\documentclass[usegraphicx,usenatbib,useAMS]{mn2e}
\usepackage{times}
  \let\url\relax
\def\apj{{ApJ}}
\def\apjs{{ApJS}}
\def\apjl{{ApJL}}
\def\aap{{\em A.\&A}}

\def\mnras{{MNRAS}}

\def\eg{{\textit{e.g.}}}

\def\araa{{\em Ann.\ Rev.\ Astron.\& Astrophys.\ }}
\def\pasj{{PASJ}}
\def\prd{{Phys Rev D}}
\newcommand{\be}{\begin{equation}}
\newcommand{\ba}{\begin{eqnarray}}
\newcommand{\ee}{\end{equation}}
\newcommand{\ea}{\end{eqnarray}}  

\def\lesssim{\mathrel{\hbox{\rlap{\hbox{\lower4pt\hbox{$\sim$}}}\hbox{$<$}}}}
\def\gtrsim{\mathrel{\hbox{\rlap{\hbox{\lower4pt\hbox{$\sim$}}}\hbox{$>$}}}}

\def\araa{{\em Ann.\ Rev.\ Astron.\& Astrophys.\ }}
\def\gtsima{$\; \buildrel > \over \sim \;$}
\def\ltsima{$\; \buildrel < \over \sim \;$}
\def\gsim{\lower.5ex\hbox{\gtsima}}
\def\lsim{\lower.5ex\hbox{\ltsima}}
\def\simgt{\lower.5ex\hbox{\gtsima}}
\def\simlt{\lower.5ex\hbox{\ltsima}}
\def\simpr{\lower.5ex\hbox{\prosima}}
\def\la{\lsim}
\def\ga{\gsim}

\def\simless{\mathbin{\lower 3pt\hbox
   {$\rlap{\raise 5pt\hbox{$\char'074$}}\mathchar''7218$}}}   
\def\simgreat{\mathbin{\lower 3pt\hbox
   {$\rlap{\raise 5pt\hbox{$\char'076$}}\mathchar''7218$}}}   

\begin{document}

\title  [Cosmological Radiative Transfer Comparison II] {Cosmological
         Radiative Transfer Comparison Project II: The 
	 Radiation-Hydrodynamic Tests} 
 
\author [I. T. Iliev, et al.]{Ilian~T.~Iliev$^{1,2,3}$\thanks{e-mail: I.T.Iliev@sussex.ac.uk}, 
         Daniel Whalen$^{4}$, Garrelt Mellema$^{5}$, Kyungjin Ahn$^{6,7}$, Sunghye Baek$^8$, 
         \newauthor Nickolay~Y.~Gnedin$^9$, Andrey~V.~Kravtsov$^{10}$, Michael Norman$^{11}$,  
         Milan Raicevic$^{12}$,  \newauthor Daniel R. Reynolds$^{13}$, 
	 Daisuke Sato$^{14}$, Paul~R.~Shapiro$^6$, Benoit Semelin$^7$, Joseph Smidt$^{15}$, 
         \newauthor Hajime Susa$^{16}$, Tom Theuns$^{12,17}$, Masayuki Umemura$^{14}$ \\
 $^1$    Astronomy Centre, Department of Physics \& Astronomy, Pevensey II Building, 
	 University of Sussex, Falmer, Brighton BN1 9QH, United Kingdom\\    
 $^2$ 	 Universit\"at Z\"urich, Institut f\"ur Theoretische Physik,
         Winterthurerstrasse 190, CH-8057 Z\"urich, Switzerland\\
 $^3$    Canadian Institute for Theoretical Astrophysics, University
         of Toronto, 60 St. George Street, Toronto, ON M5S 3H8, Canada\\
 $^4$    T-2 Nuclear and Particle Physics, Astrophysics, and Cosmology, 
         Los Alamos National Laboratory, Los Alamos, NM 87545, U.S.A.\\
 $^{5}$ Dept. of Astronomy and Oskar Klein Centre, AlbaNova, Stockholm University, SE-10691 
         Stockholm, Sweden\\
 $^6$    Department of Earth Science Education, Chosun University, Gwangju 501-759, Korea\\
 $^7$    Department of Astronomy, University of Texas, Austin, TX 78712-1083, U.S.A.\\
 $^8$    LERMA, Observatoire de Paris, 77 av Denfert Rochereau, 75014 Paris, France\\
 $^9$    Fermilab, MS209, P.O. 500, Batavia, IL 60510, U.S.A.\\
 $^{10}$ Dept. of Astronomy and Astrophysics, Center for Cosmological Physics,
         The University of Chicago, Chicago, IL 60637, U.S.A.\\
 $^{11}$ Center for Astrophysics and Space Sciences, University of California,
         San Diego, 9500 Gilman Drive, La Jolla, CA 92093-0424, U.S.A.\\
 $^{12}$ Institute for Computational Cosmology, Durham University, Durham, United Kingdom\\
 $^{13}$ Department of Mathematics, 208 Clements Hall, Southern Methodist University,
         Dallas, TX 75275, USA \\
 $^{14}$ Center for Computational Sciences, University of Tsukuba, Tsukuba,
         Ibaraki 305-8577, Japan\\
 $^{15}$ Department of Physics and Astronomy, 4129 Frederick Reines Hall, UC Irvine,
         Irvine, CA  84602, U.S.A. \\
 $^{16}$ Department of Physics, Konan University, Kobe, Japan\\
 $^{17}$ Department of Physics, University of Antwerp, Campus Groenenborger, Groenenborgerlaan B-171, B2020 Antwerp, Belgium}
\date{\today} \pubyear{2008} \volume{000}
\pagerange{1} \twocolumn \maketitle
\label{firstpage}

\begin{abstract}
  The development of radiation hydrodynamical methods that are able to follow 
  gas dynamics and radiative transfer self-consistently is key to the solution 
  of many problems in numerical astrophysics. Such fluid flows are highly complex, 
  rarely allowing even for approximate analytical solutions against which 
  numerical codes can be tested. An alternative validation procedure is to 
  compare different methods against each other on common problems, in order 
  to assess the robustness of the results and establish a range of validity 
  for the methods.  Previously, we presented such a comparison for a set of 
  pure radiative transfer tests (i.e. for fixed, non-evolving density fields). 
  This is the second paper of the Cosmological Radiative Transfer (RT)  
  Comparison Project, in which we compare 9 independent RT codes directly 
  coupled to gasdynamics on 3 relatively simple astrophysical hydrodynamics 
  problems: (5) the expansion of an H~II region in a uniform medium; (6) an 
  ionization front (I-front) in a $1/r^2$ density profile with a flat core, 
  and (7), the photoevaporation of a uniform dense clump. Results show a 
  broad agreement between the different methods and no big failures, indicating
  that the participating codes have reached a certain level of maturity and
  reliability. However, many details still do differ, and virtually every code  
  has showed some shortcomings and has disagreed, in one respect or another, 
  with the majority of the results. This underscores the fact that no method 
  is universal and all require careful testing of the particular features 
  which are most relevant to the specific problem at hand.
\end{abstract}

\begin{keywords}
  H II regions---galaxies:high-redshift---intergalactic medium---cosmology:
  theory---radiative transfer--- methods: numerical
\end{keywords}

\section{Introduction}
The transfer of ionizing radiation through optically-thick media is a key
process in many astrophysical phenomena. Some examples include cosmological
reionization \citep[$\eg$][]{2000ApJ...535..530G,2001MNRAS.321..593N,
2002ApJ...572..695R,2003MNRAS.344..607S,2003MNRAS.343.1101C,
2006MNRAS.369.1625I,2007ApJ...657...15K}, star formation
\citep[$\eg$][]{2005ApJ...623..917H,2006MNRAS.368.1885I,2006ApJ...645...55R,
  2006ApJ...645L..93S,2007MNRAS.375..881A,wn08a}, radiative feedback in 
molecular clouds \citep{2006ApJ...647..397M,2007ApJ...668..980M,
2007MNRAS.375.1291D,2007MNRAS.377..535D,2007ApJ...671..518K,2009MNRAS.393...21G}, 
and planetary nebulae \citep[$\eg$][]{1998A&A...331..335M,2003A&A...405..189L}. 
In some of these problems, fast, R-type I-fronts
predominate. Those fronts propagate faster than the hydrodynamic response of
the gas, so gas motions do not affect the I-front evolution. In these
cases the radiative transfer could be done on a fixed density field (or a
succession of such fields), and dynamic coupling to the gas is generally
not required. However, the majority of astrophysical and cosmological 
applications involve slow, D-type I-fronts (or a combination of R-type and 
D-type, as we describe in detail in section \ref{sec:T5}), so the radiative 
transfer and gasdynamics should be directly coupled and evolved simultaneously. 
Until recently, self-consistent radiation hydrodynamical codes for radiative 
transport have been rare, but this unsatisfactory situation is now rapidly 
changing due to the development of a number of such codes using a variety of 
numerical approaches.

A number of radiative transfer methods have been developed in recent years,
both stand-alone and coupled to hydrodynamics. High computational costs 
necessitate the usage of various approximations. Thus, it is of prime 
importance to validate the numerical methods developed and to evaluate 
their reliability and accuracy.  Tests with either exact or good approximate 
analytical solutions should always be the first choice for code testing.  
Extensive test suites of radiation hydrodynamical I-front transport in a 
variety of stratified media with good approximate analytical solutions do 
exist \citep{1990ApJ...349..126F,2006ApJS..162..281W} and are stringent 
tests of coupling schemes between radiation, gas, and chemistry. However, 
an alternative and complementary approach is to compare a variety of methods 
on a set of well-defined problems in astrophysical settings.  This is the 
approach we have taken in this project.

Our aim is to determine the type of problems the codes are (un)able to solve, 
to understand the origin of any differences inevitably found in the results, 
to stimulate improvements and further developments of the existing codes and, 
finally, to serve as a benchmark for testing future algorithms. All test 
descriptions, parameters, and results can be found at the project website: \\
{$\url  http://www.cita.utoronto.ca/\!\sim\!iliev/rtwiki/doku.php$}.

The first paper of this comparison project discussed the results from fixed
density field tests \citep[][hereafter Paper I]{comparison1}, i.e. without
any gas evolution. We found that all participating codes are able to track 
I-fronts quite well, within $\sim$10\% of each other. Some important differences 
also emerged, especially in the derived temperatures and spectral hardening. We 
found that some of these differences were due to variations in microphysics 
(chemical reaction rates, heating/cooling rates and photoionization cross-sections), 
while others were due to the method itself, $\eg$ how the energy equation is solved, 
how many frequency bins are used for the spectral evolution, etc. We concluded that 
the tested radiative transfer methods are producing reliable results overall, but 
that not all methods are equally appropriate for any given problem, especially in 
cases when obtaining precise temperatures and spectral features is important.

We now extend our previous work by considering a set of radiation hydrodynamical 
tests. In the spirit of Paper I, we have chosen a set of test problems which are 
relatively simple, so as to be most inclusive given the current limitations of 
the available codes ($\eg$ 1-D or 2-D vs. 3-D codes). At the same time, our tests 
consider problems of astrophysical importance, and cover a wide variety of 
situations that test the attributes of each method, including its radiative 
and hydrodynamic components and their coupling.

\begin{table*}
\caption{Participating codes and their current features.}
\label{summary}
\begin{tabular}{lllll}
\hline
Code       & Grid                   & Parallelization        & hydro method          & rad. transfer method \\ 
\hline\\
Capreole+$C^2$-Ray  & fixed         & shared/distributed     & Eulerian, Riemann solver        & short-characteristics ray-tracing         \\
TVD+$C^2$-Ray  & fixed         & shared/distributed     & Eulerian, TVD solver        & short-characteristics ray-tracing         \\
HART       & AMR                    & shared/distributed     & Eulerian, Riemann solver        & Eddington tensor moment                \\
RSPH       & particle-based         & distributed            & SPH                   & long-characteristics ray-tracing \\
ZEUS-MP    & fixed                  & distributed            & Eulerian              & 3-D ray-tracing \\
RH1D       & sph. Lagrangian        & no                     & Lagrangian            & 1-D ray-tracing            \\
Coral      & AMR                    & no                     & Eulerian, flux-vector splitting & short-characteristics ray tracing\\ 
LICORICE   & AMR                    & shared                 & SPH                   & Monte-Carlo ray-tracing\\
Flash-HC   & AMR                    & distributed            & Eulerian, PPM         & Hybrid characteristics ray-tracing \\
Enzo-RT   & fixed                  & distributed            & Eulerian, PPM         & Flux-limited diffusion\\
\hline\\
\end{tabular}
\end{table*}

\begin{figure}
\begin{center}
  \includegraphics[width=3.5in]{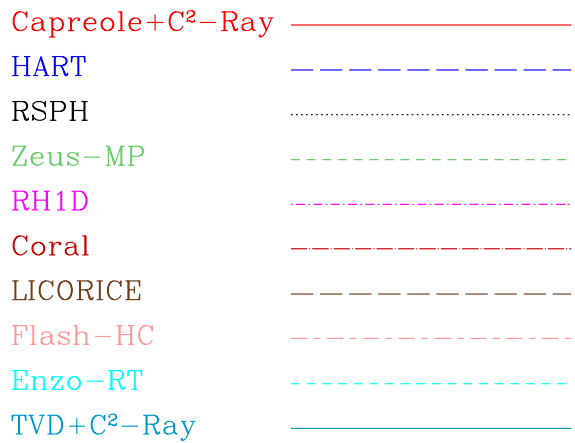} \vspace{-1in}
\caption{Legend for the line plots.
\label{legend_fig}}
\vspace{-0.5cm}
\end{center}
\end{figure}

The efficiency, optimization and performance of the codes are very important,
especially for the most complex and computationally-intensive problems.
However, there are a number of complications, which we discussed in Paper I, 
preventing us from doing such testing in a meaningful way at present. We 
therefore leave it for future work.

All test results for this study had to be supplied on a regular Cartesian
grid of $128^3$ computational cells. This relatively modest resolution was 
chosen in the interests of inclusivity, so that even codes which are not 
yet fully optimized in terms of either computations or memory can 
participate in the comparison. We note that production runs at present are 
typically run at $256^3$ or better resolution. Codes which utilize Adaptive 
Mesh Refinement (AMR) grids or particles have been requested to run the 
problem at the resolution which approximates as closely as possible the 
fixed-grid one for fair comparison. Their results have then been 
interpolated onto a regular grid for submission. 

\section{The Codes}
\label{codes_sect}

In this section we briefly describe the nine radiative transfer codes 
participating in this stage of the comparison project, with references 
to more detailed method papers if available.  Details of the codes and 
their basic features and methods are summarized in Table~\ref{summary}. 
Figure~\ref{legend_fig} provides a legend allowing the reader to identify 
which line corresponds to which code in the figures throughout the paper. 
The images we present are identified in the corresponding figure caption.

\subsection{Capreole$+C^2$-Ray and TVD$+C^2$-Ray (G. Mellema, I. Iliev, 
P. Shapiro, M. Alvarez)}

C$^2$-Ray \citep{methodpaper} is a grid-based short characteristics 
\citep[$\eg$][]{1999RMxAA..35..123R} ray-tracing code which is 
photon-conserving and causally traces the rays away from the ionizing 
sources up to each cell. Explicit photon-conservation is assured by 
taking a finite-volume approach when calculating the photoionization 
rates, and by using time-averaged optical depths. The latter property 
allows for integration time steps that are much larger than the ionization 
time scale, which results in a considerable speed-up of the calculation and
facilitates the coupling of the code to gasdynamic evolution. The code 
and the various tests performed during its development are described in 
detail in \citet{methodpaper}.

The frequency dependence of the photoionization rates and photoionization
heating rates are dealt with by using frequency-integrated rates, stored as 
functions of the optical depth at the ionization threshold. In its current 
version the code includes only hydrogen and no helium, although it could be 
added in a relatively straightforward way. 

The transfer calculation is done using short characteristics, where
the optical depth is calculated by interpolating values of grid cells
lying along the line-of-sight towards the source. Because of the causal 
nature of the ray-tracing, the calculation cannot easily be parallelized 
through domain decomposition. However, using OpenMP and MPI the code is 
efficiently parallelized over the sources and grid octants 
\citep{2008arXiv0806.2887I}. The code has been applied for large-scale 
simulations of cosmic reionization and its observability
\citep{2006MNRAS.369.1625I,21cmreionpaper,selfregulated,kSZ,pol21,cmbpol,wmap3}
on grid sizes up to $406^3$ and up to $\sim10^6$ ionizing sources, running
on up to 10,240 computing cores.

There are 1D, 2D and 3D versions of the code that are available. It was 
developed 
to be directly coupled with hydrodynamics calculations. The large time steps 
allowed for the radiative transfer enable the use of the hydrodynamic time 
step for evolving the combined system. The $C^2$-Ray radiative transfer and
nonequilibrium chemistry code has been coupled to several different
gasdynamics codes, utilizing both fixed and adaptive grids. The tests in this
project were mostly performed with the version coupled to the hydrodynamics 
code Capreole developed by Garrelt Mellema and based on Roe's approximate 
Riemann solver. The first gasdynamic application of our code is presented in 
\citet{2006ApJ...647..397M}. Additionally, one of the tests has also been run 
with C$^2$-Ray coupled to a different hydro solver, namely the TVD method of 
\citet{2004NewA....9..443T} (see Test 6 below).

\subsection{Hydrodynamic Adaptive Refinement Tree (HART) (N. Gnedin,
  A. Kravtsov)}

The Hydrodynamic Adaptive Refinement Tree (HART) code is an implementation of
the AMR technique and uses a combination of
multi-level particle-mesh and shock-capturing Eulerian methods for simulating
the evolution of the dark matter particles and gas, respectively. High dynamic
range is achieved by applying adaptive mesh refinement to both gas dynamics
and gravity calculations.

The code performs refinements locally on individual cells, and cells are
organized in refinement trees \citep{khokhlov98}. The data structure is
designed both to reduce the memory overhead for maintaining a tree and to
fully eliminate the neighbor search required for finite-difference operations.
All operations, including tree modifications and adaptive mesh refinement, can
be performed in parallel. The advantage of the tree-based AMR is its ability
to control the computational mesh on the level of individual cells. This
results in a very efficient and flexible (and thus highly adaptive) refinement
mesh which can be easily built and modified and, therefore, effectively
match the complex geometry of cosmologically interesting regions: filaments,
sheets, and clumps.  Several refinement criteria can be combined with
different weights allowing for a flexible refinement strategy that can be
tuned to the needs of each particular simulation.  The adaptive refinement in
space is accompanied by a temporal refinement (smaller time steps on meshes of
higher resolutions).

The ART code was initially developed by A.~Kravtsov in collaboration with
A.~A.~Klypin and A.~M.~Khokhlov \citep{kravtsov_etal97,kravtsov99,kravtsov_etal02}. 
N.~Gnedin joined the HART code development team in the spring of 2003 and has 
adopted the OTVET algorithm for modeling 3D radiative transfer for the ART mesh 
structure and implemented a non-equilibrium chemical network and cooling 
\citep[e.g.][]{2008arXiv0810.4148G}.

\subsection{RSPH (H. Susa, M. Umemura, D. Sato)}

The Radiation-SPH (RSPH) scheme is specifically designed to investigate the 
formation and evolution of first-generation objects at $z \ga 10$, where the 
radiative feedback from various sources plays important roles. The code can 
compute the fraction of chemical species e$^-$, H$^+$, H, H$^-$, H$_2$, and 
H$_2^+$ by fully implicit time integration. It also can deal with multiple 
sources of ionizing radiation, as well as with Lyman-Werner band photons.

Hydrodynamics is calculated by the smoothed particle hydrodynamics (SPH) method.
It uses the version of SPH by \citet{1993ApJ...406..361U} with the modification
according to \citet{1993A&A...268..391S}, and adopts the particle
resizing formalism by \citet{2000MNRAS.319..619T}. The present version 
does not use the so-called entropy formalism \citep{2002MNRAS.333..649S}. The 
non-equilibrium chemistry and radiative cooling for primordial gas are calculated 
using the code developed by \citet{2000MNRAS.317..175S}, where H$_2$ cooling and 
reaction rates are taken from \citet{1998A&A...335..403G}.

As for the photoionization process, the on-the-spot approximation is employed
\citep{1978ppim.book.....S}, meaning that the transfer of ionizing photons
directly from the source is solved, but diffuse photons are not transported. 
Instead, it is assumed that recombination photons are absorbed in the same zone 
from which they are emitted. Due to the absence of the source term in this 
approximation, the radiation transfer equation becomes very simple. Solving the 
transfer equation reduces to the easier problem of assessing the optical depth 
from the source to every SPH particle.

The optical depth is integrated utilizing the neighbour lists of SPH particles.  
It is similar to the code described in \citet{2004ApJ...600....1S}, but can 
now also deal with multiple point sources. In the new scheme fewer grid points 
are created on the light ray than in its predecessor. Instead, just one grid 
point per SPH particle is created in the particle's neighborhood. The 
'upstream' particle for each SPH particle on its line of sight to the source 
is then found. Then the optical depth from the source to the SPH particle is 
obtained by summing up the optical depth at the 'upstream' particle and the 
differential optical depth between the two particles.

The code is parallelized with the MPI library. The computational domain is
divided by the Orthogonal Recursive Bisection method.  The parallelization 
method for radiation transfer is similar to the Multiple Wave Front method 
developed by \citet{2001MNRAS.321..593N} and \citet{2006A&A...448..731H}, 
but it is adapted to fit the SPH code as described in 
\citep{2006PASJ...58..445S}.

The code computes self-gravity using a Barnes-Hut tree, which is parallelized 
as well. A Tree-GRAPE version of the code has also been developed. This code 
has been applied to radiative feedback in primordial star formation 
\citep{2006ApJ...645L..93S,2007ApJ...659..908S,2009MNRAS.tmp..445H}, as well 
as the regulation of star formation in forming galaxies by ultraviolet 
background \citep{2008ApJ...684..226S}.

\subsection{ZEUS-MP  (D. Whalen, J. Smidt, M. Norman)}

ZEUS-MP solves explicit finite-difference approximations to Euler's equations 
of fluid dynamics self-consistently with a 9-species primordial gas reaction 
network (${\rm H}$, ${\rm H^{+}}$, ${\rm He}$, ${\rm He^{+}}$, ${\rm He^{++}}$, 
${\rm H^{-}}$, ${\rm H_{2}}$, ${\rm H_{2}^{+}}$ and ${\rm  e}$) and ray-tracing 
radiative transfer, which is used to compute the radiative rate coefficients 
required by the network and the gas energy equation.  Our method is described 
in detail elsewhere \citep{2006ApJS..162..281W,wn08a}; here, we review 
multifrequency upgrades to the radiative transfer and improvements to the 
subcycling scheme \citep{wn08b}.

The ZEUS-MP RT module evaluates radiative rate coefficients by solving the 
static equation of transfer in flux form.  To obtain the total rate coefficient 
$k$ for a zone we sum the $k_{\nu}$ computed for a given binned photon emission 
rate over all energies by looping the solution to the transfer equation over 
them.  In tests spanning 40 to 2000 energy bins, good convergence is found with 
120 bins, 40 bins spaced evenly in energy from 0.755 eV to 13.6 eV and 80 bins
that are logarithmically-spaced from 13.6 eV to 90 eV.

Successive updates to the reaction network and gas energy are performed
over the minimum of the chemical time 
\begin{equation}
t_{chem} = 0.1 \, \displaystyle\frac{n_{e} + 0.001 n_{H}}{{\dot{n}}_{e}}.
\end{equation}
and the photoheating/cooling time \vspace{0.05in}
\begin{equation}
t_{hc} = 0.1 \displaystyle\frac{e_{gas}}{{\dot{e}}_{ht/cool}} \vspace{0.05in}
\end{equation}
until the larger of these two times has been crossed, at which point full
hydrodynamical updates of gas densities, energies, and velocities are 
performed.  These times are global minima for the entire grid.  Chemical 
times are defined in terms of electron flow to accommodate all chemical
processes rather than just ionizations or recombinations.  Adopting the 
minimum of the two times for chemistry and gas energy updates enforces 
accuracy in the reaction network when $t_{chem}$ becomes greater than 
$t_{hc}$ (in relic H II regions, for example).  

ZEUS-MP is now fully parallelized for three-dimensional applications.  We 
have updated the H and He recombination and cooling rates responsible for 
some minor departures between ZEUS-MP and the other codes in Paper I in the 
temperature structure of H II regions, and now use the most recent data from 
\citet{1994MNRAS.268..109H} and \citet{1998MNRAS.297.1073H}.  Our code has 
been validated with stringent tests of R-type and D-type I-fronts in a 
variety of stratified media \citep{1990ApJ...349..126F,2006ApJS..162..281W} 
and applied to both cosmological and astrophysical problems, such as the 
breakout of UV radiation from primordial star-forming clouds \citep{
2004ApJ...610...14W}, the formation of dynamical instabilities in galactic 
H II regions \citep{wn08a}, the circumstellar environments of gamma-ray 
bursts \citep{2008ApJ...682.1114W}, the photoevaporation of cosmological 
minihalos by nearby primordial stars \citep{wet08b}, and Pop III supernovae 
explosions in cosmological H II regions \citep{wet08c}.
 
\subsection{RH1D (K. Ahn, P. Shapiro)}

RH1D is a 1D, Lagrangian, spherically-symmetric, radiation-hydrodynamics code
for a two-component gas of baryons and collisionless dark matter coupled by
gravity \citep{2007MNRAS.375..881A}. For the baryonic component, the Euler
equations and the equation of state are solved, together with multi-frequency,
multi-species radiative transfer equations and a reaction network with nine
primordial species (${\rm H}$, ${\rm H^{+}}$, ${\rm He}$, ${\rm He^{+}}$,
${\rm He^{++}}$, ${\rm H^{-}}$, ${\rm H_{2}}$, ${\rm H_{2}^{+}}$ and ${\rm
  e}$).  Dark matter dynamics, governed by the collisionless Boltzmann
equations, takes a simplified form in spherical symmetry. The code solves 
an effective set of Euler equations for a dark matter fluid, based upon the 
{}``fluid approximation'' of dark matter dynamics for a spherically symmetric 
system with an isotropic velocity dispersion, derived and justified elsewhere
\citep{2005MNRAS.363.1092A}. These effective Euler equations are identical to
those for an inviscid, ideal gas with a ratio of specific heats $\gamma=5/3$.

The Euler equations are solved using the so-called {}``leap-frog'' method,
where the Lagrangian position (radius) and velocity (radial velocity) are
staggered in time to achieve a second-order accuracy in time steps, both for
baryonic and dark matter fluid. The usual artificial viscosity scheme is used
to capture shocks. We typically adopt a few thousand uniformly spaced bins in
radius. Non-equilibrium rate equations for the nine primordial species are 
solved using the backward differencing scheme of
\citet{1997NewA....2..209A}. For ${\rm H^{-}}$ and ${\rm H_{2}^{+}}$, due to
their relatively fast reaction rates, the equilibrium values may be used.

Radiative transfer is performed by ray-tracing, taking account of the optical
depth to bound-free opacity of H I, He I, He II, ${\rm H^{-}}$, and ${\rm
  H_{2}}$, as well as bound-free and dissociation opacity of ${\rm
  H_{2}^{+}}$. The optical depth to the Lyman-Werner band photons of ${\rm
  H_{2}}$, which are capable of dissociating ${\rm H_{2}}$, is treated using a
pre-calculated self-shielding function by \citet{1996ApJ...468..269D}, which
is determined by the ${\rm H_{2}}$ column density and gas temperature. Diffuse
flux is not explicitly calculated, but is accounted for implicitly by adopting
case B recombination rates.  The radiative reaction rates are calculated using
a photon-conserving scheme, which enables the code to treat optically-thick
shells ($\eg$  \citealt{1999MNRAS.309..287R}; \citealt{1999ApJ...523...66A}).
A wide range of radiation frequency (energy), $h\nu\sim[0.7\,-\,7000]\,{\rm
  eV}$, is covered by a few hundred, logarithmically spaced bins, together
with additive, linearly spaced bins where radiative cross sections change
rapidly as frequency changes. For each frequency and species, the
corresponding radiative reaction rate is calculated, then summed over
frequency to obtain the net radiative reaction rate.

The radiative transfer scheme is able to treat 1) an internal point source, 2)
an external, radially-directed source, and 3) an external, isotropic background.
The transfer for (1) and (2) is 1D, performed along the radial direction only.
For (3), the transfer is 2D in nature, and at each point the mean intensity is
required to calculate the radiative rates, which involves an angle
integration. The radiative transfer calculation is performed for each
pre-selected angle ($\theta$, measured from the radial direction), and then
the angle integral is calculated using the Gaussian quadrature method.

The code adopts a very stringent time step criterion for accuracy.  The
minimum of dynamical, sound-crossing, cooling/heating, and species change time
scales, which is multiplied by a coefficient smaller than unity ($\sim0.1$),
is chosen as the time step. All the Euler equations and rate equations are
solved with this time step, which makes the whole calculation self-consistent.
This code has been tested extensively and used to study the radiative feedback
effects by the first stars on their nearby minihalos
\citep{2007MNRAS.375..881A}.

\subsection{Coral (I. Iliev, A. Raga, G. Mellema, P. Shapiro)}

CORAL is a 2-D, axisymmetric Eulerian fluid dynamics AMR code 
\citep[see][and references therein for detailed
description]{1998A&A...331..335M,2004MNRAS.348..753S}. It solves the Euler
equations in their conservative finite-volume form using the second-order
method of van~Leer flux-splitting, which allows for correct and precise
treatment of shocks. The grid refinement and de-refinement criteria are based
on the gradients of all code variables. When the gradient of any variable is
larger than a pre-defined value the cell is refined, and when the criterion
for refinement is not met the cell is de-refined.

The code follows, by a semi-implicit method, the non-equilibrium chemistry of
multiple species (H, He, C II-VI, N I-VI, O I-VI, Ne I-VI, and S II-VI) and
the corresponding cooling \citep{1997ApJS..109..517R,1998A&A...331..335M}, as
well as Compton cooling. The photoheating rate is the sum of the
photoionization heating rates for H~I, He~I and He~II. For computational
efficiency all heating and cooling rates are pre-computed and stored in
tables.  The microphysical processes -- chemical reactions, radiative
processes, transfer of radiation, heating and cooling -- are implemented
though the standard approach of operator-splitting (i.e. solved at each
time-step, side-by-side with the hydrodynamics and coupled to it through the
energy equation).  The latest versions of the code also include the effects of
an external gravity force.

Currently the code uses a black-body or power-law ionizing source spectrum,
although any other spectrum can be accommodated. Radiative transfer of the
ionizing photons is treated explicitly by taking into account the bound-free
opacity of H and He in the photoionization and photoheating rates. The
photoionization and photoheating rates of H~I, He~I and He~II are pre-computed
for the given spectrum and stored in tables vs. the optical depths at the
ionizing thresholds of these species, which are then used to obtain the total
optical depths.  The code correctly tracks both fast (by evolving on an
ionization timestep, $\Delta t\sim\dot{n_{\rm H}}/n_{\rm H}$) and slow
I-fronts.

The code has been tested extensively and has been applied to many
astrophysical problems, $\eg$ photoevaporation of clumps in planetary nebulae
\citep{1998A&A...331..335M}, cosmological minihalo photoevaporation during
reionization \citep{2004MNRAS.348..753S,2005MNRAS...361..405I}, and studies of
the radiative feedback from propagating ionization fronts on dense clumps in
damped Lyman-$\alpha$ systems \citep{2006MNRAS.368.1885I}.

\subsection{LICORICE: LIne COntinuum Radiative tranfer Integrated
 Computing Engine (S. Baek, B. Semelin, F. Combes)} 

The LICORICE code has three main components: TreeSPH to compute gravity 
and hydrodynamics, continuum radiative transfer with hydrogen and helium 
ionization physics, and Lyman-alpha line transfer. The latter is not 
relevant to this comparison and has been described elsewhere. The ionizing 
continuum transfer has been described in details in \citet{2009A&A...495..389B}.
   
The current version of LICORICE does not include H$_2$ formation, or diffuse 
radiation from recombinations, but they will be incorporated in the future. 
LICORICE uses SPH particles for the gas dynamics and an adaptive grid for 
the radiative transfer. Physical quantities are interpolated from one to the 
other as required.  

The fluid dynamics are followed using a TreeSPH method. The implementation 
is described in detail in \citet{2002A&A...388..826S} and 
\citet{2005A&A...441...55S}. Since there are many varieties of SPH, we 
summarize the main features of our algorithm here. We use a 
spherically-symmetric spline-smoothing kernel and 50 neighbours to compute 
the SPH quantities using an arithmetic average between the neighbours of 
the smoothing length $h$ and the simple viscosity scheme by 
\citet{1992ARA&A..30..543M}. 

For the tests in this paper we implemented transmissive boundary conditions. 
This was achieved as follows: for each SPH particle within a distance of the 
simulation box boundary smaller than its smoothing length $h$, we create a
symmetrical 'ghost' particle on the other side of the boundary. All 
physical quantities for this ghost particle are equal to those of the initial 
particle, including the velocity. The ghost particles are used as neighbours 
to compute the SPH quantities of real particles. The ghost particles are 
erased and recreated at each time step. 

The continuum radiative transfer is solved using a Monte Carlo approach
similar to the one employed in the CRASH code \citep{2003MNRAS.345..379M}. 
Here we summarize only the differences between LICORICE and CRASH.
We compute the gas density at each particle's position with the SPH
smoothing kernel, and physical quantities such as ionization fraction and
temperature are updated according to these particle densities. The density 
field is generally smooth, but may sometimes show spurious fluctuations if 
the particle number density changes sharply. This is a well known but 
unavoidable problem with SPH.

The radiation field is discretized into photon packets and propagated 
through cells along directions chosen at random. The cells form an adaptive 
grid which is derived from the tree structure of the particle distribution. 
Our adaptive grid is built to keep the number of particles in each cell 
within a given range (1 to 8 and 1 to 1 ranges have been used). This yields 
greater resolution in the denser regions. The adaptive grid also requires 
fewer cells than a fixed grid to best sample a given inhomogeneous particle 
distribution, thus saving both memory and CPU time.

The time step for updating physical quantities within a cell is also 
adaptive. We update the physical quantities for all cells and particles 
after the propagation of the number of photon packets corresponding to an
integration time $dt$. However, if the number of accumulated photons in a
cell during this integration time is greater than a pre-set limit ($\eg$
10\% of the total number of neutral hydrogen atoms in the cell), we update 
the physical quantities in this cell with a time step $dt^{\prime} < dt$ 
corresponding to the time elapsed since the last update.

The test results are interpolated from the particle distribution onto 
the $128^3$ uniform Cartesian grids required in this study. Currently, the 
dynamical part of the code is parallelized for both shared and distributed 
memory architectures using OpenMP and MPI, while the radiative transfer is 
parallelized with OpenMP only. The code can now handle $256^3$ particles, 
to be increased to $512^3$ in the near future. We note that compared to a 
uniform grid with the same number of cells, the SPH Lagrangian approach 
results in higher resolution in the dense regions, but lower resolution in 
more diffuse regions.  
 
\begin{figure*}
\begin{center}
  \includegraphics[width=2.3in]{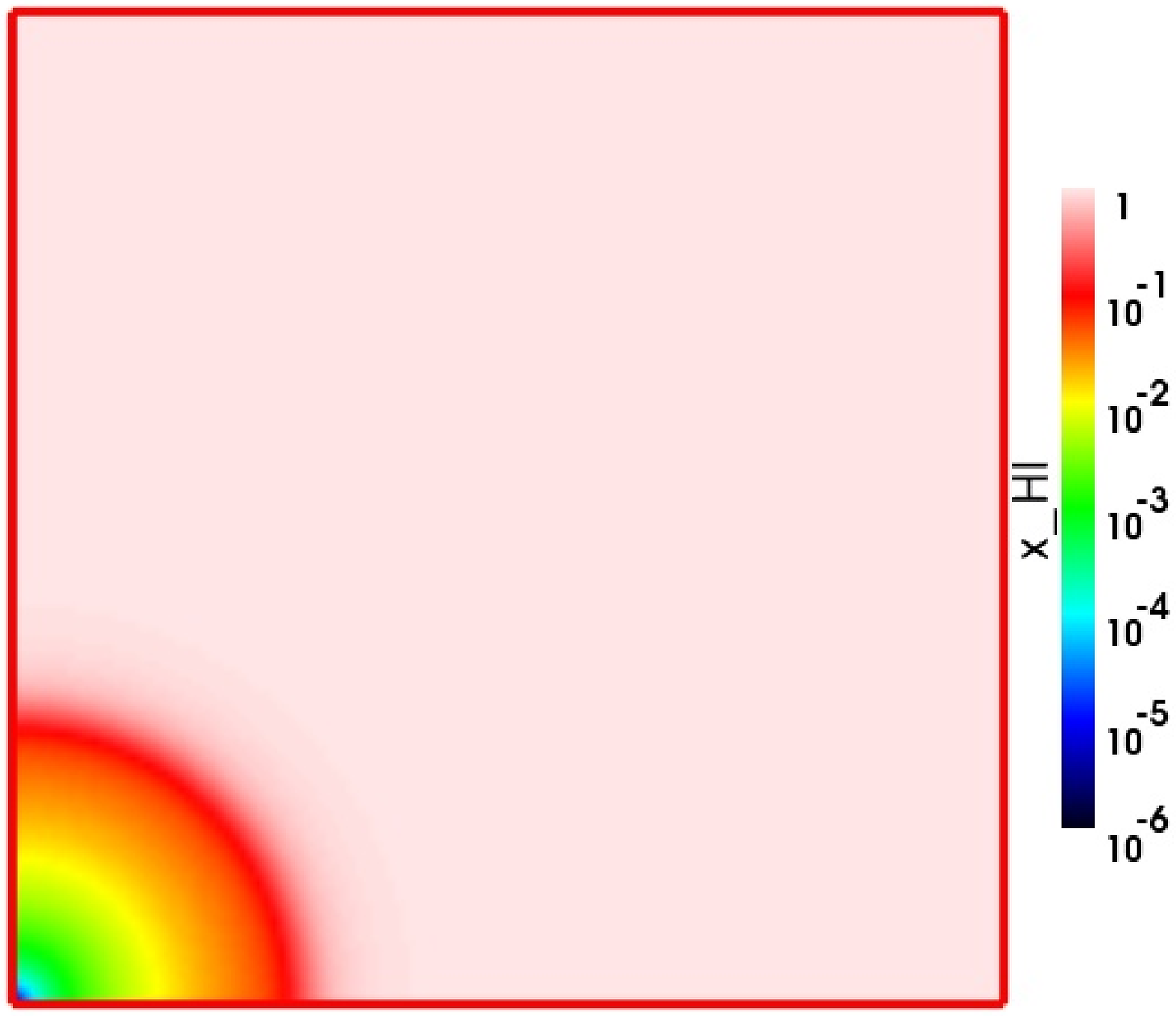}
  \includegraphics[width=2.3in]{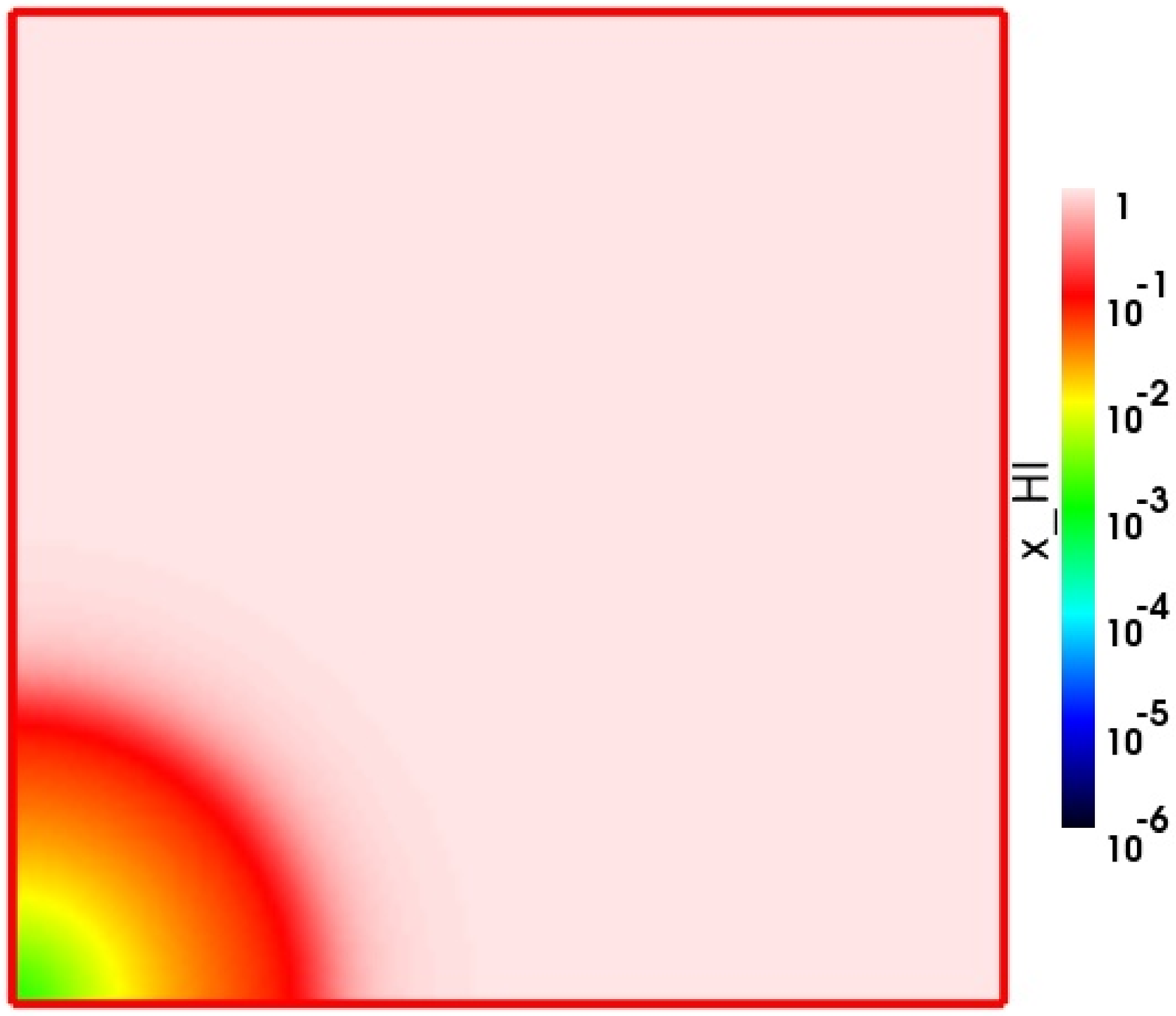}
  \includegraphics[width=2.3in]{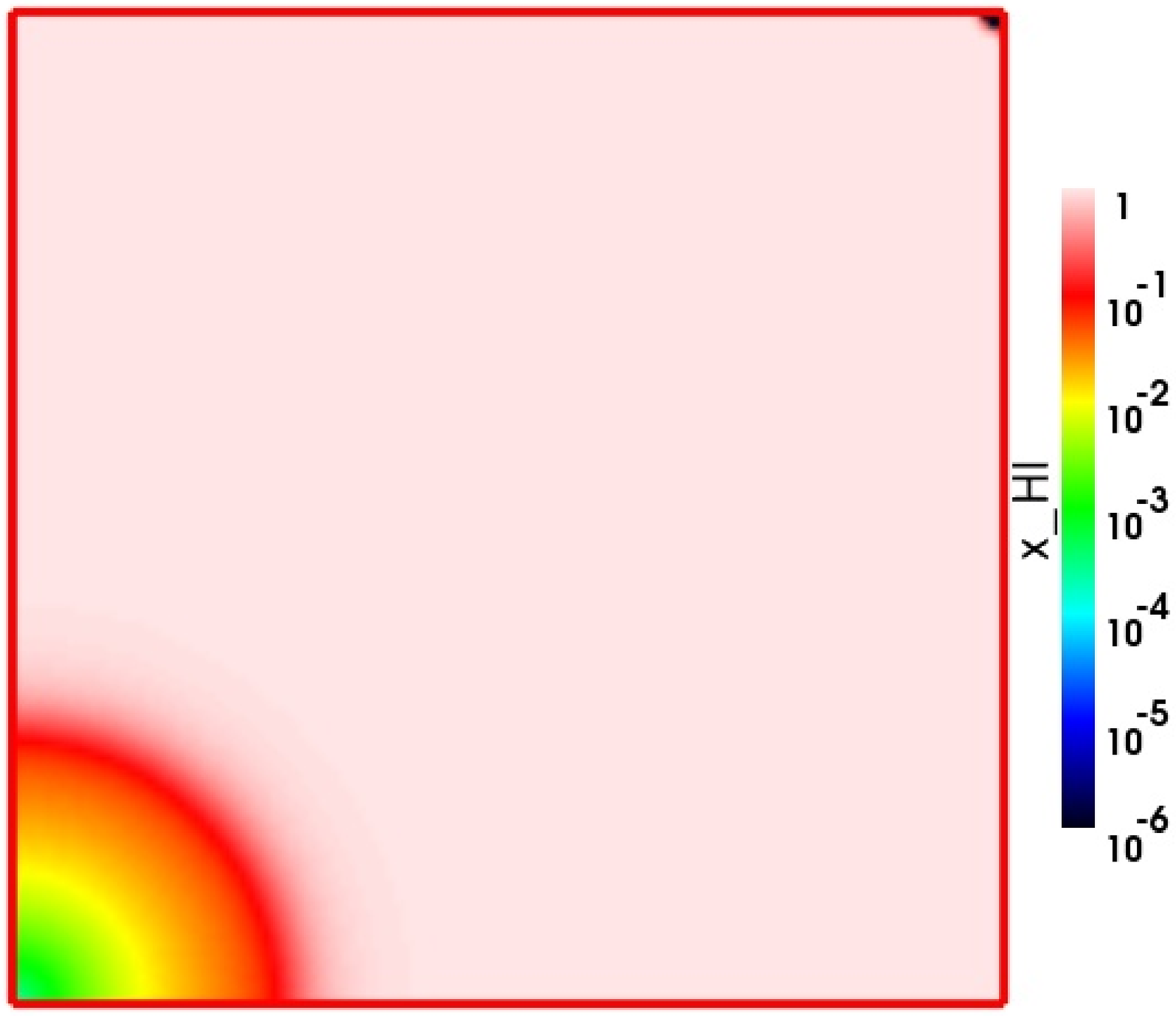}
  \includegraphics[width=2.3in]{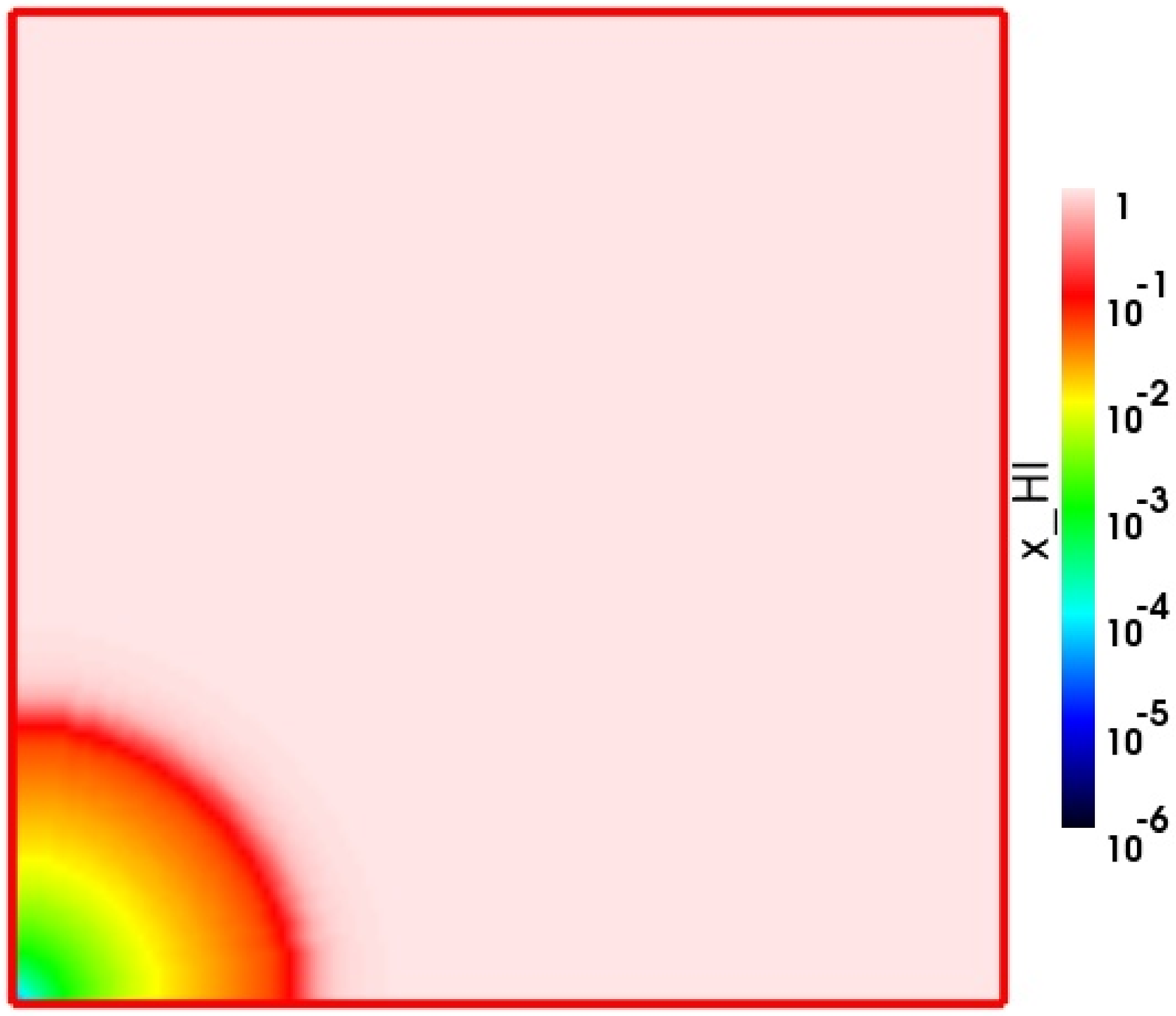}
  \includegraphics[width=2.3in]{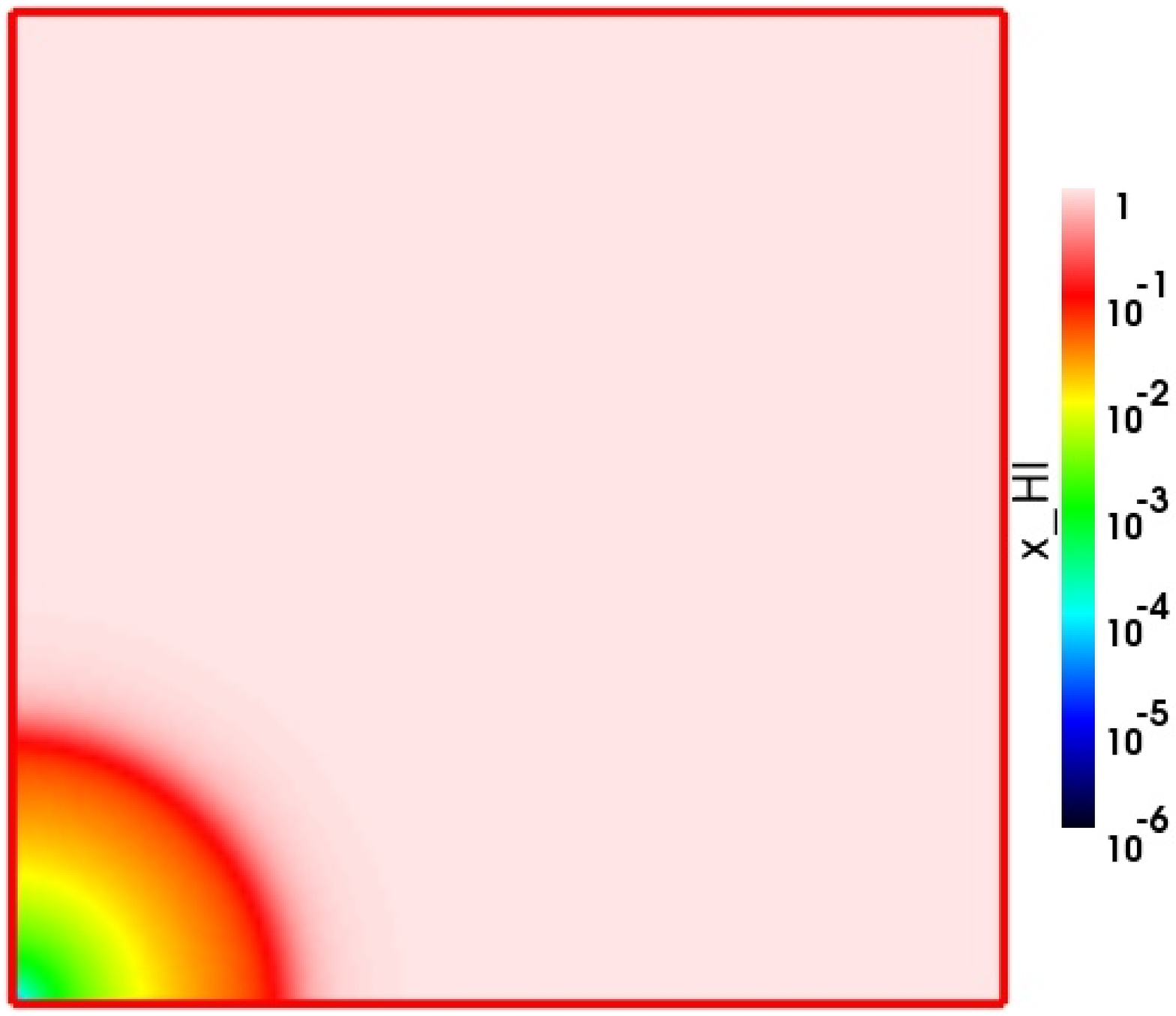}
  \includegraphics[width=2.3in]{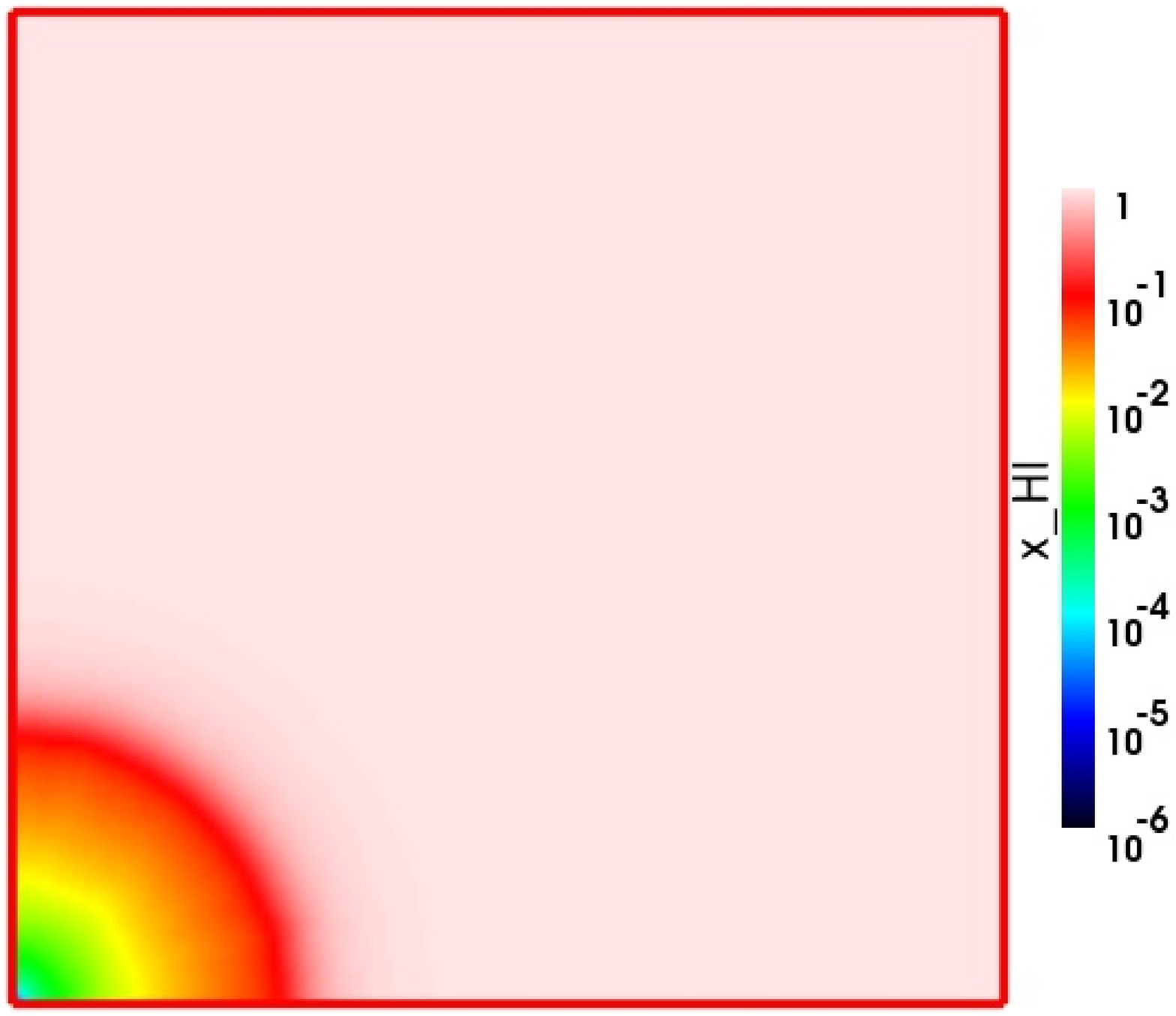}
  \includegraphics[width=2.3in]{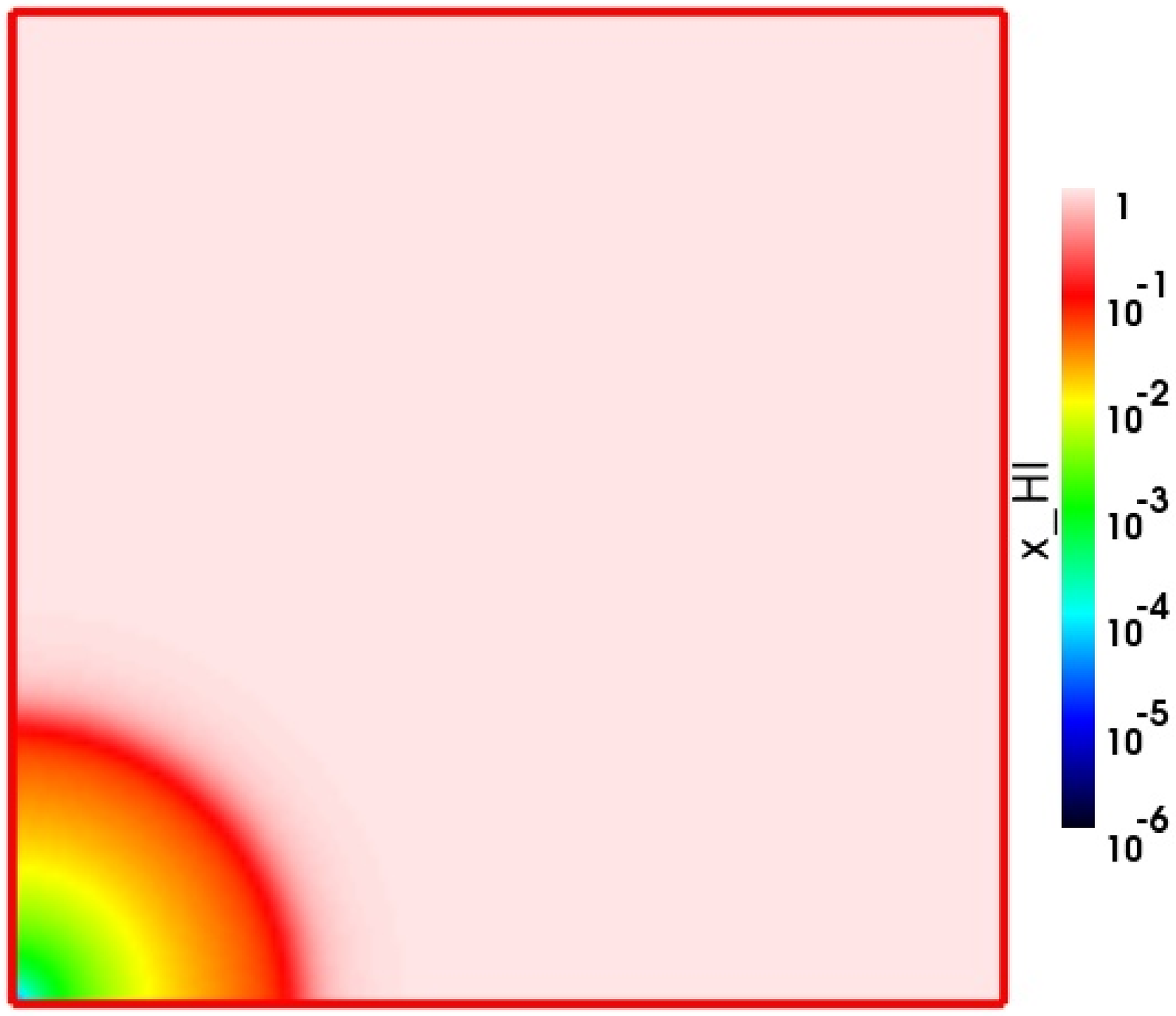}
  \includegraphics[width=2.3in]{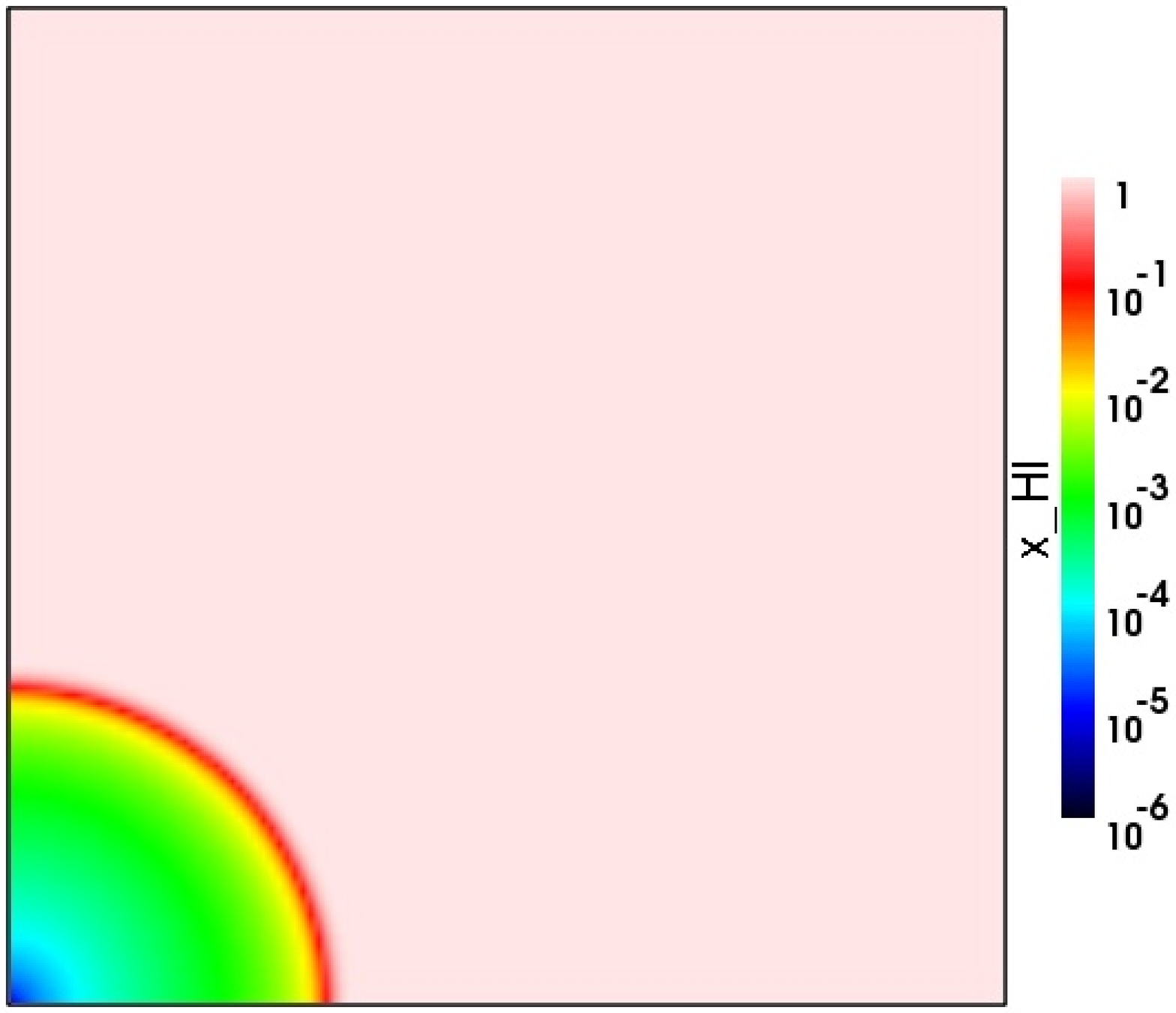}
\caption{Test 5 (H~II region expansion in an initially-uniform 
  gas): Images of the H~I fraction, cut through the simulation volume at
  coordinate $z=0$ at time $t=100$ Myr for (left to right and top to bottom)
  Capreole+$C^2$-Ray, HART, RSPH, ZEUS-MP, RH1D, LICORICE, Flash-HC and Enzo-RT.
\label{T5_images3_HI_fig}}
\end{center}
\end{figure*}

\subsection{Flash-HC: Hybrid Characteristics (T. Theuns, M. Raicevic, 
E.-J. Rijkhorst)}

The Hybrid Characteristics (HC) method \citep{Rijkhorst2005, Rijkhorst2006}
is a three-dimensional ray-tracing scheme for parallel AMR codes. It combines 
elements of long and short characteristics, using the precision and 
parallelizability of the former with efficient execution through interpolation 
of the latter.  It has been implemented into the Flash-HC AMR code 
\citep{Fryxell2000}, enabling simulations of radiation hydrodynamics problems 
with point sources of radiation. The public version of the Flash code (which
does not currently include this radiative transfer module) can be downloaded 
from\\\indent {$\url http://flash.uchicago.edu/website/home/$}.

\begin{figure*}
\begin{center}
  \includegraphics[width=2.3in]{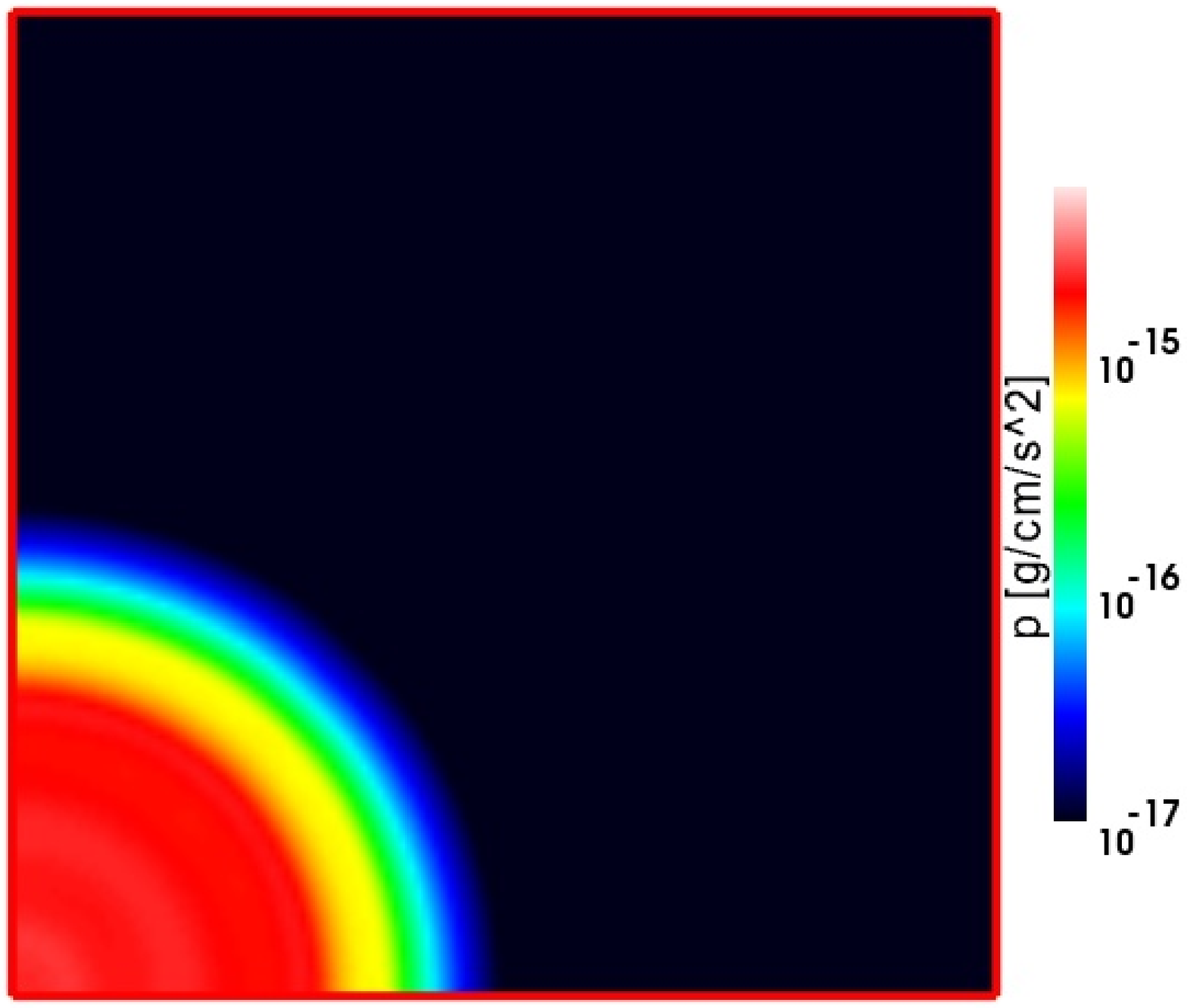}
  \includegraphics[width=2.3in]{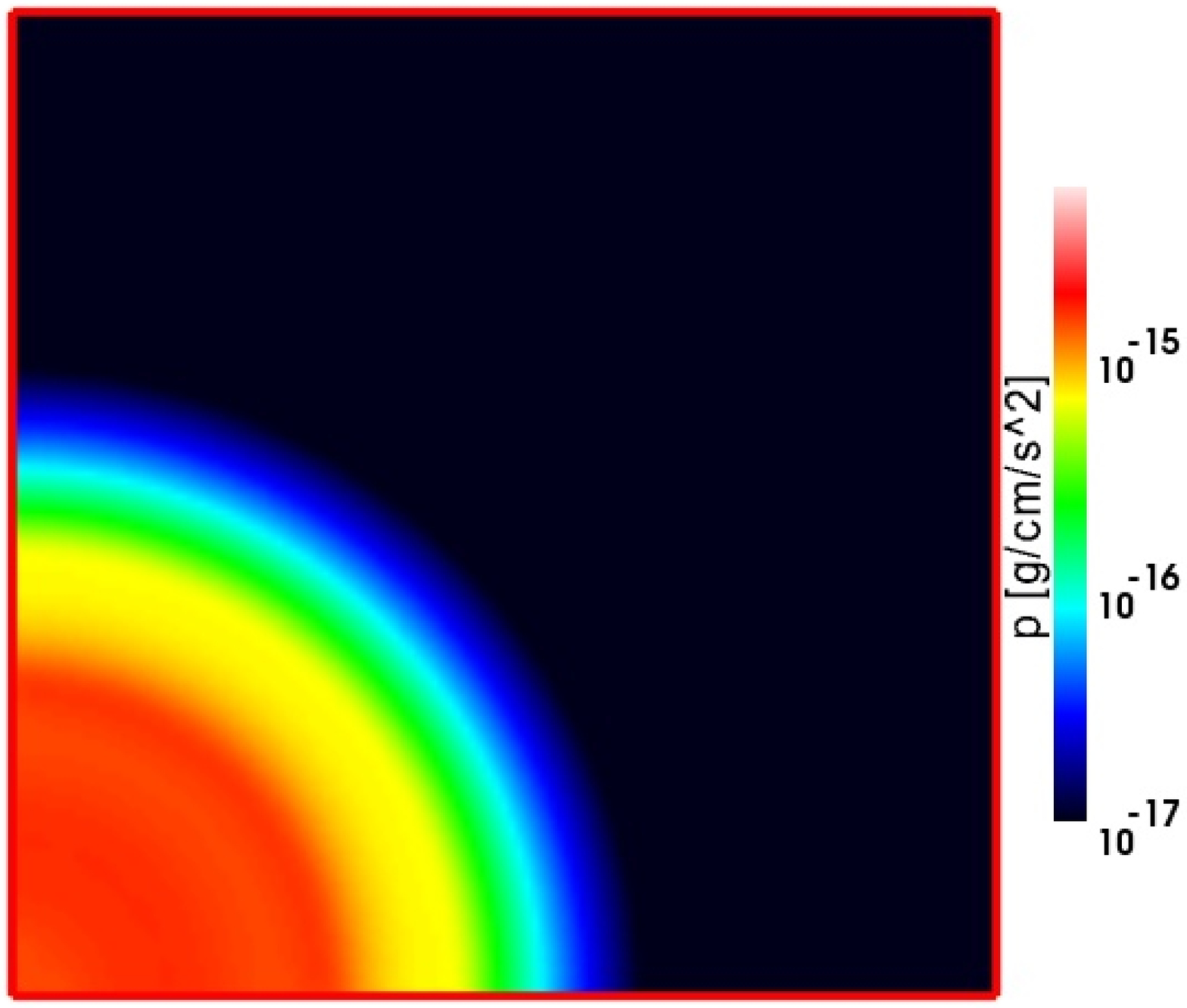}
  \includegraphics[width=2.3in]{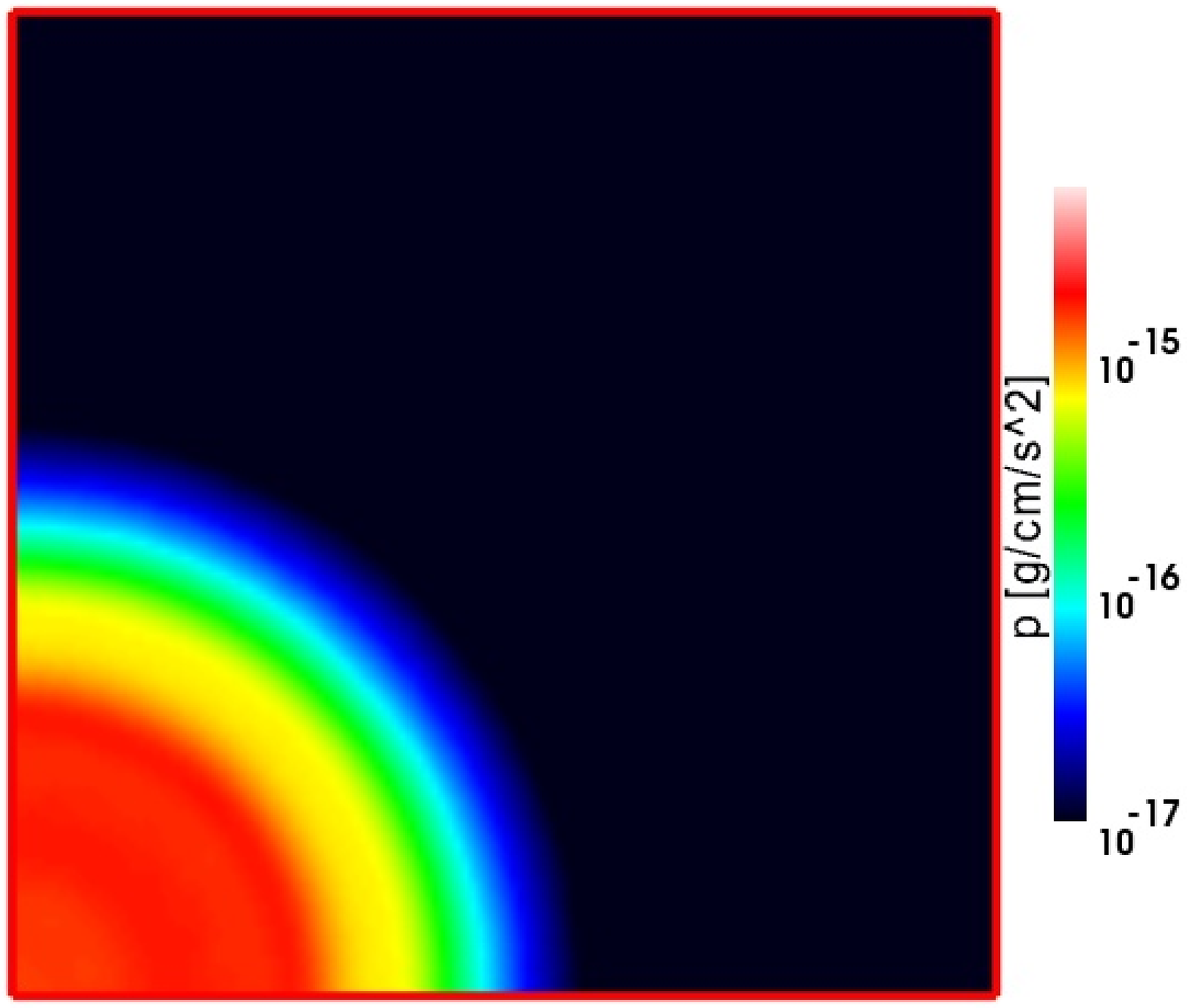}
  \includegraphics[width=2.3in]{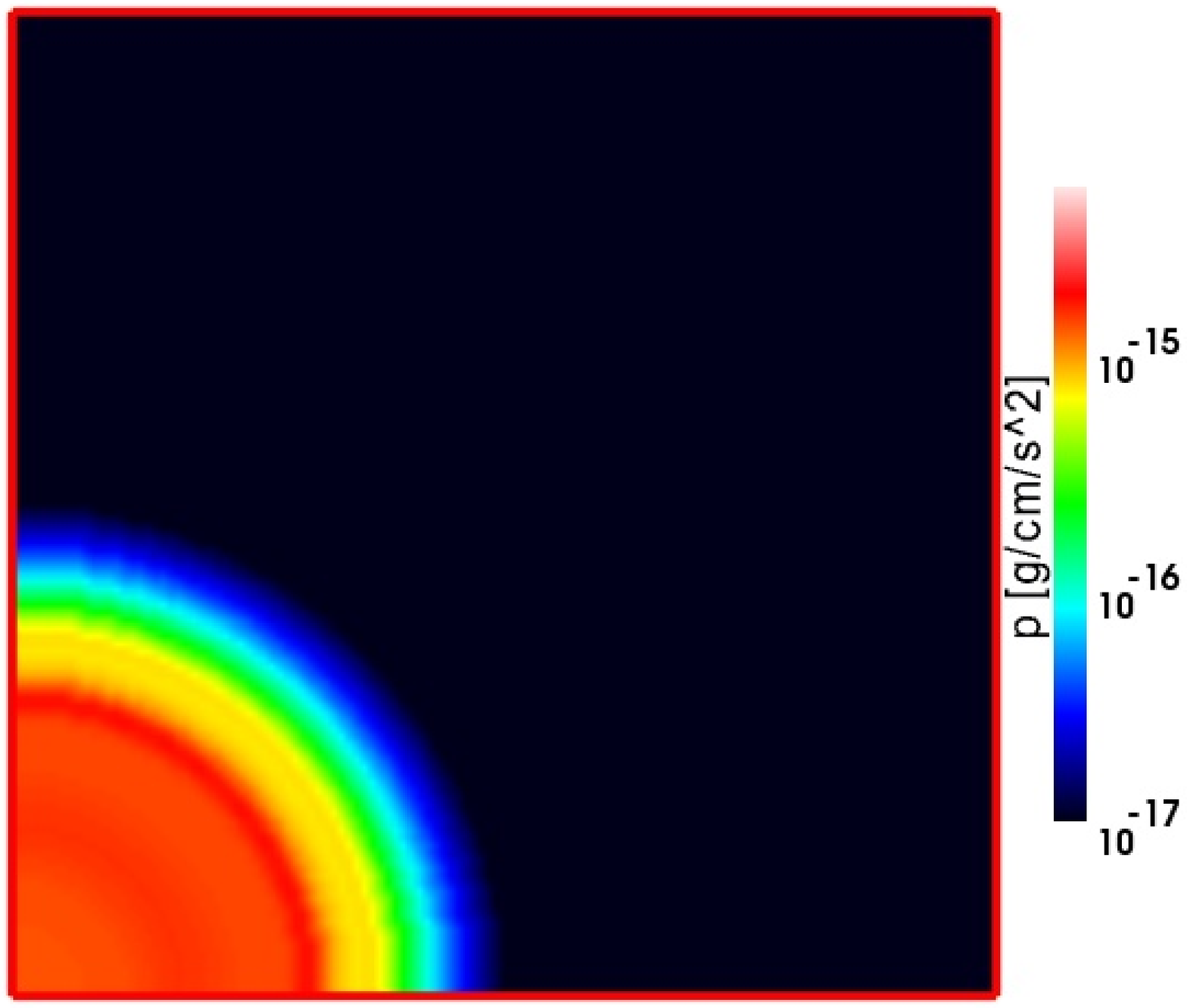}
  \includegraphics[width=2.3in]{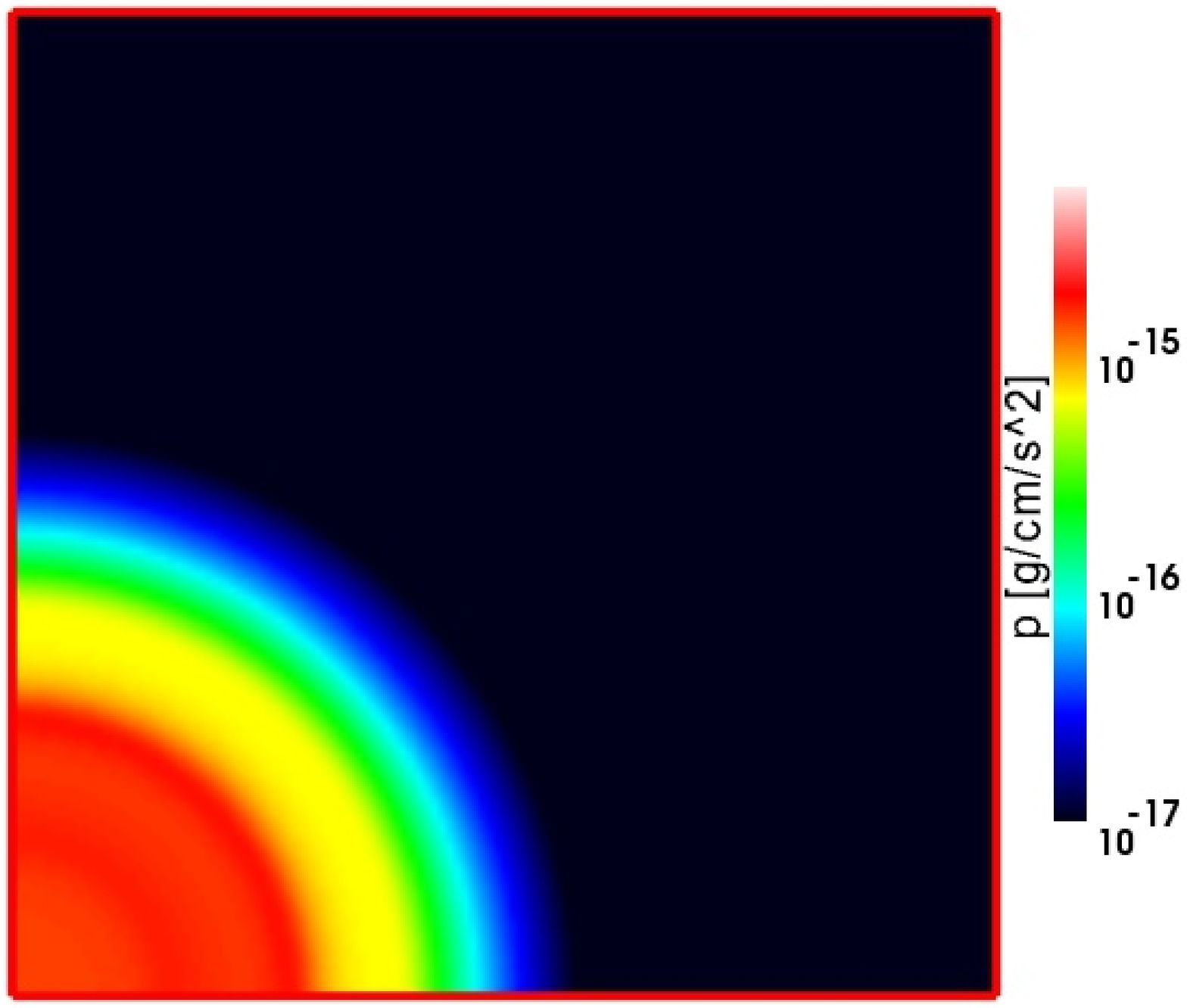}
  \includegraphics[width=2.3in]{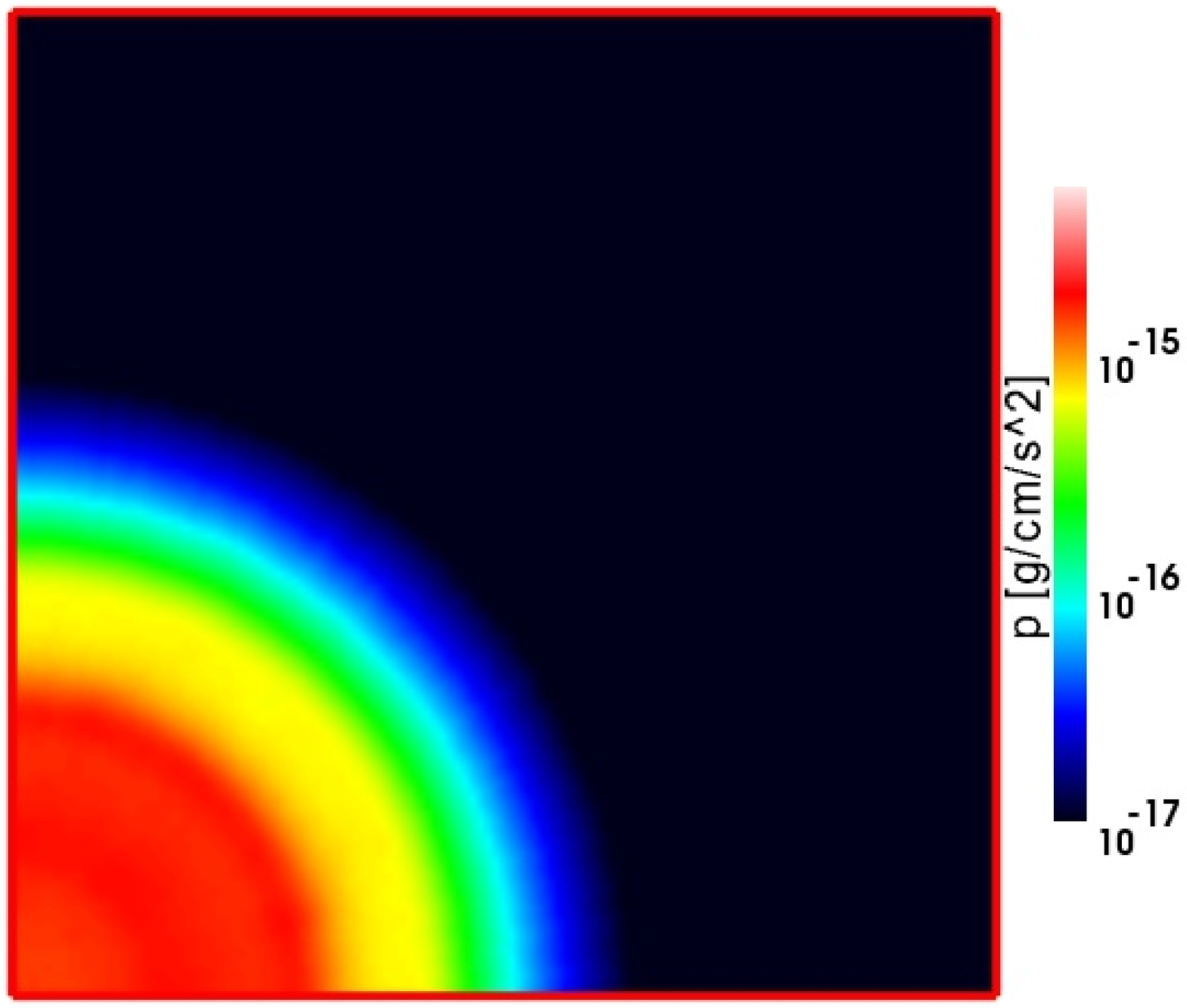}
  \includegraphics[width=2.3in]{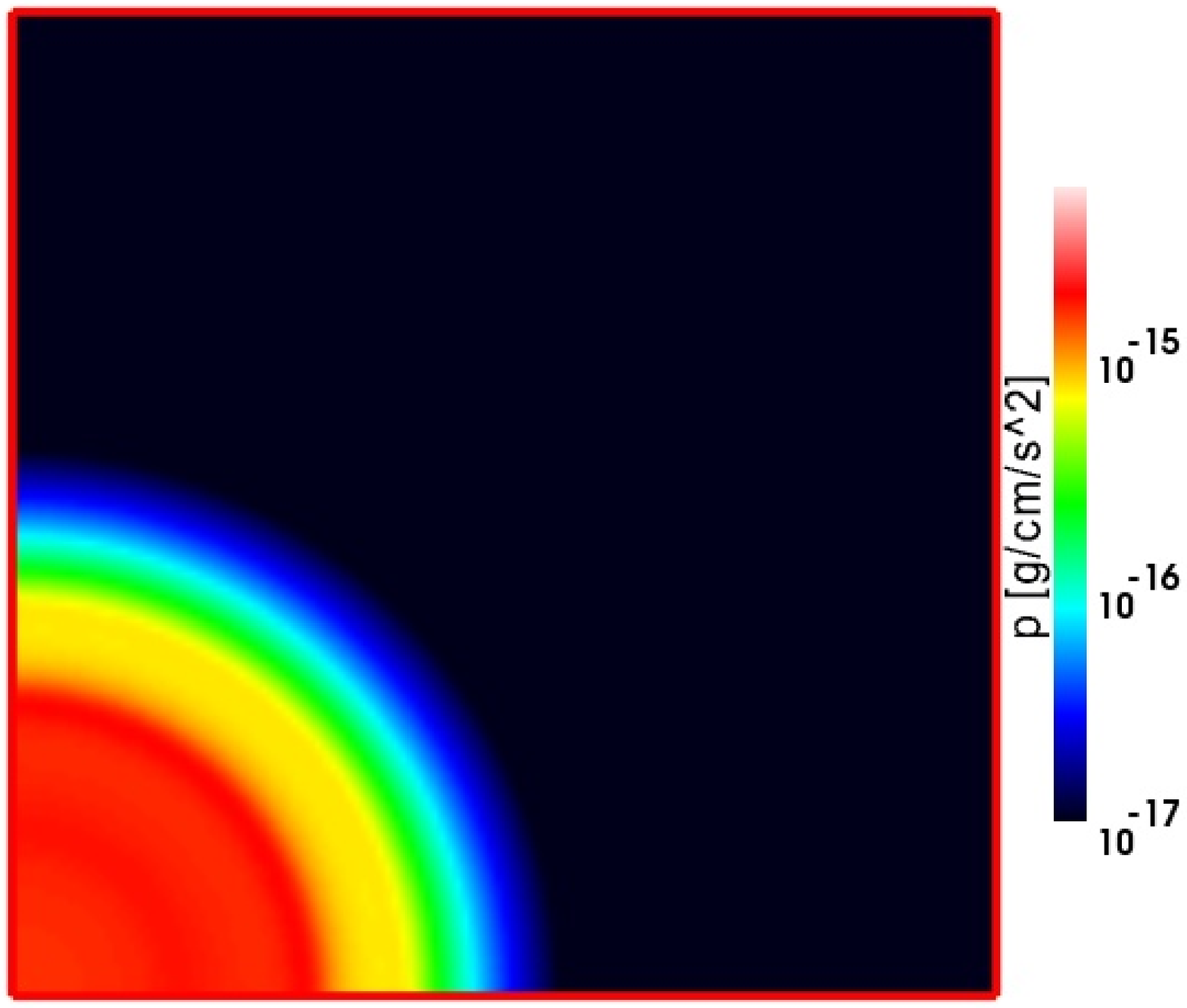}
  \includegraphics[width=2.3in]{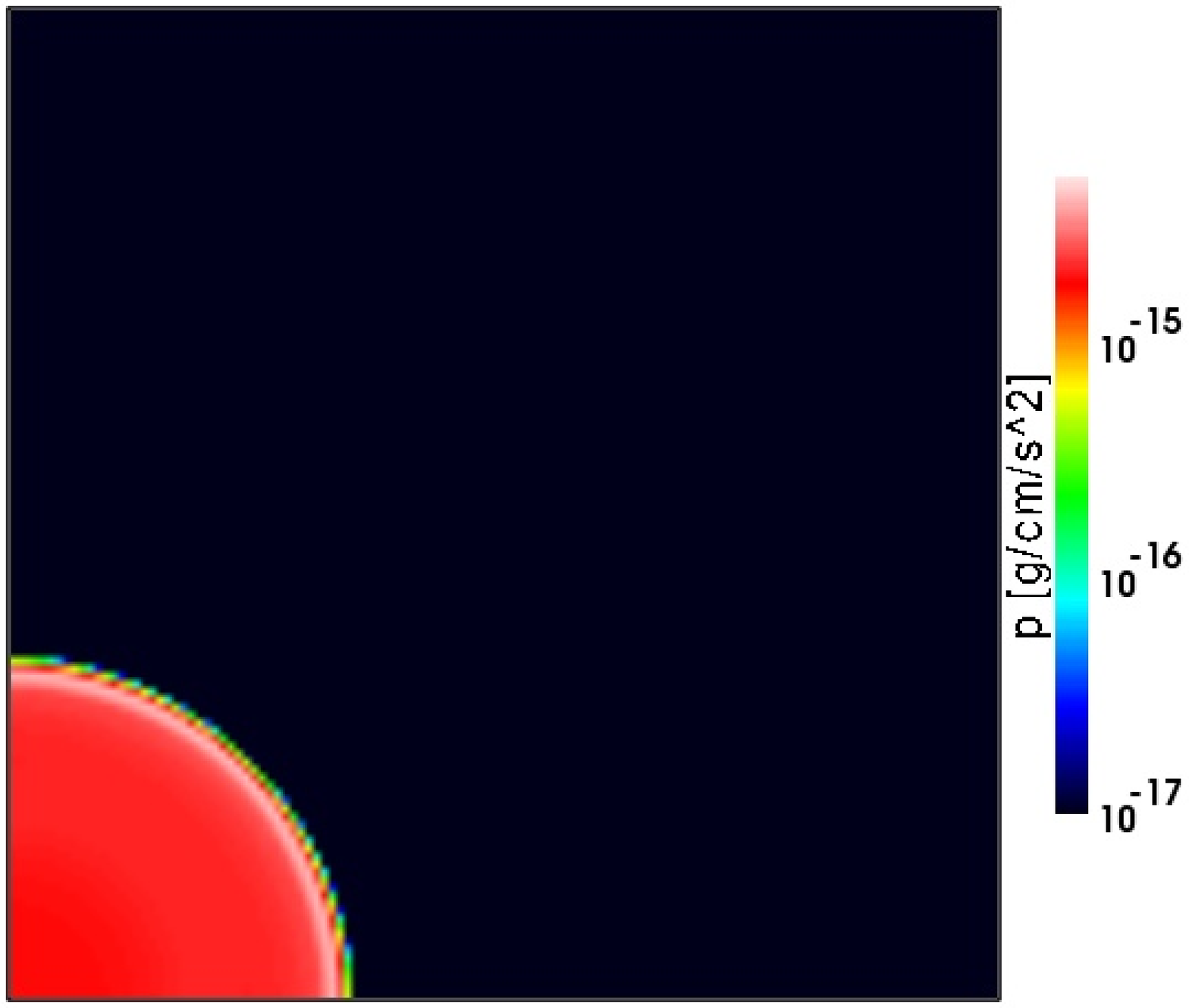}
\caption{Test 5 (H~II region expansion in an initially-uniform 
  gas): Images of the pressure, cut through the simulation volume at
  coordinate $z=0$ at time $t=100$ Myr for (left to right and top to bottom)
  Capreole+$C^2$-Ray, HART, RSPH, ZEUS-MP, RH1D, LICORICE, Flash-HC and Enzo-RT.
\label{T5_images3_p_fig}}
\end{center}
\end{figure*}

The block-structured AMR grid used in Flash-HC is distributed over
processors using a space-filling curve. Parallel ray tracing requires
each ray to be split in the independent sections where the ray
traverses the blocks held by a given processor. First, every processor
traces rays on its local blocks in directions which start from the
source, and end in the corners of each cell on the faces of the (cubic)
block. Since rays cross several blocks, interpolation is used to
assemble a ray from local block contributions. However, because some of
these blocks will be held by other processors, local column densities
need to be exchanged in one global communication. Note that only face
values are exchanged.  Finally, local and imported column densities are
combined using interpolation to assemble the complete ray. At the end
of this parallel operation, each cell has the total column density to
the source along a ray that traverses all intervening cells at the full
resolution of the AMR grid. Interpolation coefficients are chosen such
that the exact solution for the column density is obtained for a
uniform density distribution. Even in a non-uniform density
distribution, for example $1/r^2$, the differences between the value
of the correct column density and that obtained using HC is typically
less than half a percent.

\begin{figure*}
\begin{center}
  \includegraphics[width=2.3in]{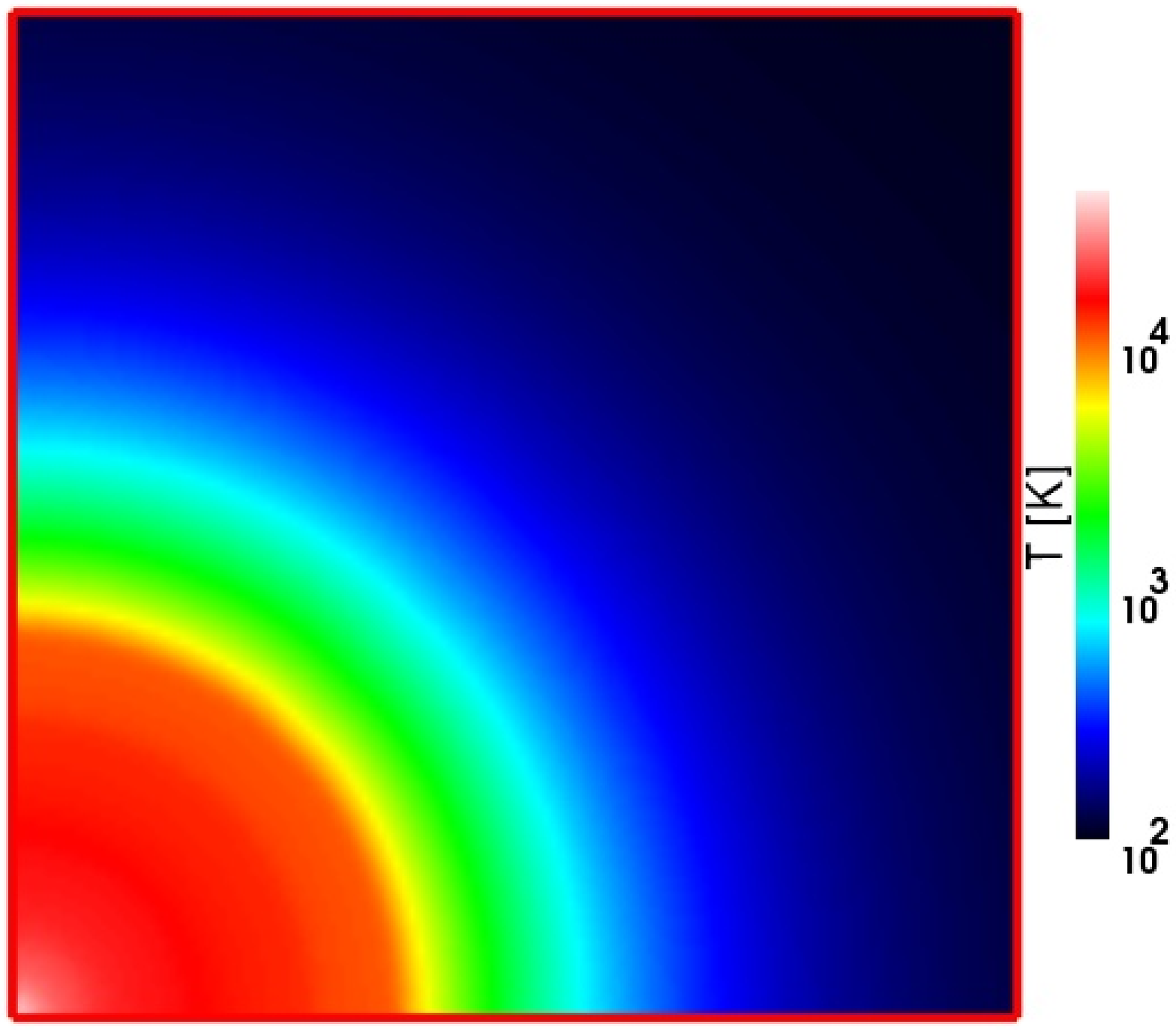}
  \includegraphics[width=2.3in]{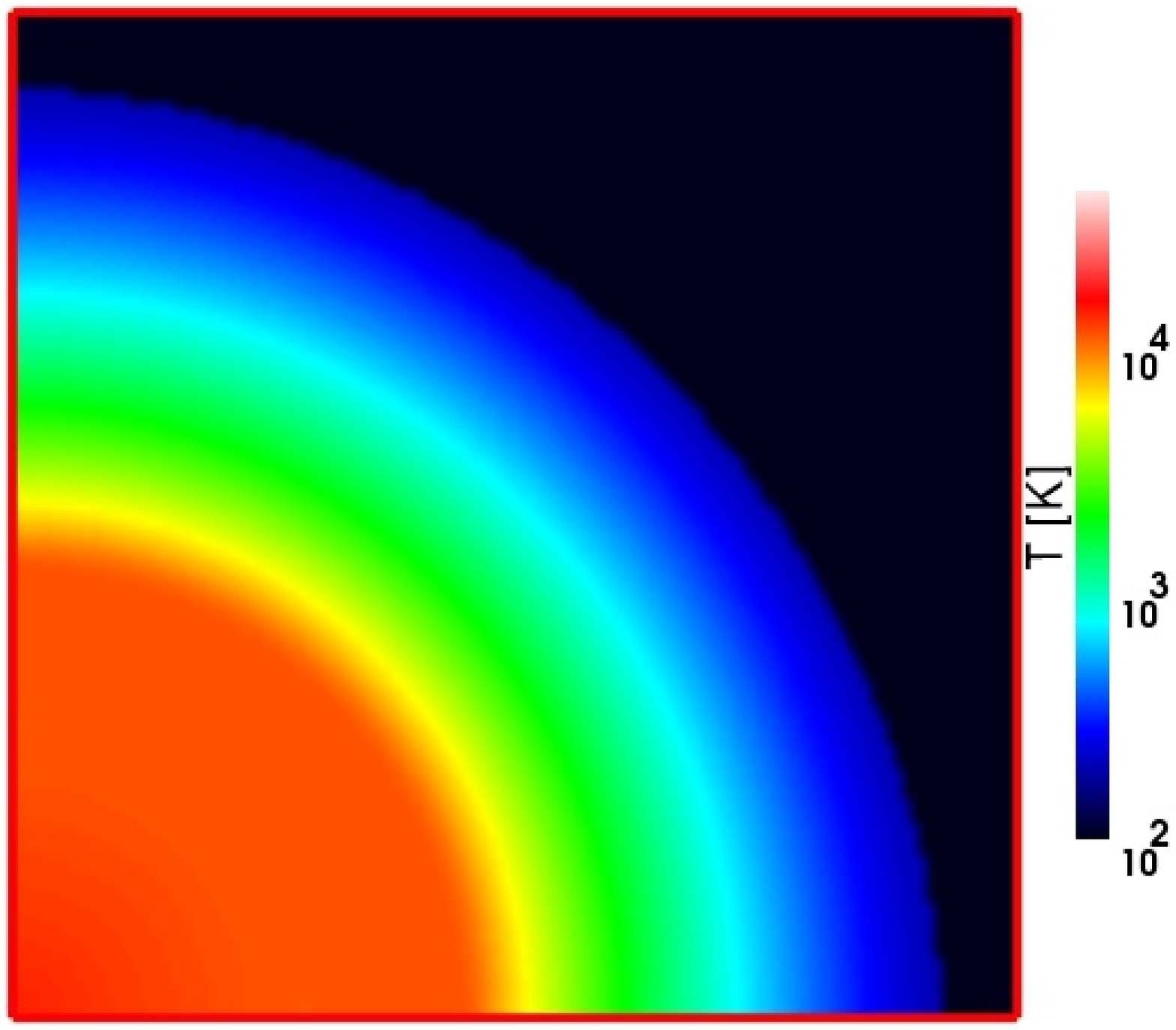}
  \includegraphics[width=2.3in]{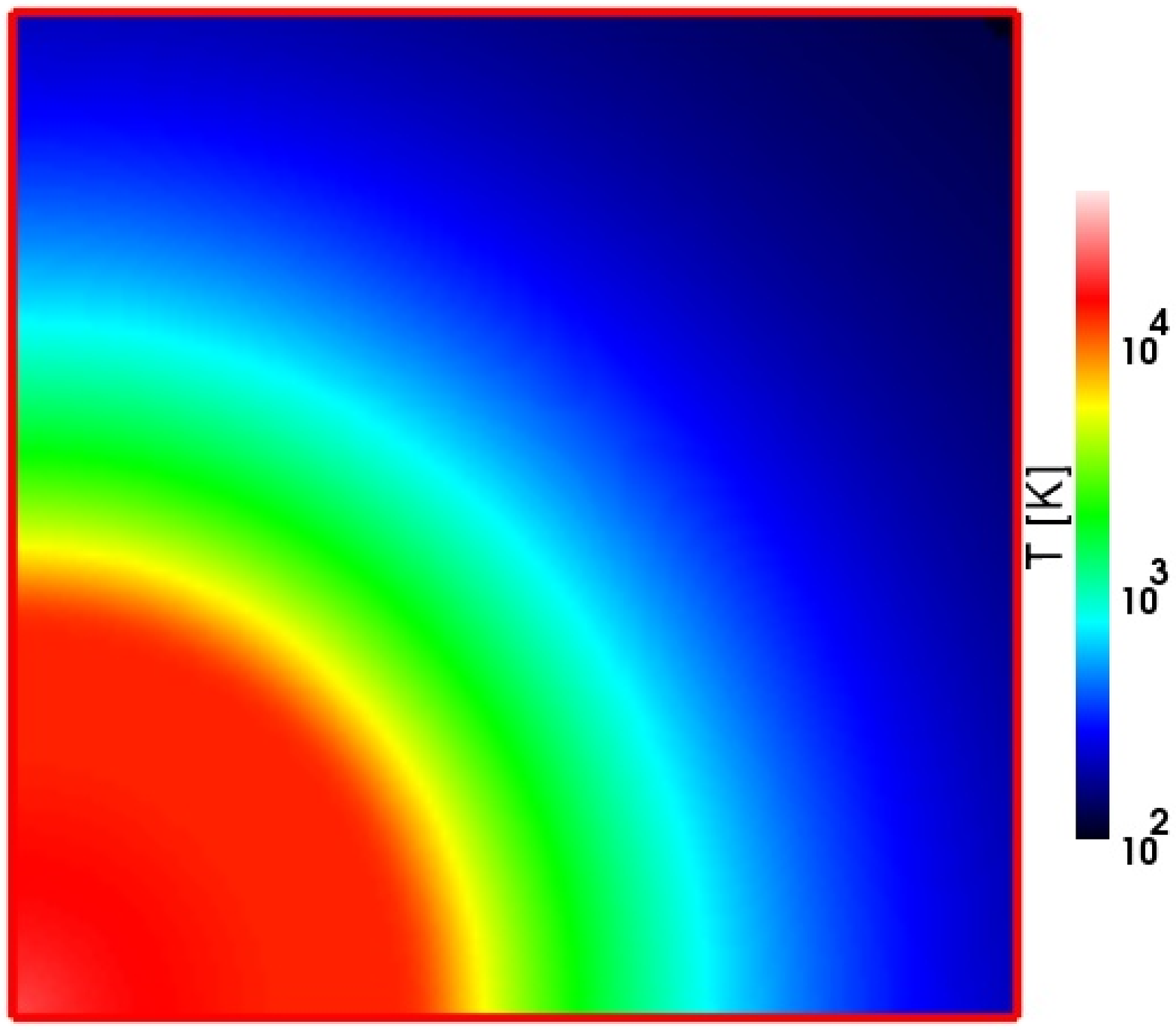}
  \includegraphics[width=2.3in]{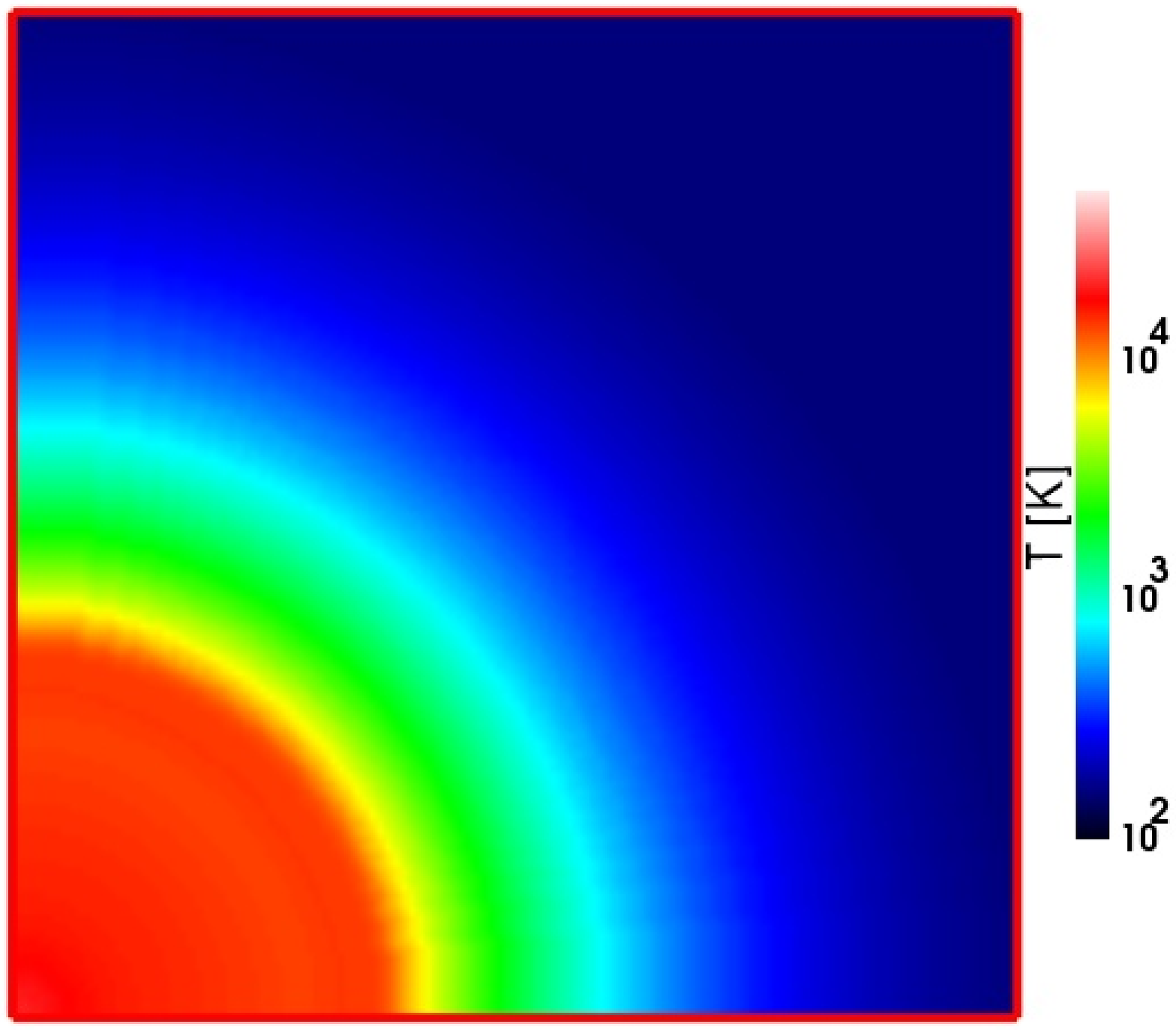}
  \includegraphics[width=2.3in]{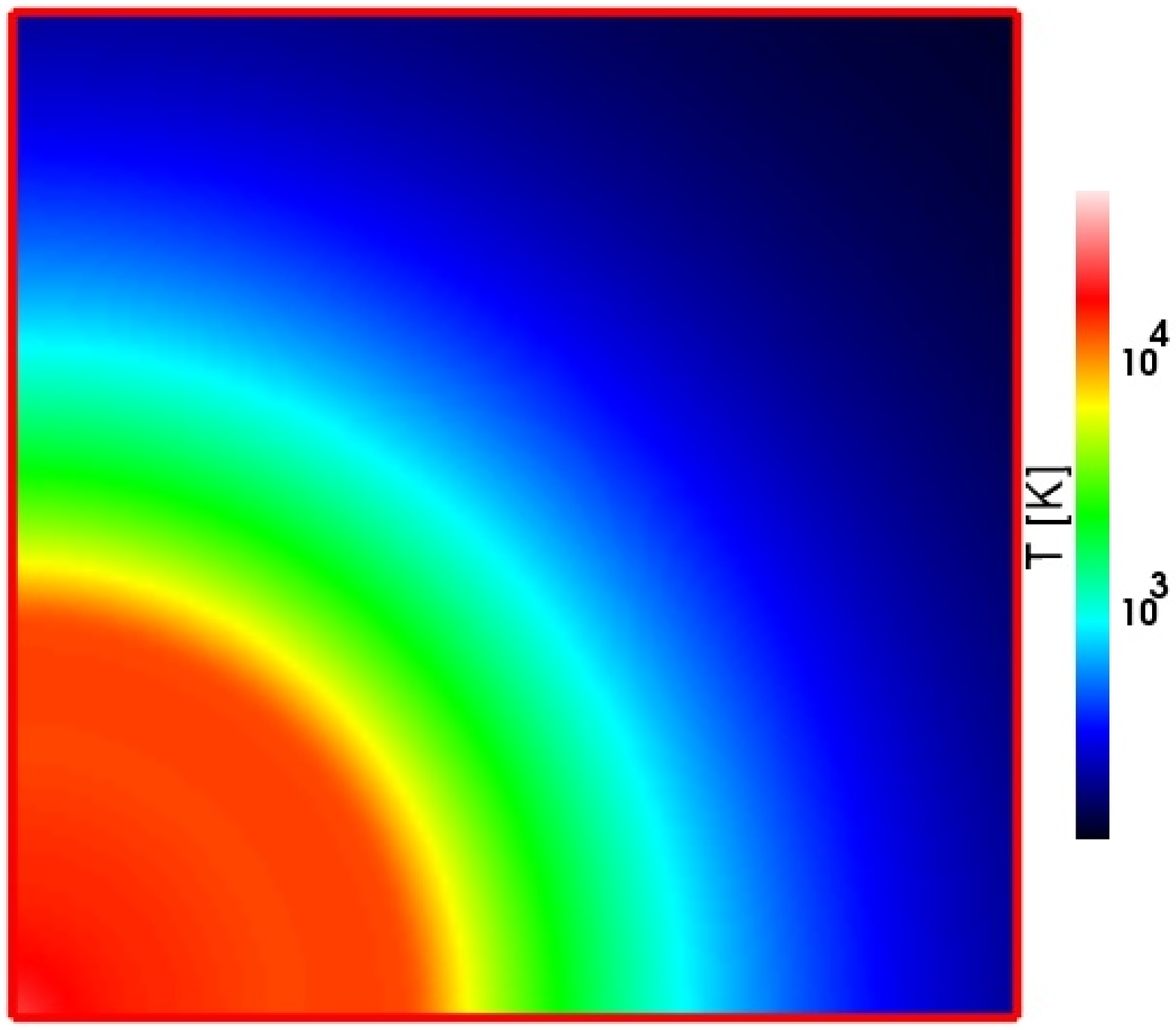}
  \includegraphics[width=2.3in]{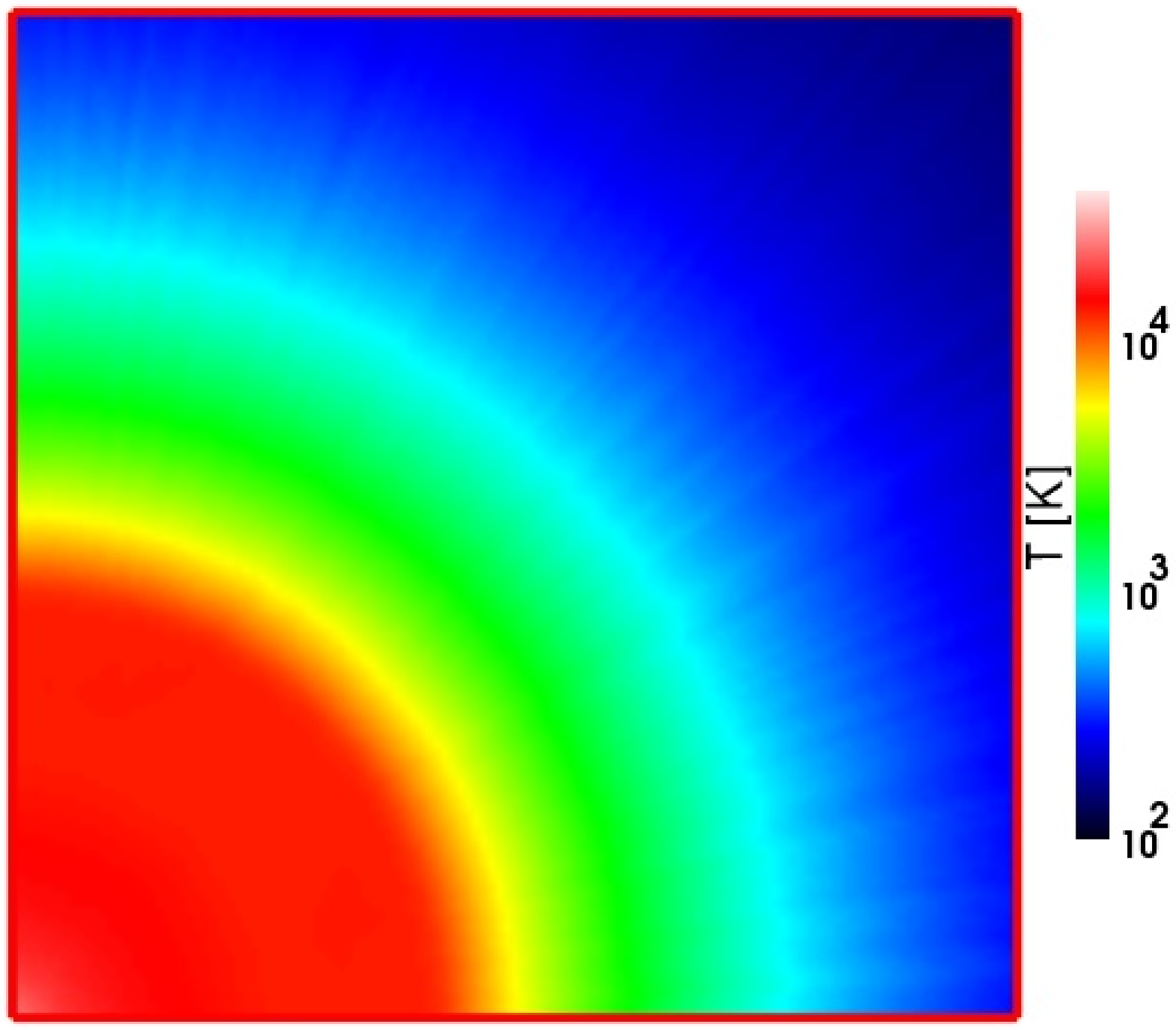}
  \includegraphics[width=2.3in]{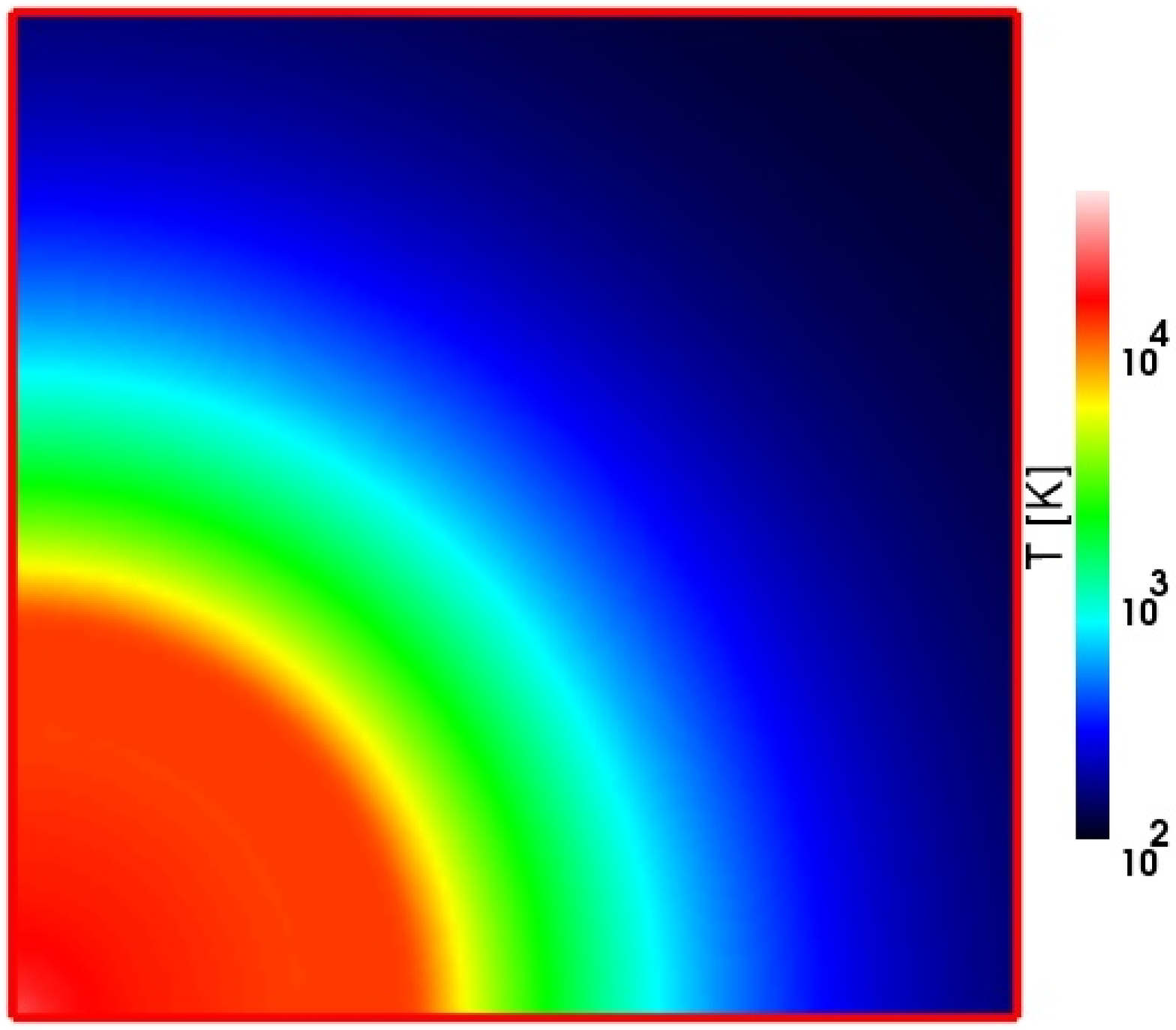}
  \includegraphics[width=2.3in]{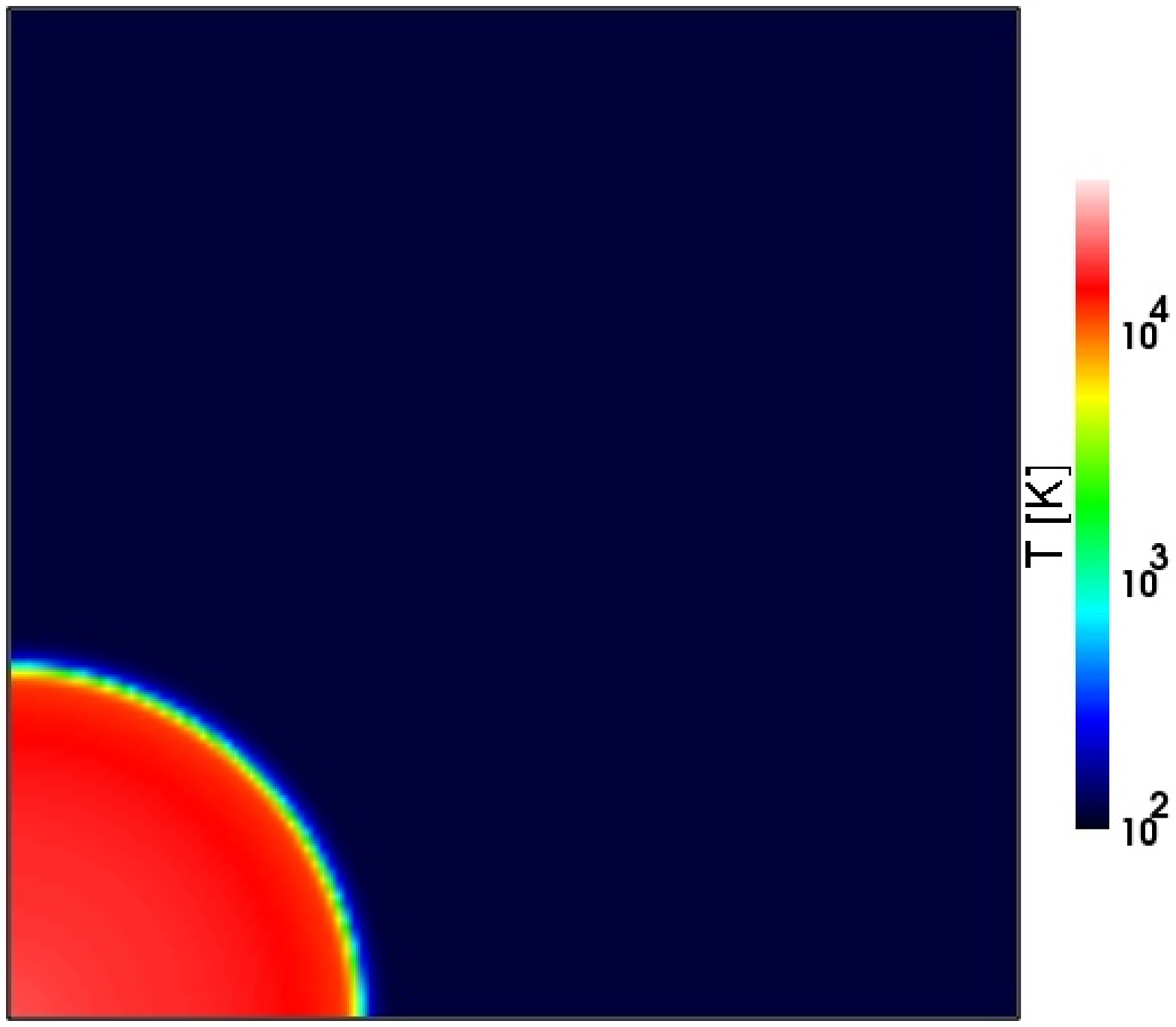}
\caption{Test 5 (H~II region expansion in an initially-uniform 
  gas): Images of the temperature, cut through the simulation volume at
  coordinate $z=0$ at time $t=100$ Myr for (left to right and top to bottom)
  Capreole+$C^2$-Ray, HART, RSPH, ZEUS-MP, RH1D, LICORICE, Flash-HC and Enzo-RT.
\label{T5_images3_T_fig}}
\end{center}
\end{figure*}

Recent improvements introduced since Paper~I
include the implementation of a fully photon conserving chemistry solver,
taking into account the effects of both spatial and temporal
discretization \citep{AbelNormanMadau1999, methodpaper}. This
implementation employs the Livermore Solver for Ordinary Differential
Equations \citep[LSODE][]{Hindmarsh1980}, which, although more computationally
intensive than the original solver used in DORIC \citep{Frank1994},
 eliminates the need for an independent radiative transfer time
step irrespective of the ionization front type and guarantees 
correct front positions and ionization heating. Details of the scheme
will be presented elsewhere. Additional functionality allows
for a radiation source outside of the computational volume, a feature
used in Test 7 to approximate a parallel ionization front.

The parallel scaling of HC was examined in \citet{Rijkhorst2005} and Paper I; 
the algorithm scales well for $\sim$ 100 processors 
on a SGI Altix, and $\sim$ 1000 processors on a IBM Blue Gene/L systems.  
The algorithm scales linearly with the number of sources. The photon-conserving 
RT and chemistry upgrades should not affect HC's scaling.

\subsection{Enzo-RT (D.R. Reynolds, M.L. Norman, J.C. Hayes, P. Paschos)}

\begin{figure*}
\begin{center}
  \includegraphics[width=2.3in]{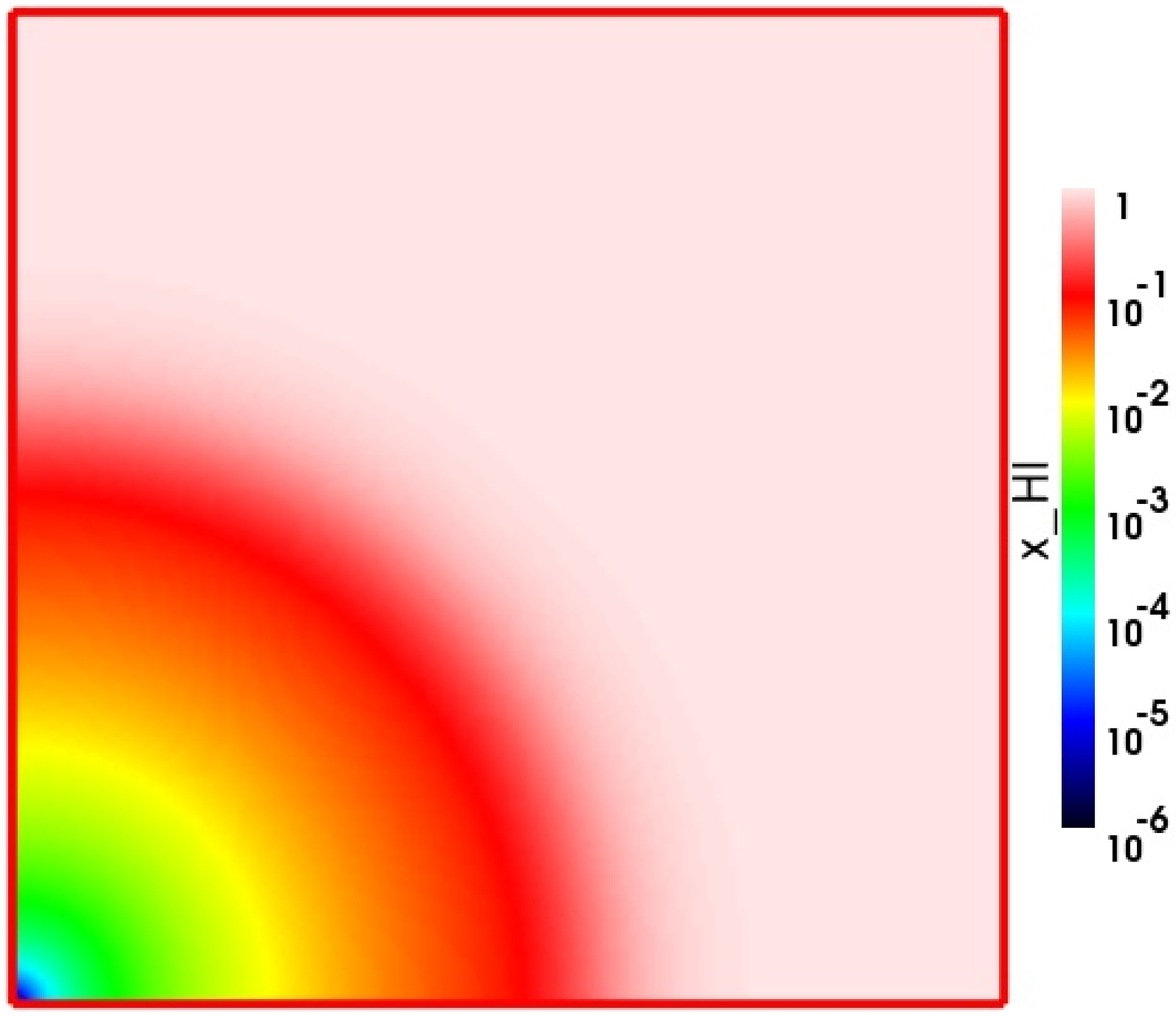}
  \includegraphics[width=2.3in]{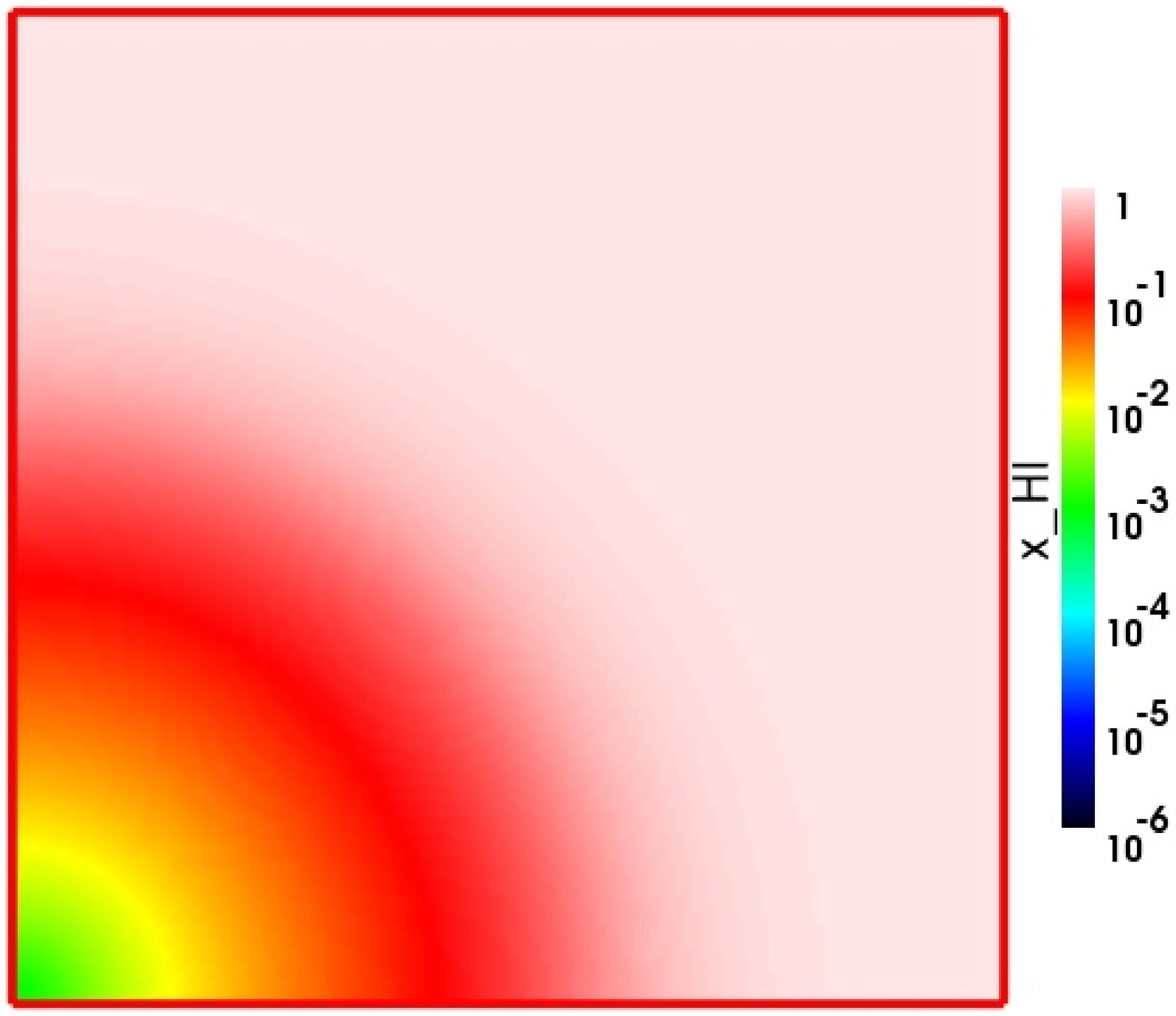}
  \includegraphics[width=2.3in]{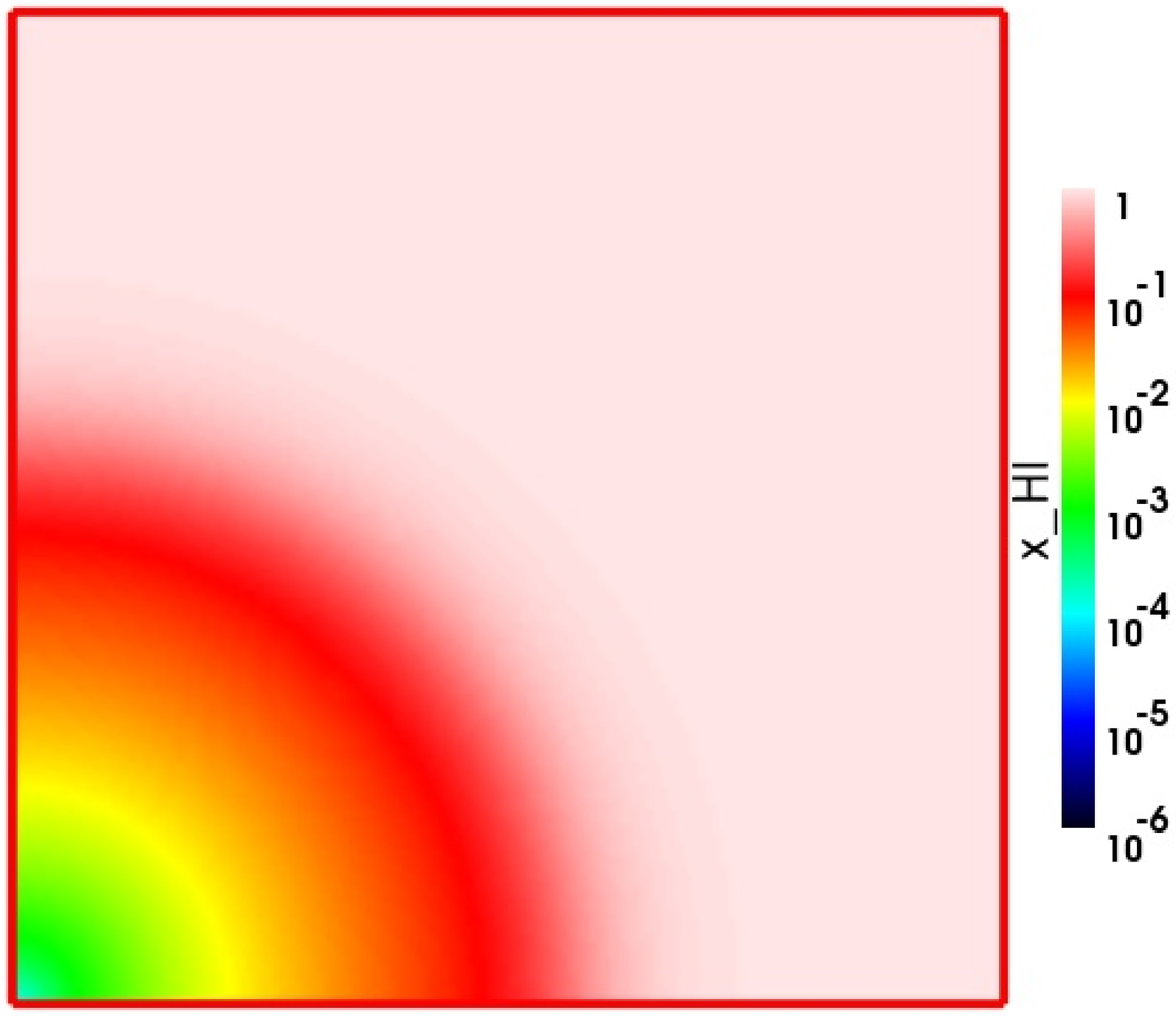}
  \includegraphics[width=2.3in]{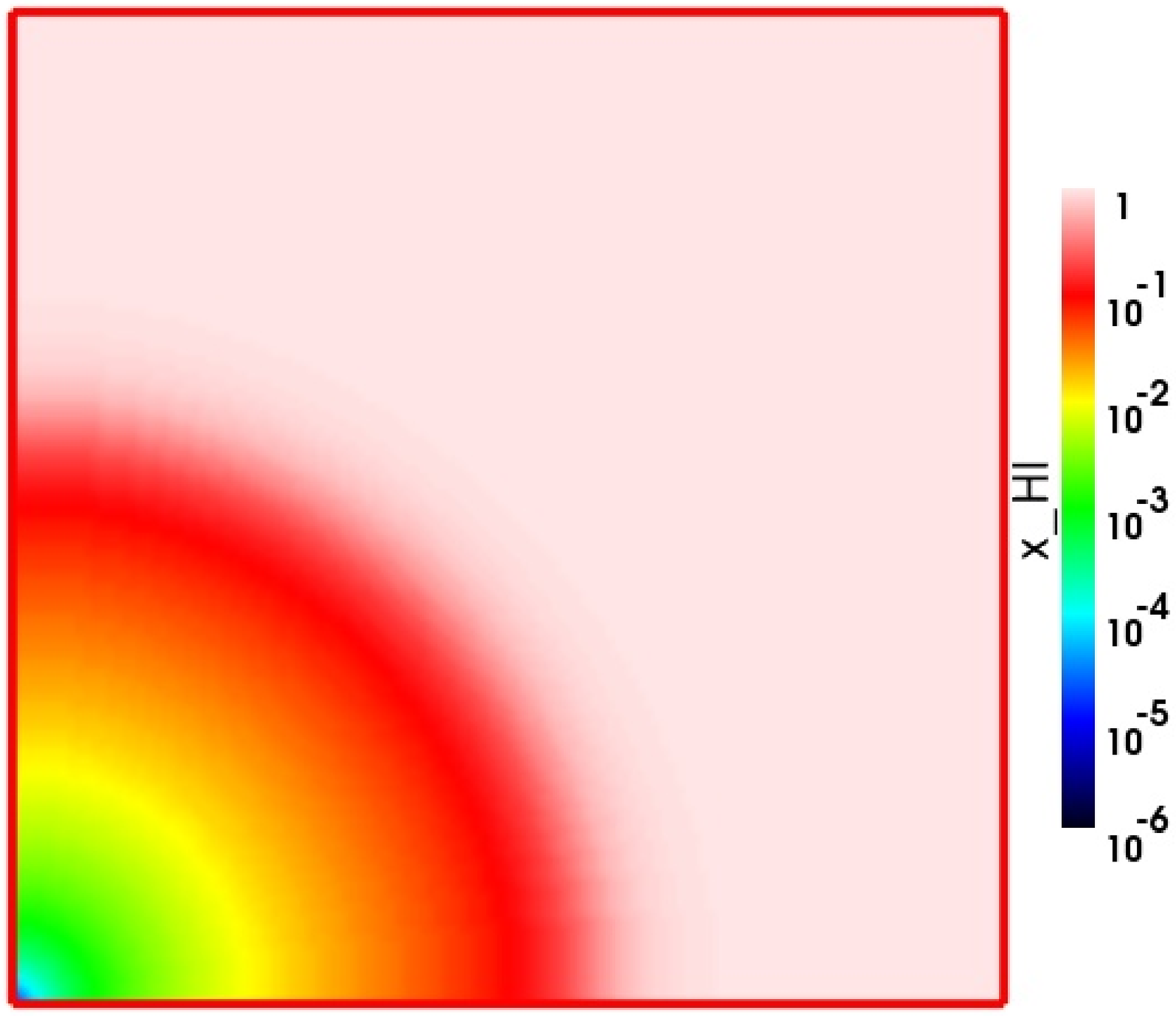}
  \includegraphics[width=2.3in]{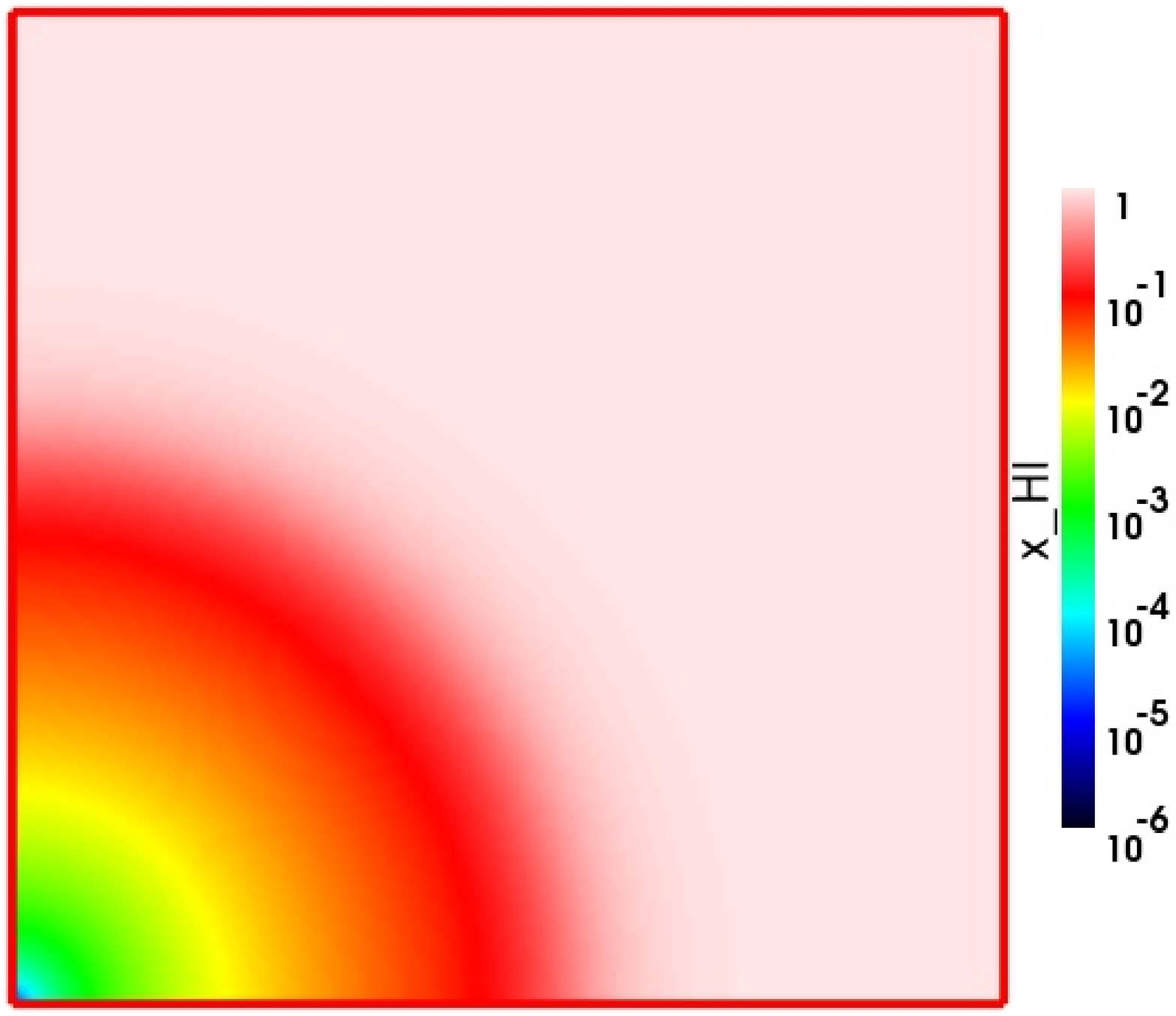}
  \includegraphics[width=2.3in]{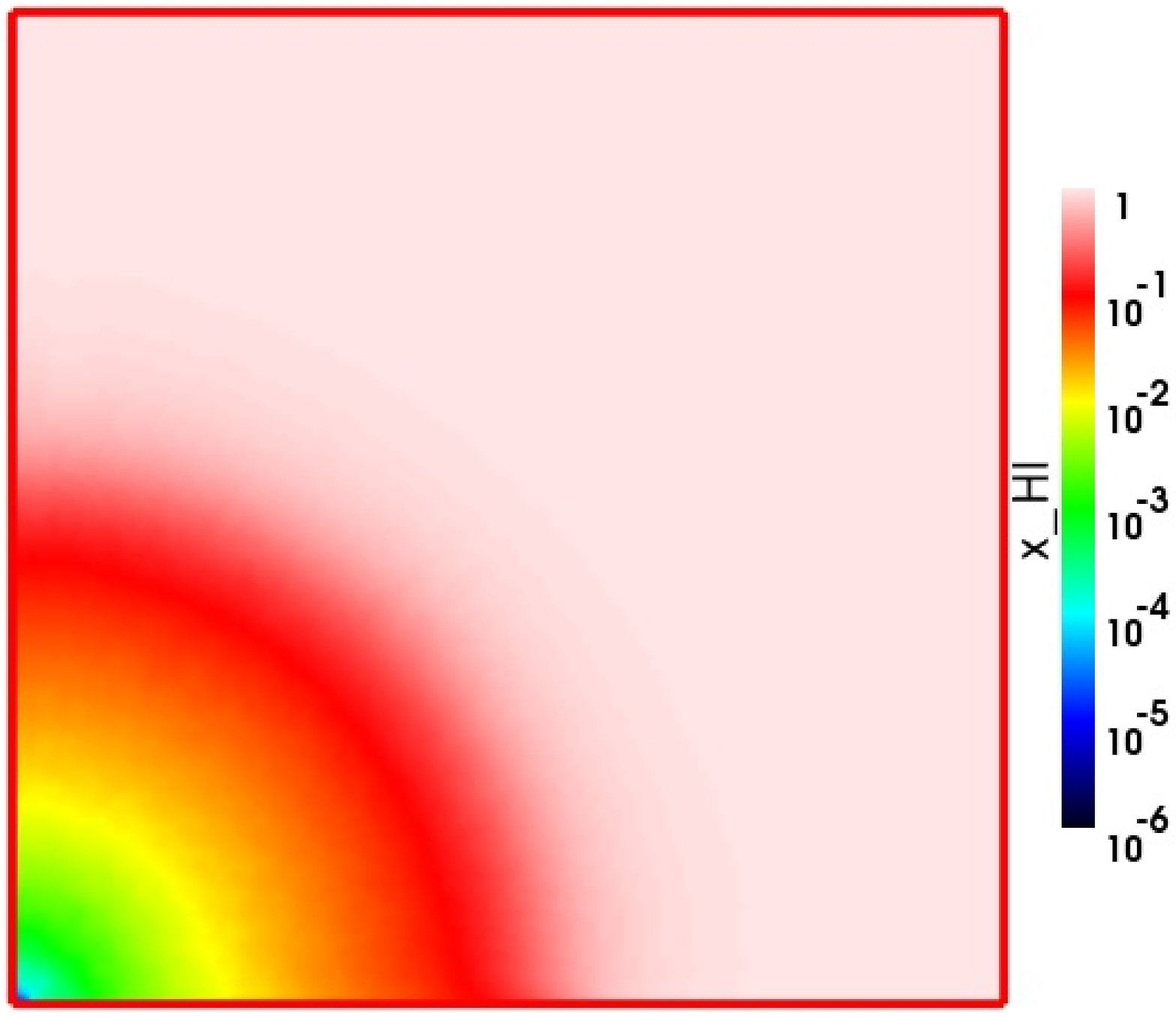}
  \includegraphics[width=2.3in]{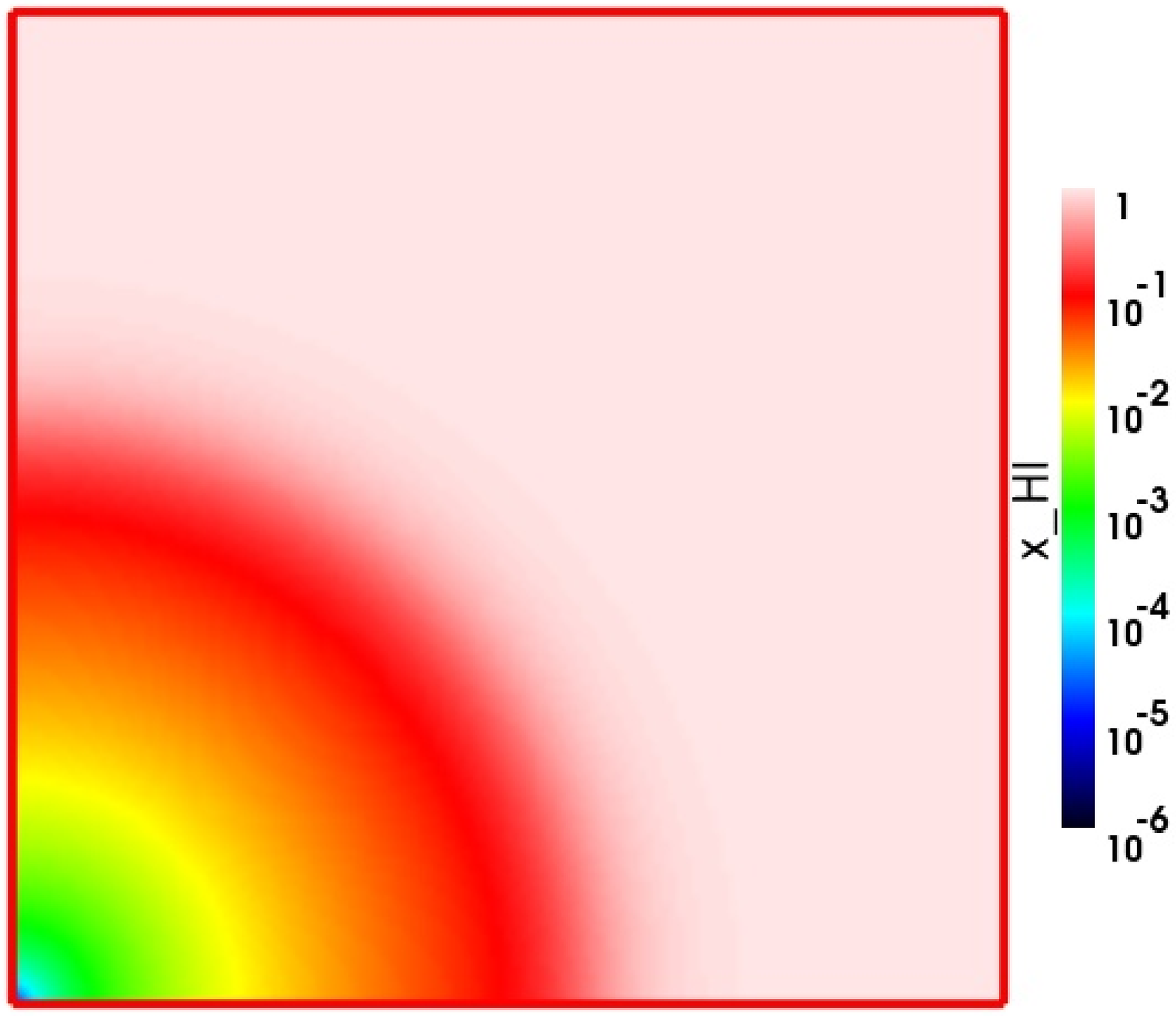}
  \includegraphics[width=2.3in]{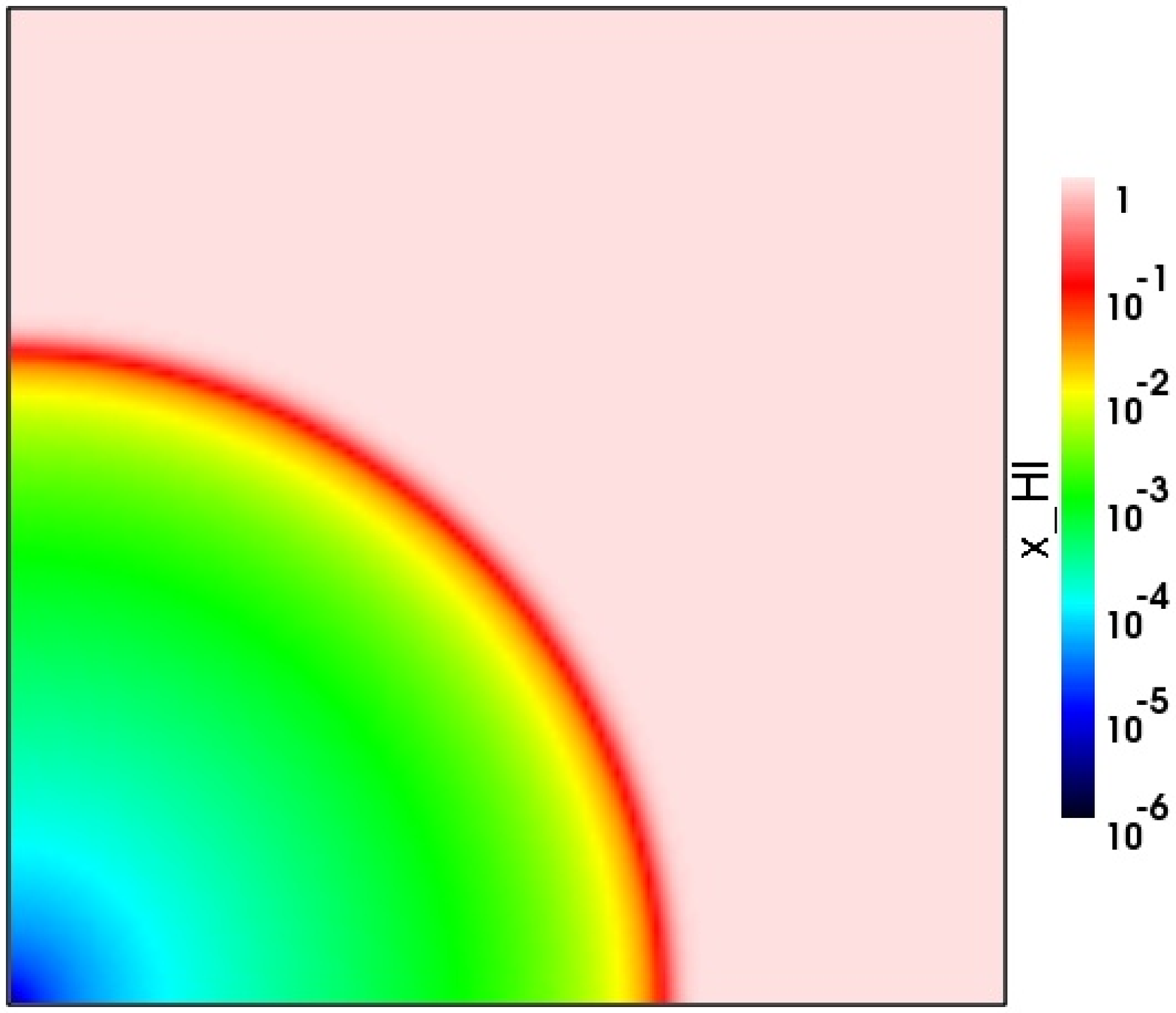}
\caption{Test 5 (H~II region expansion in an initially-uniform 
  gas): Images of the H~I fraction, cut through the simulation volume at
  coordinate $z=0$ at time $t=500$ Myr for (left to right and top to bottom)
  Capreole+$C^2$-Ray, HART, RSPH, ZEUS-MP, RH1D, LICORICE, Flash-HC and Enzo-RT.
\label{T5_images5_HI_fig}}
\end{center}
\end{figure*}

The Enzo-RT code is an extension to the freely-available Enzo 
code\footnote{\url{http://lca.ucsd.edu/portal/software/enzo}} that 
self-consistently 
incorporates coupled radiation transport and chemical ionization 
kinetics within Enzo's formulation for cosmological hydrodynamics 
on AMR meshes \citep{2007arXiv0705.1556N}. In Enzo-RT we 
approximate the radiation transport processes using a single 
integrated radiation energy density in each spatial cell that is 
propagated with flux-limited diffusion on a finite-volume mesh.  
The radiation field is implicitly coupled in time to a 
multi-species chemical reaction network.  This implicit radiation 
chemistry system is then coupled in an operator-split fashion with 
Enzo's cosmological hydrodynamics solver, which utilizes the Piecewise 
Parabolic Method for the advection of matter and gas energy \citep{
1984JCoPh..54..174C}.  The coupled algorithm, along with a suite of 
verification tests, is fully described in \cite{2009arXiv0901.1110R}.

The frequency dependence of the photoionization rates is treated
by integrating a prescribed radiation frequency spectrum,
typically chosen to be either monochromatic, blackbody, or a $\left(
\frac{\nu}{\nu_0}\right)^{-\beta}$ power law.  This integration is 
performed upon initialization of the solver and the integrated rates 
are re-used throughout the simulation. In the current version of the 
code, only a single radiation profile is allowed, although this 
formulation may be easily extended to allow for multifrequency 
calculations. 

\begin{figure*}
\begin{center}
  \includegraphics[width=2.3in]{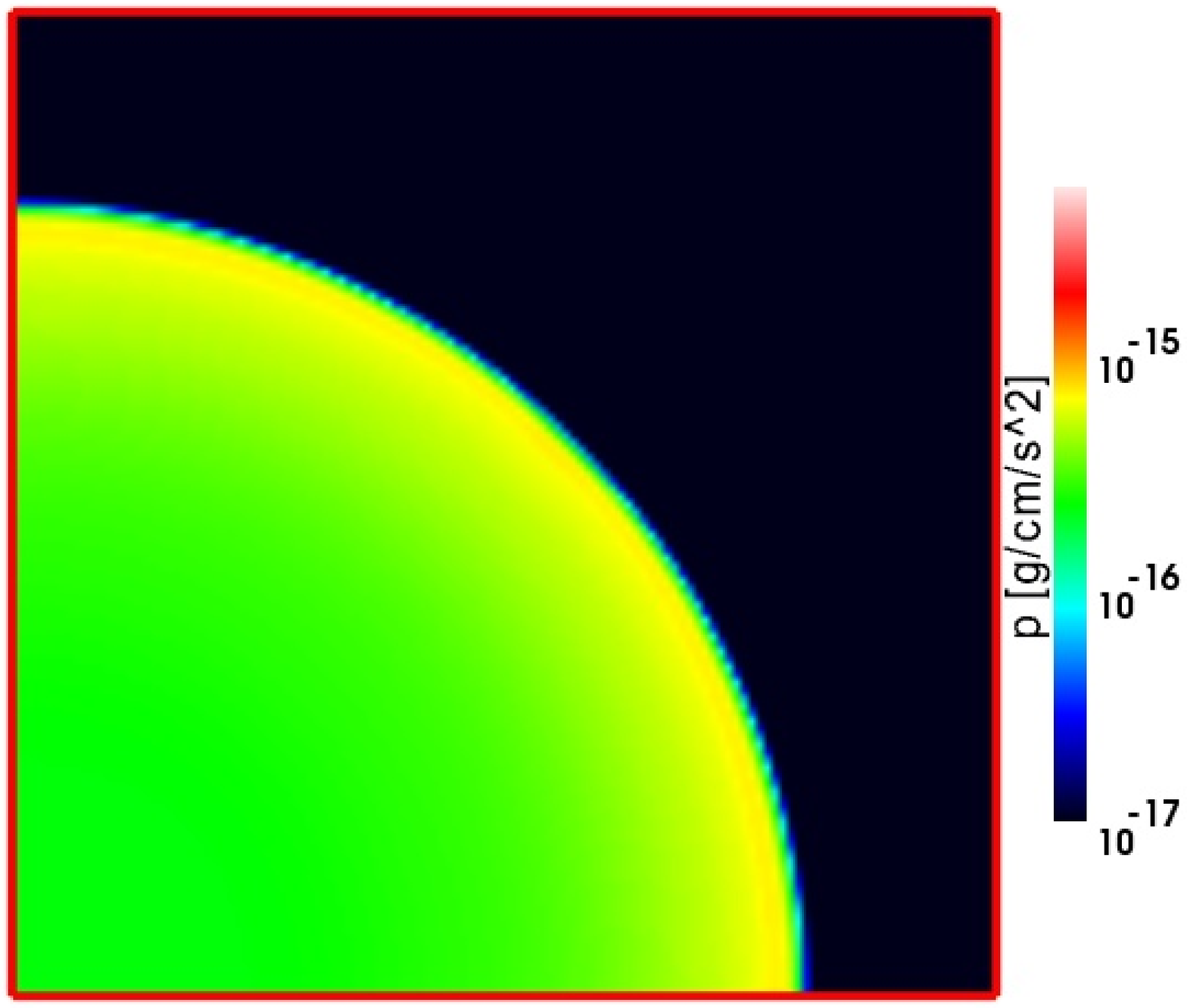}
  \includegraphics[width=2.3in]{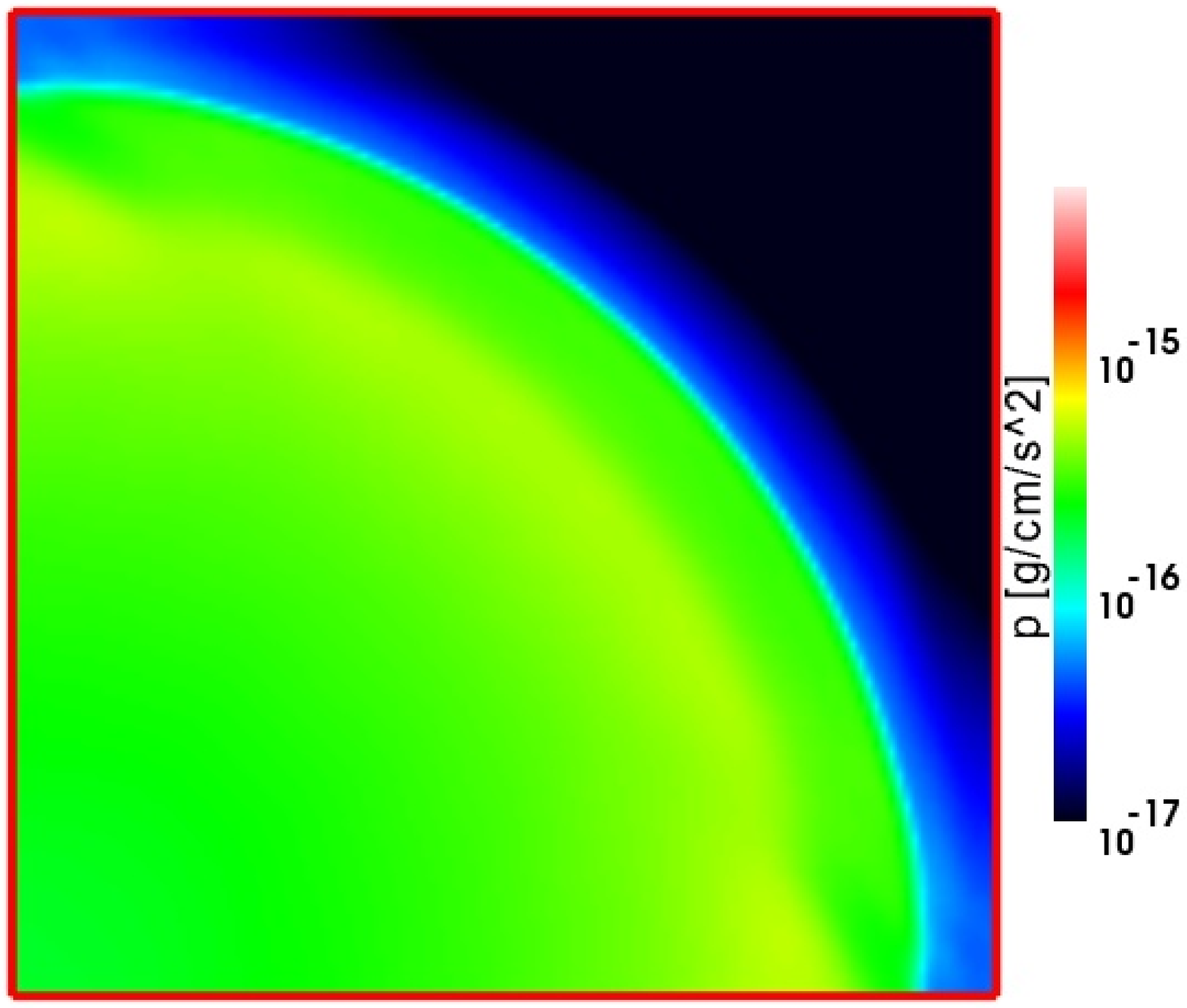}
  \includegraphics[width=2.3in]{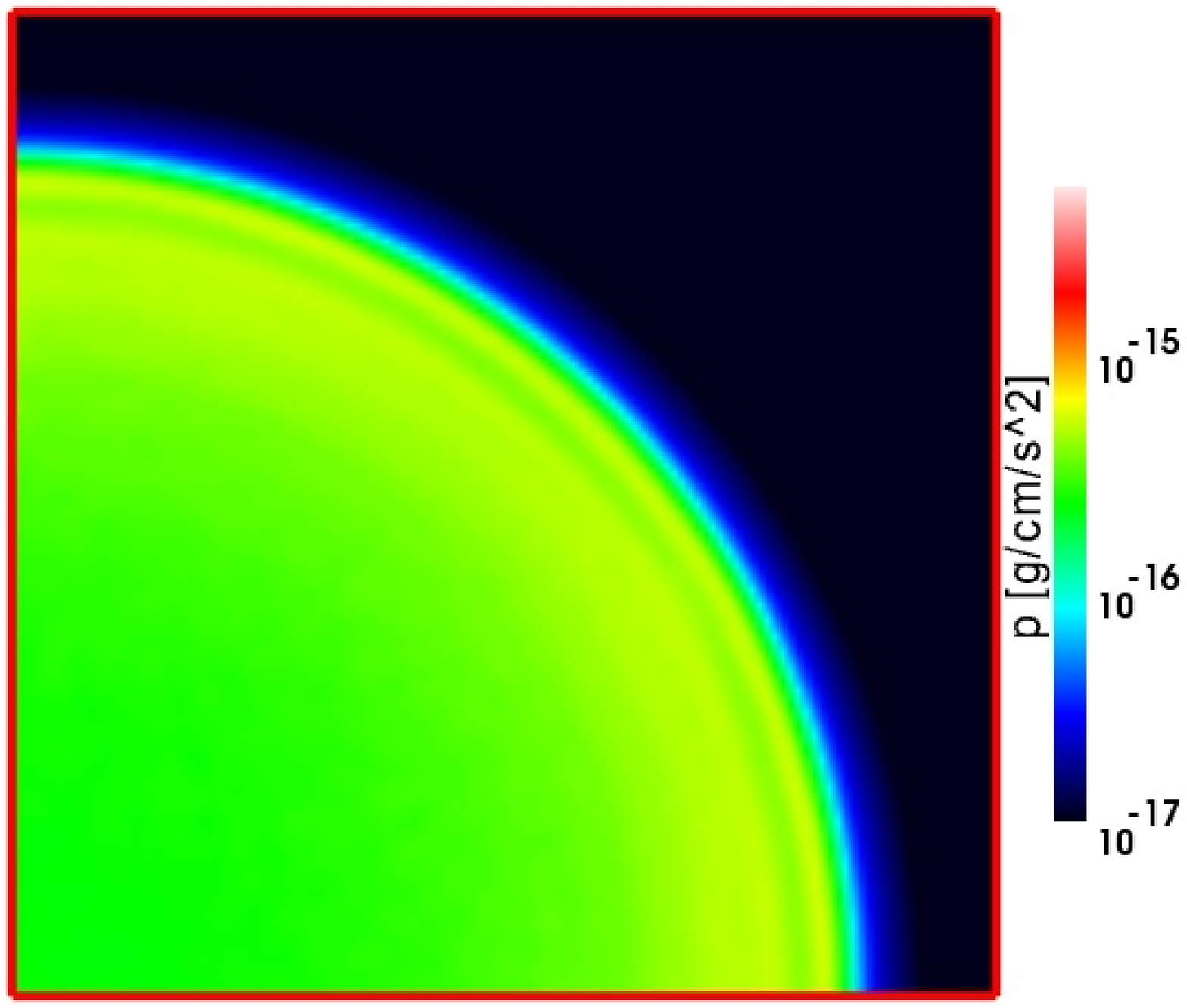}
  \includegraphics[width=2.3in]{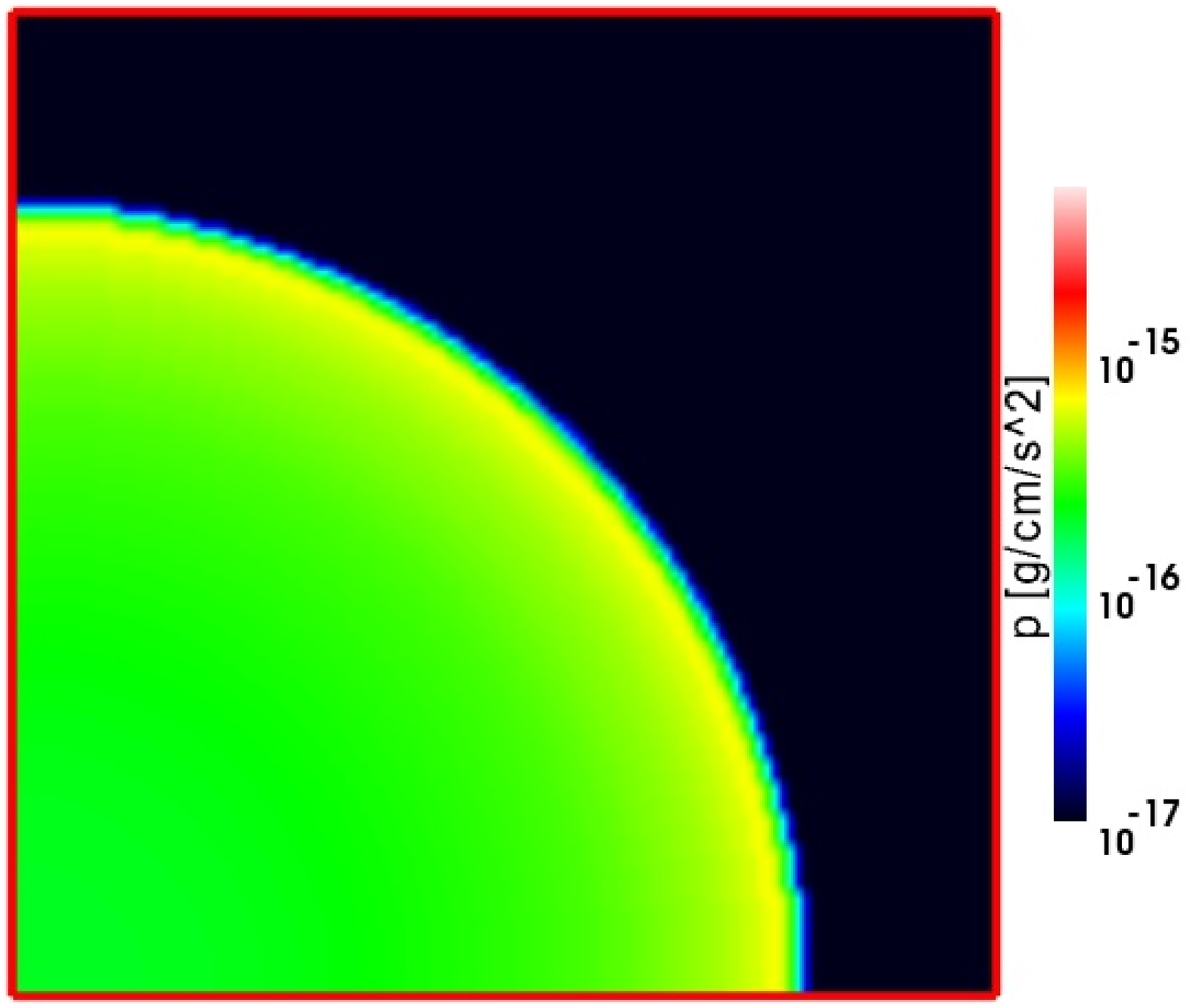}
  \includegraphics[width=2.3in]{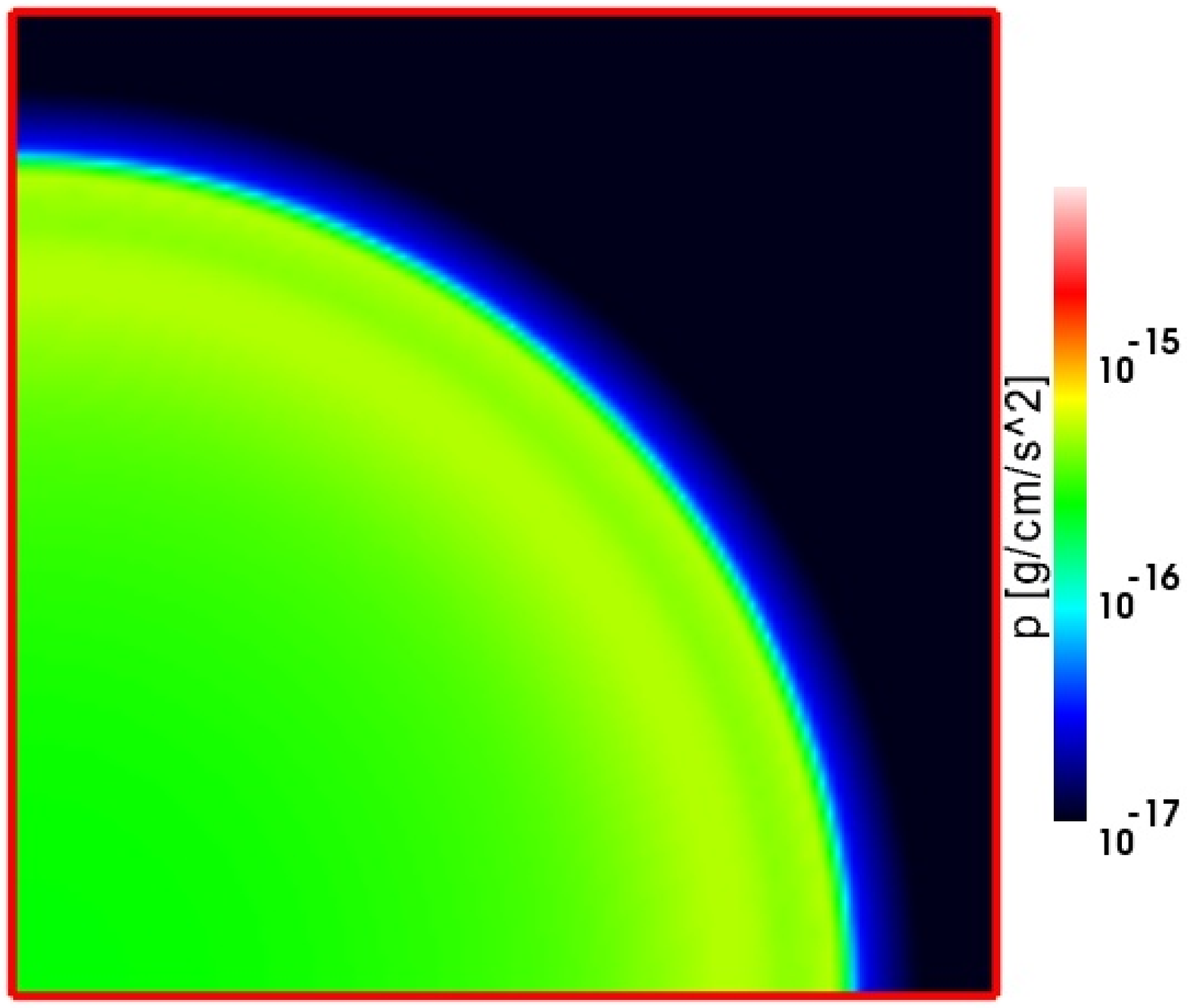}
  \includegraphics[width=2.3in]{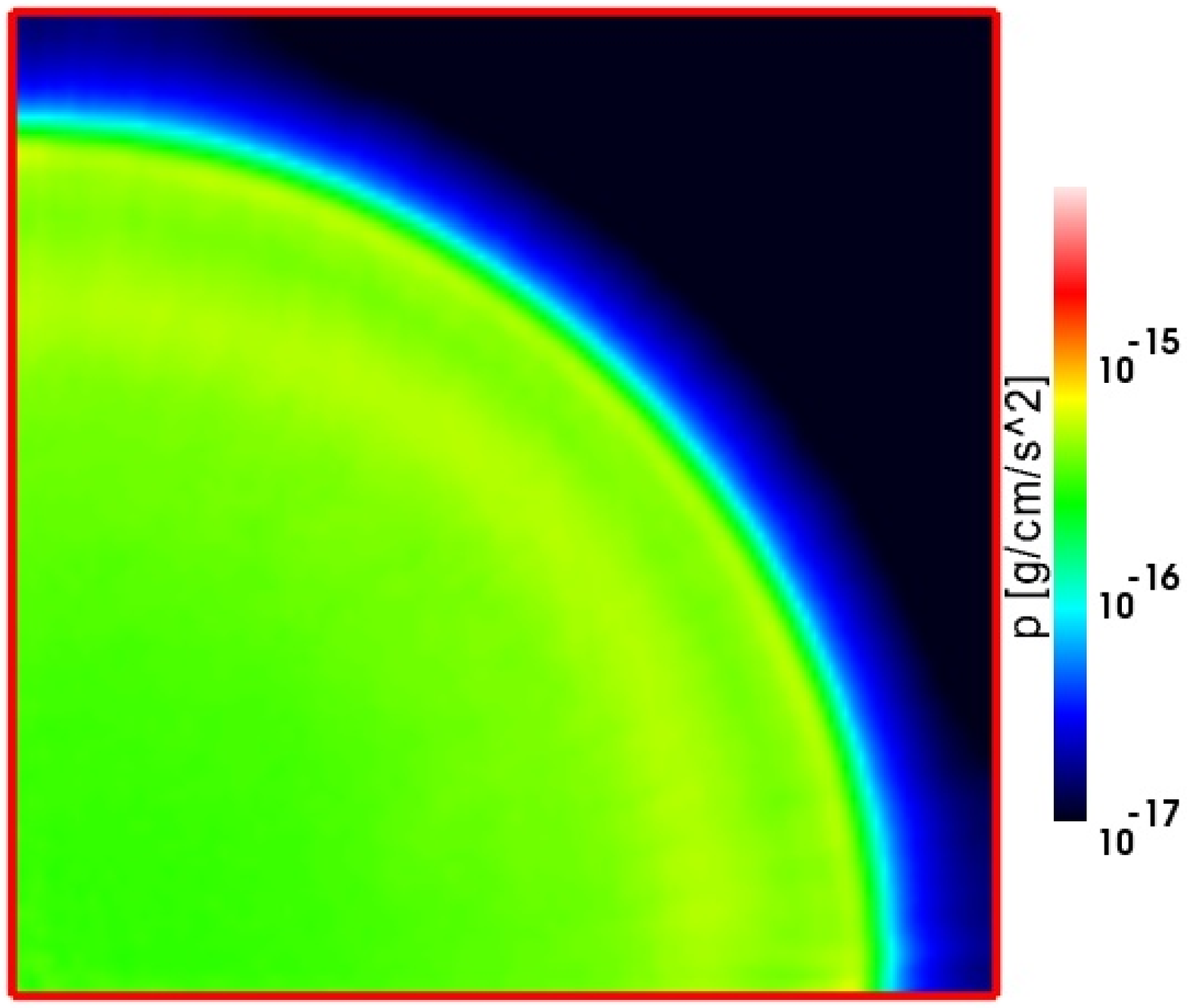}
  \includegraphics[width=2.3in]{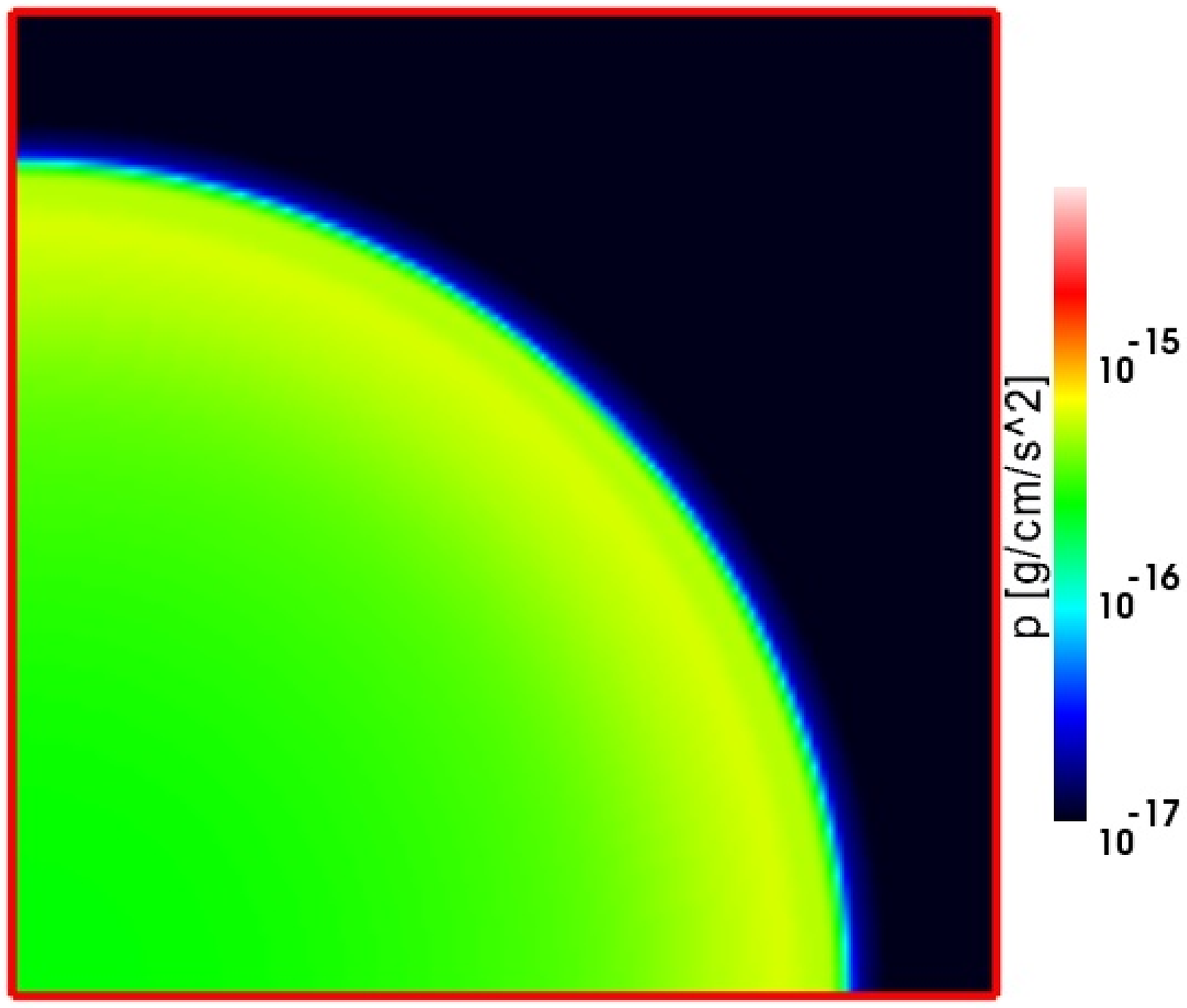}
  \includegraphics[width=2.3in]{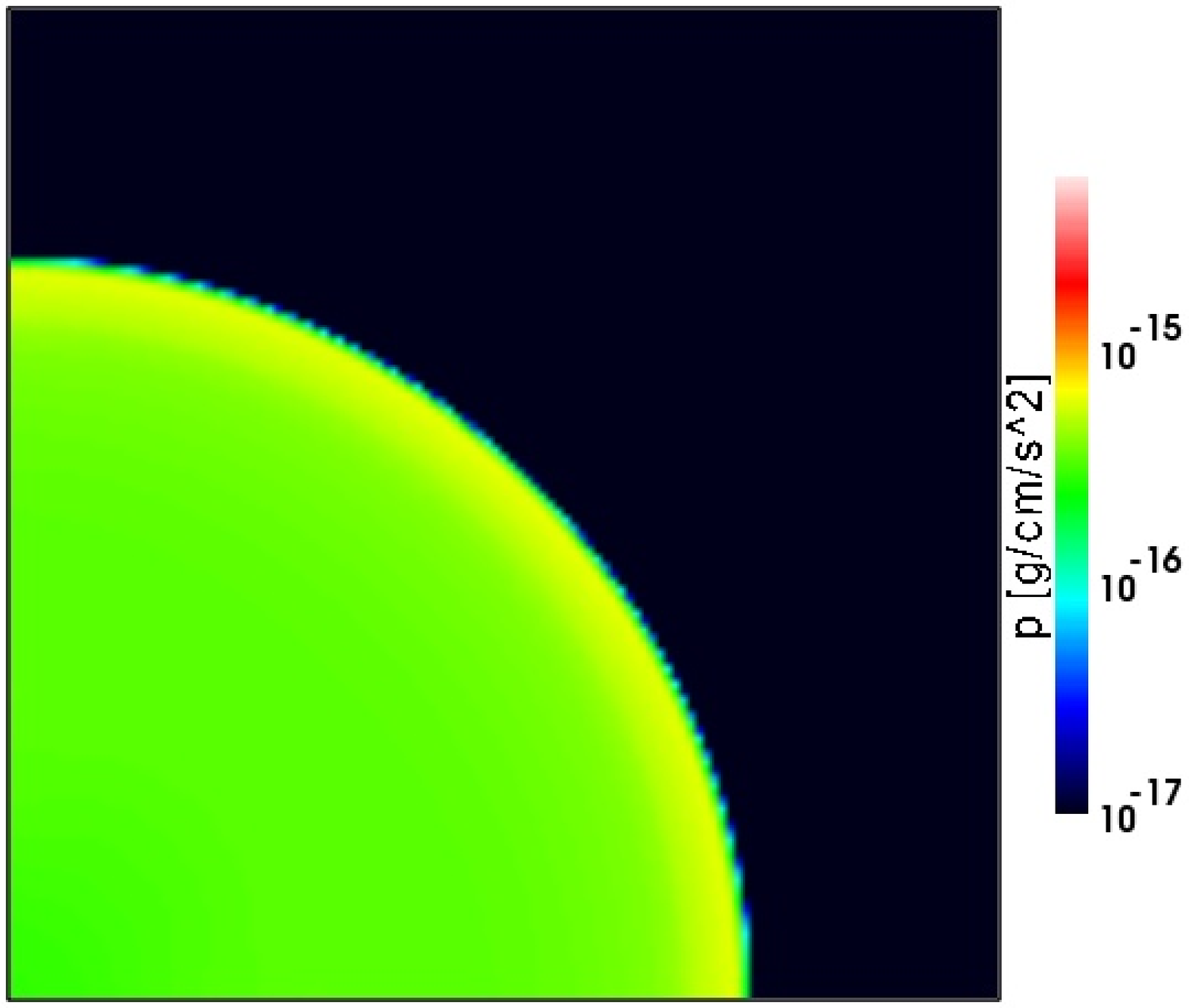}
\caption{Test 5 (H~II region expansion in an initially-uniform 
  gas): Images of the pressure, cut through the simulation volume at
  coordinate $z=0$ at time $t=500$ Myr for (left to right and top to bottom)
  Capreole+$C^2$-Ray, HART, RSPH, ZEUS-MP, RH1D, LICORICE, Flash-HC and Enzo-RT.
\label{T5_images5_p_fig}}
\end{center}
\end{figure*}

The solver for propagating radiation throughout the domain follows a
standard flux-limited diffusion model, in which the radiation flux
${\bf F}$ is approximated by
\begin{equation}
  {\bf F} = -\frac1a D\nabla E.
\end{equation}
Here $E$ is the radiation energy density, and the flux-limiter $D$ 
smoothly connects the limiting cases of (nearly) isotropic and 
free-streaming radiation:
\begin{equation}
  D(E) = diag(D_1(E), D_2(E), D_3(E)),
\end{equation}
where
\begin{equation}
  D_i(E) = \frac{c(2\kappa_T + R_i)}{6\kappa_T^2 + 3\kappa_TR_i +
    R_i^2}, \;\; i=1,\ldots,3,
\end{equation}
and $R_i = |\partial_i E|/E$, $c$ is the speed of light, and
$\kappa_T$ is the opacity.

\begin{figure*}
\begin{center}
  \includegraphics[width=2.3in]{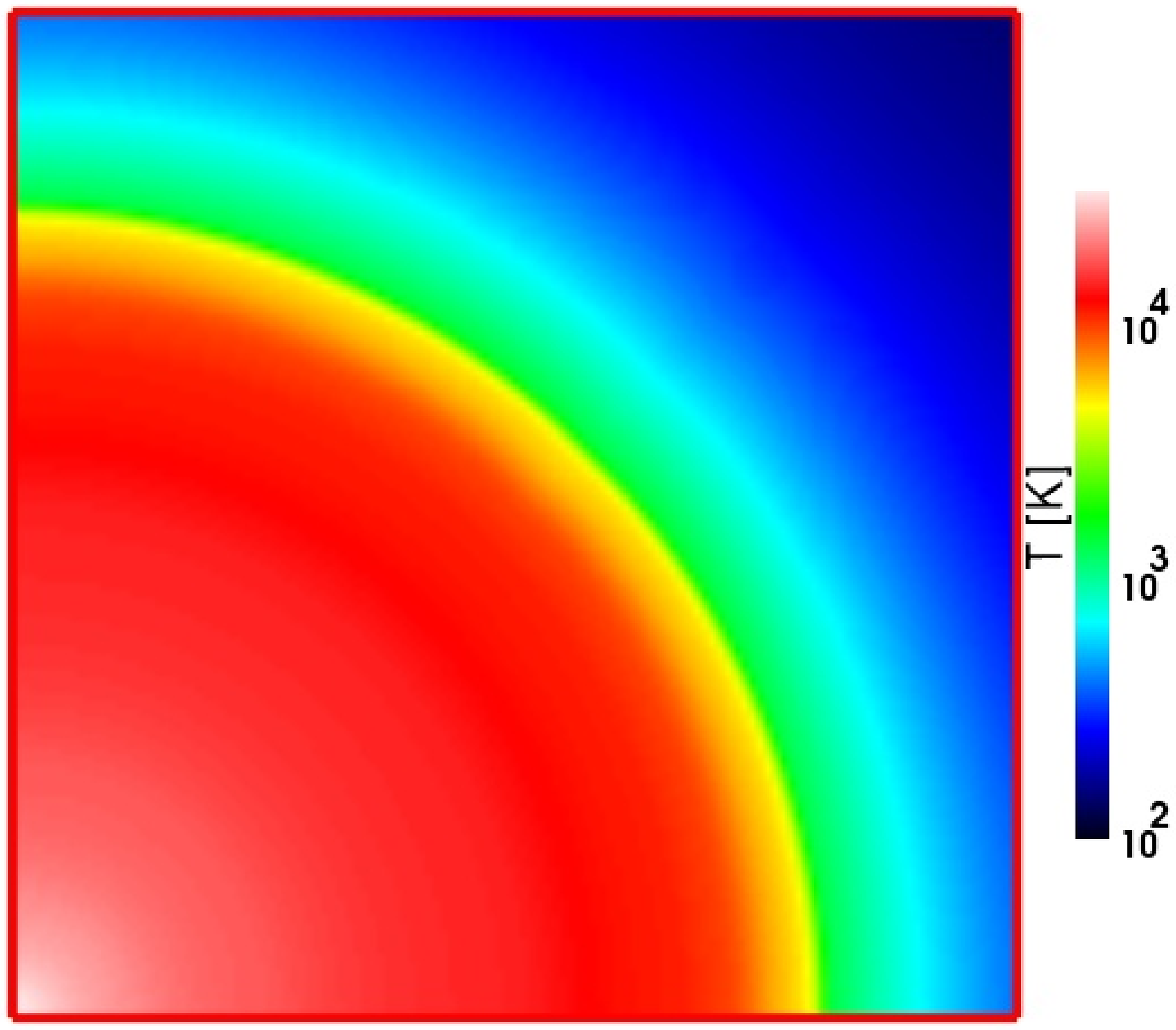}
  \includegraphics[width=2.3in]{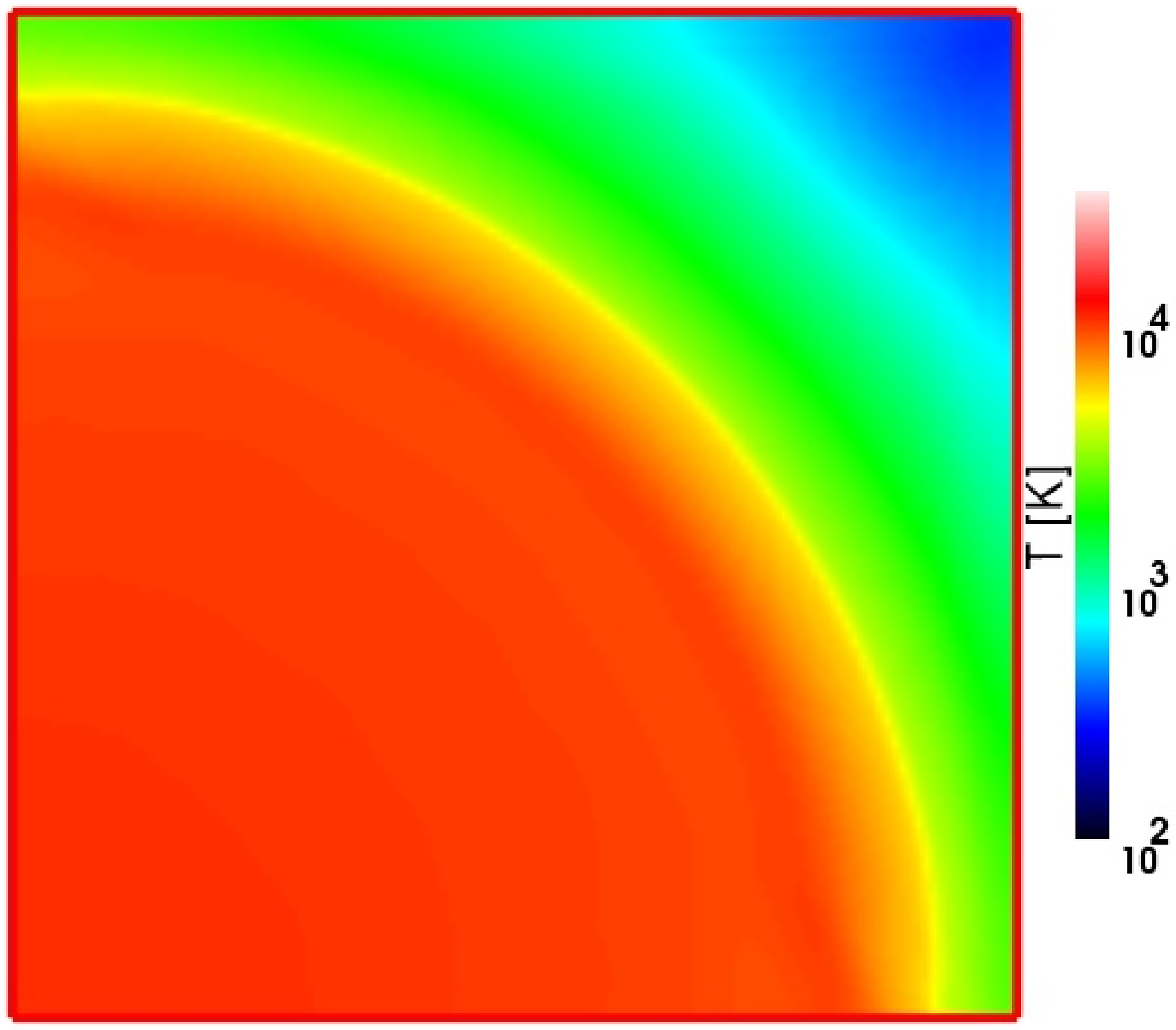}
  \includegraphics[width=2.3in]{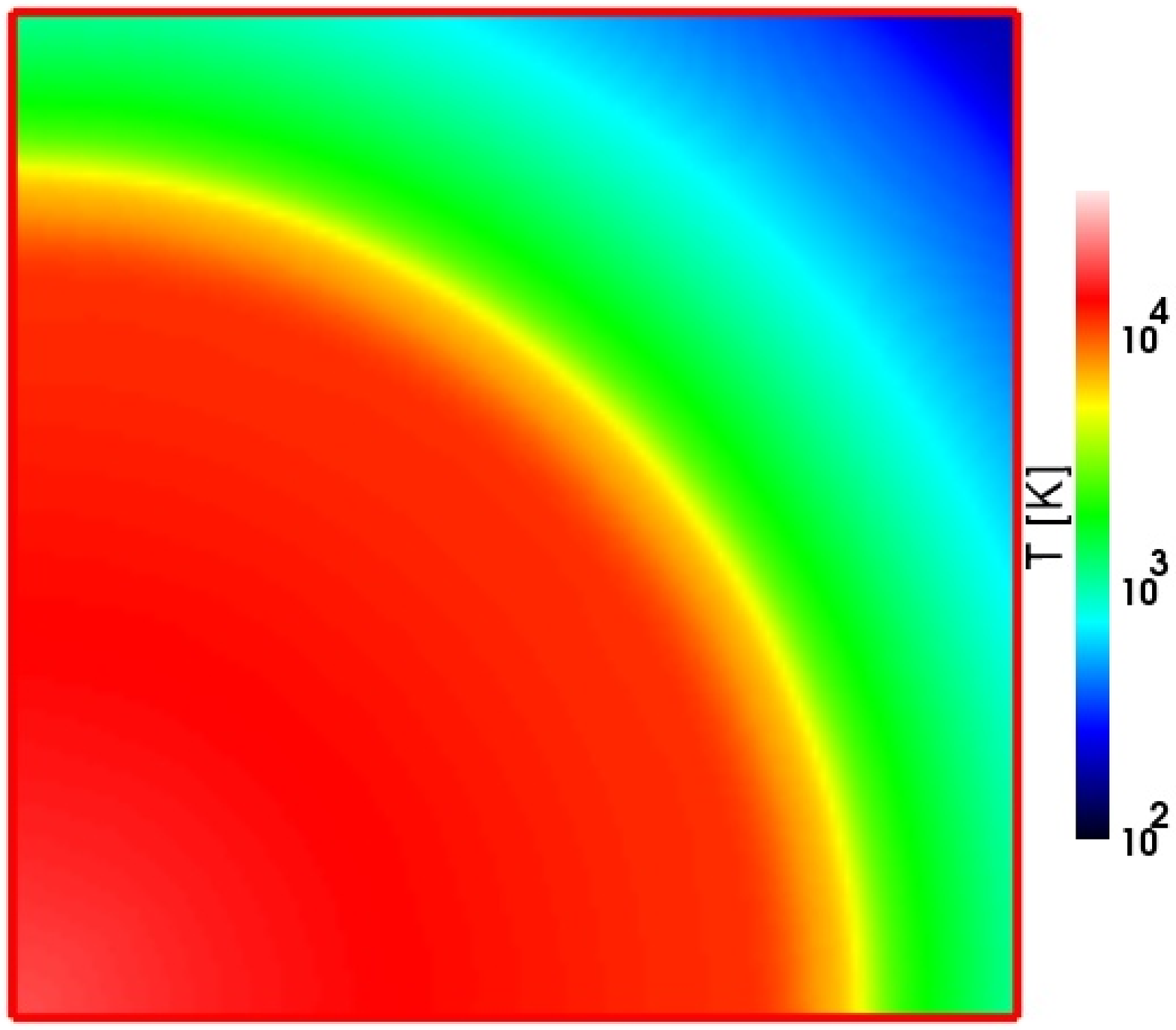}
  \includegraphics[width=2.3in]{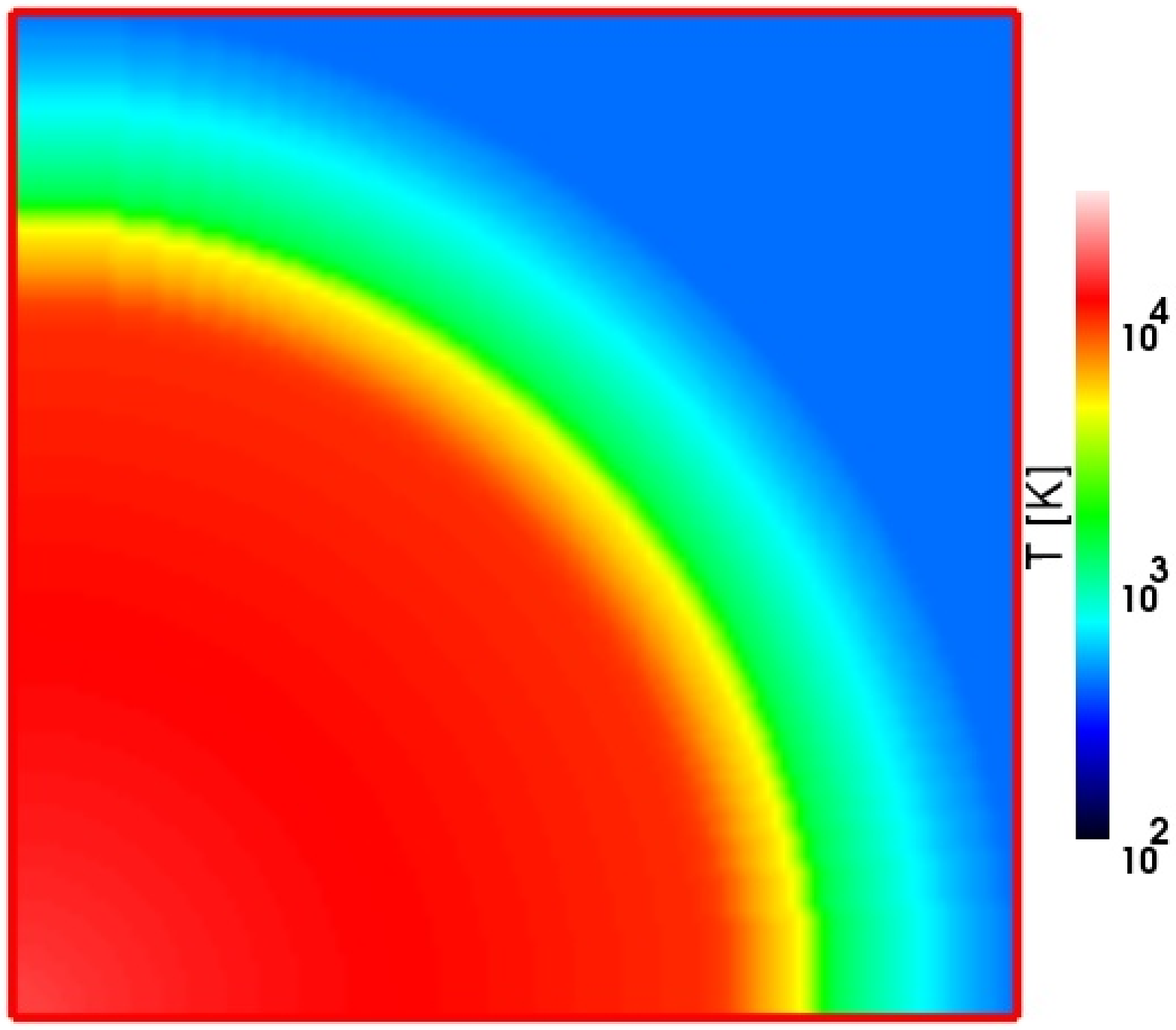}
  \includegraphics[width=2.3in]{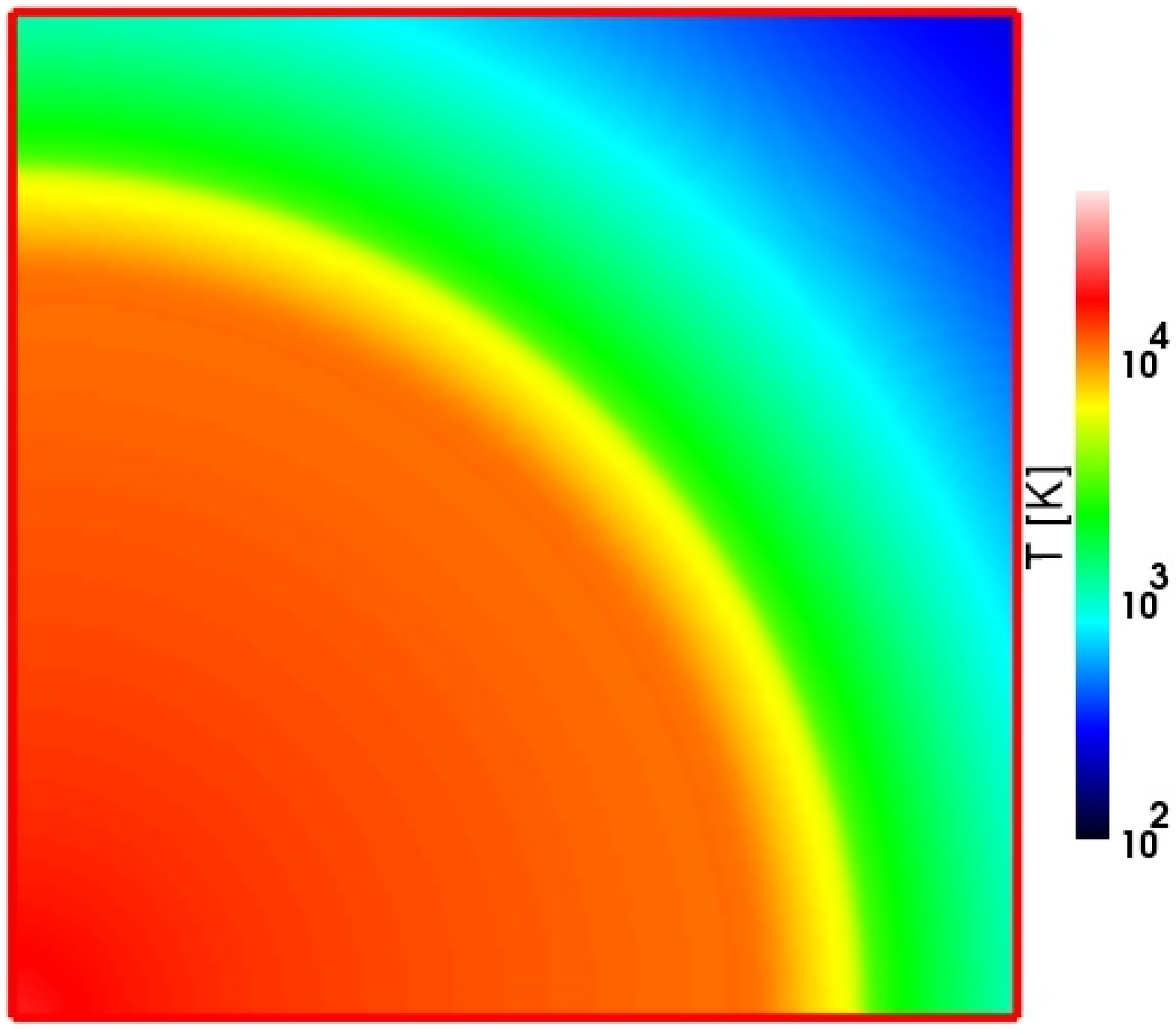}
  \includegraphics[width=2.3in]{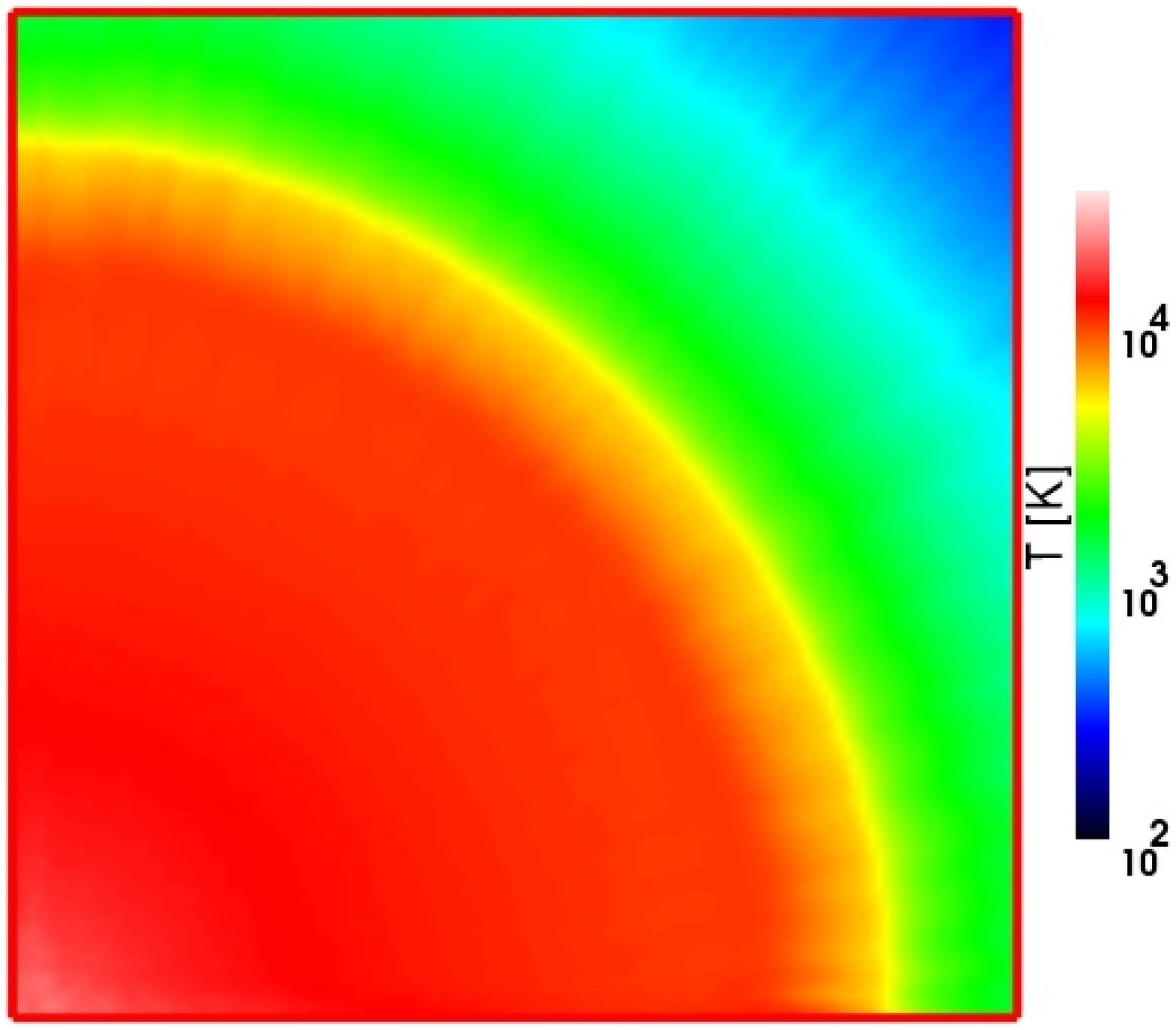}
  \includegraphics[width=2.3in]{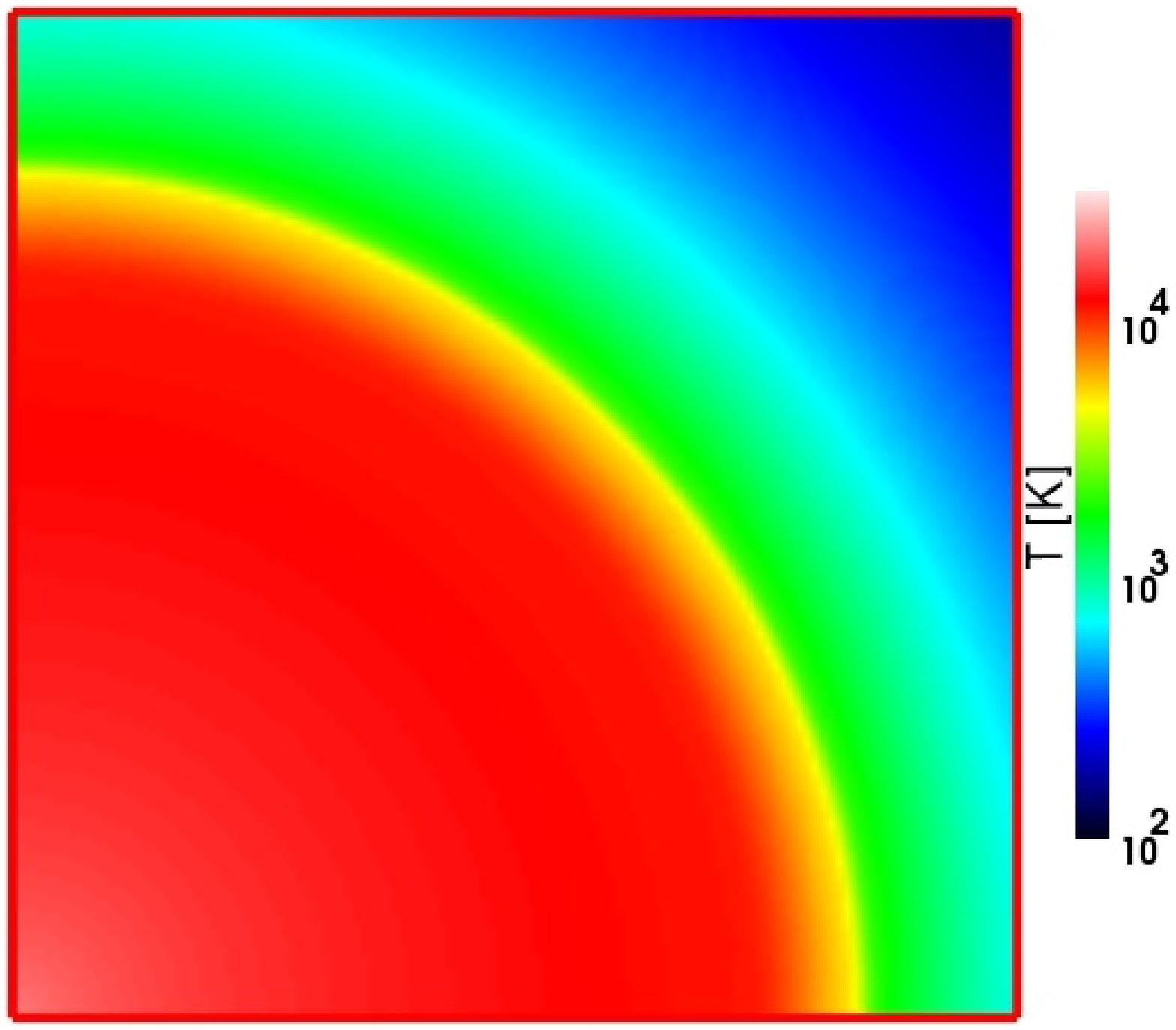}
  \includegraphics[width=2.3in]{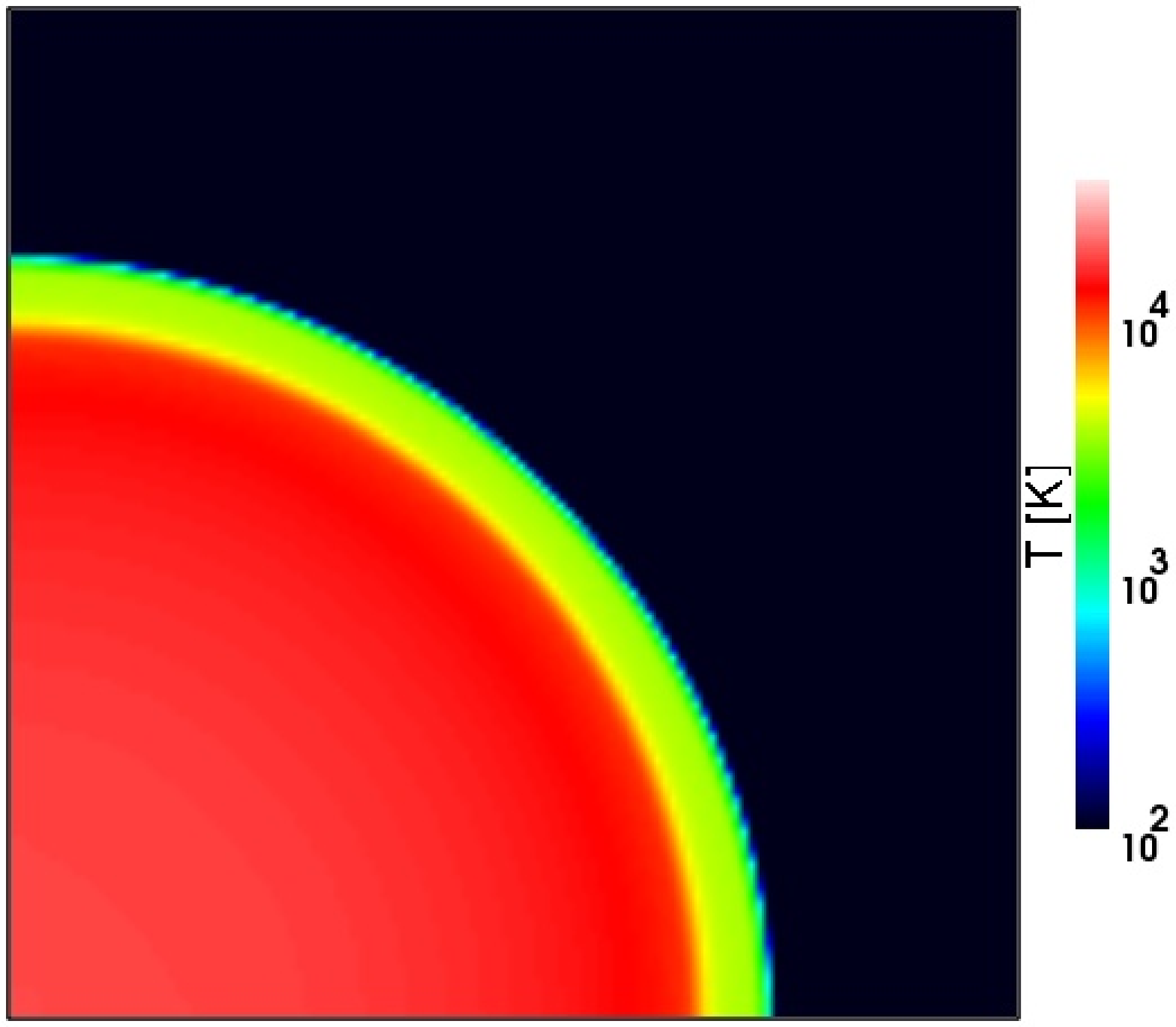}
\caption{Test 5 (H~II region expansion in an initially-uniform 
  gas): Images of the temperature, cut through the simulation volume at
  coordinate $z=0$ at time $t=500$ Myr for (left to right and top to bottom)
  Capreole+$C^2$-Ray, HART, RSPH, ZEUS-MP, RH1D, LICORICE, Flash-HC and Enzo-RT.
\label{T5_images5_T_fig}}
\end{center}
\end{figure*}

The coupled implicit radiation-chemistry system further includes a gas
energy feedback field, which allows us to self-consistently heat and
cool the gas in an operator-split manner, capturing all of
the stiff components involved in radiation transport, primordial 
chemistry and thermal heating/cooling in a tightly-coupled implicit
system. Enzo's explicit Eulerian hydrodynamics solver and its parallel 
implementation have been exhaustively described elsewhere \citep{
2007arXiv0705.1556N}. Parallelism of the coupled implicit system
follows a standard domain-decomposition approach and is solved using
state-of-the-art Newton-Krylov-Multigrid solvers \citep{KnollKeyes2004}, 
potentially allowing scalability of the algorithm to up to tens of 
thousands of processors.

While Enzo allows for spatial adaptivity through structured adaptive 
mesh refinement (SAMR), our initial implementation of Enzo-RT is 
currently limited to uniform grids in 1-, 2- or 3-dimensions, although 
their upgrade to AMR is under development. Extensions of this approach to 
variable Eddington tensors, multigroup flux-limited diffusion, or 
multigroup variable Eddington tensors are easily accommodated within 
our implicit formulation and are planned as future extensions. 
One benefit is that the timestep is 
independent of grid resolution, at least for the radiation solve. Another 
advantage of our approach is that by defining radiation as a 
field variable, scalability with respect to the number of point sources 
ceases to be an issue.  Instead, scalability is dictated by the underlying 
linear system solver, which for the case of multigrid is optimal.  

\section{Radiation Hydrodynamics Tests: Description}

For simplicity and inclusivity (since currently not all codes have 
implemented helium or metals chemistry and cooling) all tests assume 
the gas to be composed of pure hydrogen.

\subsection{Test 5: Classical H~II Region Expansion} 
\label{sec:T5}

\begin{figure*}
\begin{center}
  \includegraphics[width=2.3in]{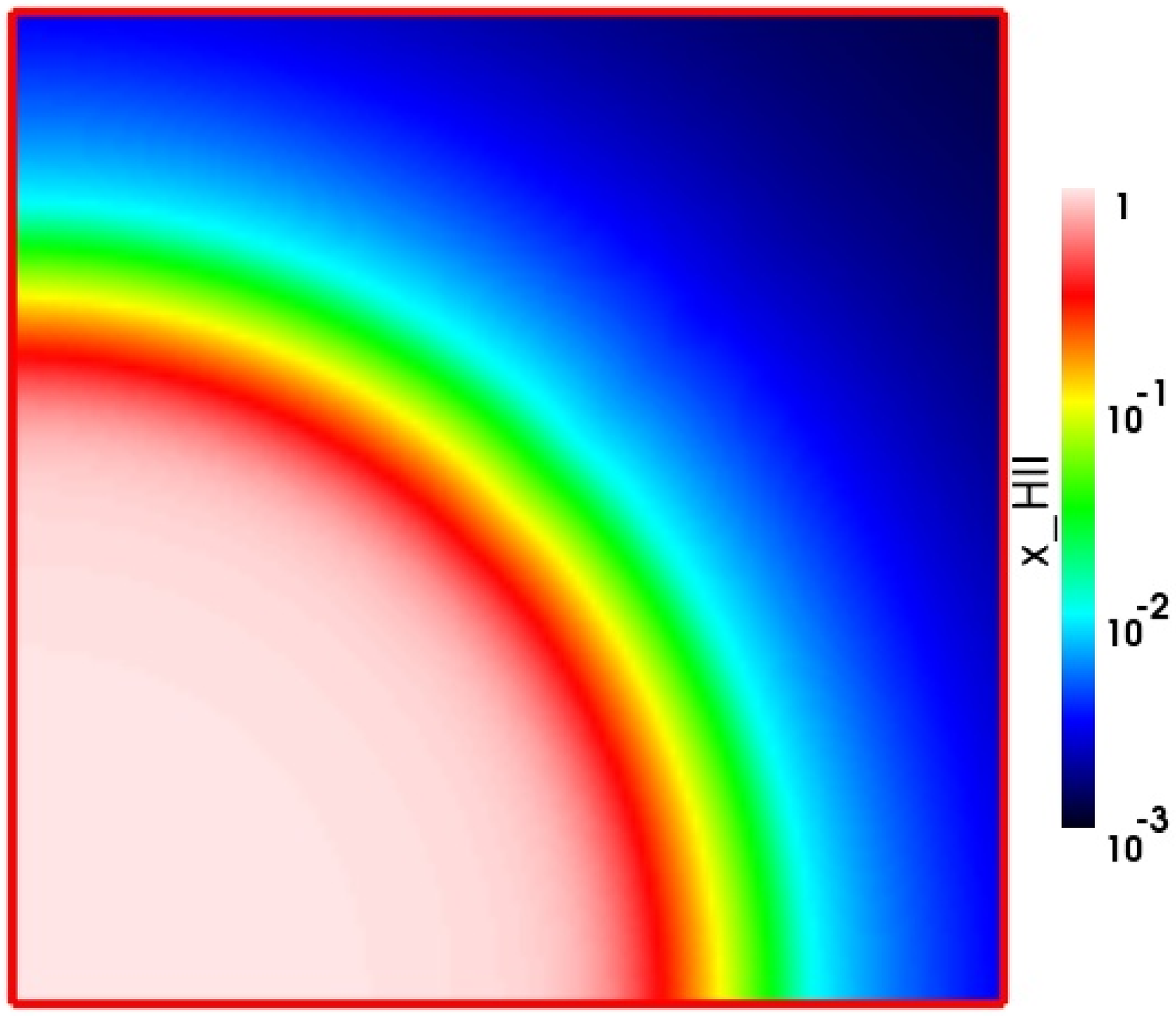}
  \includegraphics[width=2.3in]{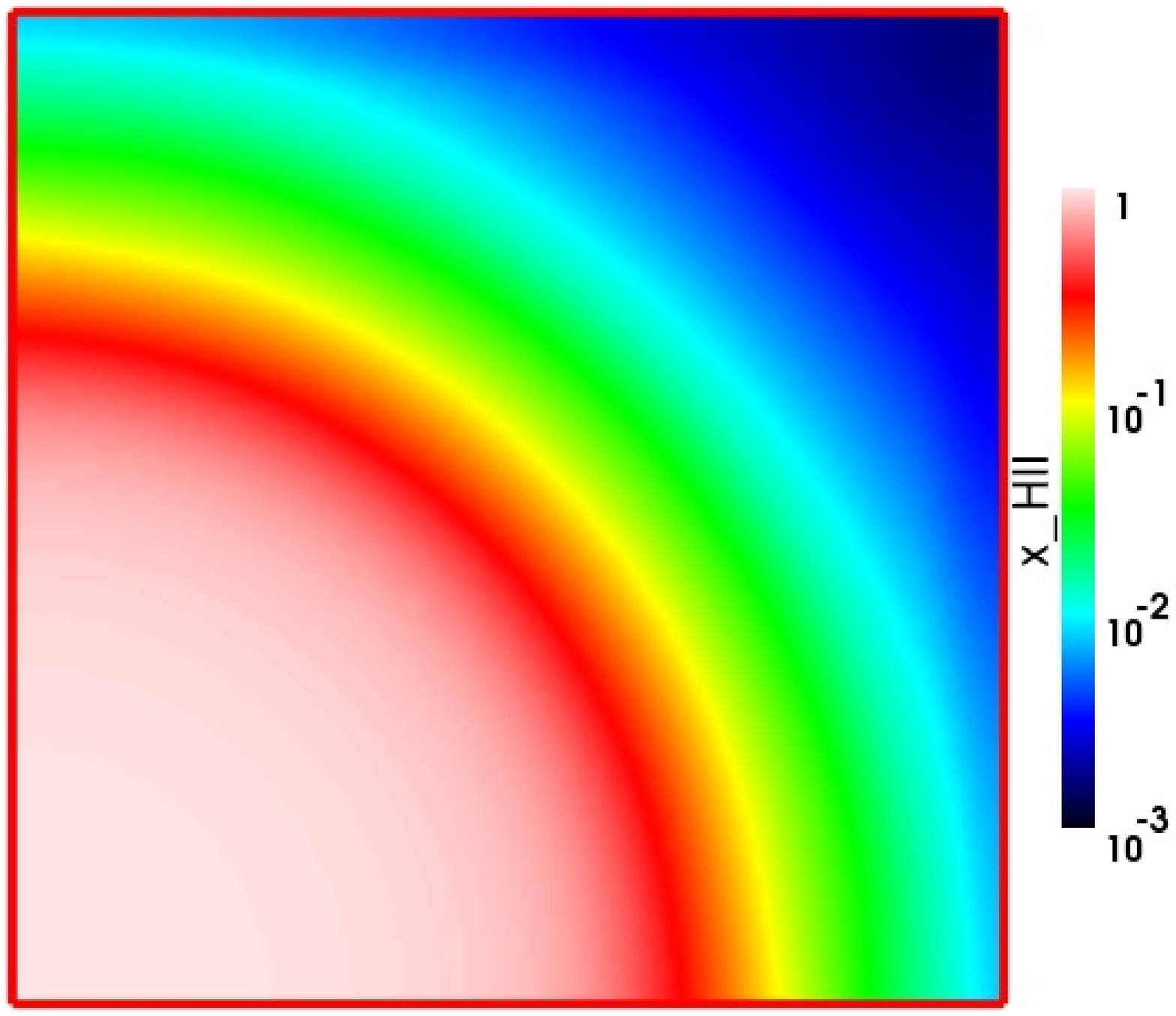}
  \includegraphics[width=2.3in]{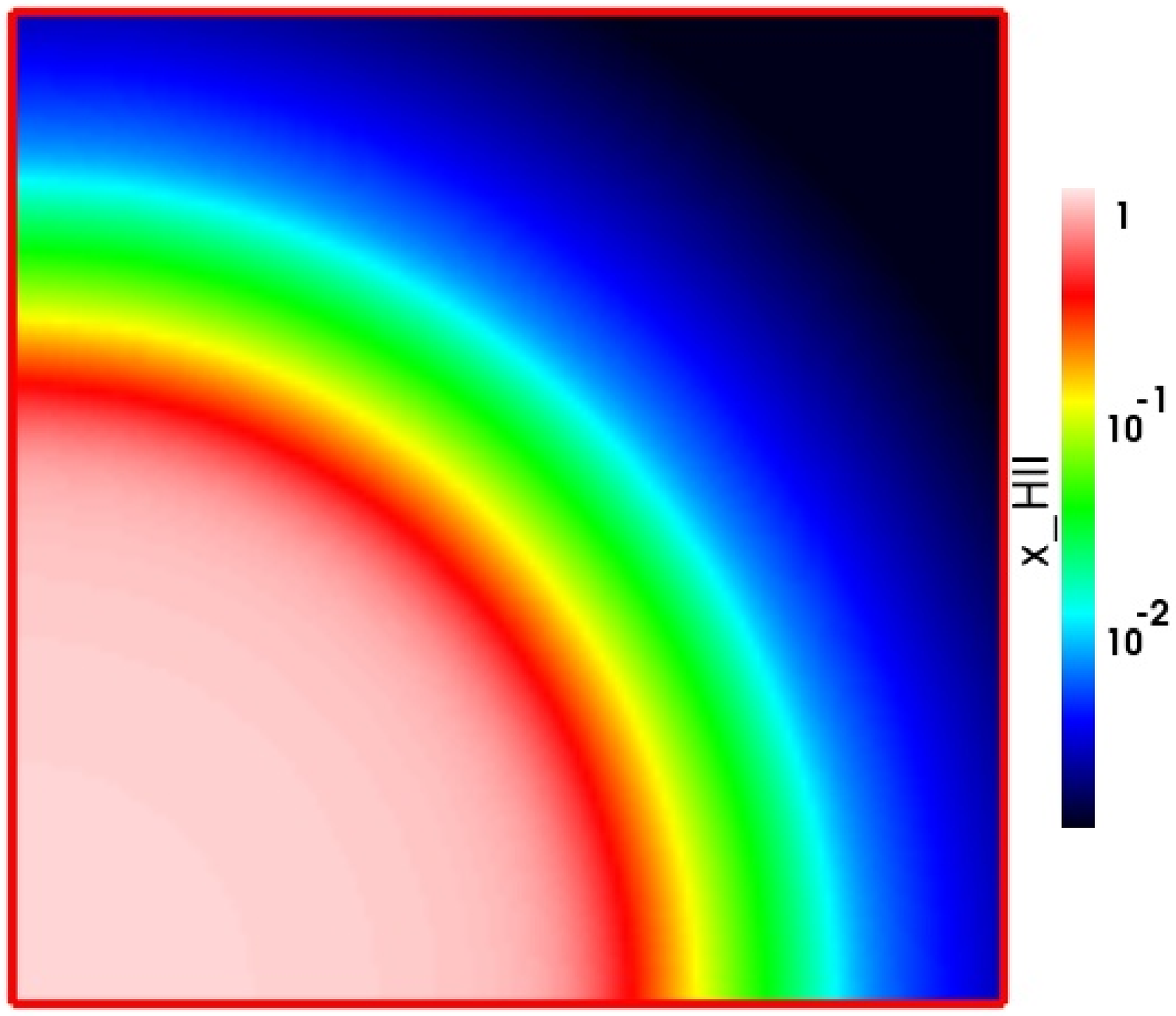}
  \includegraphics[width=2.3in]{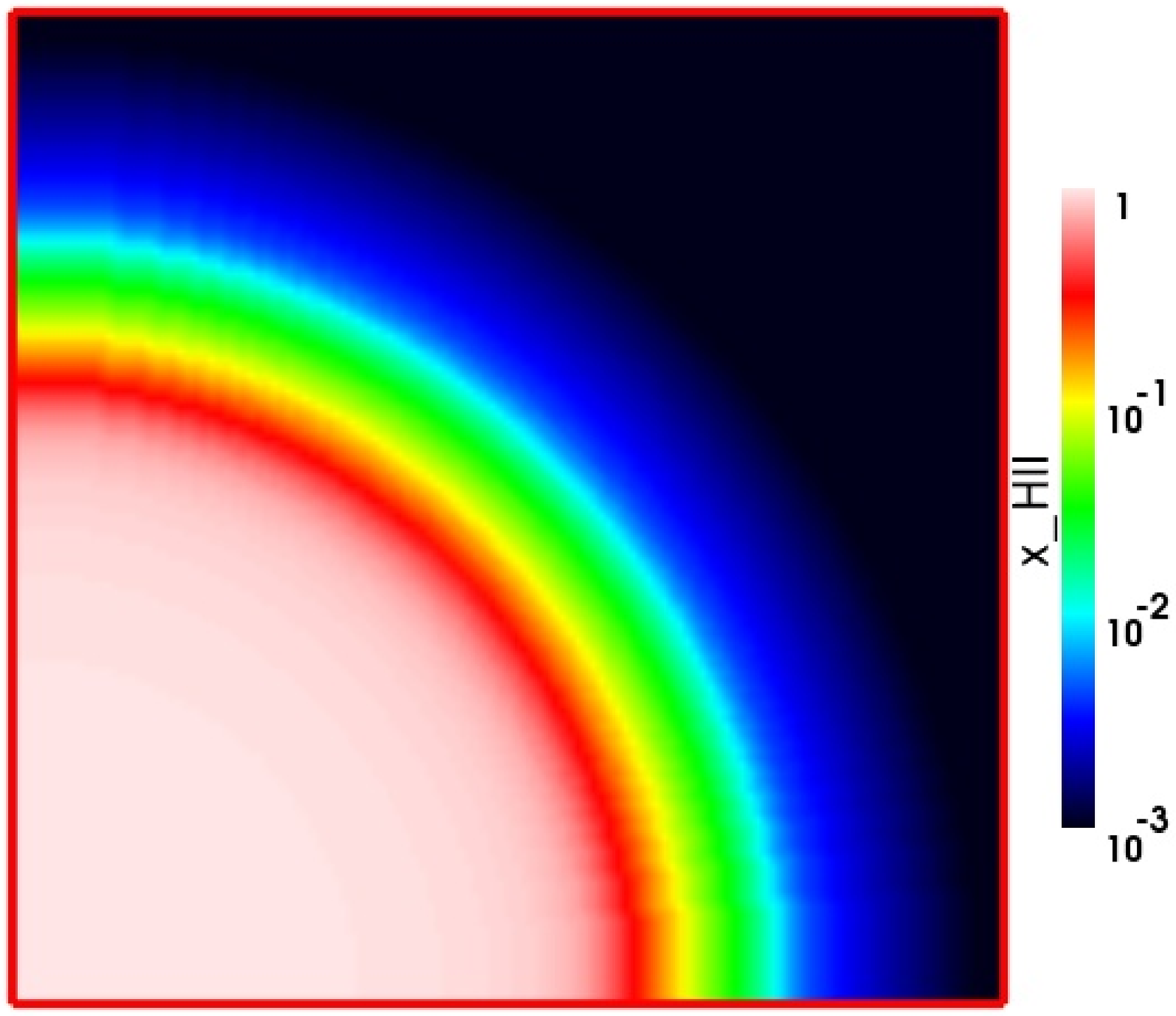}
  \includegraphics[width=2.3in]{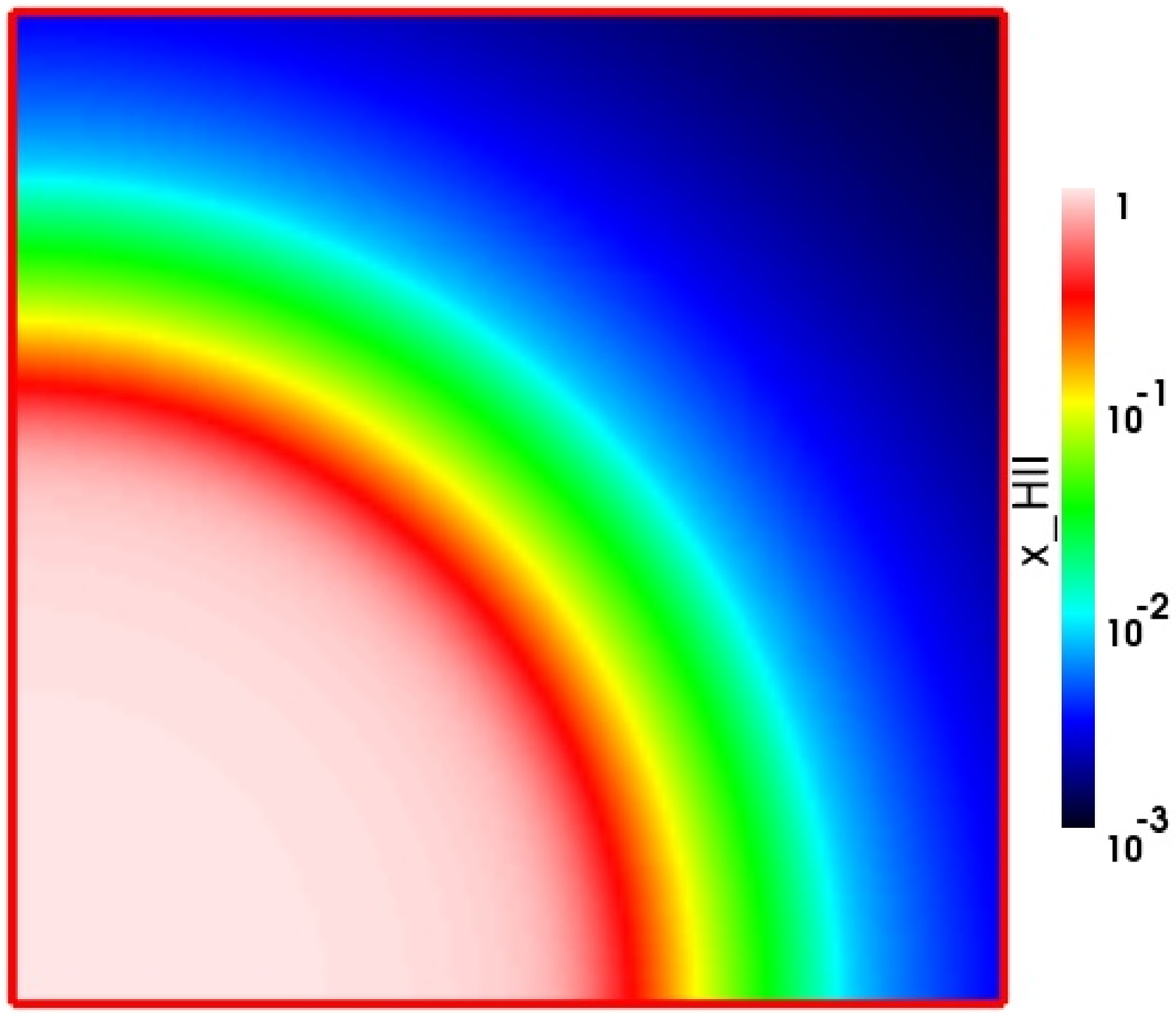}
  \includegraphics[width=2.3in]{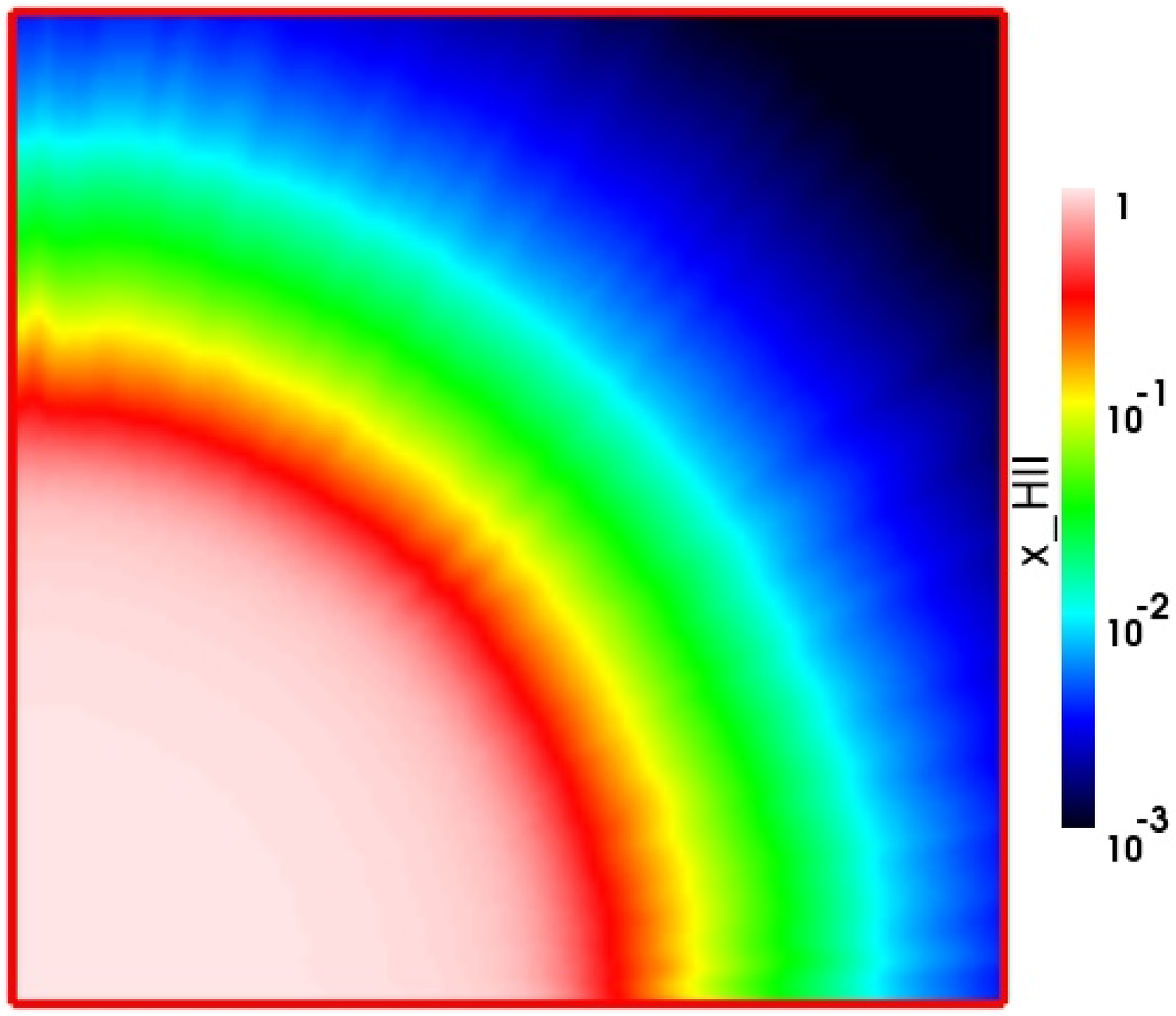}
  \includegraphics[width=2.3in]{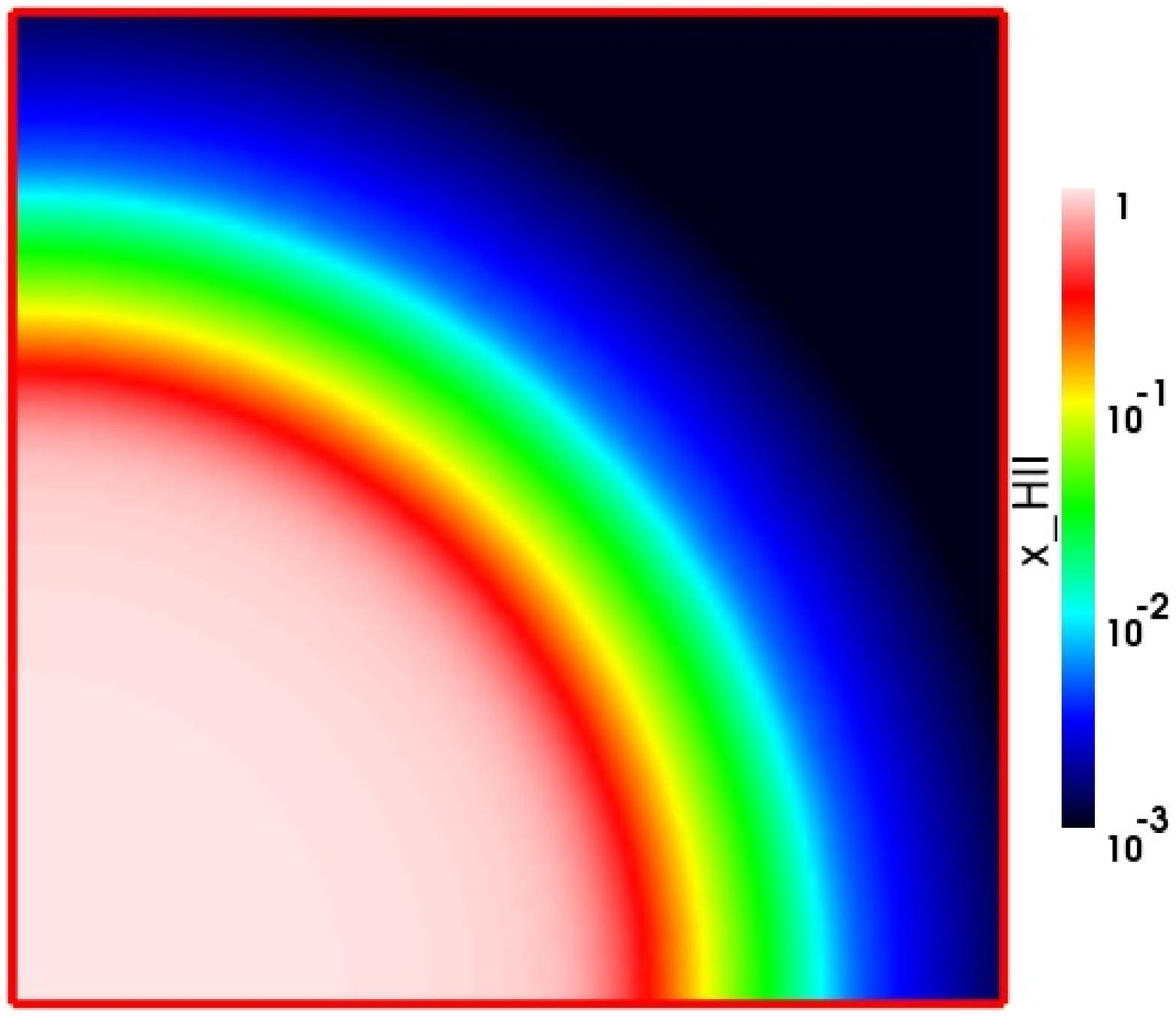}
  \includegraphics[width=2.3in]{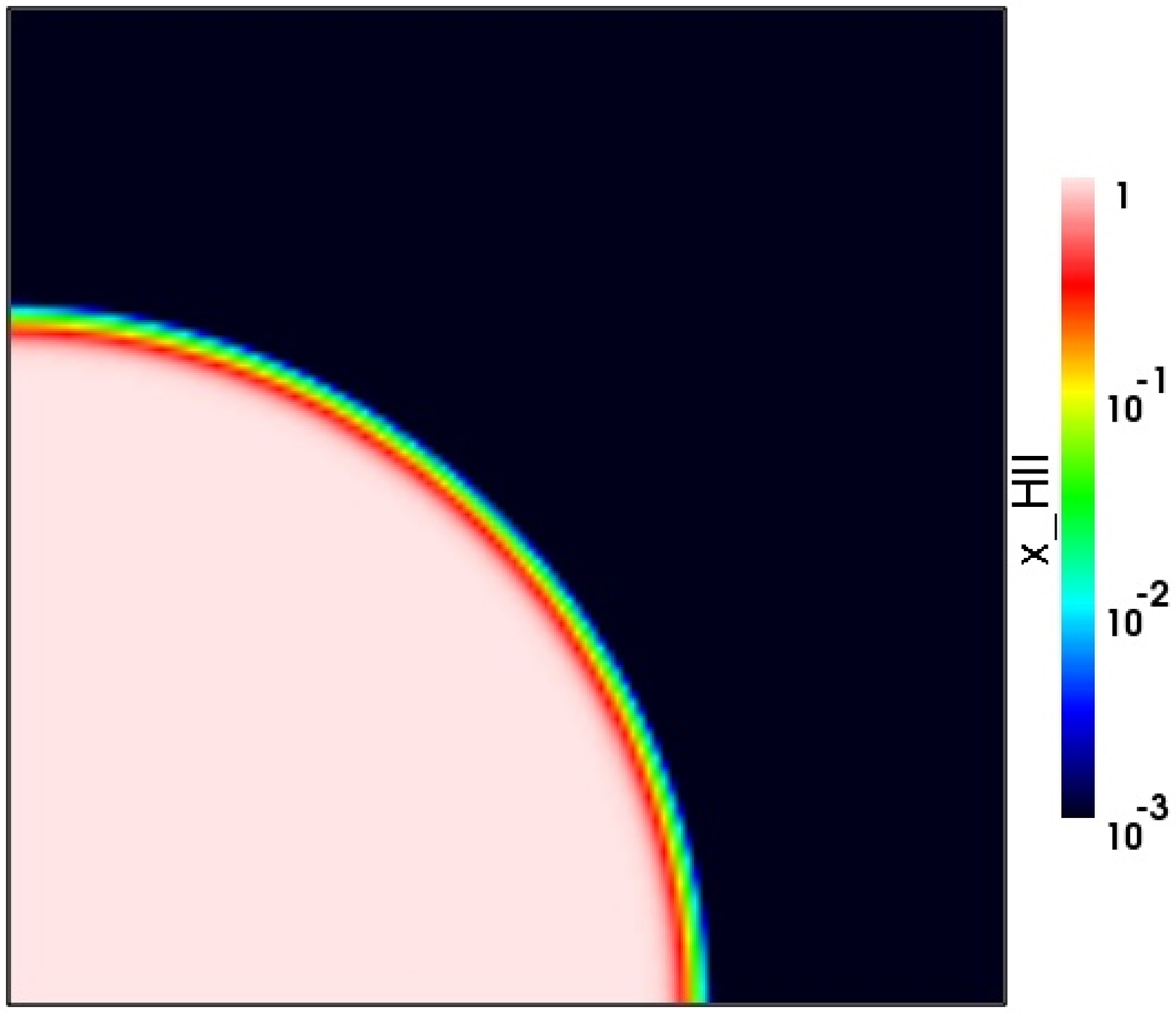}
\caption{Test 5 (H~II region expansion in an initially-uniform 
  gas): Images of the H~II fraction, cut through the simulation volume at
  coordinate $z=0$ at time $t=500$ Myr for (left to right and top to bottom)
  Capreole+$C^2$-Ray, HART, RSPH, ZEUS-MP, RH1D, LICORICE, Flash-HC and Enzo-RT.
\label{T5_images5_HII_fig}}
\end{center}
\end{figure*}

Test 5 is the classical problem of the expansion of an I-front due to a point 
source in an initially uniform-density medium. In general, I-fronts are 
classified according to their speed with respect to the gas and the change 
in gas density through the 
I-front \citep[c.f.][]{1965ARA&A...3...47K,1978ppim.book.....S}. There are two 
critical speeds: R-critical, defined as $v_R=2c_{s,I,2}$, and D-critical, given 
by $v_D=c_{s,I,2}-(c_{s,I,2}^2-c_{s,I,1}^2)^{1/2}\approx c_{s,I,1}^2/(2c_{s,I,2})$, 
where $c_{s,I,1}=(p_1/\rho_1)^{1/2}$ and $c_{s,I,2}=(p_2/\rho_2)^{1/2}$ are the 
{\it isothermal} sound speeds in the gas ahead of and behind the I-front, 
respectively. Note that in the test the gas is {\it not} assumed to be isothermal. 
The velocity of the I-front is given by the jump condition $v_I=F/n$ (which 
guarantees photon 
conservation), where $n$ is the number density of the neutral gas entering the 
front and $F$ is the flux of ionizing photons at the I-front transition (which 
is attenuated due to absorptions in the gas on the source side). We note that 
this jump condition is modified significantly for I-fronts moving with 
relativistic speeds with respect to the gas \citep{2006ApJ...648..922S}.
This can occur in a number of astrophysical and cosmological environments.
However, we do not consider such cases here since currently few radiative
transfer codes (and no radiation hydrodynamics codes, to our knowledge) are
able to handle such relativistic I-fronts.

When $v_{\rm I}\geq v_R$ ($\eg$ close to the source, where the flux $F$ is
large) the I-front is R-type (R-critical when $v_{\rm I}=v_R$).  R-type 
I-fronts always move supersonically with respect to the neutral gas ahead, 
while with respect to the ionized gas the front can move either subsonically 
(strong R-type, highly compressive, but generally irrelevant to H II regions 
since it means that the isothermal sound speed behind the front is lower than 
the one ahead of it), or supersonically (weak R-type, resulting in only slight 
compression of the gas moving through the front). When $v_{\rm I}\leq v_D$, the 
I-front is D-type (D-critical in the case that $v_{\rm I}=v_D$). The gas passing 
through this type of I-front always expands, and the front is subsonic with 
respect to the gas beyond. With respect to the ionized gas, the I-front can 
again be either supersonic (strong D-type), or subsonic (weak D-type). When 
$v_D< v_{\rm I}<v_R$ (sometimes referred to as an M-type I-front) the I-front 
is necessarily led by a shock which compresses the gas entering the I-front 
sufficiently to slow it down and guarantee that $v_{\rm I}\leq v_D$.

\begin{figure*}
\begin{center}
  \includegraphics[width=2.3in]{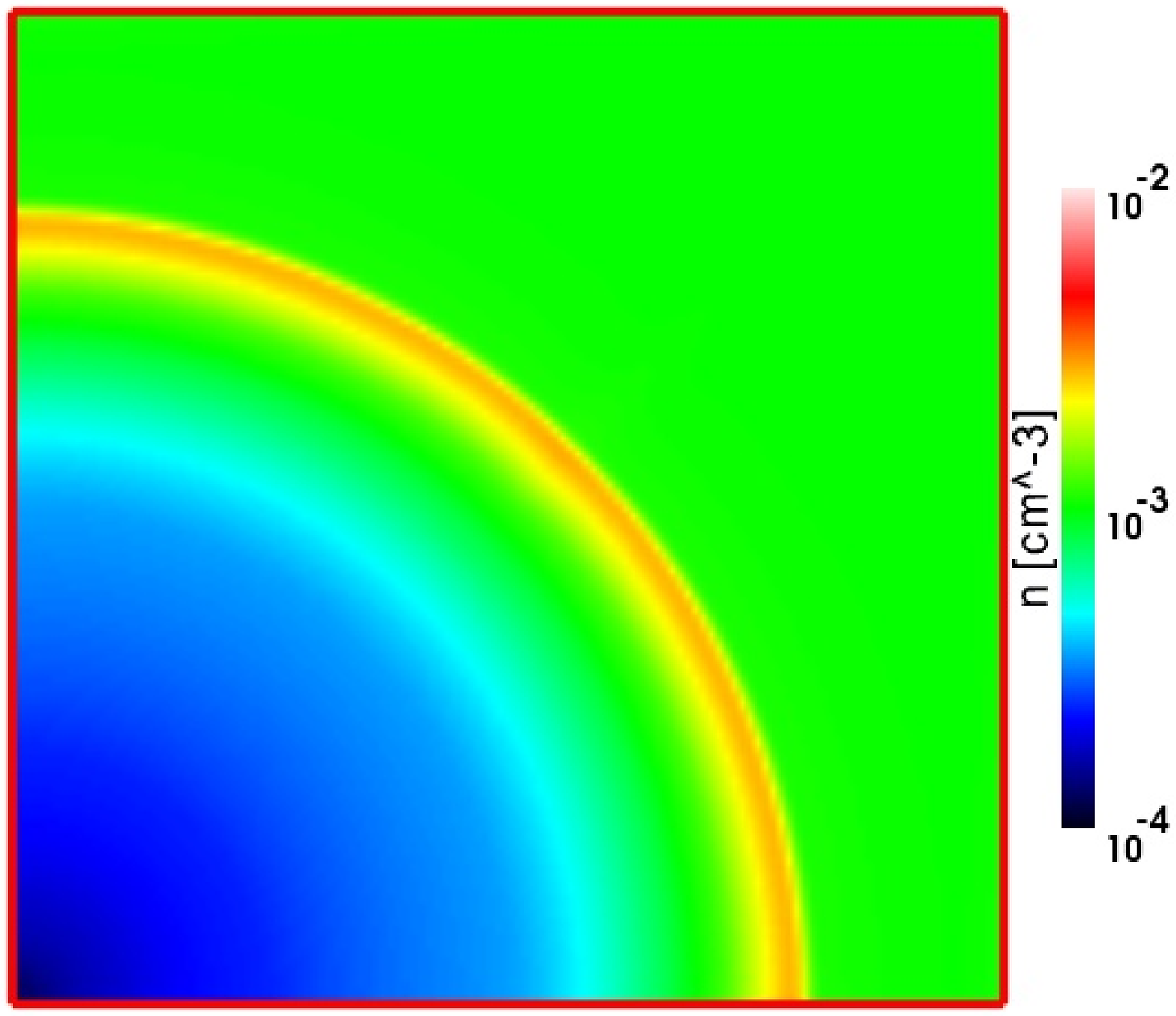}
  \includegraphics[width=2.3in]{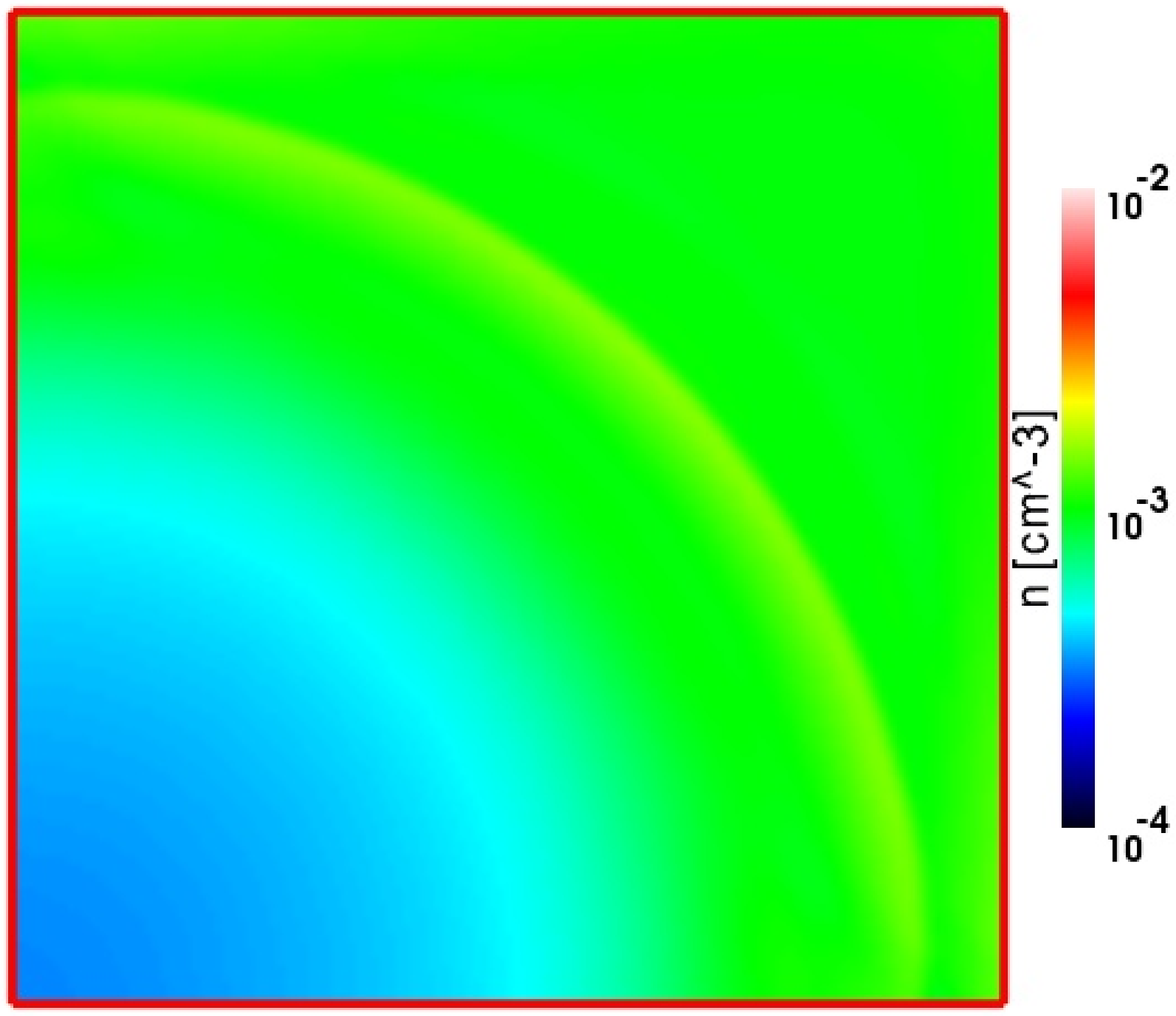}
  \includegraphics[width=2.3in]{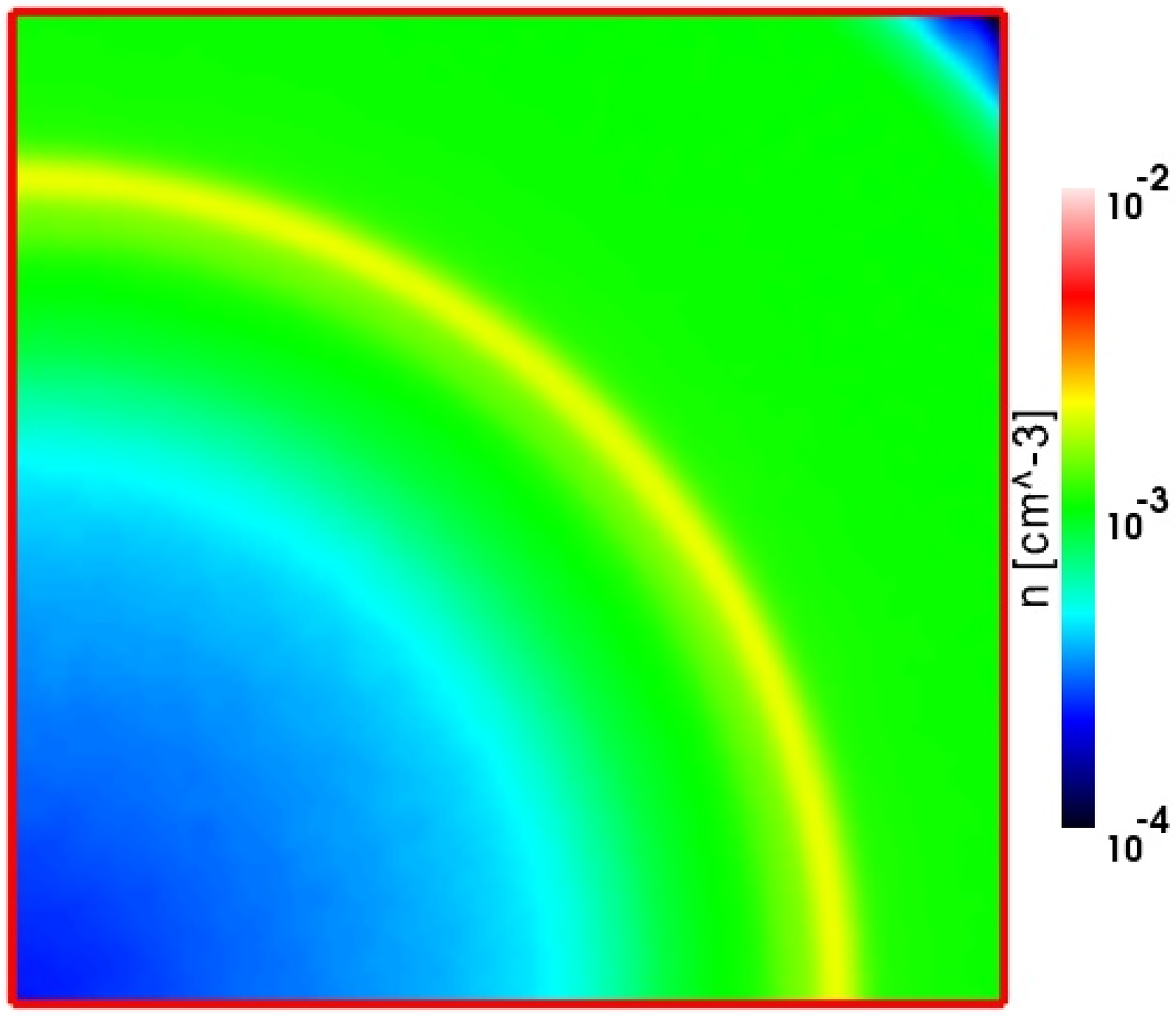}
  \includegraphics[width=2.3in]{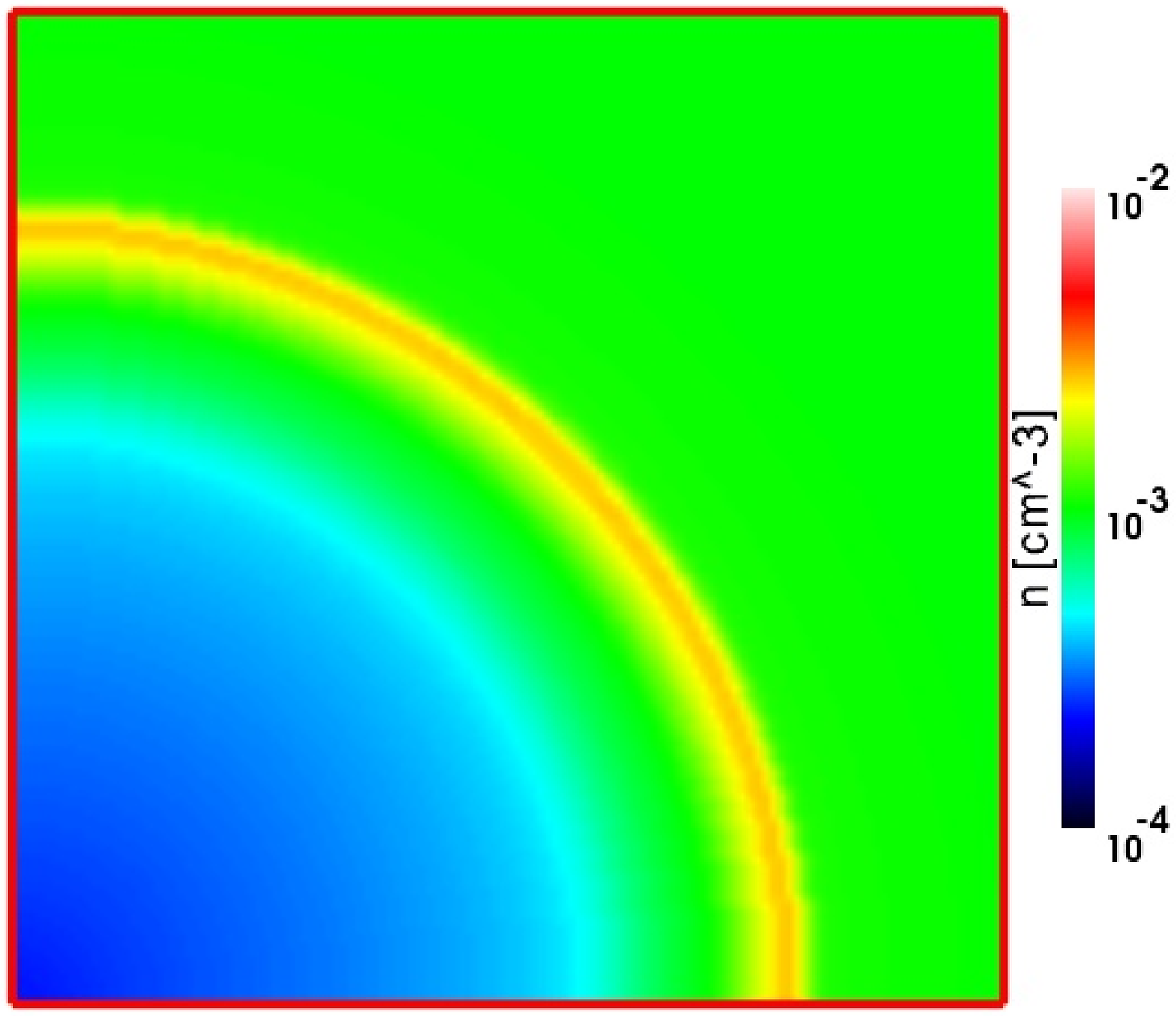}
  \includegraphics[width=2.3in]{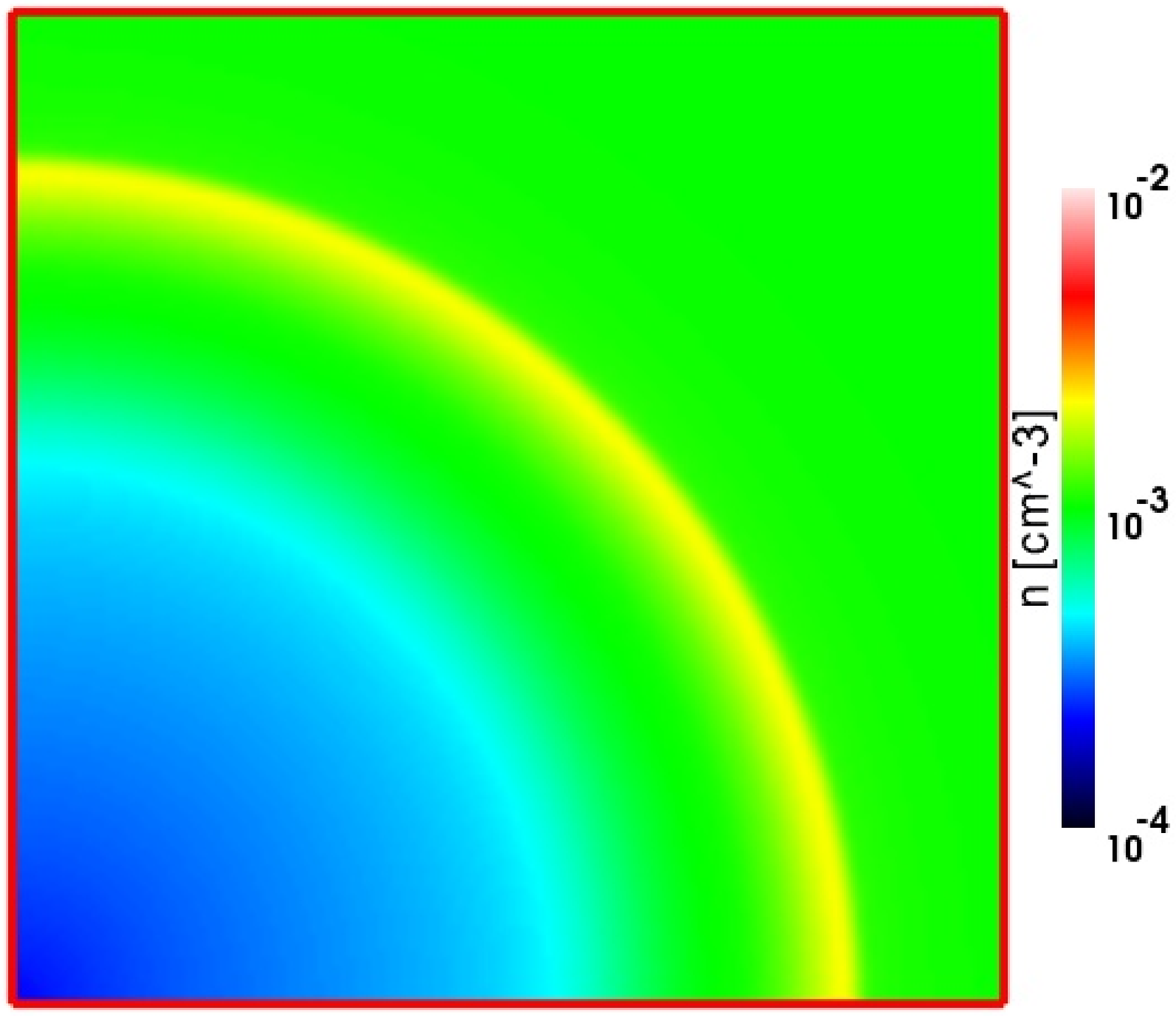}
  \includegraphics[width=2.3in]{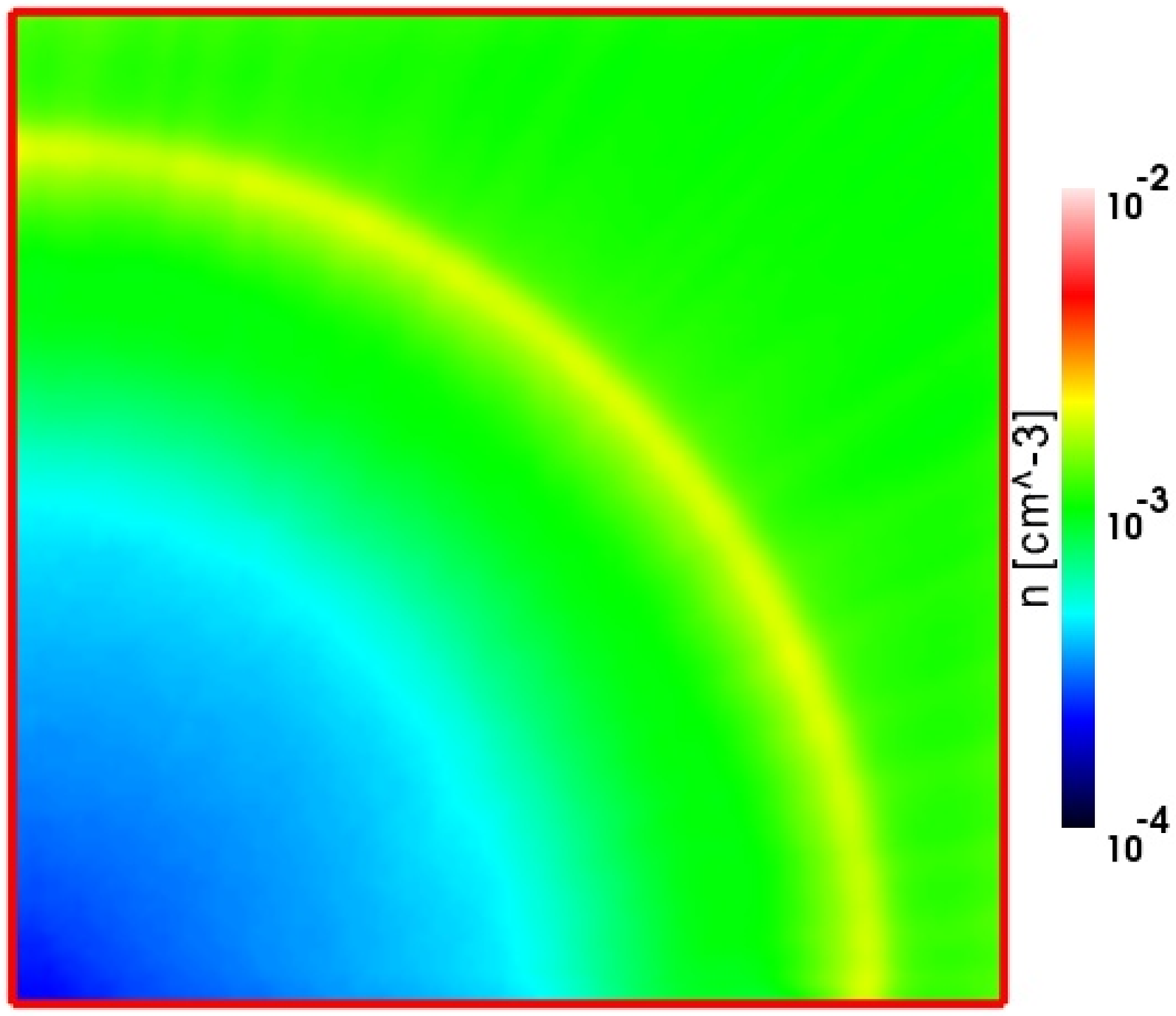}
  \includegraphics[width=2.3in]{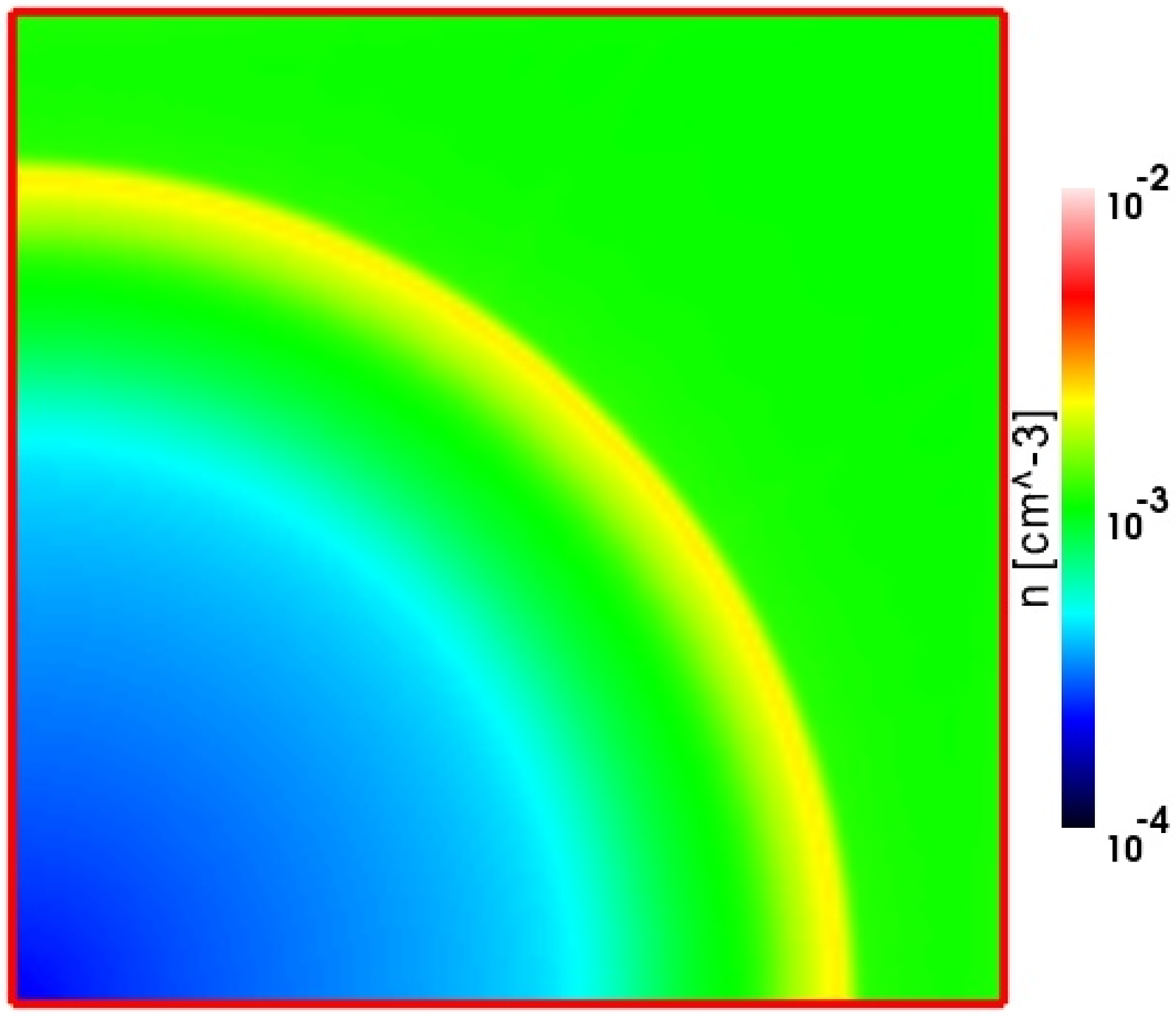}
  \includegraphics[width=2.3in]{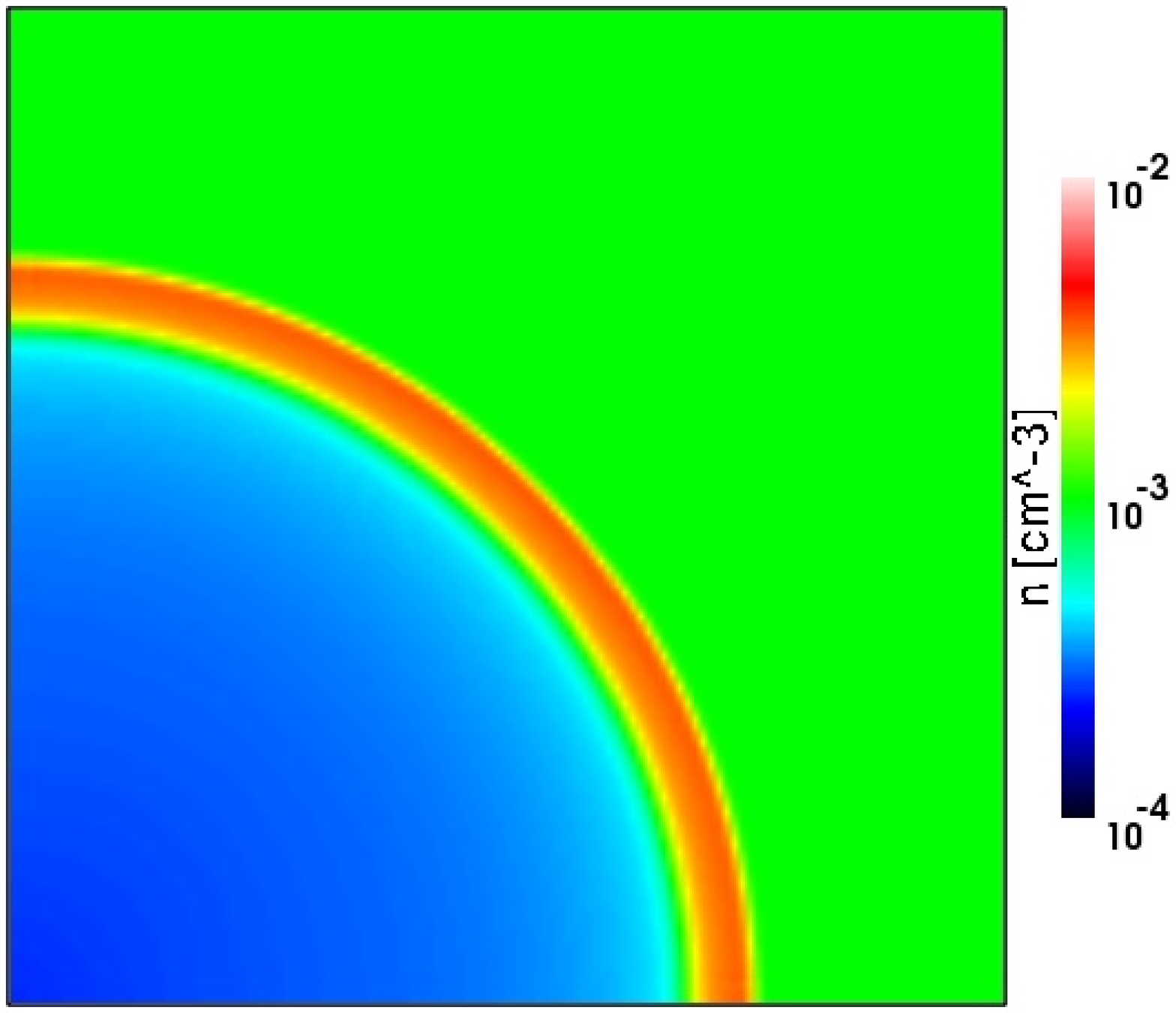}
\caption{Test 5 (H~II region expansion in an initially-uniform 
  gas): Images of the gas number density, cut through the simulation volume at
  coordinate $z=0$ at time $t=500$ Myr for (left to right and top to bottom)
  Capreole+$C^2$-Ray, HART, RSPH, ZEUS-MP, RH1D, LICORICE, Flash-HC and Enzo-RT.
\label{T5_images5_n_fig}}
\end{center}
\end{figure*}

In a static medium with number density $n_{\rm H}$ and constant ionized gas 
temperature $T$, the evolution of the I-front radius $r_{\rm I}$ and velocity 
$v_{\rm I}$ for a point source emitting $\dot{N}_\gamma$ ionizing photons per 
second are given by
\ba 
r_{\rm I}&=&r^0_{\rm S}\left[1-\exp(-t/t_{\rm rec})\right]^{1/3}\,\label{strom0}\\
\rm v_I&=&\frac{r_{\rm S}}{3t_{\rm rec}}\frac{\exp{(-t/t_{\rm rec})}} 
    {\left[1-\exp(-t/t_{\rm rec})\right]^{2/3}}\,,\label{strom01}
\ea 
where
\begin{equation}
r^0_{\rm S}=\left[{3\dot{N}_\gamma\over 4\pi \alpha_B(T)n_{\rm H}^2}\right]^{1/3}\,,
\label{strom_rad}
\end{equation}
the Str\"omgren radius (assuming full ionization), which is reached when
the number of recombinations in the ionized volume per unit time exactly
balances the number of ionizing photons emitted by the source per unit time.
This final static stage is commonly referred to as Str\"omgren sphere. 
The recombination time is given by
\begin{equation}
t_{\rm rec}=\left[\alpha_B(T) n_{\rm H}\right]^{-1}\,.
\end{equation}
Here $\alpha_B(T)$ is the case B recombination coefficient of hydrogen at
temperature $T$ in the ionized region.

In reality, the ionized gas is not static and its much higher pressure than 
that of the ambient medium causes it to expand outward beyond the Stromgren 
radius. Analytical models predict that in this phase the I-front radius 
evolves approximately according to \citep[c.f.][]{1978ppim.book.....S} 
\be 
r_I=r^0_{\rm S}\left(1+\frac{7c_st}{4r^0_{\rm S}}\right)^{4/7},
\label{rI_p_driven} 
\ee 
where $r^0_S$ is the Str\"omgren radius and $c_s$ is the sound speed in the 
ionized gas. The expansion finally stalls when a pressure equilibrium is 
reached. The predicted final H~II region radius is 
\be 
r_f=\left(\frac{2T}{T_e}\right)^{2/3}r^0_S,
\ee 
where $T$ is the temperature inside the H~II region and $T_e$ is the
external temperature. In reality, the evolution is more complicated,
with non-uniform temperatures inside the H~II region, broadened I-fronts due
to pre-heating by energetic photons, etc. Furthermore, equation~\ref{rI_p_driven}  
describes correctly only in the purely pressure-driven, late-time
evolution, but not the transition from fast, R-type to D-type I-front.
These analytical solutions should therefore only be considered to be 
guidelines for the expected behaviour, not as exact solutions for this problem.

The numerical parameters for Test 5 are as follows: computational box size
$L=15$~kpc, initial gas number density $n_H=10^{-3}$ cm$^{-3}$, initial 
ionization fraction $x=0$, constant ionizing photon emission rate 
$\dot{N}_\gamma=5\times10^{48}\,\rm s^{-1}$, initial gas velocity zero and 
initial gas temperature $T_e=100$~K. The radiation source is at the 
$(x_s,y_s,z_s) = (0,0,0)$ corner of the computational box. For reference, 
if we assume that the temperature of the ionized gas is $T=10^4$~K, and that 
the recombination rate is given by $\alpha_B(T)=2.59\times10^{-13}\rm\,cm^{3}s^{-1}$, 
we find $t_{\rm rec}=3.86\times10^{15}\,\rm s=122.4$~Myr, $r_S=5.4$~kpc,
and $r_f\approx185$~kpc. This rough final pressure-equilibrium radius is 
thus well outside of our computational volume, which was instead chosen to 
resolve the more physically-interesting transition from R-type to D-type, 
which occurs around $r^0_S$. Boundary conditions are reflective for the 
boundaries which contain the origin (where the ionizing source is positioned) 
and transmissive for the other boundaries. The ionizing spectrum is that of a 
$10^5$~K  black body, as expected for a massive, metal-free Pop~III star. 
Hydrogen line cooling, recombinational cooling, and bremsstrahlung cooling 
are all included, but not Compton cooling. The simulation running time is 
$t_{\rm sim}=500$ Myr $\approx4\,t_{\rm rec}$.  The required outputs are 
the neutral fraction of hydrogen, gas pressure, temperature and Mach number 
on the entire grid at $t=10,30,100,200,$ and 500 Myr, and the I-front position 
(defined as the position where the neutral fraction is 50\%) and I-front 
velocity vs. time along the $x$-axis.

\subsection{Test 6: H~II region expansion in $1/r^2$ density profile}

Test 6 is the propagation of an I-front created by a point source at the 
center of a spherically-symmetric, steeply-decreasing power-law density 
profile with a small flat central core of gas number density $n_0$ and 
radius $r_0$: 
\[ n_{H}(r) = \left\{ 
\begin{array}{ll}
  n_{0}                 & \mbox{if $r \leq r_{0}$} \\
  n_{0}(r/r_{0})^{-2}   & \mbox{if $r \geq r_{0}$}
                          \end{array}
                  \right.\vspace{0.1in} \]
\begin{figure*}
\begin{center}
  \includegraphics[width=2.3in]{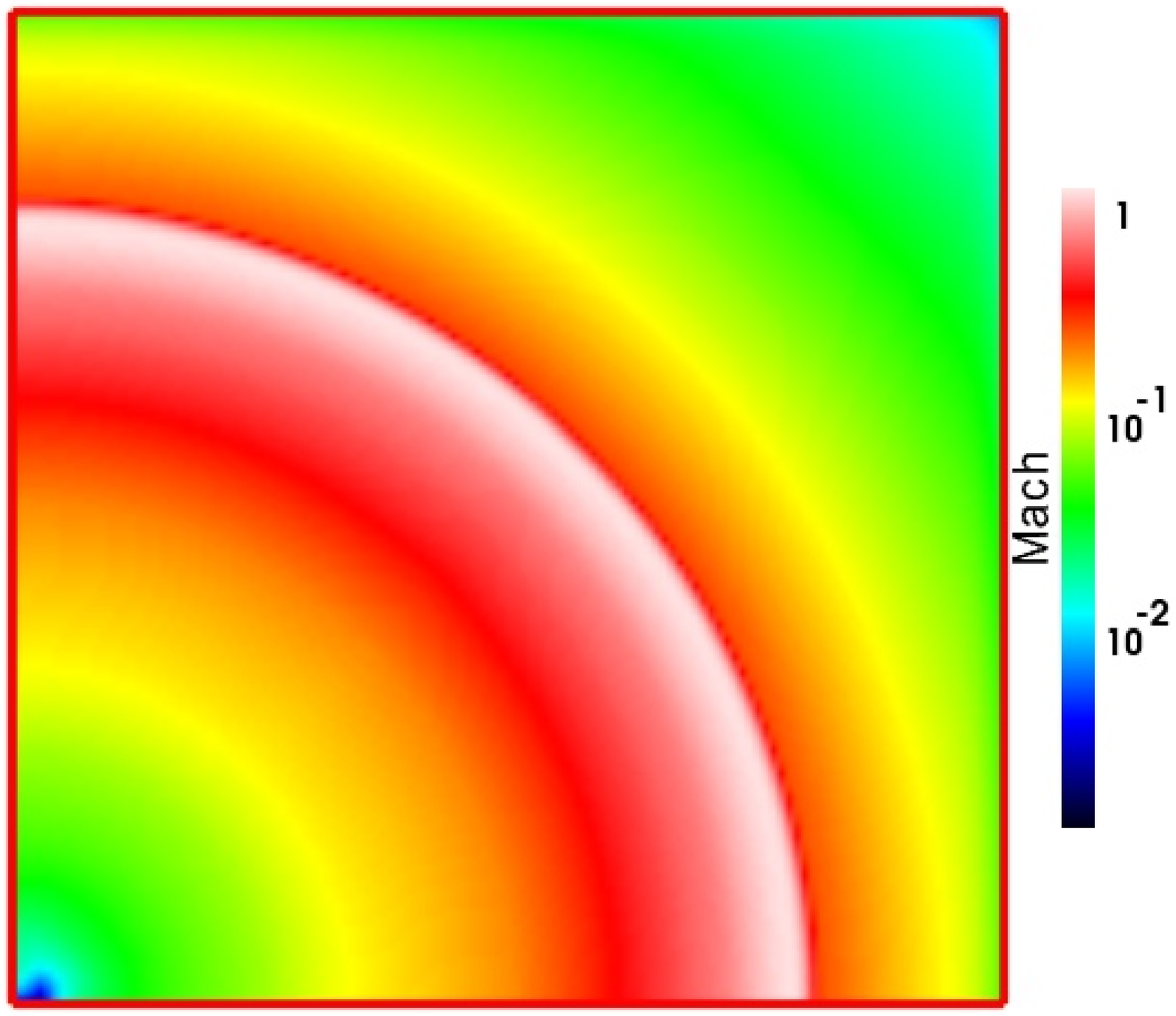}
  \includegraphics[width=2.3in]{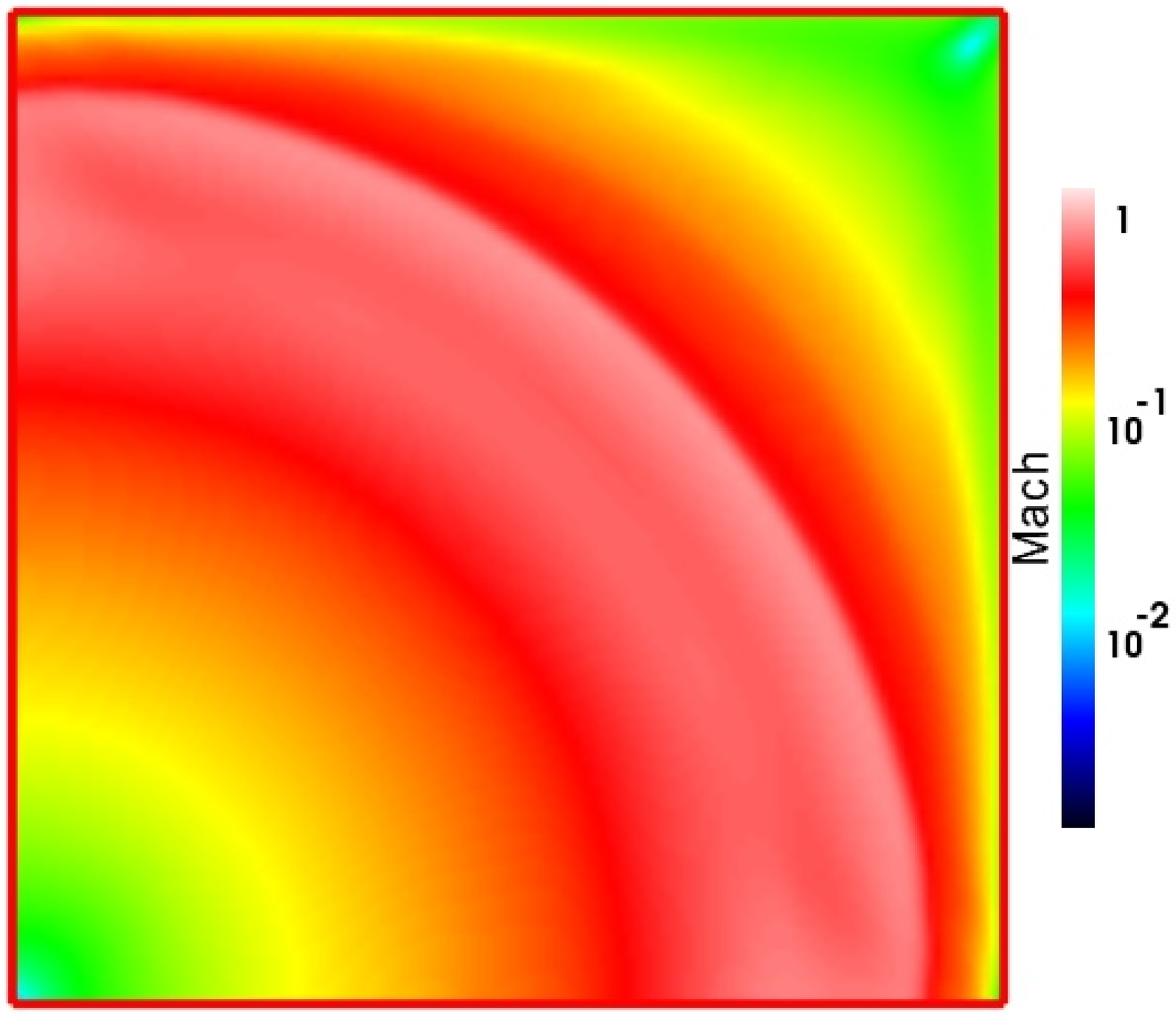}
  \includegraphics[width=2.3in]{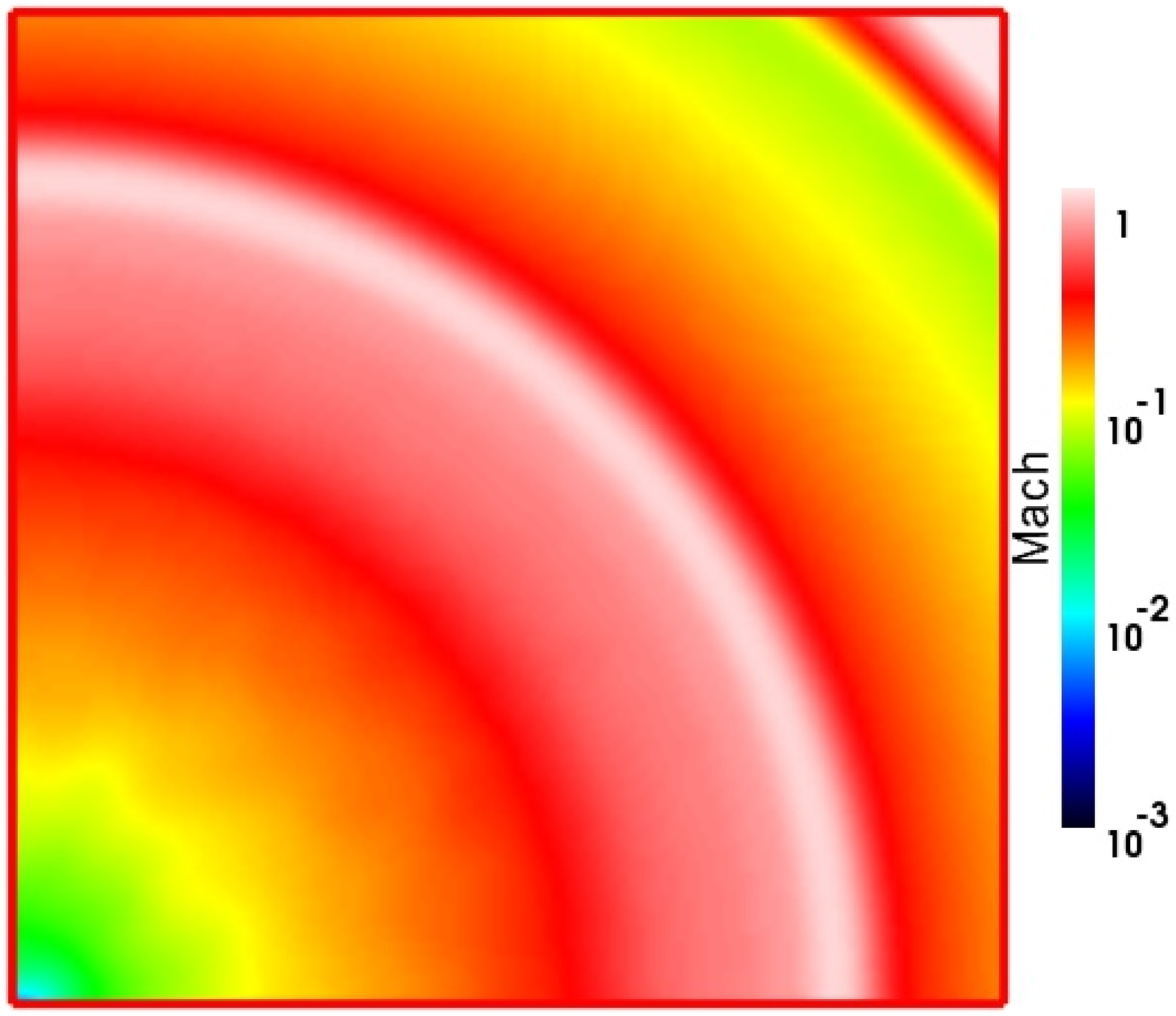}
  \includegraphics[width=2.3in]{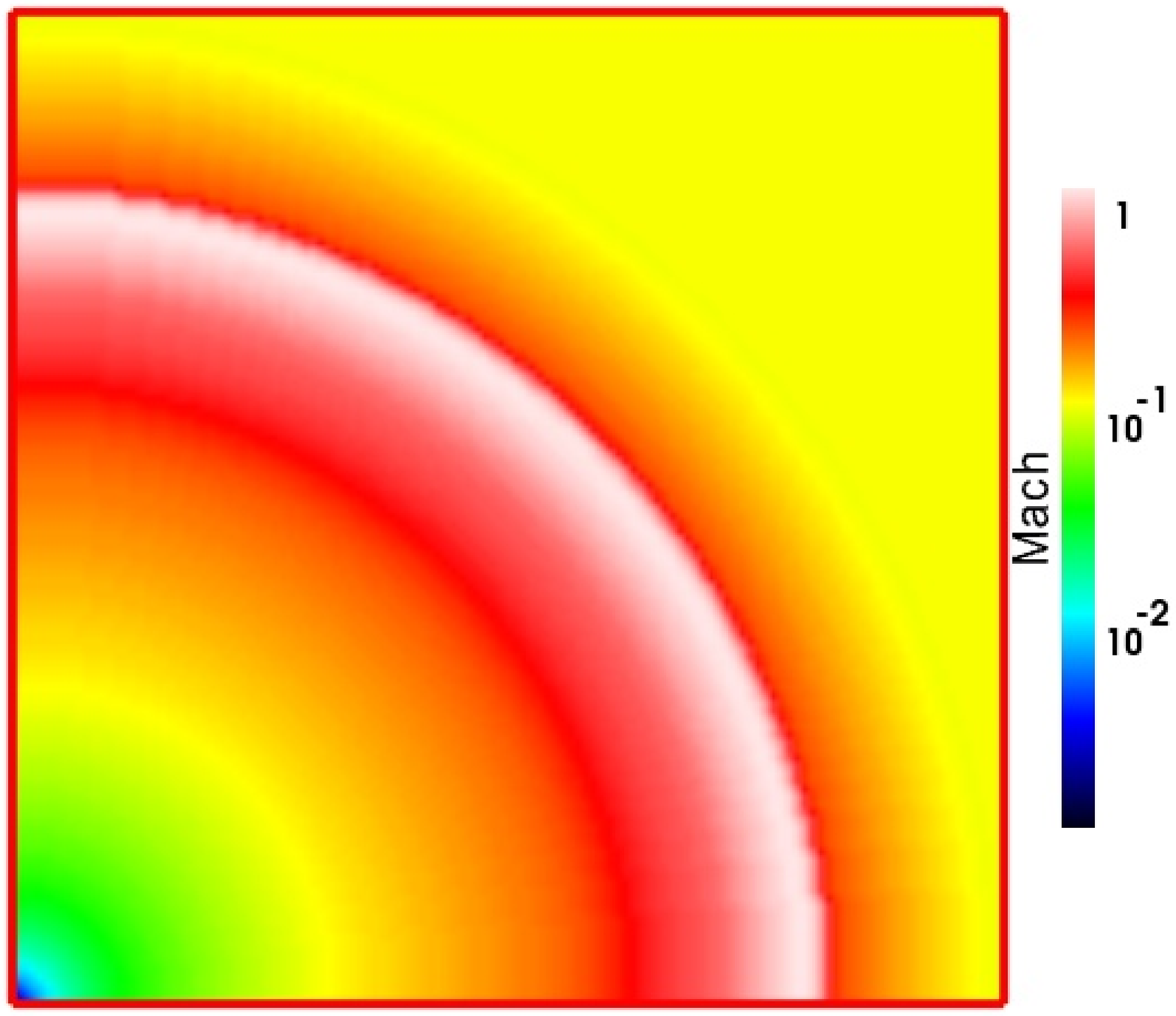}
  \includegraphics[width=2.3in]{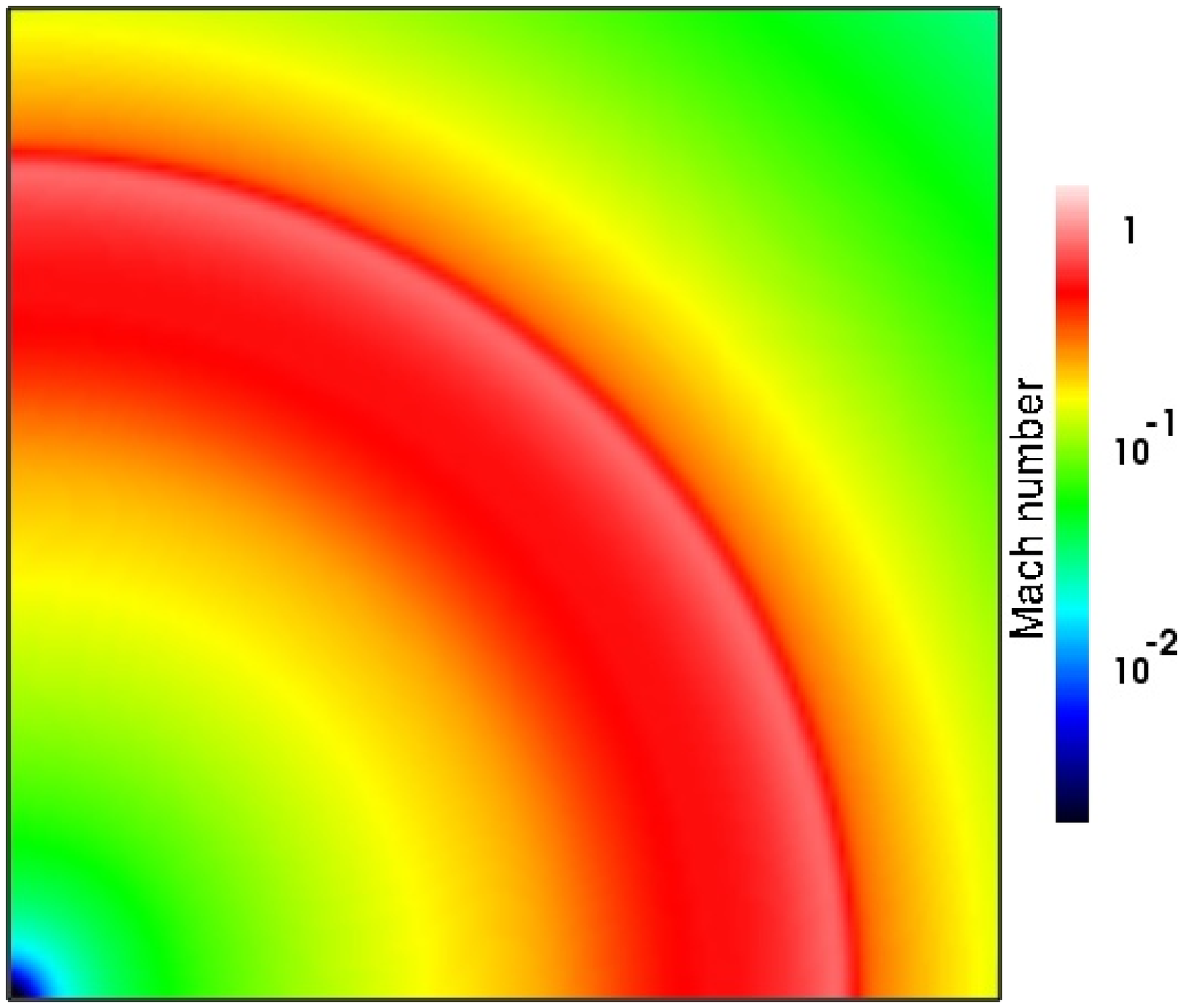}
  \includegraphics[width=2.3in]{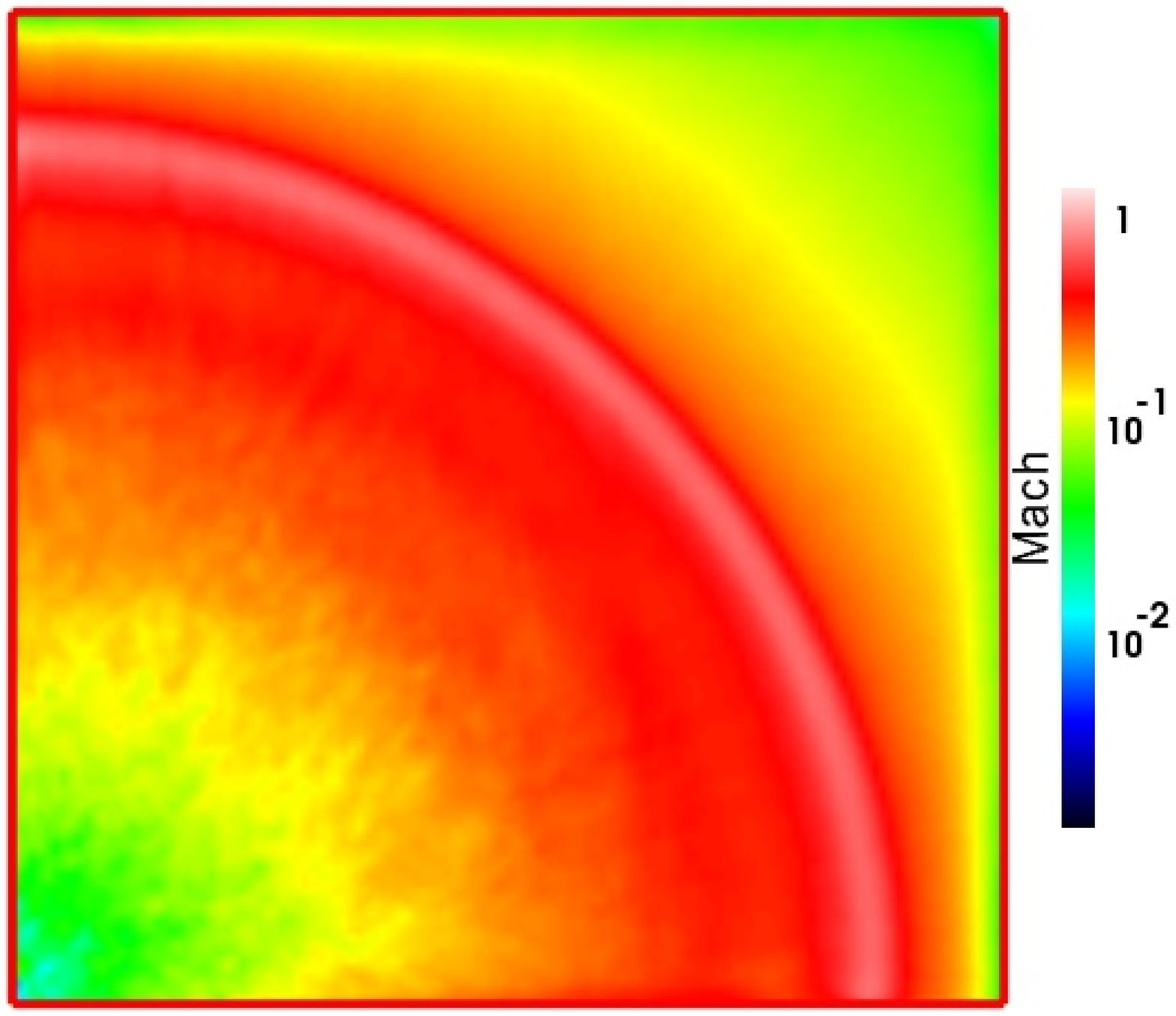}
  \includegraphics[width=2.3in]{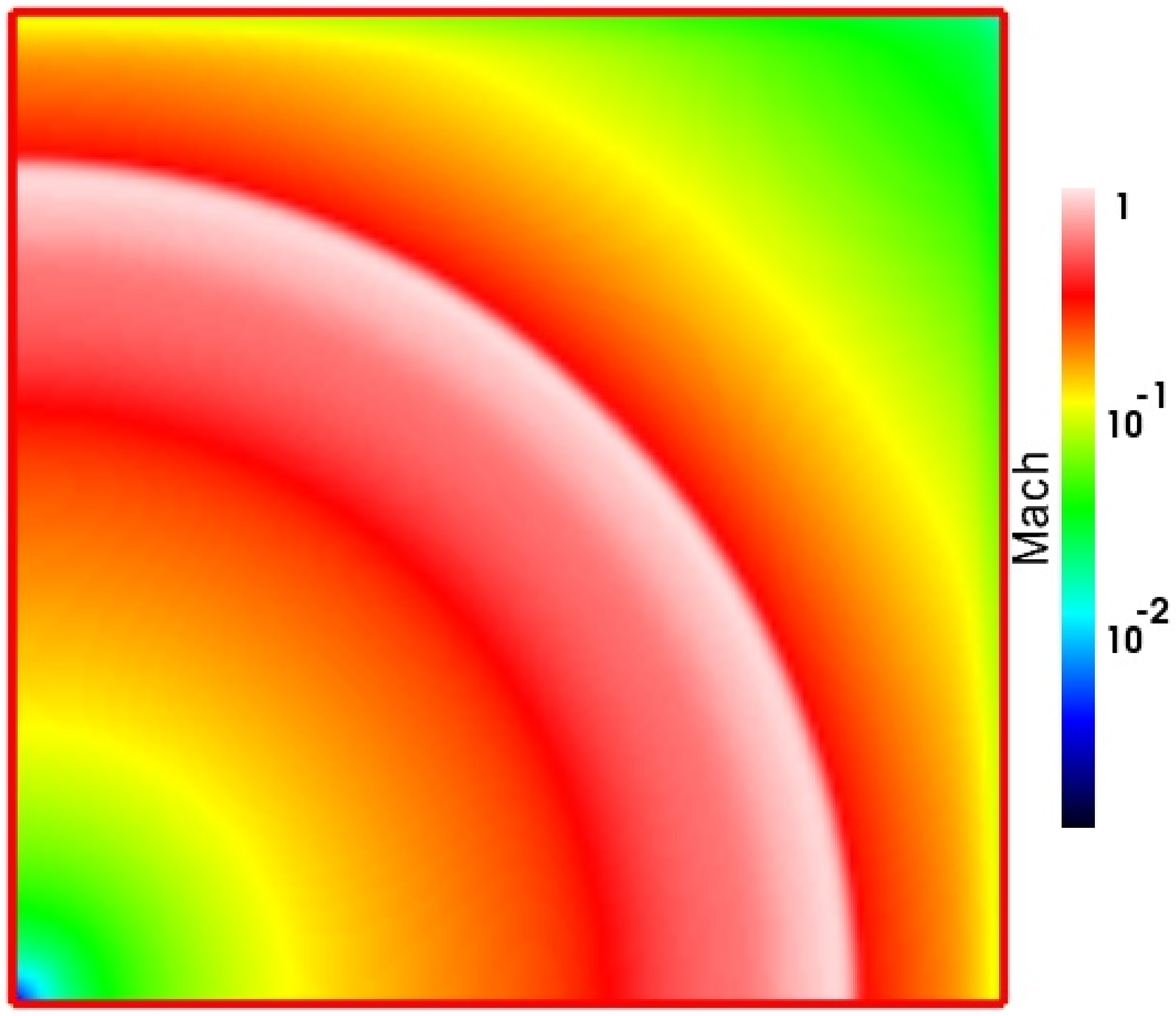}
  \includegraphics[width=2.3in]{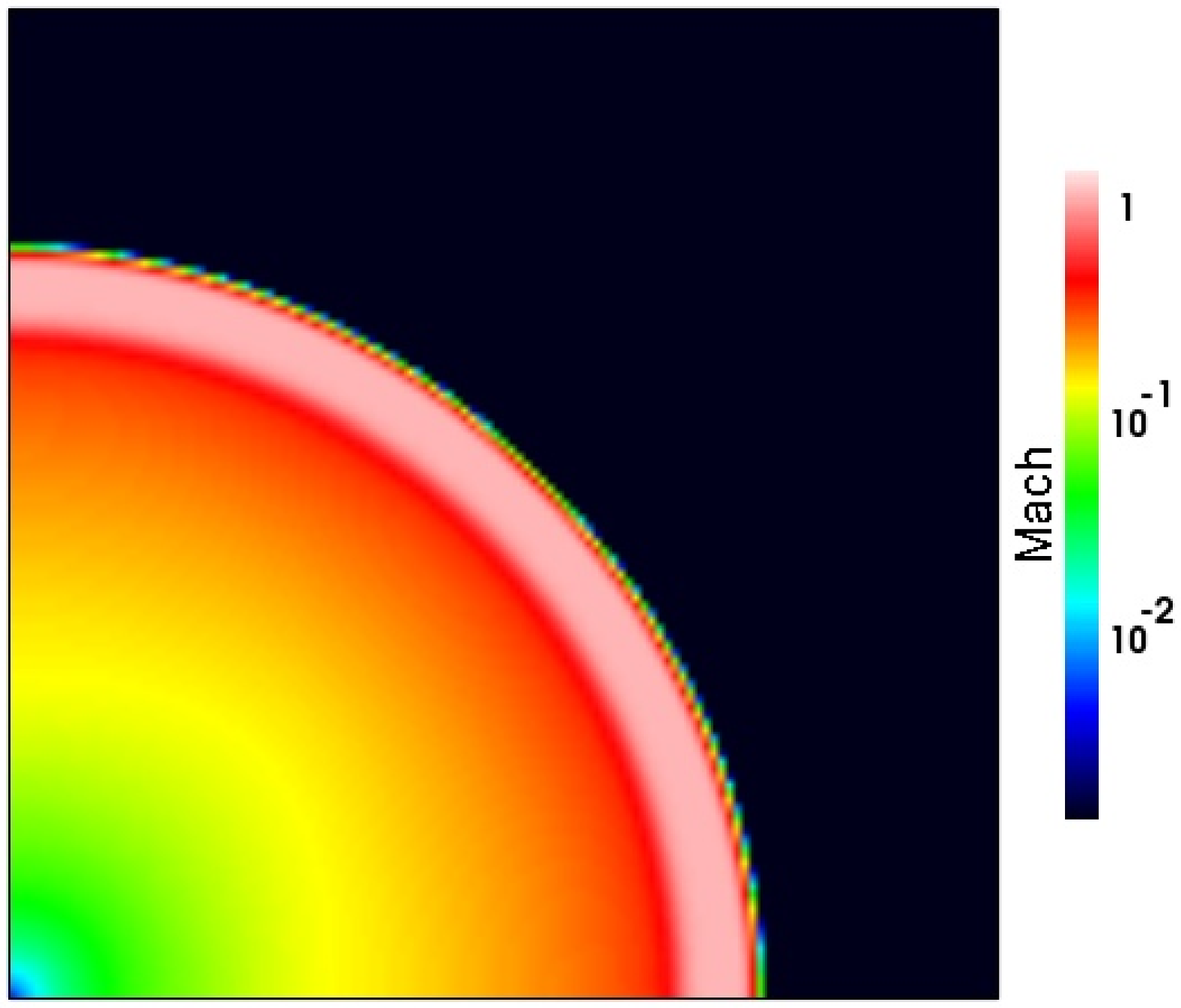}
\caption{Test 5 (H~II region expansion in an initially-uniform 
  gas): Images of the Mach number, cut through the simulation volume at
  coordinate $z=0$ at time $t=500$ Myr for (left to right and top to bottom)
  Capreole+$C^2$-Ray, HART, RSPH, ZEUS-MP, RH1D, LICORICE, Flash-HC and Enzo-RT.
\label{T5_images5_M_fig}}
\end{center}
\end{figure*}
For a static-density medium the evolution of the I-front within 
the flat-density core is described by equations~\ref{strom0} and 
\ref{strom01}. In this case, if the Str\"omgren radius associated 
with the core density $n_0$, 
$r_{S,0}=[3\dot{N}_{\gamma}/(4\pi\alpha_B(T)n_0^2)]^{1/3}$, is 
smaller than $r_0$, the front will come to a halt within the core. 
If, instead, $r_{S,0}>r_0$, the front escapes the core and propagates 
into the stratified envelope. Thereafter, the I-front position and 
velocity as a function of time have complex analytical forms for an 
arbitrary source fluxes and densities \citep{methodpaper}. A simple 
solution exists for the special case of the central ionizing source 
rate of photon emission $\dot{N}_\gamma =16\pi r_0^3n_0^2\alpha_B/3$, 
in which case the I-front radius upon leaving the core is
\begin{equation}
r_I=r_0(1+2t/t_{\rm rec,core})^{1/2},
\end{equation}
where $t_{\rm rec,core}$ is the recombination time in the core
\citep{methodpaper}. Similar solutions exist also when the I-front 
is moving relativistically \citep{2006ApJ...648..922S}. 

The propagation of an I-front in $r^{-2}$ density profiles with 
full gas dynamics does not have an exact analytical solution, but 
has been well studied with both semianalytical and numerical methods 
\citep{1990ApJ...349..126F}. If $r_{S,0}<r_0$ then the I-front 
converts to D-type within the core, but starts to re-accelerate upon 
entering the steep density gradient. Numerical simulations indicate 
that in density profiles approximating those of galactic molecular 
cloud cores or cosmological minihalos at high redshift, the I-front 
remains D-type for the lifetime of typical UV sources \citep{wn08b}.  
If $r_{S,0}$ is instead equal to or greater than $r_0$, the I-front 
may briefly convert to D-type, but then rapidly reverts to R-type and
flash-ionizes the cloud on timescales shorter than the dynamical time 
of the gas. Now completely ionized and nearly isothermal, strong pressure 
gradients form wherever there are steep density gradients, the sharpest 
of which are found at the edge of what was once the edge of the core.  
These pressure gradients drive the gas outward into the ionized cloud 
forming a shock that moves with a roughly constant velocity in $r^{-2}$ 
density profiles \citep{1990ApJ...349..126F}.

\begin{figure*}
\begin{center}
  \includegraphics[width=2.3in]{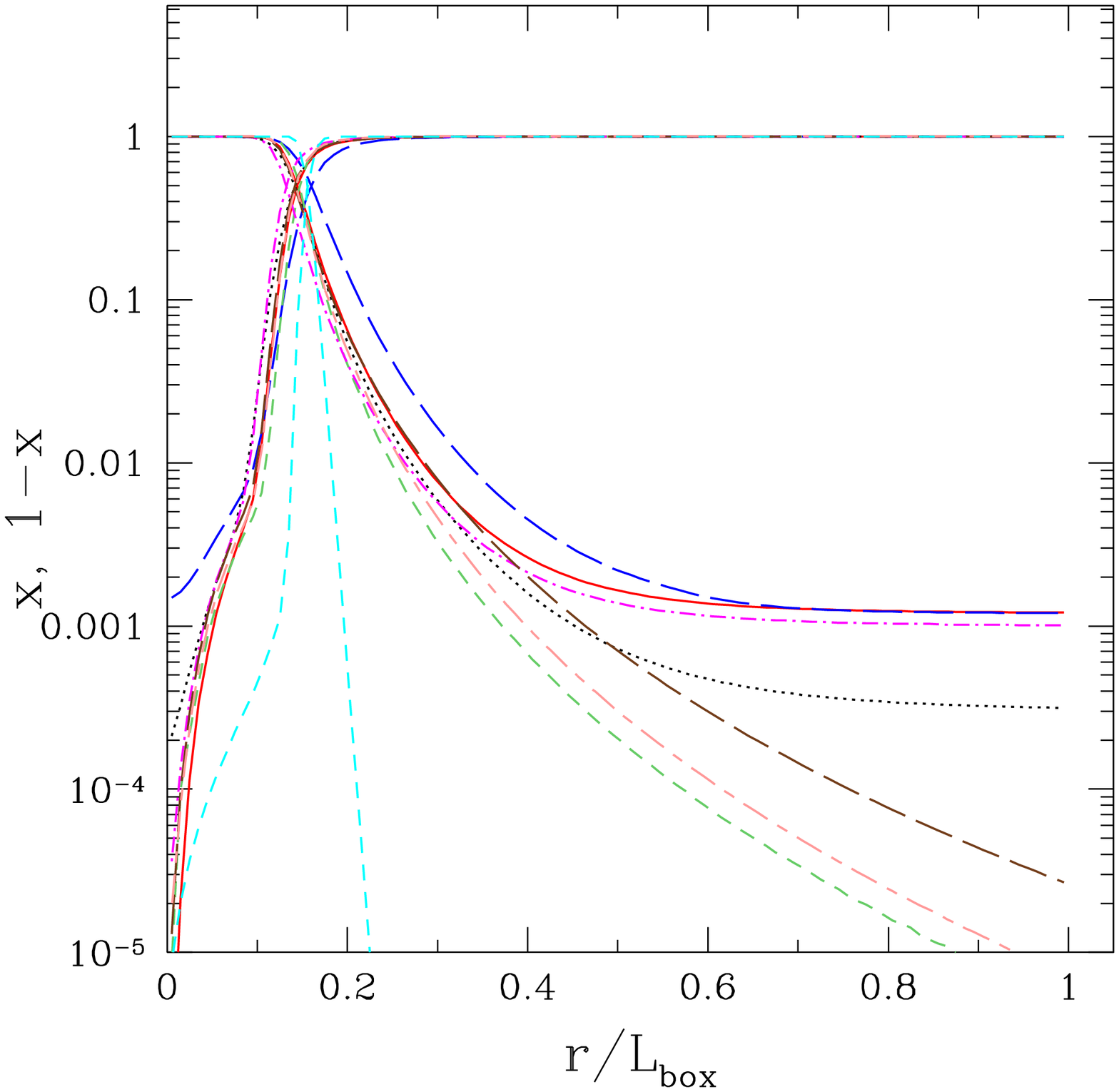}
  \includegraphics[width=2.3in]{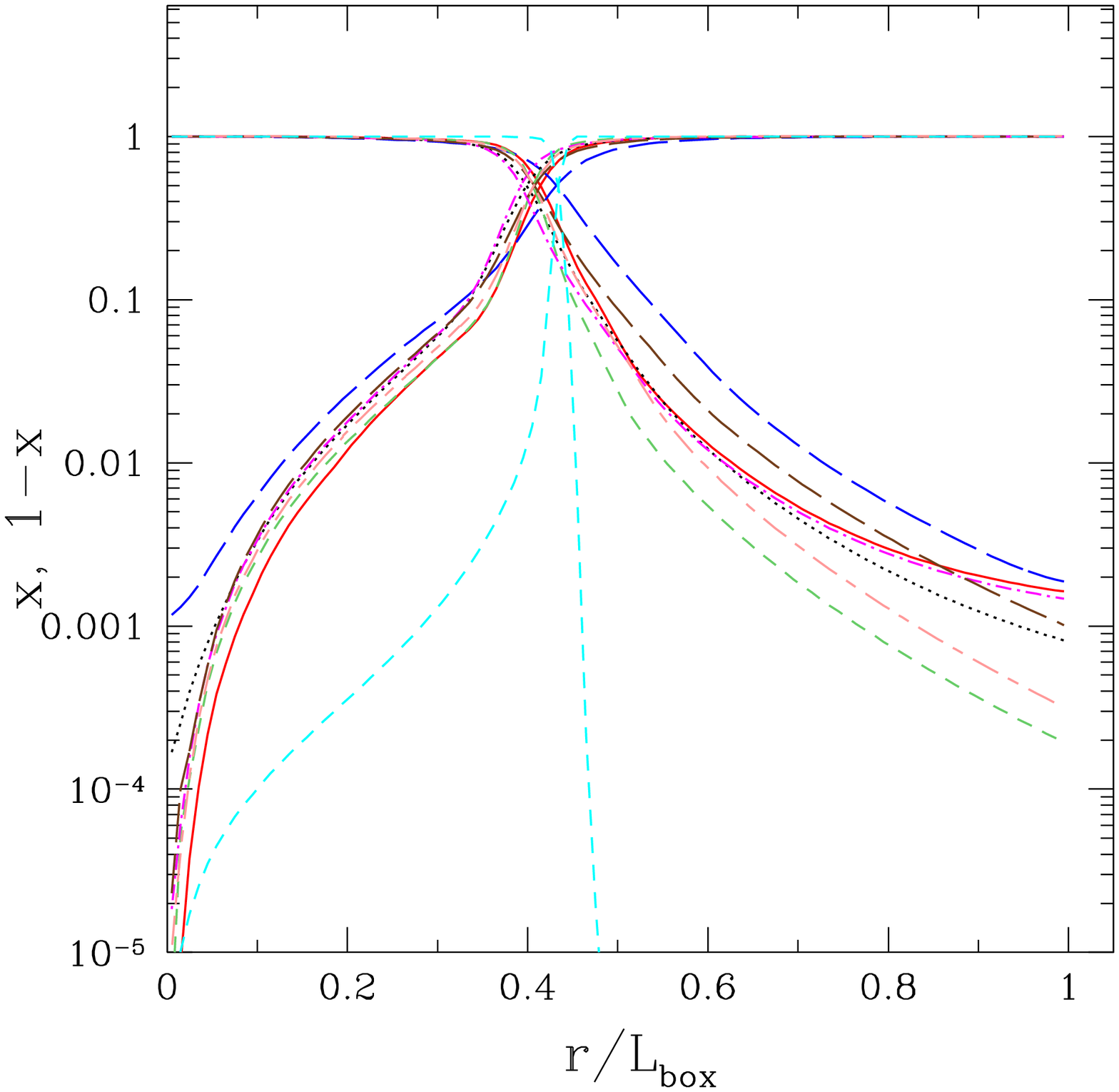}
  \includegraphics[width=2.3in]{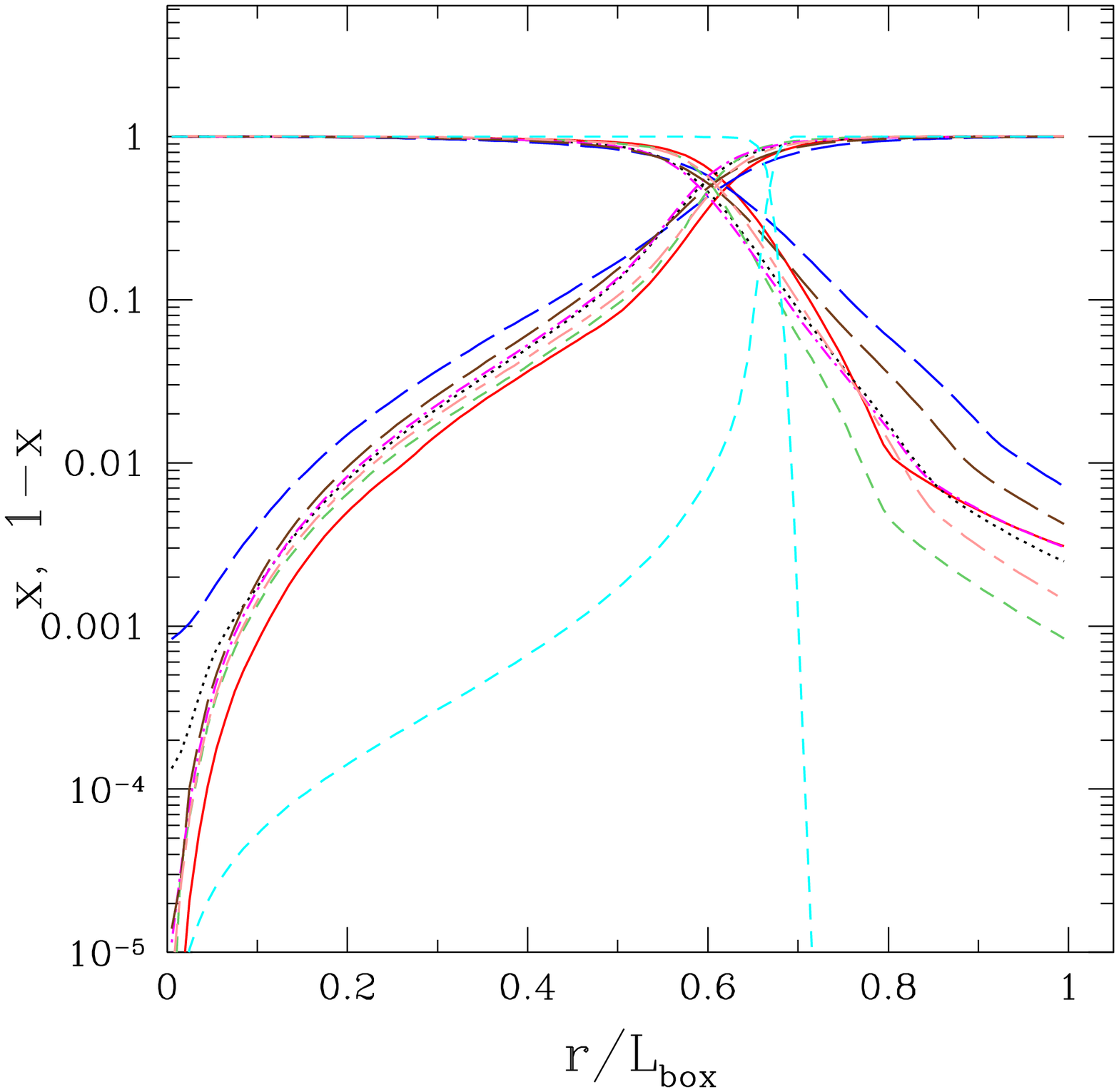}
\caption{Test 5 (H~II region expansion in an initially-uniform 
  gas): Spherically-averaged profiles for ionized fractions $x$ 
  and neutral fractions $x_{\rm HI}=1-x$ at times $t=10$ Myr, 200 Myr and 500
  Myr vs. dimensionless radius (in units of the box size). 
\label{T5_profs_fig}}
\end{center}
\end{figure*}

\begin{figure*}
\begin{center}
  \includegraphics[width=2.3in]{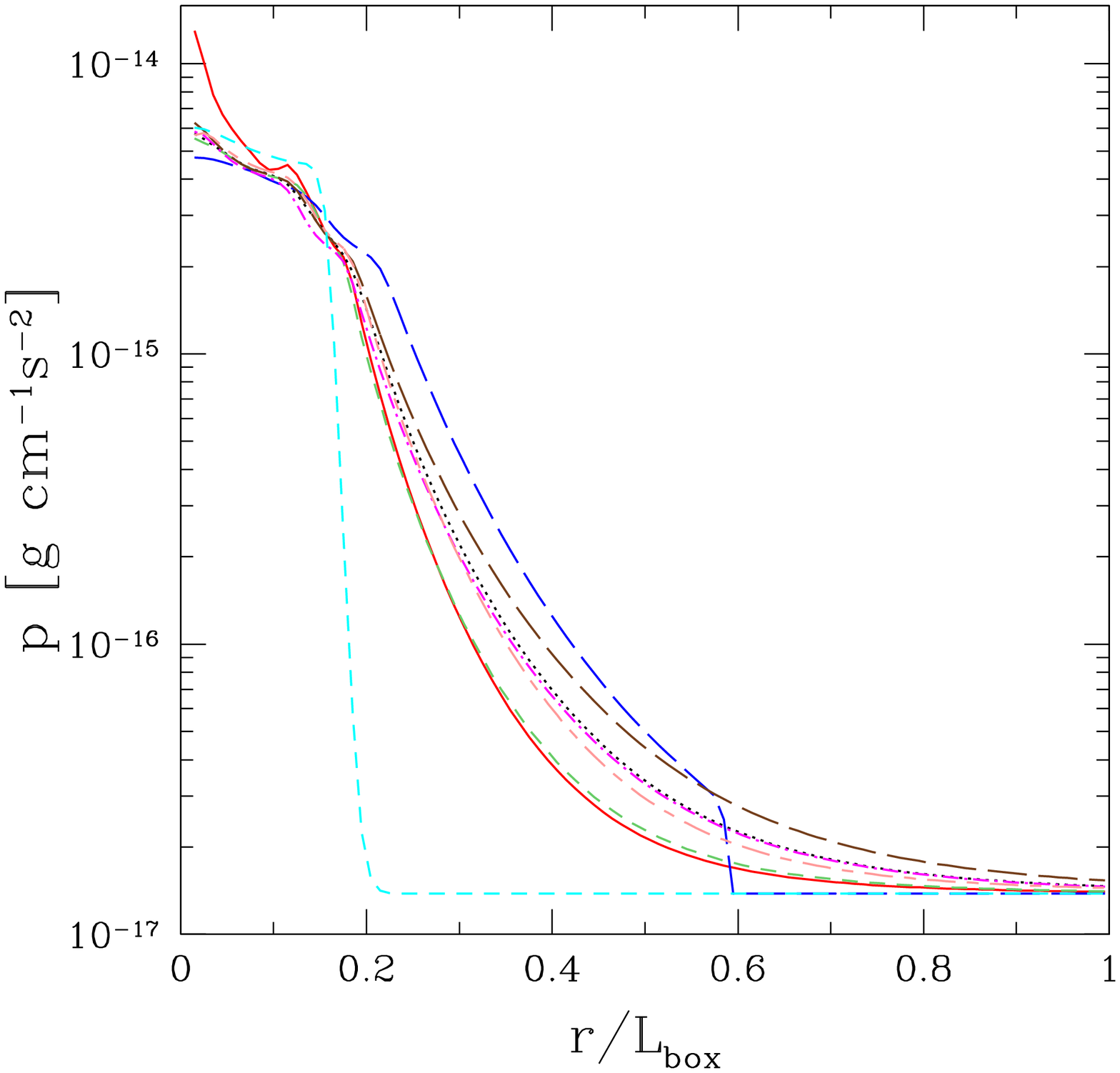}
  \includegraphics[width=2.3in]{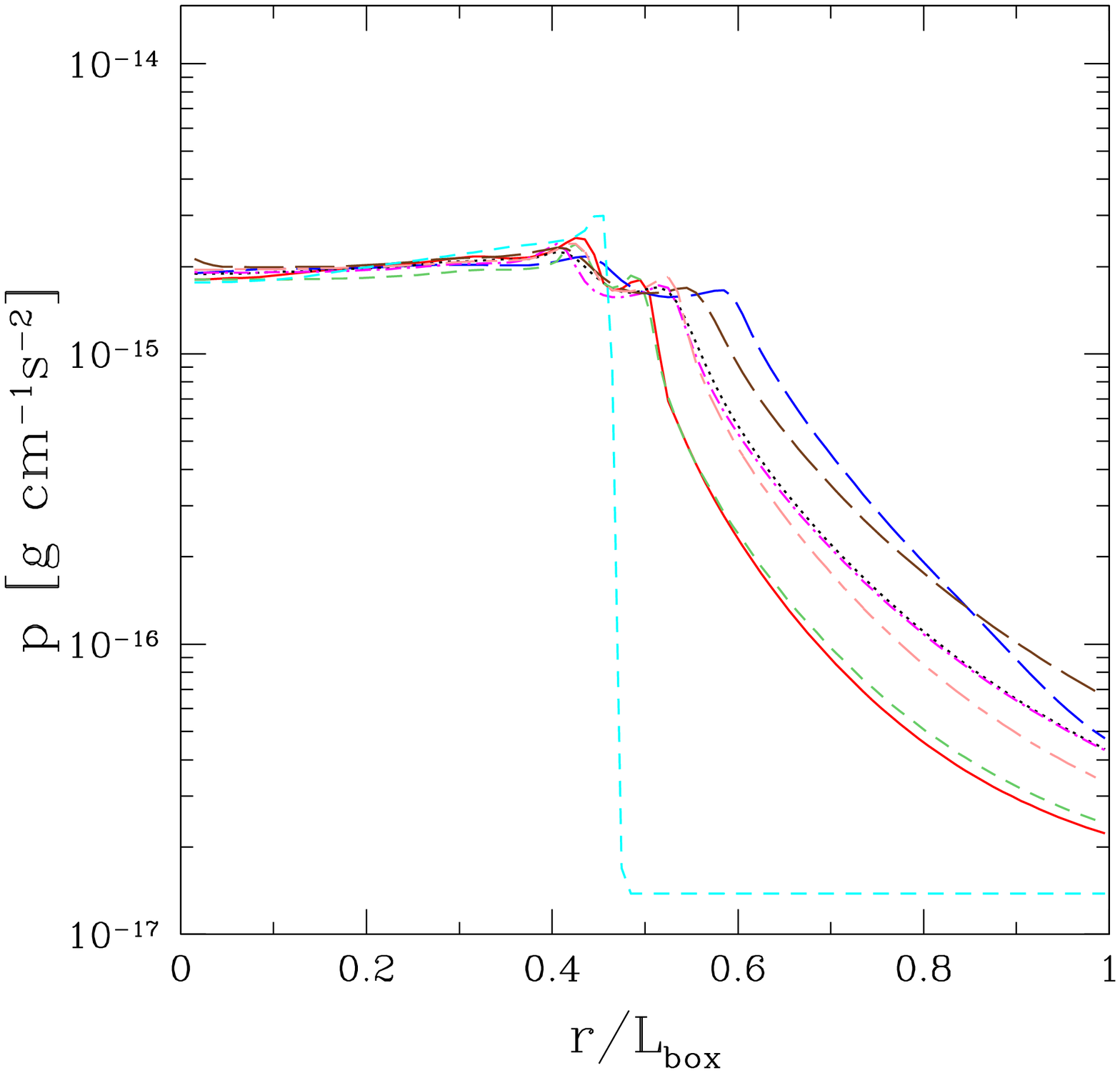}
  \includegraphics[width=2.3in]{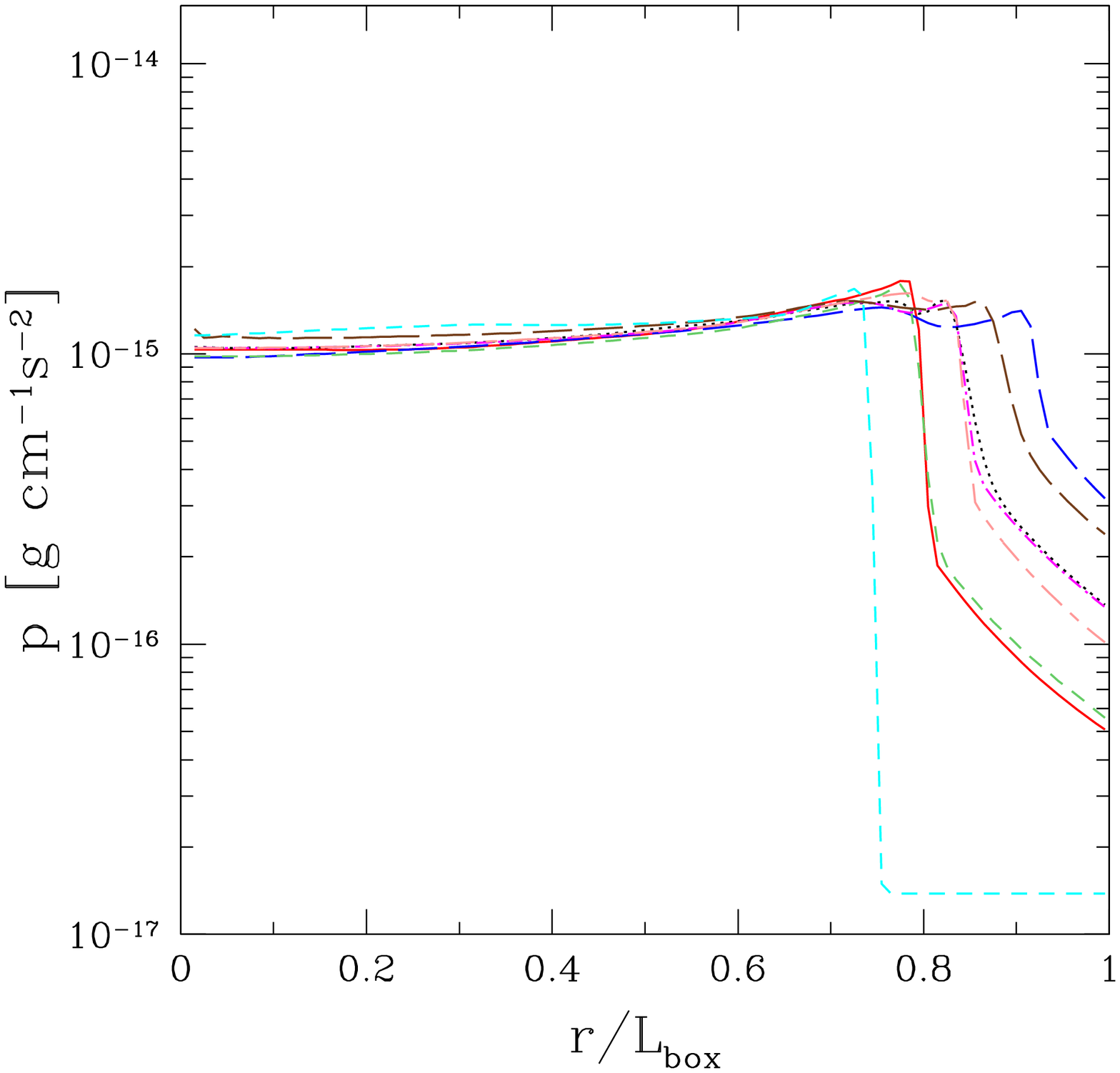}
\caption{Test 5 (H~II region expansion in an initially-uniform 
  gas): Spherically-averaged profiles for pressure, $p$, at times $t=10$~Myr, 
200 Myr and 500 Myr vs. dimensionless radius (in units of the box size). 
\label{T5_profsp_fig}}
\end{center}
\end{figure*}

\begin{figure*}
\begin{center}
  \includegraphics[width=2.3in]{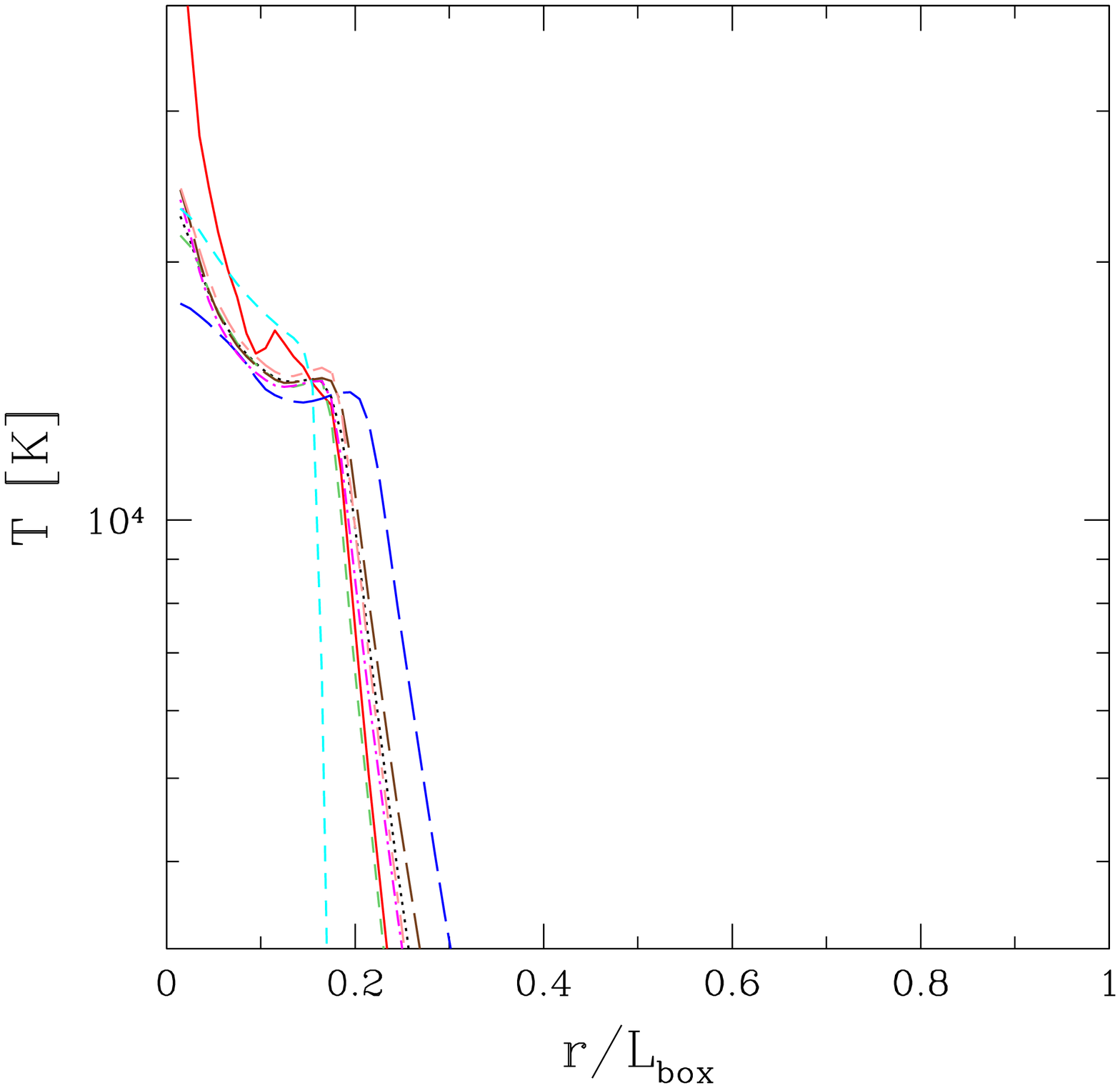}
  \includegraphics[width=2.3in]{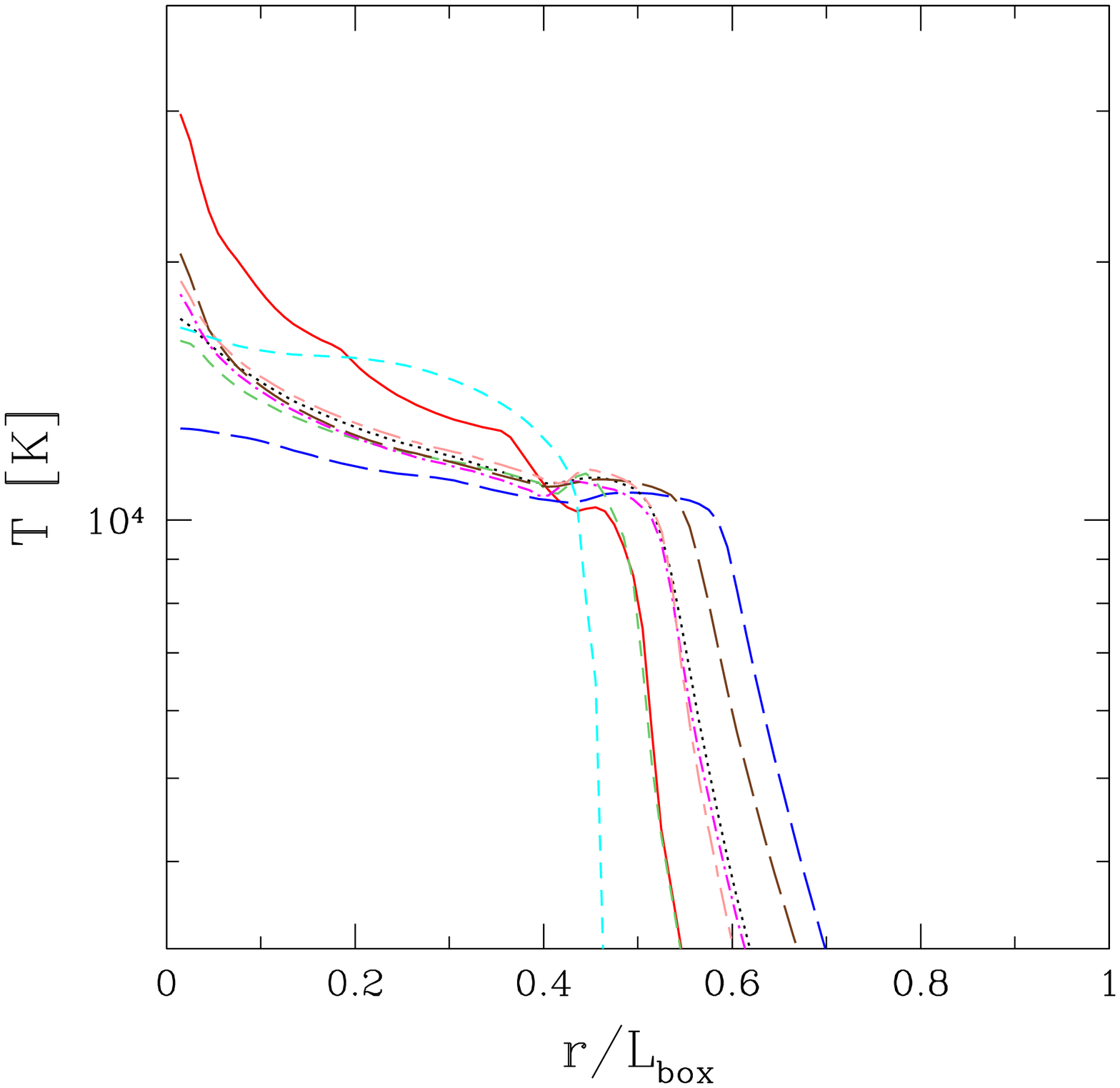}
  \includegraphics[width=2.3in]{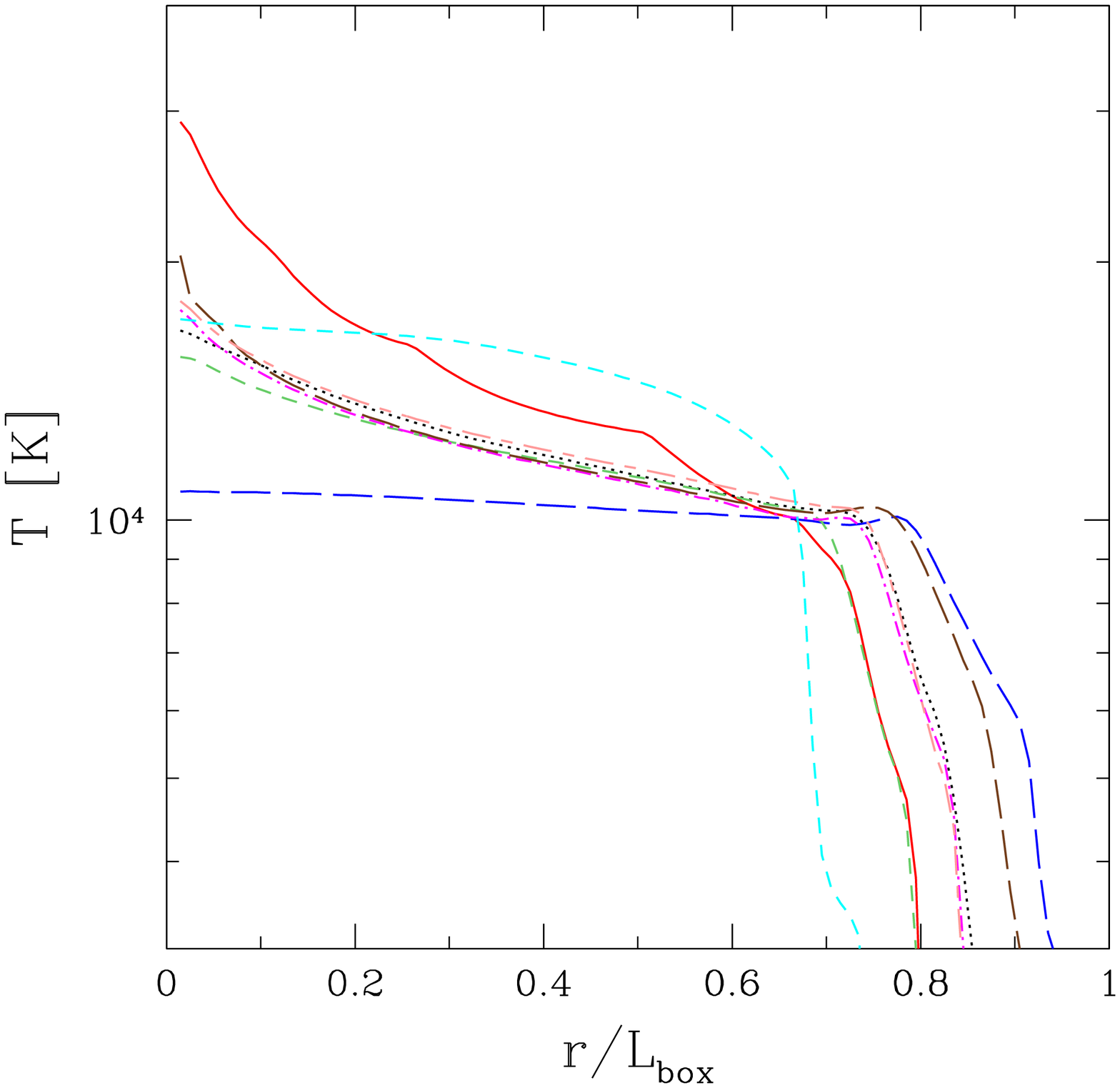}
\caption{Test 5 (H~II region expansion in an initially-uniform 
  gas): Spherically-averaged profiles for temperature at times $t=10$~Myr, 
 200 Myr and 500 Myr vs. dimensionless radius (in units of the box size). 
\label{T5_profsT_fig}}
\end{center}
\end{figure*}

In Test 6 we examine the former case, in which the initial Str\"omgren
radius is smaller than the core radius. The aim of this test is to study 
the initial transition of the I-front from R-type to D-type and back to 
R-type over a fairly restricted range of radii, rather than its long-term 
behavior thereafter. Accordingly, we adopt the following numerical 
parameters: computational box length $L=0.8$ kpc, $n_0=3.2$ cm$^{-3}$, 
$r_0=91.5$ pc, zero initial ionization fraction, ionizing photon emission
rate $\dot{N}_\gamma=10^{50}$ photons\,$s^{-1}$ and initial temperature 
$T=100$~K. The source position is at the corner of the computational volume 
$(x_s,y_s,z_s) = (0,0,0)$. Boundary conditions are reflective for the 
boundaries which contain the origin and transmissive for the other boundaries.
For these parameters the I-front changes from 
R-type to D-type inside the core. Once the front reaches the core edge it 
will accelerate as it propagates down the steep density slope. The initial
recombination time inside the core (assuming ionized gas temperature 
$T=10^4$~K) is $t_{\rm rec,core}=0.04$~Myr. The ionizing spectrum is again 
that of a $10^5$~K black body, as expected for a massive, metal-free 
Pop~III star. Hydrogen line cooling, recombinational cooling, and 
bremsstrahlung cooling are all included, but 
again not Compton cooling. For simplicity, gravitational forces are ignored 
and no hydrostatic equilibrium is imposed on the cloud. Unlike in Test 5,  
left on their own the pressure forces will accelerate gas outward in this 
density-stratified cloud, albeit those forces are much inferior than the 
stronger ones due to pressure from the photoheated gas. The running time 
is $t_{\rm  sim}=75$ Myr. The required outputs are neutral fraction of 
hydrogen, gas number density, temperature and Mach number on the grid at 
times $t=1, 3, 10, 25$ and 75 Myr, and the I-front position (as defined in 
Test 5) and velocity vs. time along the $x$-axis.

\subsection{Test 7: Photoevaporation of a dense clump}

In Test 7, a plane-parallel I-front encounters a uniform spherical clump 
in a constant background density field.  This problem has been studied in
many contexts, $\eg$ in relation to the photoevaporation of dense clumps 
in planetary nebulae \citep{1998A&A...331..335M}. Depending on the assumed
parameters the clump may either initially trap the I-front, or be 
flash-ionized without ever trapping the I-front, the so-called 'cloud-zapping' 
regime \citep[c.f.][]{1989ApJ...346..735B}. The condition for an I-front to be 
trapped by a dense clump with number density $n_H$ can be derived by 
defining a ``Str\"omgren length'', $\ell_S(r)$, at a given impact parameter 
$r$ using equations (\ref{strom0}) and (\ref{strom01}), and solving them 
for each impact parameter \citep{2004MNRAS.348..753S}. We can then define 
the ``Str\"omgren number'' for the clump as 
$L_S\equiv2r_{\rm  clump}/\ell_S(0)$, where $r_{\rm clump}$ is the clump 
radius and $\ell_S(0)$ is the Str\"omgren length for zero impact parameter.  
If $L_S>1$, then the clump is able to trap the I-front, while if $L_S<1$, 
the I-front quickly ionizes the clump and is never trapped. 
For a uniform clump equation~(\ref{strom_rad}) reduces to 
\ba
\ell_S&=&\frac{F}{\alpha_H^{(2)} n_H^2},
\label{l_S_uniform}
\ea
and $L_S$ becomes
\begin{equation}
L_S=\frac{2r_{\rm clump}\alpha_H^{(2)} n_H^2}{F}.
\end{equation}

\begin{figure*}
\begin{center}
  \includegraphics[width=2.3in]{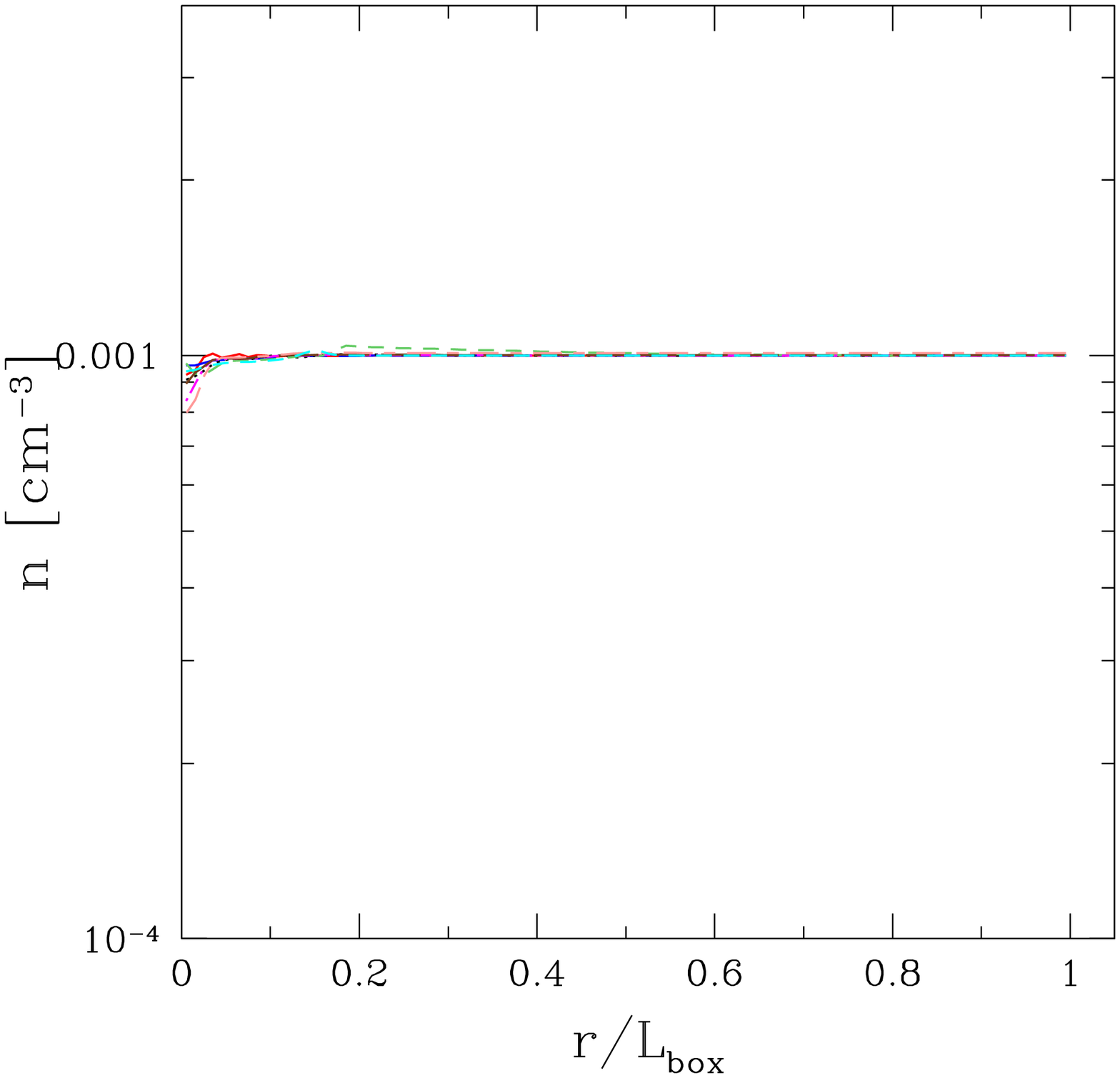}
  \includegraphics[width=2.3in]{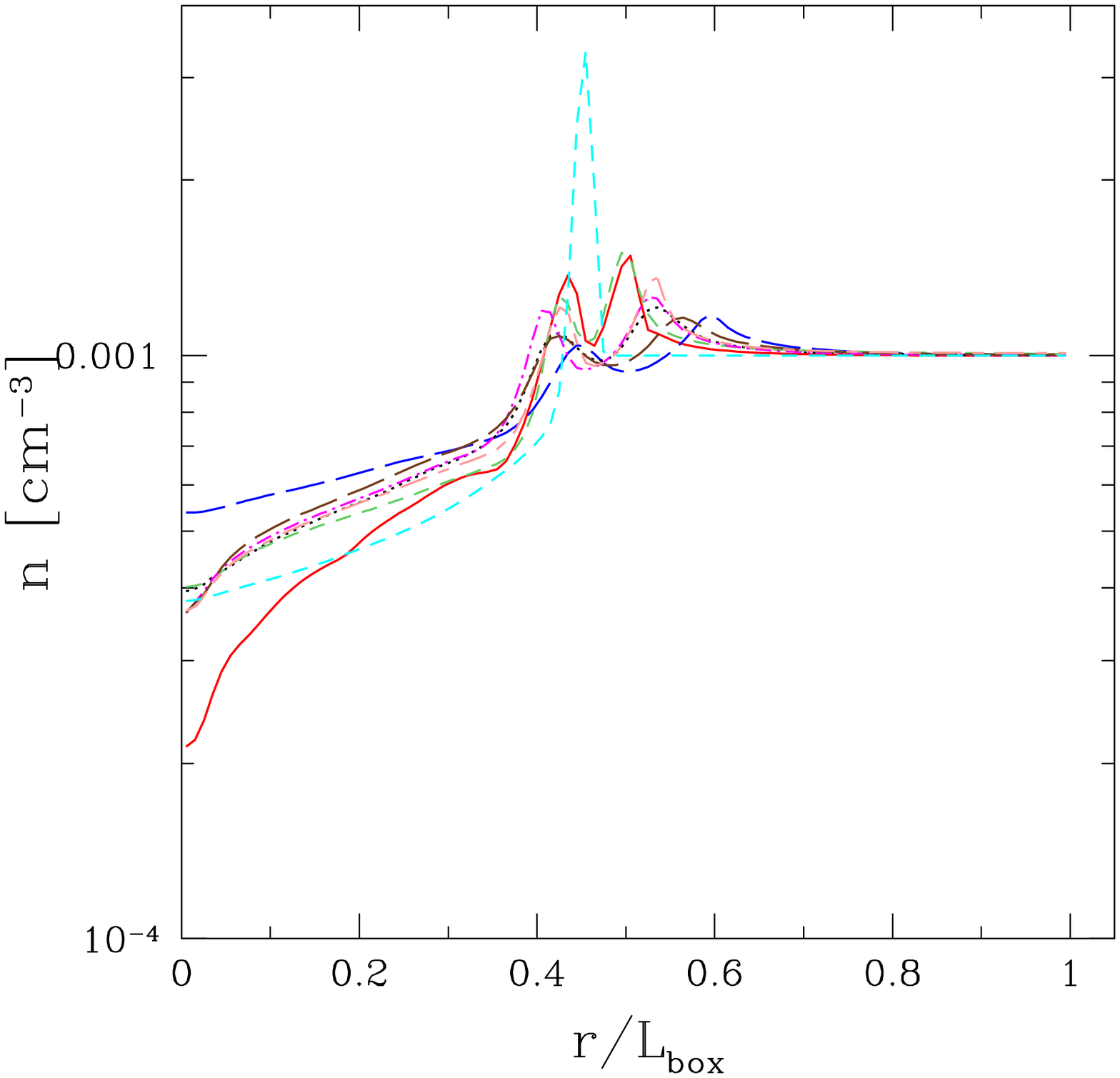}
  \includegraphics[width=2.3in]{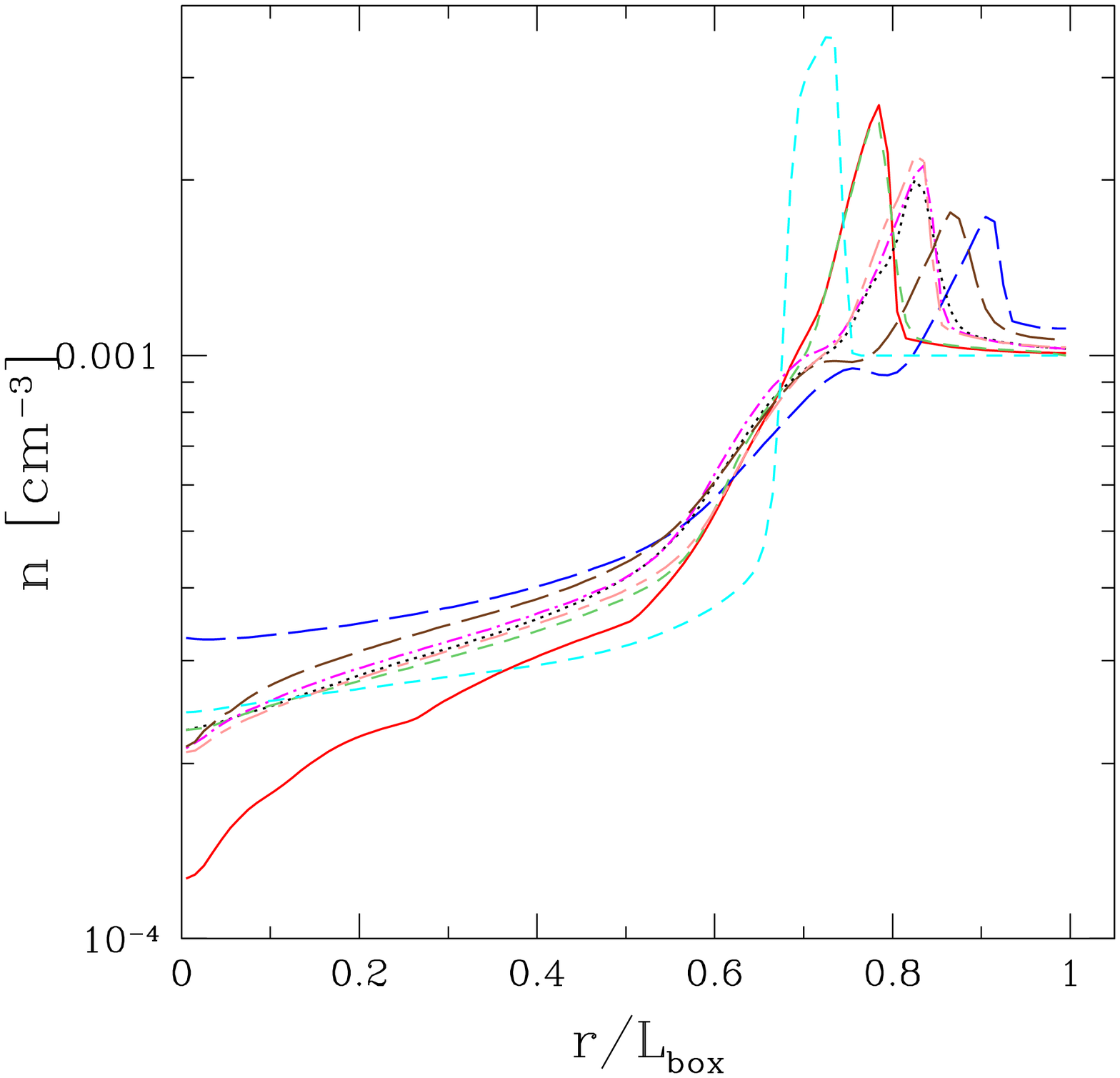}
\caption{Test 5 (H~II region expansion in an initially-uniform 
  gas): Spherically-averaged profiles for the hydrogen number density, $n$,
  at times $t=10$ Myr, 200 Myr and 500 Myr vs. dimensionless radius (in units 
  of the box size). 
\label{T5_profsn_fig}}
\end{center}
\end{figure*}

\begin{figure*}
\begin{center}
  \includegraphics[width=2.3in]{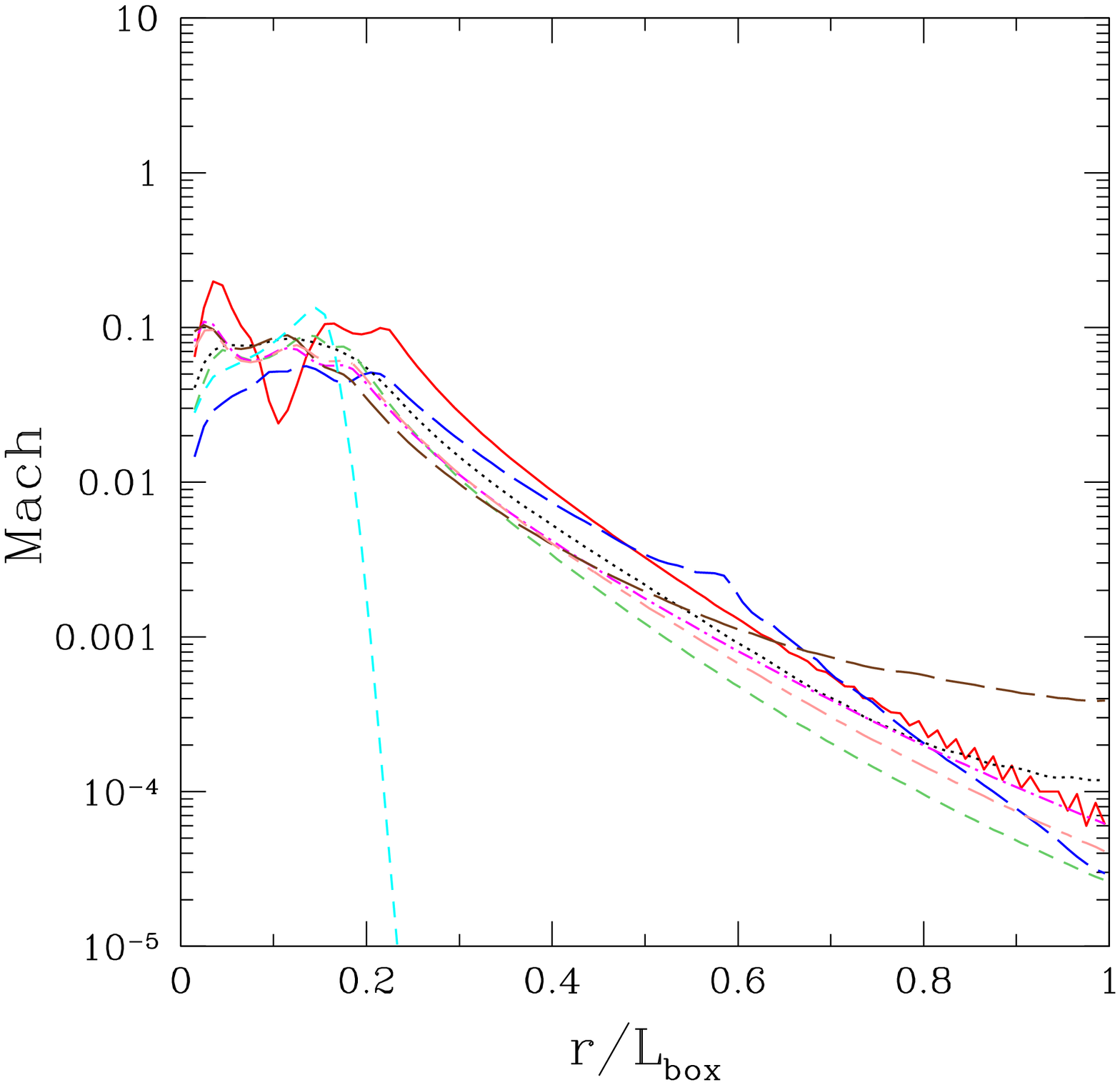}
  \includegraphics[width=2.3in]{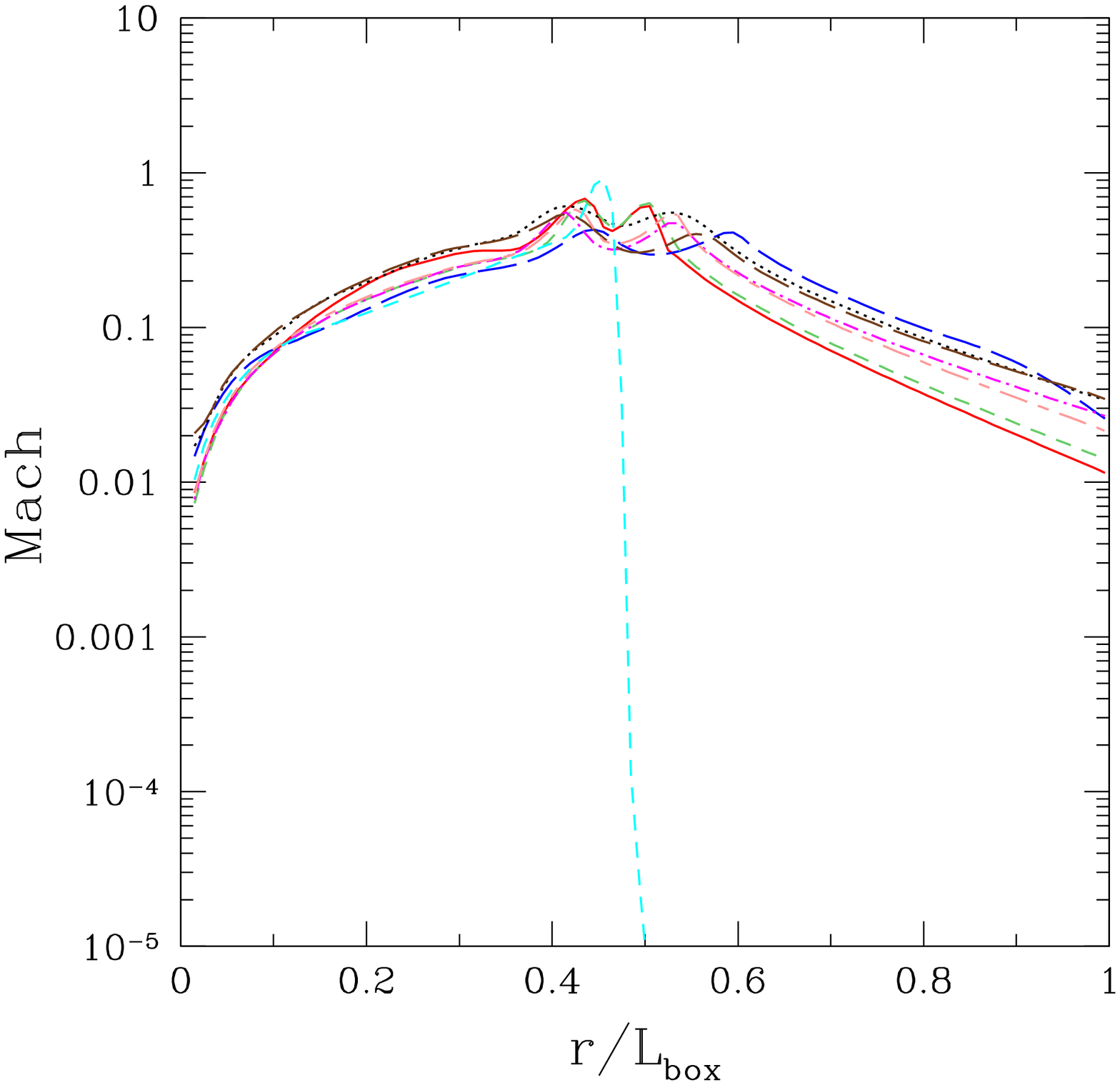}
  \includegraphics[width=2.3in]{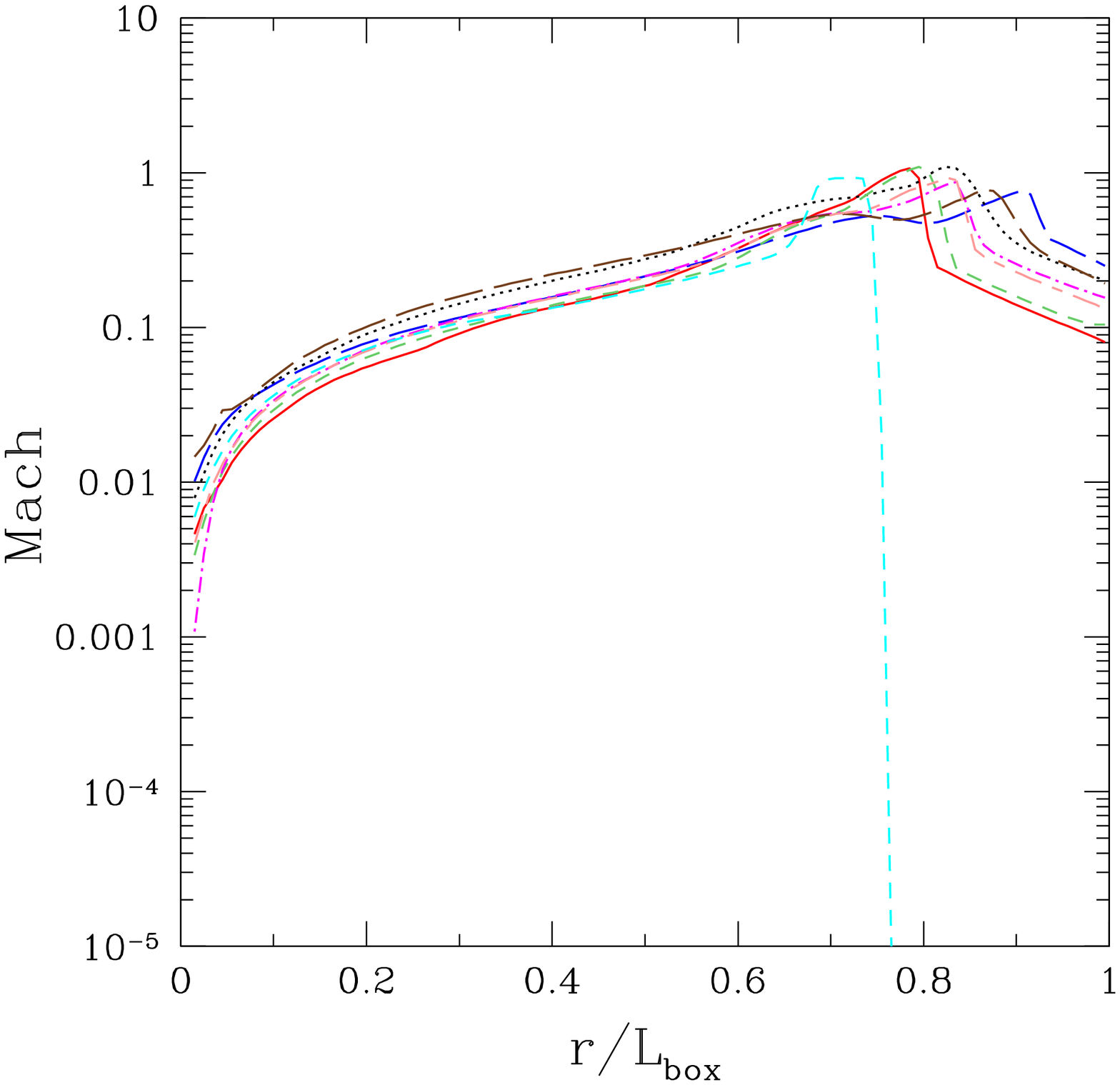}
\caption{Test 5 (H~II region expansion in an initially-uniform 
  gas): Spherically-averaged profiles for the flow Mach number, $M$,
  at times $t=10$ Myr, 200 Myr and 500 Myr vs. dimensionless radius (in units 
  of the box size). 
\label{T5_profsm_fig}}
\end{center}
\end{figure*}

The numerical parameters for Test 7 are the same as for Test 3 in Paper~I: 
a constant ionizing photon flux of $F=10^6\rm\,s^{-1}cm^{-2}$ is incident 
at $y=0$, the ambient hydrogen gas number density and temperature are 
$n_{\rm out}=2\times10^{-4}\rm\, cm^{-3}$ and $T_{\rm out}=8,000$ K, 
respectively, while the initial clump density and temperature are 
$n_{\rm clump}=200\,n_{\rm out}=0.04\rm\, cm^{-3}$ and $T_{\rm clump}=40$ K.  
These parameters ensure that outward pressures in the clump balance those 
from the hot gas so that the clump is initially in pressure equilibrium 
with the surrounding medium. The column density of the clump is sufficient 
to trap the I-front and compel its transition to D-type, in contrast to 
the less interesting for us ``cloud-zapping'' regime in which the front 
flash-ionizes the cloud and remains R-type throughout. The computational 
box size is $x_{\rm  box}=6.6$ kpc, the radius of the clump is 
$r_{\rm clump}=0.8$ kpc, and its center is at $(x_c,y_c,z_c)=(5,3.3,3.3)$~kpc 
= $(97,64,64)$ cells. Hydrogen line cooling, recombinational cooling, and 
bremsstrahlung cooling are included, but not Compton cooling. Boundary 
conditions are transmissive for all grid boundaries.

With hydrodynamics the evolution beyond the trapping phase proceeds very 
differently from the static Test 3 in Paper I. As the heated and ionized 
gas is evaporated and expands towards the source, its recombination rate 
falls and it attenuates the ionizing flux less. As a consequence, the 
I-front slowly consumes the clump until it photoevaporates completely. The 
required outputs are H~I fraction, gas pressure, temperature, and Mach 
number at times $t=1,5,10,25$ and 50 Myr and the position and velocity 
of the I-front along the axis of symmetry.

\section{Results}

\subsection{Test 5}

\begin{figure*}
\begin{center}
  \includegraphics[width=3.4in]{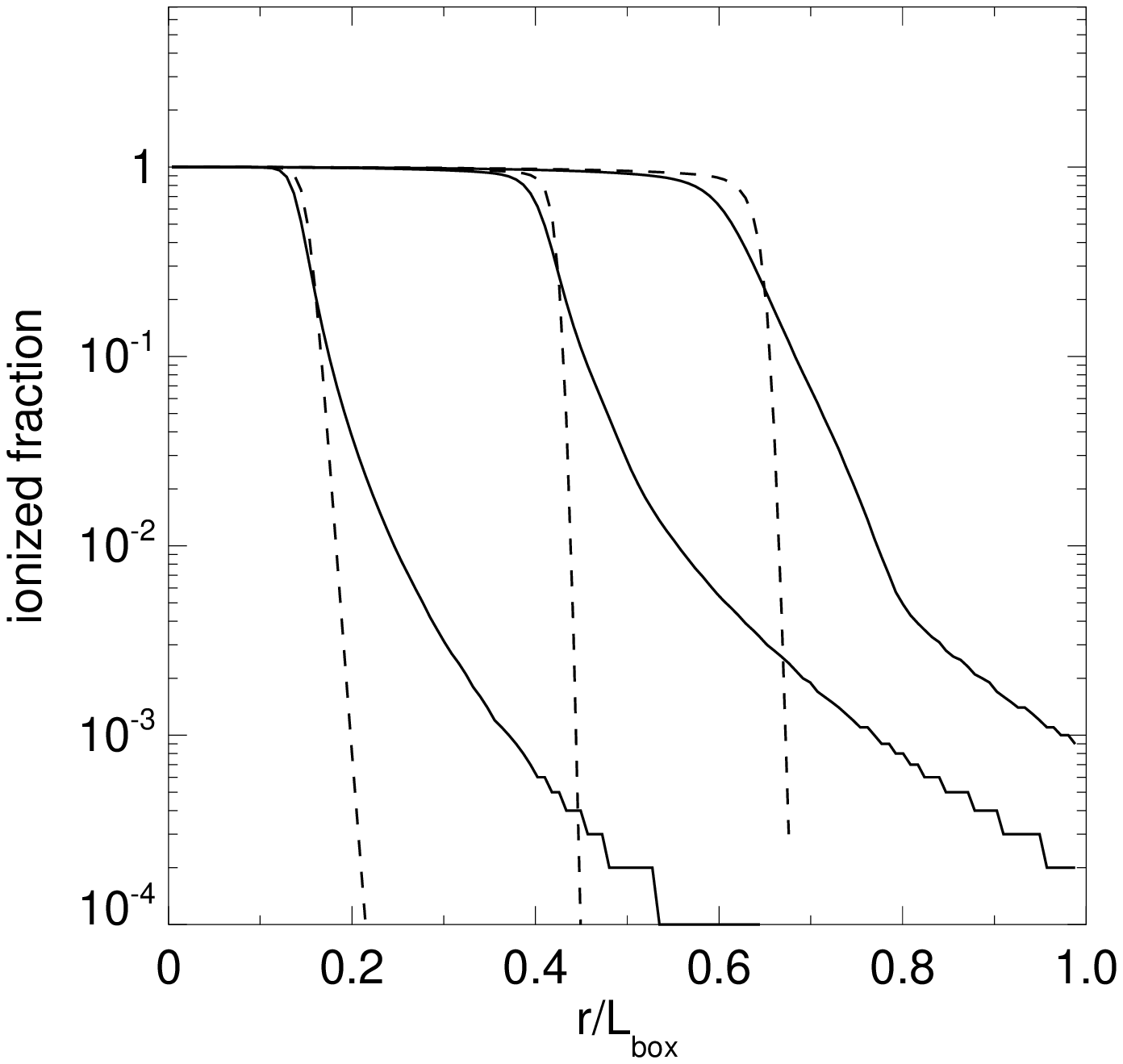}
  \includegraphics[width=3.4in]{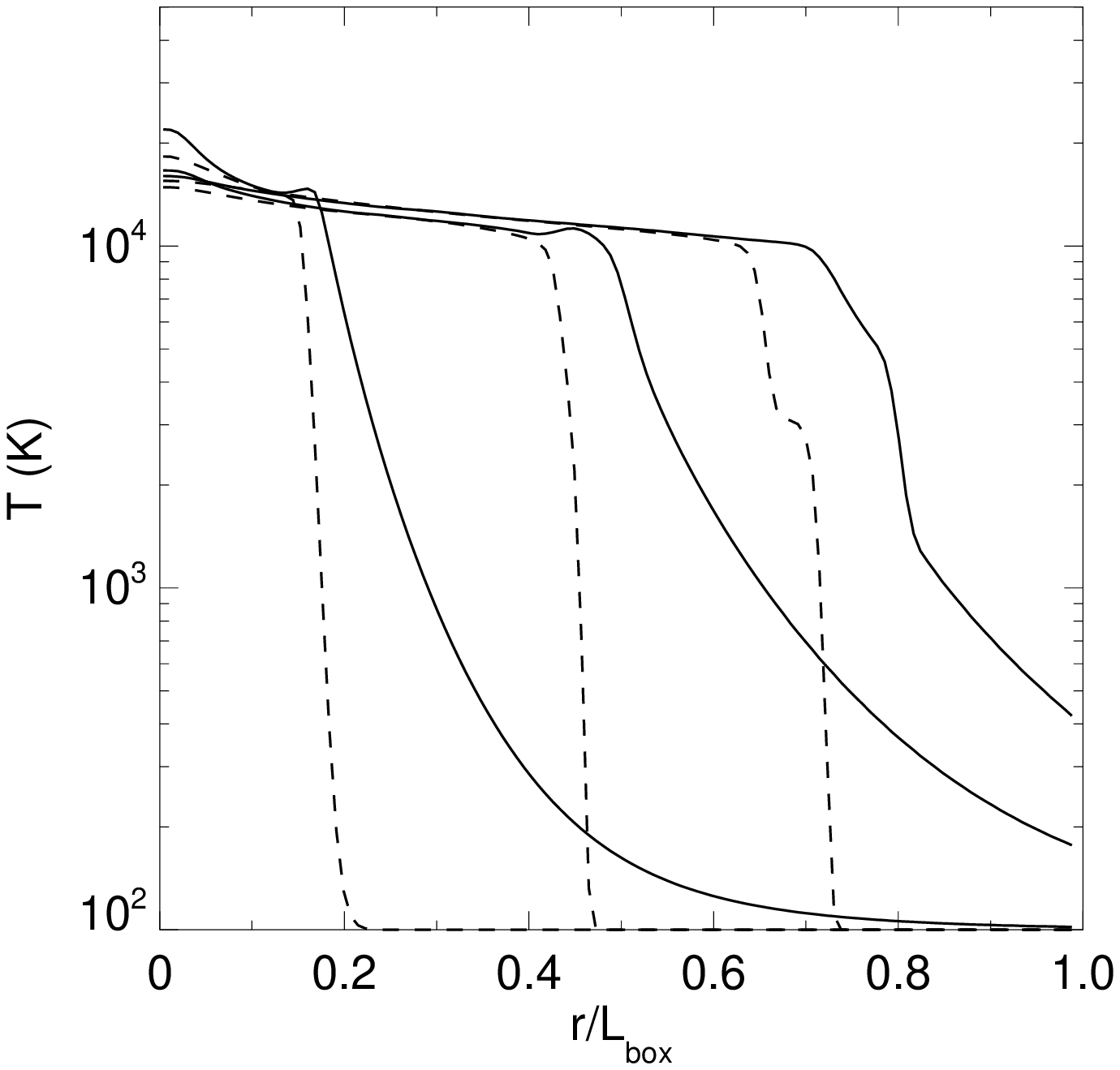}
\caption{ZEUS-MP Test 5 ionized fraction (left) and temperature (right) profiles
 with monoenergetic photons (dashed) and a 10$^5$ K blackbody spectrum (solid) at 
 times $t=10$ Myr (left pairs), 200 Myr (central pairs) and 500 Myr (right pairs)
 vs. dimensionless radius (in units of the box size). 
\label{ifrac_comparison}}
\end{center}
\end{figure*}

\begin{figure*}
\begin{center}
  \includegraphics[width=3.4in]{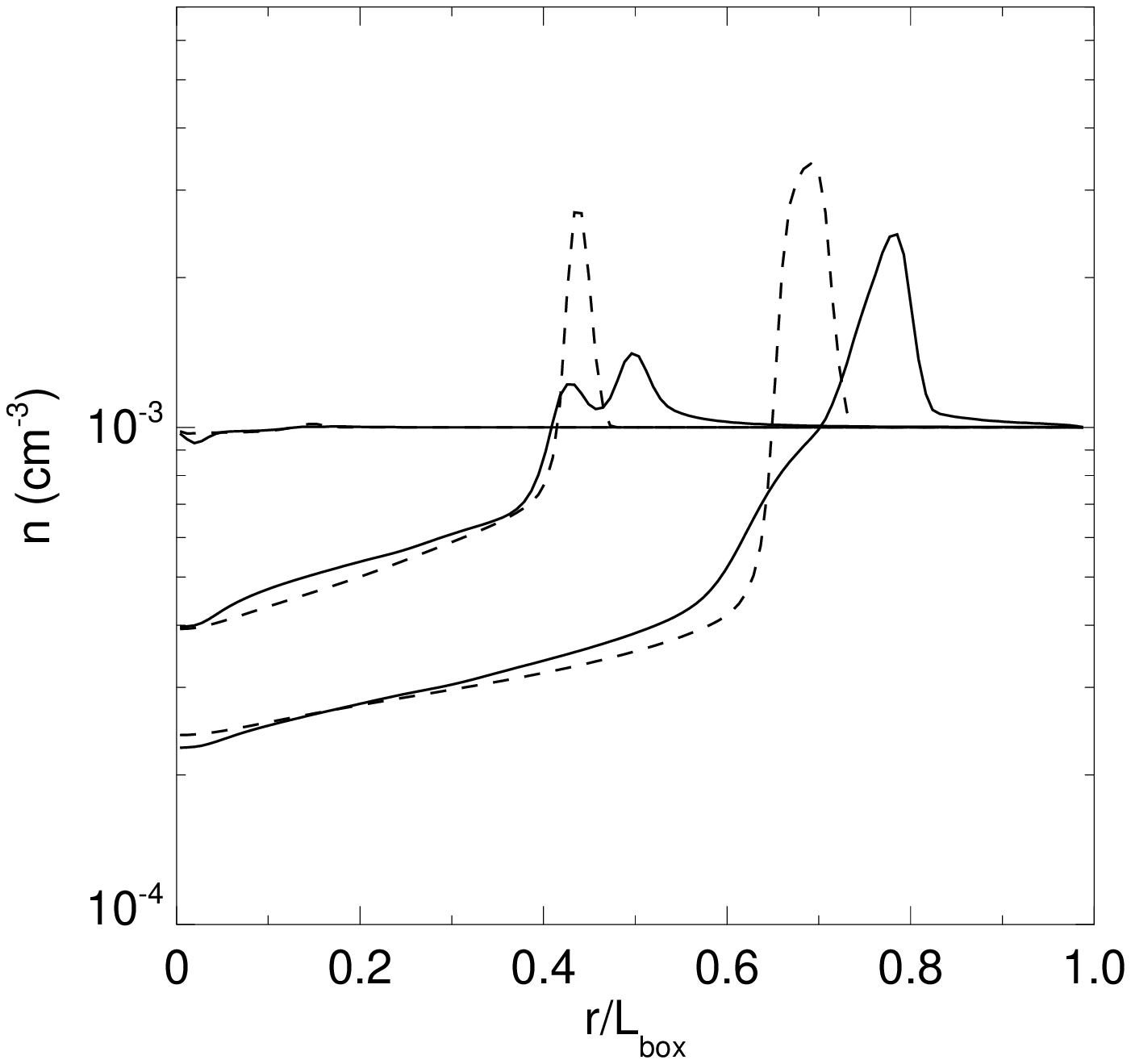}
  \includegraphics[width=3.4in]{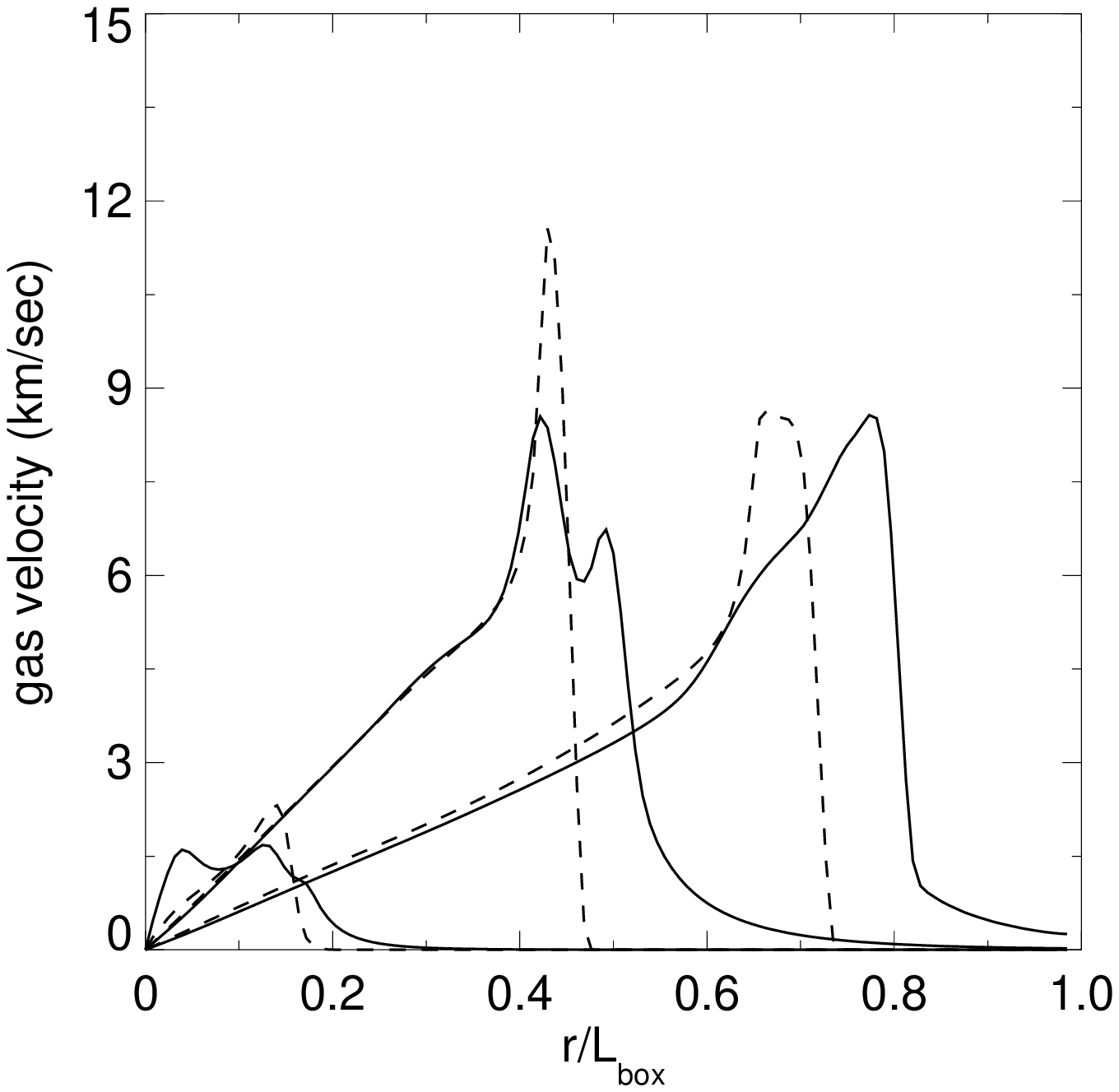}
\caption{ZEUS-MP Test 5 density (left) and gas velocity (right) profiles with 
 monoenergetic photons (dashed) and a 10$^5$ K blackbody spectrum (solid) at 
 times $t=10$ Myr (left pairs), 200 Myr (central pairs) and 500 Myr (right pairs)
 vs. dimensionless radius (in units of the box size). 
\label{dens_comparison}}
\end{center}
\end{figure*}

We start by comparing the fluid flow and ionization structure at two 
characteristic stages of the evolution, at $t=100$~Myr, of order of one 
recombination time, which is the start of the I-front conversion from R- 
to D-type, shortly before the initial Str\"omgren radius is reached, and 
at $t=500$~Myr, corresponding to a few recombination times, when the 
I-front is D-type preceded by a shock. In Figures~\ref{T5_images3_HI_fig}
- \ref{T5_images3_T_fig} and \ref{T5_images5_HI_fig} - \ref{T5_images5_T_fig} 
we show image cuts at coordinate $z=0$ of the neutral hydrogen fraction, 
pressure, and temperature at 100 Myr and at 500 Myr, respectively, while in 
Figures~\ref{T5_images5_HII_fig} and \ref{T5_images5_n_fig} we show the 
ionized fraction and number density at 500 Myr.  We note here that unlike 
the other simulations which are fully 3-D in both the hydrodynamic and the 
radiative transfer treatment, the RH1D results are 1-D spherically-symmetric
Lagrangian profiles mapped onto the 3-D Cartesian grid required in this
study. 

With a few exceptions, discussed below, all results exhibit reasonably 
good agreement throughout the flow evolution. As we found also for the 
static tests in Paper I, the majority of the differences are a consequence 
of the different handling of the energy equation and the hard photons with 
long mean free paths. These variations yield different spatial structures 
in the temperatures (Figures~\ref{T5_images3_T_fig}, \ref{T5_images5_T_fig} 
and \ref{T5_profsT_fig}) and ionized fractions in the gas just ahead of the 
I-front (which are dictated by the hard photons and non-equilibrium chemistry, 
Figures~\ref{T5_images5_HII_fig} and \ref{T5_profs_fig}), but very similar 
ionization profiles inside the H~II region (which sees the whole spectrum of 
photons and is mostly chemically equilibrated, Figures~\ref{T5_images3_HI_fig}, 
\ref{T5_images5_HI_fig} and \ref{T5_profs_fig}).
   
Regardless of the variations in the temperature and ionization profiles among 
the codes, the overall differences in I-front position and velocity are very 
modest, of order only a few percent, with the exception of Enzo-RT and, to a 
lesser extent, HART.  The hydrodynamical profiles also cluster fairly closely 
together.  The codes basically agree on the temperature structure of the 
evolving H II region over time except for HART, which predicts flat, lower 
temperatures at later times, and C$^2$-Ray, which yields higher ionized gas 
temperatures close to the ionizing source due to its simplified method for 
handling the energy, and again Enzo-RT because of its monochromatic spectrum. 
The reason for the sharp drop in pressure at 0.6 $L_{box}$ at 10 Myr in the 
HART results is unclear.

Apart from the differences discussed above, there are several features of 
the HART, LICORICE and Enzo-RT methods worth noting. The OTVET moment radiative
transfer method used in HART is somewhat diffusive, as was already noted in
Paper~I, which results in thicker I-front and less sharp flow features 
overall. There are some radial striations visible in the LICORICE results, 
especially in the temperature images that are reminiscent of those observed
in the CRASH code results in Test 2 of Paper~I. Since LICORICE adopts the 
Monte Carlo radiative transfer found in the original version of CRASH, the 
radial artifacts in its temperatures are similarly due to the noise in that 
version's energy sampling scheme, which has been corrected in the latest 
release of the CRASH code \citep{2009MNRAS.393..171M}. The wall effects in 
the upper left and lower right corners of the box in the HART pressure and 
Mach number images reflect the fact that mirror rather that transmissive 
boundary conditions were utilized. This is due to the natively-periodic nature
of the OTVET method, which demands special handling in order to run the 
non-periodic test problems in this comparison. The LICORICE, and to a lesser
extent the RSPH Mach numbers exhibit a somewhat grainy structure deep
inside the H~II region not visible in the other quantities. The origin of 
these features is likely due to the low SPH resolution in the evacuated 
interior of the H II region, which is nearly an order of magnitude lower in 
density than its surroundings. The difference in the degree of graininess 
between the two SPH codes may in part be due to how each code's particle 
data was mapped onto the Cartesian grid. The origin of the third outermost 
band in the RSPH Mach numbers, which is not present in those of the other 
codes, is likely due to the utilization of a larger box in that calculation
compared to the other cases, which changes the flow boundary conditions.  

\begin{figure}
\begin{center}
  \includegraphics[width=3.5in]{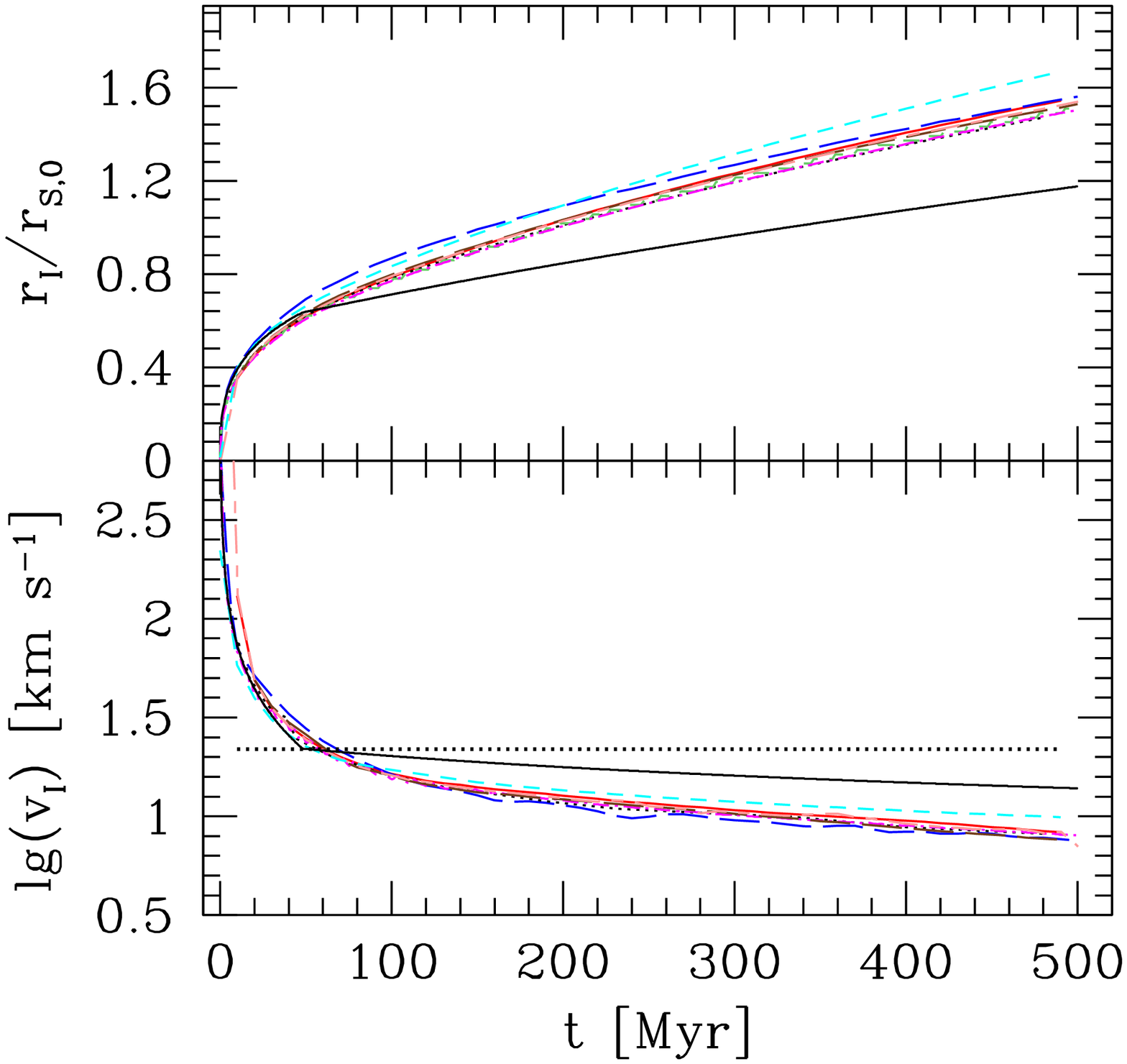}
\caption{Test 5 (H~II region gasdynamic expansion in an initially-uniform
  gas): The evolution of the position and velocity of the I-front. Solid 
  lines show the approximate analytical solution as described in the text.
  Dotted horizontal line indicates the approximate value of $v_{R}$.
\label{T5_Ifront_evol_fig}}
\end{center}
\end{figure}

Several important questions arise in this Test.  First, does the broadening 
of the front by high-energy photons from a hard UV source alter its radius 
as a function of time in comparison to a monochromatic front with the same
average ionized gas temperatures?  This issue is key because it determines
if the extensive approximate analytical solutions to hydrodynamical I-front
transport that exist in the literature apply to ionization fronts in which 
there is spectral hardening due to hard UV sources.  Second, how does the 
penetration of hard photons into the dense shocked neutral gas ahead of the 
I-front alter its structure and flow?  Third, how do these changes to the 
shocked flow alter its own rate of advance and that of the front?  Finally, 
what are the origin of the distinctive double peaks in density, velocity 
and Mach number in the full spectrum I-fronts at intermediate times, and 
why are they absent in the Enzo profiles?

We cannot resort to comparison of the present code results alone to resolve these 
questions because they are all are multifrequency in nature except for Enzo, 
and even Enzo integrates over the blackbody spectrum to implement the grey 
approximation to radiation transport.  These issues can only be settled by 
comparing the multifrequency I-front in Test 5 to a monochromatic one whose 
photon energy has been adjusted to yield the same average ionized gas 
temperature as for the 10$^5$ K blackbody spectrum.  This guarantees that 
any discrepancy in position between the two fronts will be due only to the 
broadening of the front and its modifications to the shocked flow just 
beyond it, not to differences in the average sound speed within the H II 
regions, which is primarily what determines the rate of advance of the 
I-front when it is D-type.  This approach also ensures that any variations 
in the structure of the shocked flow between the two I-fronts are due to 
spectral hardening only, since both are being driven by the same ionized gas 
pressure. 

To investigate these points and determine the origin of some of the features 
in the hydrodynamic profiles in Figures~\ref{T5_profs_fig} -\ref{T5_profsm_fig}, 
we performed two fiducial runs of Test 5 with ZEUS-MP.  The first was with the 
original 10$^5$ K blackbody spectrum and the second was with monoenergetic 
photons at 17.0 eV.  Both had the same ionizing photon rate $\dot{N}_\gamma=5
\times10^{48}\,\rm s^{-1}$.  The 17.0 eV monochromatic photons establish the 
same average
ionized gas temperature as in the multifrequency H II region in ZEUS-MP.  We 
show ionized fractions, temperatures, velocities, and densities for the two 
runs at $t=$10, 200, and 500 Myr in Figures \ref{ifrac_comparison} and 
\ref{dens_comparison}.  The broadening of the I-front in the multifrequency 
calculation is apparent at all three times in the ionized fractions, becoming 
greater as the front expands.  In contrast, the monoenergetic I-front remains
sharp, intersecting the multifrequency front at very nearly the same ionized
fraction at all three radii.  Except for small differences in the elevated 
values near the source, the two H II regions exhibit nearly identical 
temperatures out to where the full-spectrum I-front broadens.  Out to this 
same radius the density and velocity profiles are also nearly identical. 

The ionization profiles demonstrate that the position of the I-front as a
function of time is not signficantly altered by either the broadening of the 
front or the partial ionization and heating of the dense shocked gas in front
of it by high-energy photons, at least for the radii considered in this problem.  
This affirms that the global dynamics of the D-type I-front depend primarily 
on the temperature (and hence sound speed) of the ionized gas.  Past tests by 
one of us of D-type I-fronts in $r^{-2}$ density gradients confirm that this 
holds well beyond the R to D transition surveyed in this Test.  Likewise, 
these two models demonstrate that the ionized flow within the H II region is 
also mostly unchanged by spectral hardening over the radii enclosed by the 
computational box. However, here it is important to distinguish between the 
motion of the I-front and that of the shocked flow it drives. The latter is 
dramatically altered by spectral hardening as we discuss below.

The simple-structured 3000 K layer of shocked gas driven by the monochromatic 
I-front is split into the double-peaked structure in the full-spectrum front 
at $t=200$~Myr and is present in the profiles of all the codes except Enzo, as 
shown in Figures~\ref{T5_profsn_fig} and \ref{T5_profsm_fig}.  This feature 
is transient and disappears by $t=500$~Myr.  Its origin is the heating of the 
dense shell by the high-frequency photons.  At 200 Myr, they partially ionize
the base of the shocked shell: ionized fractions of 10\% or more extend out to 
0.5 $L_{box}$. The high frequency tail of the spectrum cannot maintain large 
ionized fractions in this layer but does effectively deposit heat there, as 
evidenced by the rise in temperature at 0.45 $L_{box}$, which is positioned 
approximately in the valley between the two peaks in the density and velocity 
profiles. This energy ablates the lower layer of the dense shocked shell, 
driving both inward and outward photoevaporative flows in the frame of the 
shock that split the density and velocity peaks into two smaller ones. The 
forward flow accelerates the gas in the outer peak to 7 km s$^{-1}$ by 200 Myr. 
However, pressure gradients from the ionized interior of the H~II region drive 
the inner peak to higher velocities that cause it to later overtake the forward 
peak (500 Myr). At this distance from the central source, high energy photons do 
heat the base of the shocked shell but not to sufficient temperatures to create 
backflow, as seen in the disappearance of the temperature bump that was present 
at 200 Myr. However, they do smear out the sharp interface between the ionized 
and shocked gas temperatures that is present in the monoenergetic front at 500 
Myr.  In contrast, monoenergetic photons result in much simpler structure in 
both the front and the dense shell at 200 and 500 Myr.  At 500 Myr, $\sim$15,000~K 
ionized gas drives a clearly defined shocked shell and there are no ablation 
flows.  The absence of backflows allow peak gas velocities to reach higher 
values in the shell at intermediate times than in the hard spectrum case. 

What effect does pre-heating by hard photons leaking ahead of the I-front have
on the propagation of the shocked flow?  It weakens the shock, as evidenced by
the smaller density jump, lowering the density compression there and thus
enlarging its detachment from the I-front.  This can be clearly seen by comparing 
the position of the shock for the two I-fronts in Figure \ref{dens_comparison}.  
Thus, while the positions of the two fronts are nearly identical, the shocked 
flow of the multifrequency front is well ahead of that of the monochromatic one. 
It is clear from this comparison that multifrequency photon transport, or the 
use of lookup tables of ionizing rates as a function of optical depth as a proxy, 
is necessary to capture the correct structure of I-fronts and shocks driven by 
high-temperature UV sources.

\begin{figure}
\begin{center}
  \includegraphics[width=3.5in]{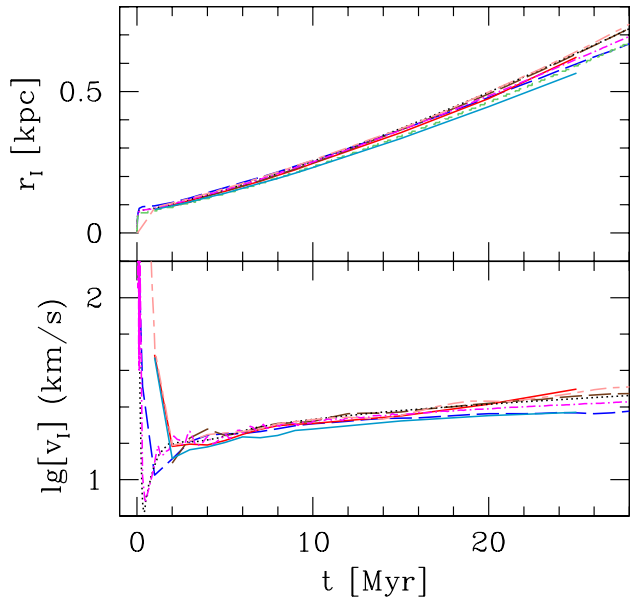}
\caption{Test 6 (H~II region gasdynamic expansion down a power-law initial
  density profile): The evolution of the position and velocity of the I-front.
\label{T6_Ifront_evol_fig}}
\end{center}
\end{figure}

Having isolated the effect of spectral hardening on both the dynamics of 
the I-front and on the shocked neutral flows it drives, we can now disentangle
the sometimes competing effects that account for the differences that do exist
among the hydrodynamic profiles in Test 5, both among the codes with 
multifrequency physics and between those codes and Enzo.  
Setting aside the Enzo results for the moment, the minor spread in I-front 
position in the other codes can now be traced to the variations of their H II 
region temperatures. This is primarily due to how each code handles the energy 
equation.  Furthermore, it is now clear that (1) the origin of the temperature 
bumps in the shocked gas at intermediate times; (2) the double peaks in the 
densities, velocities, and Mach numbers; and (3) shocks that are more fully 
detached from the I-front are all a consequence of spectral hardening.  

The profiles that are most distinct from the rest are those of Enzo-RT.  
This can now be understood to be due to the lack of spectral hardening in this 
code.  Although it carries out an integration over the 10$^5$ K spectrum to
compute the photoionization cross sections, Enzo-RT it employs a grey 
approximation for the photoionization cross-section, i.e. a cross-section
independent of the frequency and is therefore essentially monochromatic in 
its current form. No hard photons 
means no pre-heating ahead of the front and also a much sharper I-front.  This 
lack of pre-heating in turn means that, as in the ZEUS-MP monochromatic results 
above, the Enzo-RT shock is stronger than the others, as evidenced by its higher 
Mach number, density compression and pressure jump.  Because the Enzo I-front 
remains sharp, its shocked flow also exhibits the single peak associated with 
monoenergetic I-fronts.  Because it is stronger, the shock in the Enzo profiles 
propagates somewhat more slowly, lagging behind those of the other codes.  On 
the other hand, the Enzo-RT I-front actually leads the others, almost coinciding 
with the shock.  This is a consequence of the greater average ionized gas 
temperatures in its H II region in comparison to the others, a result of its 
integration over the blackbody spectrum.  The Enzo-RT temperature profile is 
similar to the flat profile found by HART, but at a higher value. Its origin is 
unknown but could be associated with its flux-limited diffusion radiative transfer, 
which shares some similarities with the OTVET method in HART.

Except for the variations noted above, the codes agree well on I-front 
position and velocity but are at variance with the analytical solution,
leading it by roughly 30\%. This is in part because the solution plotted 
in Figure 17 is for fixed temperatures of 15,000 K behind and 1000 K ahead
of the I-front (allowing for some pre-heating by hard photons, which alters
the values of $v_R$ and $v_D$), while in reality the profiles exhibit a
complex temperature structure. Furthermore, as we discussed above, the 
analytical solution describes well the early and late evolution, but not 
the intermediate one. The H II region radial evolution at late times does 
not exactly match the asymptotic $t^{4/7}$ slope predicted for self-similar 
flows but approaches it at larger radii.  This is to be expected since the 
box size was chosen to enclose only the transition of the I-front from R-type 
to D-type, during which the assumption of self-similarity is not satisfied.  
The codes asymptotically approach the expected solution as the fronts grow in 
radius.

\begin{figure*}
\begin{center}
  \includegraphics[width=2.3in]{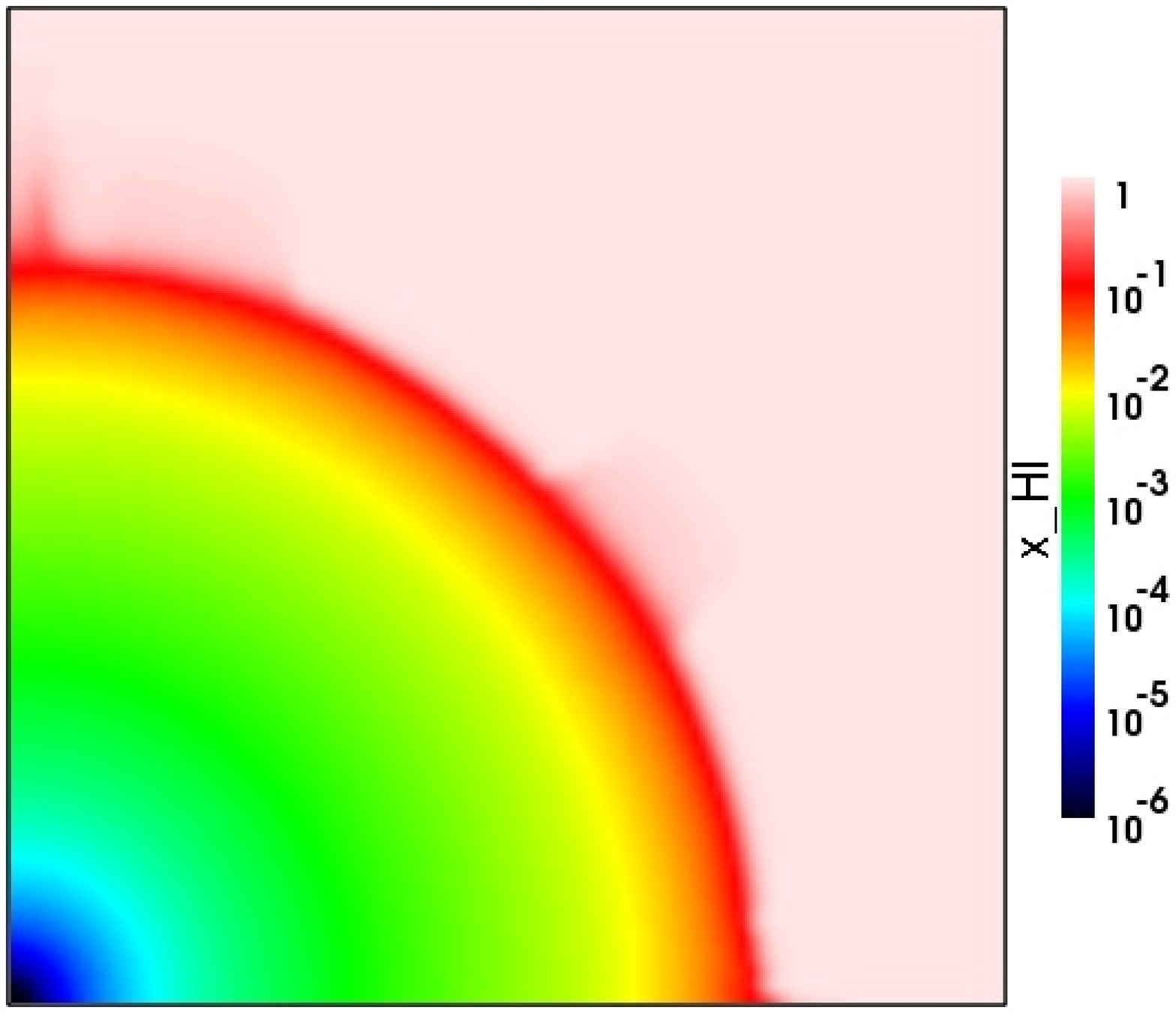}
  \includegraphics[width=2.3in]{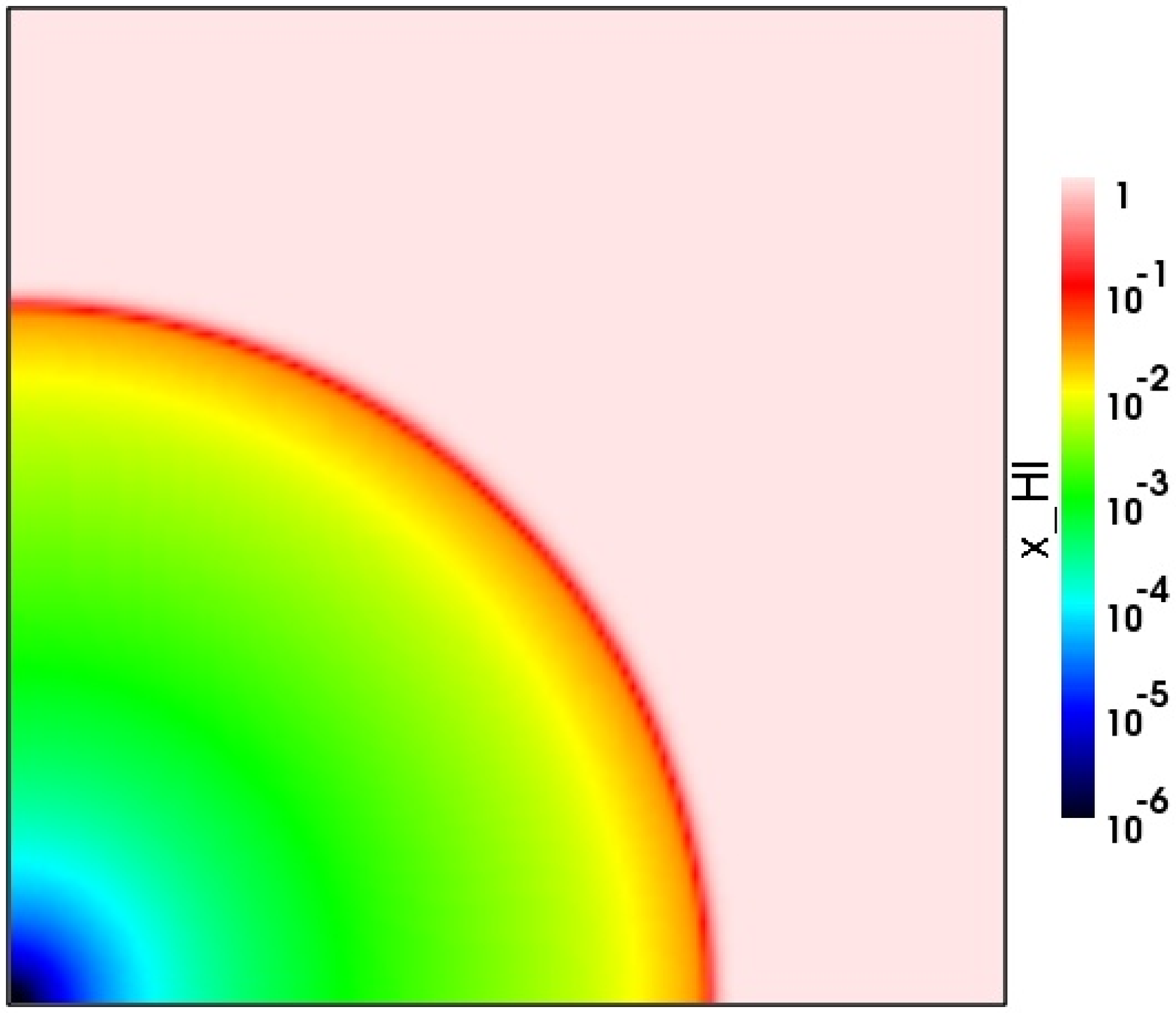}
  \includegraphics[width=2.3in]{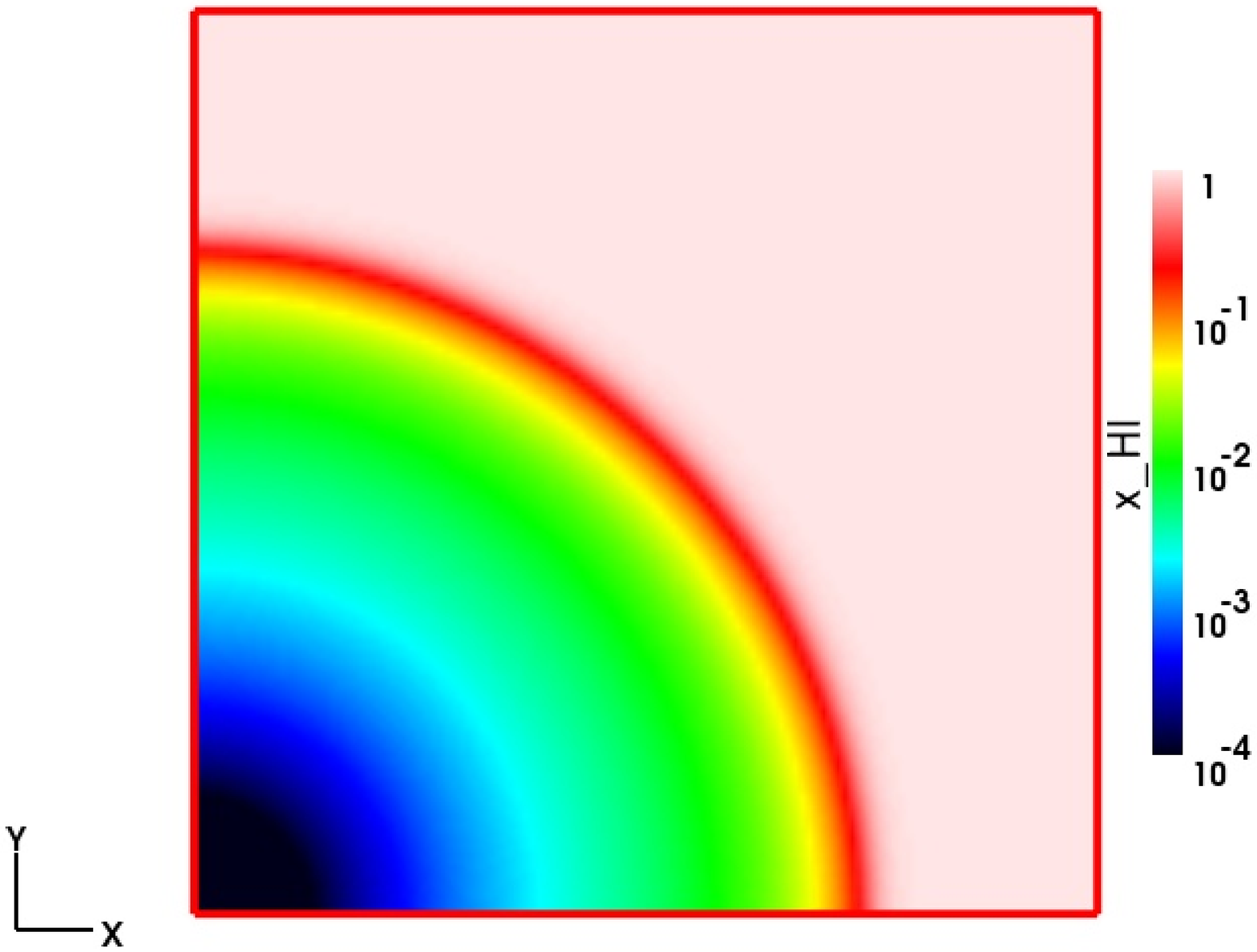}
  \includegraphics[width=2.3in]{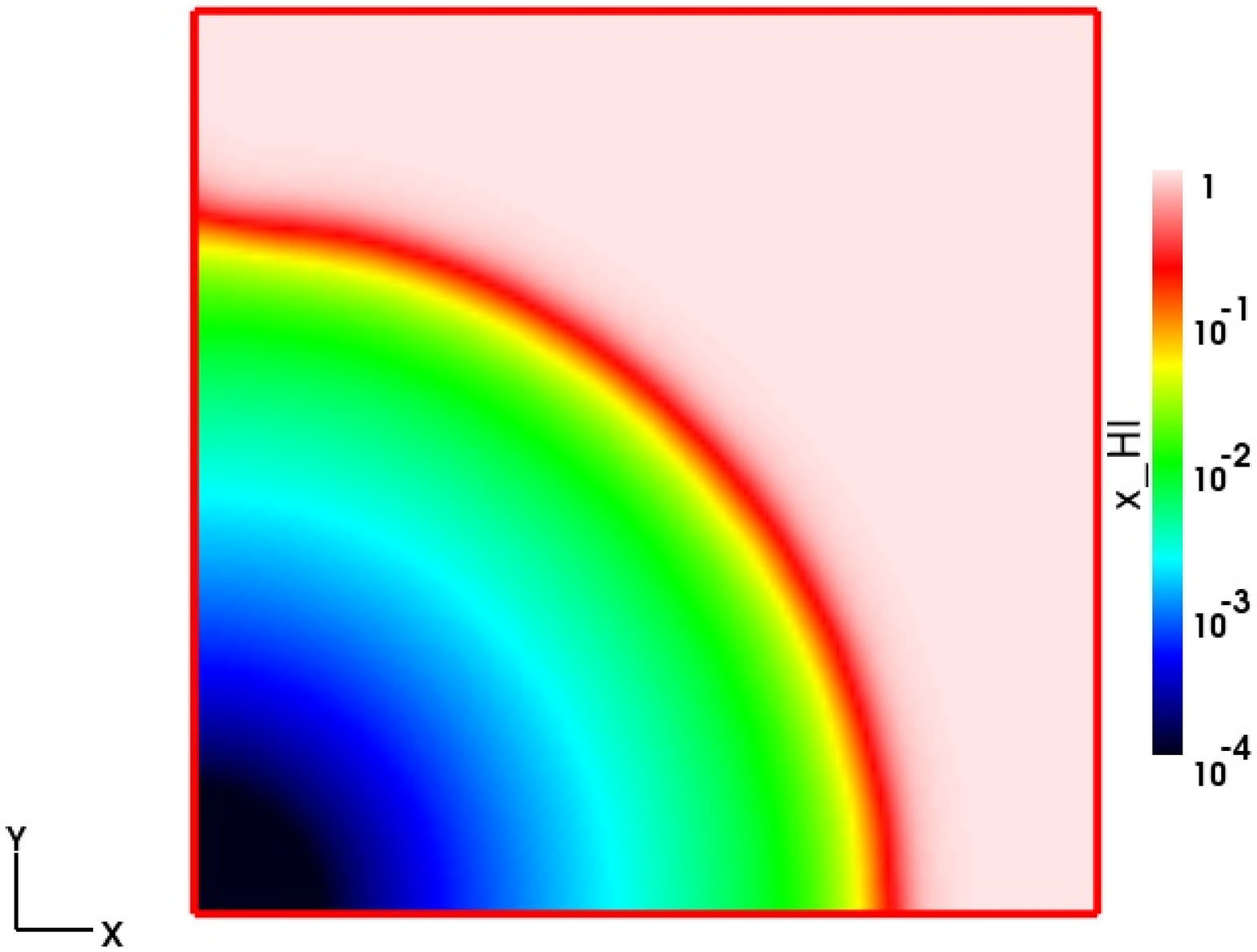}
  \includegraphics[width=2.3in]{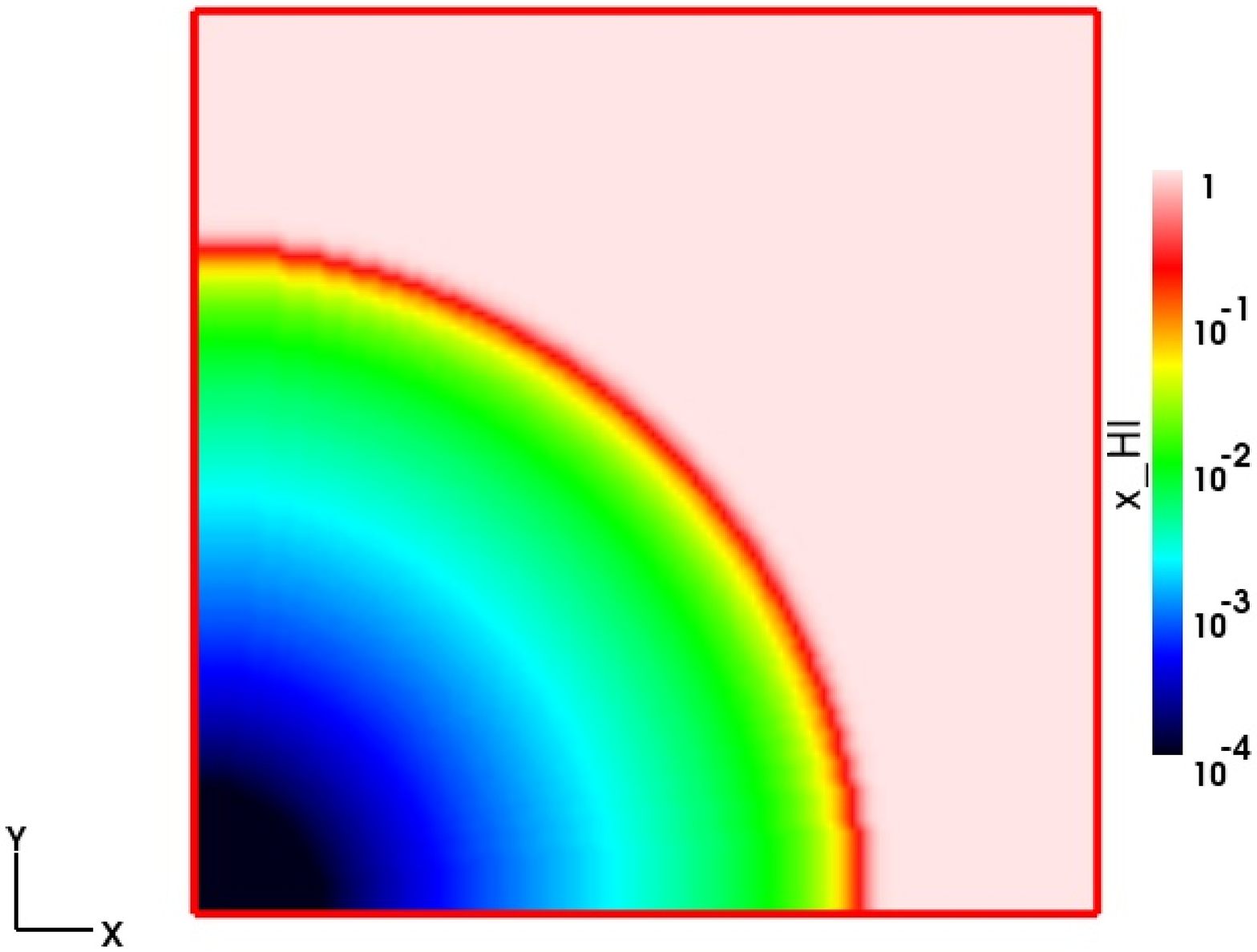}
  \includegraphics[width=2.3in]{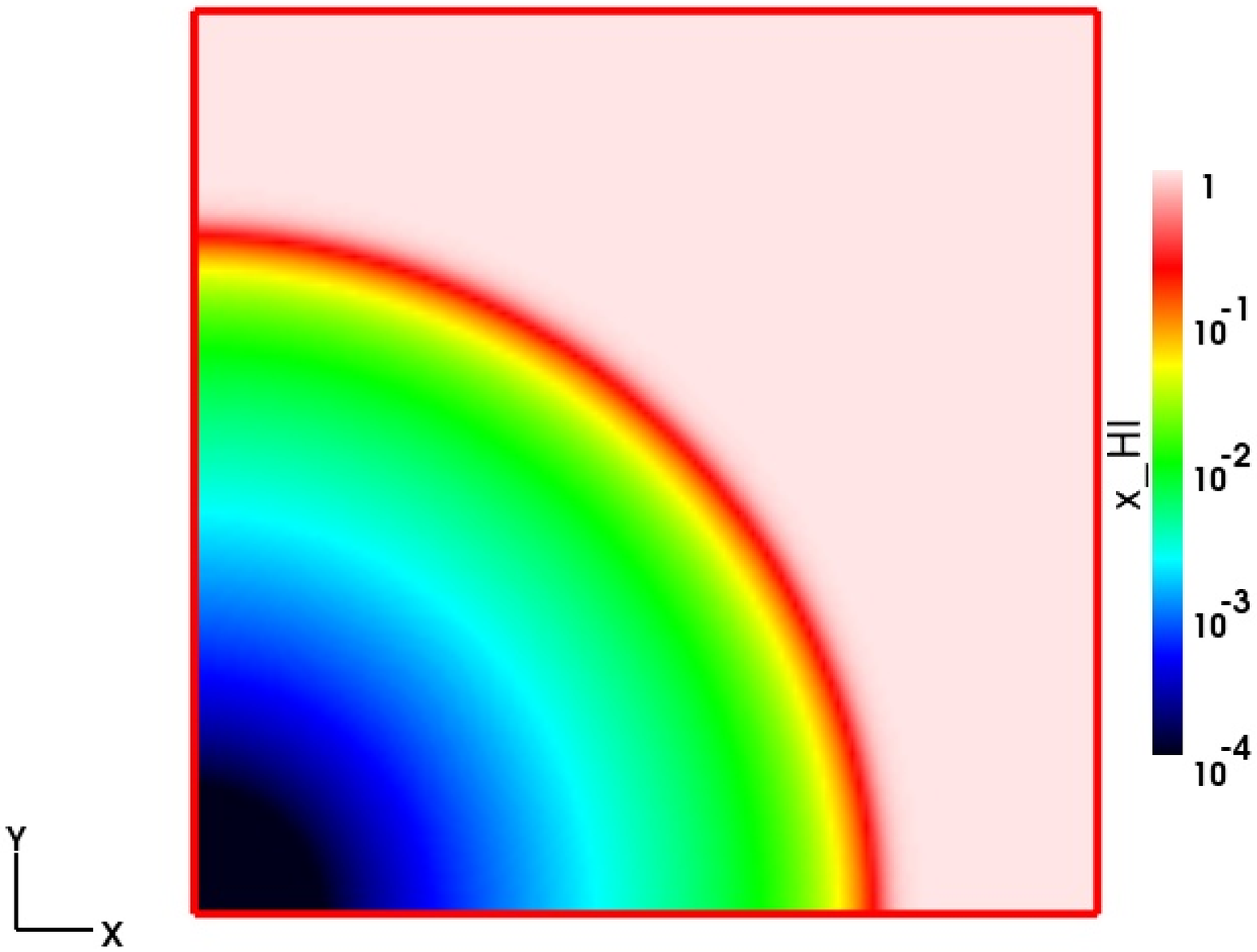}
  \includegraphics[width=2.3in]{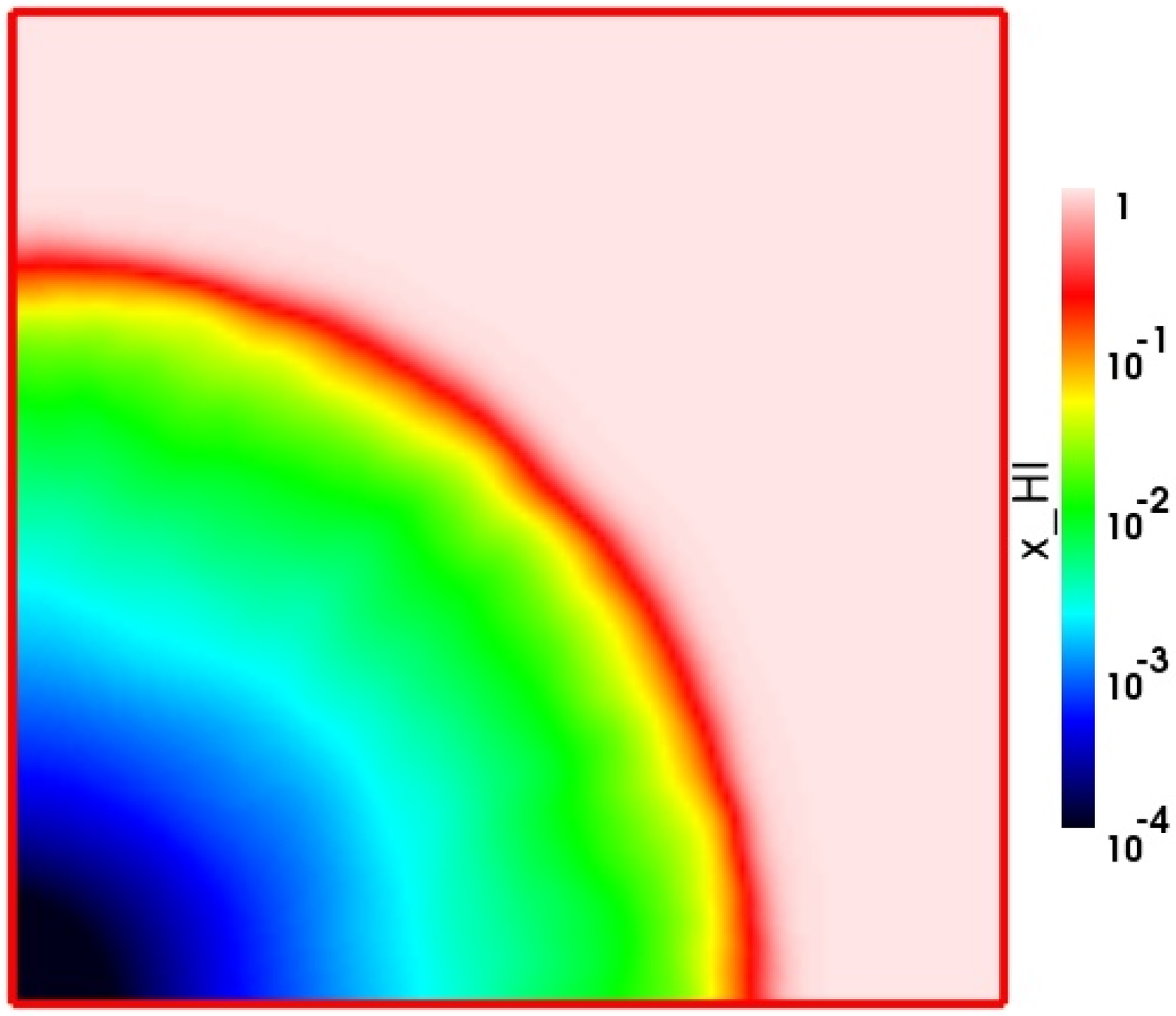}
  \includegraphics[width=2.3in]{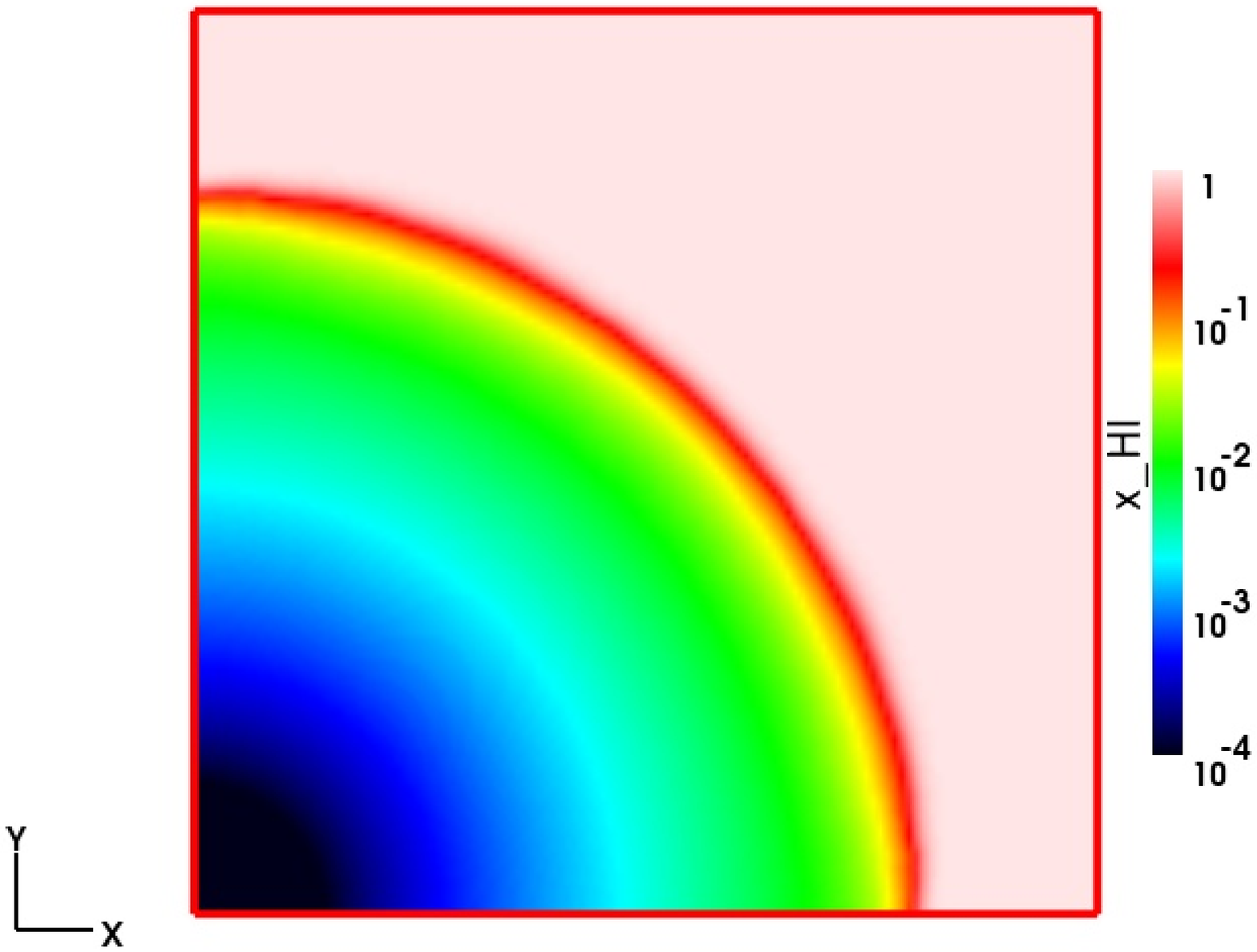}
\caption{Test 6 (H~II region gasdynamic expansion down a power-law initial
  density profile): Images of the H~I fraction, cut through the simulation
  volume at coordinate $z=0$ at time $t=25$ Myr for (left to right and top 
  to bottom) Capreole+$C^2$-Ray, TVD+$C^2$-Ray, HART, RSPH, ZEUS-MP, RH1D, 
  LICORICE, and Flash-HC.
\label{T6_images4_HI_fig}}
\end{center}
\end{figure*}

Finally, we note that the I-fronts in this test are dynamically stable.  
If H$_2$ cooling, LW photodissociation, and self-shielding to LW photons 
had been included, violent hydrodynamical instabilities mediated by H$_2$ 
cooling might have erupted in the fronts after becoming D-type, as 
explained in greater detail in Test 6 below.  Line cooling in H  
alone appears to be unable to incite such instabilities \citep{wn08a}.


\subsection{Test 6} 

\begin{figure*}
\begin{center}
  \includegraphics[width=2.3in]{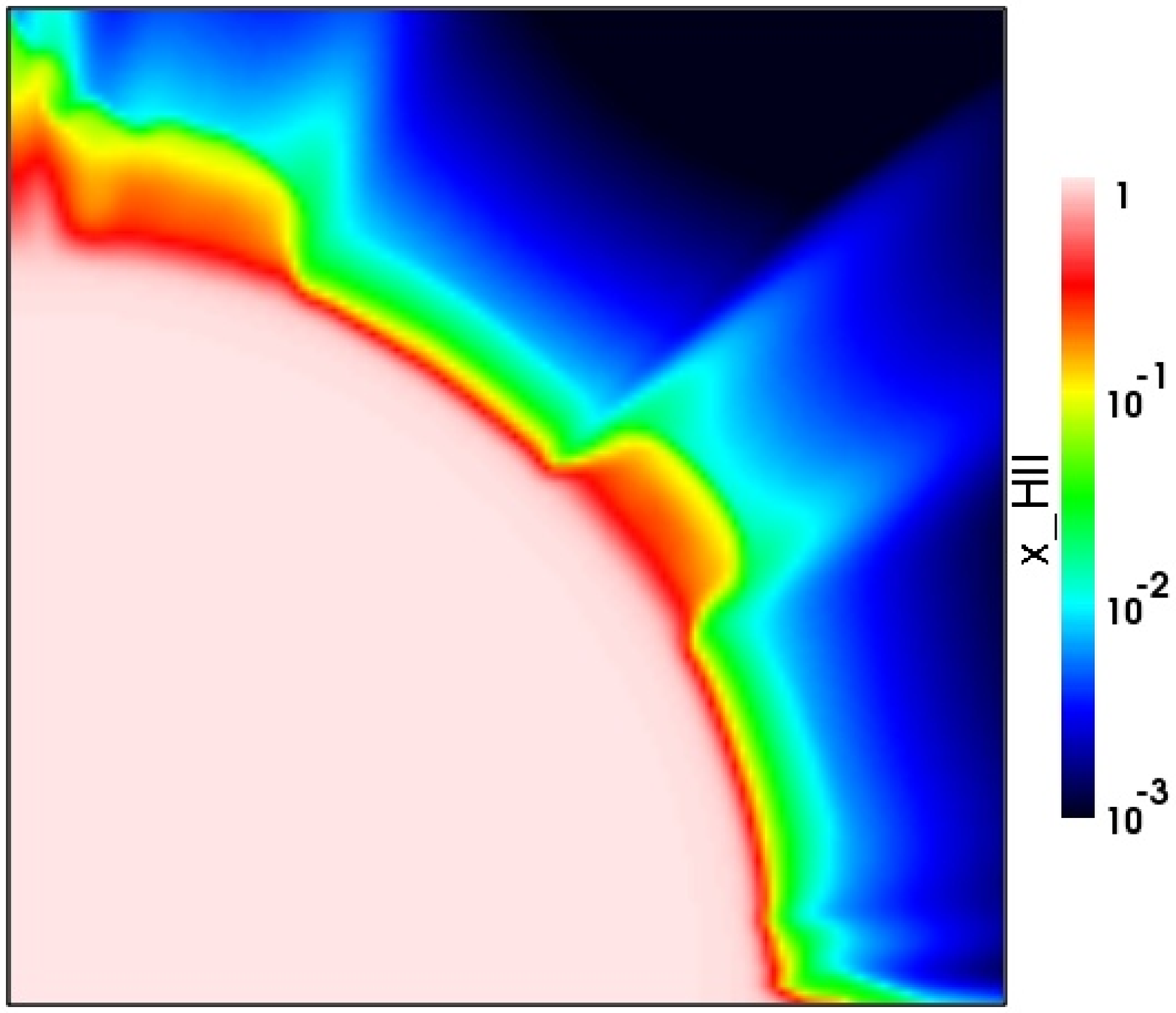}
  \includegraphics[width=2.3in]{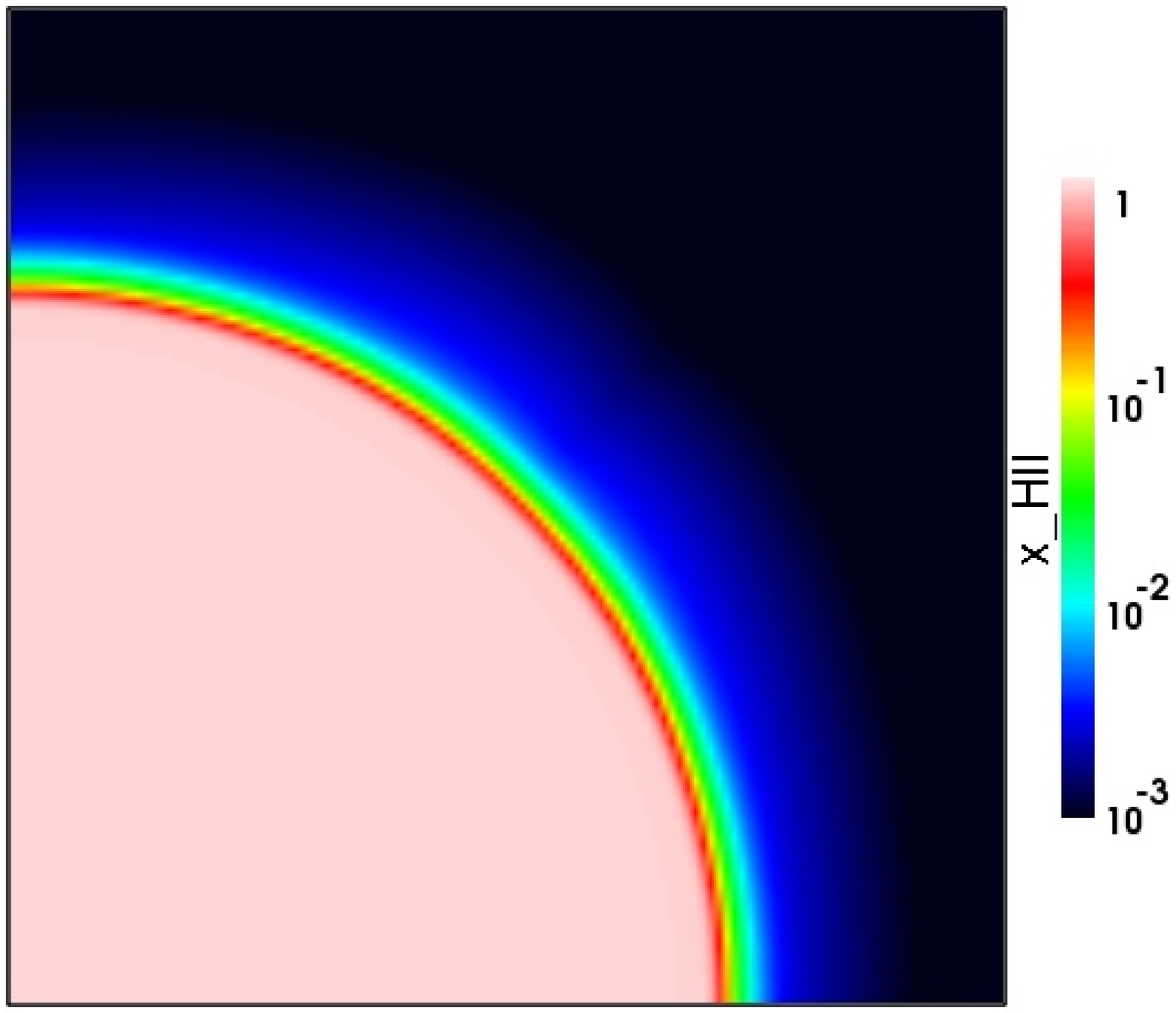}
  \includegraphics[width=2.3in]{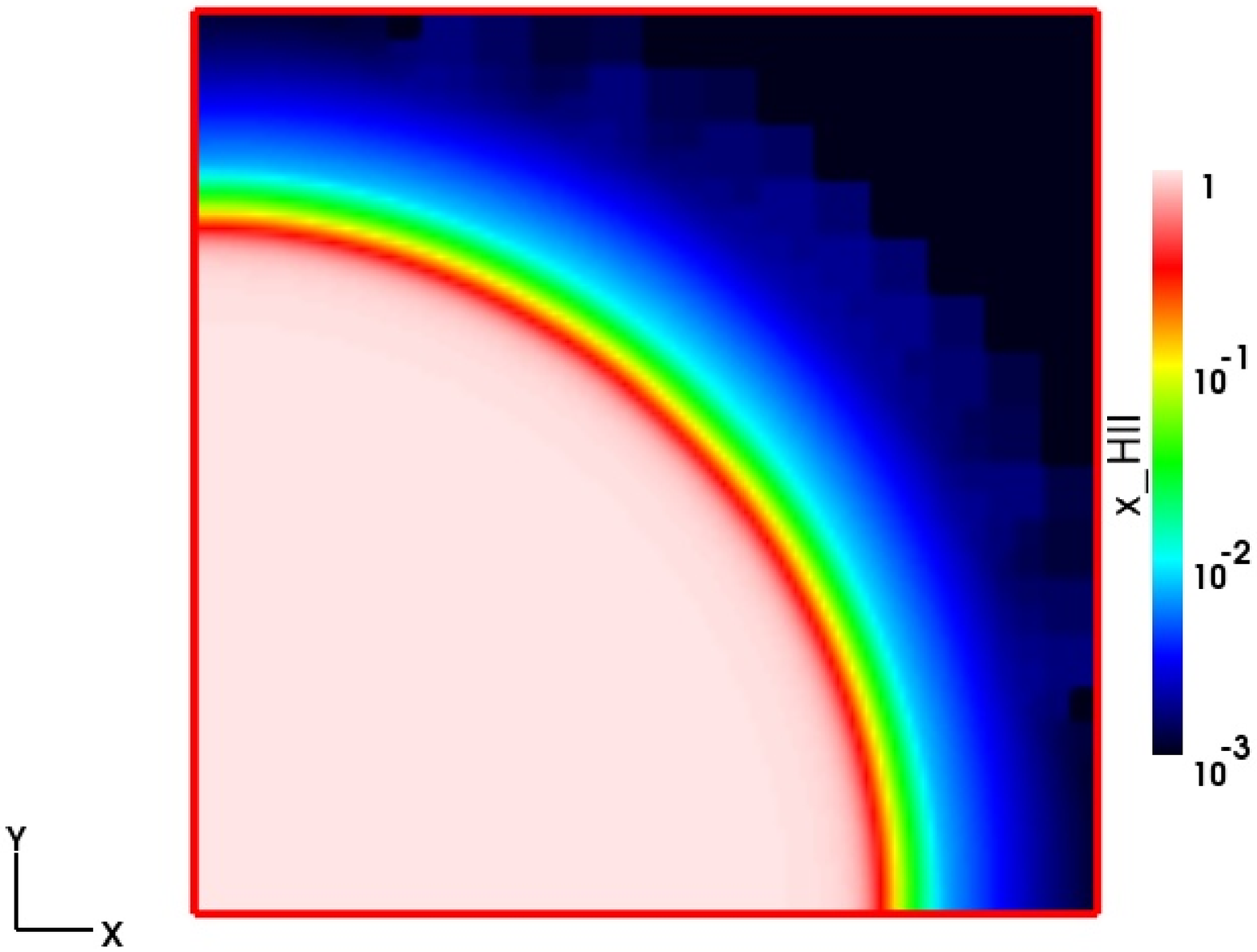}
  \includegraphics[width=2.3in]{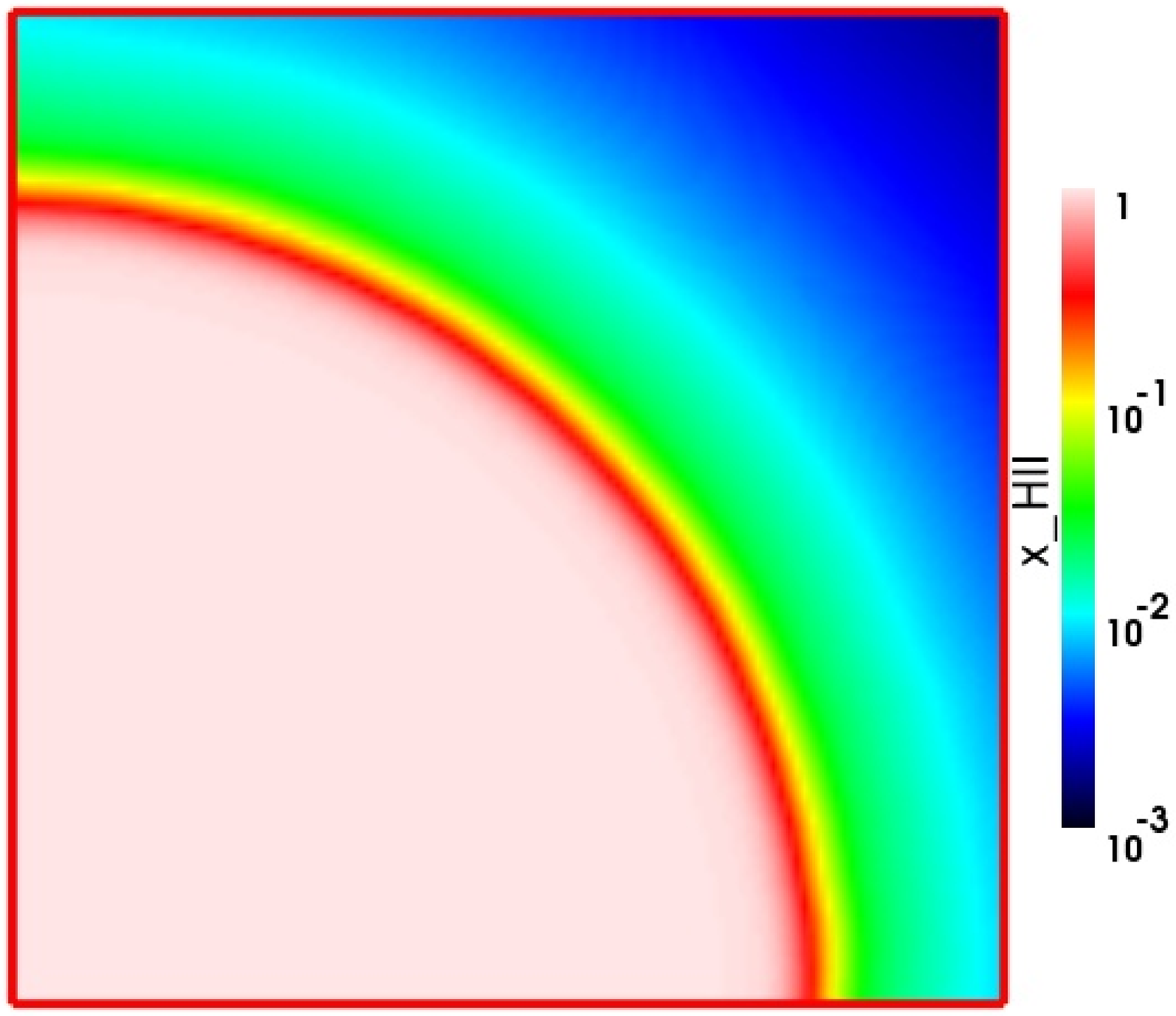}
  \includegraphics[width=2.3in]{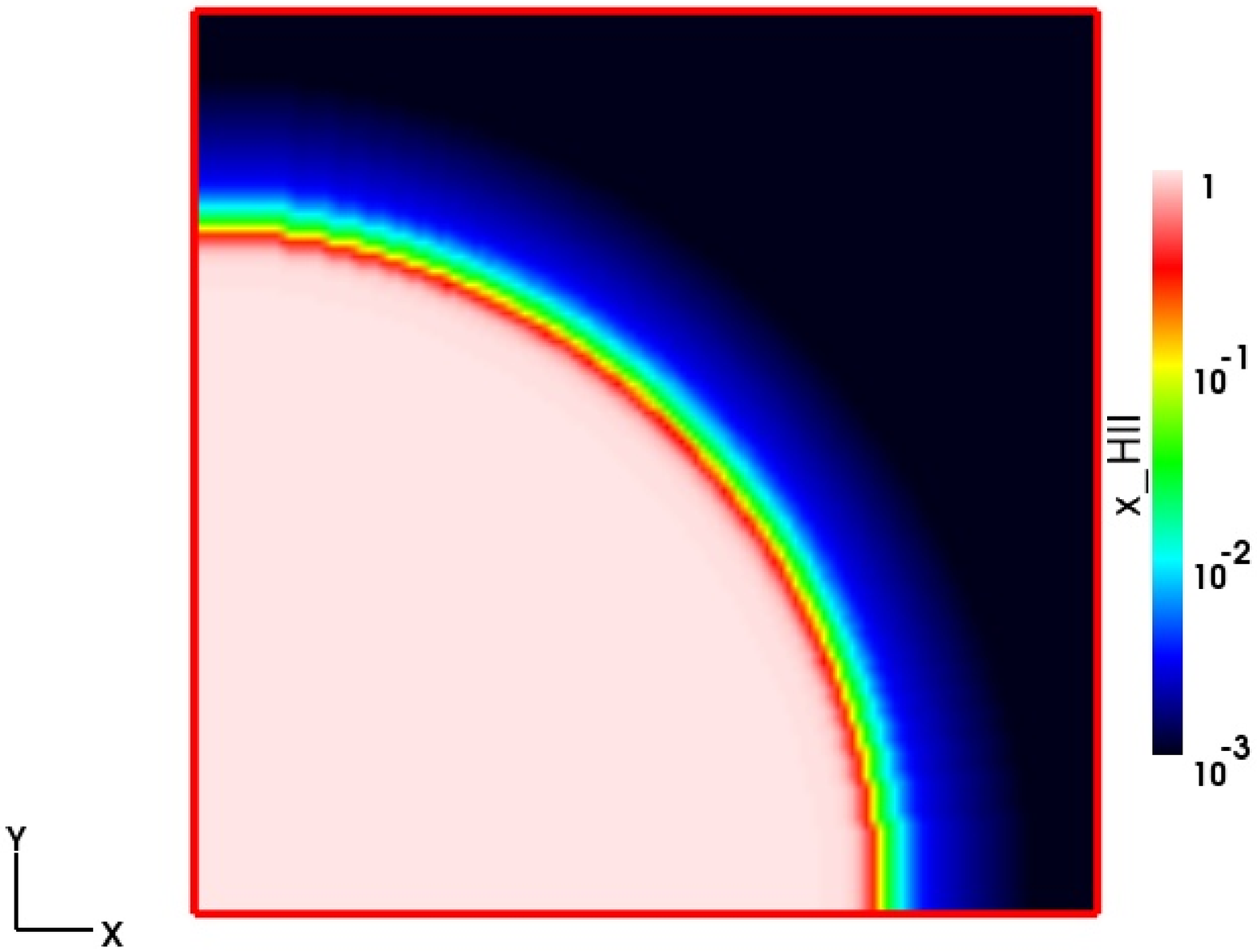}
  \includegraphics[width=2.3in]{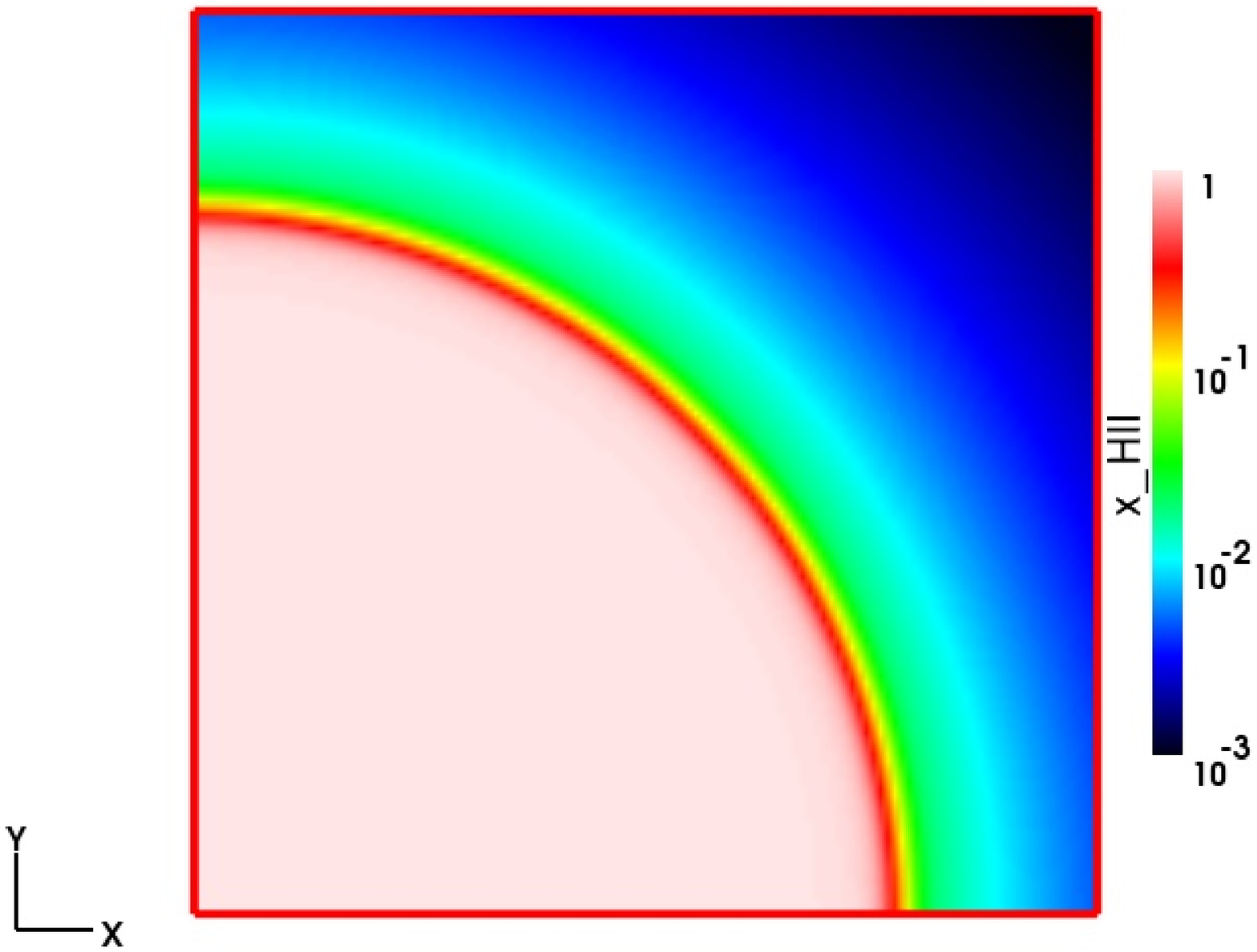}
  \includegraphics[width=2.3in]{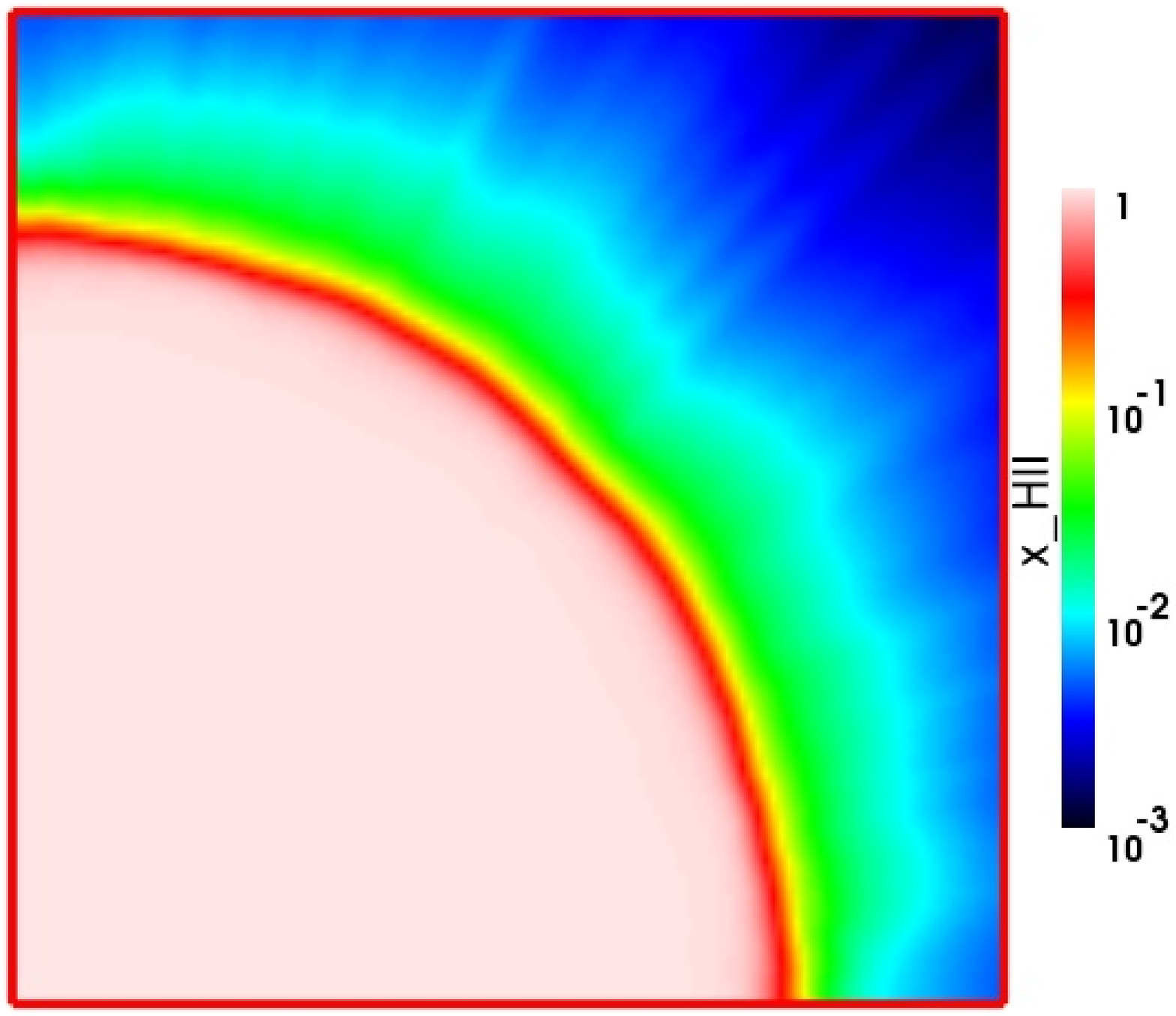}
  \includegraphics[width=2.3in]{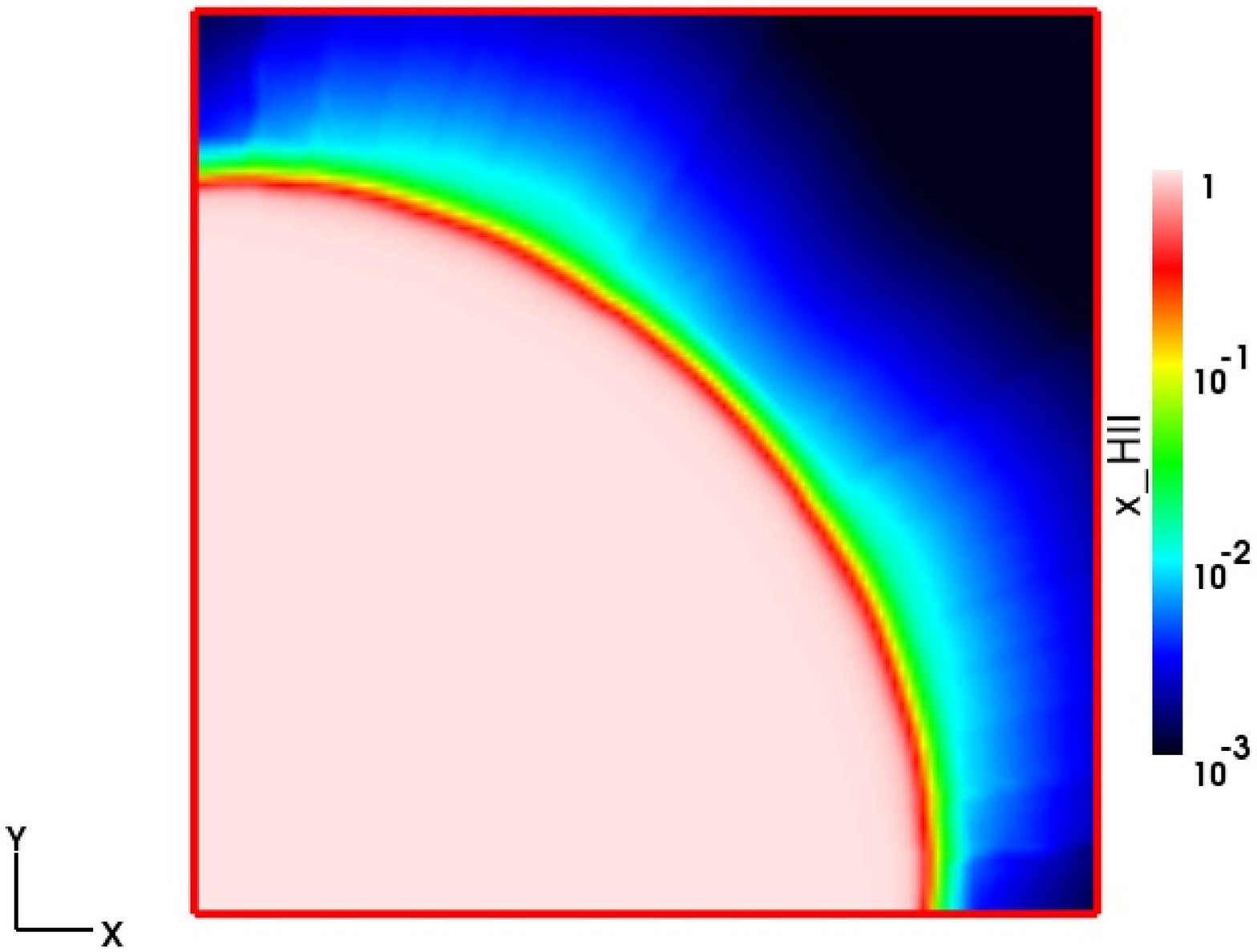}
\caption{Test 6 (H~II region gasdynamic expansion down a power-law initial
  density profile): Images of the H~II fraction, cut through the simulation
  volume at coordinate $z=0$ at time $t=25$ Myr for (left to right) and top 
  to bottom) Capreole+$C^2$-Ray, TVD+$C^2$-Ray, HART, RSPH, ZEUS-MP, RH1D, 
  LICORICE, and Flash-HC.
\label{T6_images4_HII_fig}}
\end{center}
\end{figure*}

We start our analysis with a head-to-head comparison of the evolution of 
the position and velocity of the I-front, plotted in 
Figure~\ref{T6_Ifront_evol_fig}. In the velocity plot we clearly see the
evolution stages outlined in the discussion of Test 6 above. Initially, 
while it is still within the density core the I-front moves very fast 
(is of R-type), but precipitously slows down as it approaches its 
Str\"omgren radius (whose precise value is temperature-dependent, but is
slightly smaller than the core radius chosen here). The fast R-type phase 
is over within a fraction of a Myr, after which the expansion becomes 
pressure-driven, and the front itself converts to a D-type led by a shock. 
The I-front speed reaches a minimum of just a few kilometers per second,
well below $v_R$ - the critical velocity defined in \S~\ref{sec:T5}. We 
note that although some of the results appear to never show I-front 
velocities below $\sim10$~km/s, this is in fact due to insufficient number 
of early-time snapshots being saved in the I-front evolution data. 
For this reason the short transition stage does not appear in some of 
the plotted results, and this does not imply any problem with the codes.
The later-time evolution is not affected, as long as the actual 
time-stepping in the code is sufficiently fine to properly follow the 
early evolution. Once out of the core, the I-front re-accelerates as it 
descends the steep $r^{-2}$ density gradient, eventually reaching speeds 
of $25-28$ km/s. These speeds never surpass $v_R$ and therefore the front 
remains D-type until leaving the computational volume at $t\sim30$~Myr 

\begin{figure*}
\begin{center}
  \includegraphics[width=2.3in]{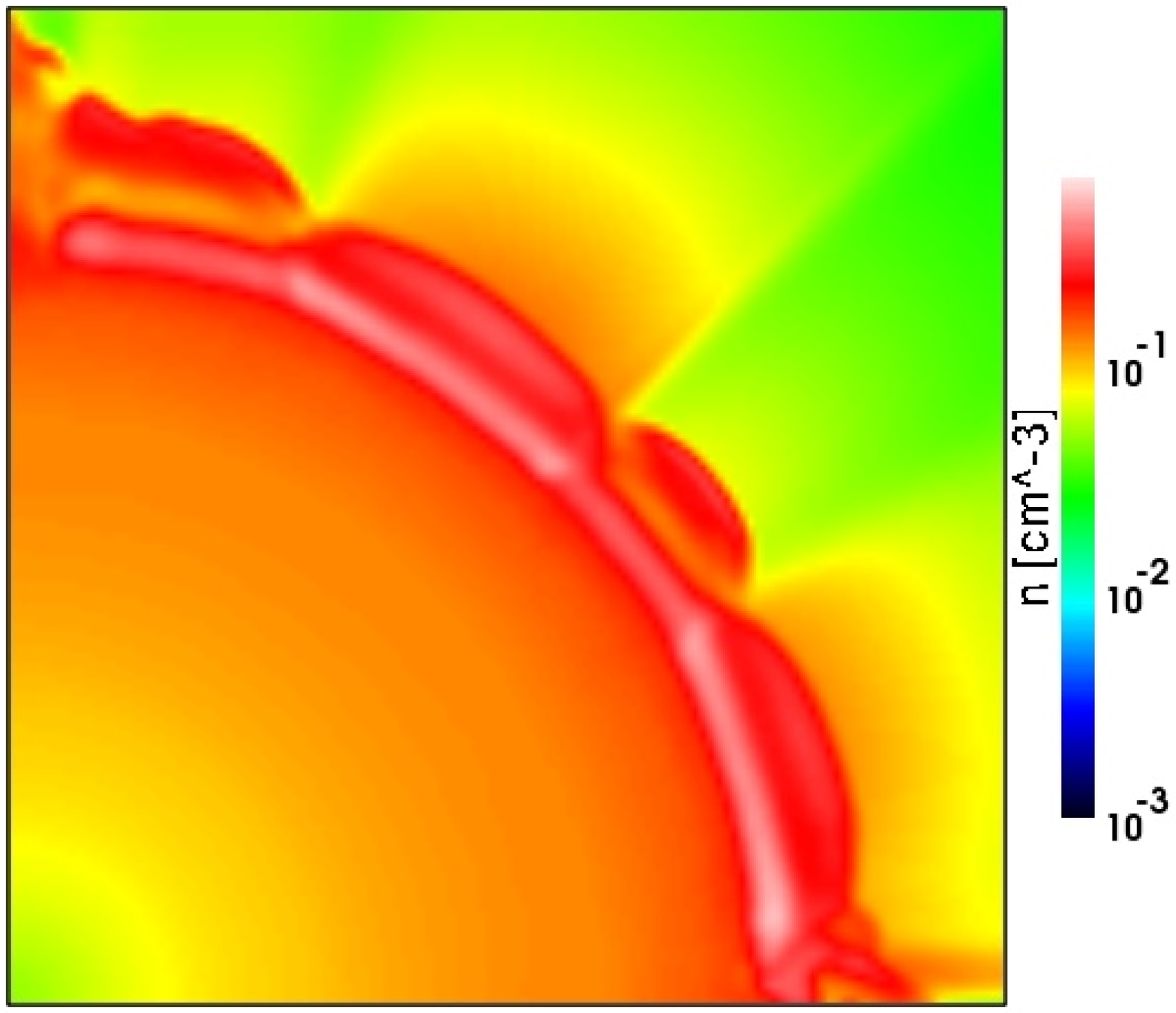}
  \includegraphics[width=2.3in]{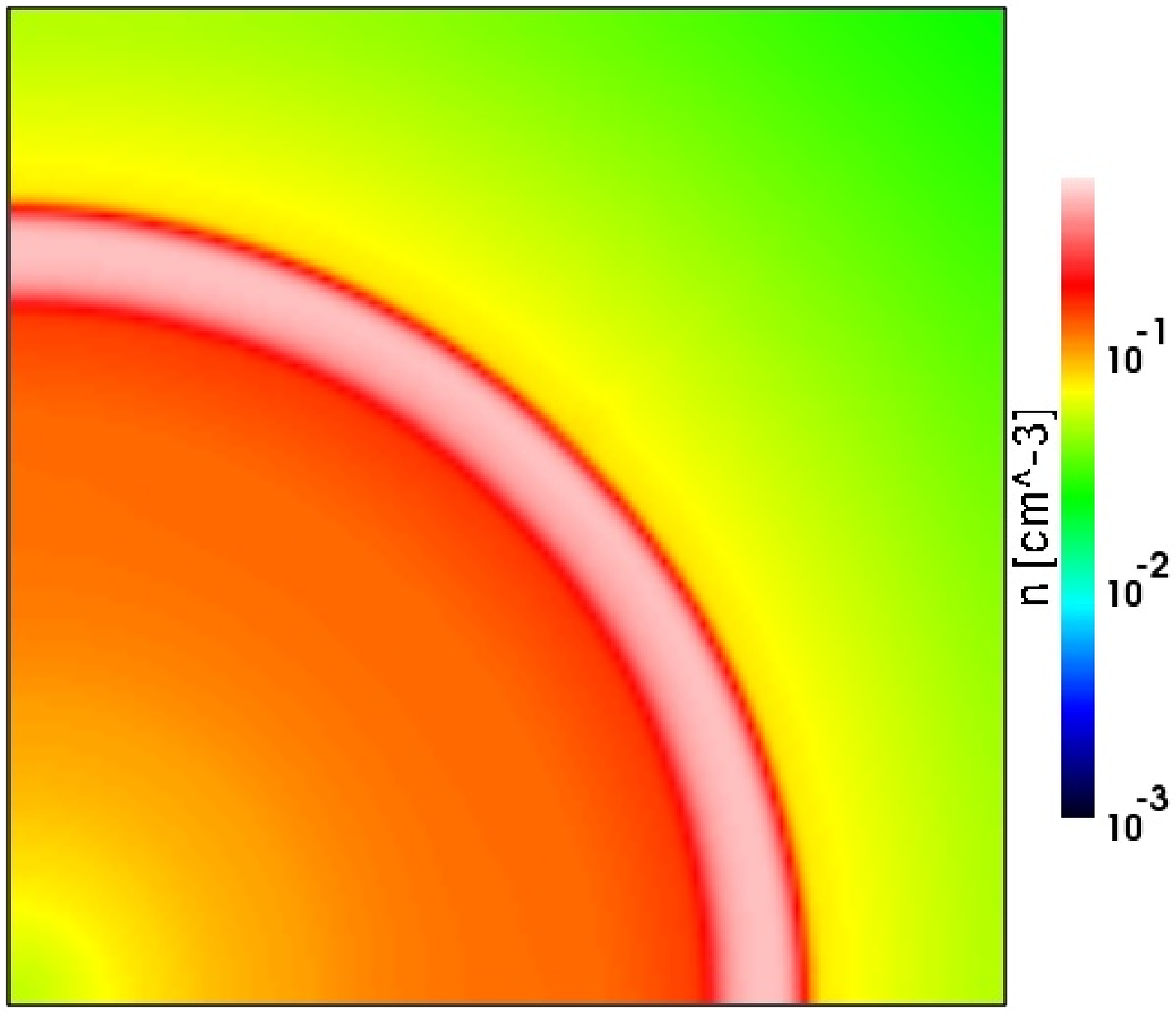}
  \includegraphics[width=2.3in]{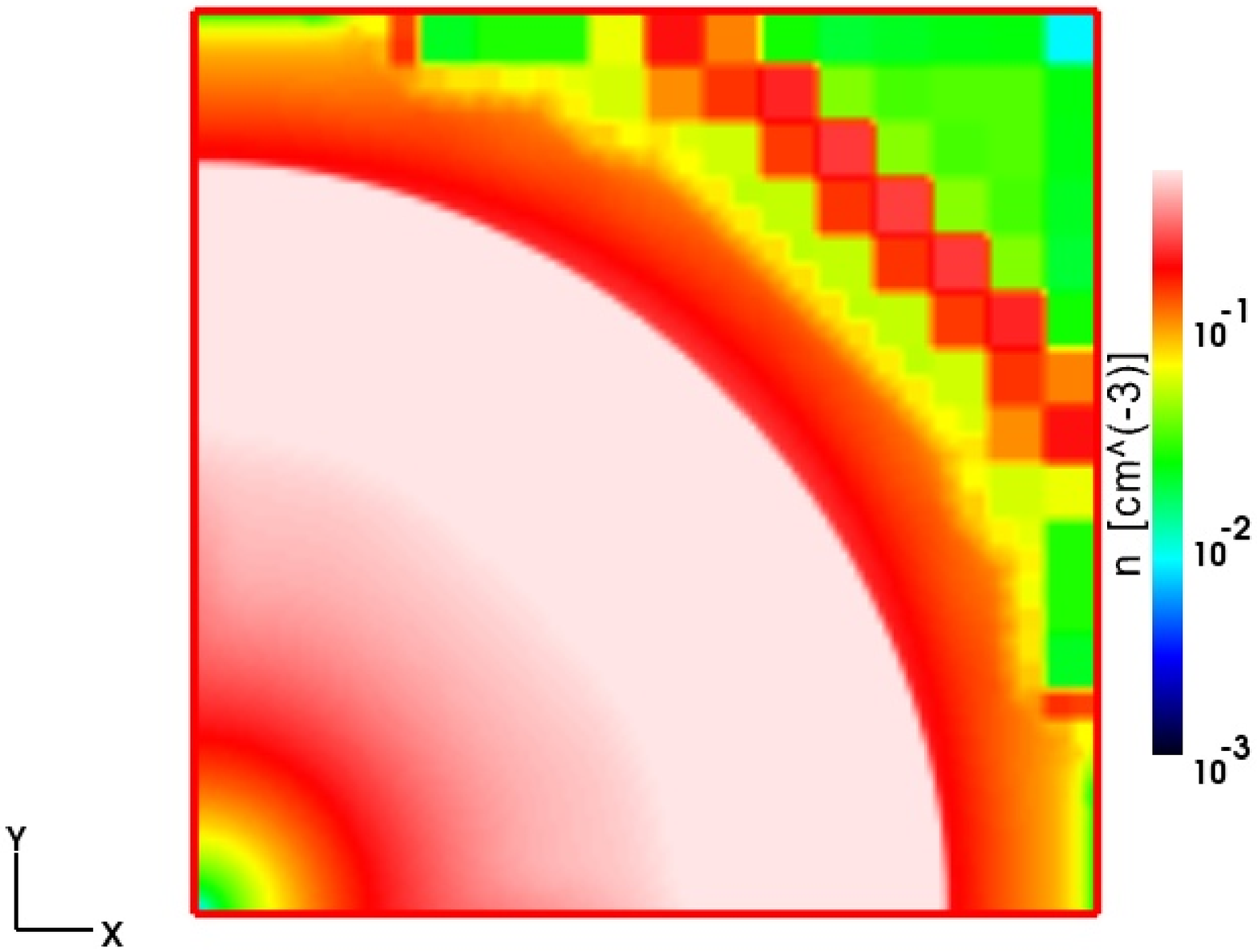}
  \includegraphics[width=2.3in]{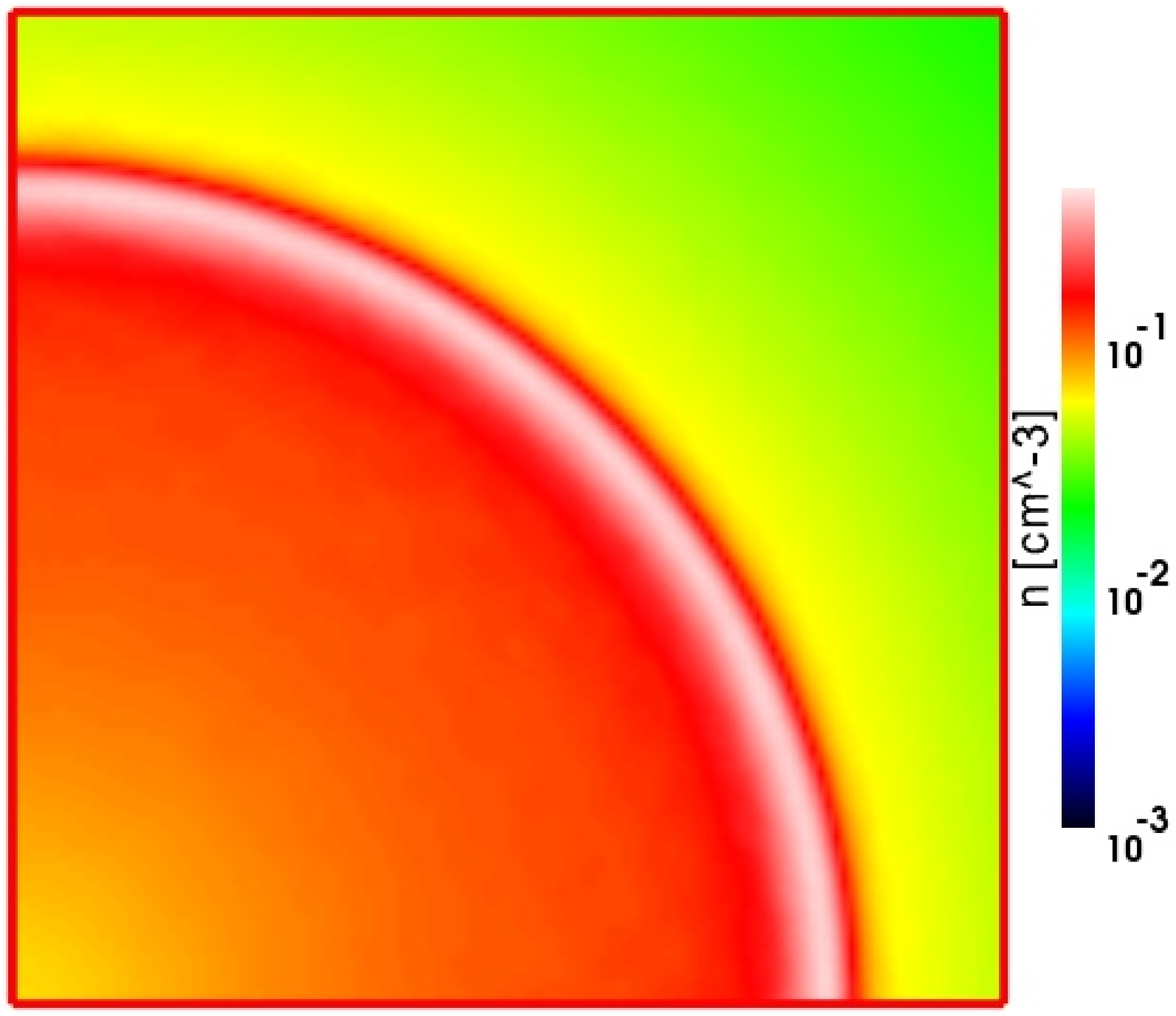}
  \includegraphics[width=2.3in]{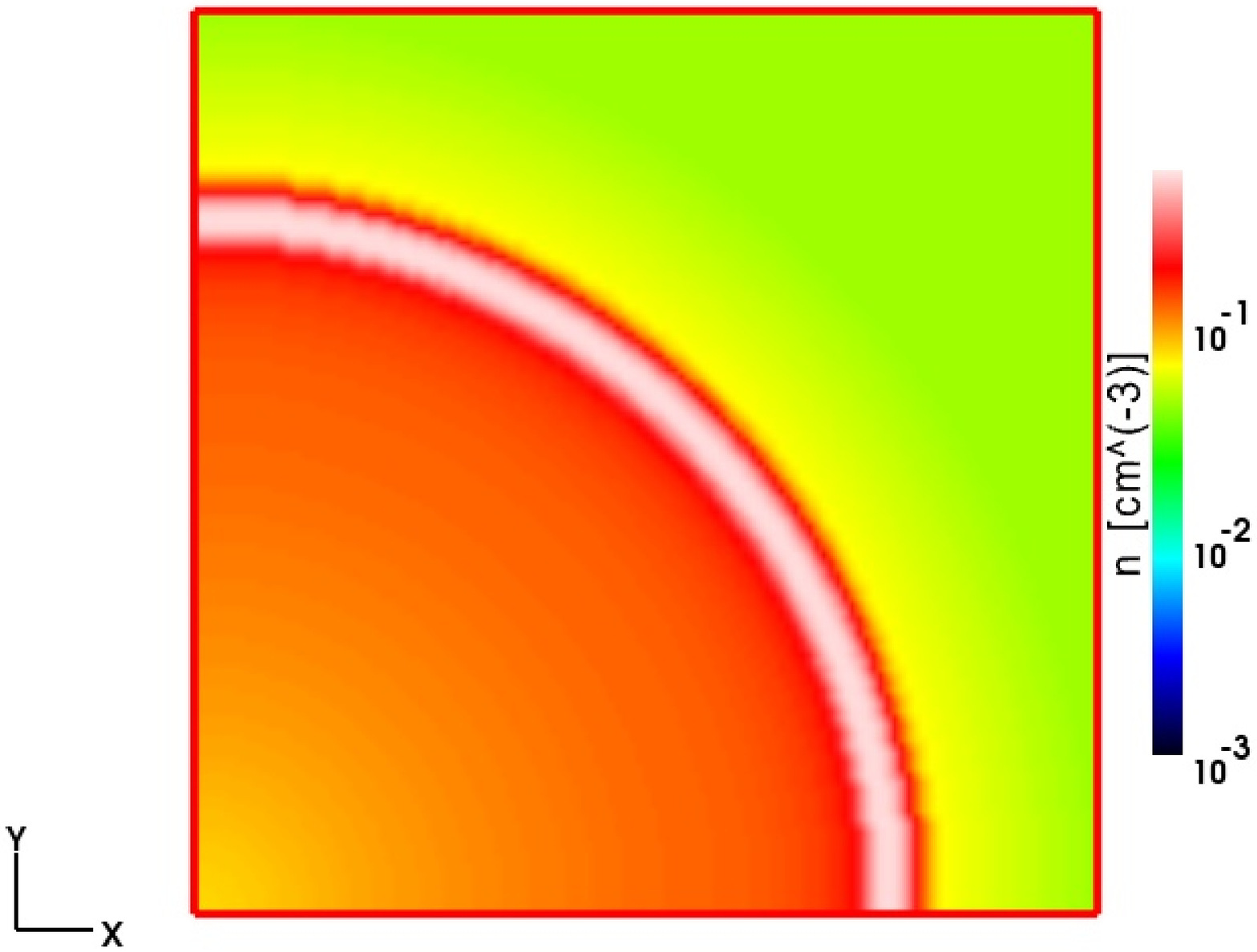}
  \includegraphics[width=2.3in]{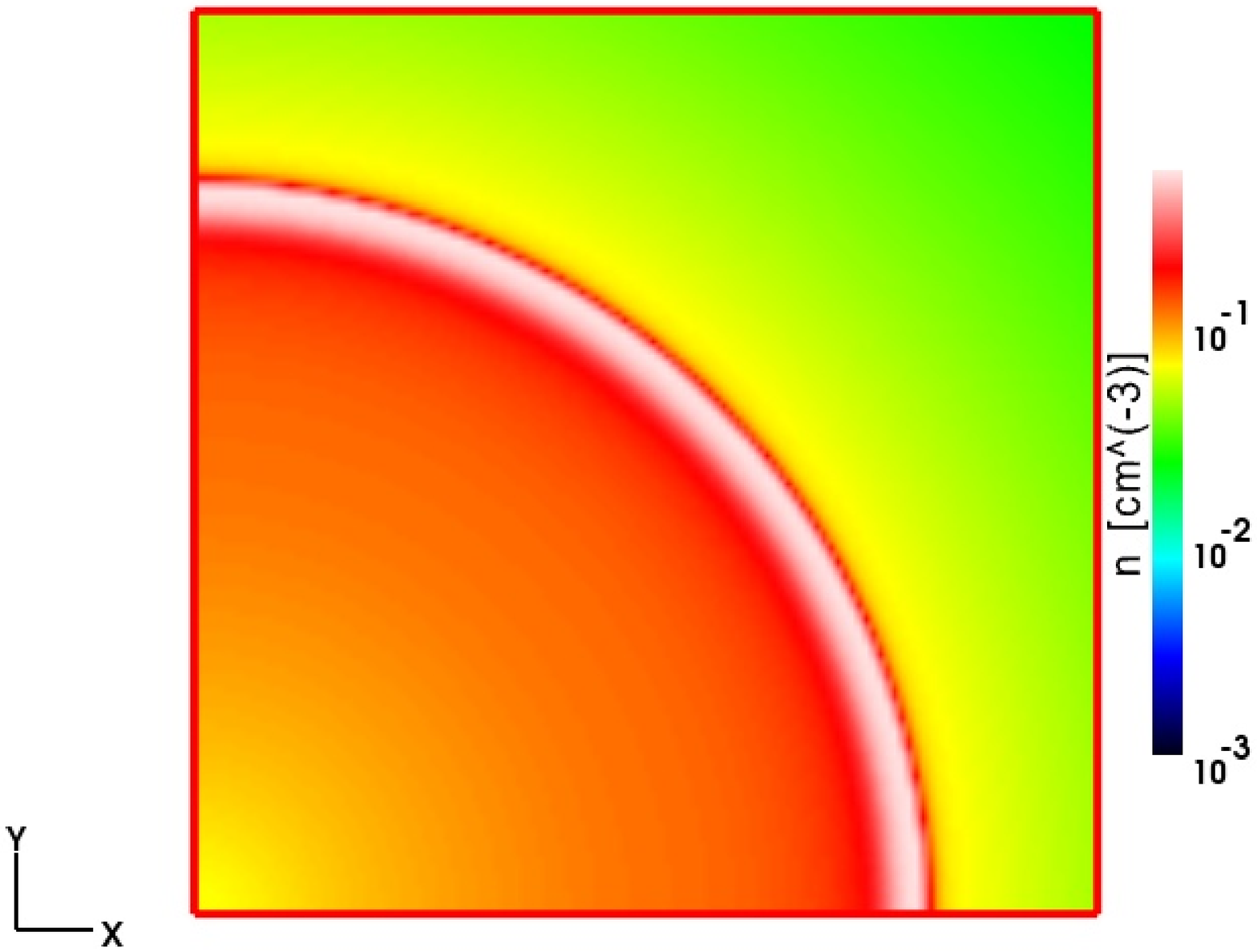}
  \includegraphics[width=2.3in]{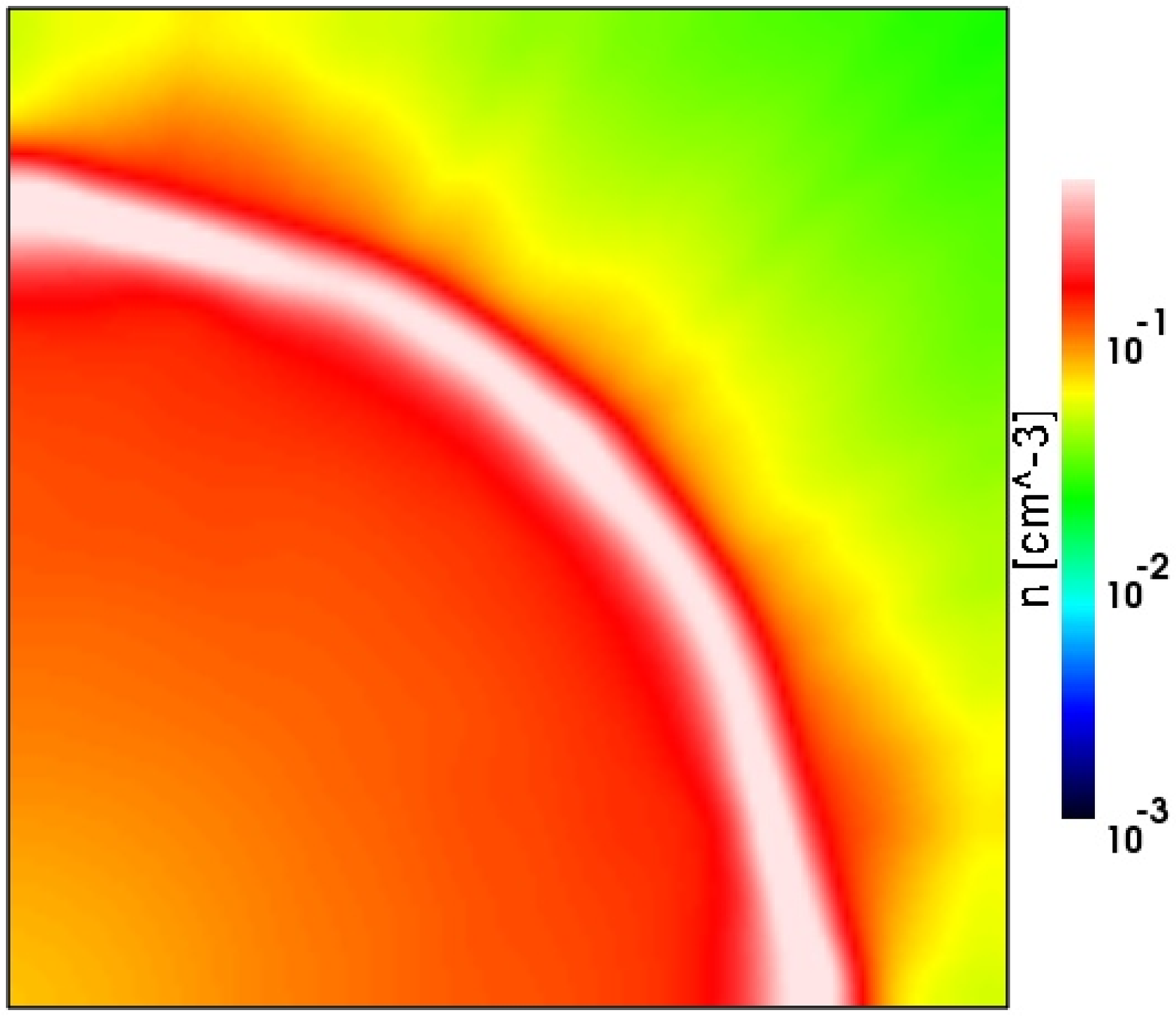}
  \includegraphics[width=2.3in]{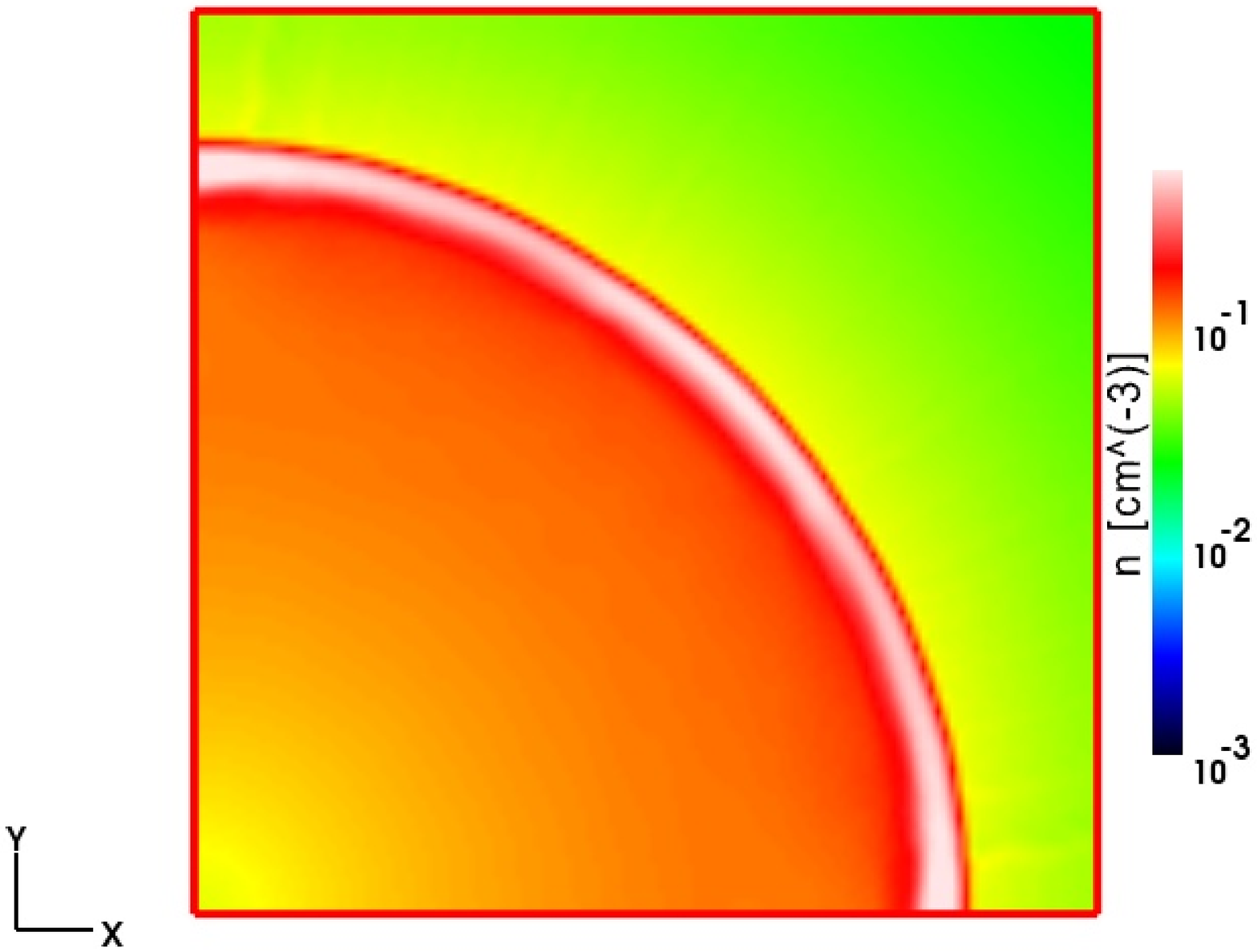}
\caption{Test 6 (H~II region gasdynamic expansion down a power-law initial
  density profile): Images of the density, cut through the simulation
  volume at coordinate $z=0$ at time $t=25$ Myr for (left to right and top 
  to bottom) Capreole+$C^2$-Ray, TVD+$C^2$-Ray, HART, RSPH, ZEUS-MP, RH1D, 
  LICORICE, and Flash-HC.
\label{T6_images4_n_fig}}
\end{center}
\end{figure*}

All codes agree on the later-time, pressure-driven expansion, both 
qualitatively (a slow, D-type I-front, preceded by a relatively weak 
shock, as we shall see below), and quantitatively. There are some modest 
differences in the I-front speed, $\sim10$\% or less between cases, which 
results in I-front positions whose spread grows with time, but never 
exceeds $\sim5-7\%$.

\begin{figure*}
\begin{center}
  \includegraphics[width=2.3in]{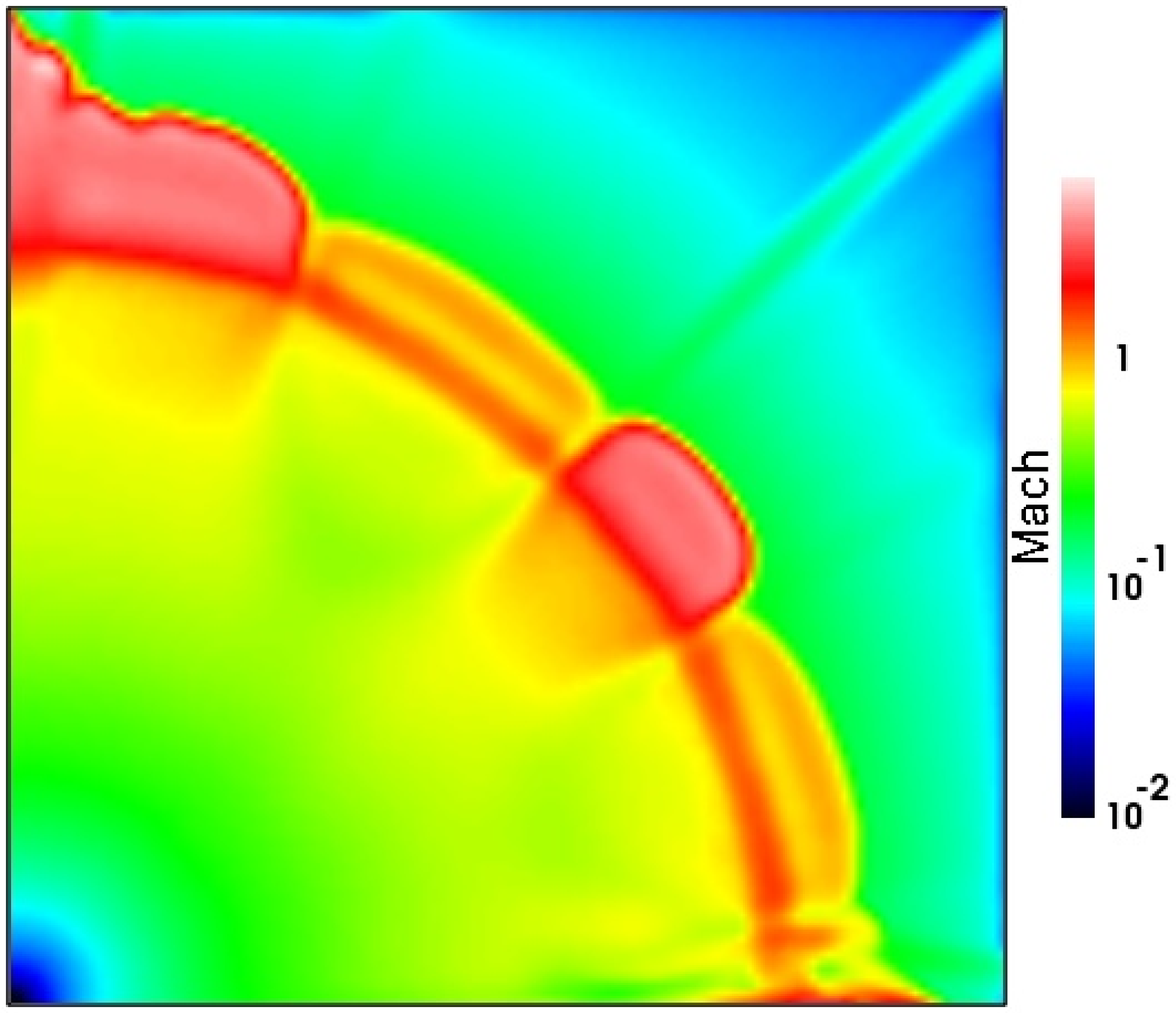}
  \includegraphics[width=2.3in]{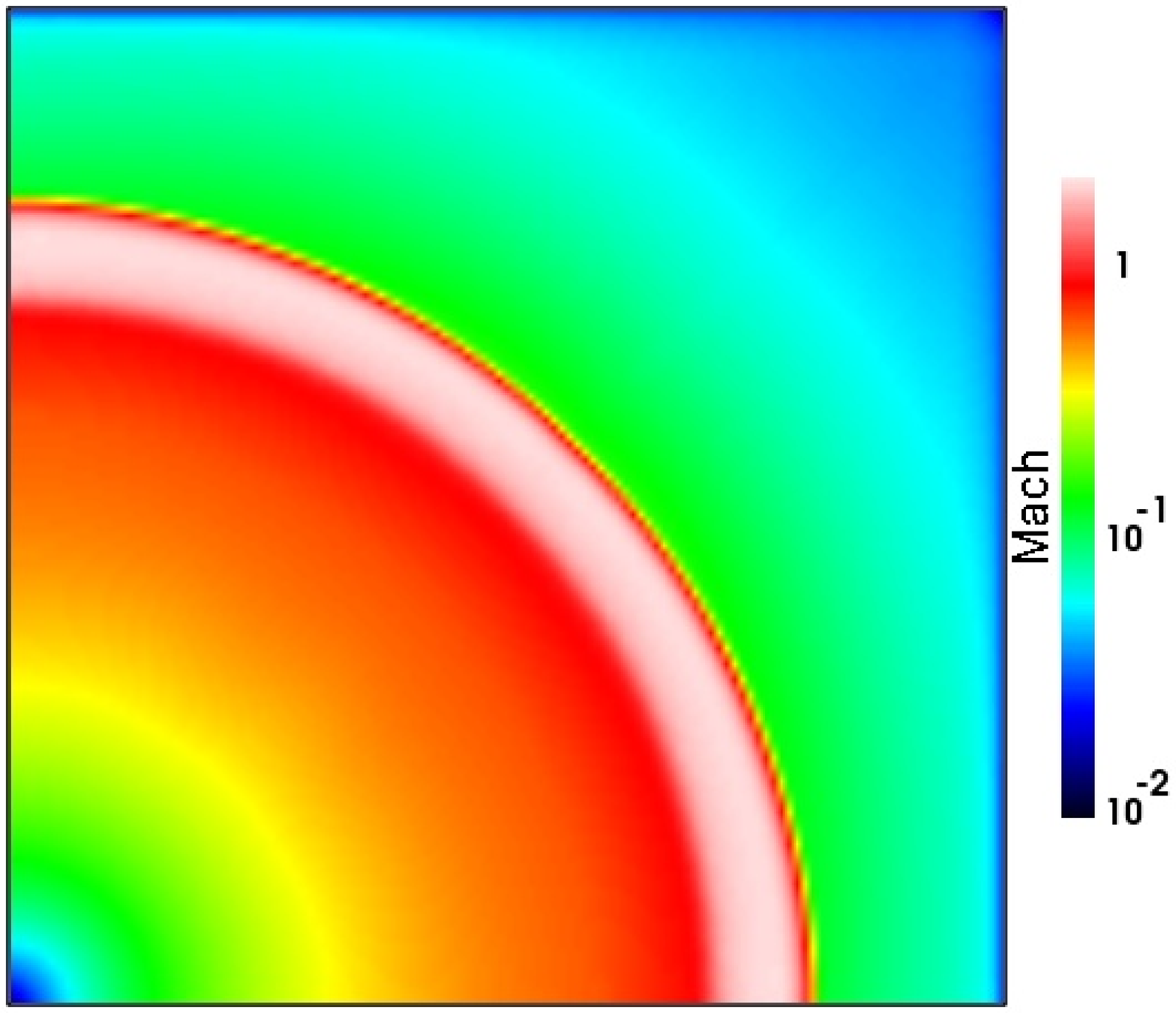}
  \includegraphics[width=2.3in]{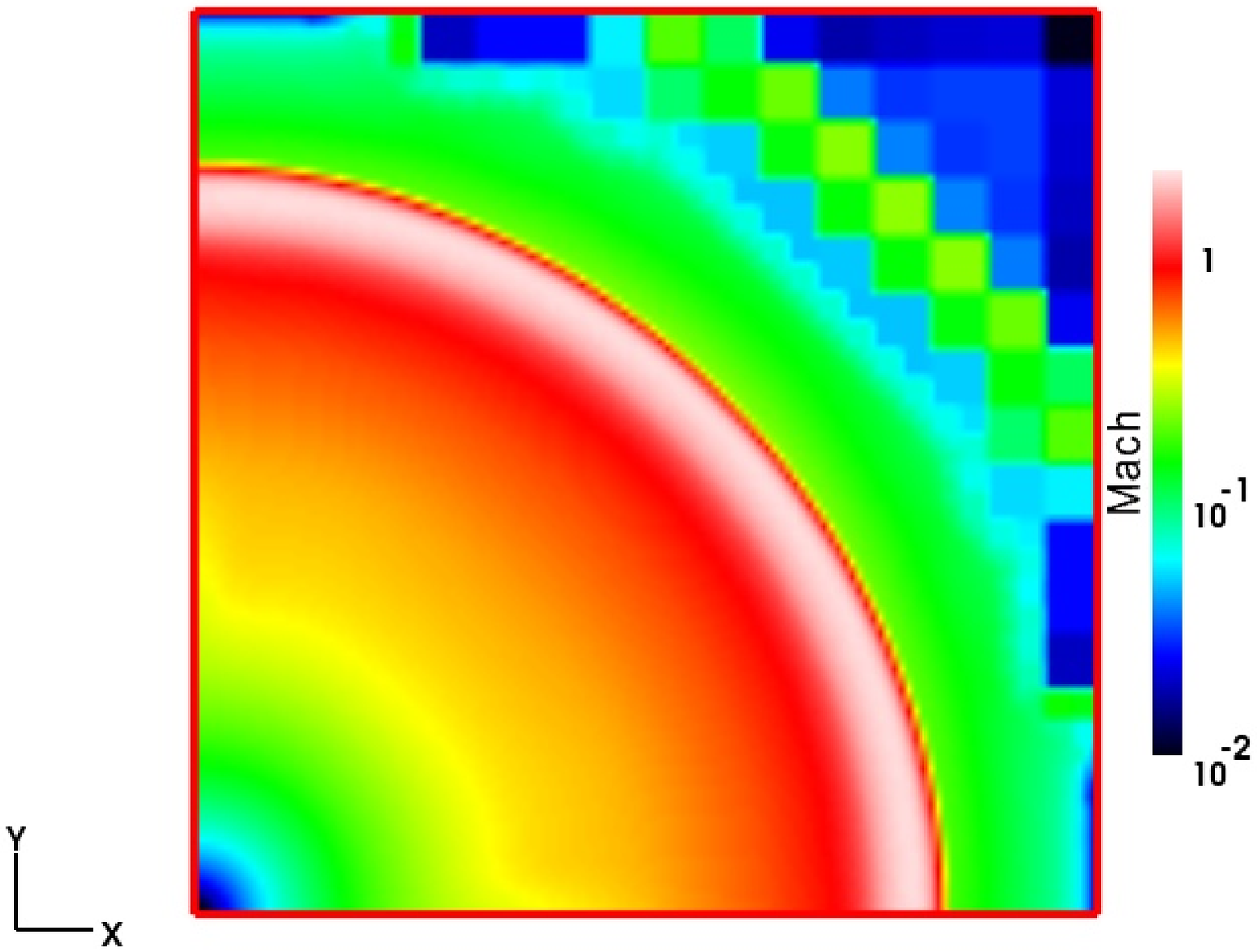}
  \includegraphics[width=2.3in]{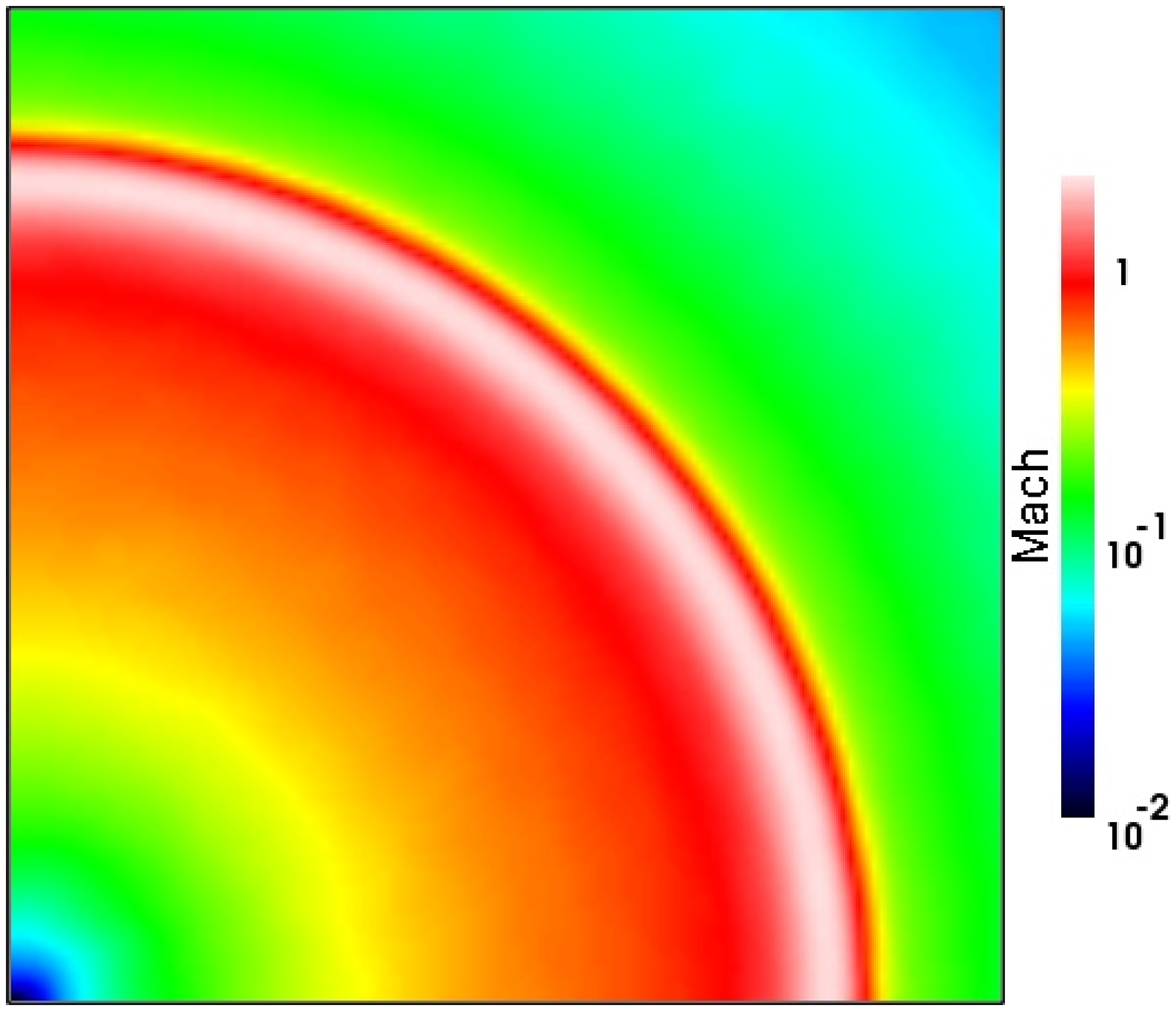}
  \includegraphics[width=2.3in]{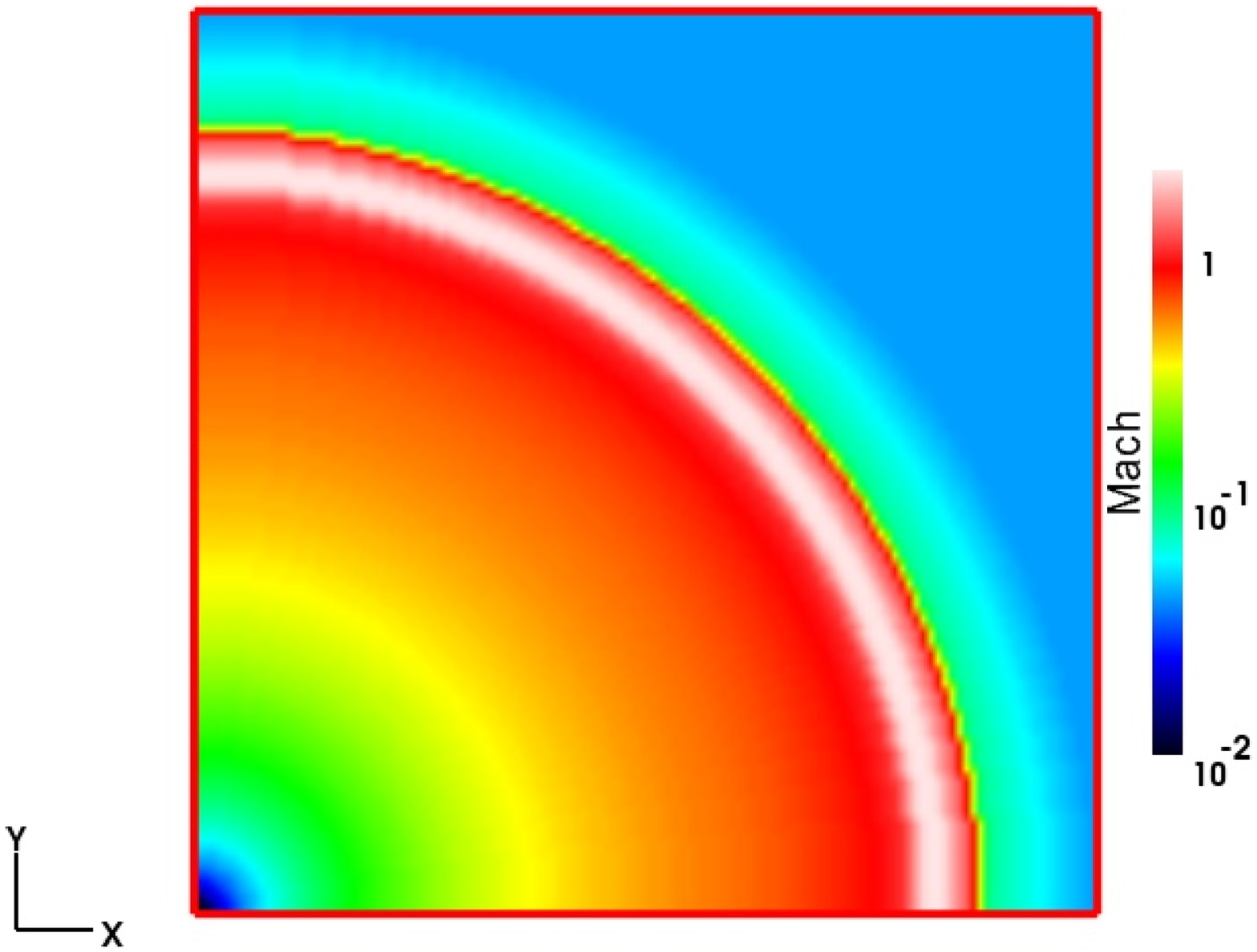}
  \includegraphics[width=2.3in]{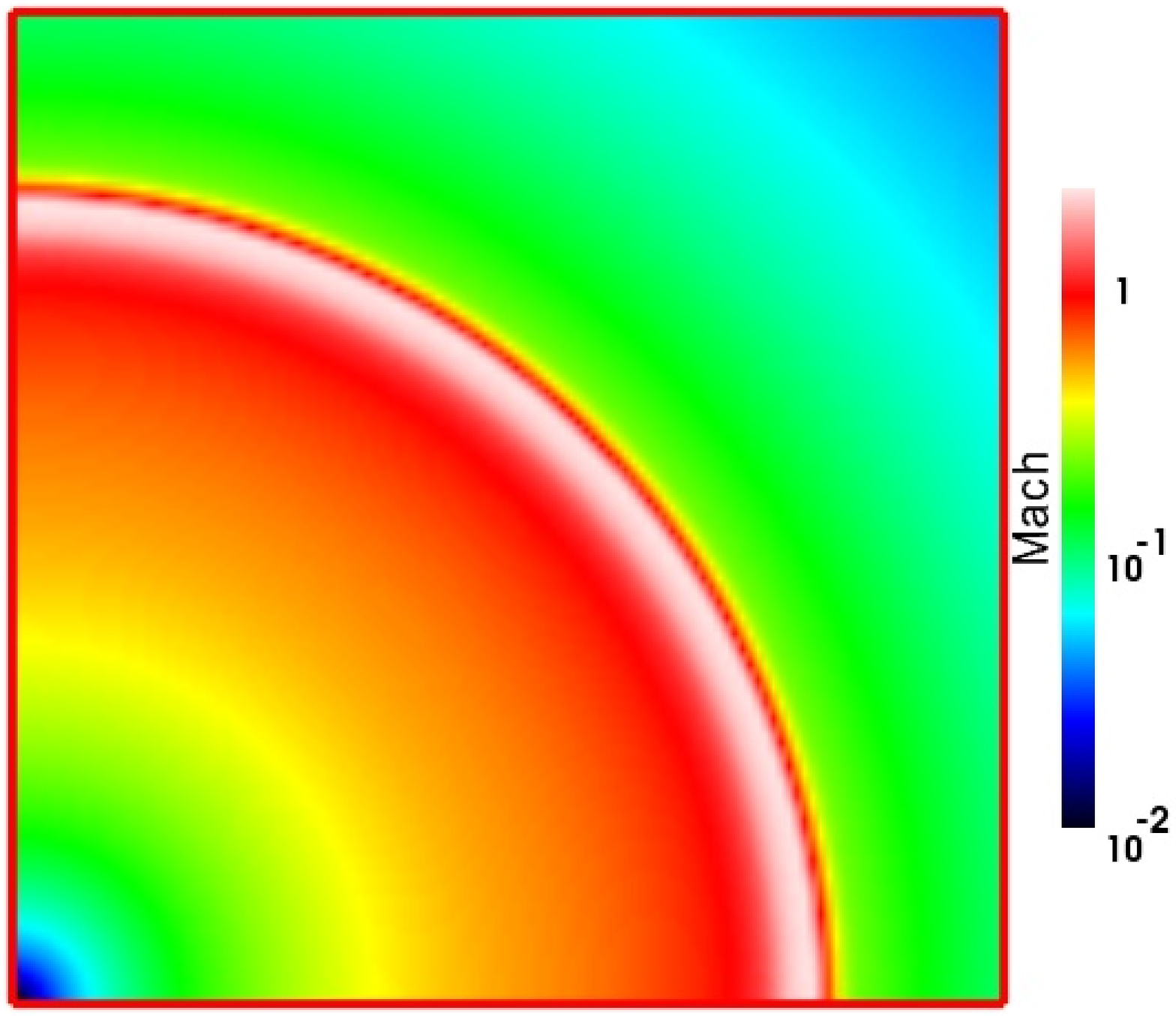}
  \includegraphics[width=2.3in]{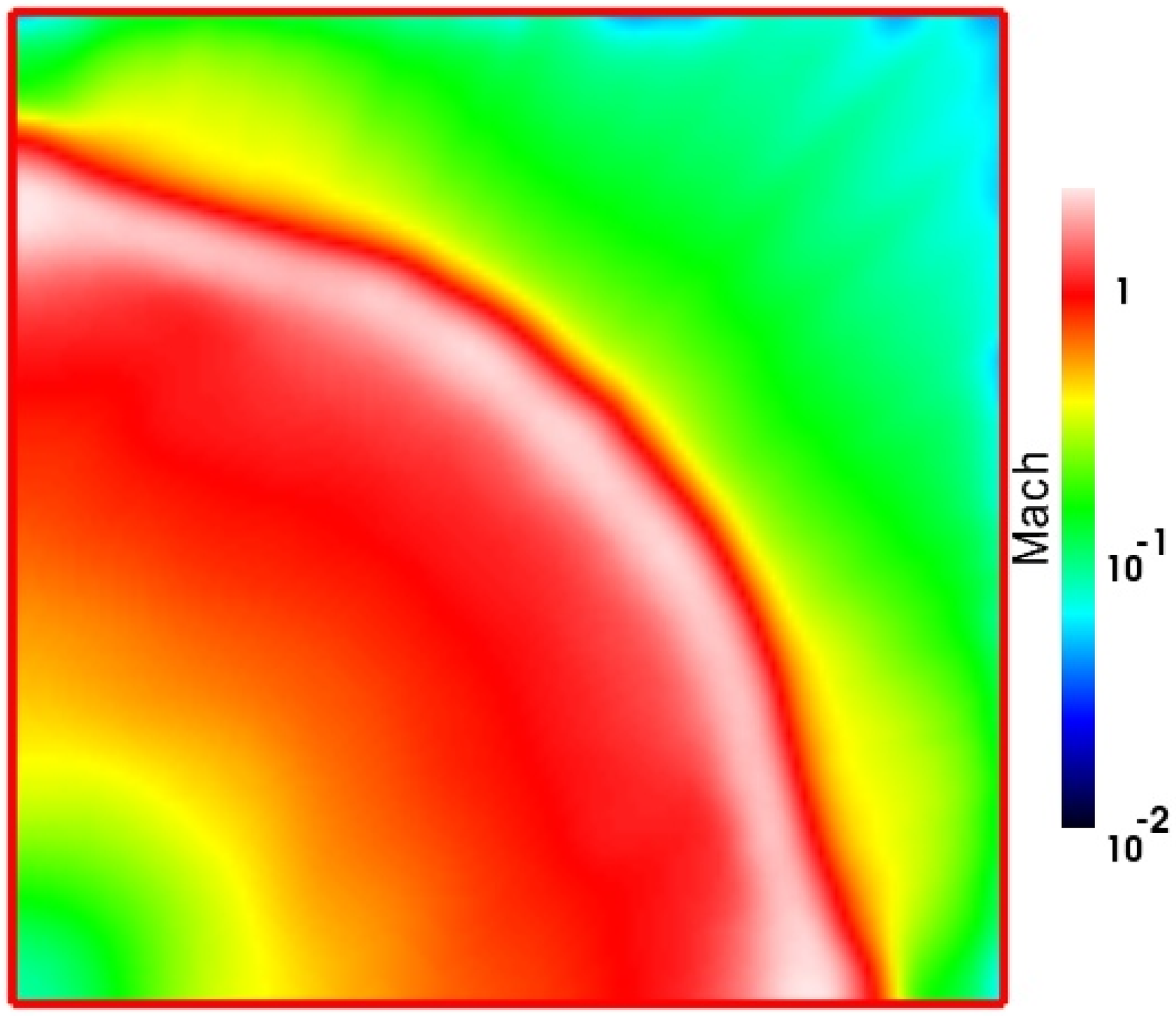}
  \includegraphics[width=2.3in]{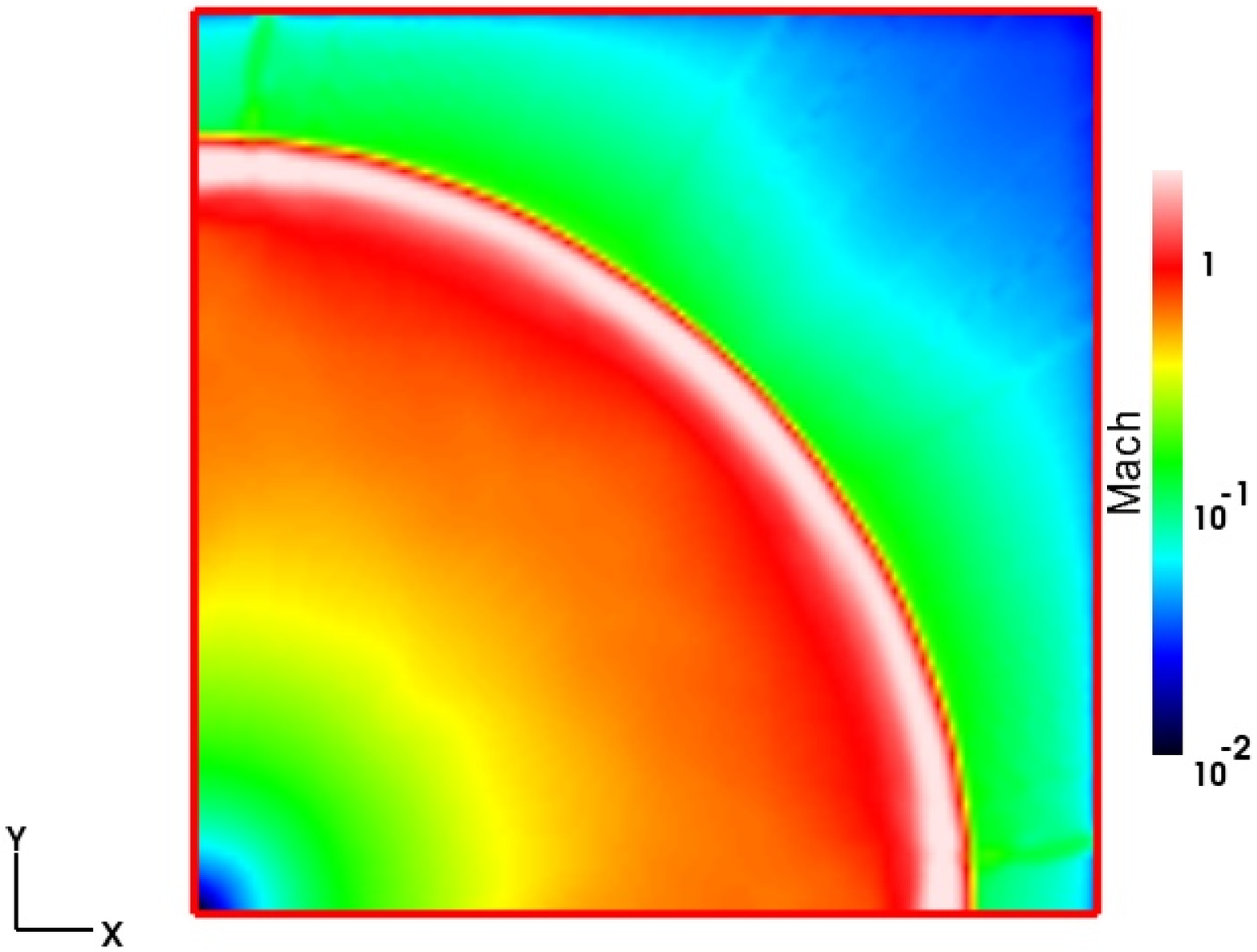}
\caption{Test 6 (H~II region gasdynamic expansion down a power-law initial
  density profile): Images of the Mach number, cut through the simulation
  volume at 
  coordinate $z=0$ at time $t=25$ Myr for (left to right) and top to bottom)
  Capreole+$C^2$-Ray, TVD+$C^2$-Ray, HART, RSPH, ZEUS-MP, RH1D, LICORICE, and 
Flash-HC.
\label{T6_images4_M_fig}}
\end{center}
\end{figure*}
\begin{figure*}
\begin{center}
  \includegraphics[width=2.3in]{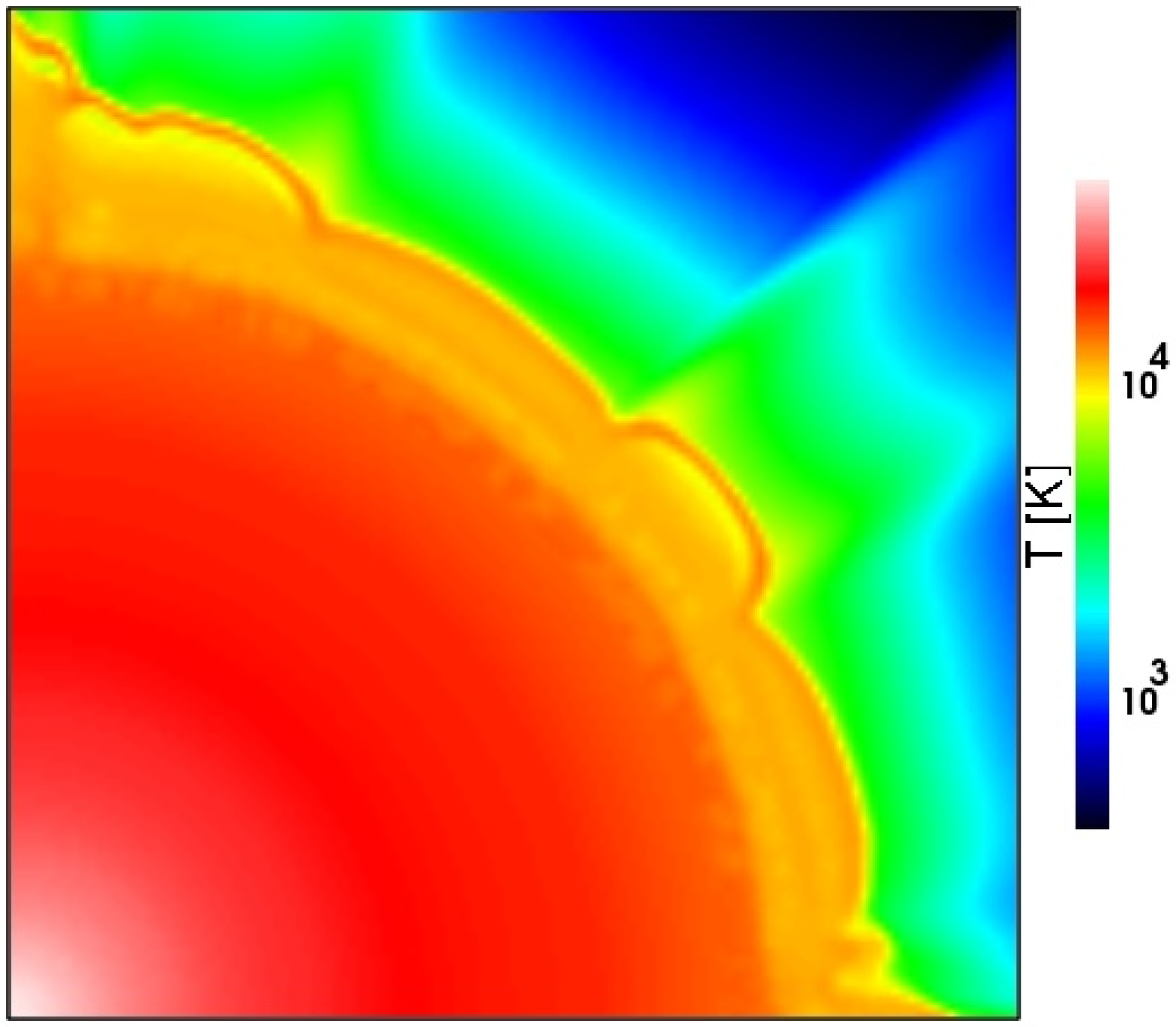}
  \includegraphics[width=2.3in]{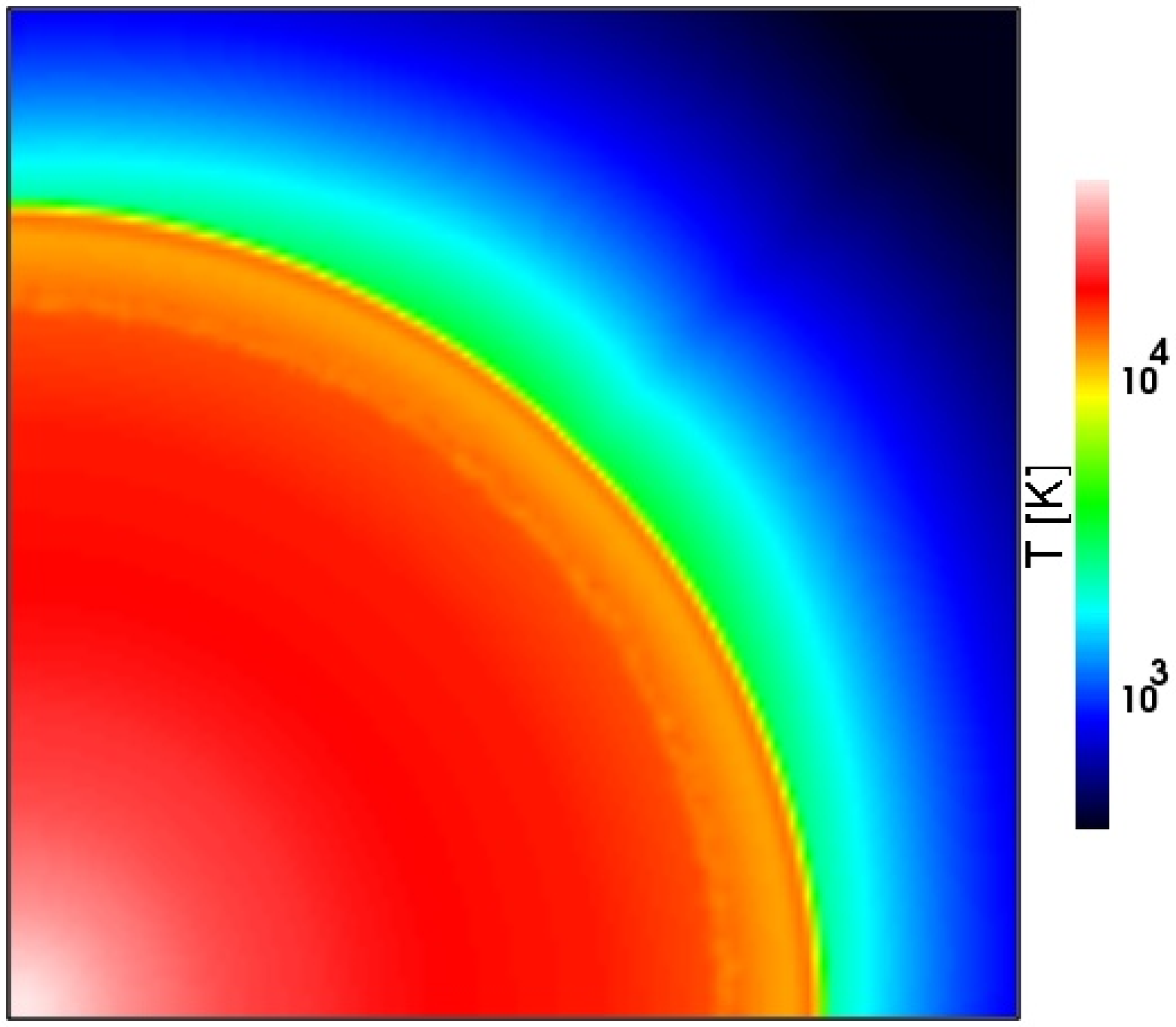}
  \includegraphics[width=2.3in]{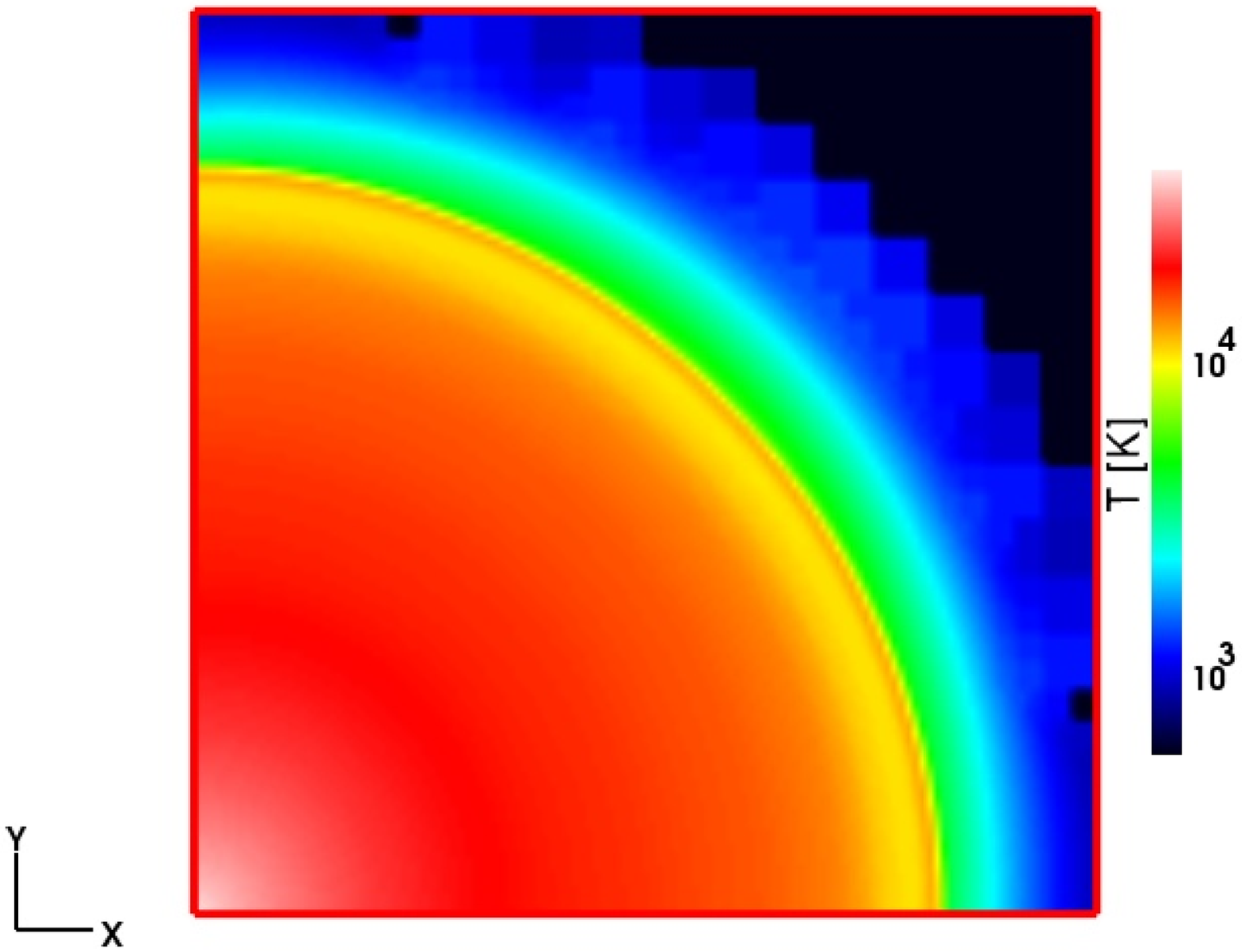}
  \includegraphics[width=2.3in]{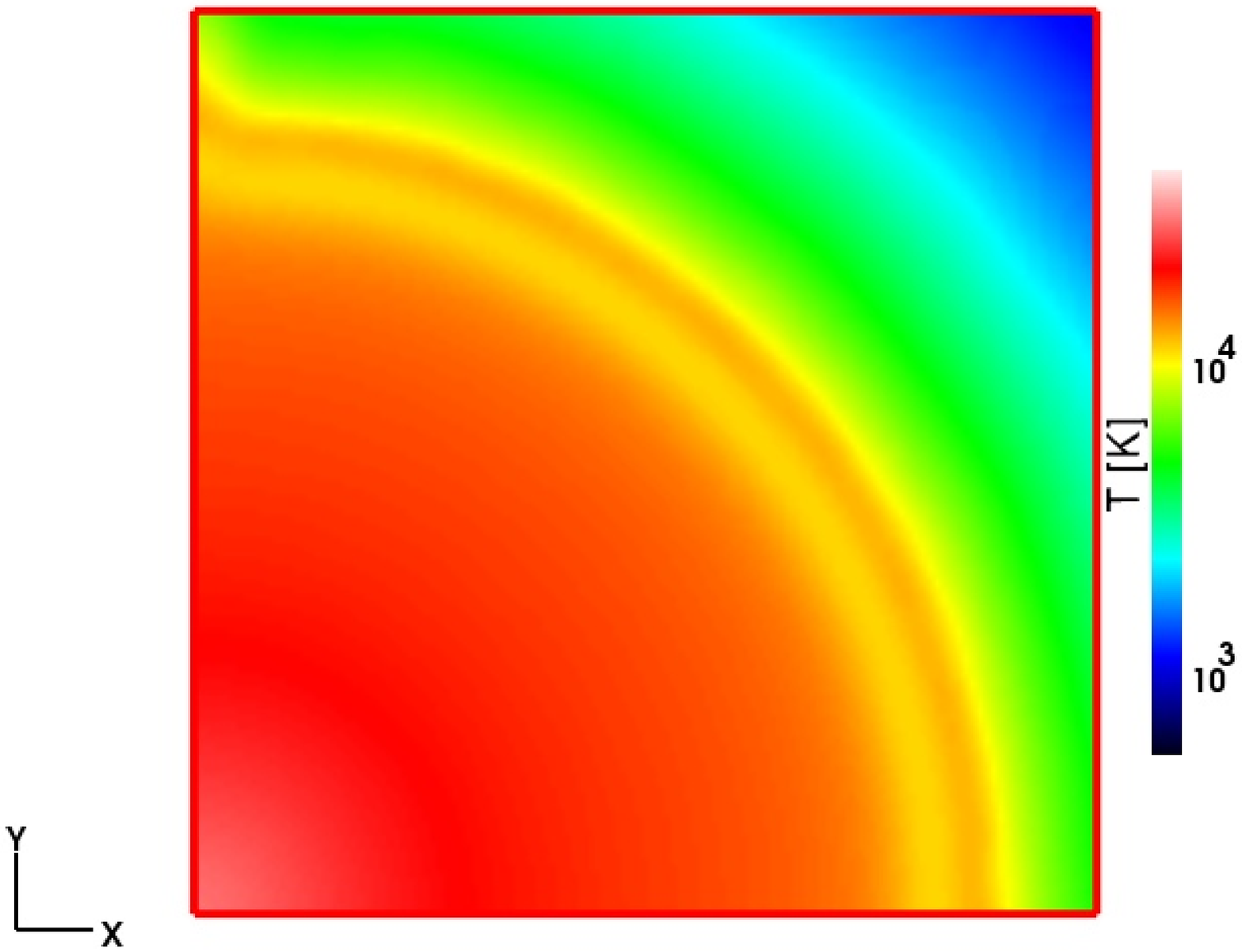}
  \includegraphics[width=2.3in]{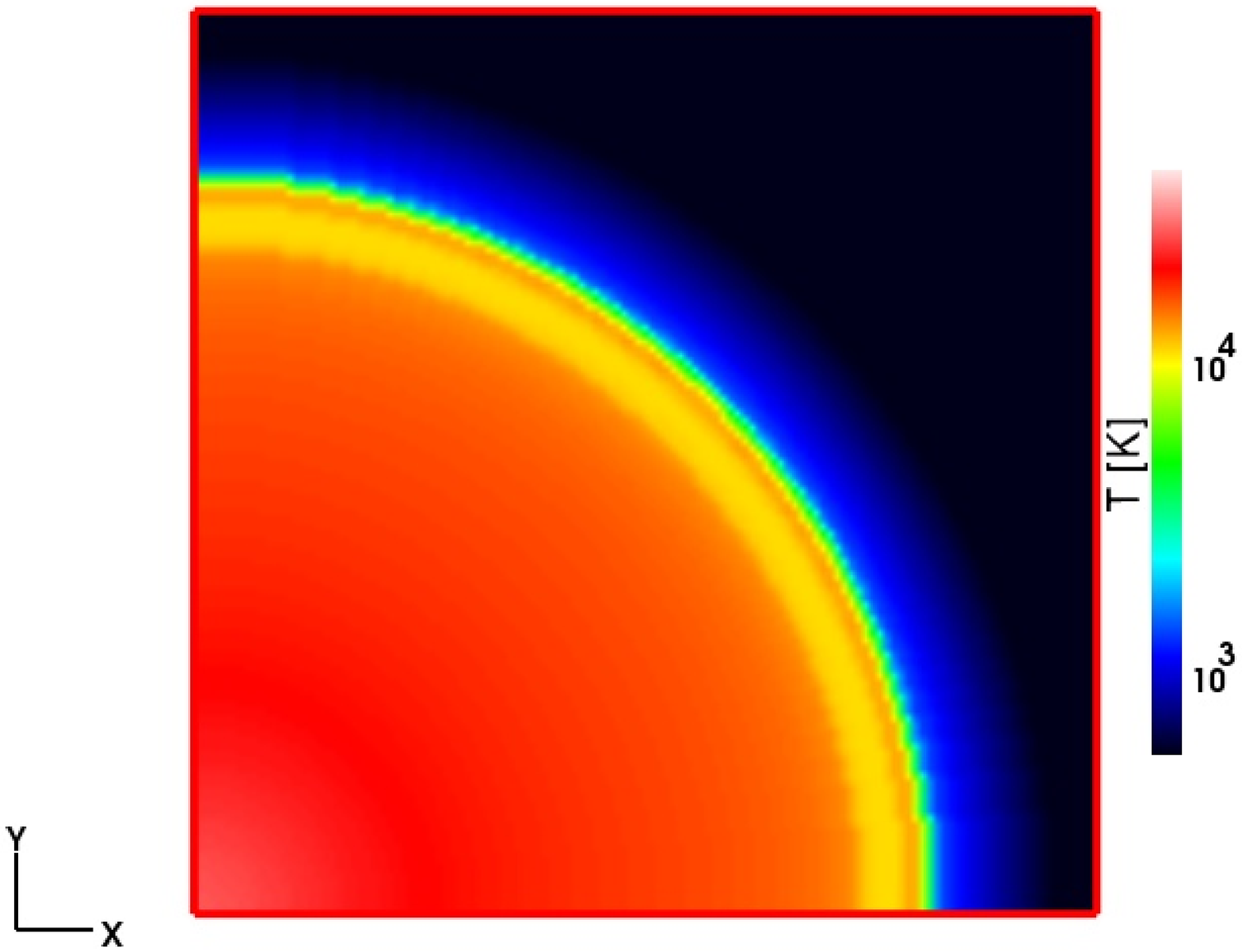}
  \includegraphics[width=2.3in]{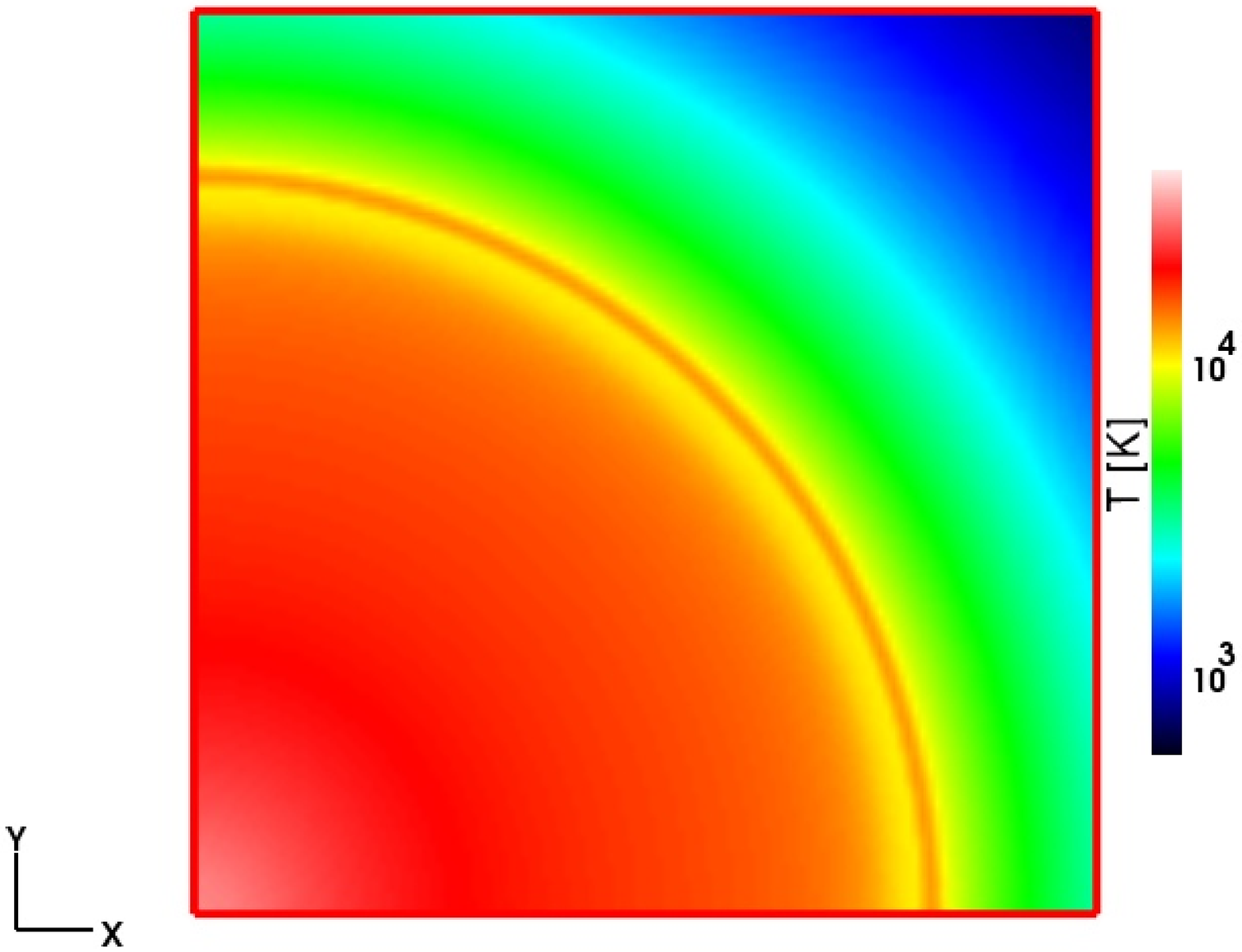}
  \includegraphics[width=2.3in]{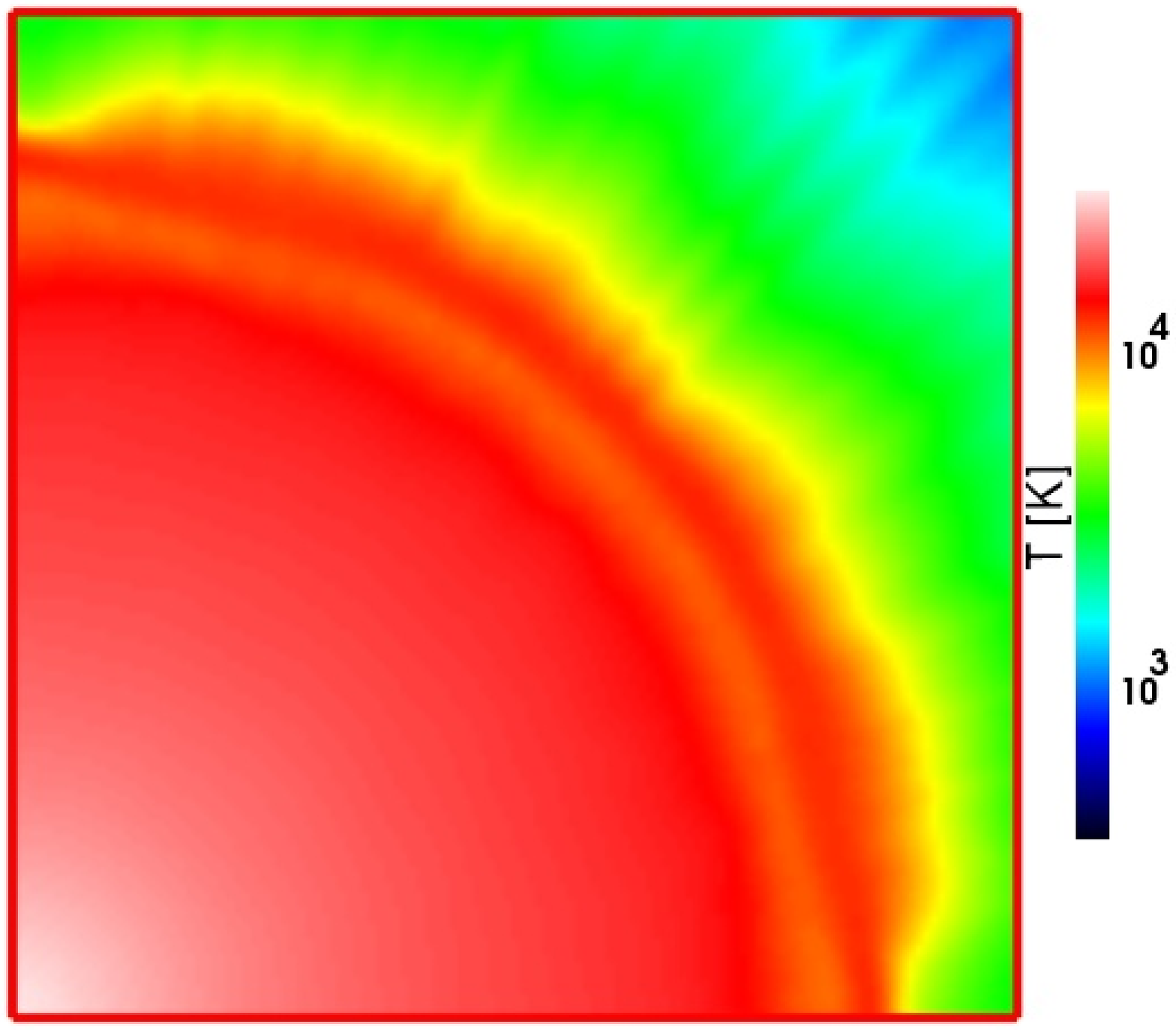}
  \includegraphics[width=2.3in]{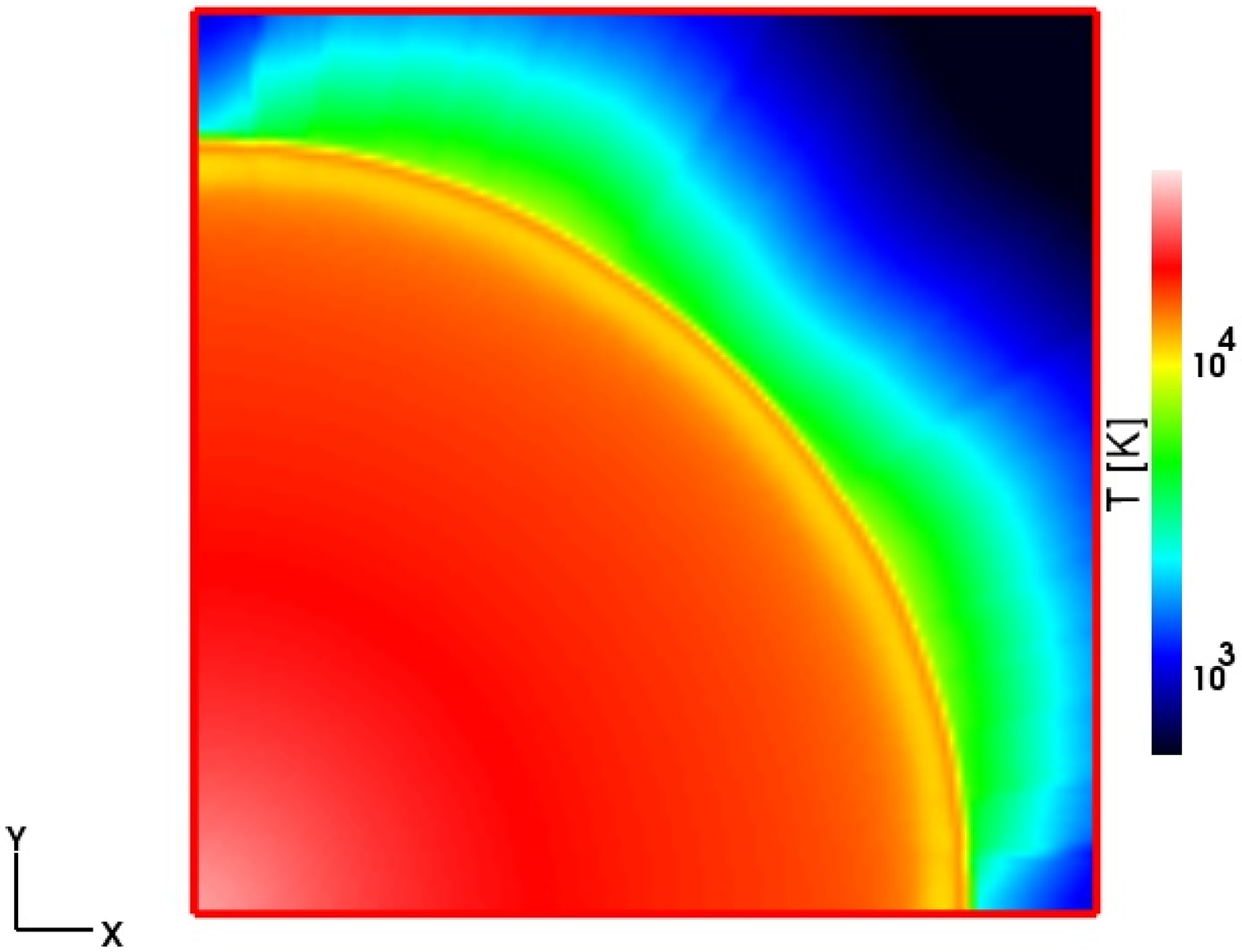}
\caption{Test 6 (H~II region gasdynamic expansion down a power-law initial
  density profile): Images of the temperature, cut through the simulation
  volume at coordinate $z=0$ at time $t=25$ Myr for (left to right and top 
  to bottom) Capreole+$C^2$-Ray, TVD+$C^2$-Ray, HART, RSPH, ZEUS-MP, RH1D, 
  LICORICE, and Flash-HC.
\label{T6_images4_T_fig}}
\end{center}
\begin{center}
  \includegraphics[width=2.3in]{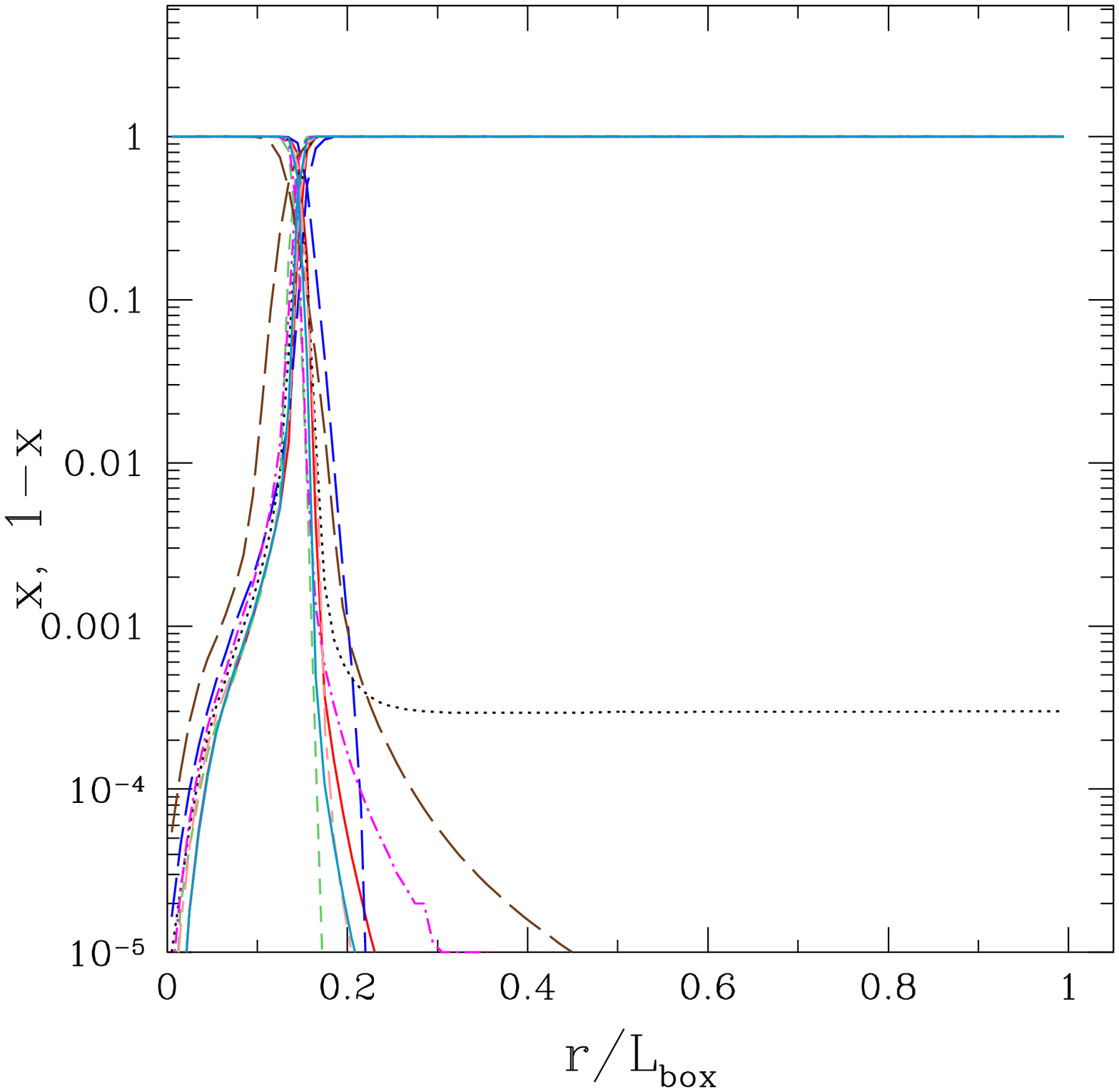}
  \includegraphics[width=2.3in]{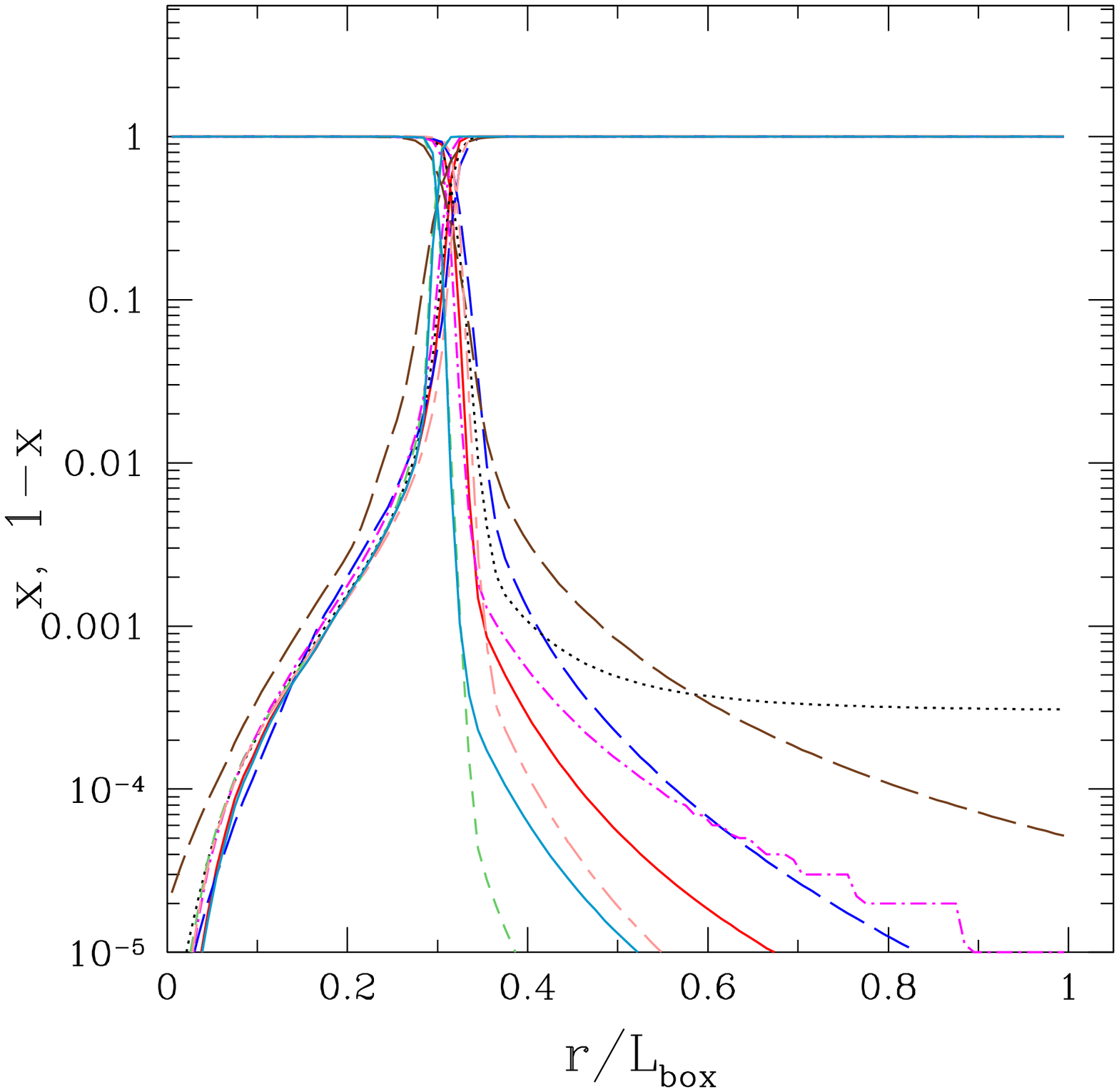}
  \includegraphics[width=2.3in]{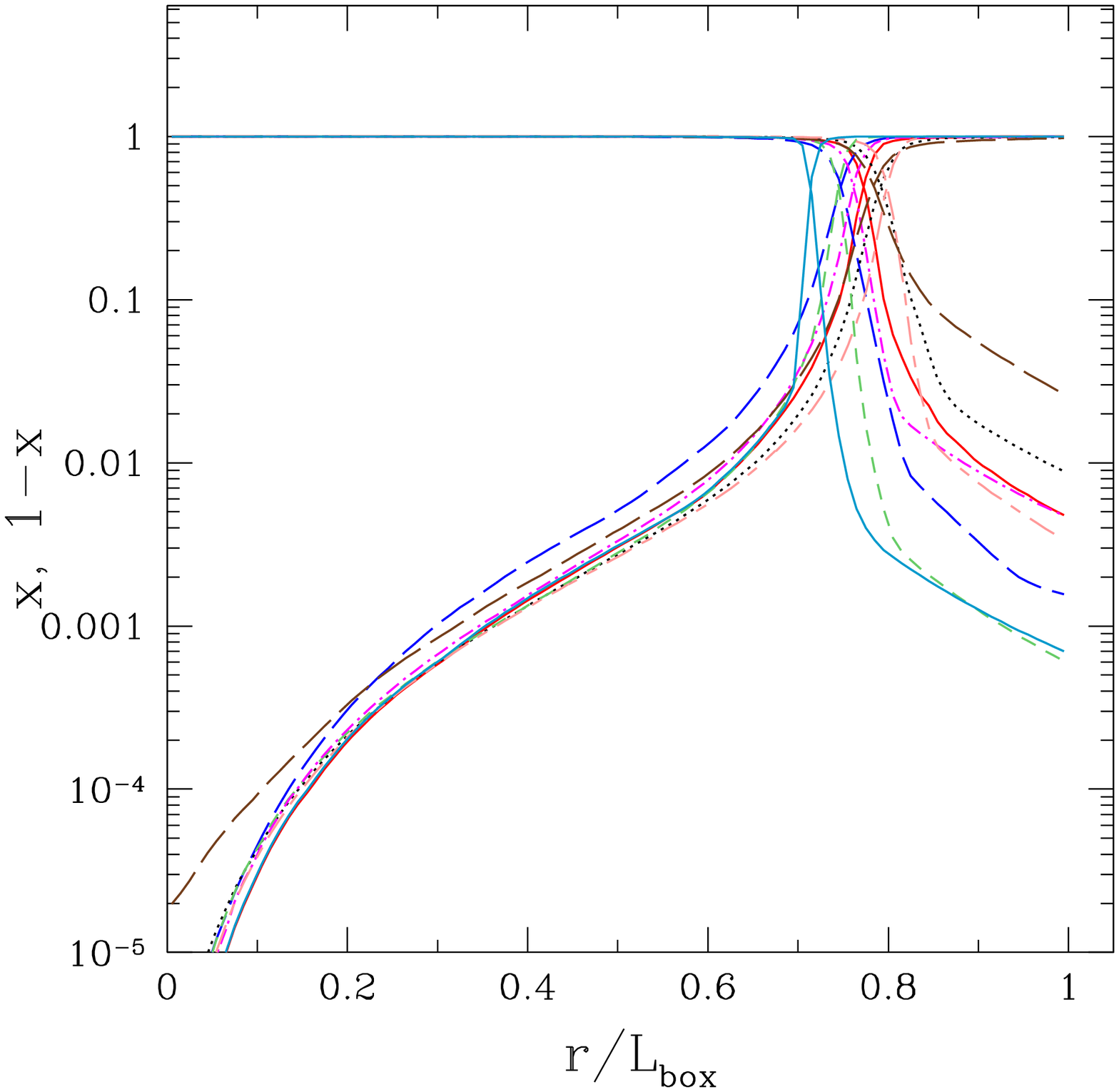}
\caption{Test 6 (H~II region gasdynamic expansion down a power-law initial
  density profile): Spherically-averaged profiles for ionized fractions $x$ 
  and neutral fractions $x_{\rm HI}=1-x$ at times $t=3$ Myr, 10 Myr and 25 Myr
  vs. dimensionless radius (in units of the box size). 
\label{T6_profs_fig}}
\end{center}
\end{figure*}

Next we turn our attention to the overall structure of the fluid flow and 
ionization, shown in 2-D cuts along the x-y plane of the H~I and H~II fractions, 
density, temperature and Mach number at time $t=25$~Myr in 
Figures~\ref{T6_images4_HI_fig} - \ref{T6_images4_T_fig}. We again remind the 
reader that unlike the other simulations which are fully 3-D in both the hydrodynamic and 
the radiative transfer treatment, the RH1D results are 1-D spherically-symmetric
Lagrangian profiles mapped onto the 3-D Cartesian grid required in this
study.  There is good 
agreement between the results, in terms of the positions of the I-front 
and the shock, the size of the growing H~II region and its ionization, density 
and temperature structures. There are some differences in the level of hard photon
penetration ahead of the I-front and the temperature distribution, similar to 
the ones we observed in Test 5 and Paper I. 

However, in this test also a new kind of difference shows up, namely the 
appearance of instabilities near the I-front. Such instabilities occur for 
several of the codes, and their nature varies between codes. In the cases 
of C$^2$-Ray+Capreole and LICORICE, the instabilities are clearly visible 
in the ionized fractions, temperatures, densities, and Mach numbers, while 
in Flash-HC they are mostly visible in the temperatures and ionized fractions.  
RSPH exhibits a minor anomaly in only the temperature at 25 Myr. The RH1D data 
cannot exhibit such instabilities because they are 1D spherical polar coordinate 
profiles mapped onto the 3D Cartesian grid mandated for this test. The ZEUS-MP 
profiles, which were computed on a 3D spherical polar coordinate grid and then
mapped onto 3D Cartesian coordinates, manifest no instabilities in any of the 
profiles, and the HART results do not show them either. Are these instabilities 
physical or numerical?

\begin{figure*}
\begin{center}
  \includegraphics[width=2.3in]{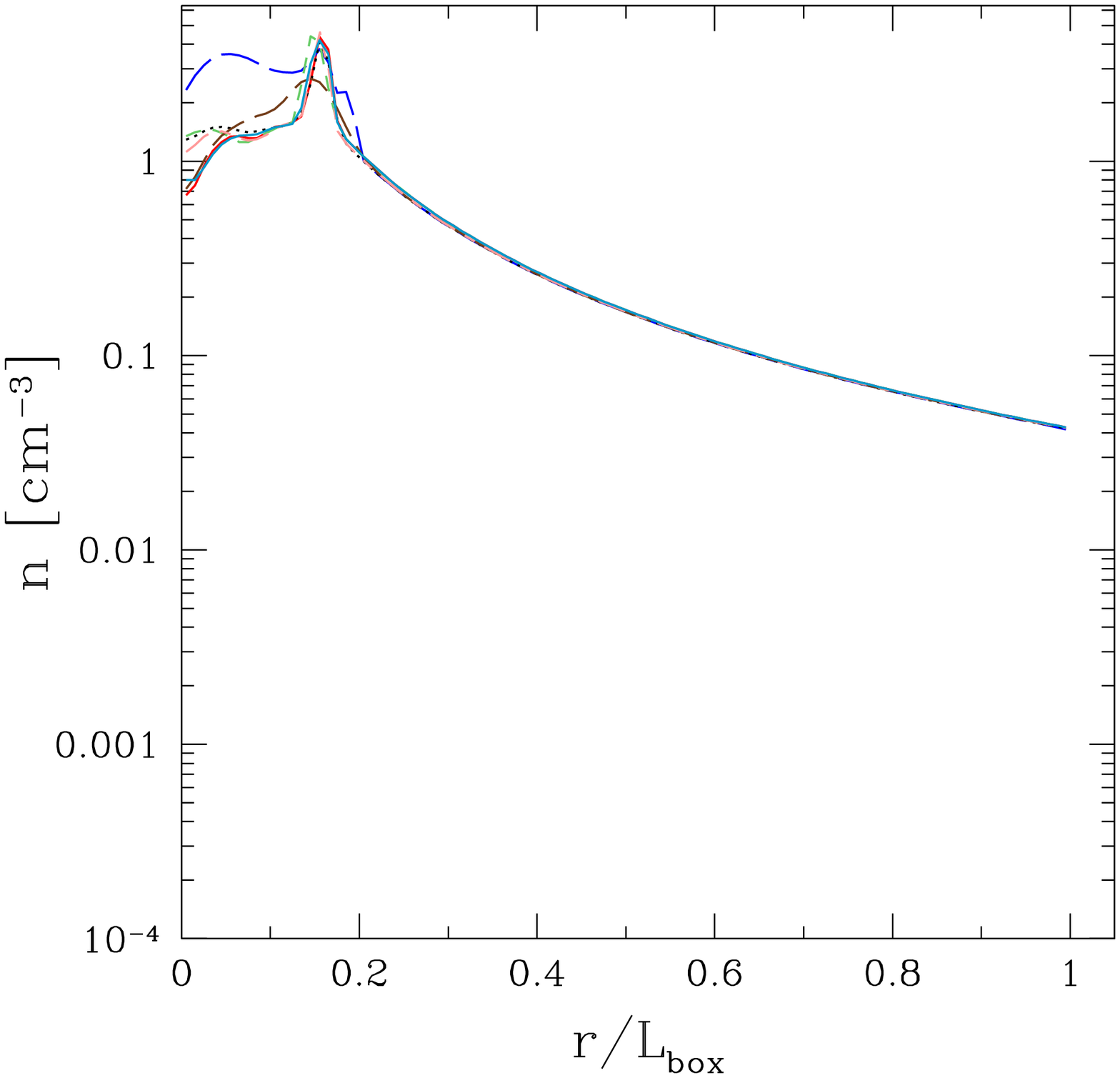}
  \includegraphics[width=2.3in]{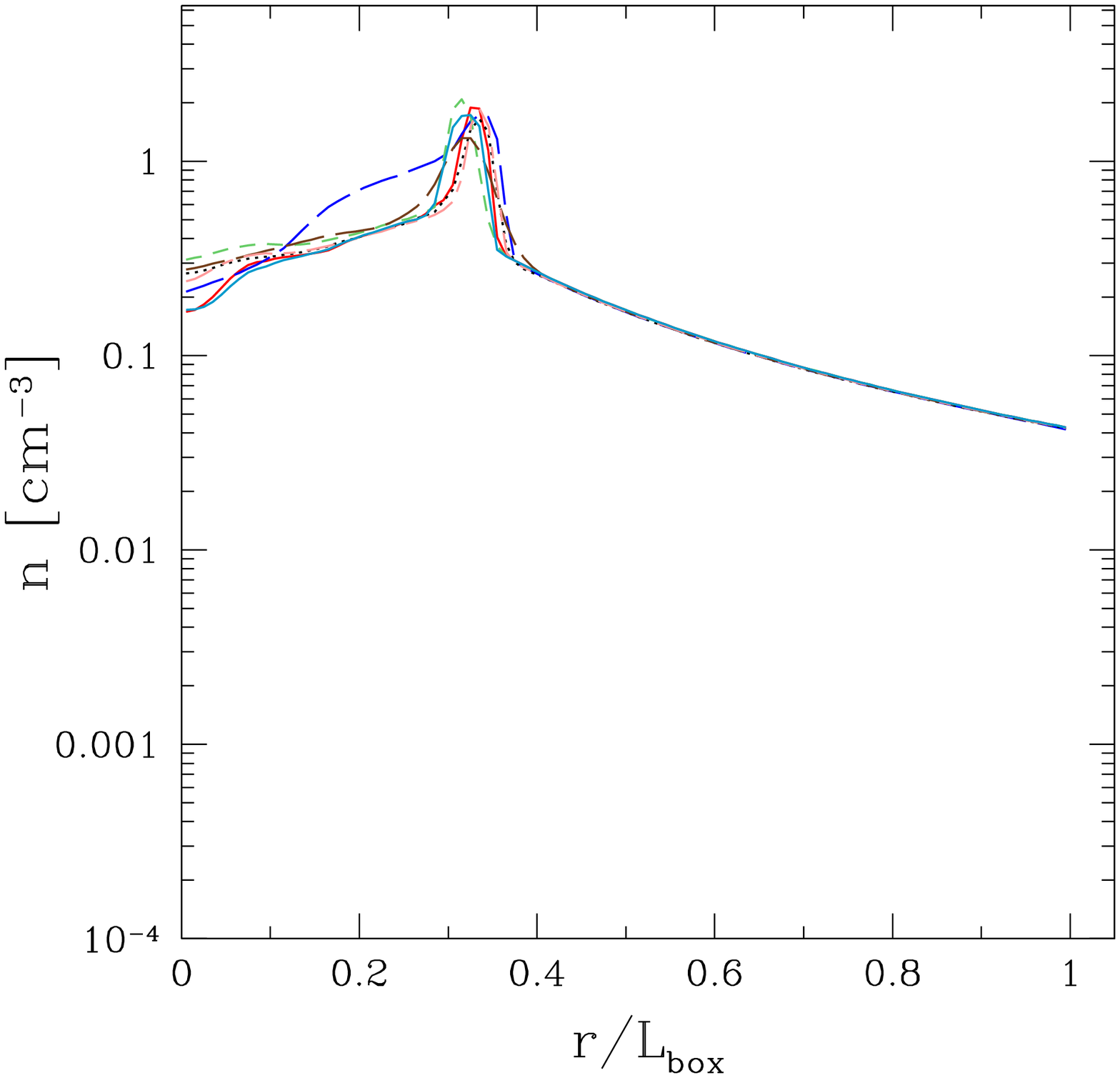}
  \includegraphics[width=2.3in]{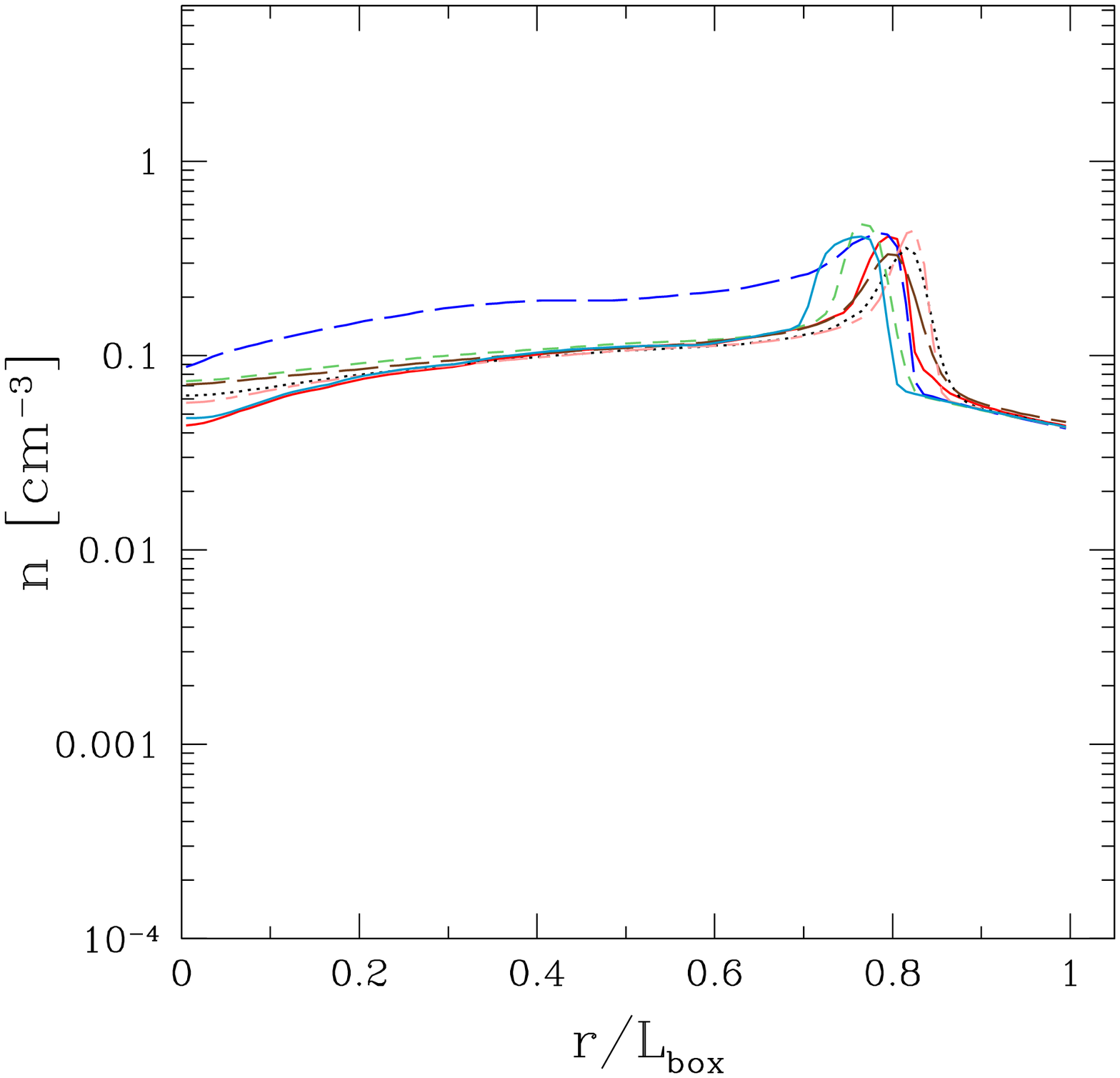}
\caption{Test 6 (H~II region gasdynamic expansion down a power-law initial
  density profile): Spherically-averaged profiles for the gas number
  density, $n$, at times $t=3$ Myr, 10 Myr and 25 Myr vs. dimensionless radius
  (in units of the box size).  
\label{T6_profsn_fig}}
\end{center}
\end{figure*}

\begin{figure*}
\begin{center}
  \includegraphics[width=2.3in]{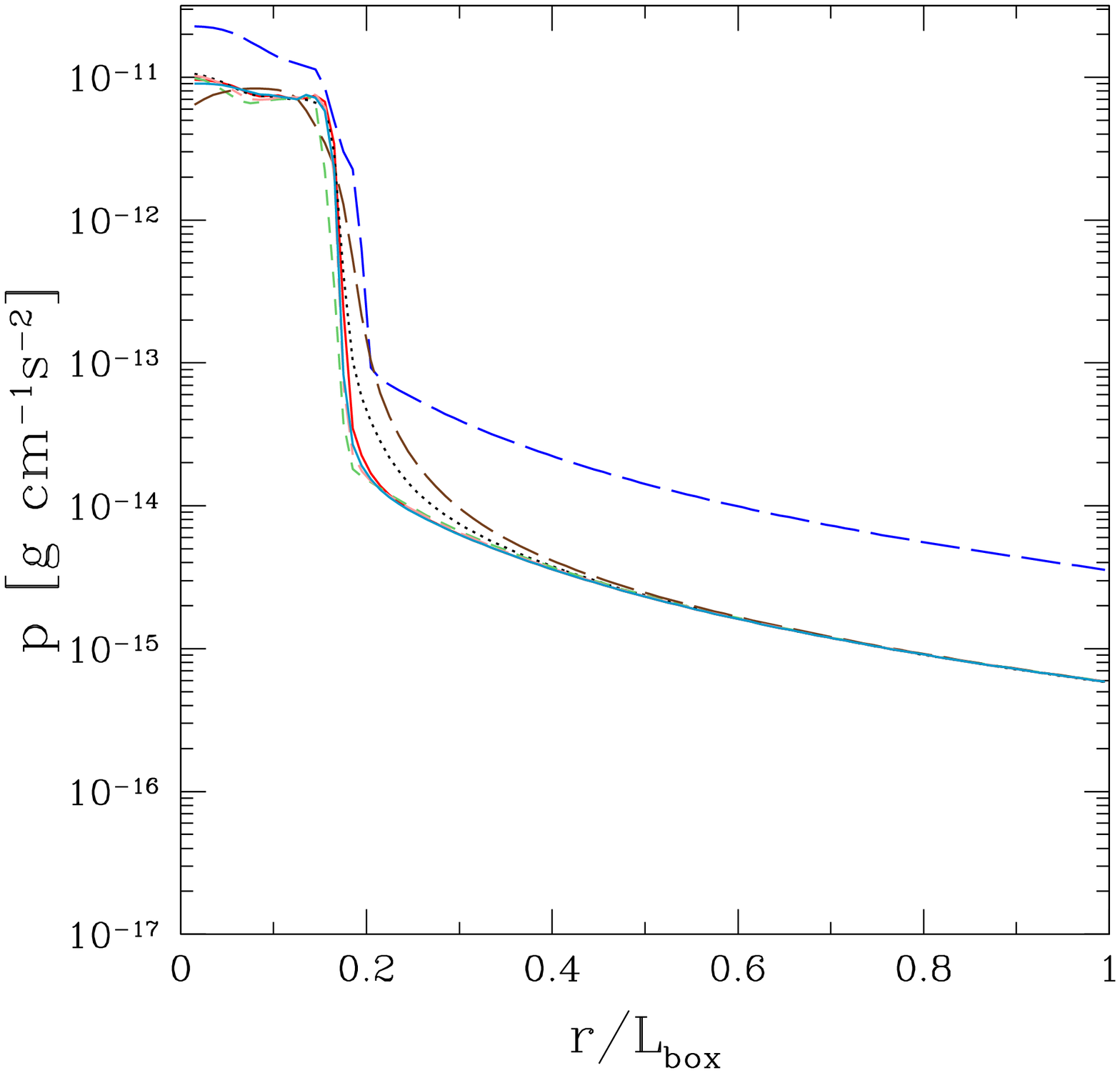}
  \includegraphics[width=2.3in]{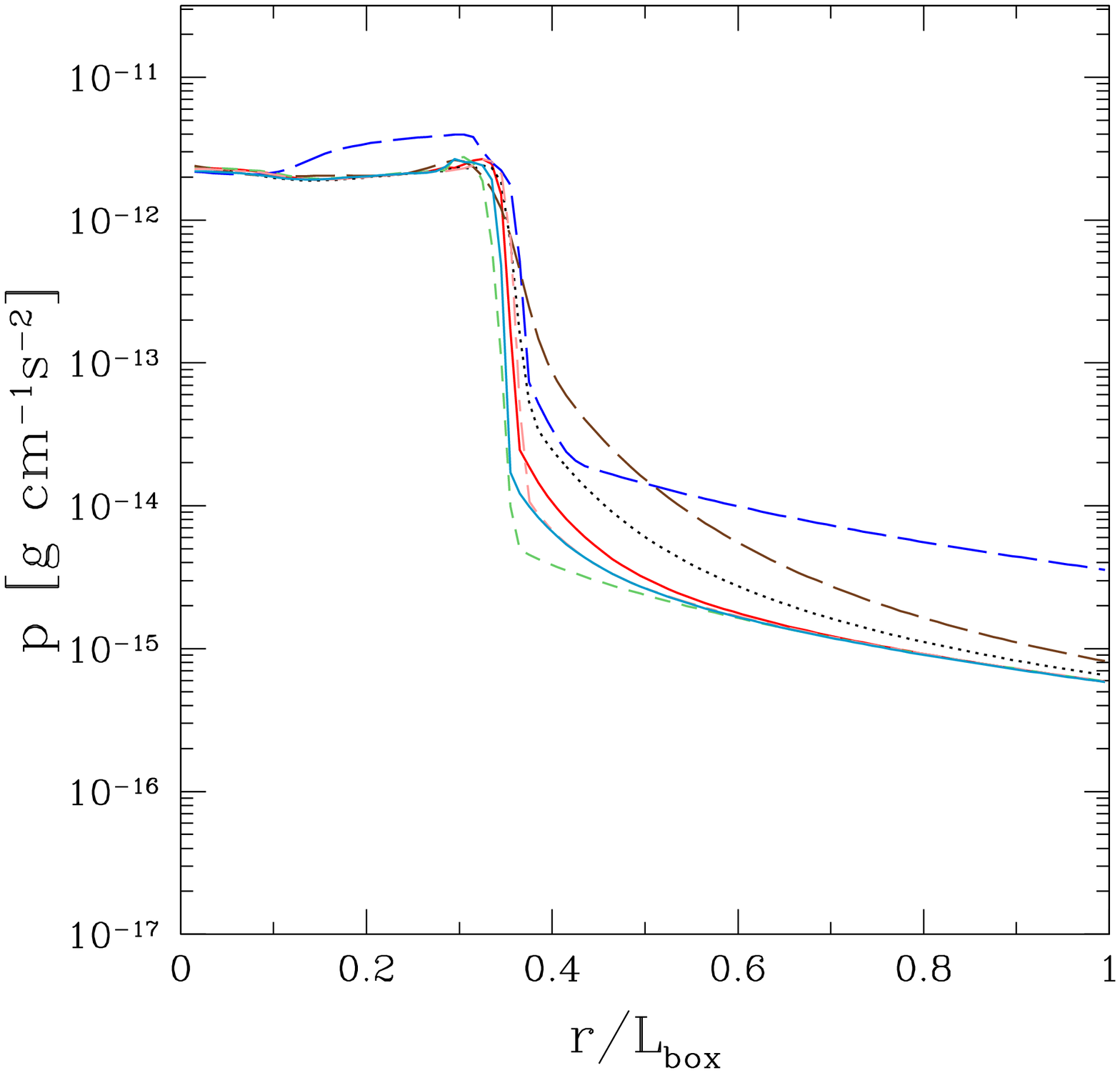}
  \includegraphics[width=2.3in]{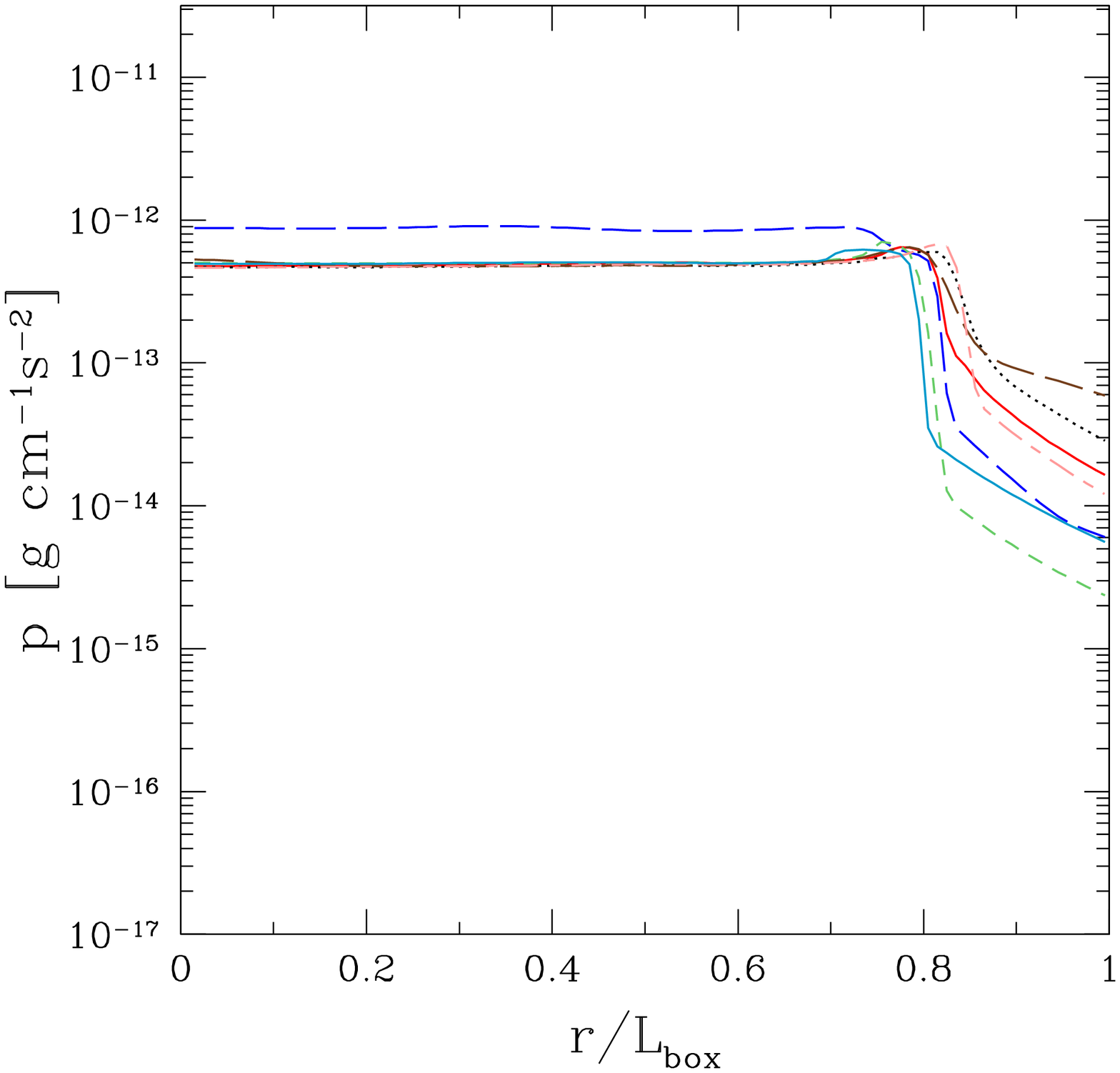}
\caption{Test 6 (H~II region gasdynamic expansion down a power-law initial
  density profile): Spherically-averaged profiles for pressure, $p$, at times
  $t=3$~Myr, 10 Myr and 25 Myr vs. dimensionless radius (in units of the box
  size).  
\label{T6_profsp_fig}}
\end{center}
\end{figure*}
\begin{figure*}
\begin{center}
  \includegraphics[width=2.3in]{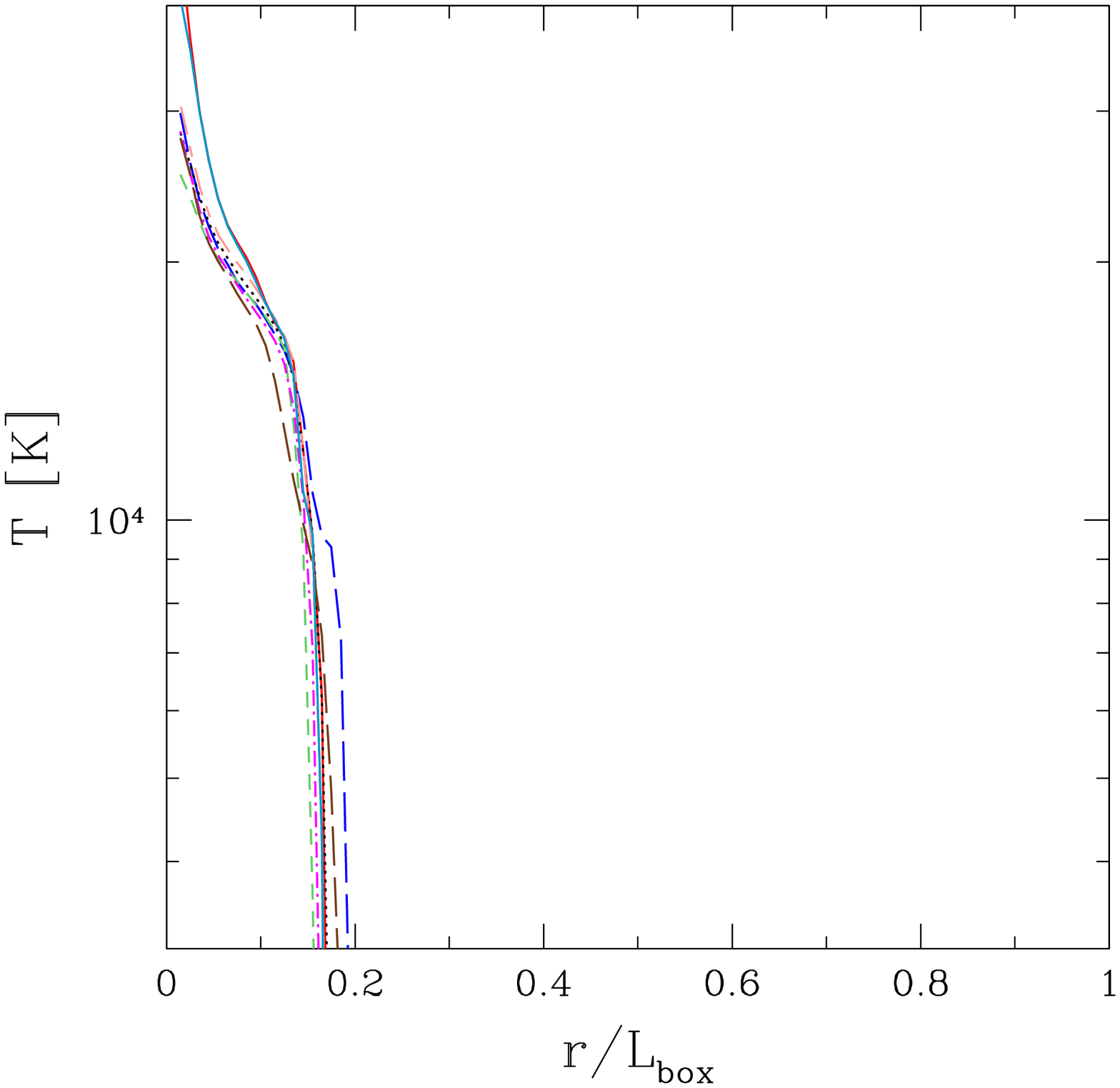}
  \includegraphics[width=2.3in]{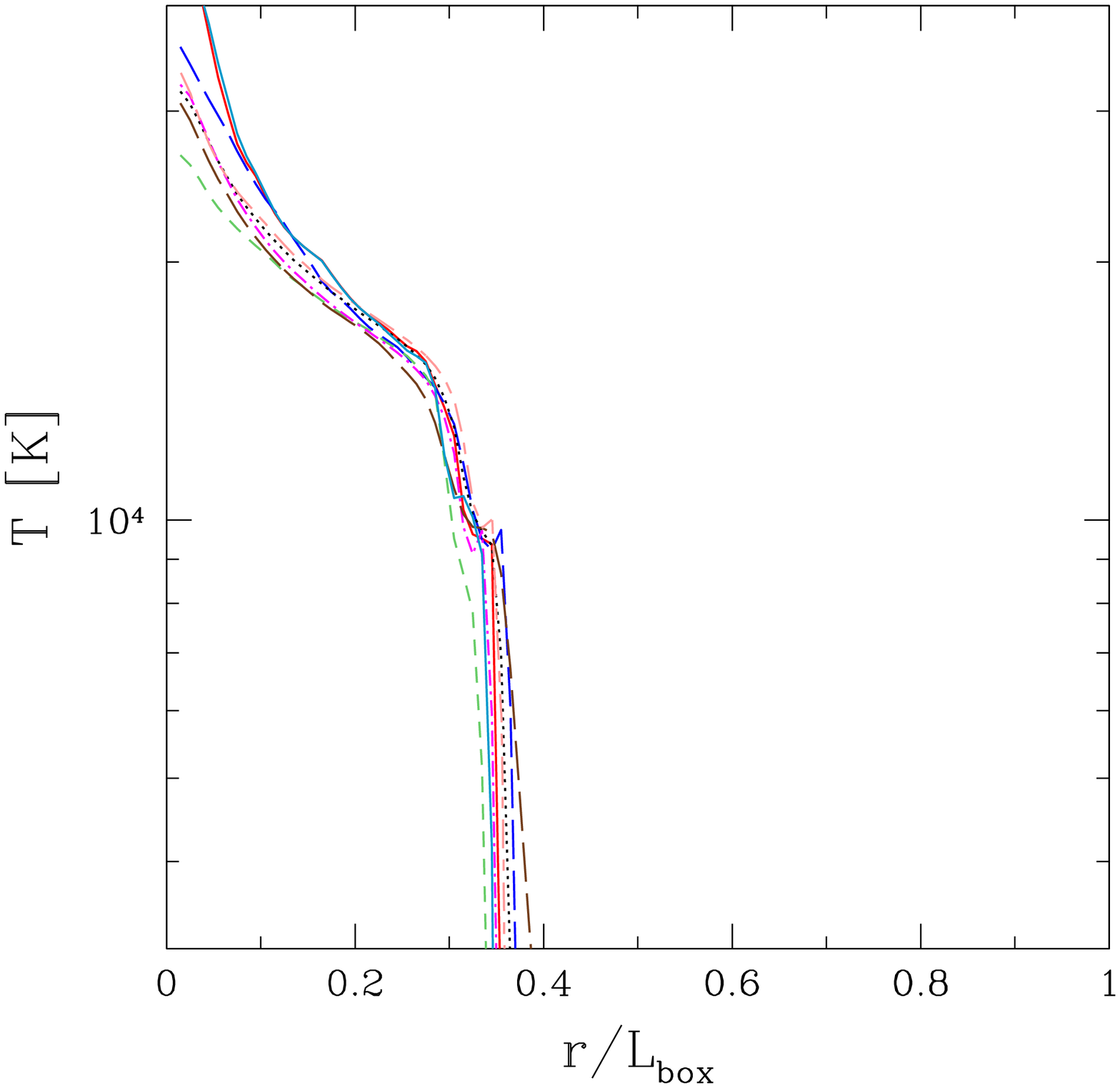}
  \includegraphics[width=2.3in]{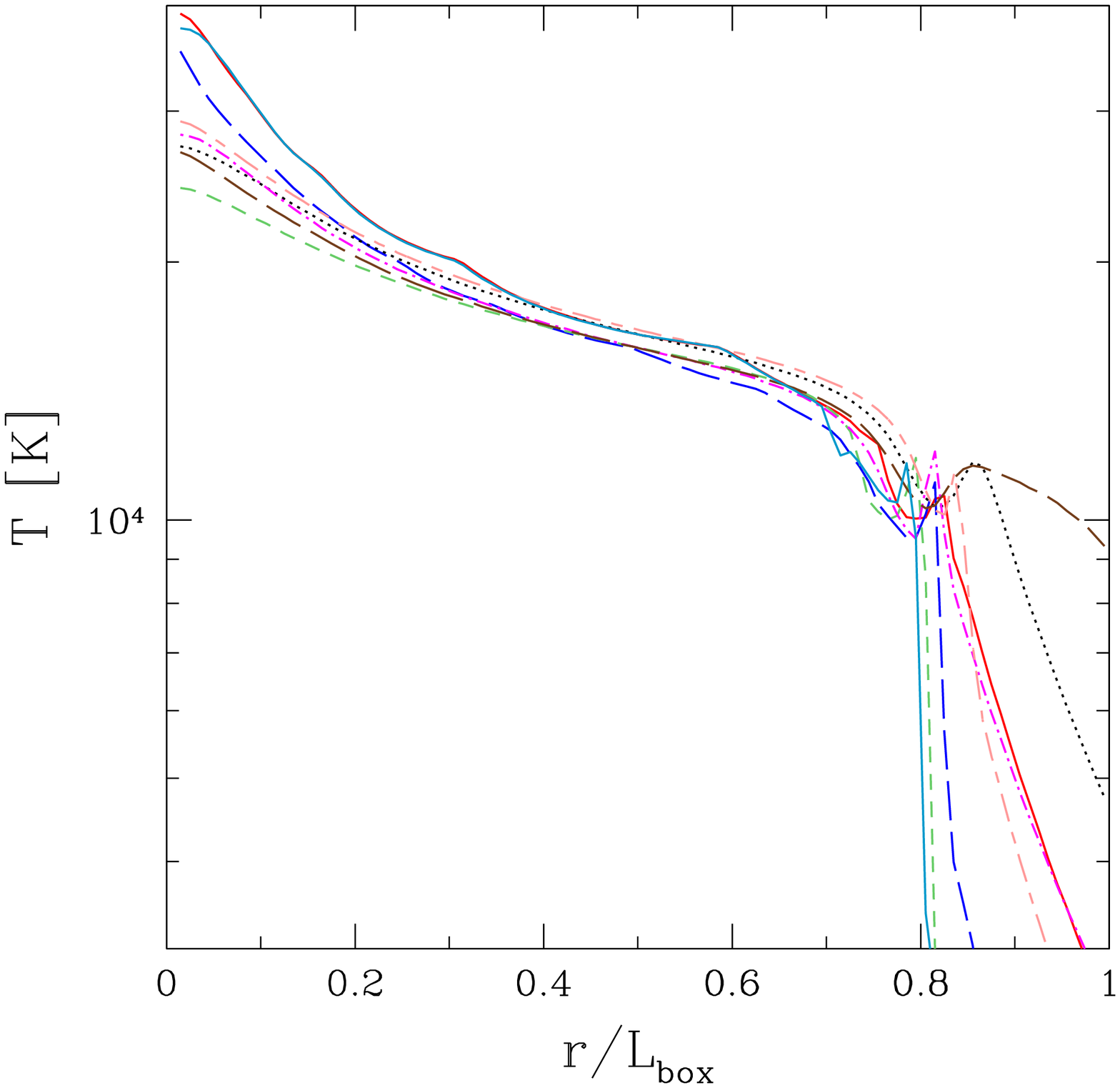}
\caption{Test 6 (H~II region gasdynamic expansion down a power-law initial
  density profile): Spherically-averaged profiles for temperature at times
  $t=3$~Myr, 10 Myr and 25 Myr vs. dimensionless radius (in units of the box
  size).  
\label{T6_profsT_fig}}
\end{center}
\end{figure*}

\begin{figure*}
\begin{center}
  \includegraphics[width=2.3in]{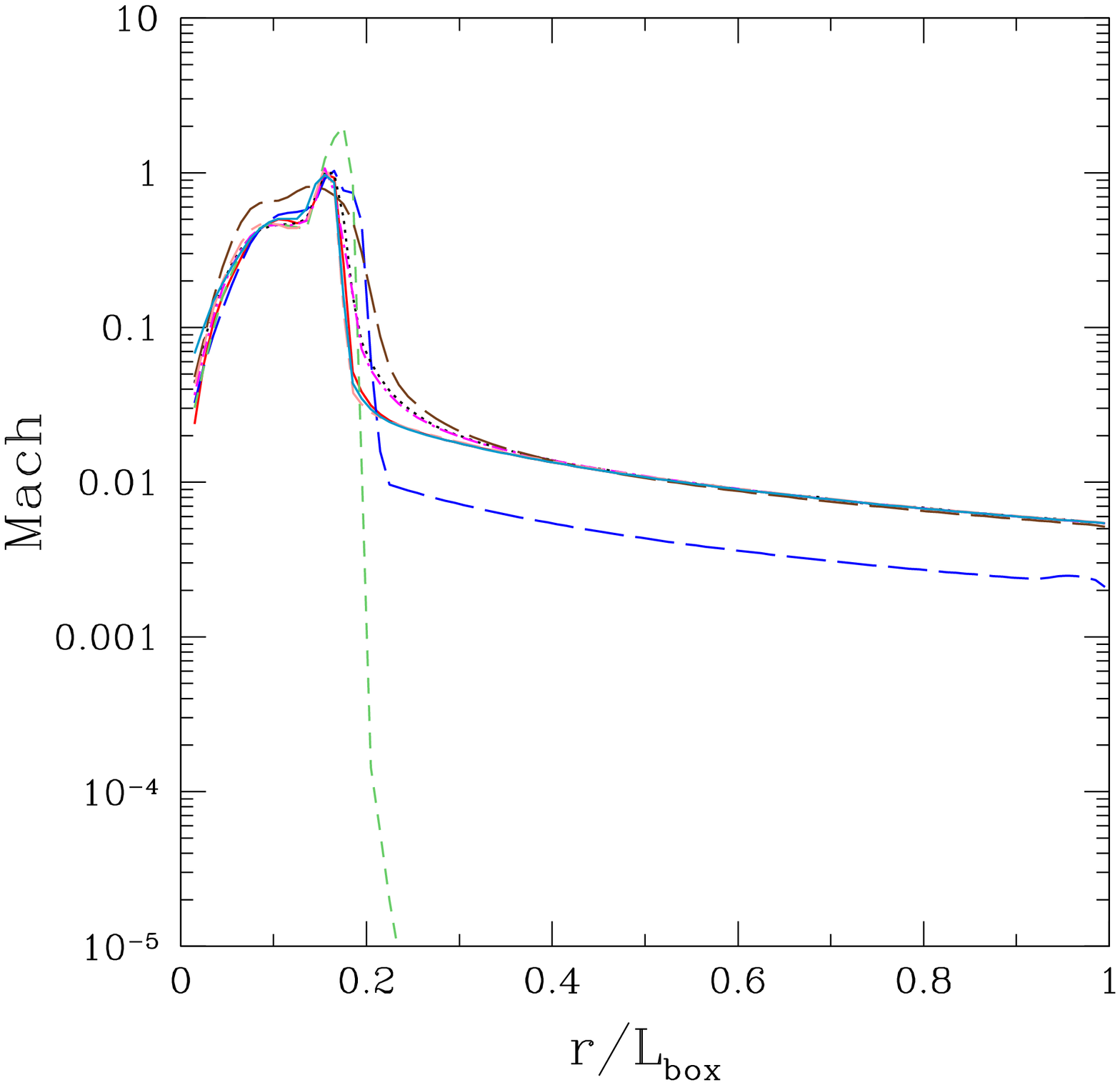}
  \includegraphics[width=2.3in]{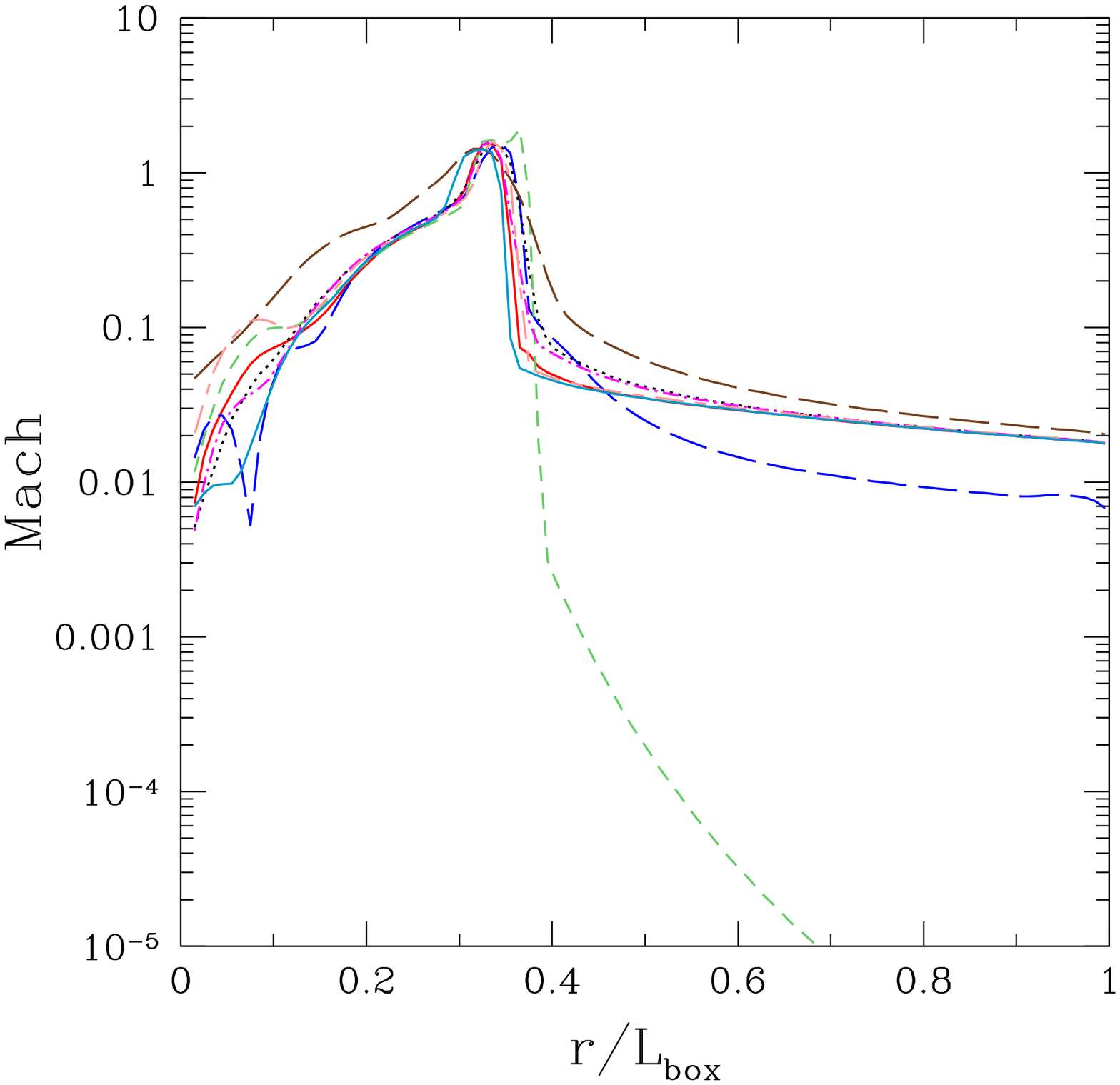}
  \includegraphics[width=2.3in]{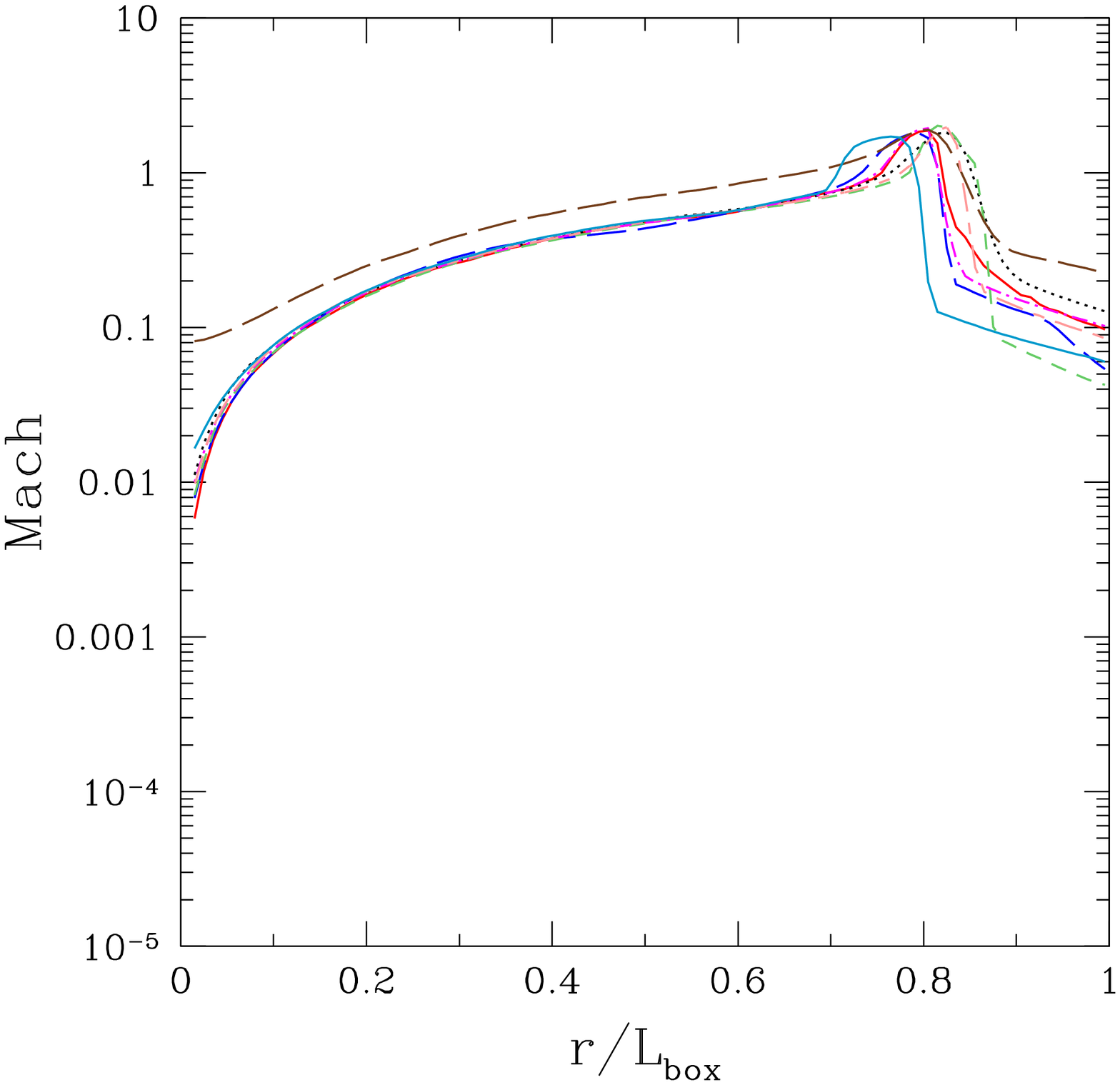}
\caption{Test 6 (H~II region gasdynamic expansion down a power-law initial
  density profile): Spherically-averaged profiles of the Mach number at times
  $t=3$~Myr, 10 Myr and 25 Myr vs. dimensionless radius (in units of the box
  size).  
\label{T6_profsm_fig}}
\end{center}
\end{figure*}

Three types of dynamical instabilities in ionization fronts have been discovered 
in the past thirty years. The first type occurs in D-type ionization fronts whose 
shocked neutral gas shells can cool efficiently by radiation \citep{g79,gsf96}.  
Cooling collapses the gas into a cold thin dense layer that is prone to 
oscillations and fragmentation \citep{v83}. Ionizing UV radiation then 
opportunistically escapes through the cracks in the shell and flares outward in 
violent instabilities. However, in the current Test only H lines can cool the 
shell, and recent numerical experiments prove that such cooling is too 
inefficient to initiate dynamical instabilities of this type \citep{wn08b}. The 
fact that thin-shell instabilities do not arise in the Test 5 profiles further 
attests to the fact that H line cooling is not responsible for the corrugations 
in the Test 6 profiles. In general, shocks that accelerate down power law density 
gradients steeper than r$^{-2}$ are also prone to Rayleigh-Taylor instabilities 
that do not require radiative cooling. They too are also capable of inciting 
violent instabilities in I-fronts, but are not relevant to the r$^{-2}$ gradients 
in the current test. Another type of instability can appear in D-type fronts when 
photons are incident to the front at oblique angles \citep{rjr02}, but they cannot 
develop in I-fronts given the imposed initial spherical symmetry of this test.

\begin{figure}
 \begin{center}
   \includegraphics[width=3.5in]{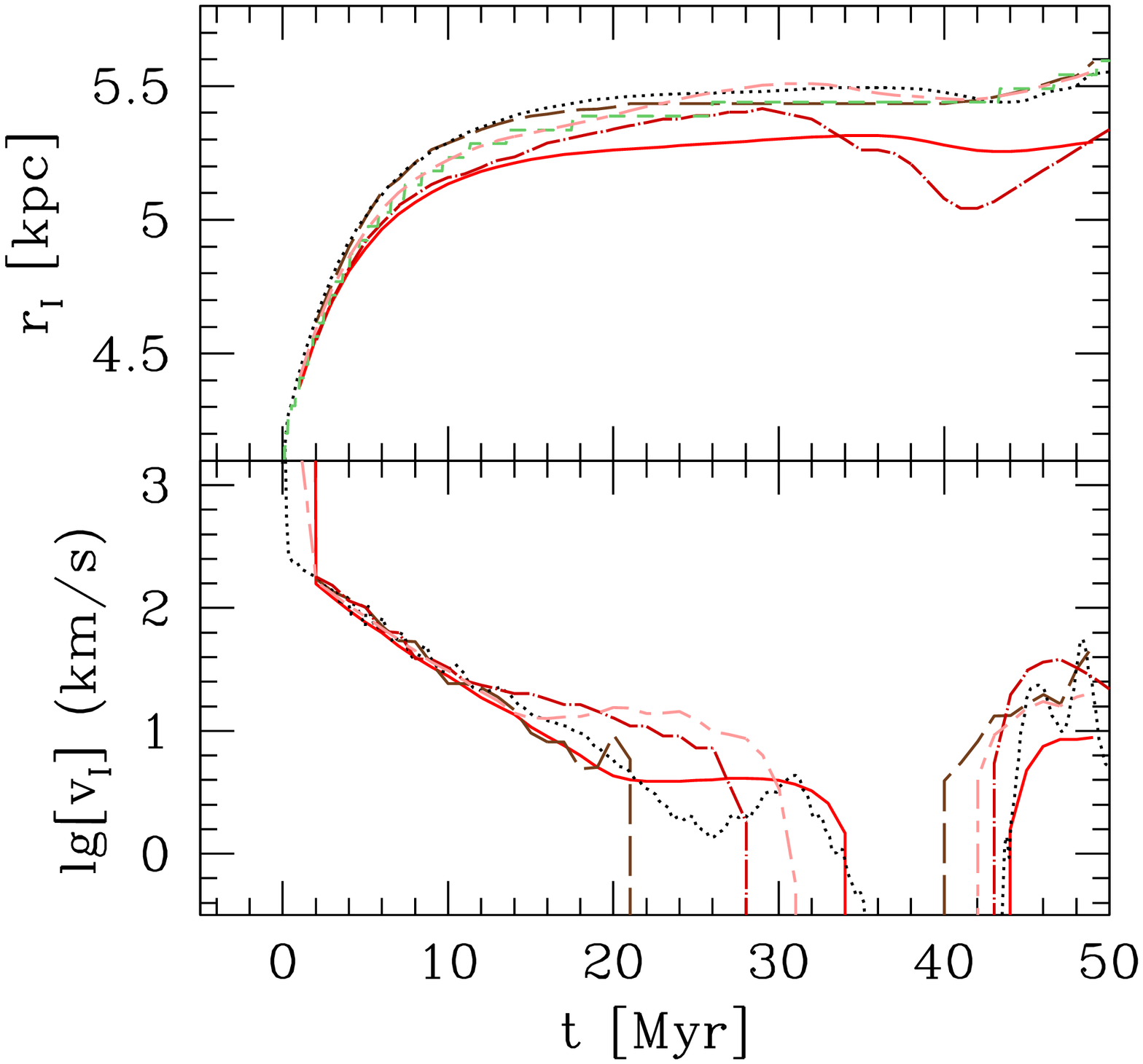}
 \caption{Test 7 (Photoevaporation of a dense clump): The evolution of the 
 position and velocity of the I-front along the axis of symmetry through the 
 centre of the clump.
 \label{T7_Ifront_evol_fig}}
 \end{center}
 \end{figure}

The third type, shadow instabilities, can appear when a density perturbation is 
advected through an R-type front, forming dimples that erupt into violent 
instabilities when the I-front becomes D-type \citep{rjr99}. Although the density 
profile in this test is radially symmetric, prescriptions for imposing 
spherically-symmetric profiles on a Cartesian mesh as a function of radius that 
are too simple can lead to minor departures from radial symmetry in densities 
between neighbor grid points. As the front crosses these mesh points it can become 
dimpled just as if real physical perturbations had traversed it. These corrugations 
would then grow into the much more prominent features visible in the 25 Myr images 
upon transformation of the front to D-type. If this were the case, they would be 
dampened by employing higher grid resolution or by a better prescription for 
smoothing densities between neighbor mesh points. We tested the latter possibility
by applying an algorithm in C$^2$-Ray that sub-sampled volumes enclosed between 
adjacent points in radius with thirty more finely subdivided shells and then 
interpolated densities accordingly between the points. This measure failed to 
alleviate the instabilities in the I-front in C$^2$-Ray, suggesting that they are 
not shadow instabilities. 

Instead, their shape and position at early times reveal 
that they are the infamous 'carbuncle' phenomenon or 'odd-even decoupling' 
\citep{quirk94}. Low-diffusion solvers, such as the Roe's approximate Riemann 
solver employed in Capreole+C$^2$-Ray and the PPM Riemann solvers in Flash-HC 
and Enzo-RT, sometimes suffer from these numerical 
instabilities.  They occur when shocks travel parallel to one of the coordinate 
axes, in this case beginning near the symmetry axes of the expanding shell. These 
instabilities are well-known and mostly understood, although their occurrence is 
not always predictable.  Although by 25 Myr most of the shell has been disrupted, 
we find that the perturbations begin near the axis and exhibit the characteristic 
morphology of the carbuncle instability. While the usual solution is to artificially 
introduce more diffusion only where it is needed, this approach did not suppress the 
phenomenon in C$^2$-Ray. If the test is instead run with C$^2$-Ray coupled to the
TVD solver of Trac and Pen \citep{2004NewA....9..443T}, which is more diffusive and 
not known to suffer from the carbuncle instability, the shell indeed remains well 
behaved. We also note that this carbuncle instability, while fairly violent, does 
not in fact affect the fluid flow or I-front propagation significantly and the
I-front position and velocity evolution discussed above and the spherically-averaged 
profiles discussed below are still in good agreement with the other results.
Therefore, effect of this is fairly modest at the early times studied here, but as the 
previously cited work demonstrates, they could become important at later times.

Another possibility is that there is a hitherto unknown breakout instability 
associated with the transition of the D-type I-front back to R-type as it 
descends the density gradient.  However, such an instability would not have
the opportunity to propagate throughout and disrupt the entire shell during 
breakout as in the C$^2$-Ray results because the transition from D-type to 
R-type is too abrupt, and, as discussed above, the I-front remains D-type within
the computational volume. Such an instability would instead be manifest as a 
premature supersonic runaway of radiation from the surface of the shell along 
certain lines of sight with little disruption of the shell itself, as observed
in the Flash-HC results. However, this does not happen in the ZEUS-MP profiles, 
which are computed on a 3D spherical coordinate grid that is naturally suited to 
spherically-symmetric density fields and on which the 'corner' effects inherent 
in Cartesian grids are absent. Furthermore, as the spherically averaged profiles 
show, the I-front in this test is not on the verge of breaking past the shock and 
becoming R-type at late times. We therefore conclude that the instability in the 
Flash-HC results is not physical and that the early breakout of radiation there 
is at least partly the result of gridding a spherical density on a Cartesian grid.
The effect of this is fairly modest, however, and does not disturb the overall
dynamics significantly.

The irregular morphology of the shell in the LICORICE results and to a small 
extent in the RSPH profile at 25 Myr is likely due to spurious fluctuations in 
the density field where the local particle number changes sharply, in this case 
in the vicinity of the dense shell.  This well-known feature of SPH, as discussed 
in section 2.7, is what probably allows radiation to preferentially advance along 
lines of sight through low-density fluctuations in the two profiles. The larger
effects for LICORICE compared to RSPH are probably due to the usage of a grid to 
perform the radiative transfer in the former. Once again, none of these effects 
appears to affect the overall evolution significantly, but they might matter in 
certain astrophysical situations. 

The HART results exhibit banding in all the profiles except for HI fraction at 
25 Myr. The origin of these features is unclear, but is possibly related to the 
much coarser AMR griding used around the outer edges. They may also be related 
to the greater diffusivity of the OTVET algorithm, although no such features 
were observed in the other tests performed with OTVET. However, the density 
profile found by HART is much flatter, with no clear dense shell swept by the
shock, in clear contrast to all other results. It is possible that the time step 
applied to the gas energy updates in HART is too coarse for I-fronts in r$^{-2}$ 
density gradients, which has been found to lead to banding in temperatures and 
densities in H II regions in stratified media \citep{1986MNRAS.221..635T,
2006ApJS..162..281W}. 

We finally note that had H$_2$ cooling been included in either this test or in 
Test 5, violent physical instabilities might have arisen in the I-front when it 
became D-type \citep{wn08b}. Hard UV spectra significantly broaden ionization 
fronts, forming regions of a few thousand K and ionized fractions of 10\% in their 
outer layers. These are prime conditions for the catalysis of H$_2$ via the H$^-$ 
channel, which forms between the I-front and the dense shocked shell when the front 
becomes D-type \citep{2002ApJ...575...49R}.  H$_2$ - H, H$_2$ - e$^-$, and H$_2$ - 
H$^+$ collision channels emit ro-vibrational lines \citep{ls83,gp98,ga08} that can 
radiatively cool the base of the shocked layer and incite dynamical instabilities 
there just as metal ions do in galactic I-fronts. These instabilities may have been
common in the early universe, such as in UV breakout from the first star-forming
clouds.  They appear if there is enough H$_2$ in the cloud to self-shield from the 
Lyman-Werner (LW) flux (11.18-13.6 eV) also being emitted by the source, which 
photodissociates molecular hydrogen \citep{1996ApJ...468..269D}. Thus, while the 
instabilities manifested by some of the codes in Test 6 are numerical, physical
ones are possible when H$_2$ cooling, LW photons, and self-shielding to LW 
radiation are properly included.  Since not all the codes contain these physical 
processes, it was not included in any of the current tests, but will be a target
for future stages of this comparison project.   

In spite of the prominence of the numerical instabilities in some of the codes,
we re-iterate that they were not catastrophic to the overall dynamics over the 
range of radii and interval of time which we study here, as shown 
by the spherically averaged hydrodynamical profiles. In Figs \ref{T6_profs_fig} 
- \ref{T6_profsm_fig} we show ionization fractions, number densities, pressures, 
temperatures and Mach numbers at 3, 10, and 25 Myr. Comparison of ionization 
fractions and densities at 3 Myr indicates that the I-front is D-type at 
$r\sim120$~pc, somewhat beyond the flat central core of the initial density 
profile. A thin layer of shocked neutral gas is visible at $r \sim$ 140 pc in 
the Mach number profile.  
Acoustic waves are evident at $r < 100$~pc within the H II region in both the 
density and Mach number plots and are consistently reproduced by all the codes.  

At this early stage, the I-front widths (customarily defined by the difference 
between the positions at which 0.1 and 0.9 ionization fractions are reached) 
vary from $\sim$ 20~pc to 40~pc. At the relatively high inner-profile density 
of $n\sim $1 cm$^{-3}$, the mean free paths (mfp) of 13.6 eV and 60 eV photons,
which roughly bracket the available energies in the $10^5$~K black-body 
spectrum used for this test, are $\sim0.05$~pc and 4~pc, respectively. The 
intrinsic width of the I-front is approximately 20 mean-free paths, or between 
1 and 80 pc. Therefore, all the codes give widths roughly consistent with the 
expected values, but somewhat on the wider side, primarily due to the 
diffusivity of some of the algorithms. In particular, LICORICE and HART have 
wider fronts, while ZEUS-MP has the narrowest one and the rest are spread 
between those two extremes. As explained above, the details of the structure 
of the I-front are interesting since they relate to the formation of molecular 
hydrogen in its outer layers \citep[e.g.][]{2002ApJ...575...49R} where hard, 
deeply penetrating photons could yield a positive feedback mechanism during 
early structure formation. The post-front (ionized) gas temperatures of the 
codes at these early times differ by at most 10\%. The low Mach numbers 
outside the H II region at early times, ranging from 0.001 to 0.01 arise 
because hydrostatic equilibrium was not imposed on the original density 
profile in any of the codes except for ZEUS-MP. Pressure forces gently 
accelerate the gas outward, but this has little effect on the late-time
evolution of the H II region. For the purposes of this comparison the lack
of initial hydrostatic equilibrium is irrelevant, as long as all 
codes start from the same initial conditions.

By 10 Myr the H II region has grown to 240 pc, with all results agreeing 
well on the I-front position. There is very little variation in the ionization 
structure inside the H~II region, with only LICORICE finding a slightly lower 
level of ionization. The differences in the pre-front ionization structures 
are much more pronounced, underlying again the variety in the treatments of
multi-frequency photons. All the codes still find postfront gas temperatures 
within 10\% of one another at most radii. The temperature profiles drop sharply
just beyond the I-front as before, but then briefly plateau at 10$^4$ K for 
$\sim$ 16 pc before falling further. This is the dense ambient neutral gas 
shell-swept up by the shock, clearly seen in the density profiles 
(Figure~\ref{T6_profsn_fig}), which 
is sufficiently hot and dense to become collisionally ionized to a small degree.  
However, the minute residual ionized fractions (10$^{-4}$ - 10$^{-3}$) in and 
beyond the shell in most of the plots occur because the I-front broadens over 
time. As more neutral gas accumulates on the shell with the expansion of the 
H~II region, its density decreases because its area grows, and its optical 
depth to photons in the high-energy tail of the spectrum decreases. ZEUS-MP 
still finds the sharpest I-front and LICORICE the thickest, with the rest 
dispersed between. No single cause can be ascribed to the moderate variation 
in I-front structure amongst the codes; for example, both ZEUS-MP and RH1D 
perform multifrequency ray-tracing radiative transfer with similar integration 
schemes and frequency binning, but RH1D has a noticeably wider I-front. 
Tabulating pre-computed frequency-dependent ionization rate integrals as a 
function of optical depth as an alternative to full multifrequency RT in 
Capreole+$C^2$-Ray leads to a somewhat different structure for the front.  
There is an unmistakable trend toward greater diffusivity with the SPH-coupled 
radiative transfer codes RSPH and LICORICE that is likely related to the inherent 
difficulty in representing low-density regions with SPH particles and the 
tendency of SPH to broaden shocks. Nonetheless, the grid-based codes and RSPH 
agree to within a few percent on the density structure of the shocked shell at 
10 Myr. LICORICE does not resolve the shell as well, but this would likely be 
remedied by using more particles to resolve the flow, or by using a more 
adaptive smoothing kernel. Overall, the codes agree reasonably well on the shock
position and the corresponding density and pressure jumps (although, as was 
discussed above, HART yields higher and more uniform density and pressure 
distributions behind the shock than the other the methods, which agree on that 
quite well among themselves).  

At 25 Myr the I-front is at 640 pc, approaching the boundary of our computational 
volume. At this stage the subsonic expansion of the front with respect to the 
sound speed in the ionized gas is evident: acoustic waves have erased density 
fluctuations up to the shocked shell in the pressure and density plots. The 
acceleration of the shock down the density gradient can be seen in the heating 
by the shock: the temperature of the dense shell is 25\% greater than at 10~Myr.  
The velocities beyond the shock are now 20\% of the sound speed of the neutral 
gas, and the peak density of the shell has fallen to 0.4 cm$^{-3}$. There is a 10\% 
variation in the position of the I-front among the codes that is not attributable
to differences in chemistry or radiative transfer because of the uniformity in
ionized gas temperatures (and therefore sound speeds).  More likely, it is due 
to the variety of hydrodynamics schemes (both grid and particle based) applied
to the models.  Apart from the diffusivity of some of the algorithms as manifest
in I-front structure, we find good agreement on the evolution of the H~II region  
in this stratified medium between all the codes. Direct multifrequency RT and 
approximations to full multifrequency transfer with precomputed ionization
integrals both yield extended I-front structures, but it is difficult to assess
which is more accurate since even the two direct methods disagree with each other 
to some degree. The disagreement between the multifrequency codes on the width
of the I-front is probably due to their discretization of the blackbody curve 
and resultant binning of ionizing photon rates according to energy, since they 
otherwise employ the same ionization cross sections and photon-conserving ray
tracing. The number of bins per decade in energy and their distribution in 
frequency can lead to different thicknesses for the front. They have a much
smaller effect on the temperature of the ionized gas, and therefore the position
of the front, because the temperature is more strongly governed by the cooling 
rates than by minor discrepancies in spectral profile.

\subsection{Test 7}

\begin{figure*}
 \begin{center}
   \includegraphics[width=2.3in]{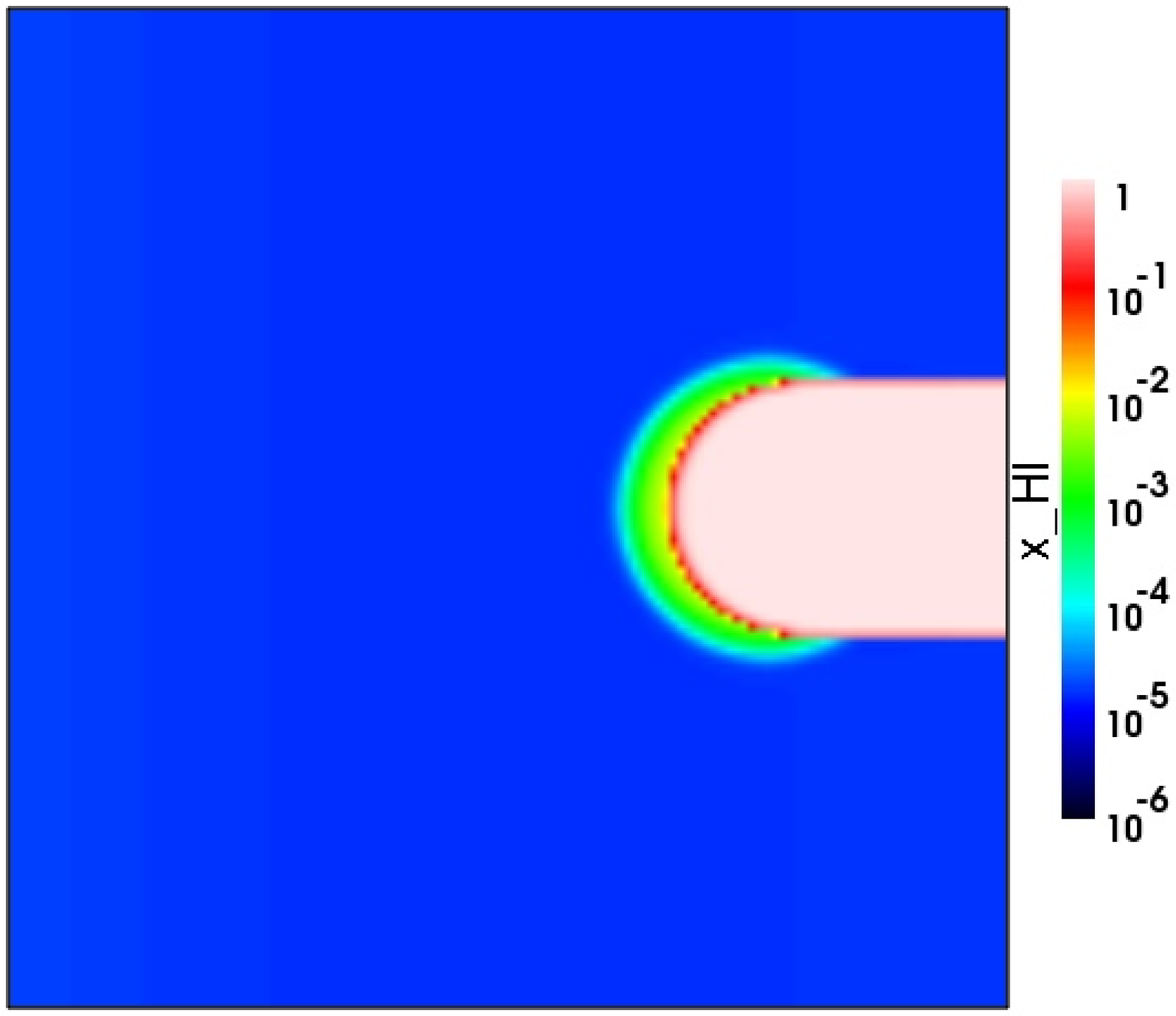}
   \includegraphics[width=2.3in]{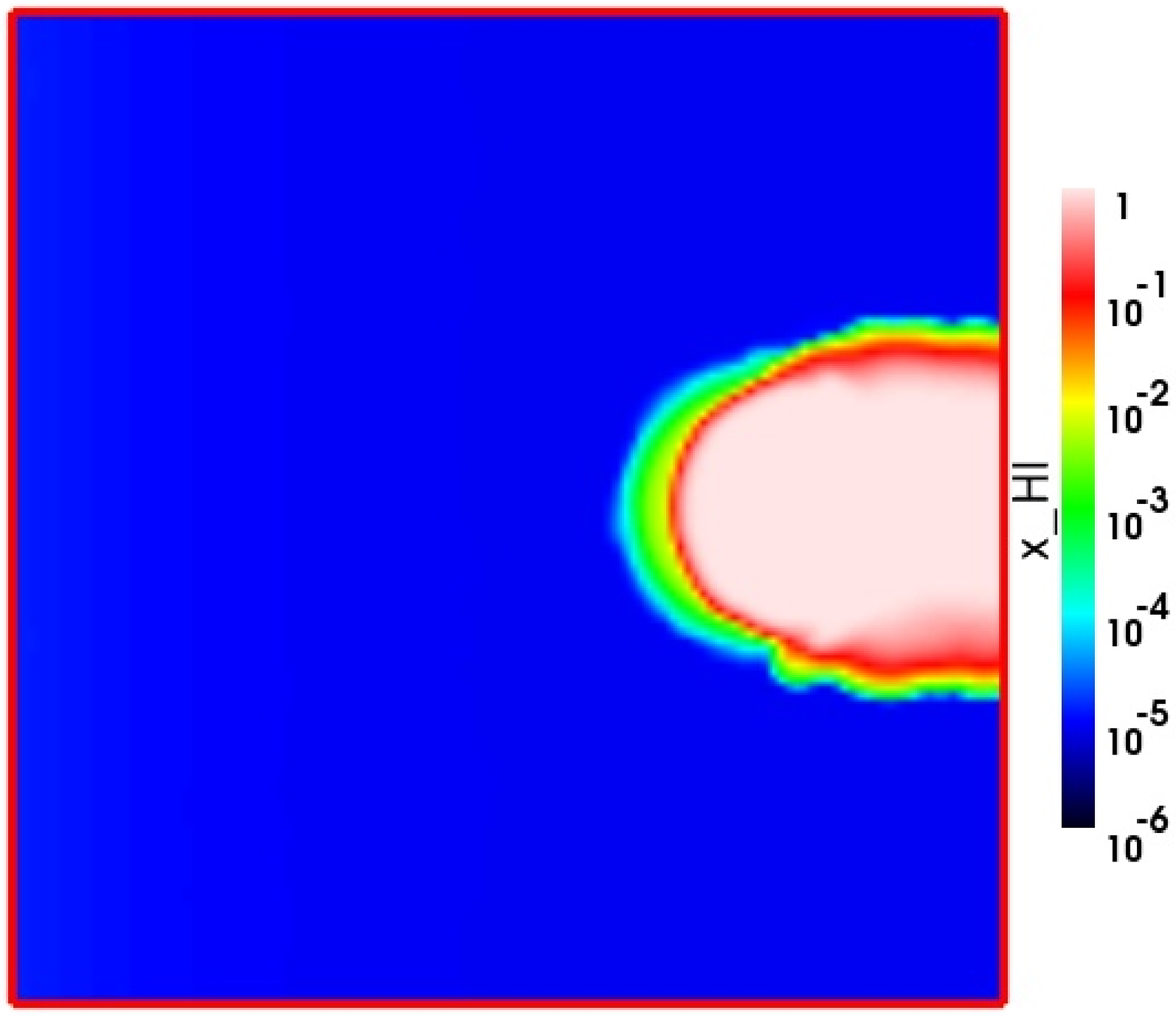}
   \includegraphics[width=2.3in]{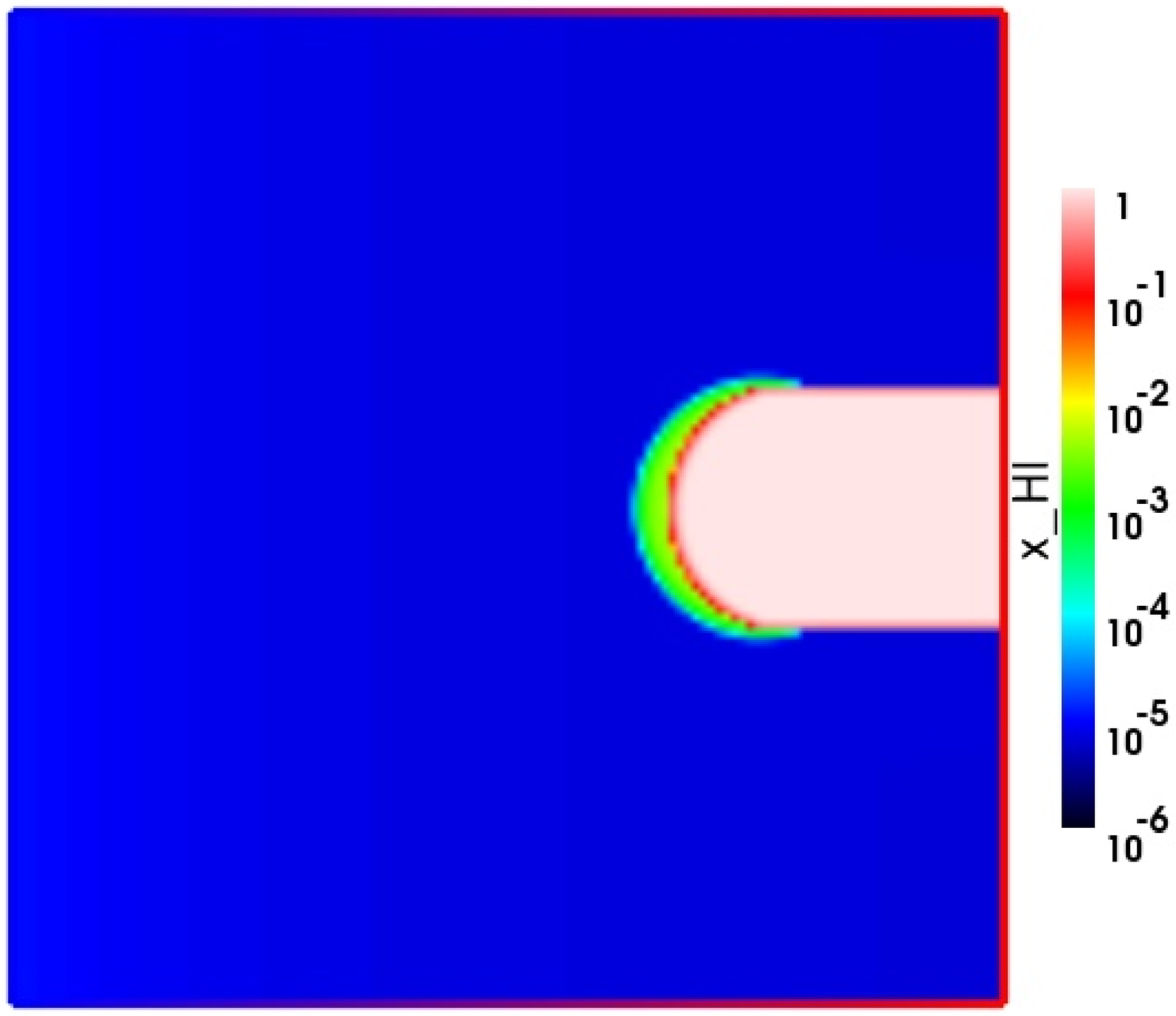}
   \includegraphics[width=2.3in]{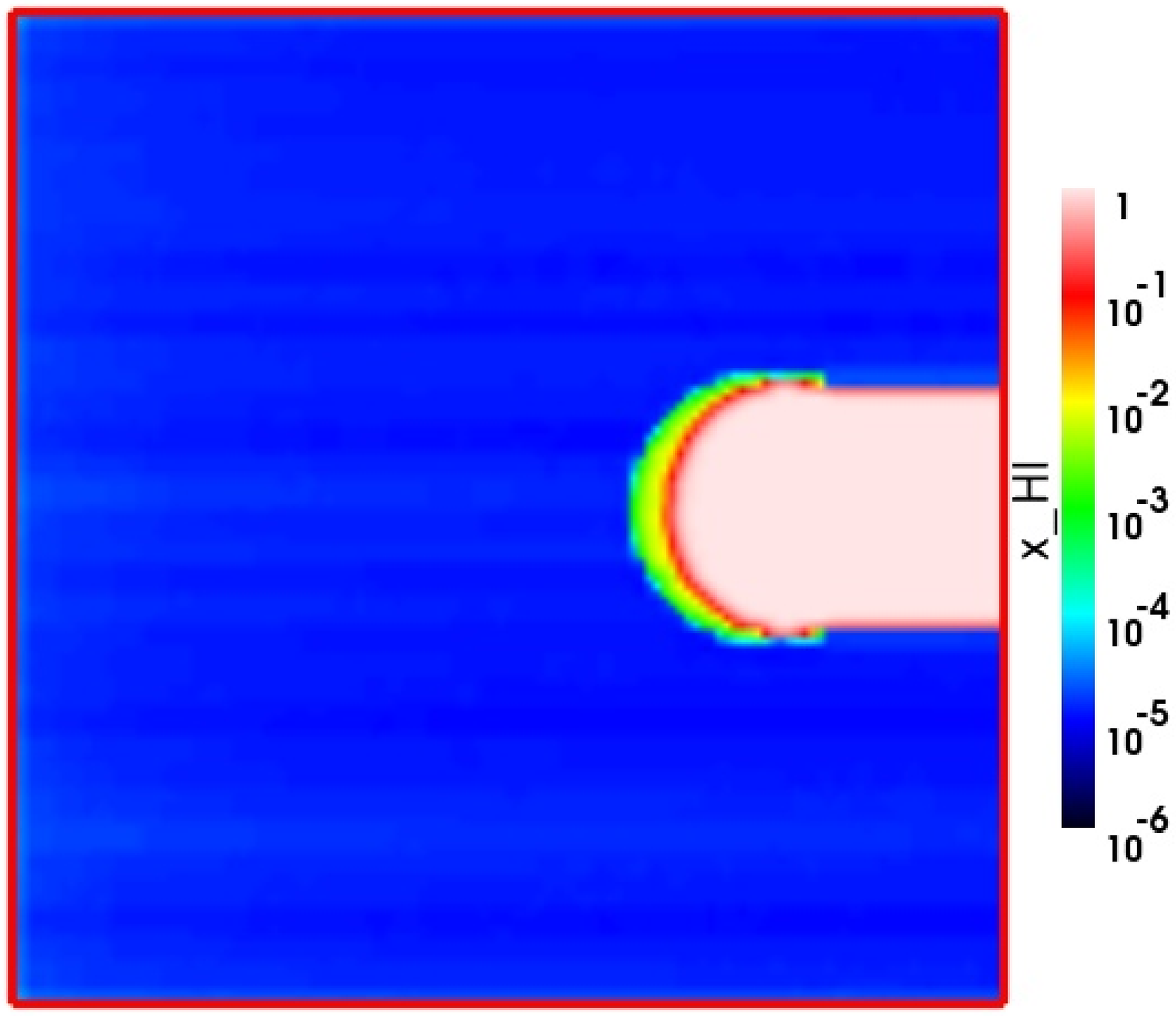}
   \includegraphics[width=2.3in]{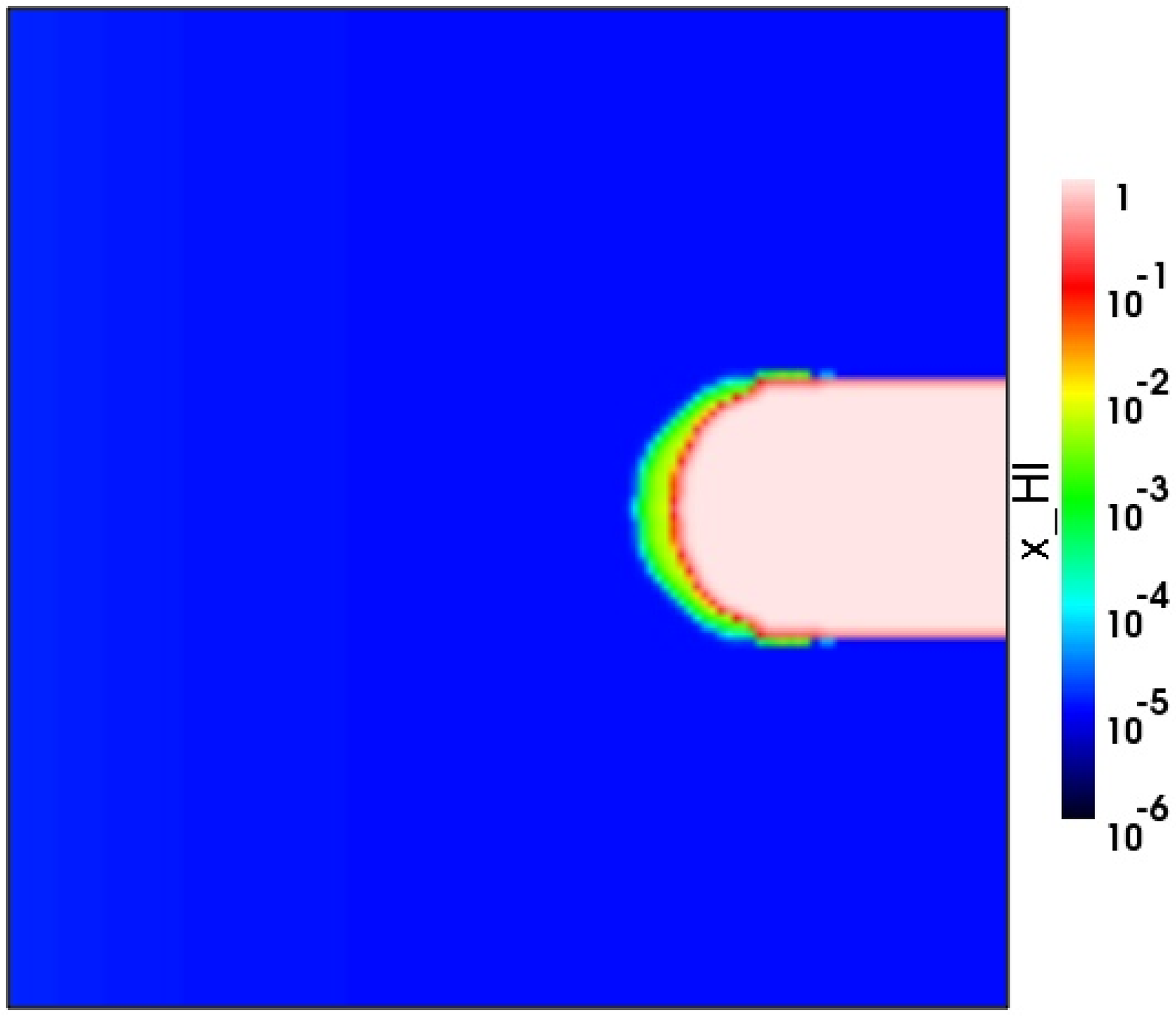}
   \includegraphics[width=2.3in]{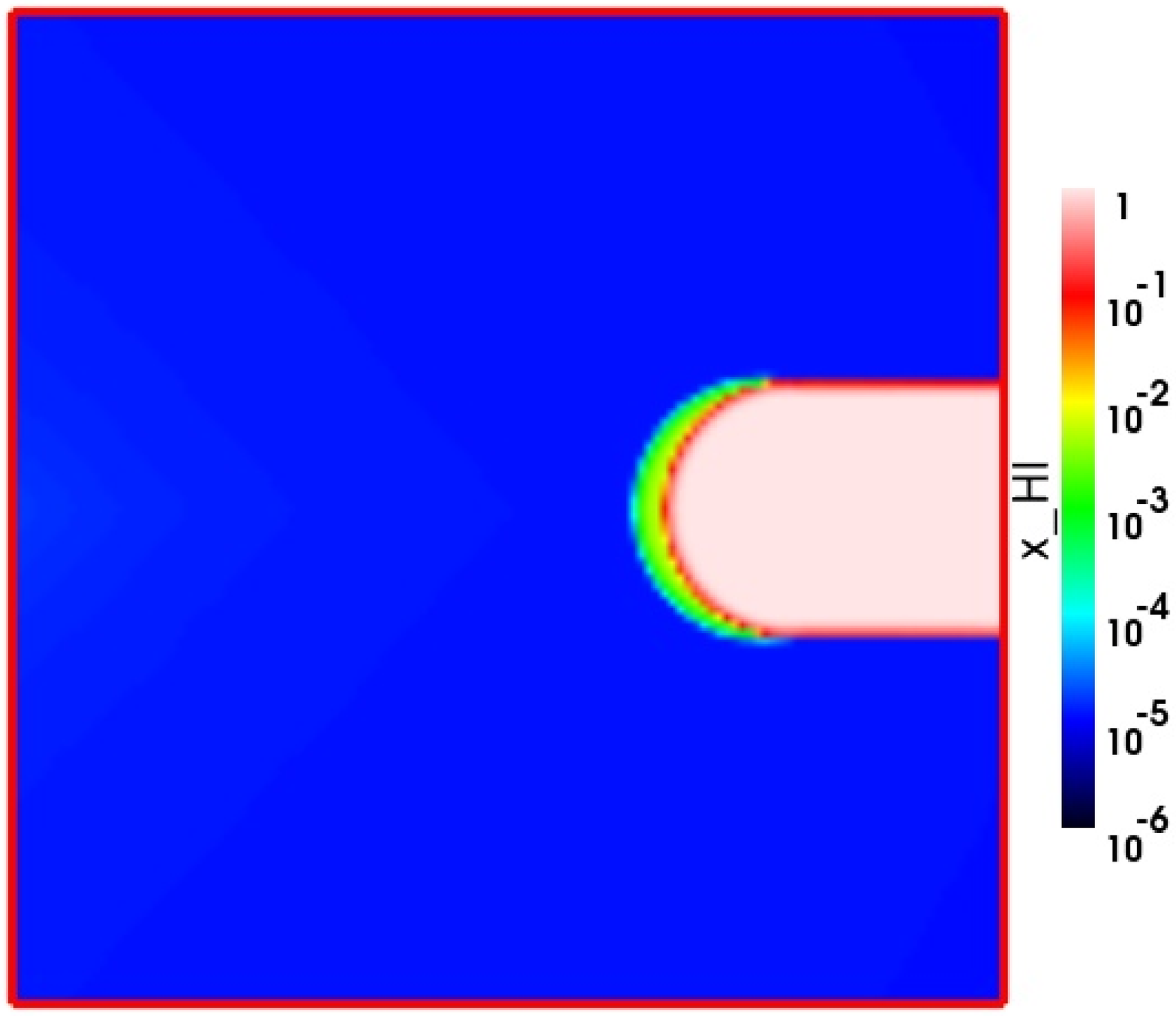}
 \caption{Test 7 (Photoevaporation of a dense clump): Images of the H~I
   fraction, cut through the simulation volume at coordinate $z=0$ at time 
   $t=1$ Myr for (left to right  and top to bottom)
   Capreole+$C^2$-Ray, RSPH, ZEUS-MP, LICORICE, Flash-HC and Coral.
 \label{T7_images1_xhi_fig}}
 \end{center}
 \end{figure*}

The evolution of the I-front along the axis of symmetry through the center of the 
dense clump is shown in Figure~\ref{T7_Ifront_evol_fig}.
The I-front starts off very fast, R-type in the low-density medium surrounding
the dense clump, but slows down quickly once it enters the high-density gas, 
which occurs in less than a Myr. Thereafter, the front slows down more gradually 
as source photons encounter more and more recombining atoms in the photoevaporative 
flow, which attenuates the flux which reaches the I-front. This initial 
trapping phase is largely over by $t=1$~Myr, yielding a thin ionized layer in the 
dense clump on the source side and a clear shadow behind, as illustrated in 
Figure~\ref{T7_images1_xhi_fig}. Due to the short evolution time, by this point the 
gas is still essentially static and Test 6 reproduces the analogous stage in Test 3
in Paper I. There are only a few modest differences between the neutral gas 
distributions. The boundary of the shadowed region "flares" especially for the 
RSPH result, primarily because their particle neighbour list based ray-tracing 
scheme inevitably introduces some "diffusion of optical depth", whereby high 
optical depth spreads through the neighbour list. Flaring of the boundary is also 
in part due to the interpolation procedures to set up the initial conditions which, 
with their sharp boundaries, are unnatural for SPH and thus difficult to represent 
well. In fact, even grid-based codes exhibit similar problems, since the low
resolution required in this Test imposes some grid artefacts on the spherical 
dense clump that are later manifest as ripples in the ablation shock of the clump 
unless a smoothing 
procedure is applied at the setup of the problem.  To minimize artificial features
in the photoevaporative flow from the clump, ZEUS-MP and $C^2$-Ray implemented the
same smoothing procedure to the clump as in Test 6 in their initial density profiles.
There is also some faint striping of the neutral fraction in the low-density region 
for the case of LICORICE, probably due to insufficient Monte-Carlo sampling, as was 
discussed earlier. However, this does not have any apparent effect on the 
photoevaporation of the clump. 

 \begin{figure*}
 \begin{center}
   \includegraphics[width=2.3in]{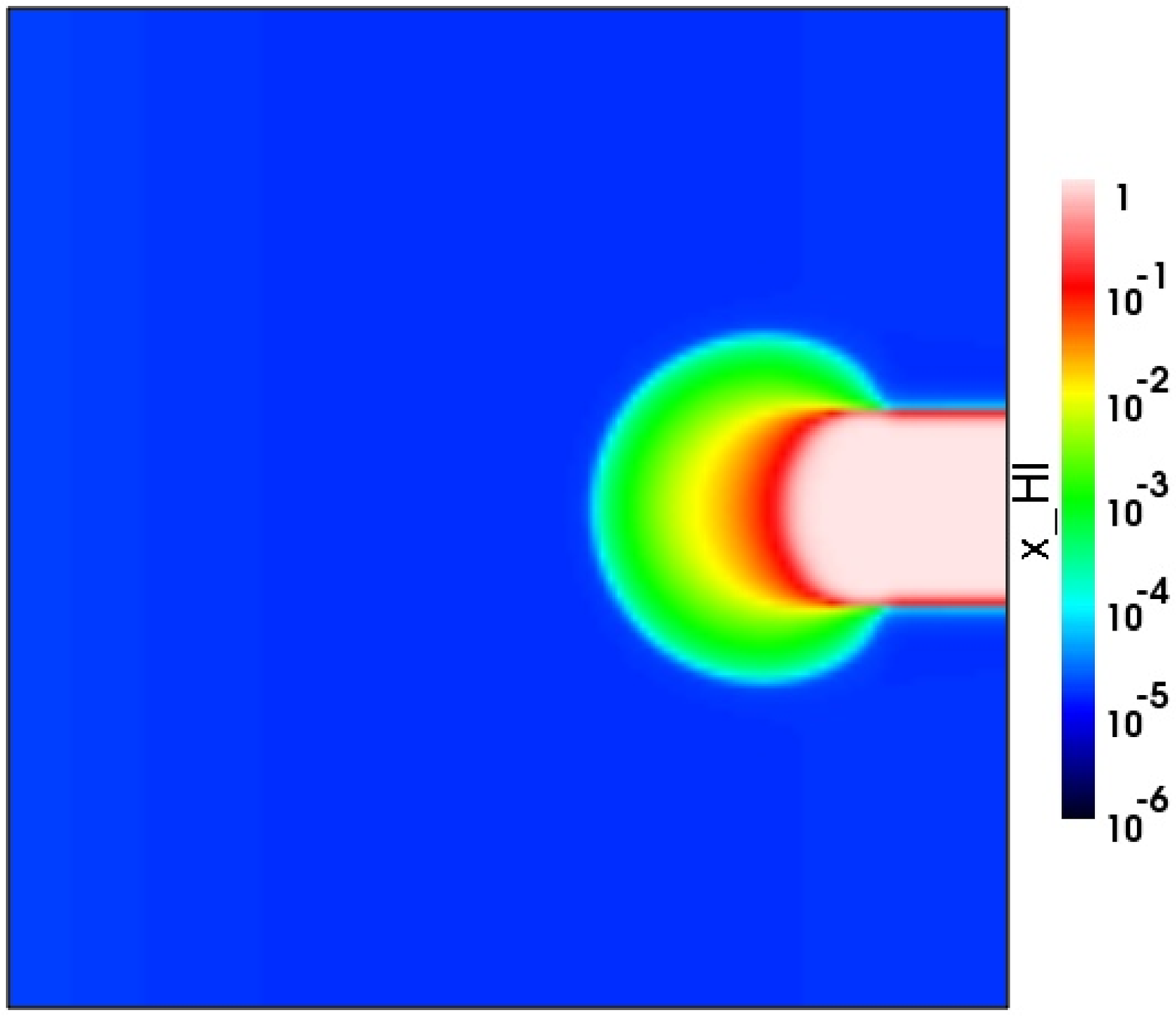}
   \includegraphics[width=2.3in]{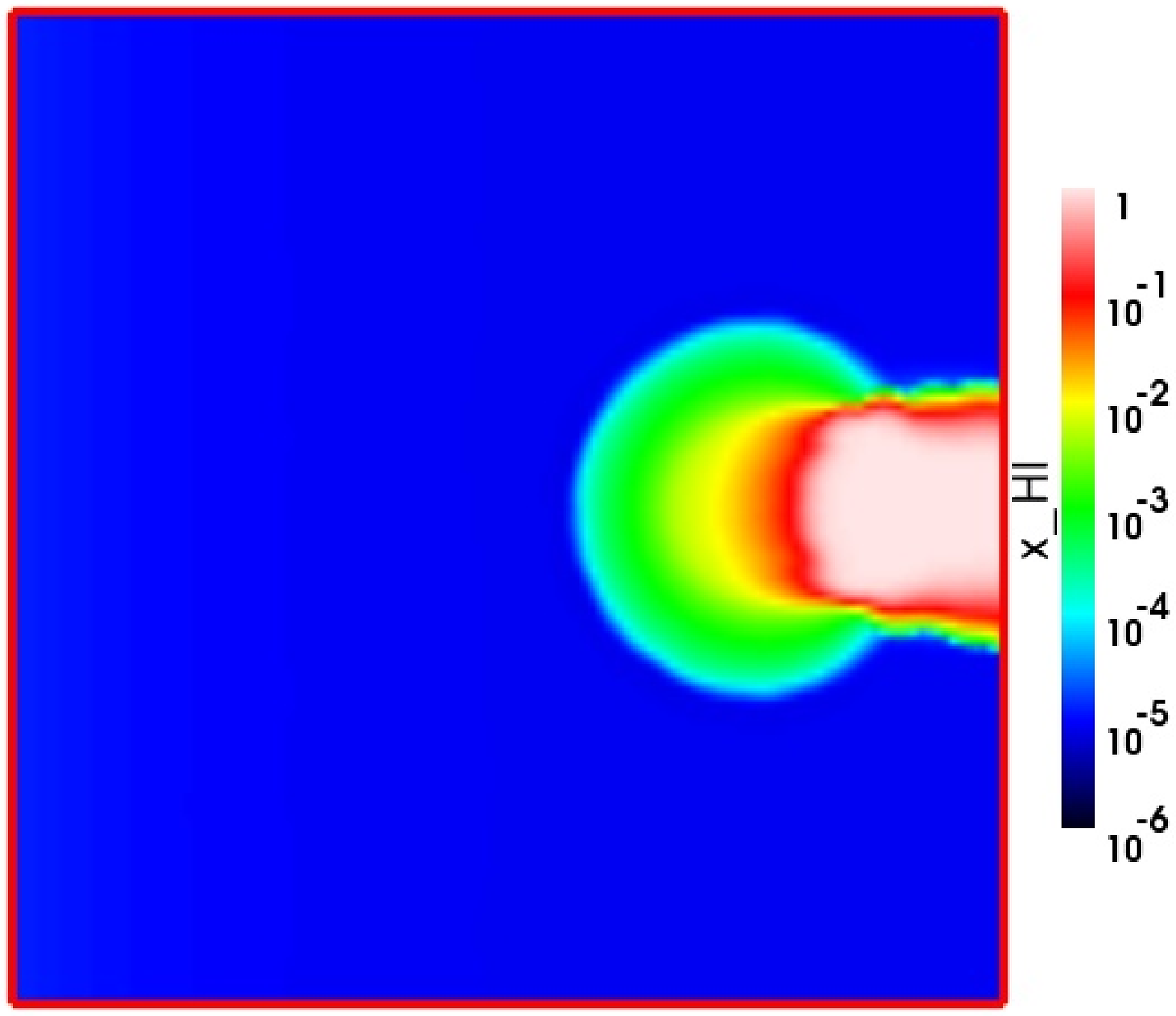}
   \includegraphics[width=2.3in]{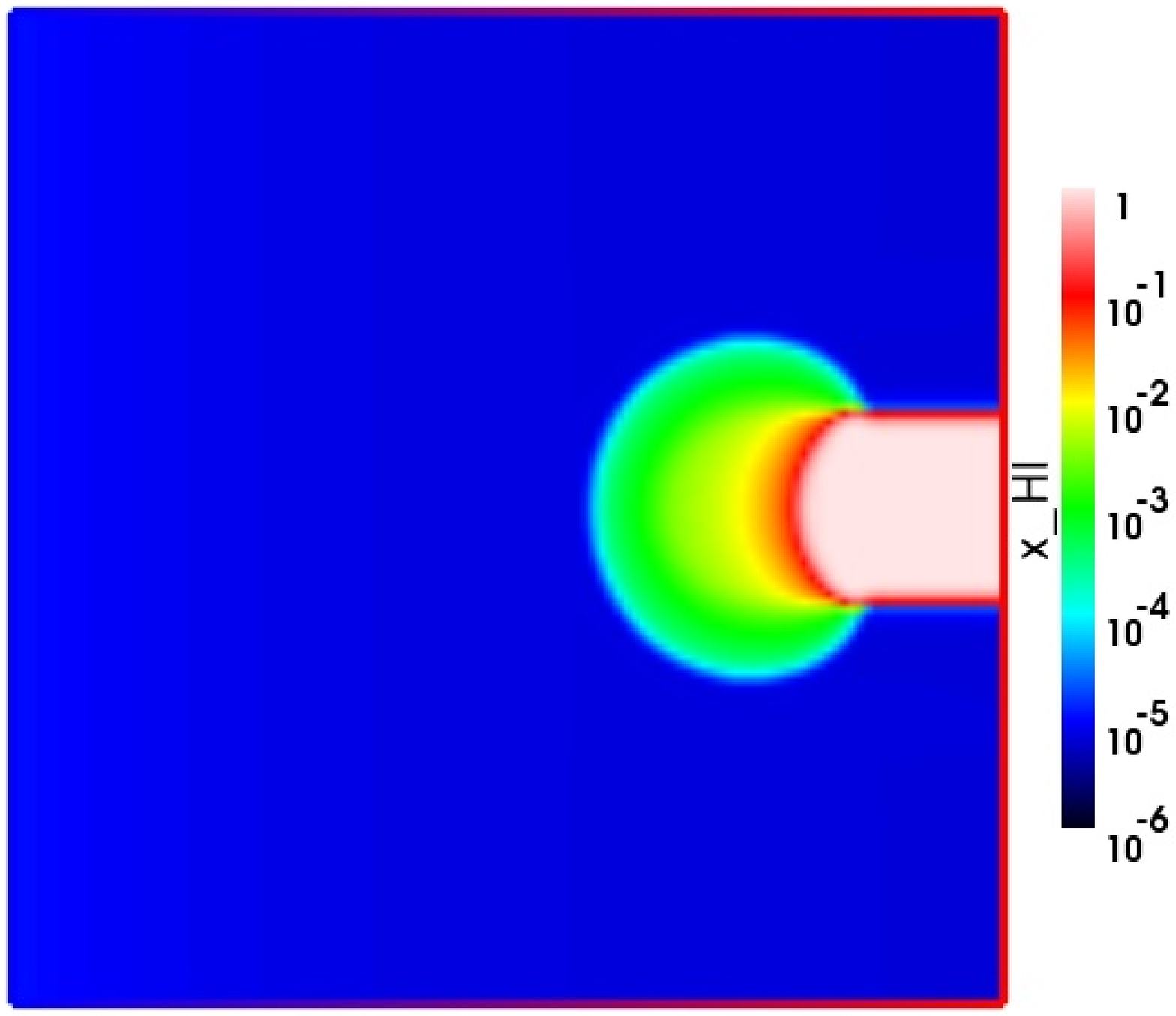}
   \includegraphics[width=2.3in]{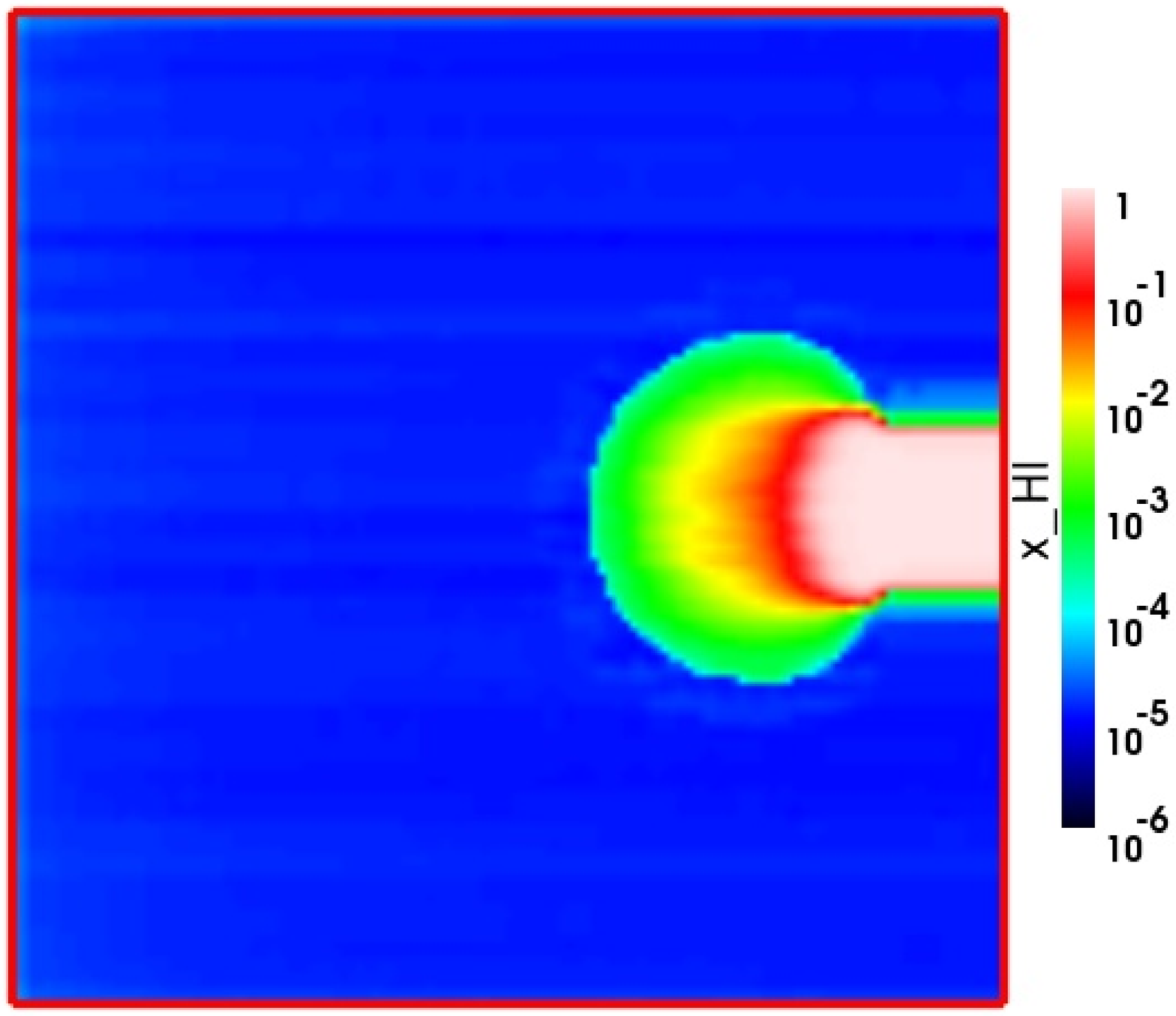}
   \includegraphics[width=2.3in]{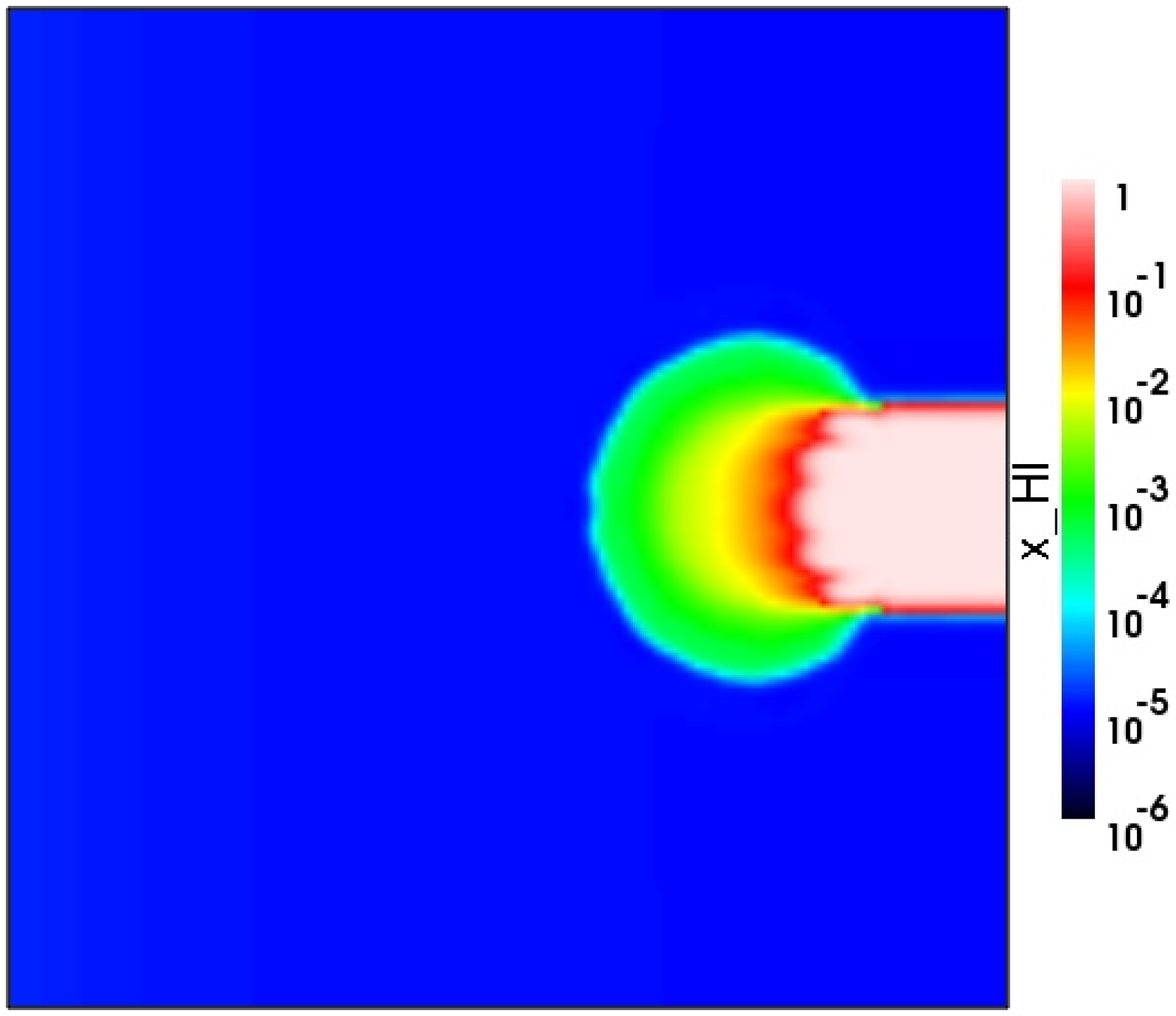}
   \includegraphics[width=2.3in]{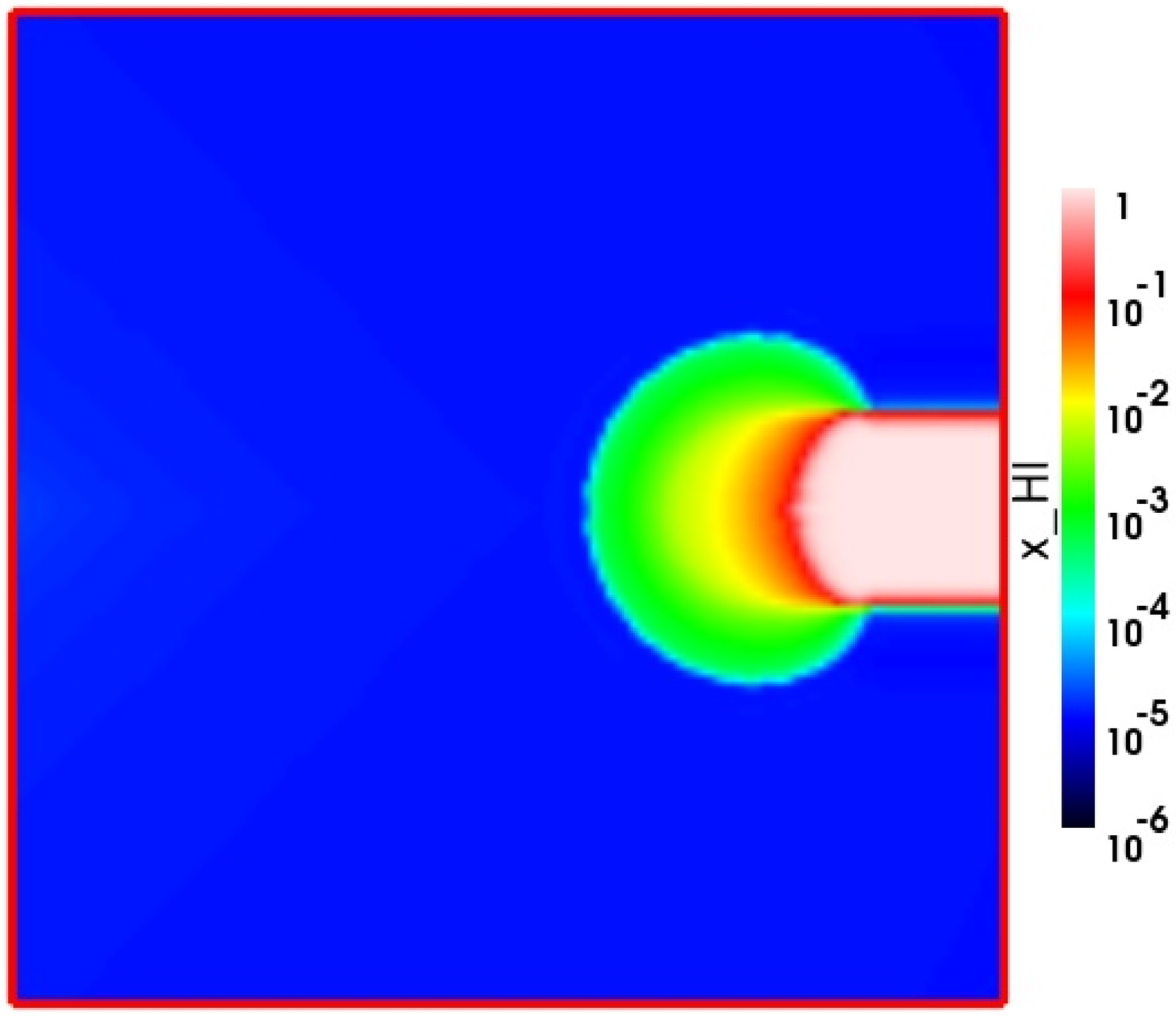}
 \caption{Test 7 (Photoevaporation of a dense clump): Images of the H~I
   fraction, cut through the simulation volume at coordinate $z=0$ at time 
   $t=10$ Myr for (left to right  and top to bottom)
   Capreole+$C^2$-Ray, RSPH, ZEUS-MP, LICORICE, Flash-HC and Coral.
 \label{T7_images3_xhi_fig}}
 \end{center}
 \end{figure*}

 \begin{figure*}
 \begin{center}
   \includegraphics[width=2.3in]{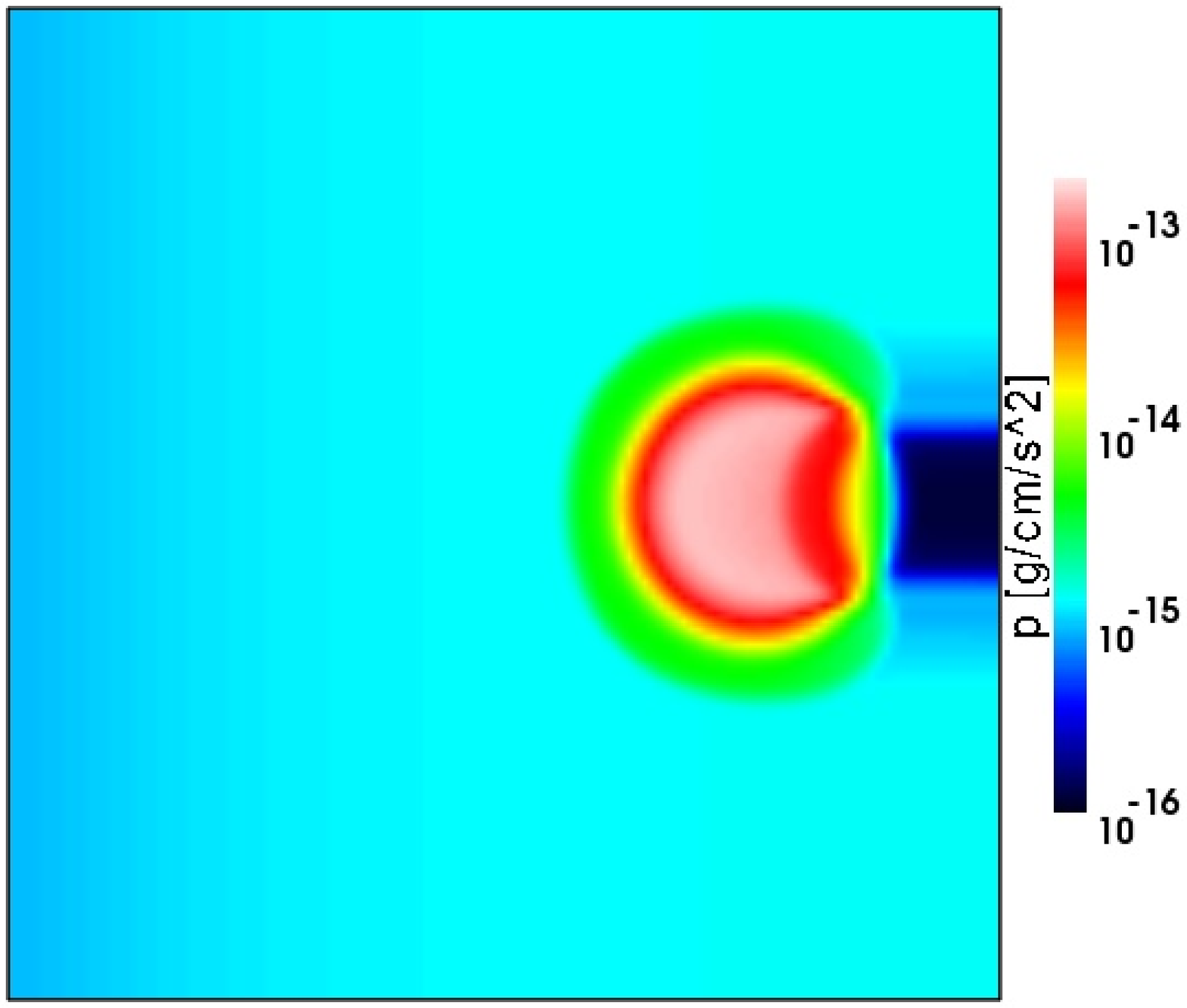}
   \includegraphics[width=2.3in]{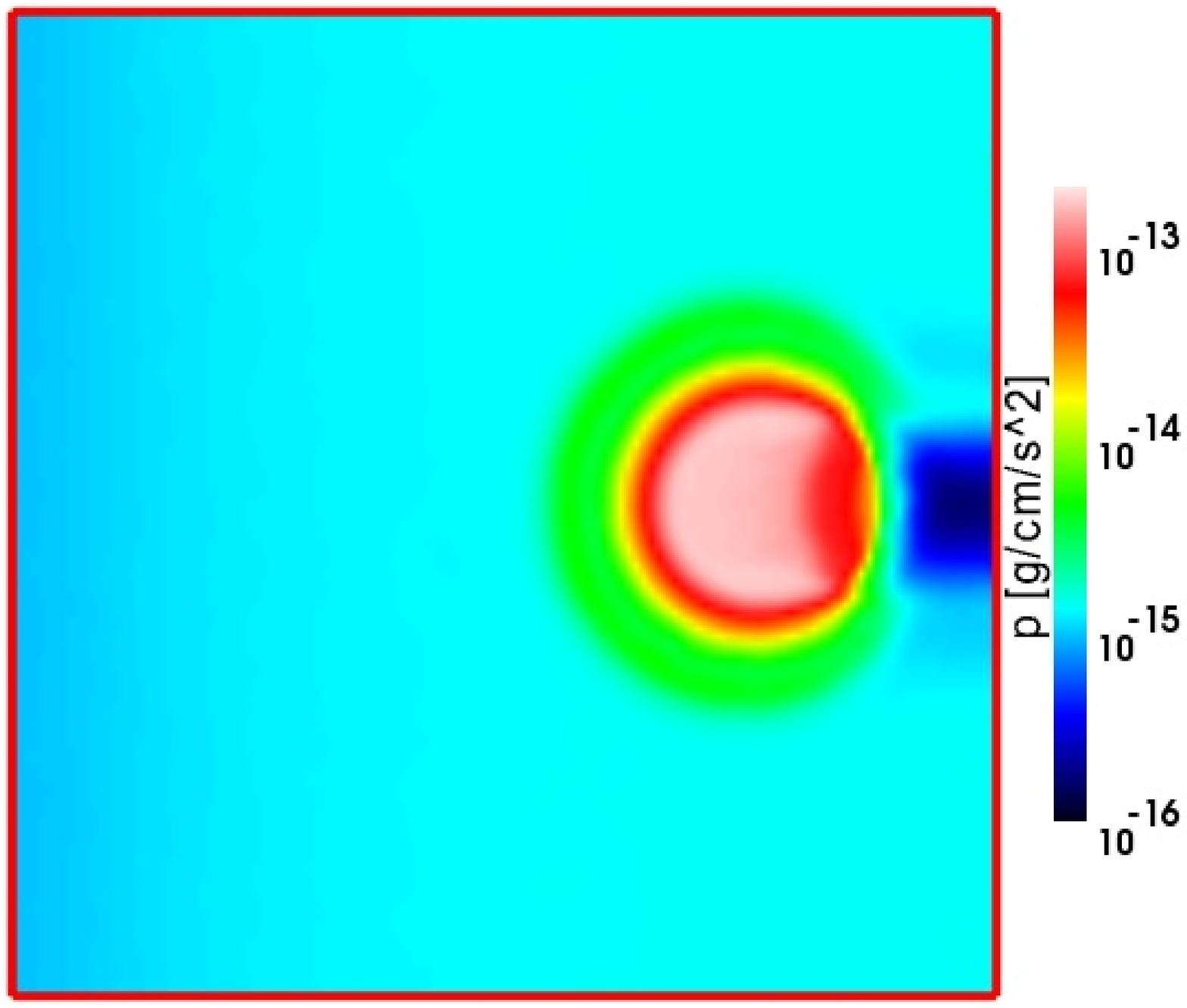}
   \includegraphics[width=2.3in]{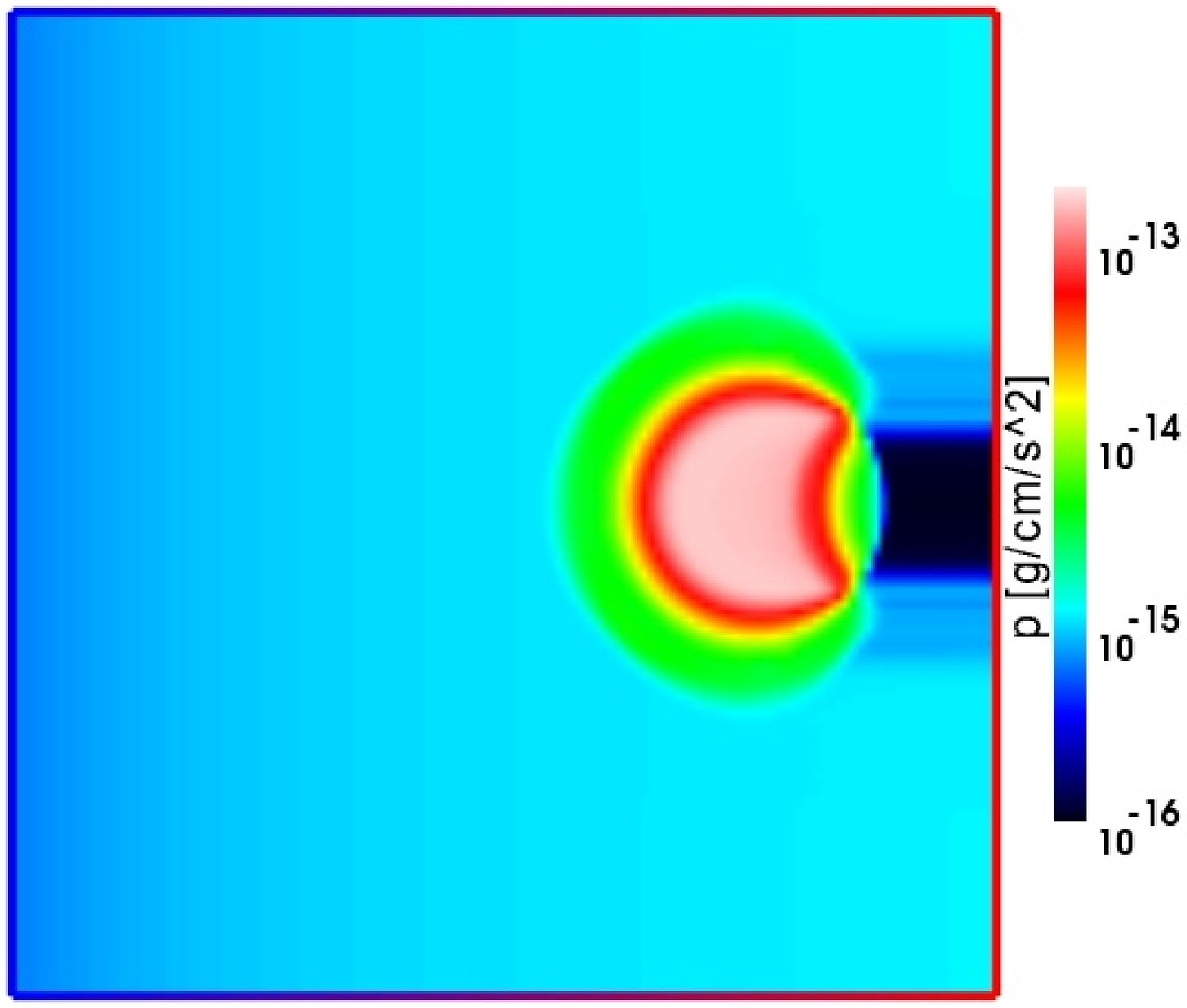}
   \includegraphics[width=2.3in]{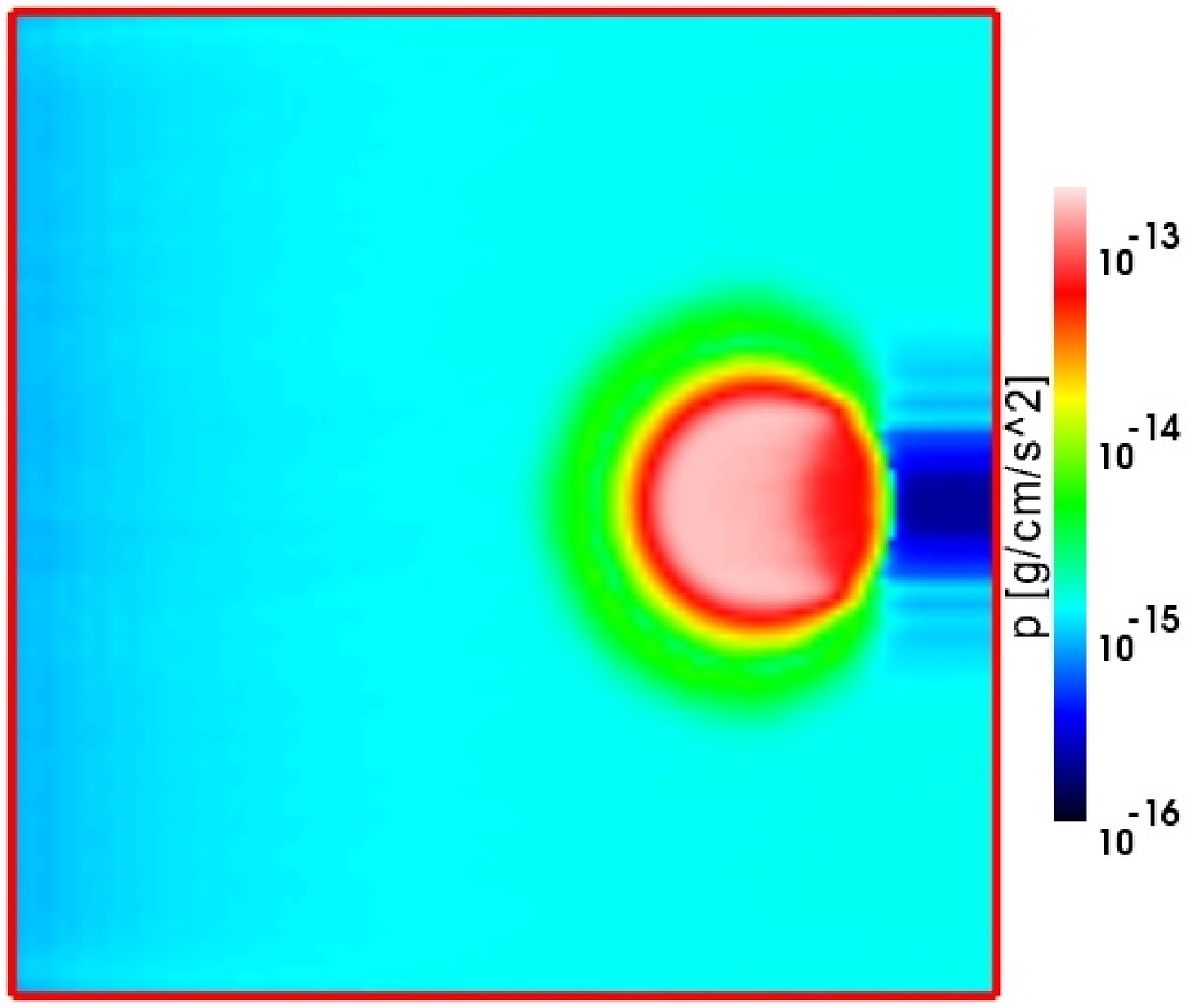}
   \includegraphics[width=2.3in]{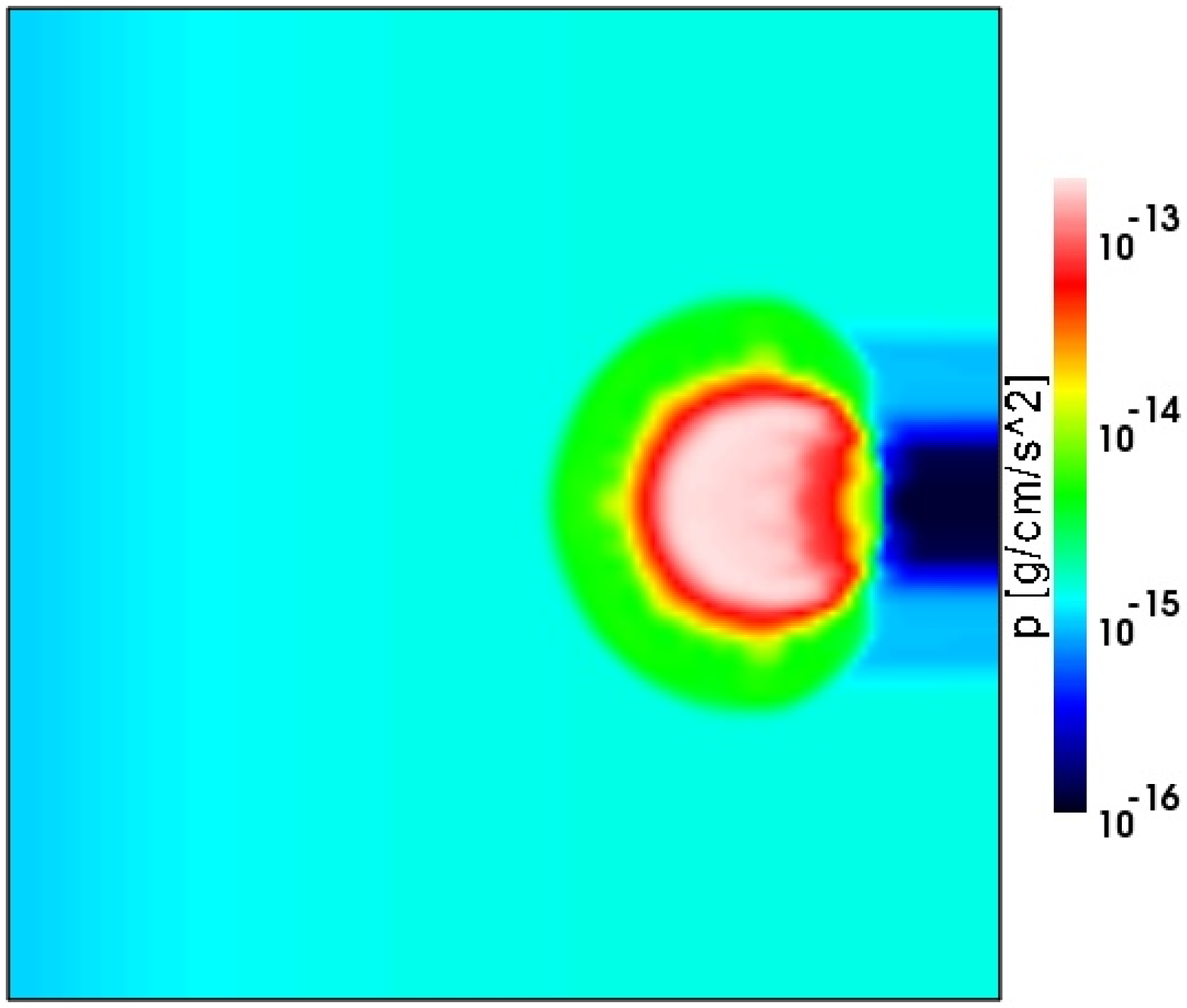}
   \includegraphics[width=2.3in]{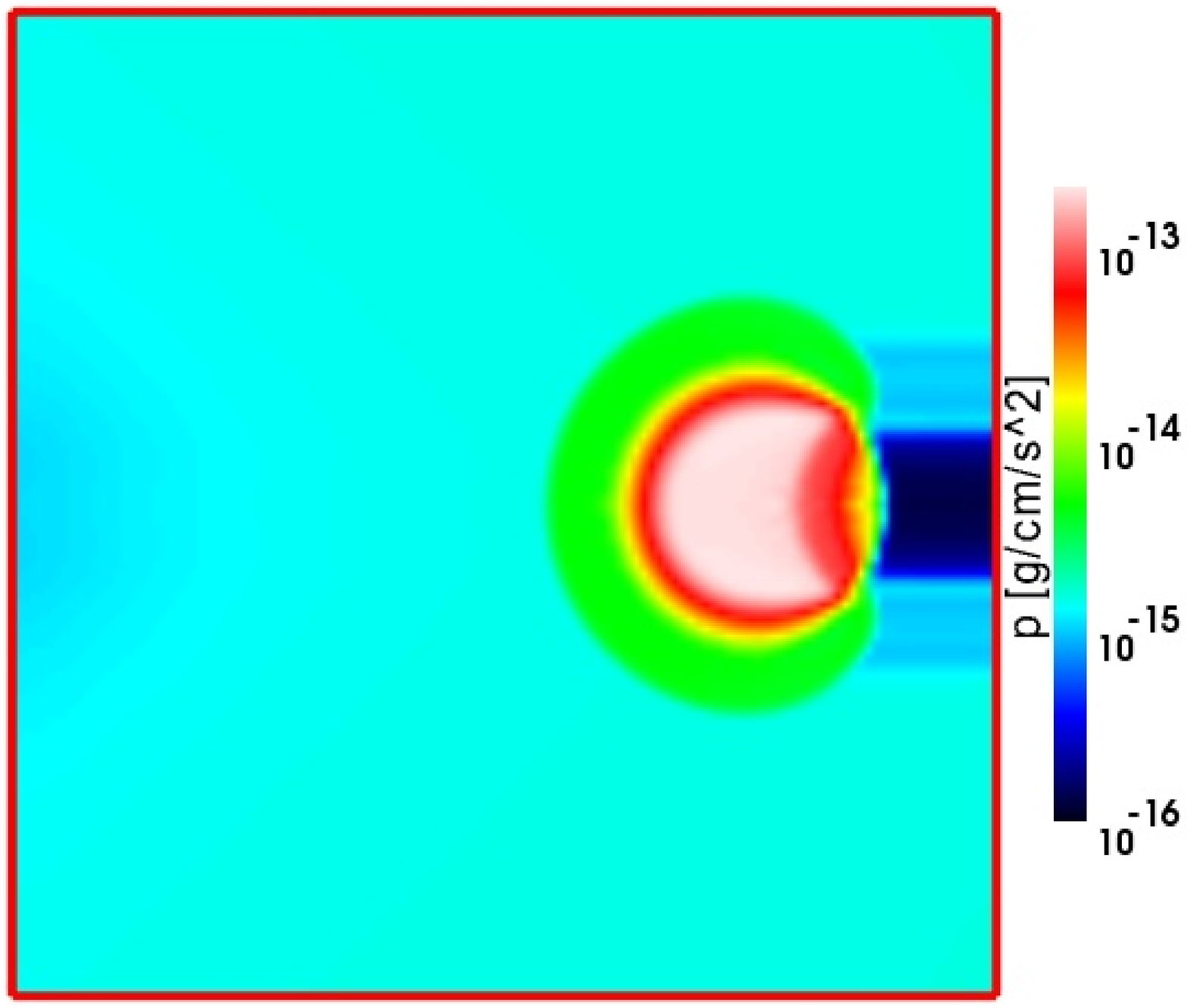}
 \caption{Test 7 (Photoevaporation of a dense clump): Images of the gas 
   pressure, cut through the simulation volume at coordinate $z=0$ at time 
   $t=10$ Myr for (left to right  and top to bottom)
   Capreole+$C^2$-Ray, RSPH, ZEUS-MP, LICORICE, Flash-HC and Coral.
 \label{T7_images3_p_fig}}
 \end{center}
 \end{figure*}

 \begin{figure*}
 \begin{center}
   \includegraphics[width=2.3in]{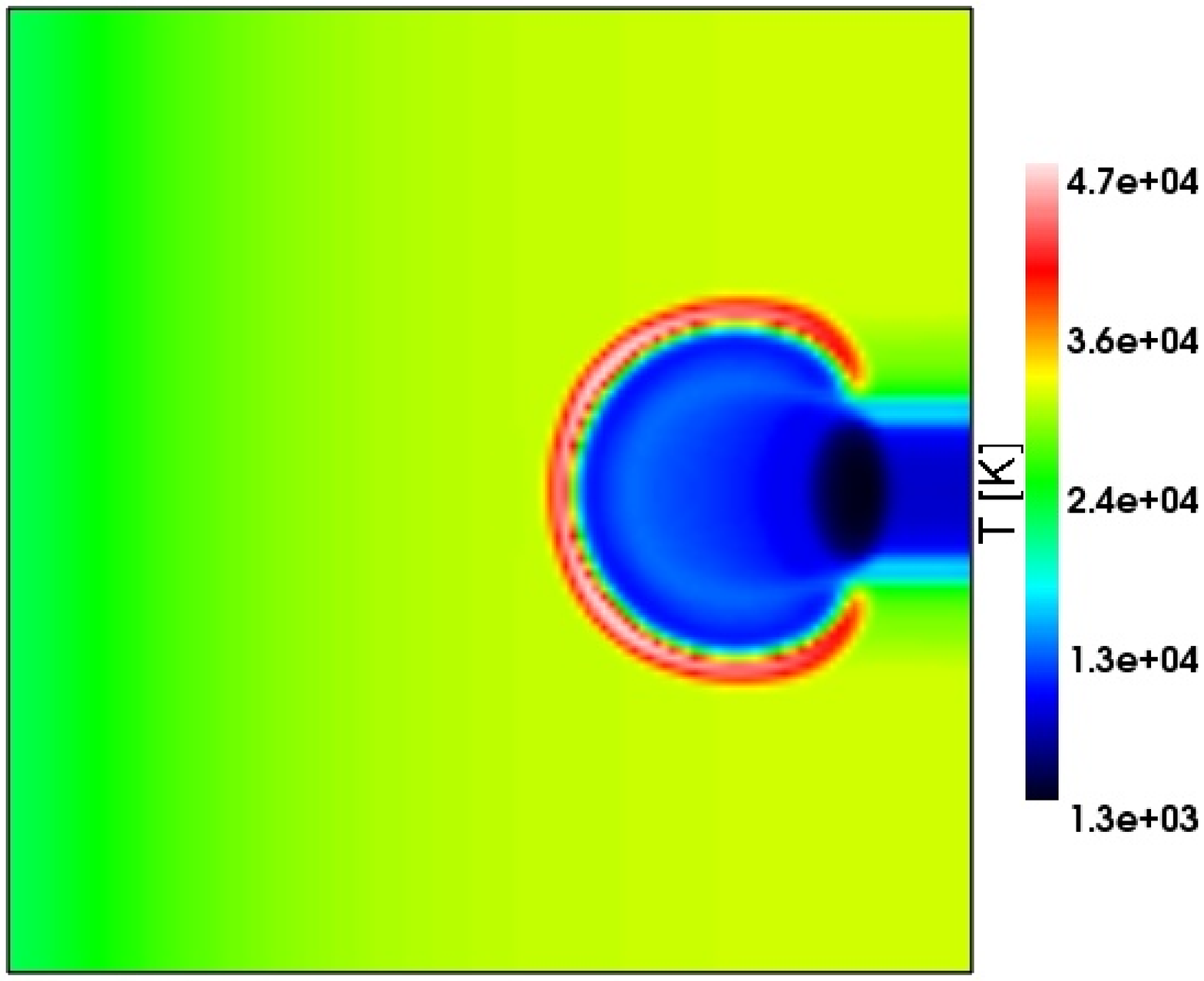}
   \includegraphics[width=2.3in]{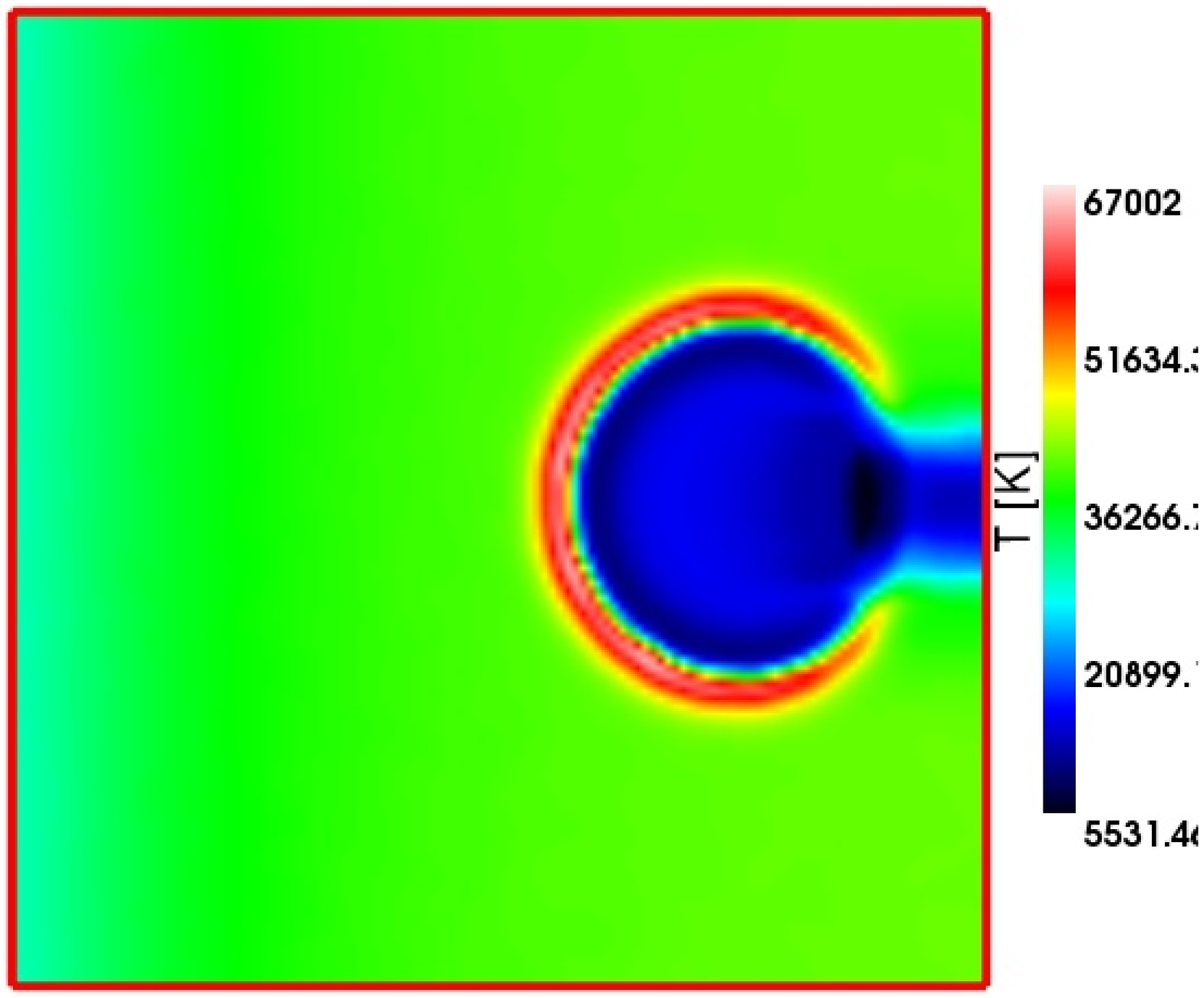}
   \includegraphics[width=2.3in]{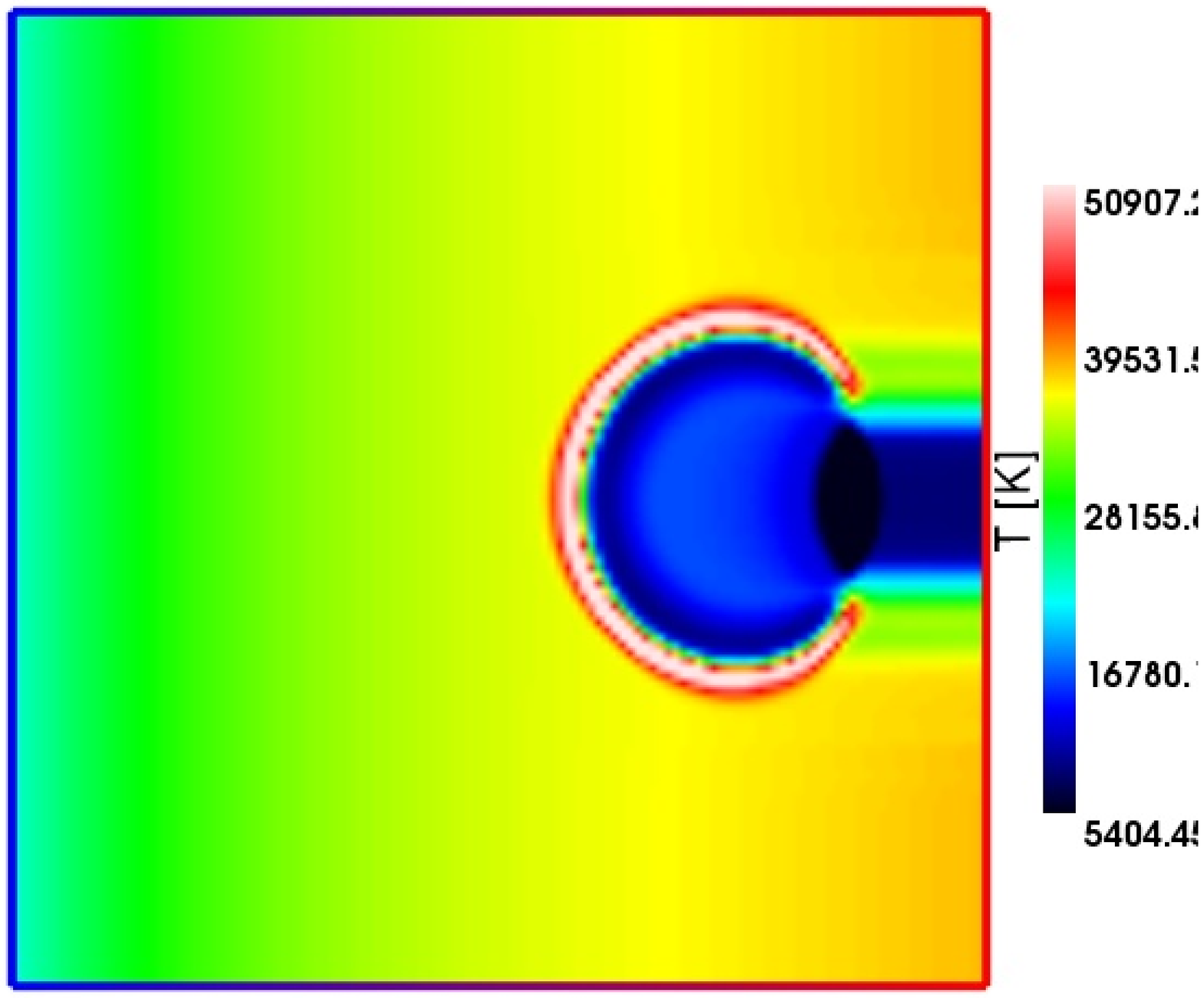}
   \includegraphics[width=2.3in]{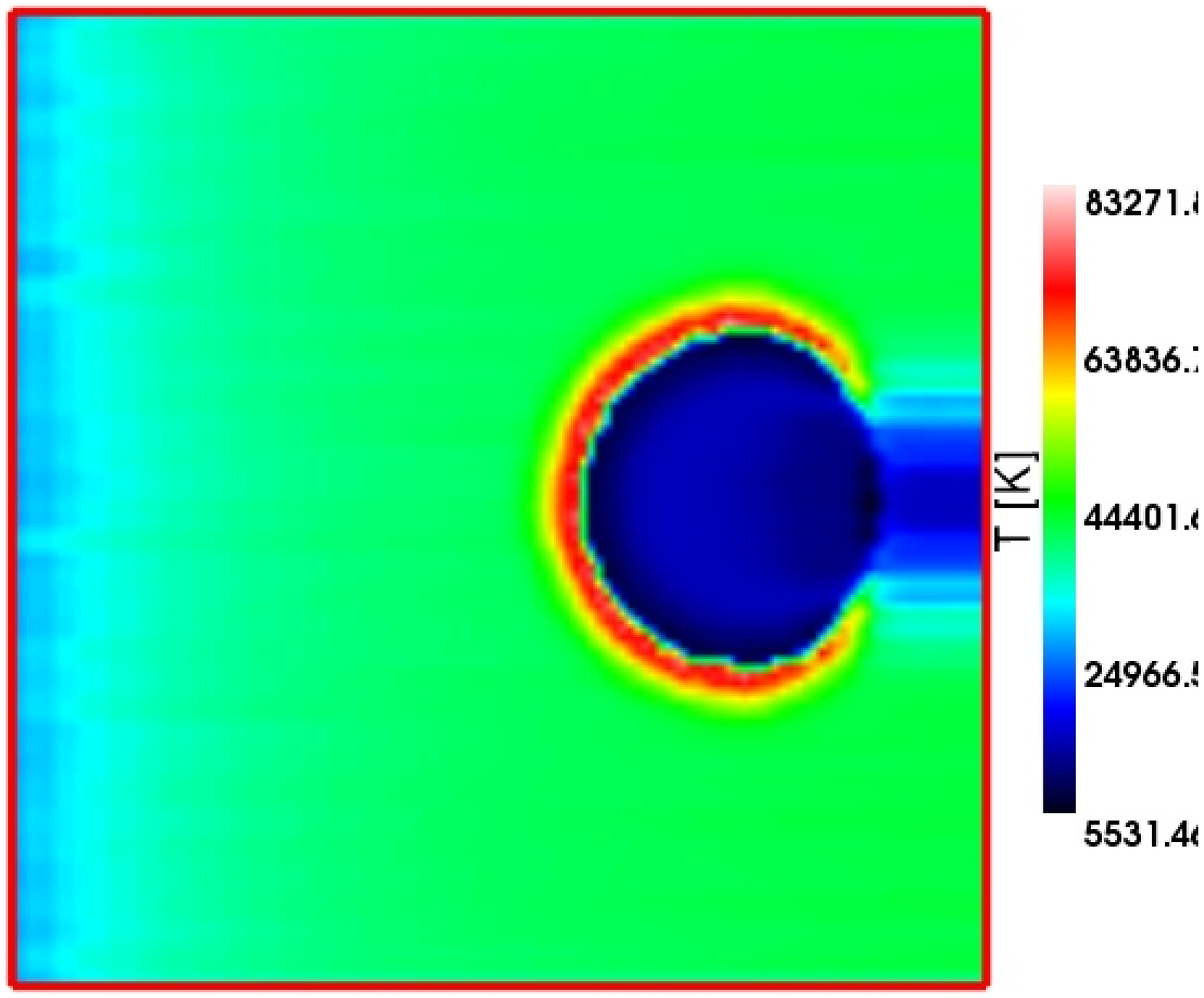}
   \includegraphics[width=2.3in]{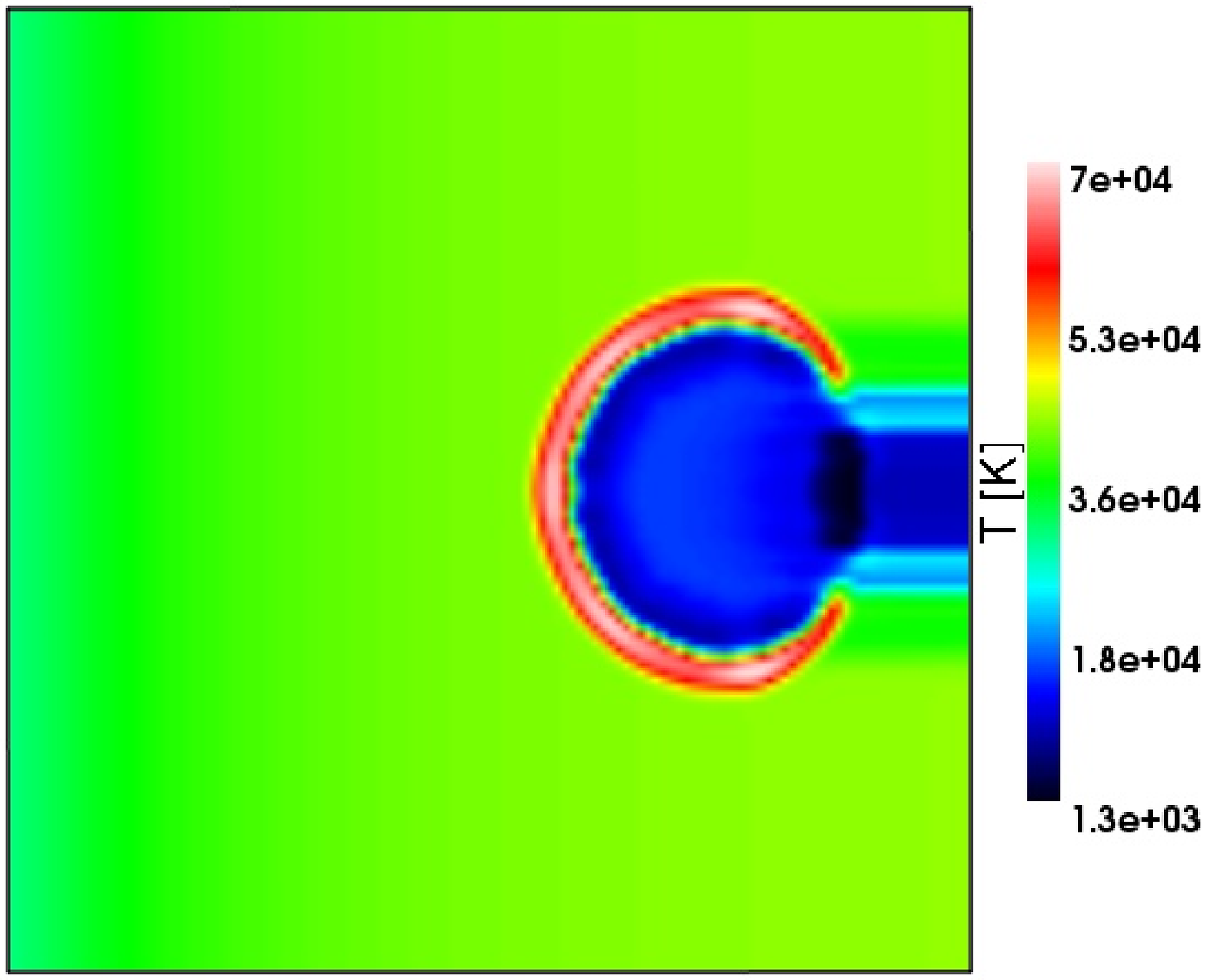}
   \includegraphics[width=2.3in]{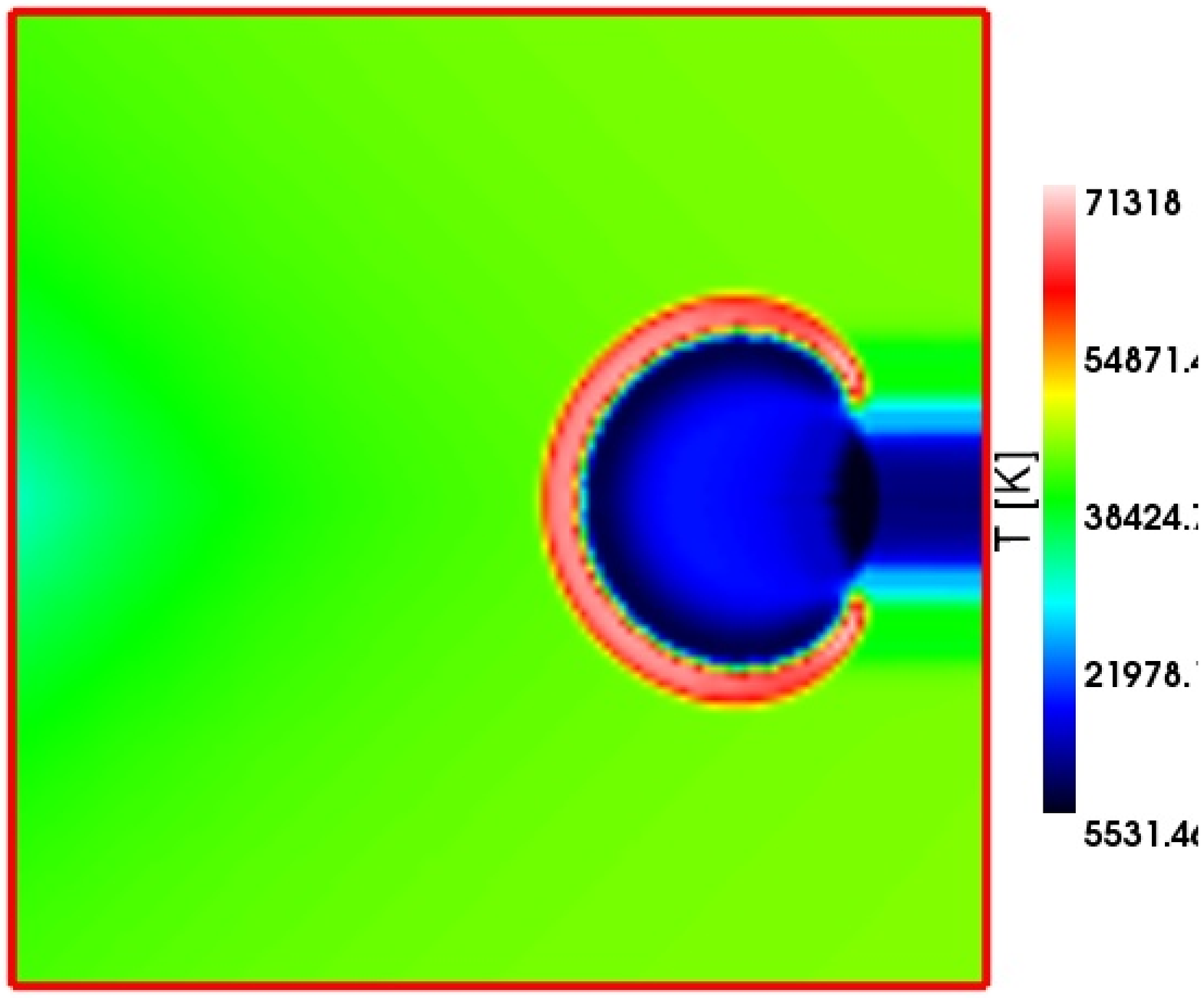}
 \caption{Test 7 (Photoevaporation of a dense clump): Images of the gas 
   temperature, cut through the simulation volume at coordinate $z=0$ at time 
   $t=10$ Myr for (left to right  and top to bottom)
   Capreole+$C^2$-Ray, RSPH, ZEUS-MP, LICORICE, Flash-HC and Coral.
 \label{T7_images3_T_fig}}
 \end{center}
 \end{figure*}
Once the front speed drops below $v_R$ a shock starts to form ahead of it, 
converting it to D-type front. The photoheated material on the source side 
starts photoevaporating, by blowing supersonic wind towards the ionizing 
source. The I-front slowly eats its way into the dense clump, as shell after 
shell of gas boils off  and joins the wind. The I-front velocity gradually 
drops to a few km/s and its position remains roughly constant. Some differences 
among the derived I-front evolution in position and velocity are observed, 
but they remain small throughout the evolution, never exceeding 10\% in terms 
of position. In Figure~\ref{T7_images3_xhi_fig} we show images of the neutral 
hydrogen fraction at $t=10$~Myr. Overall results agree fairly well, with the 
expanding wind and shadow at very similar stages of evolution. There are also 
a few, relatively minor, differences which should be noted. The RSPH result 
remains somewhat more diffuse and asymmetric then the rest, as noted above, 
but as the evolution proceeds the differences are somewhat less notable. 
However, the photoevaporation does proceed somewhat more rapidly in this case 
due to the inevitably more diffuse initial conditions. There is some leaking of 
light at the edges of the LICORICE shadow which is not seen in the other results 
and should therefore be related to the radiative transfer method employed, rather 
than to other factors, e.g. to the limb column density of the clump being small 
and allowing some light to go through. Finally, there are some uneven features 
at the edge of the shadow on the source side in the case of Flash-HC, whose origin 
is currently unclear.    

 \begin{figure*}
 \begin{center}
   \includegraphics[width=2.3in]{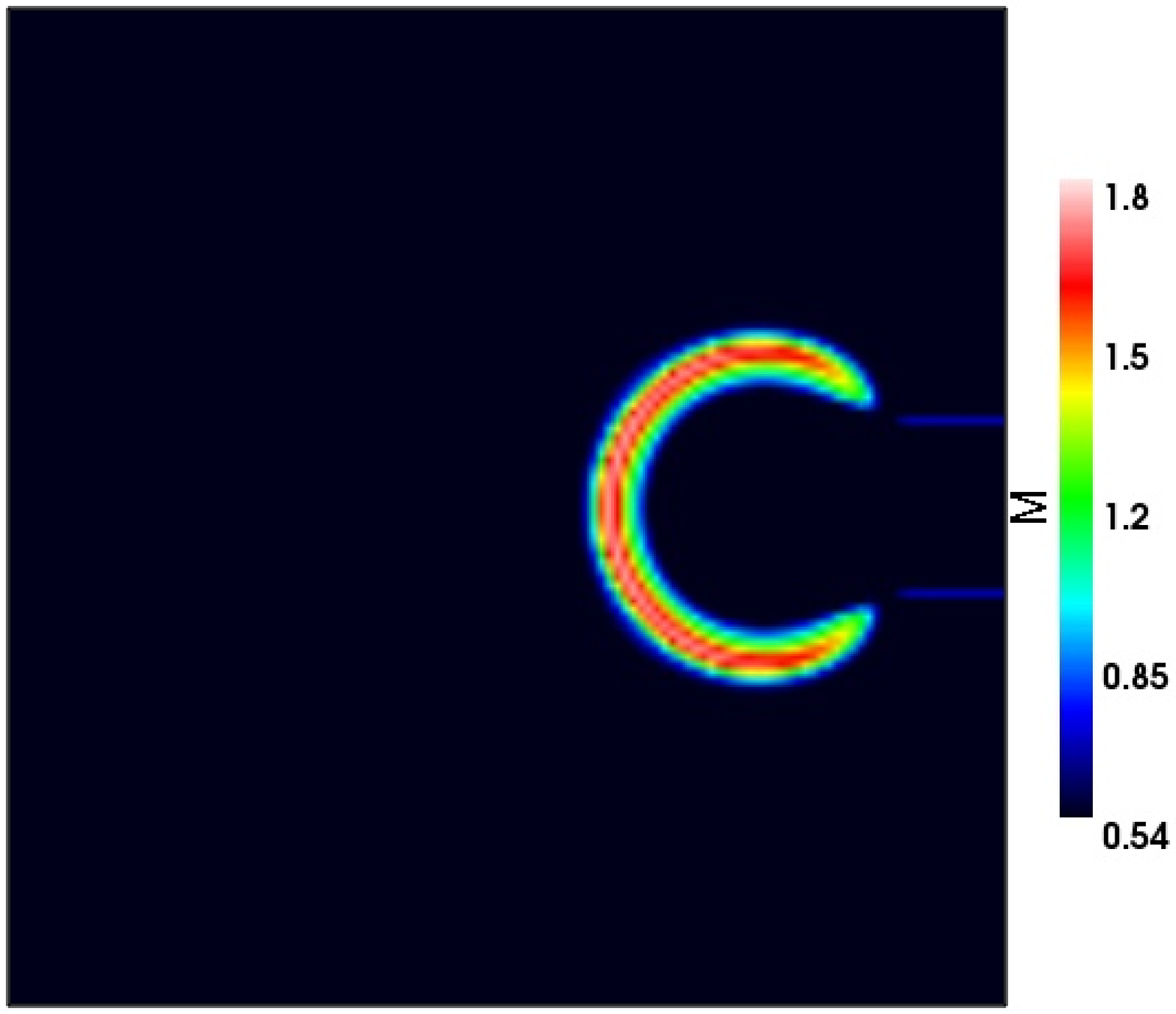}
   \includegraphics[width=2.3in]{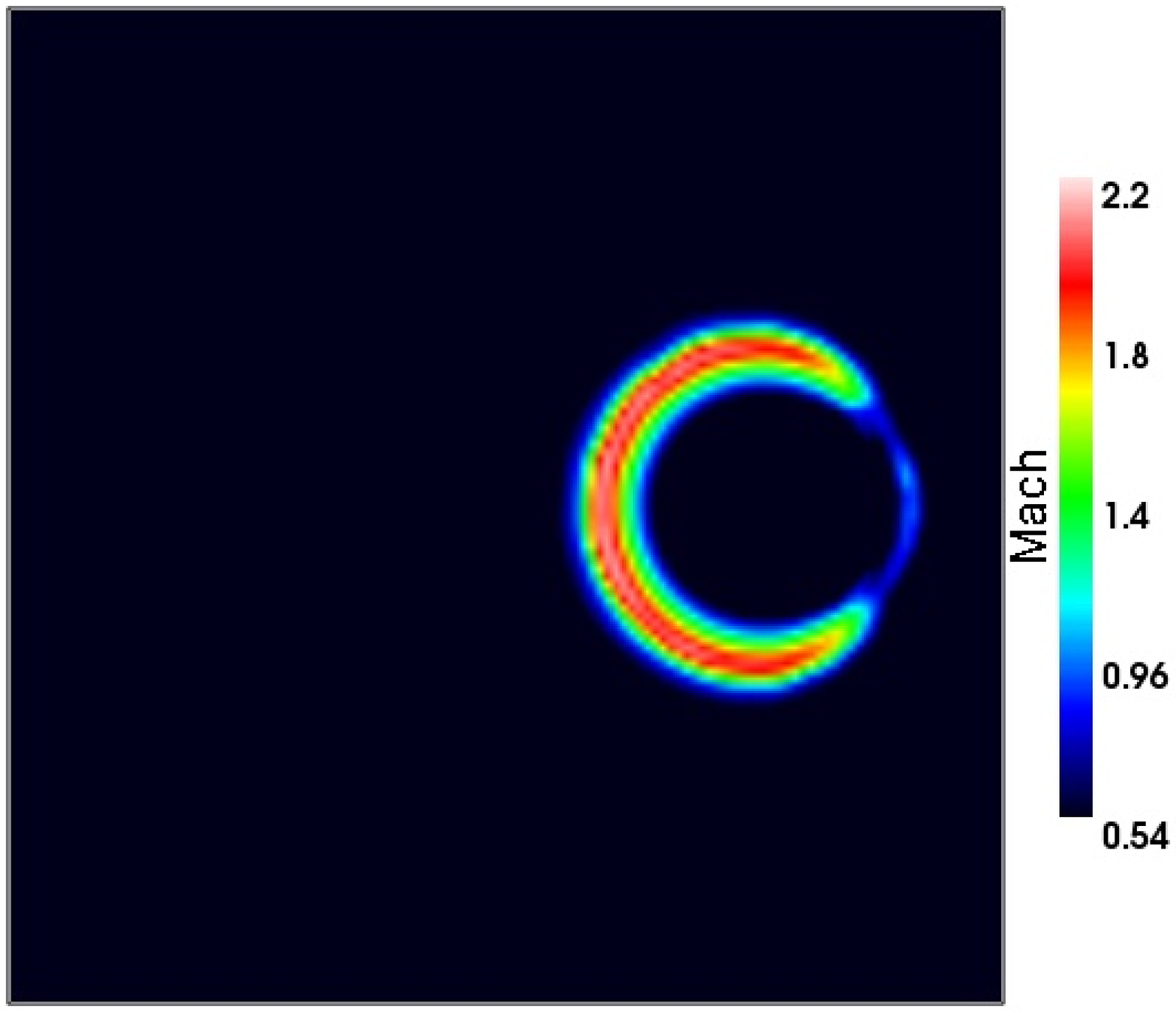}
   \includegraphics[width=2.3in]{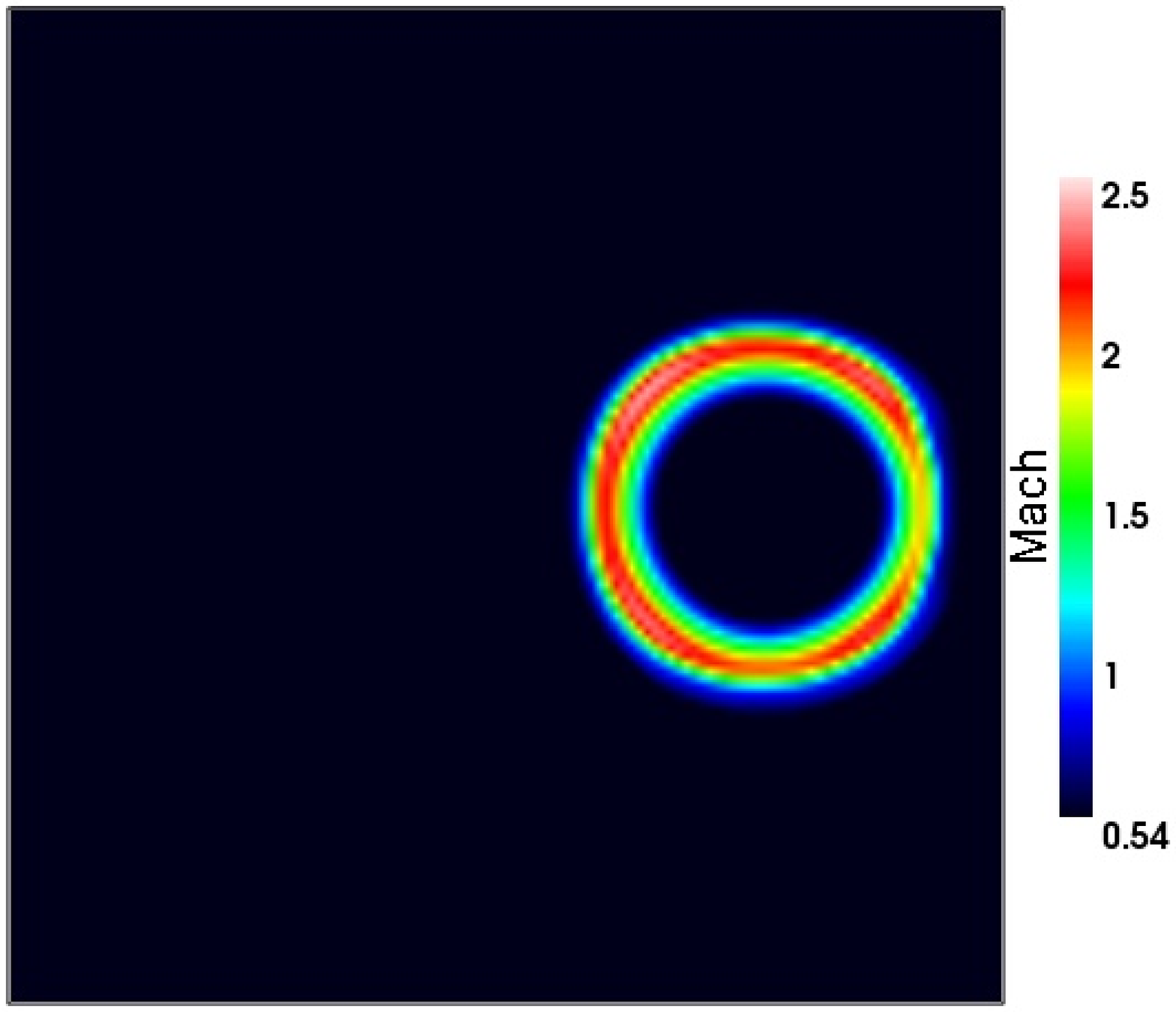}
   \includegraphics[width=2.3in]{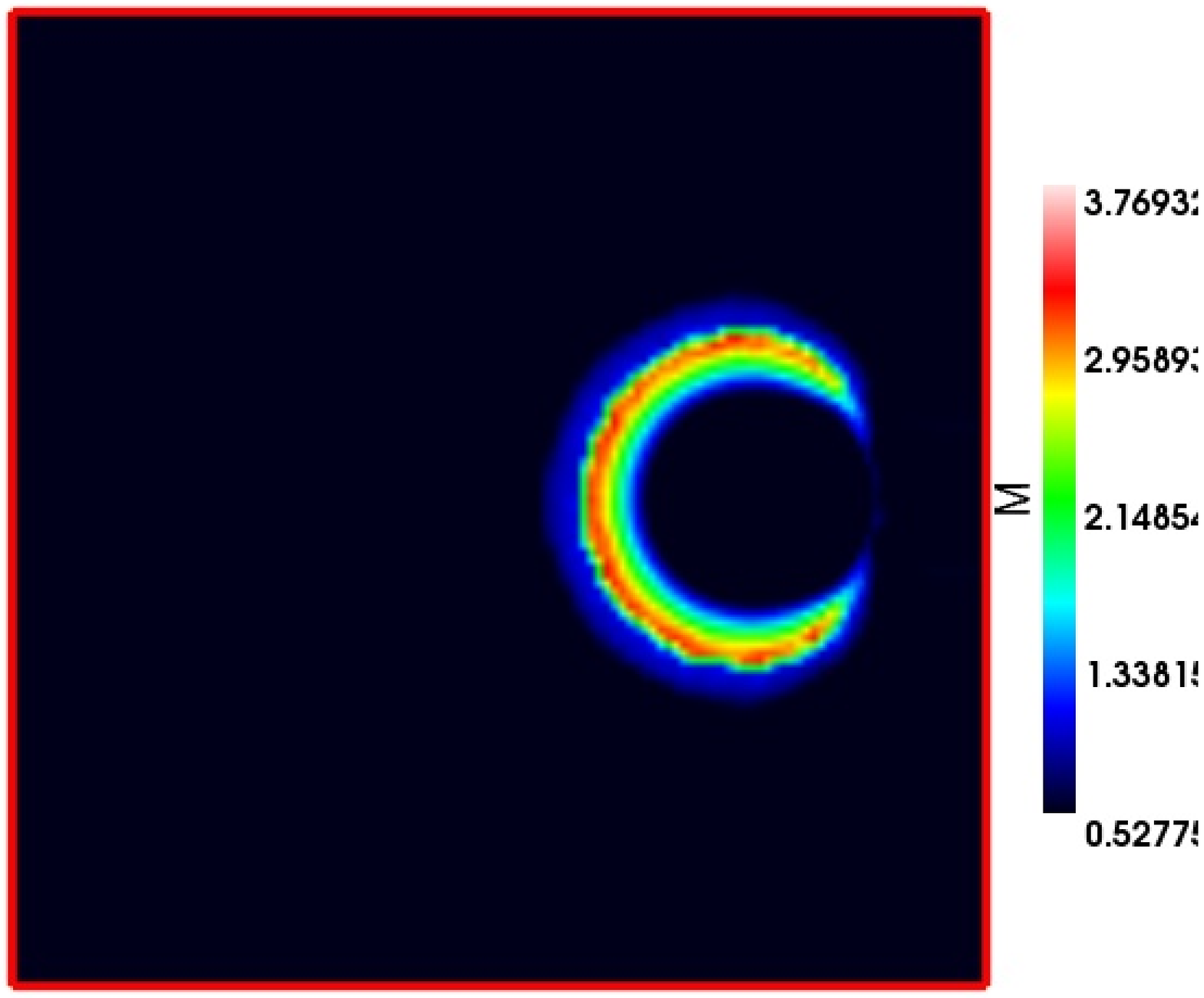}
   \includegraphics[width=2.3in]{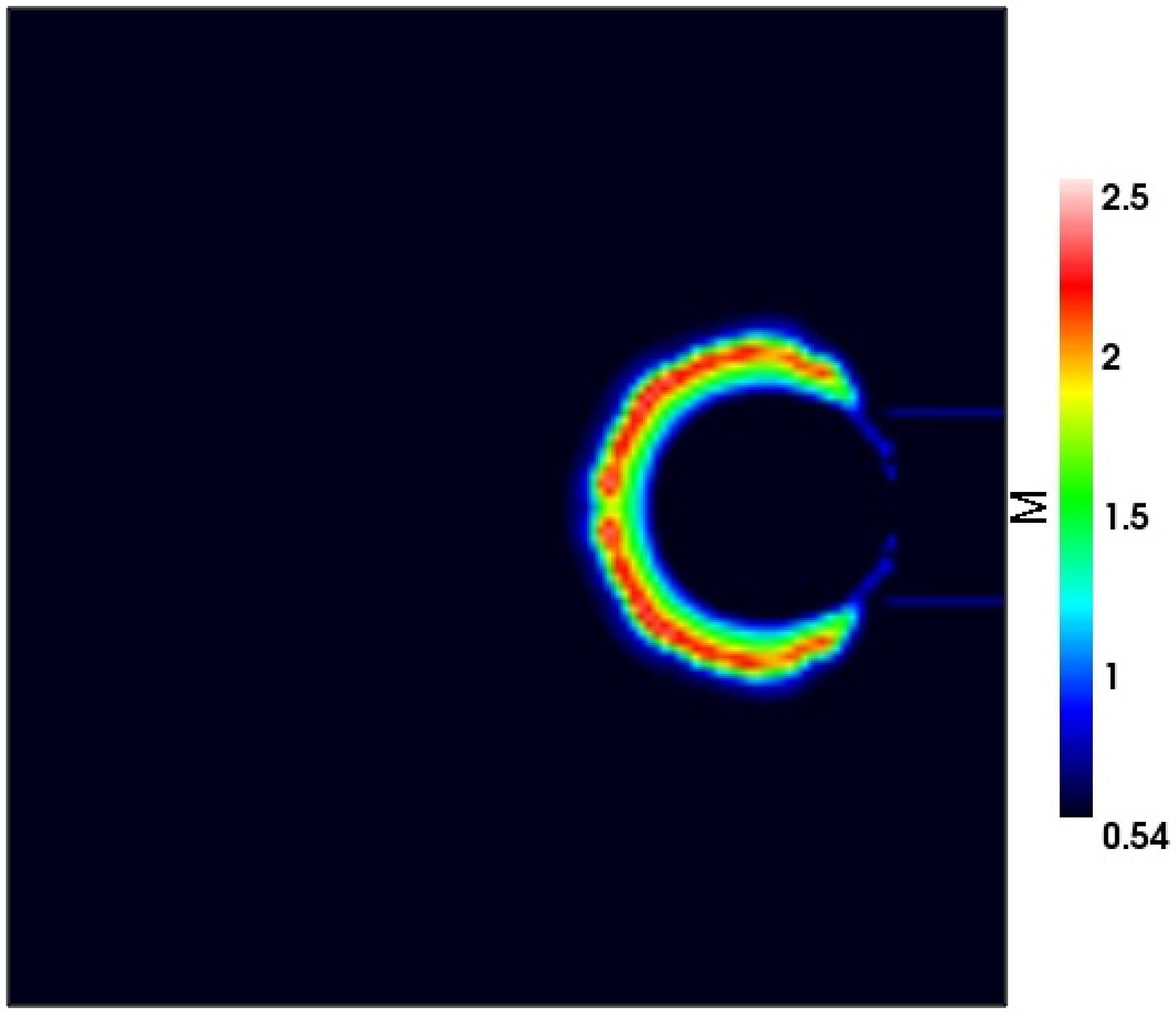}
   \includegraphics[width=2.3in]{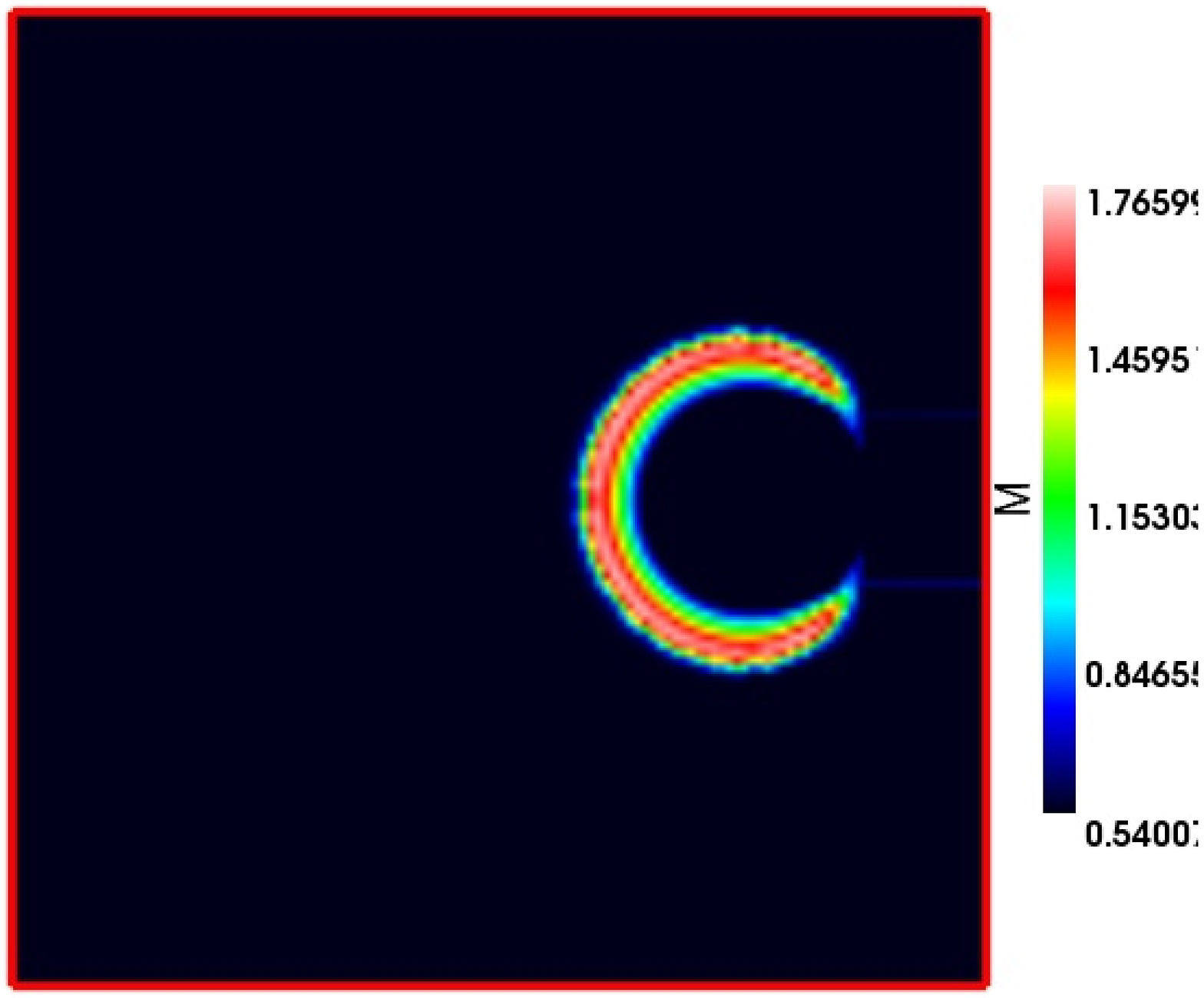}
 \caption{Test 7 (Photoevaporation of a dense clump): Images of the flow 
   Mach number, cut through the simulation volume at coordinate $z=0$ at 
   time $t=10$ Myr for (left to right  and top to bottom)
   Capreole+$C^2$-Ray, RSPH, ZEUS-MP, LICORICE, Flash-HC and Coral.
 \label{T7_images3_M_fig}}
 \end{center}
 \end{figure*}

The pressure images at $t= 10$~Myr in Figure~\ref{T7_images3_p_fig} 
essentially agree, with only minor 
morphological differences between the results. The shadow is somewhat thicker and
less squeezed at the edges for $C^2$-Ray, ZEUS-MP and Flash-HC, compared to RSPH and
LICORICE, with Coral results intermediate between the two groups. The reason for
this difference becomes apparent from the corresponding temperature images 
(Figure~\ref{T7_images3_T_fig}). In the cases of $C^2$-Ray, ZEUS-MP and Flash-HC there 
is clear temperature gradient from the edges of the shadow going inward, while
for RSPH and LICORICE this temperature gradient is largely absent. There are also 
noticeable temperature variations within the clump for $C^2$-Ray, ZEUS-MP and Flash-HC
which are less pronounced for RSPH and LICORICE. The reason for these differences
appears to be the different levels of penetration by hard photons through the
high column density material in the clump, and the corresponding varying levels
of energy deposit by those photons. The variations in evolution this introduces seem
minor in our particular test problem, but such differences might matter more in 
problems in which the precise level of the number of free electrons and the local 
temperature within dense clumps is of importance. One example may be the study of 
the the production of molecular hydrogen within dense regions irradiated by UV 
radiation, which can regulate (stimulate or suppress) local star formation 
\citep{2006MNRAS.368.1885I,wet08b}. Finally, the Mach number images at $t=10$~Myr 
are shown in Figure~\ref{T7_images3_M_fig}. The supersonic wind which starts to blow
towards the ionizing source is clearly visible, with only small differences in 
terms of the thickness of this layer and the Mach number values between the 
different runs. The only peculiarity visible here is that in the case of ZEUS-MP 
this supersonic layer is almost spherical, surrounding the clump from all sides, 
which is not seen in any of the other results. This appears to be a consequence of 
the very cold ($T<1000$~K) region remaining at the back of the clump, which is not 
present in any of the other cases (see also Figure~\ref{T7_profsT_fig}). The reason 
for this region remaining so cold in the ZEUS-MP simulation is unclear at present, 
considering that (as we discussed above) the spectrum hardening and penetration of 
hard photons through the clump and into the shadow are similar to $C^2$-Ray, 
Flash-HC and Coral and stronger than RSPH and LICORICE.

 \begin{figure*}
 \begin{center}
   \includegraphics[width=2.3in]{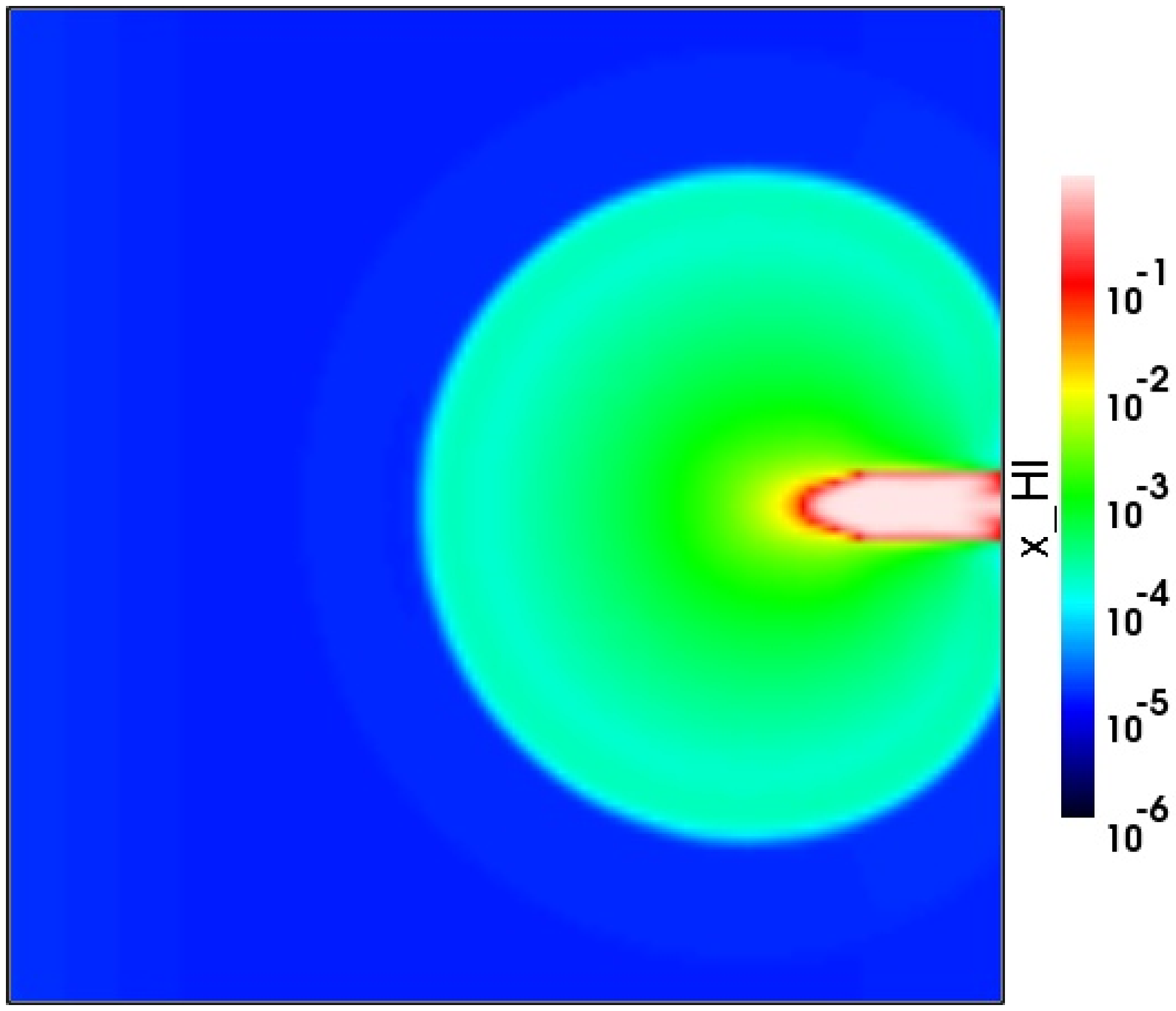}
   \includegraphics[width=2.3in]{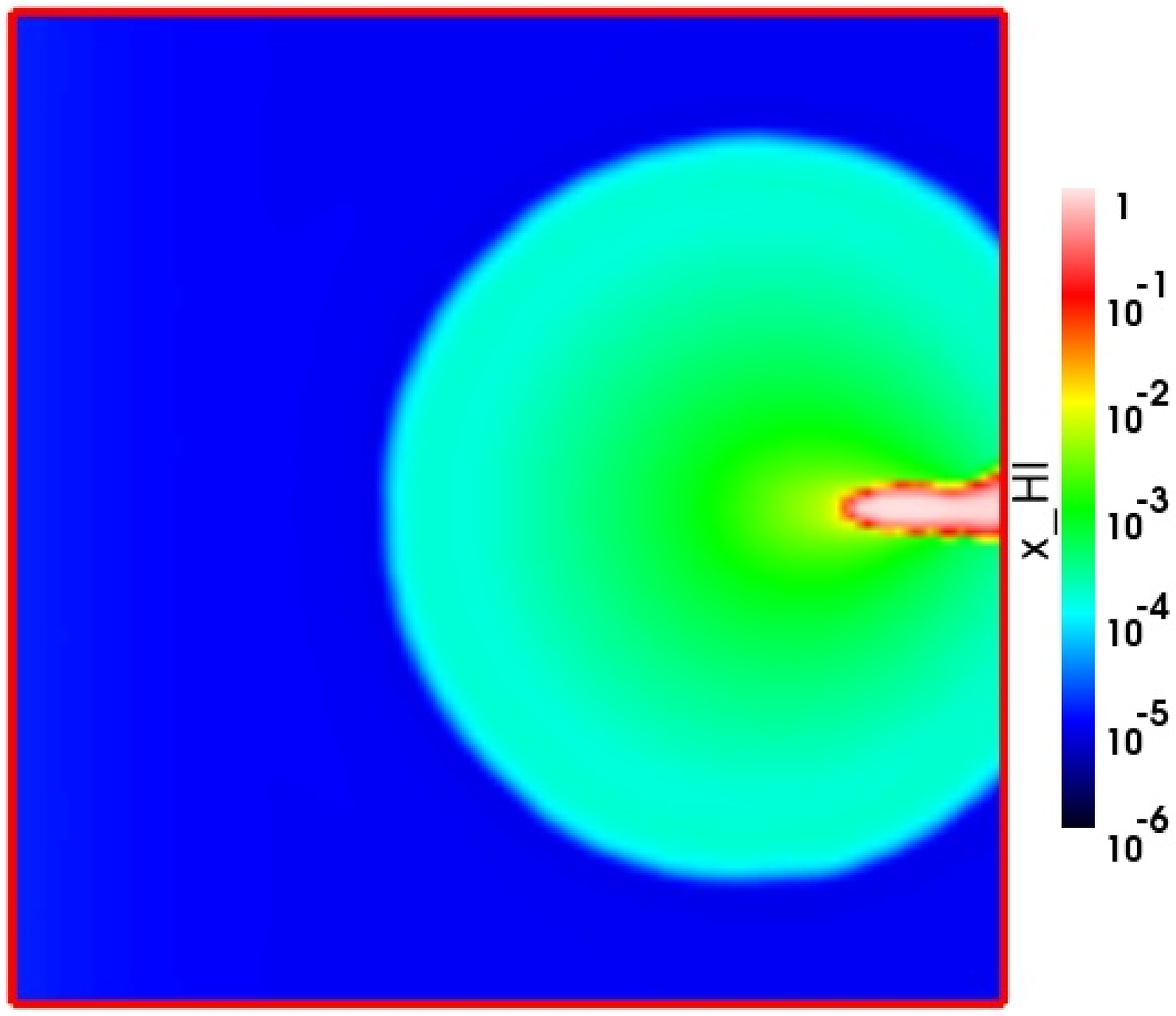}
   \includegraphics[width=2.3in]{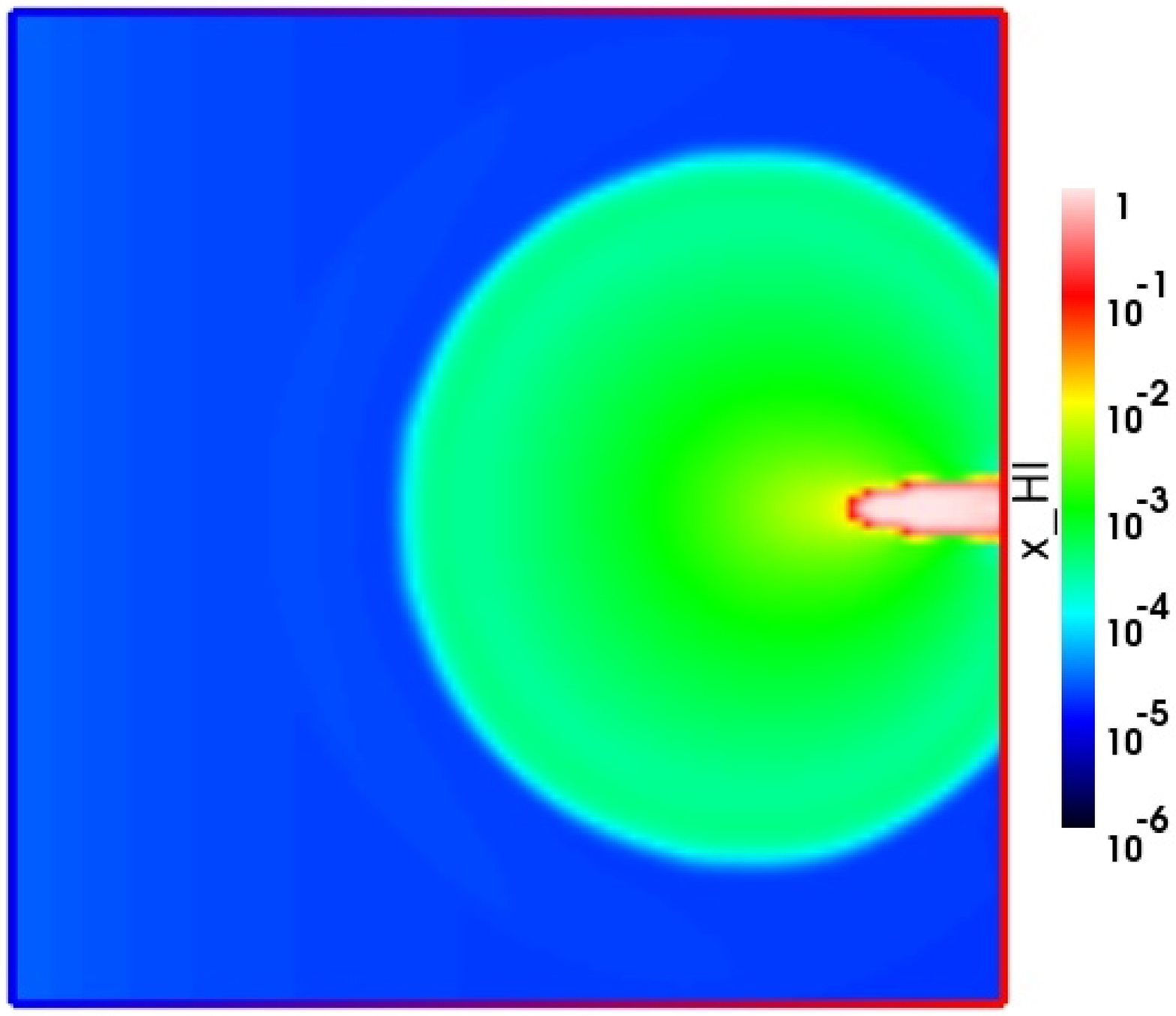}
   \includegraphics[width=2.3in]{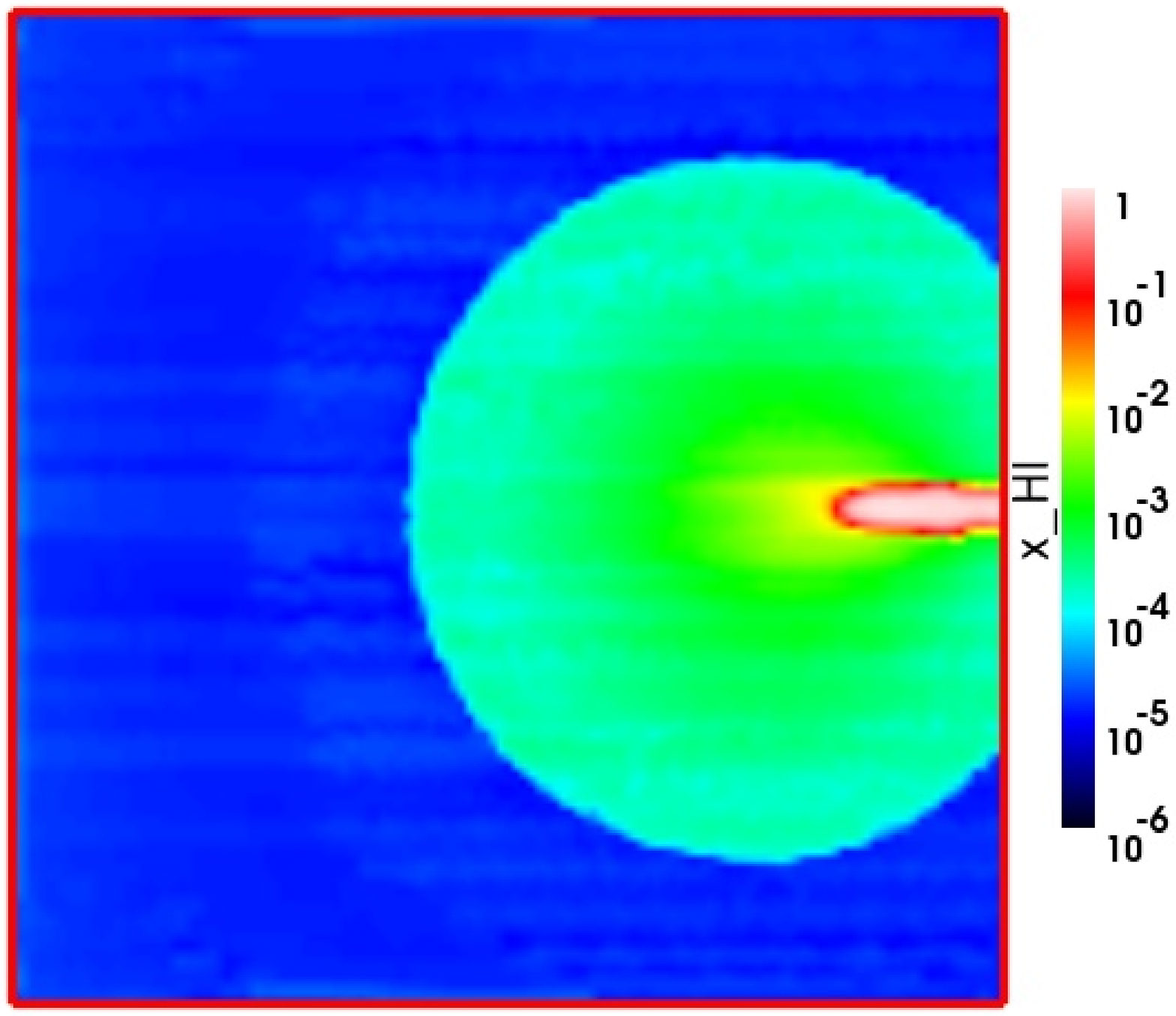}
   \includegraphics[width=2.3in]{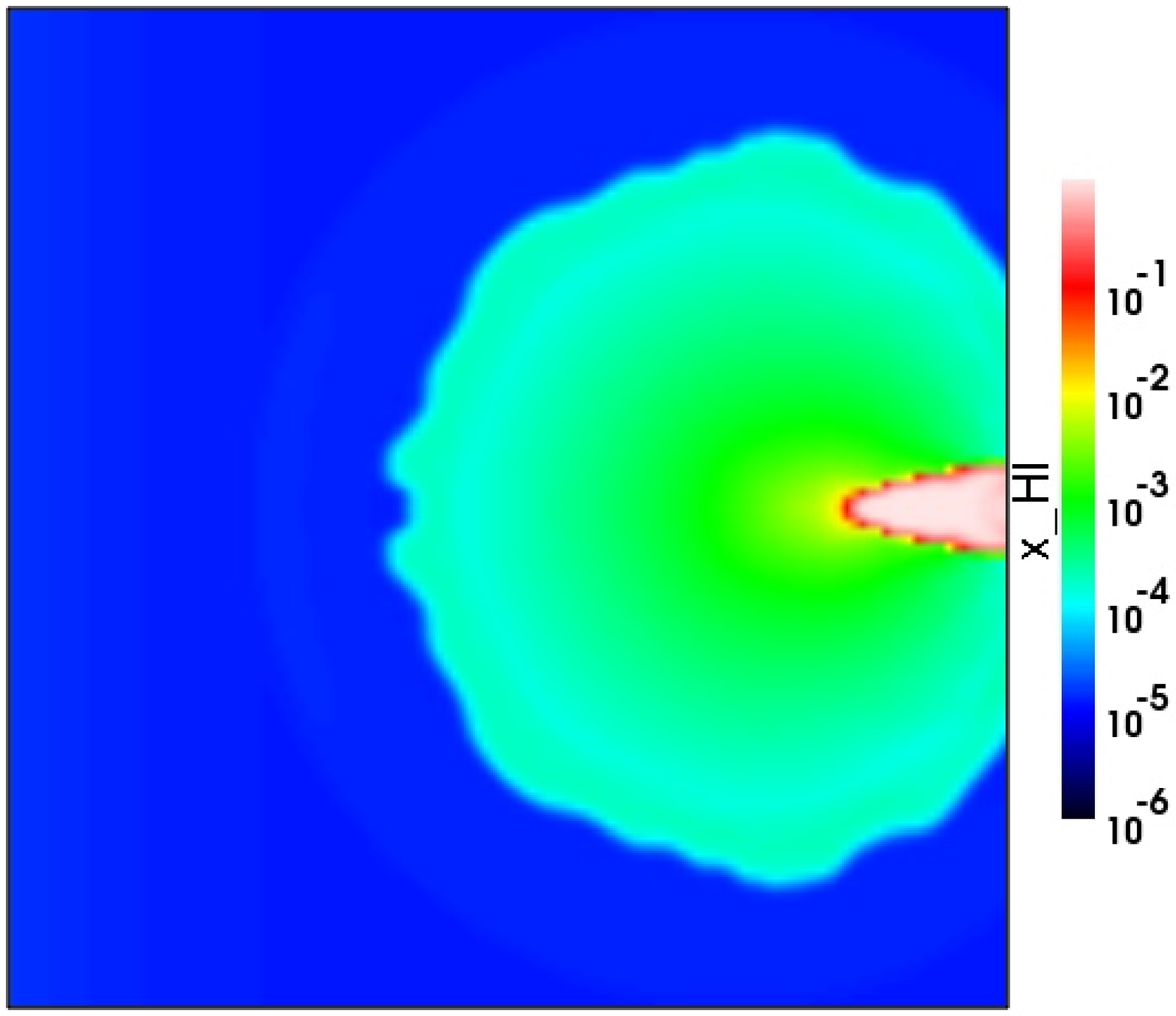}
   \includegraphics[width=2.3in]{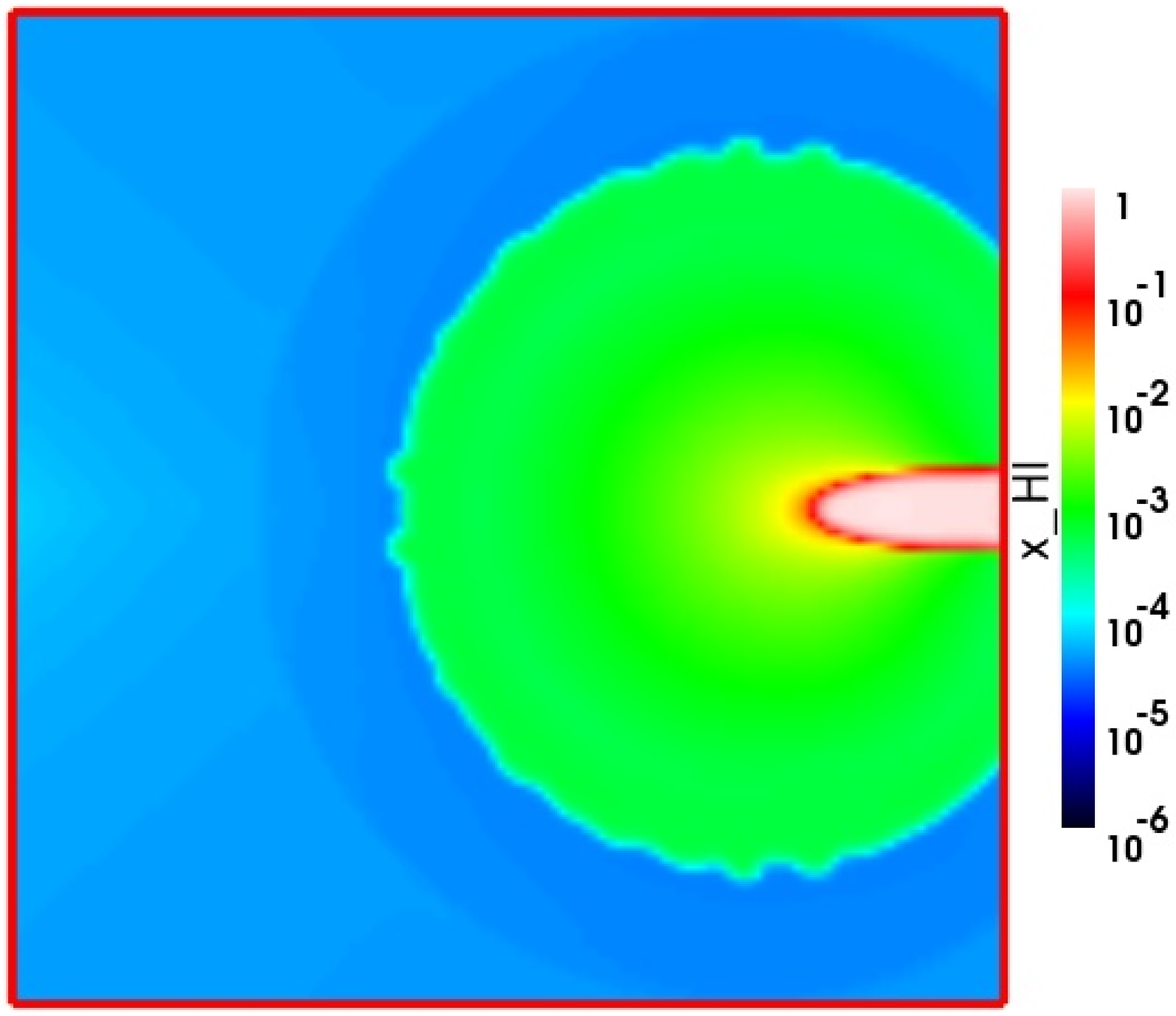}
 \caption{Test 7 (Photoevaporation of a dense clump): Images of the H~I
   fraction, cut through the simulation volume at coordinate $z=0$ at time 
   $t=50$ Myr for (left to right  and top to bottom)
   Capreole+$C^2$-Ray, RSPH, ZEUS-MP, LICORICE, Flash-HC and Coral.
 \label{T7_images5_xhi_fig}}
 \end{center}
 \end{figure*}

 \begin{figure*}
 \begin{center}
   \includegraphics[width=2.3in]{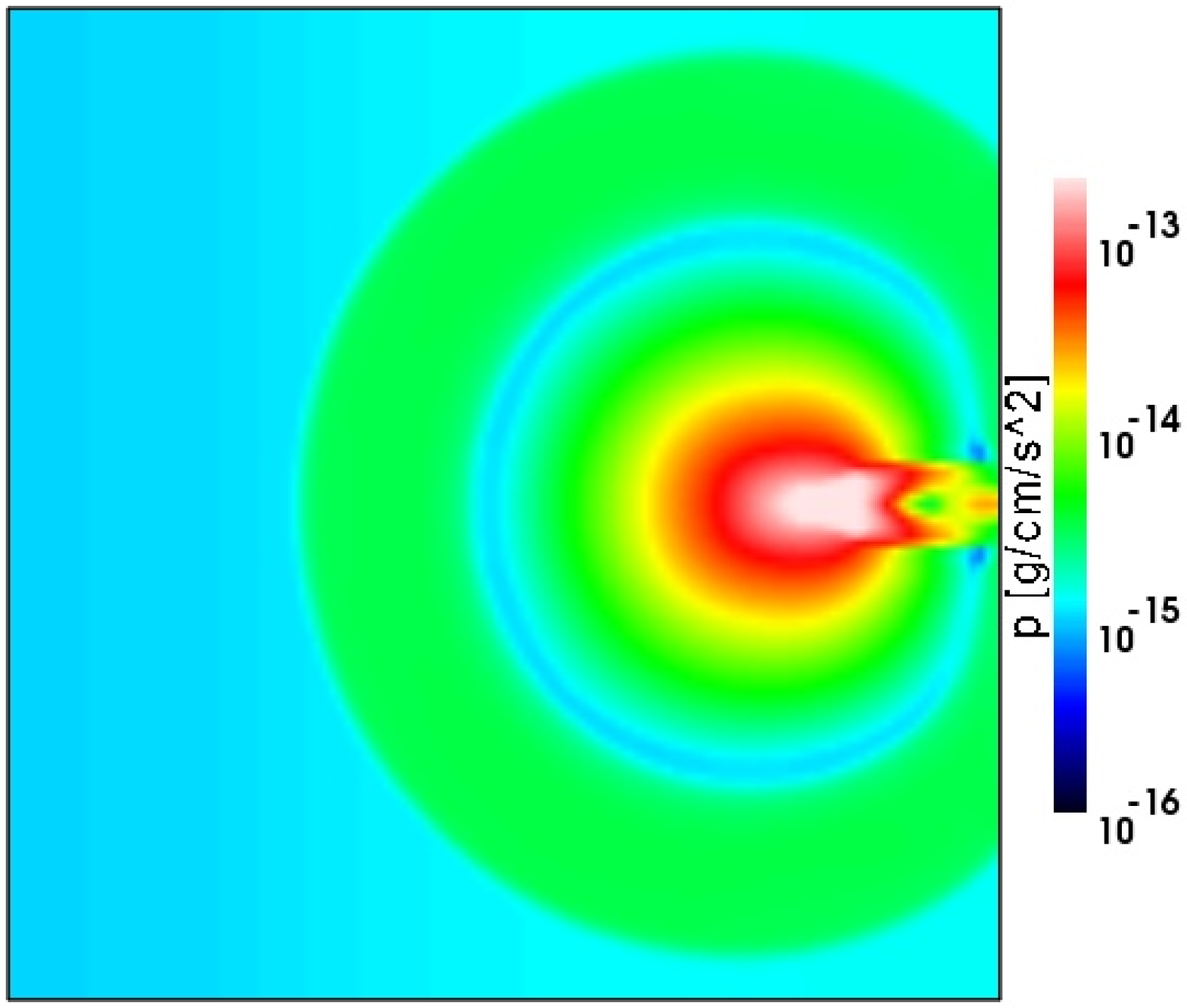}
   \includegraphics[width=2.3in]{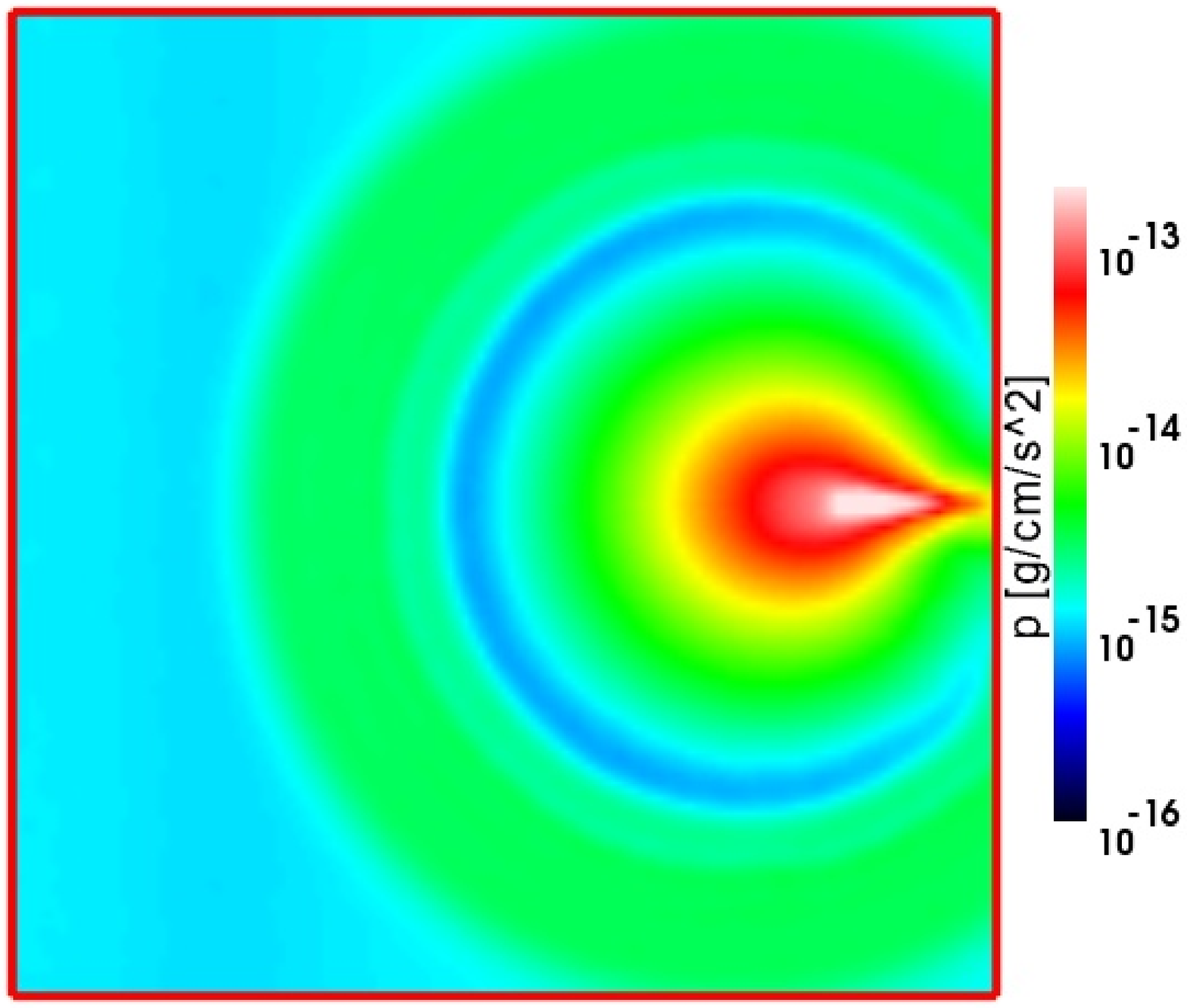}
   \includegraphics[width=2.3in]{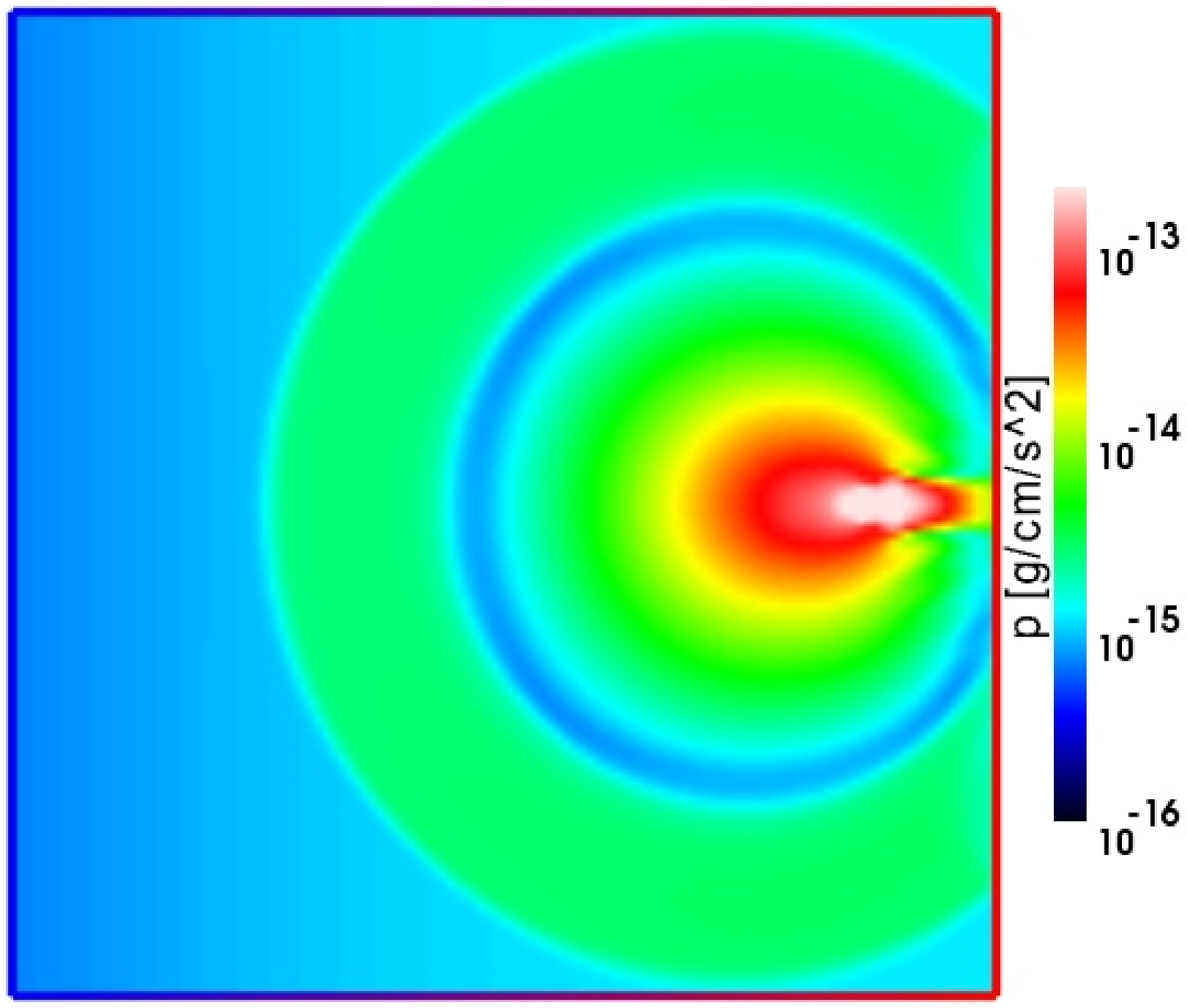}
   \includegraphics[width=2.3in]{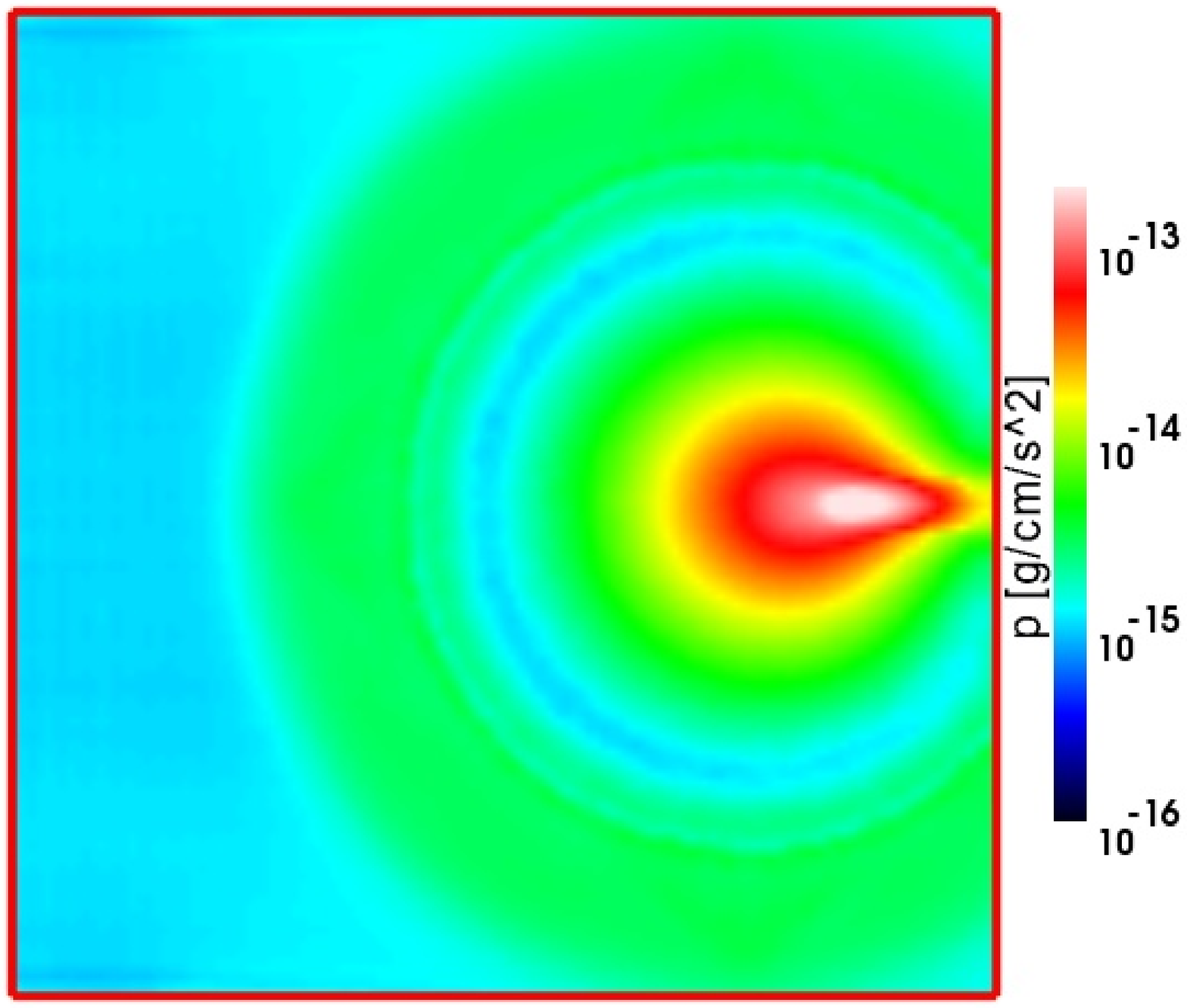}
   \includegraphics[width=2.3in]{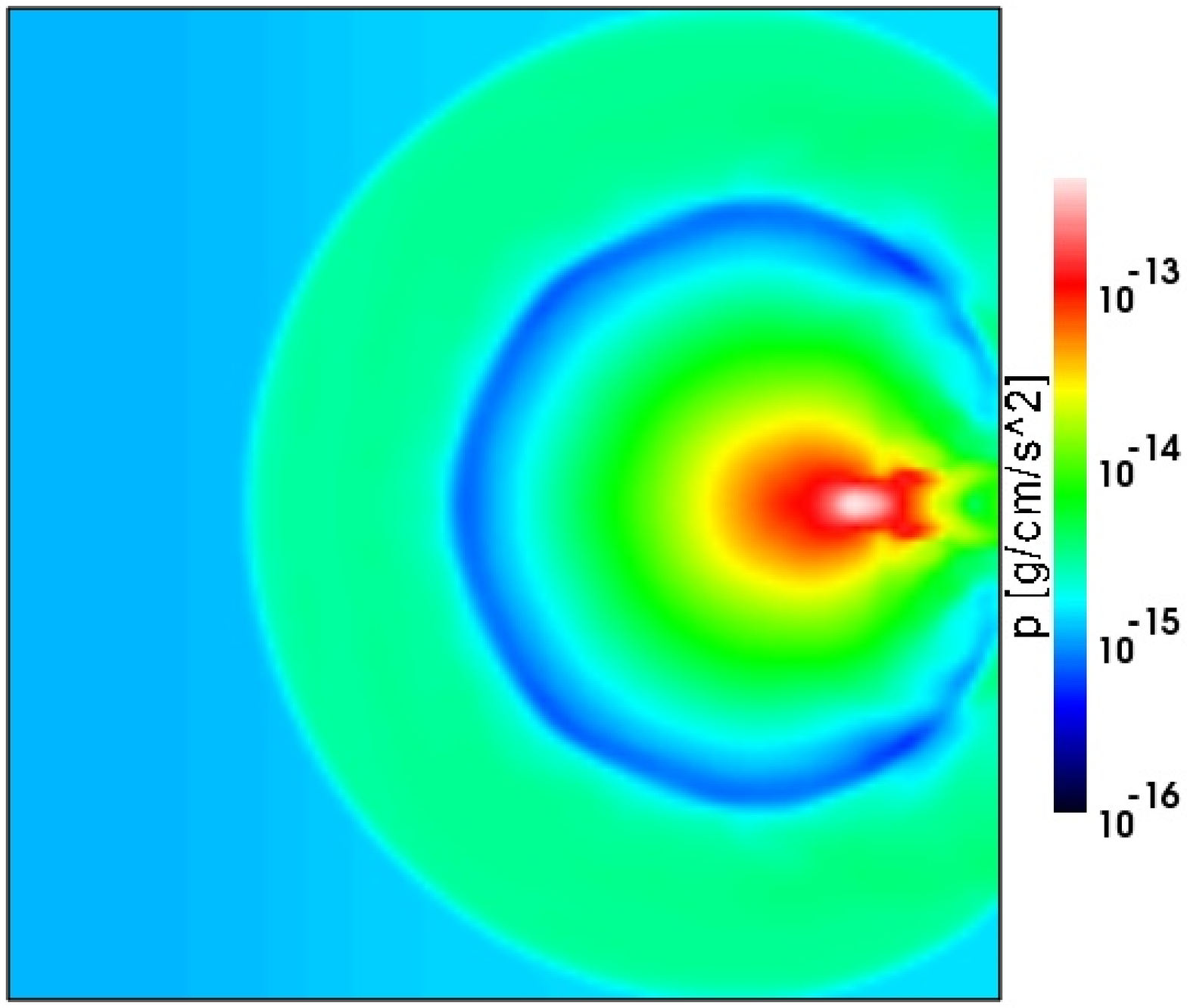}
   \includegraphics[width=2.3in]{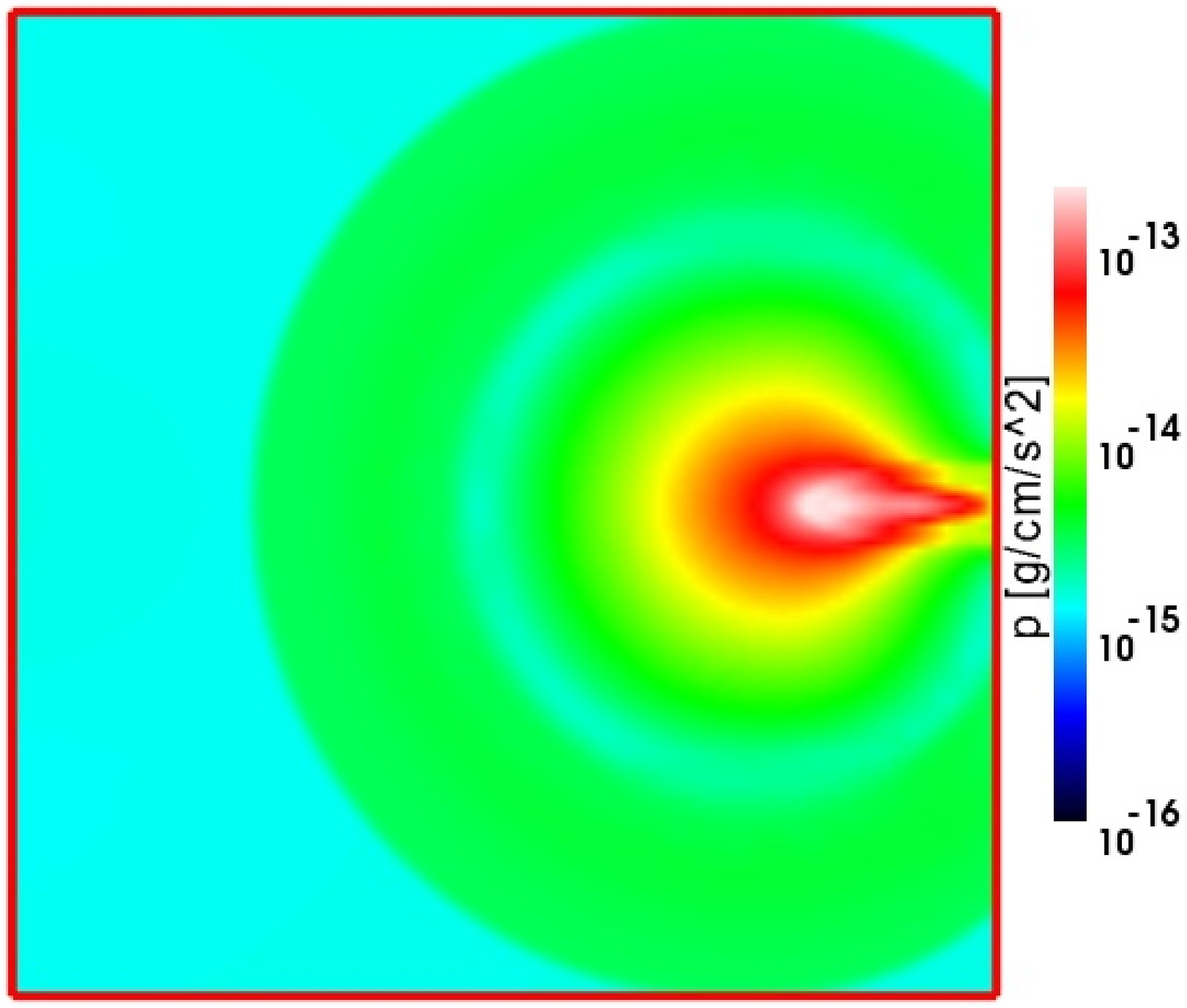}
 \caption{Test 7 (Photoevaporation of a dense clump): Images of the 
   pressure, cut through the simulation volume at coordinate $z=0$ at 
   time $t=50$ Myr for (left to right  and top to bottom)
   Capreole+$C^2$-Ray, RSPH, ZEUS-MP, LICORICE, Flash-HC and Coral.
 \label{T7_images5_p_fig}}
 \end{center}
 \end{figure*}

\begin{figure*}
 \begin{center}
   \includegraphics[width=2.3in]{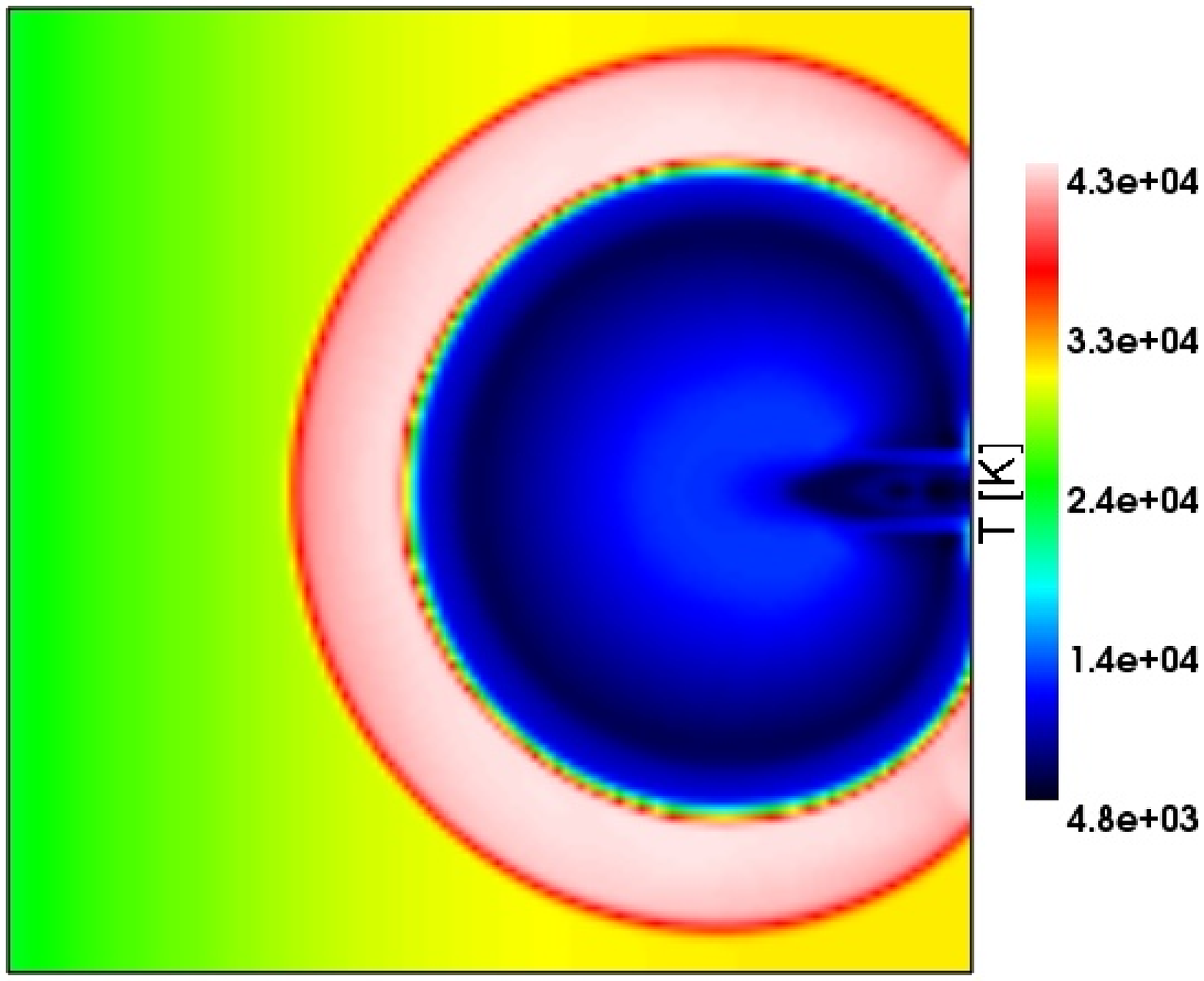}
   \includegraphics[width=2.3in]{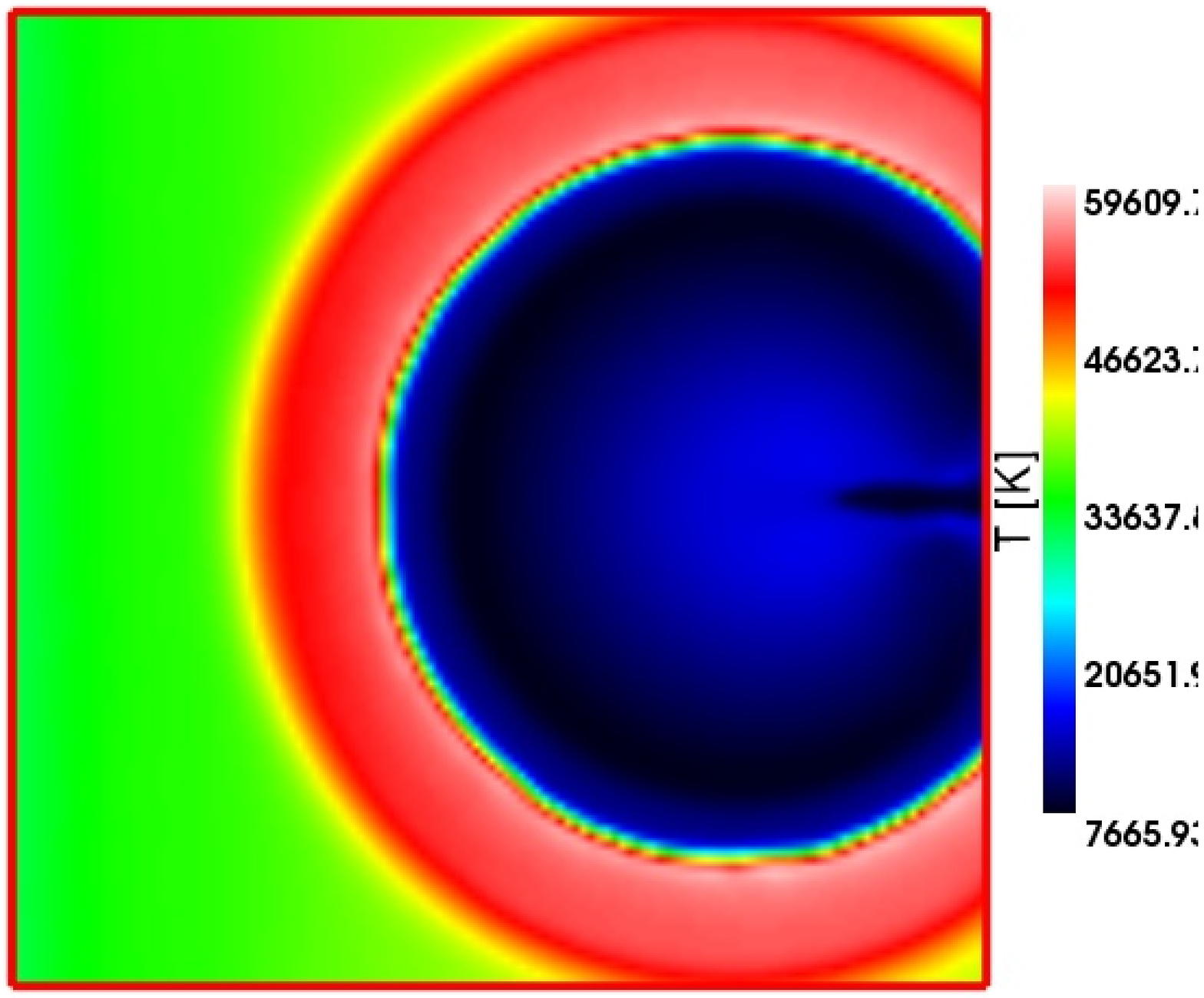}
   \includegraphics[width=2.3in]{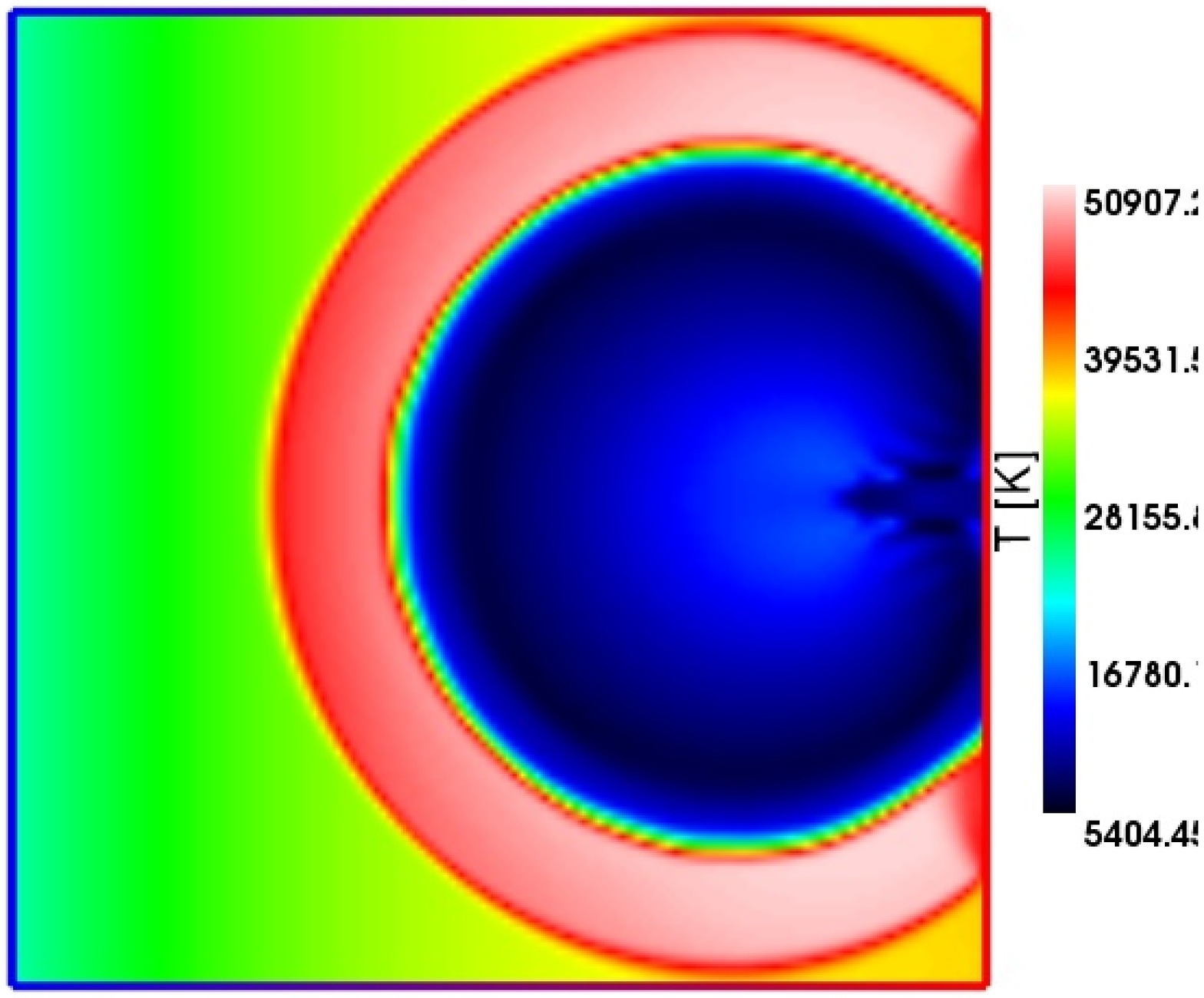}
   \includegraphics[width=2.3in]{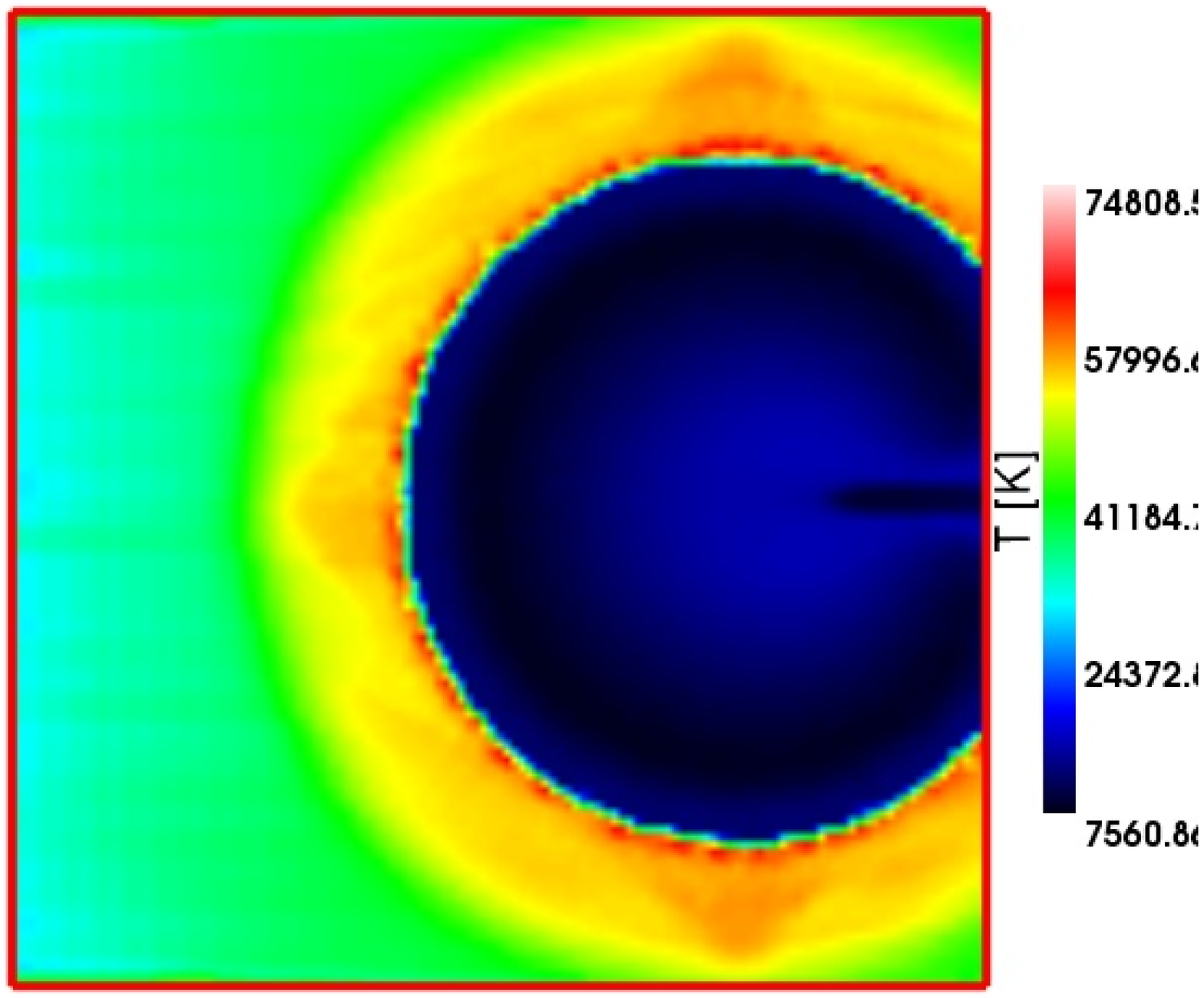}
   \includegraphics[width=2.3in]{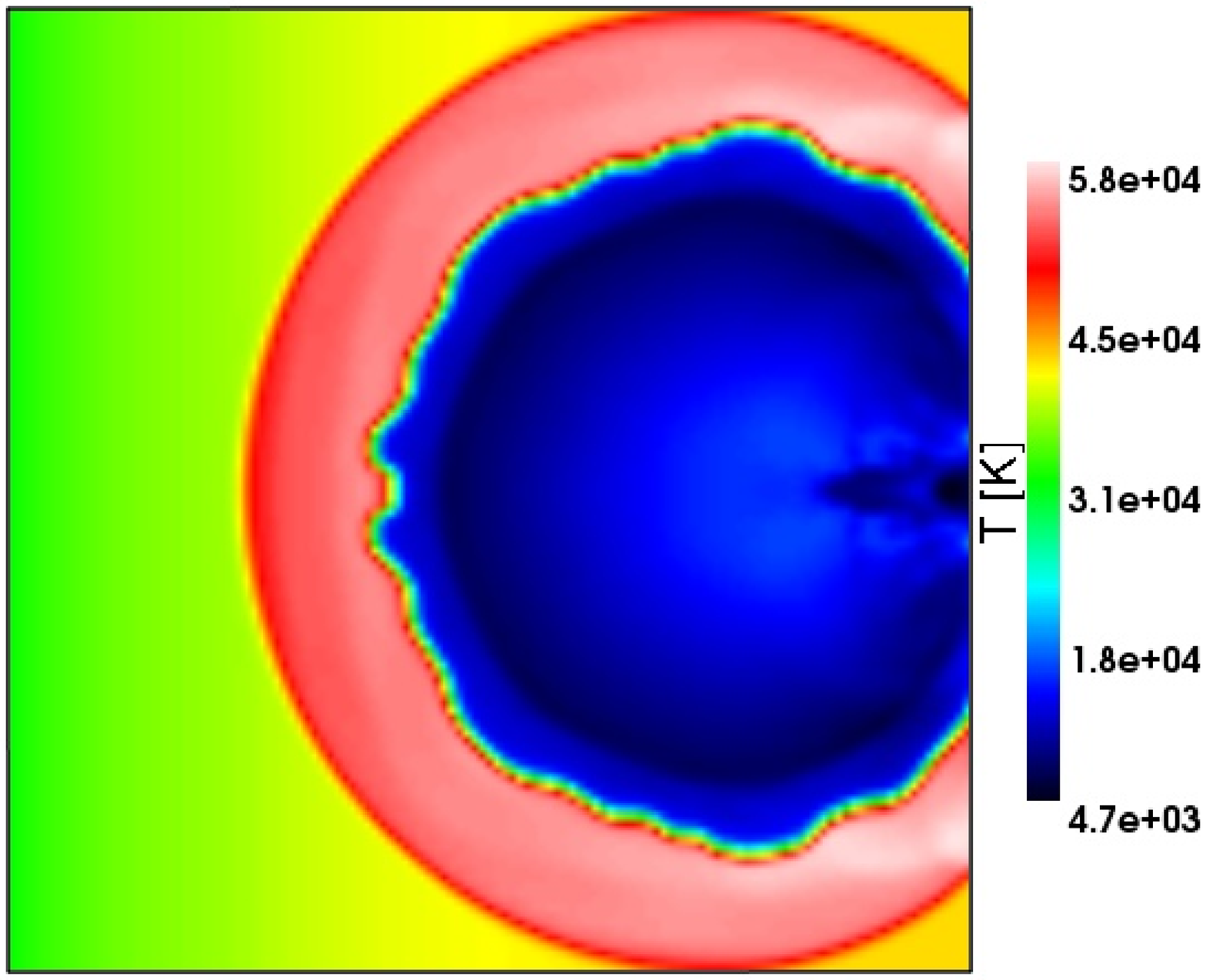}
   \includegraphics[width=2.3in]{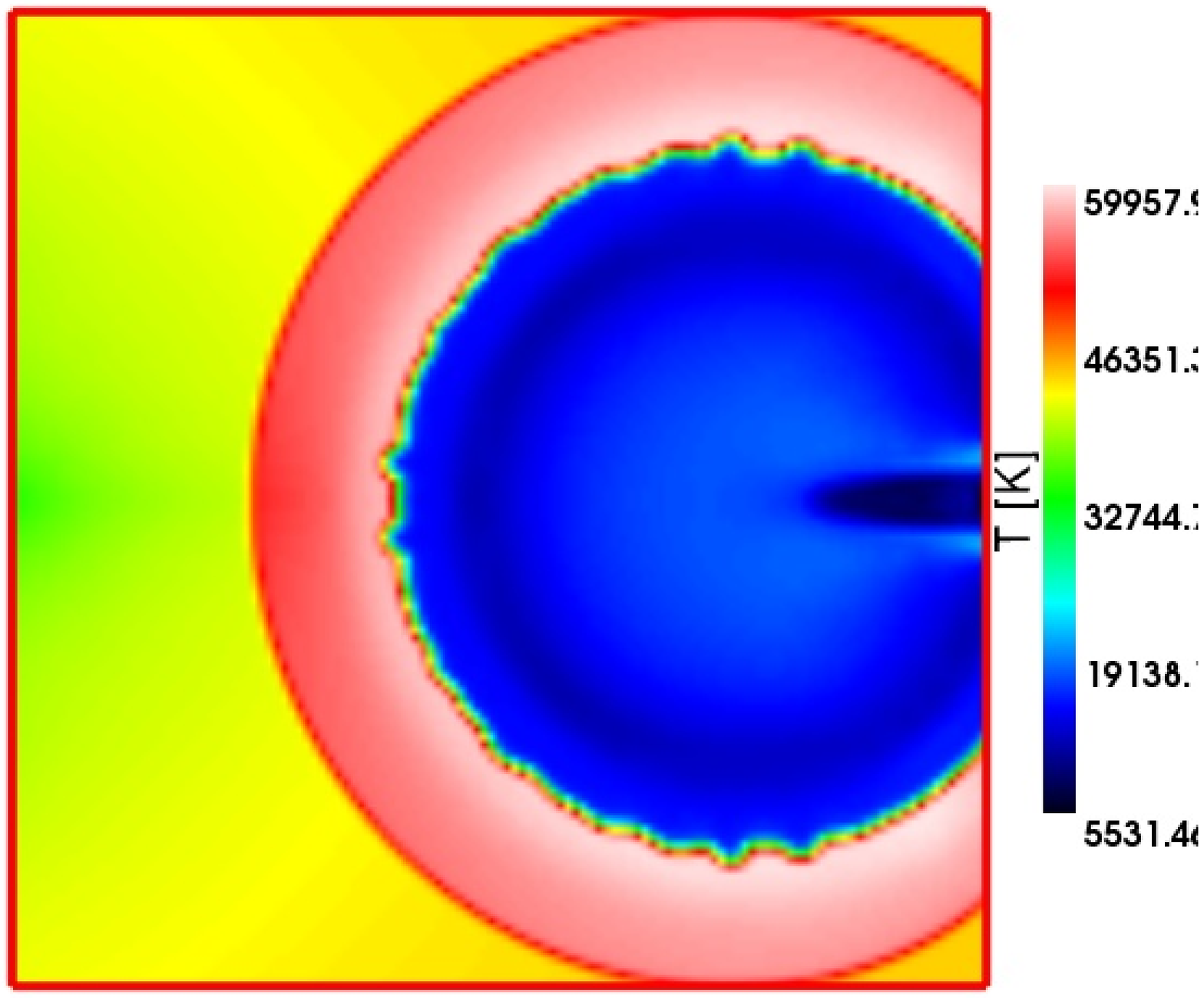}
 \caption{Test 7 (Photoevaporation of a dense clump): Images of the 
   temperature, cut through the simulation volume at coordinate $z=0$ at 
   time $t=50$ Myr for (left to right  and top to bottom)
   Capreole+$C^2$-Ray, RSPH, ZEUS-MP, LICORICE, Flash-HC and Coral.
 \label{T7_images5_T_fig}}
 \end{center}
 \end{figure*}

\begin{figure*}
 \begin{center}
   \includegraphics[width=2.3in]{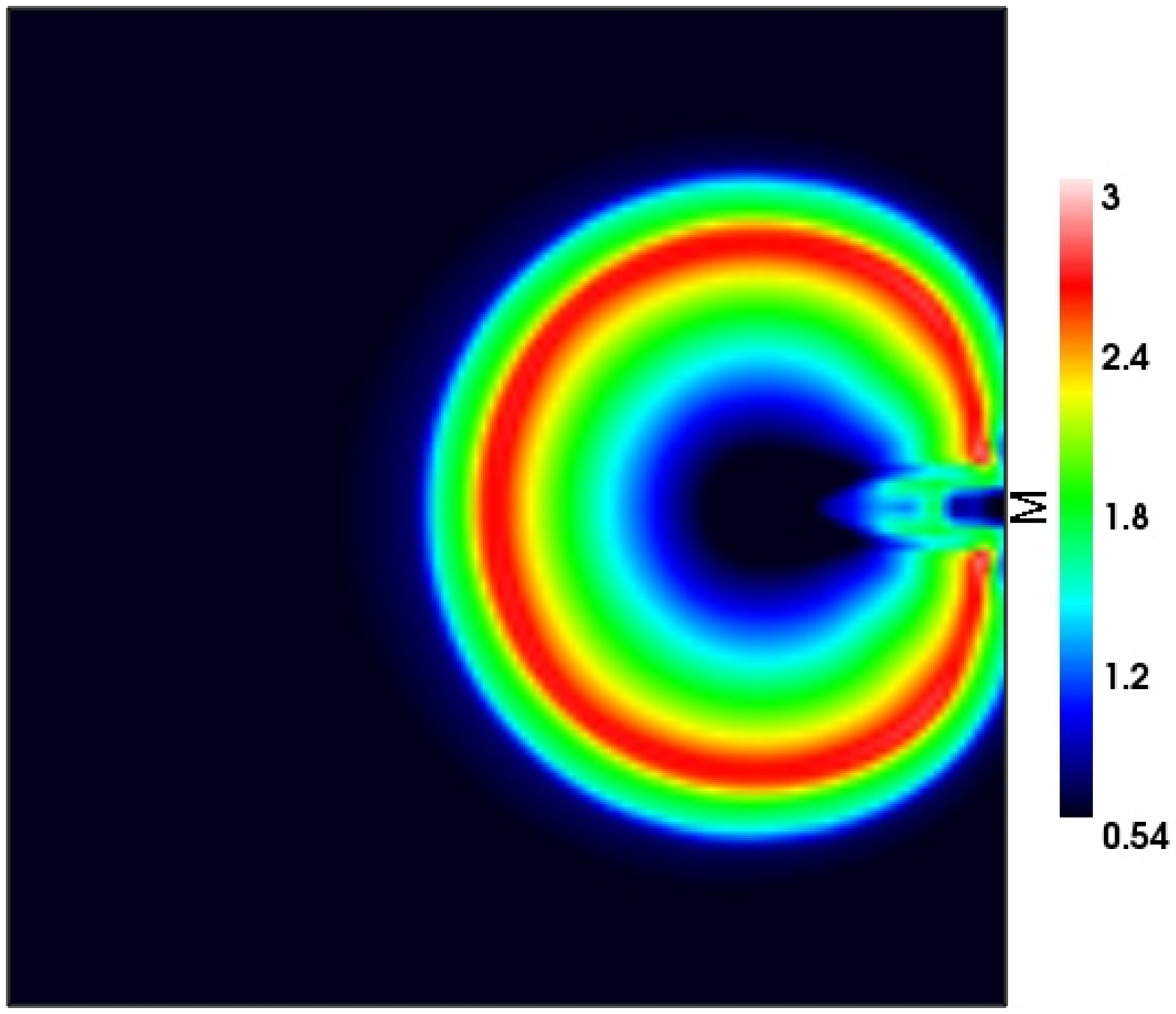}
   \includegraphics[width=2.3in]{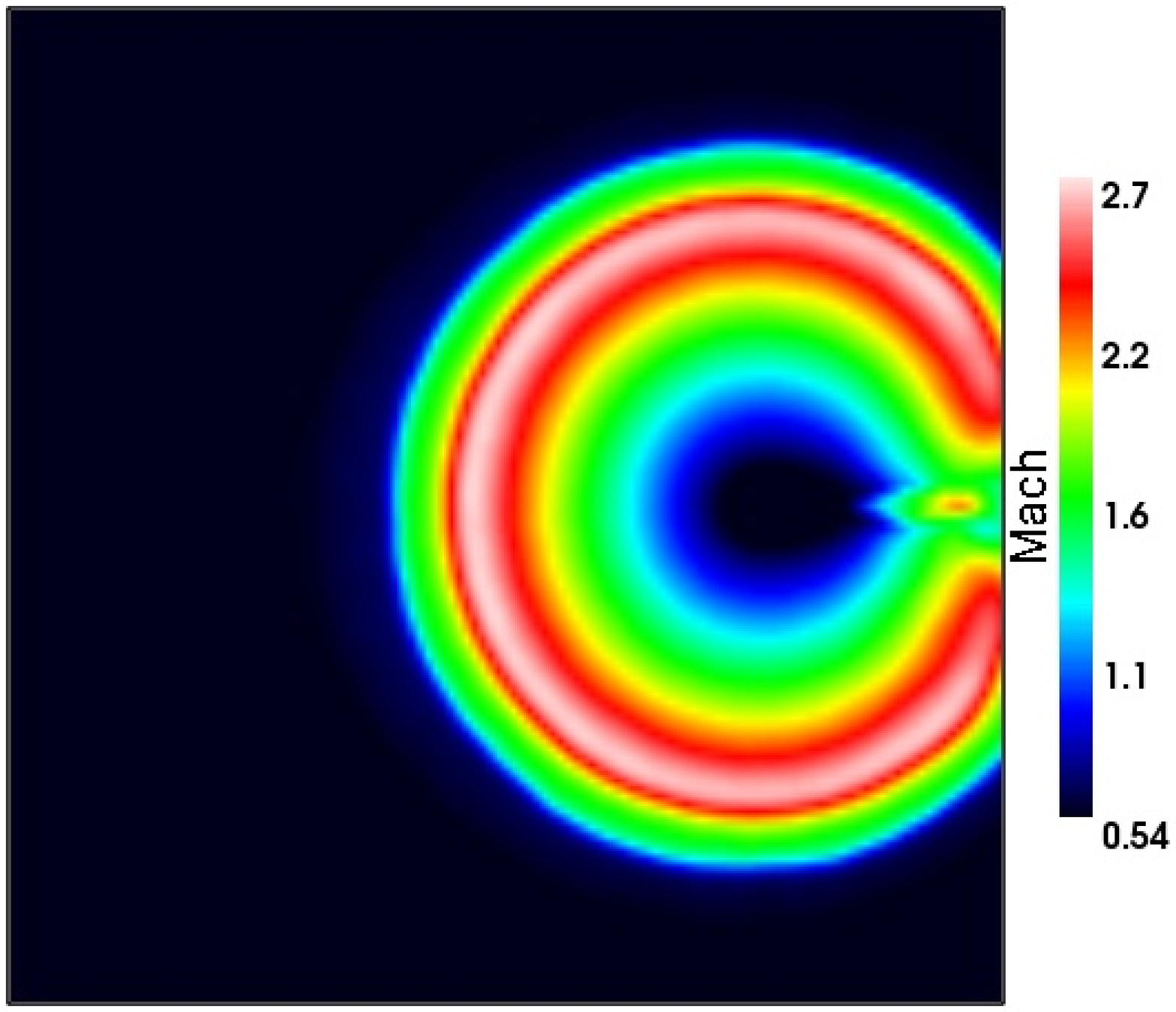}
   \includegraphics[width=2.3in]{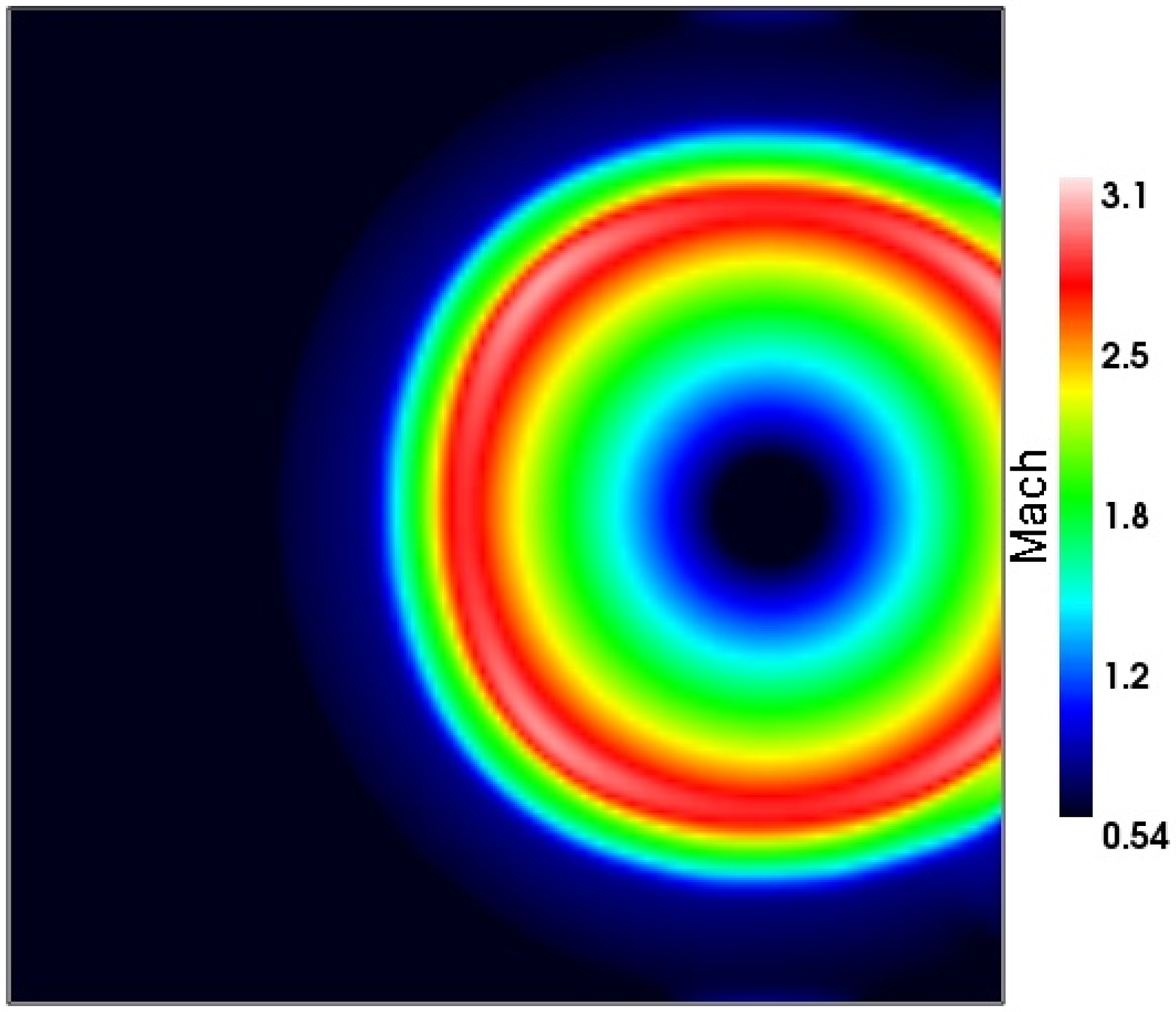}
   \includegraphics[width=2.3in]{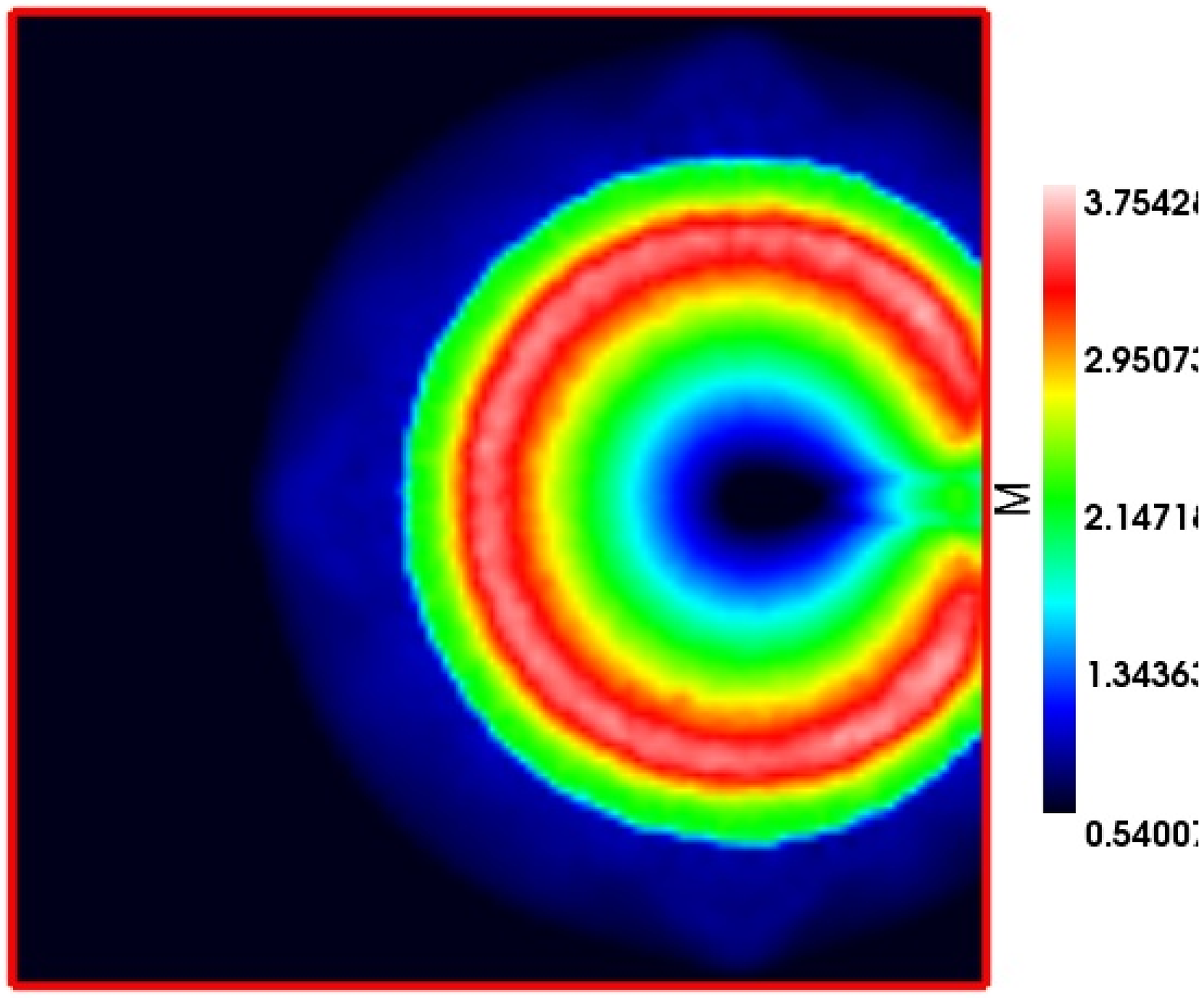}
   \includegraphics[width=2.3in]{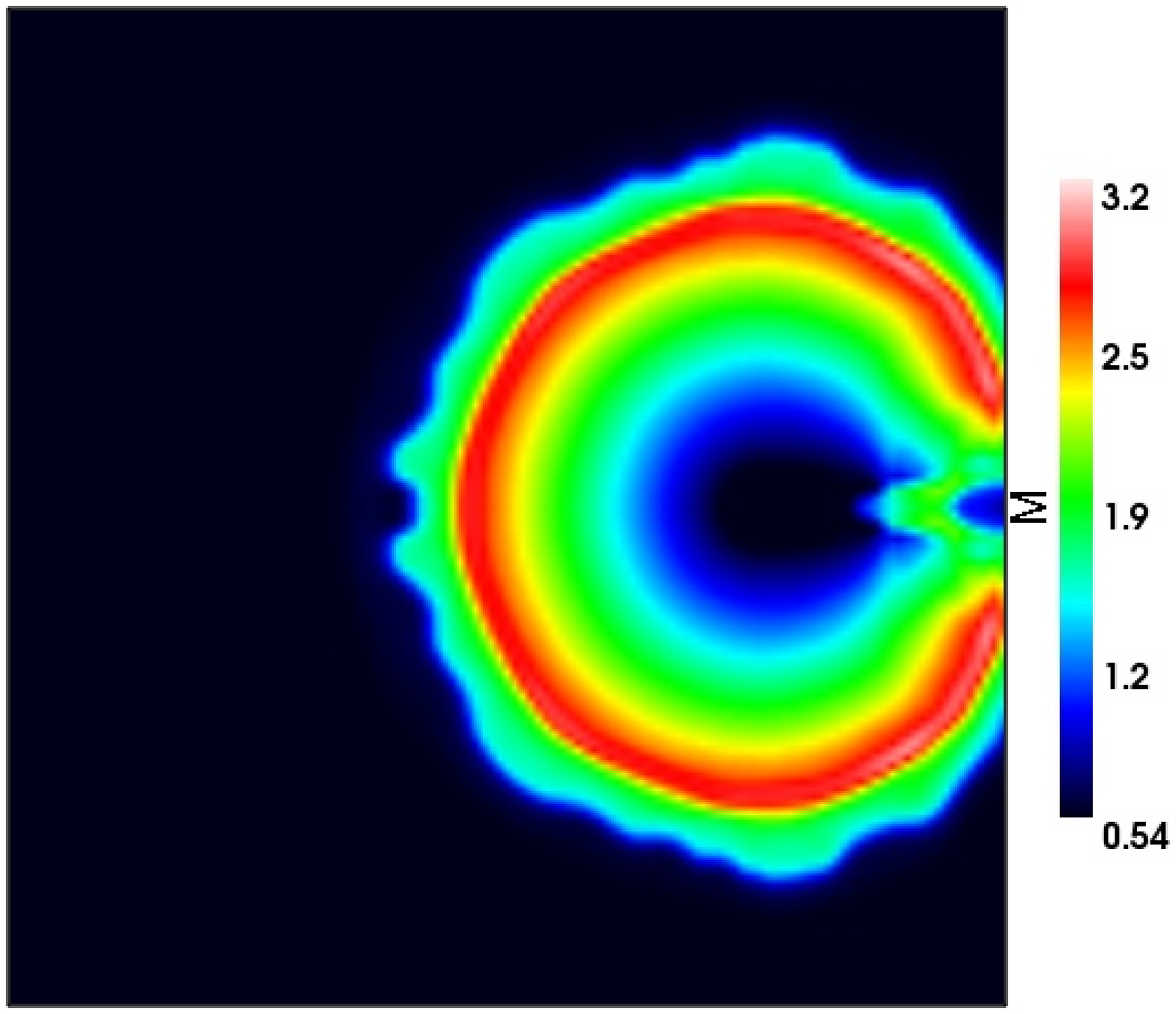}
   \includegraphics[width=2.3in]{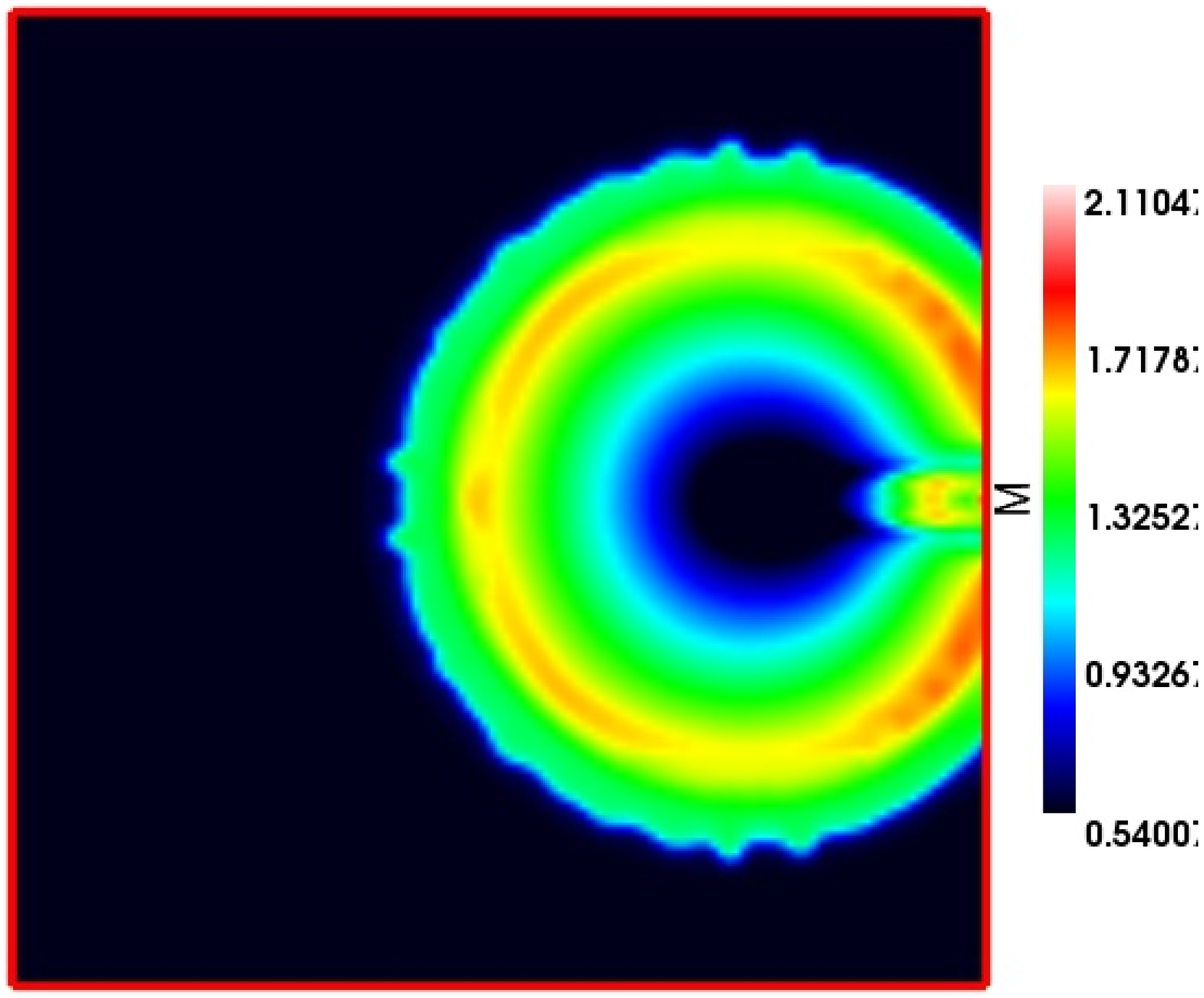}
 \caption{Test 7 (Photoevaporation of a dense clump): Images of the flow
   Mach number, cut through the simulation volume at coordinate $z=0$ at 
   time $t=50$ Myr for (left to right  and top to bottom)
   Capreole+$C^2$-Ray, RSPH, ZEUS-MP, LICORICE, Flash-HC and Coral.
 \label{T7_images5_M_fig}}
 \end{center}
 \end{figure*}

By $t=50$~Myr (Figures~\ref{T7_images5_xhi_fig}-\ref{T7_images5_M_fig}) 
the photoevaporation process is well advanced. The region swept by the 
expanding supersonic wind has grown quite large and takes up a significant 
fraction of the simulation volume. There are only modest differences in
its size between the different codes. In the case of Flash-HC and, to a 
lesser extent Coral, the edge of the expanding wind region is uneven, 
as a consequence of the grid effects in the initial conditions, as 
discussed above, when representing a spherical object on a relatively 
coarse rectangular grid with no interpolation used. These grid effects
could be seen at earlier times as well, but at a lower level. The 
overall size of the wind region is the same, however, thus this problem 
does not affect the evolution significantly. 

A small core region from the initial clump remains neutral and still 
casts a clear shadow which also remains neutral in all cases. This 
neutral region is moderately compressed by the higher external pressure 
of the ionized and heated gas surrounding it. The size of this neutral 
dense core and its shadow varies between the runs, being somewhat larger
for Capreole+C$^2$-Ray, Flash-HC and Coral than for RSPH, ZEUS-MP and LICORICE. There 
is also some 'flaring' (i.e. widening) of the shadow for Flash-HC and Coral,
probably due to the specific interpolation weighting used in the short 
characteristics methods they both employ \citep[for discussion and testing
of this see][Appendix A]{methodpaper}. 

In the pressure and temperature images shown in Figures~\ref{T7_images5_p_fig} 
and \ref{T7_images5_T_fig} we clearly see the shocked shell of gas swept up by 
the supersonic wind of evaporating clump material. The inner zone on the side 
of the clump facing the source is being evacuated and is accordingly colder due 
to adiabatic cooling, while the outer shocked shell is much hotter, with 
temperatures 
reaching 40,000-70,000 K (note the different upper limits for the temperature 
images). Some quantitative and morphological differences are easily noticed. 
The evacuated region yields a shell of low pressure whose depth varies between 
the runs by about an order of magnitude, from the very low pressure 
$\sim10^{-16}\rm g/cm/s^2$ found by Flash-HC, through the intermediate cases of 
RSPH and ZEUS-MP, to the relatively higher pressure  $\sim10^{-15}\rm g/cm/s^2$
found by Capreole+$C^2$-Ray, LICORICE and Coral. The dense, high-pressure 
central region which remains neutral and the shadow behind it show quite 
different morphologies between the runs, clearly seen in the pressure images
(less so in the temperature ones, due to the lack of color dynamic range). This 
morphology arises as a consequence of successive reflecting oblique shocks which 
form behind the evaporating clump by the interaction between the evaporative wind 
and the partly collapsed shadow squeezed inward by the high external pressure of 
the ionized region. The reason for the morphological differences between the cases 
is most probably a slight difference in the timing of these shocks for each run, 
but ascertaining this will require more detailed analysis of the evolution.    

Finally, the Mach number images at $t=50$~Myr shown in Figure~\ref{T7_images5_M_fig}
show that while the wind is clearly supersonic, with Mach numbers of a few, the 
shocked swept material moves subsonically. The peak Mach numbers vary from 2 to 3.7,
with typical peak values around 3. The shock is clearly somewhat weaker for Coral,
a consequence of this code's more diffusive hydrodynamic solver (based on van Leer 
flux splitting). All other methods, both Eulerian grid-based (Capreole+$C^2$-Ray,
ZEUS-MP and Flash-HC) or particle-based (RSPH, LICORICE) yield very similar results 
in terms of Mach number values. The only significant difference between the results
is again the more spherical high Mach number shell found by ZEUS-MP.

\begin{figure*}
\begin{center}
  \includegraphics[width=3.3in]{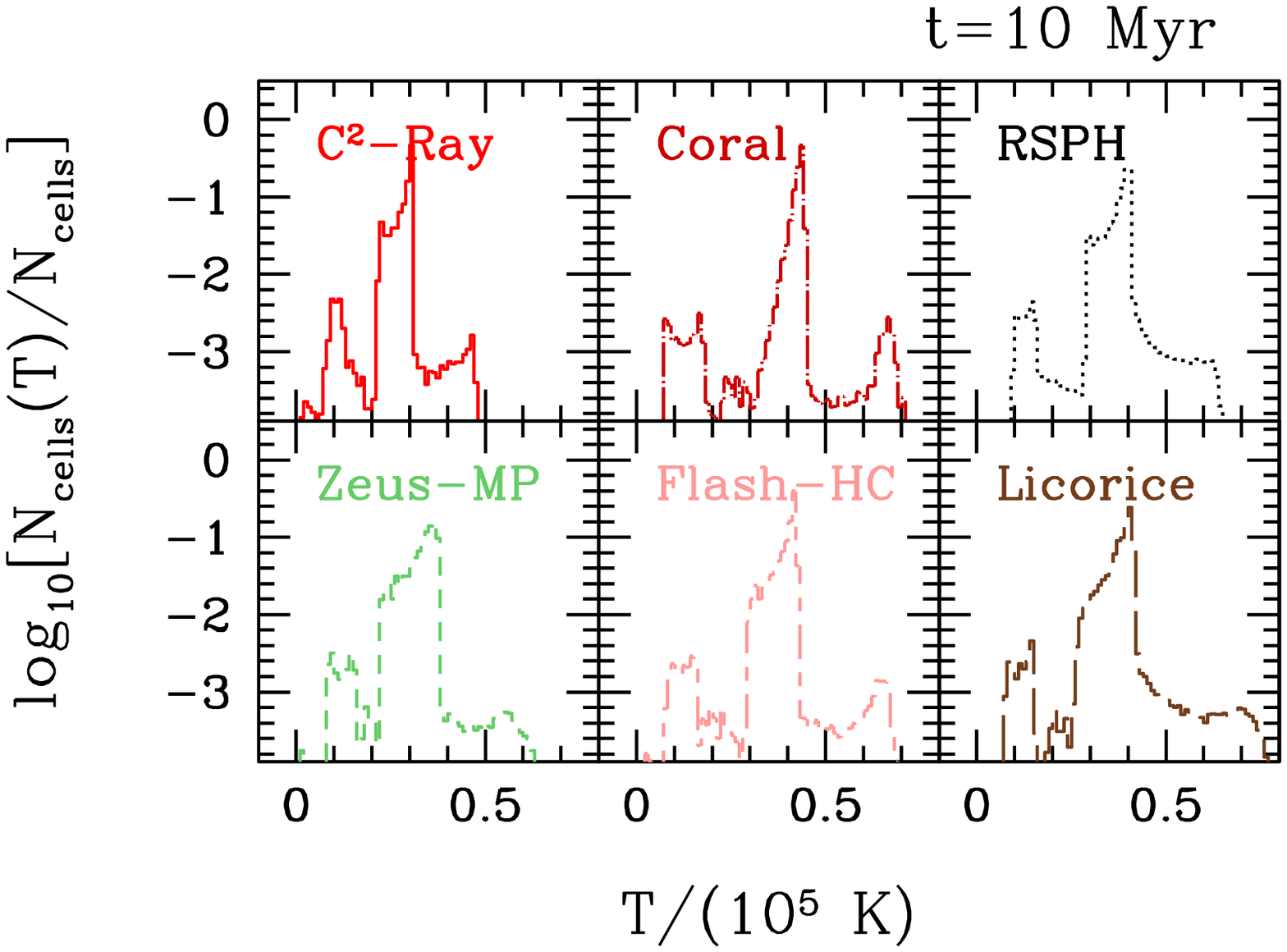}
  \includegraphics[width=3.3in]{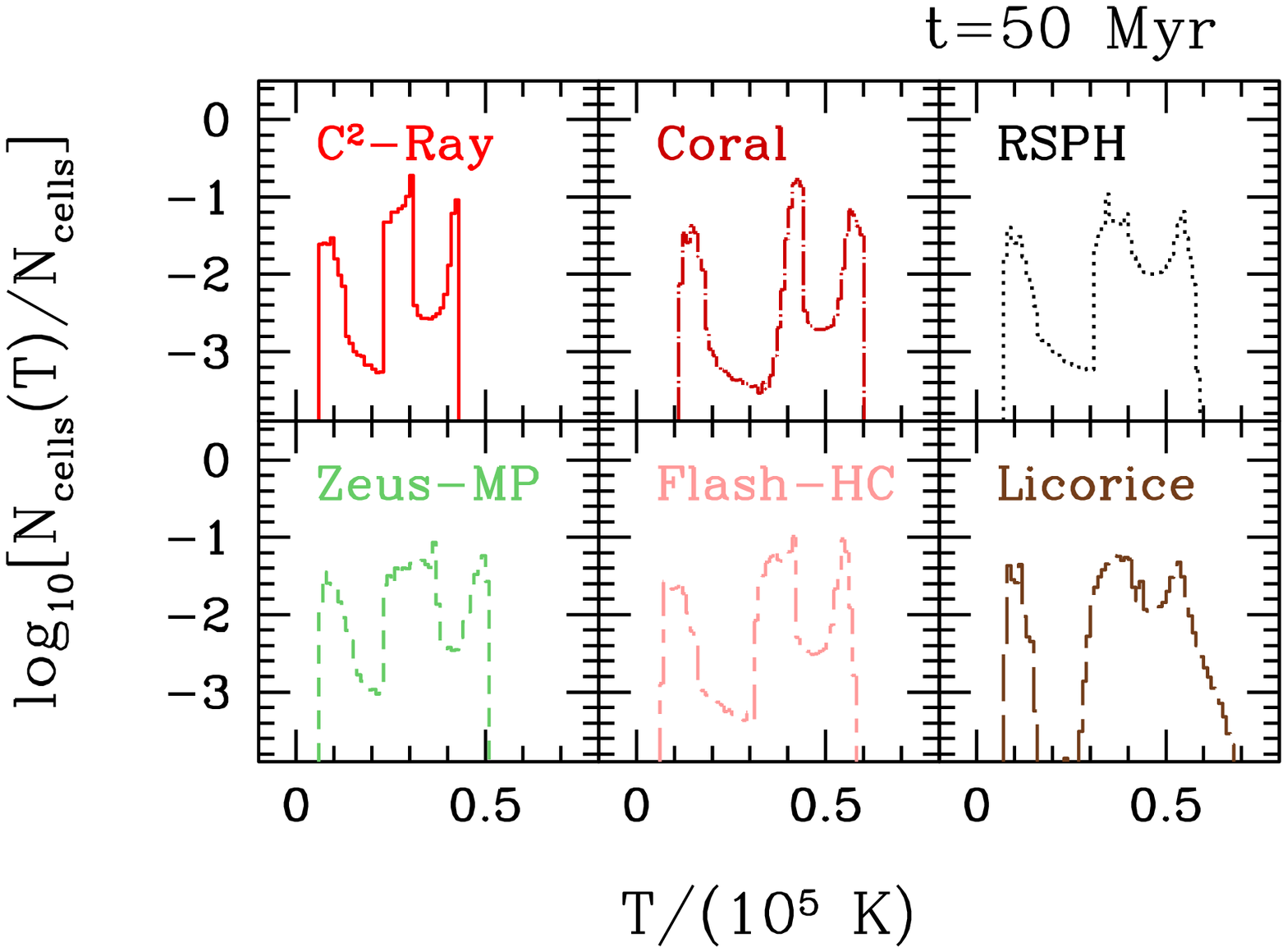}\vspace{-0.9in}
\caption{Test 7 (Photoevaporation of a dense clump): Histograms of the 
gas temperature at times 
$t=10$ and 50 Myrs.
\label{T7_histT_fig}}
\end{center}
\end{figure*}

\begin{figure*}
\begin{center}
  \includegraphics[width=3.3in]{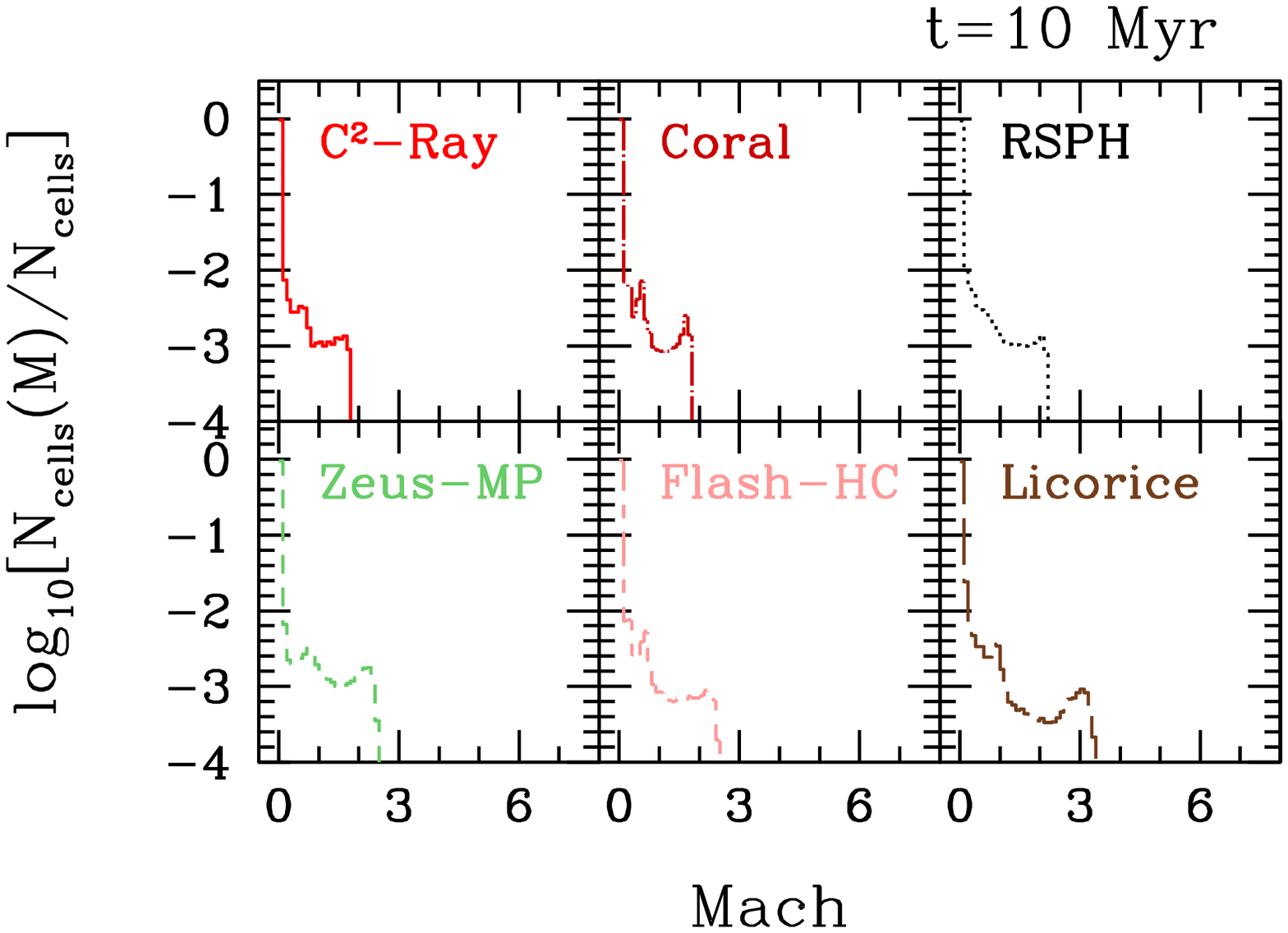}
  \includegraphics[width=3.3in]{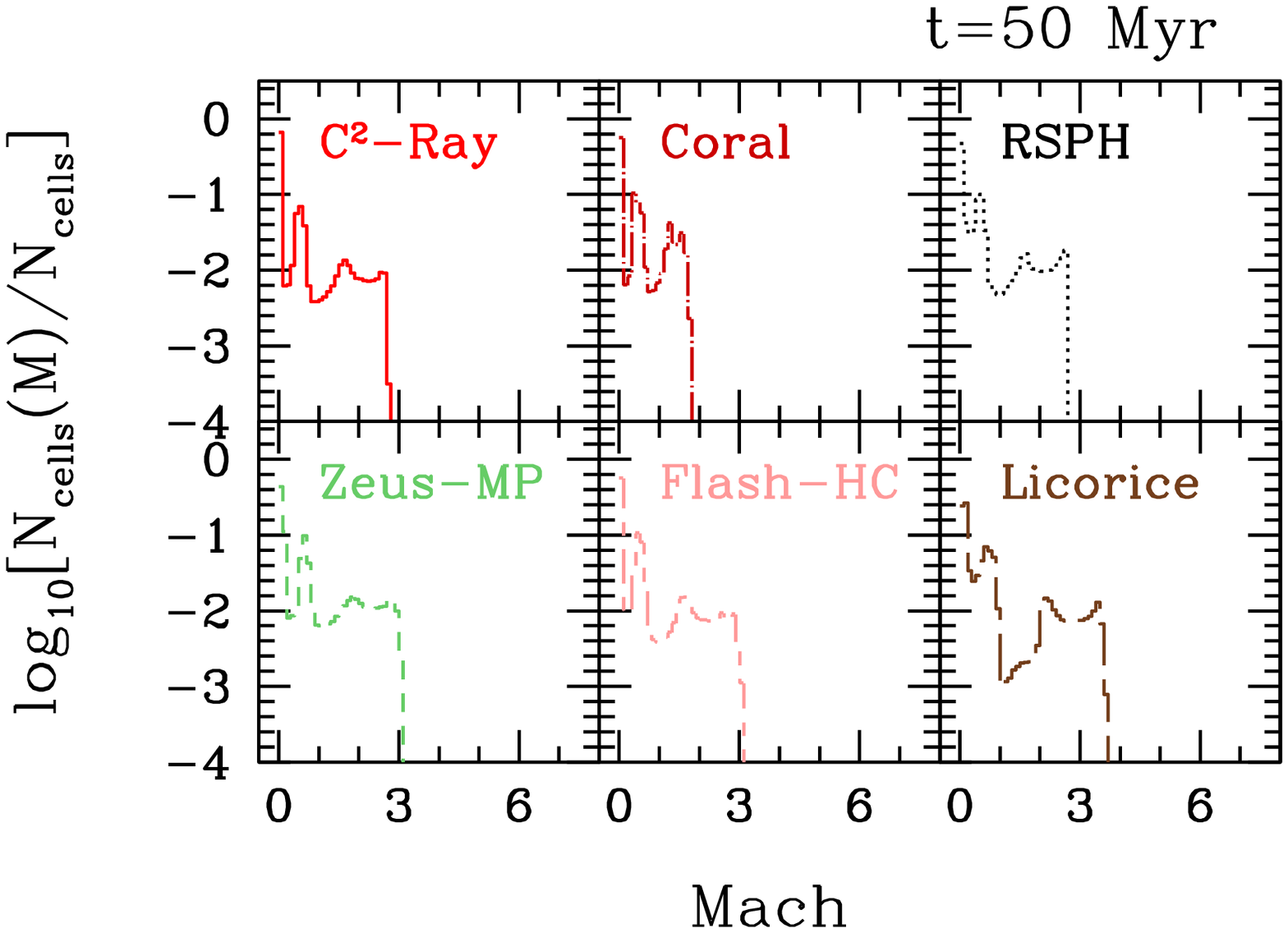}\vspace{-0.9in}
\caption{Test 7 (Photoevaporation of a dense clump): Histograms of the 
flow Mach number at times $t=10$, and 50 Myrs.
\label{T7_histM_fig}}
\end{center}
\end{figure*}

Next we turn our attention to the statistical distributions of the temperature
(shown in Figure~\ref{T7_histT_fig}) and the Mach number (in Figure~\ref{T7_histM_fig}).
We notice that three distinct temperature phases, represented by the three peaks of
the histograms, exist throughout the evolution - hot, photoionized gas with temperatures 
$T\sim25,000-45,000$~K, very hot, $T>50,000$~K, shock-heated gas and a cold phase, 
consisting in part of self-shielded gas and in part of adiabatically cooled gas behind 
the expanding supersonic wind. These three phases are observed in all cases and the 
histograms are very similar. The Mach number histograms are in good agreement as well.
For RSPH and LICORICE the hot, shocked phase is less distinct from the photoionized phase. 
The shocked gas temperature is a bit higher for LICORICE, due to the stronger shock
(evidenced by the higher peak Mach number) observed in this case. On the other hand, 
the temperatures found by Capreole+$C^2$-Ray are somewhat lower than the rest, which
is related to the more approximate treatment of the energy equation in that case. 

\begin{figure*}
\begin{center}
  \includegraphics[width=2.2in]{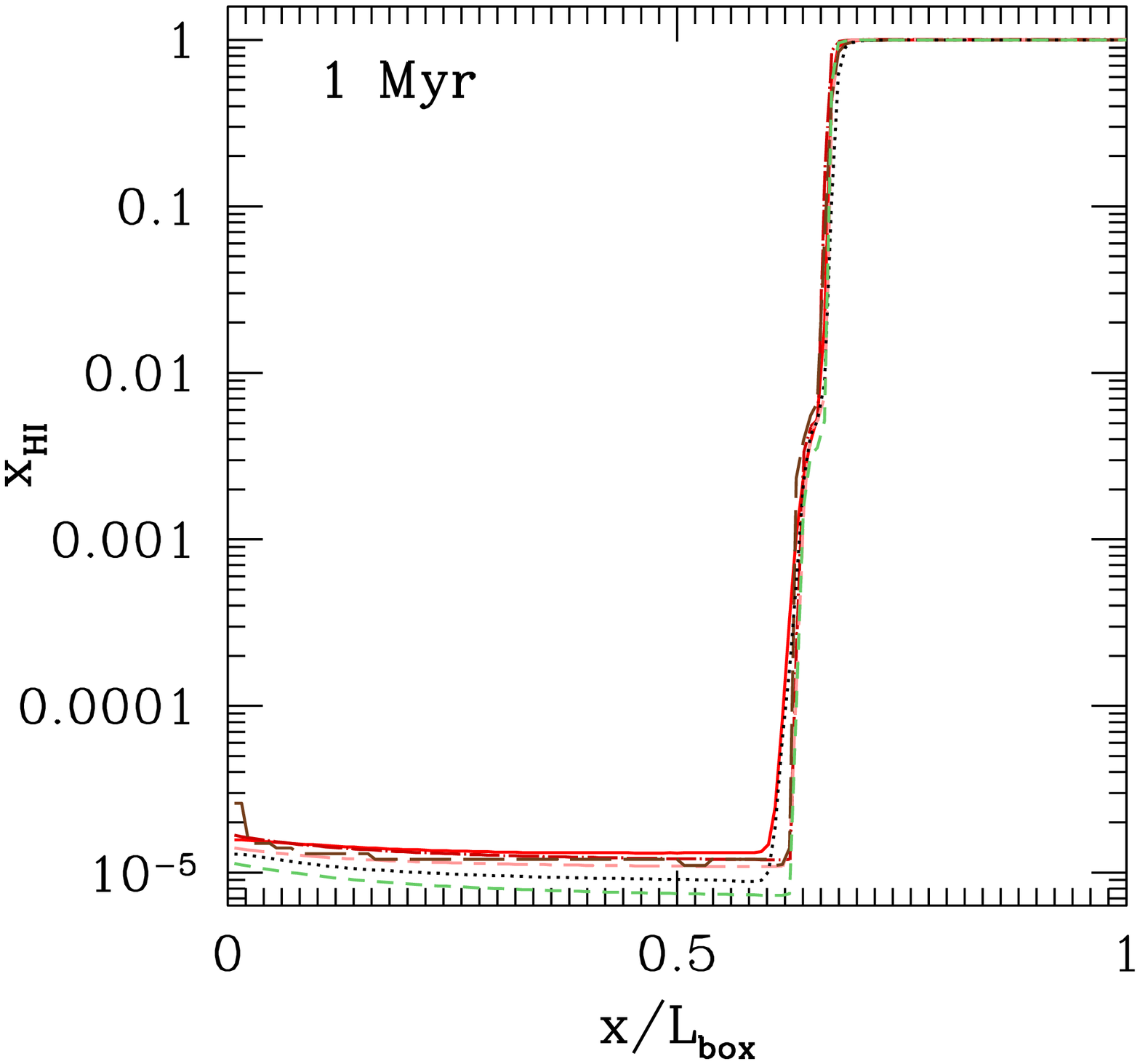}
  \includegraphics[width=2.2in]{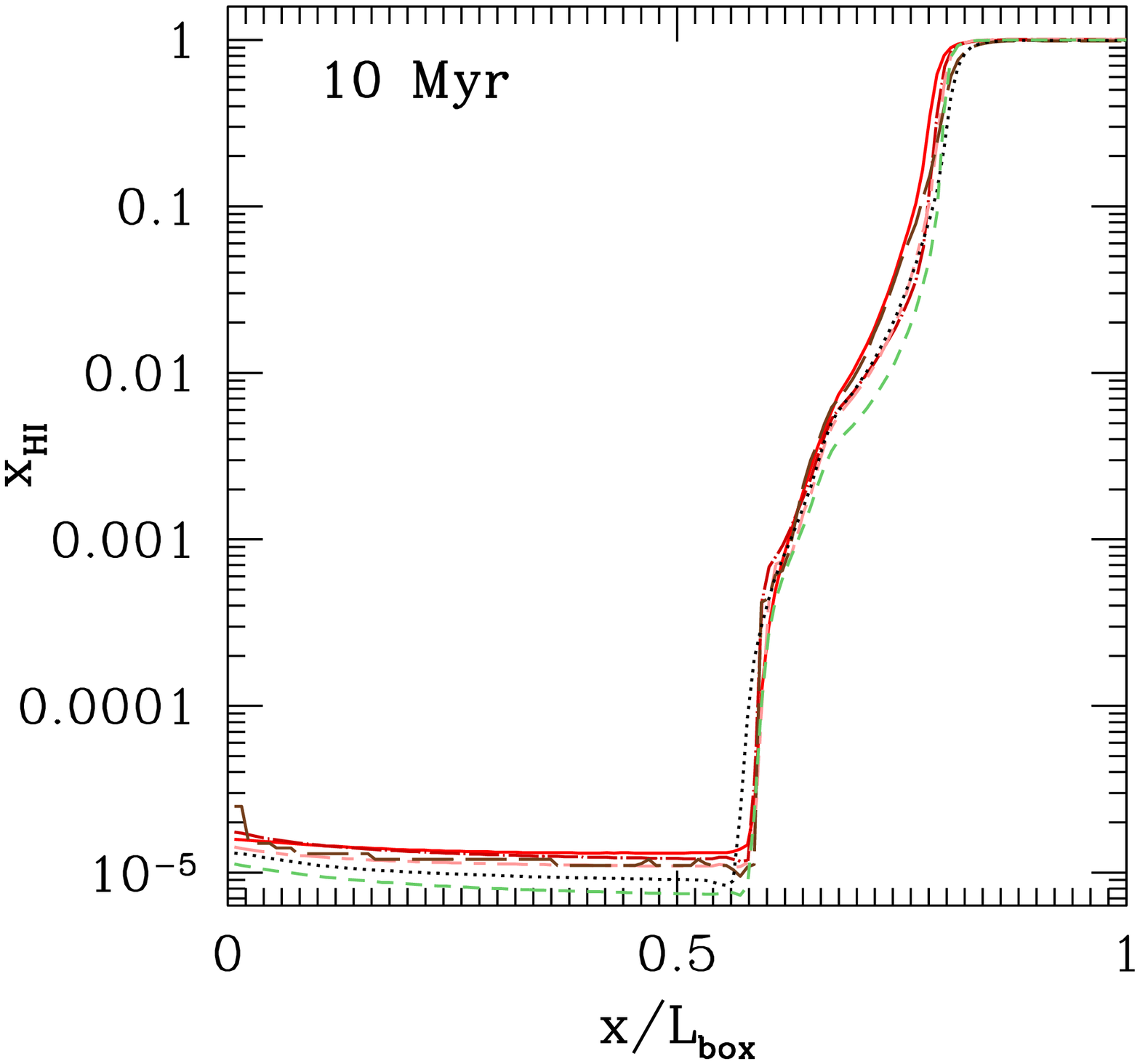}
  \includegraphics[width=2.2in]{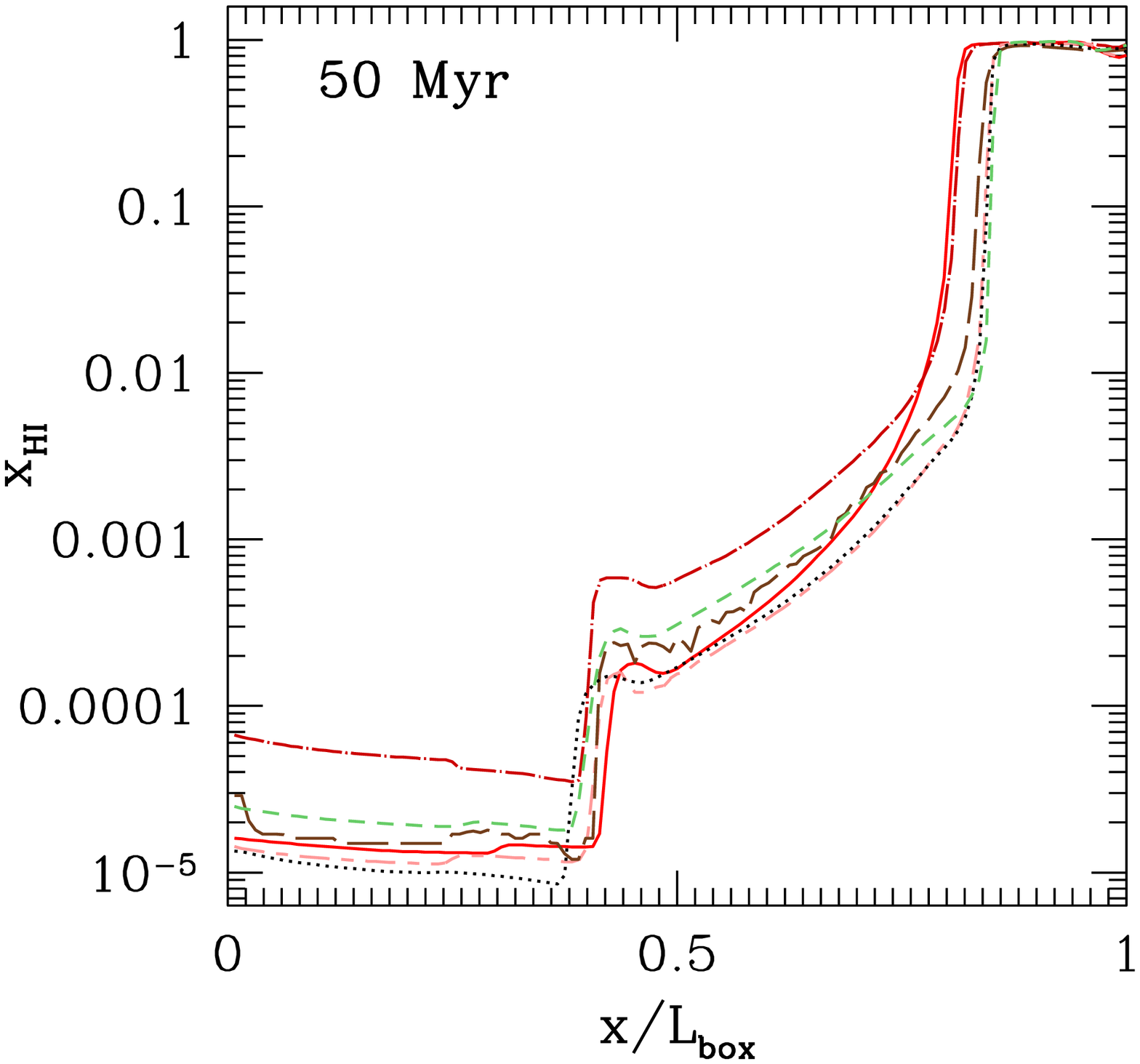}
\caption{Test 7 (Photoevaporation of a dense clump): Line cuts of the 
neutral fraction along the axis of symmetry through the centre of the 
clump at times $t=1$ Myr, 10 Myr and 50 Myr (left to right). 
\label{T7_profsHI_fig}}
\end{center}
\end{figure*}

\begin{figure*}
\begin{center}
  \includegraphics[width=2.2in]{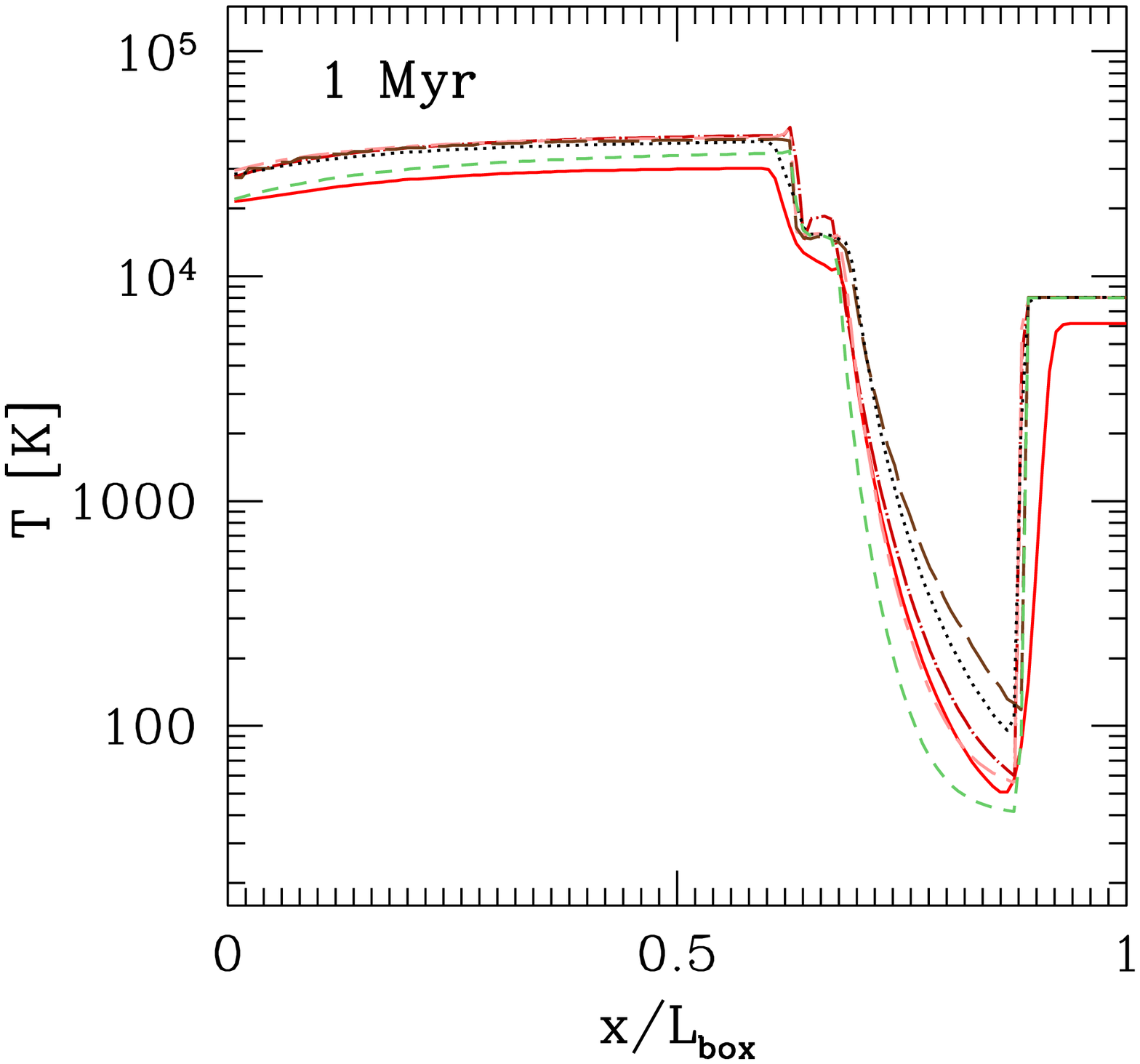}
  \includegraphics[width=2.2in]{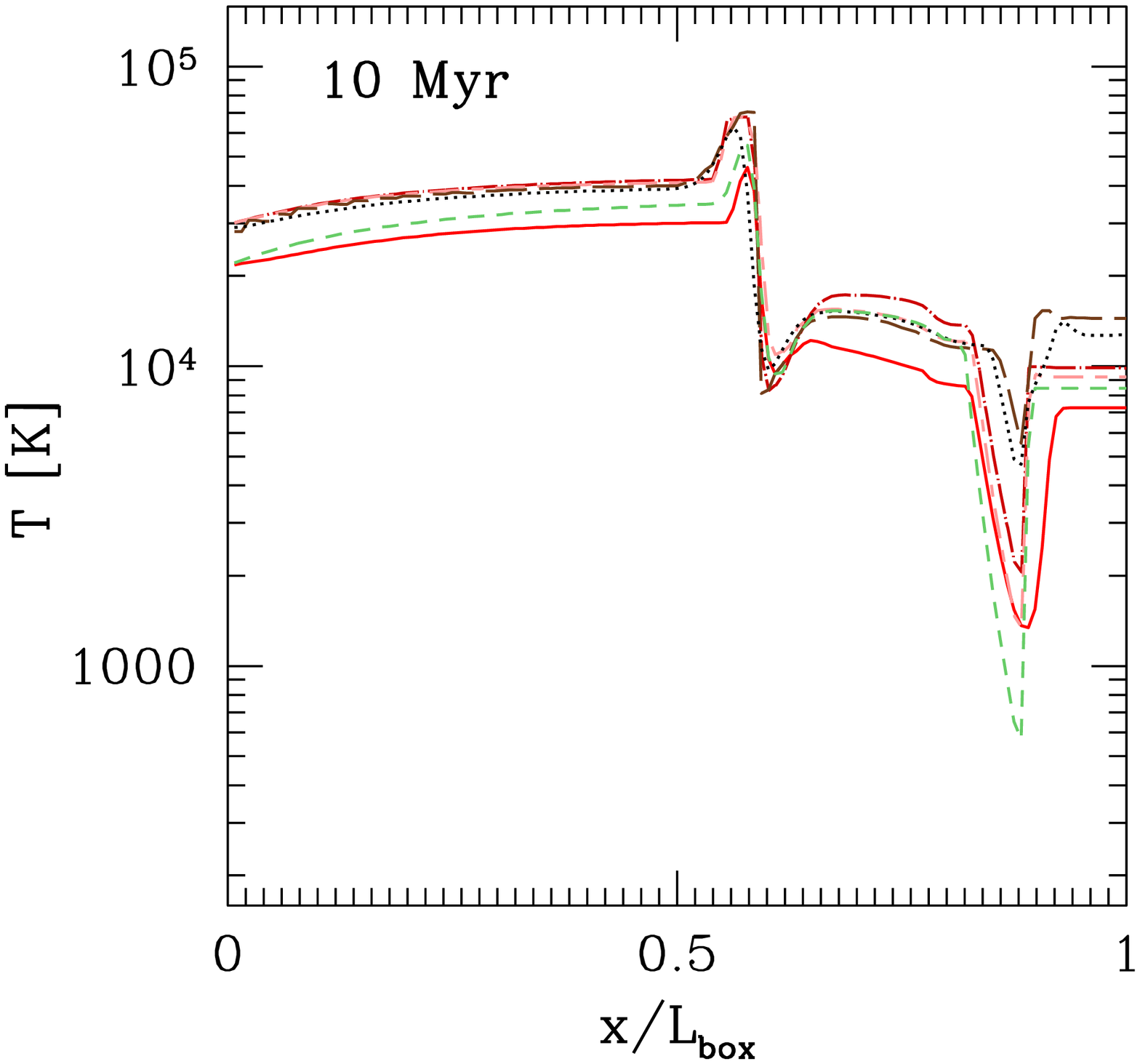}
  \includegraphics[width=2.2in]{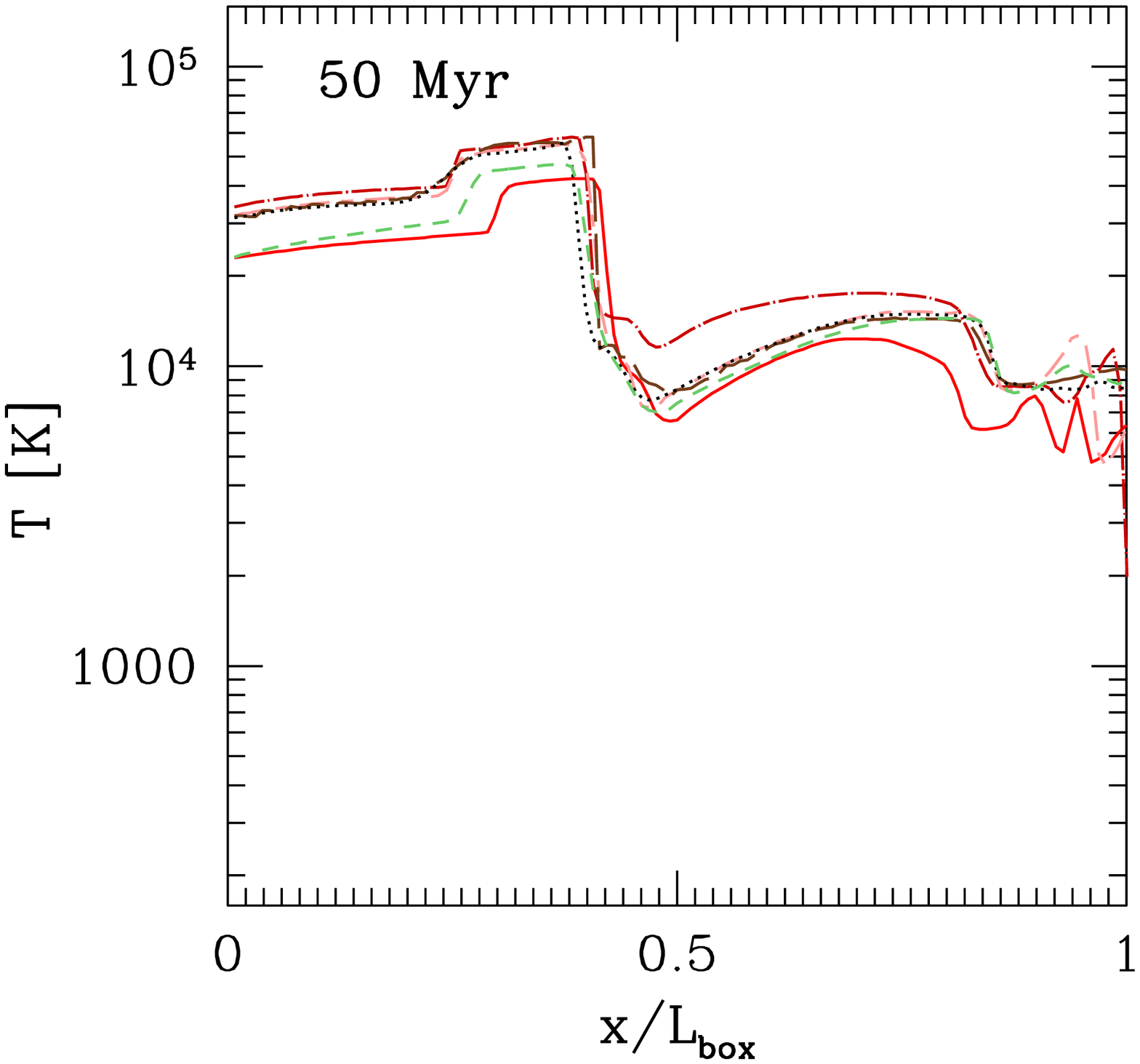}
\caption{Test 7 (Photoevaporation of a dense clump): Line cuts of the 
temperature along the axis of symmetry through the centre of the 
clump at times $t=1$ Myr, 10 Myr and 50 Myr (left to right). 
\label{T7_profsT_fig}}
\end{center}
\end{figure*}

\begin{figure*}
\begin{center}
  \includegraphics[width=2.2in]{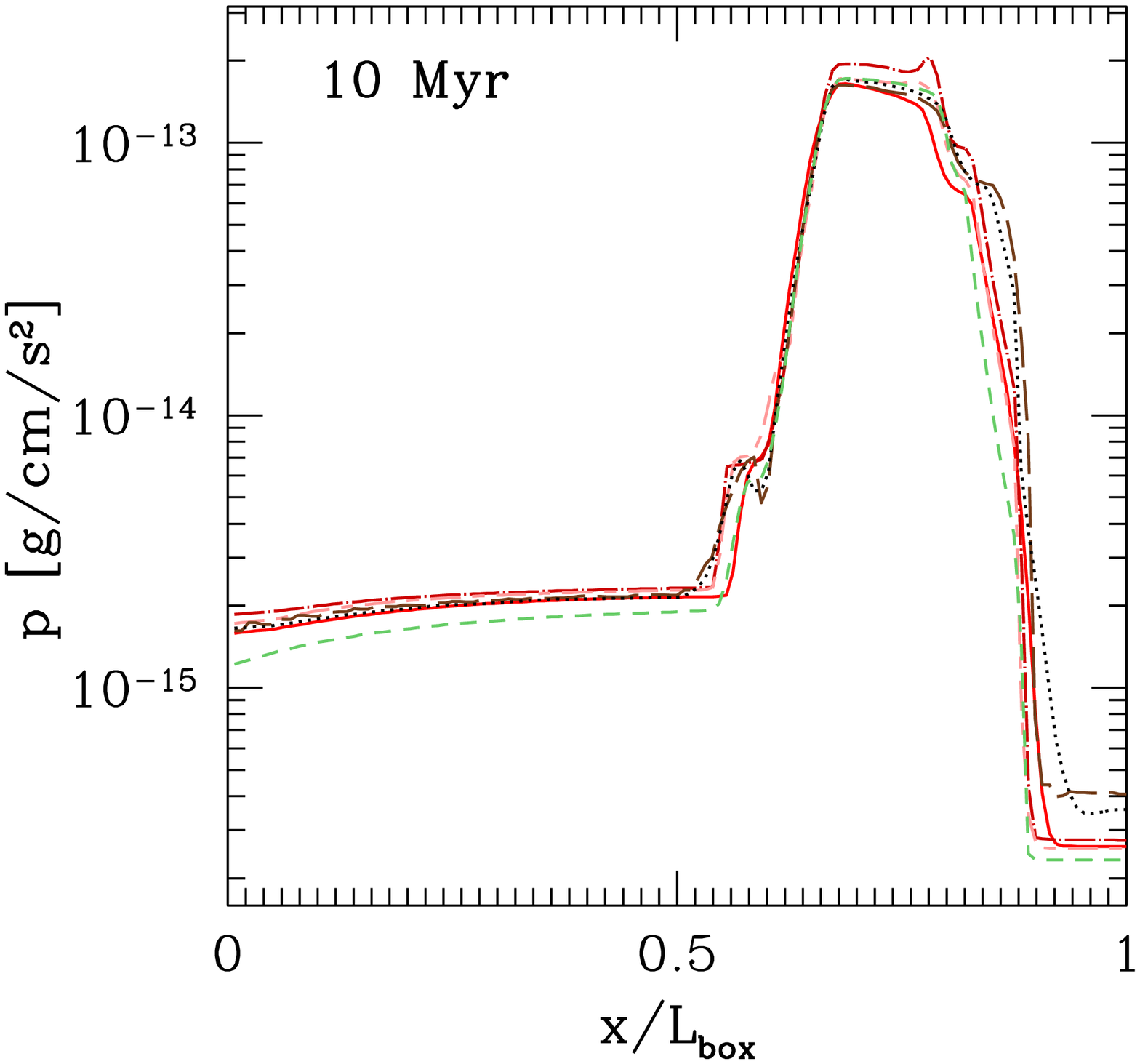}
  \includegraphics[width=2.2in]{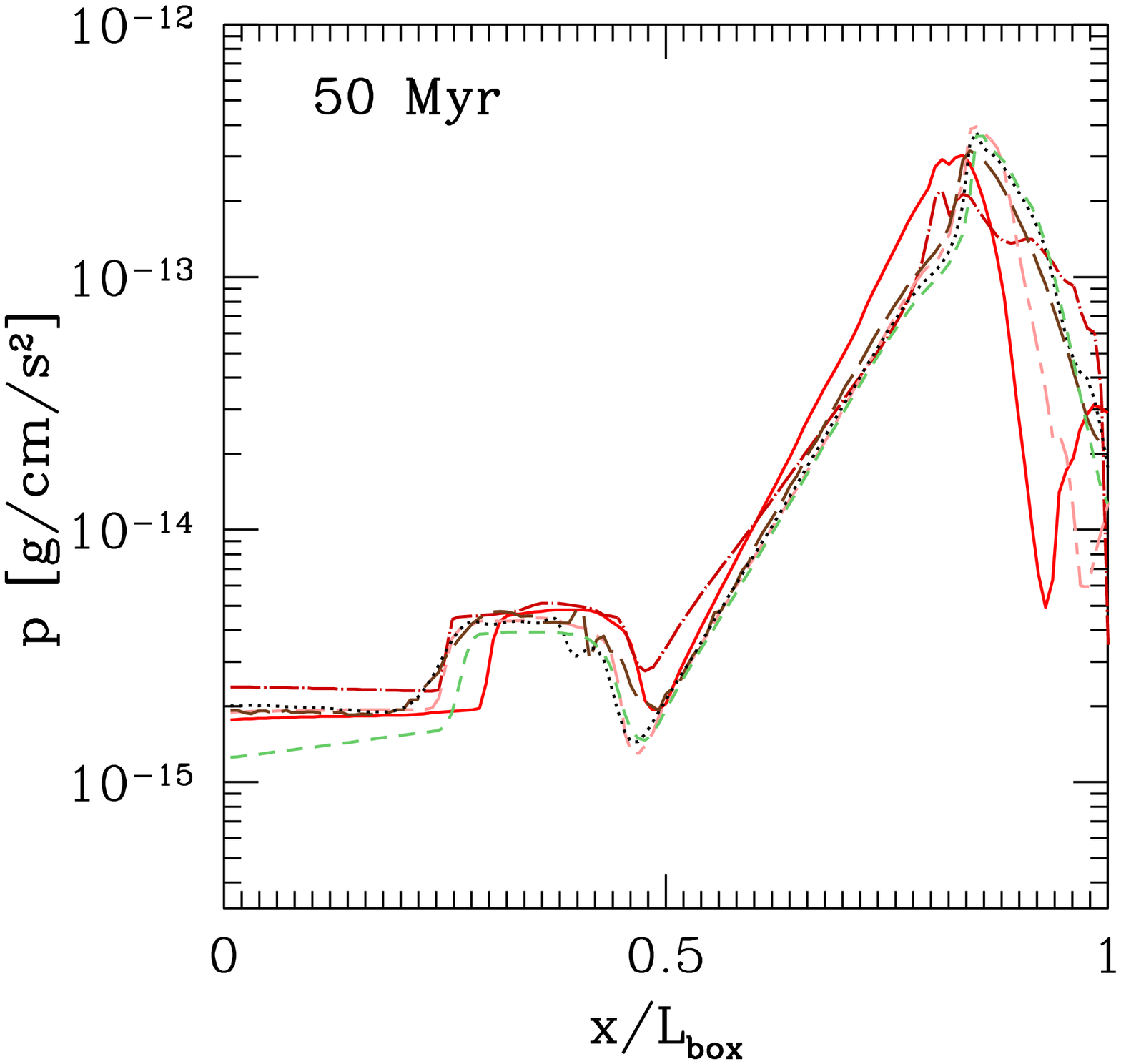}
  \includegraphics[width=2.2in]{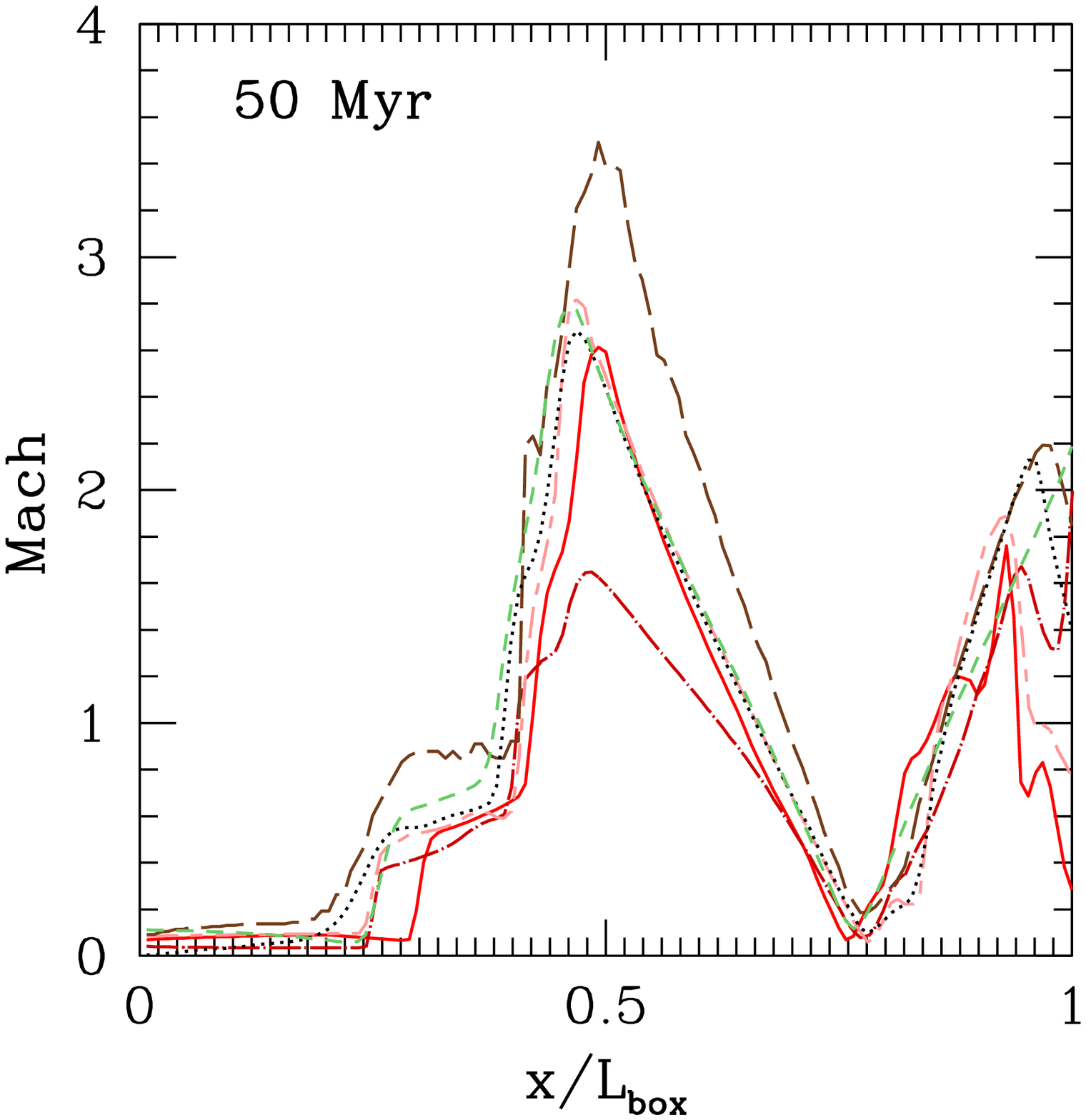}
\caption{Test 7 (Photoevaporation of a dense clump): Line cuts of the 
pressure at times $t=10$ Myr (left), and 50 Myr (centre) and of the Mach 
number at time $t=50$ Myr (right) along the axis of symmetry through the 
centre of the clump.
\label{T7_profs_p_m_fig}}
\end{center}
\end{figure*}

Finally, in Figures~\ref{T7_profsHI_fig}-\ref{T7_profs_p_m_fig} we present 
cuts along the x-axis of the neutral fraction, $x_{\rm HI}$, temperature, $T$,
pressure, $p$, and Mach number $M$ at selected times, as indicated. At early times 
($t=1-10$~Myr) all codes agree very well on both the ionization front position 
and its profile. The only modest differences are found in the semi-shielded
part of the dense gas ($x_{\rm HI}=0.01-1$), due to variations of the treatment
of hard photons, and in the low-density gas between the clump and the source,
where the neutral fractions are affected by the slightly different temperatures
found by the different methods. This is confirming the conclusions reached in
Paper I that with no (or little) gas motions any differences are due to the
treatment of the energy equation and the hard photons. The hydrodynamic 
evolution introduces some differences, particularly in the I-front position, 
but the scatter remains small.  

The temperature profiles generally agree in shape and in the position of the
flow features, the expanding wind and its leading shock. The main differences 
are in the amplitude, which varies by up to 50\%, except for the cold, shielded
gas at the back of the remaining dense clump at $t=10$~Myr (at position 
$x/L_{\rm box}\sim0.8$), where the variation between results reached an order of 
magnitude. This large variation does not affect the later-time evolution 
considerably, however. The pressure and Mach number profiles 
(Figure~\ref{T7_profs_p_m_fig}) show similar trends, with very small differences
during the early evolution, growing to somewhat larger ones at later times, but
with all prominent flow features agreeing in both nature and position.     

\subsection{Summary and Conclusions}

In this work we compared the results from 10 directly coupled hydrodynamics 
and radiative transfer codes on three test problems of astrophysical interest 
- H~II region expansion in initially uniform gas, as well as internal and 
external photoevaporation of dense clumps of galactic-like size and density. 
Our aims are to validate our codes and test their reliability. Our test 
problems, while chosen to be relatively simple and clean, nevertheless cover
a wide range of regimes applicable to photoionization-driven astrophysical 
flows, including propagation of fast (R-type) and slow (D-type) I-fronts, 
shock creation and supersonic photoevaporative winds. All the data is available 
on the Radiative Transfer Comparison Project wiki-based website, so future code 
developers can test their codes against our results.  

Overall, the agreement is quite good and all codes are generally reliable and
produce reasonable results. However the results also highlighted some important 
differences between the methods. All participating algorithms track fast, R-type 
I-fronts well, in agreement with the results we obtained in Paper I. We note 
that this is not the trivial statement that we simply reproduce our previous 
static density field results, since in this second Comparison Project phase 
there are several codes which are newly developed (RH1D, LICORICE, Enzo-RT) 
and therefore did not participate in Paper I, and even the ones 
which were present then have been further developed over the intervening 
period and are thus not identical to the versions used in Paper~I. 

Again, as we found in Paper I, the treatment of multi-frequency 
radiative transfer and particularly of the hard tail of the photon spectrum 
varies significantly among the methods and yields correspondingly large range
of temperature and ionization structure just beyond the I-front itself. We 
showed with a specific example that the 
spectral energy distribution of the ionizing source changes the I-front 
structure and shocked flow features considerably. Monochromatic light yields 
much sharper I-fronts and shocks and certain flow features like the 
double-peaked profile found in Test 5 disappear altogether.

For static density distributions the variations in the multi-frequency 
radiative transfer treatment had little effect on the I-front positions 
and propagation speeds since those are largely determined (apart from 
recombinations-related effects) by simple photon counting and balancing 
this number against the number of atoms entering the front. For 
dynamically-coupled evolution, this changes and there are significant 
feedback effects, with the radiative transfer effects affecting the gas 
dynamics and vice versa. For example, pre-heating by hard photons, or 
lack of it, can affect the dynamics significantly. More specifically, higher
pre-heating results in shocks, e.g. ones typically leading D-type I-fronts, 
which are weaker and faster-propagating, and vice-versa. The internal structure 
of such a front and the relative spacing between the shock and the I-front can 
also change considerably. Shocks created by photoheating effects tend to be 
relatively weak, with Mach numbers of a few or less. The density compression
resulting from them is strongly dependent on the pre-heating by hard photons, 
but generally did not exceed factors of 1.5-2. The profiles of the fluid 
quantities in supersonically expanding regions (e.g. the photoevaporative 
wind in Test 7) show good agreement among the different methods.

Significant differences were noted in the numerical diffusivity of the 
methods. Numerical diffusion could be due to either the radiative 
transfer method employed (e.g. the moment method OTVET used in HART), or the 
hydrodynamics (SPH in LICORICE). Higher diffusion could have notable effects
on some properties of the flow (features become smoother, high contrasts are 
diminished), but seems to have modest effects on the overall gross features 
and the basic dynamics remains largely unaffected. However, care should be 
taken when using such methods for problems in which the sharp features might 
matter, e.g. enhanced molecule formation due to shocks.

The propagation of an accelerating I-front down a steep ($1/r^2$) density 
profile proved to be a quite difficult problem and several codes
developed significant instabilities, while the rest did not. While there 
are a number of physical instabilities which can develop in similar situations,
as we discussed in some detail, in this particular case the instabilities we
observed proved to be numerical in nature. The most severe one was the 
carbuncle instability or odd-even decoupling, which in some cases affects 
low-diffusion hydrodynamic solvers (here a Roe Riemann solver). This problem
can be eliminated by either adding some artificial diffusion or using a more
diffusive hydrodynamic solver. 

In summary, we have found a considerable level of agreement between the wide 
variety of radiative transfer and hydrodynamics coupled methods participating 
in this project. The basic flow features and their evolution are reproduced 
well by all the methods. There are some variations whose origins we did our best 
to understand. The recurring differences were mostly due to the different 
treatment of the energy equation and the transfer of multi-frequency radiation. 
There were also some problems specific to certain methods which we discussed 
in detail. While none of the codes gave any obviously unphysical or incorrect 
results and all largely agreed with each other, some of the methods were clearly 
less suited for certain problems. No method is universally applicable to all
astrophysical situations and every one of the participating codes showed some
behaviour discrepant with the majority in one respect or another. Care should
therefore be taken in applying any given algorithm to a new type of problem and
detailed testing is always advised. 

\section*{Acknowledgments} 
This study was supported in part by Swiss National Science 
Foundation grant 200021-116696/1, the U.S. Department of Energy 
at Los Alamos National Laboratory under Contract No. 
DE-AC52-06NA25396, research funds from Chosun University, 
NSF grant AST 0708176, NASA grants NNX07AH09G and NNG04G177G, 
Chandra grant SAO TM8-9009X, and Swedish Research Council 
grant 60336701. The work with RSPH was supported in part by 
the {\it FIRST} project based on Grants-in-Aid for Specially 
Promoted Research by MEXT (16002003), JSPS Grant-in-Aid for 
Scientific Research (S) (20224002) and Inamori foundation. 
The Flash code was developed by the DOE-supported ASC / Alliance 
Center for Astrophysical Thermonuclear Flashes at the University 
of Chicago. MLN and DRR acknowledge partial support from NSF 
Grant AST-0808184.


\begin{thebibliography}{97}
\expandafter\ifx\csname natexlab\endcsname\relax\def\natexlab#1{#1}\fi

\bibitem[{{Abel} {et~al.}(1999{\natexlab{a}}){Abel}, {Norman}, \&
  {Madau}}]{1999ApJ...523...66A}
{Abel} T., {Norman} M.~L., {Madau} P., 1999{\natexlab{a}}, \apj, 523, 66

\bibitem[{{Abel} {et~al.}(1999{\natexlab{b}}){Abel}, {Norman}, \&
  {Madau}}]{AbelNormanMadau1999}
---, 1999{\natexlab{b}}, \apj, 523, 66

\bibitem[{{Ahn} \& {Shapiro}(2005)}]{2005MNRAS.363.1092A}
{Ahn} K., {Shapiro} P.~R., 2005, \mnras, 363, 1092

\bibitem[{{Ahn} \& {Shapiro}(2007)}]{2007MNRAS.375..881A}
---, 2007, \mnras, 375, 881

\bibitem[{{Anninos} {et~al.}(1997){Anninos}, {Zhang}, {Abel}, \&
  {Norman}}]{1997NewA....2..209A}
{Anninos} P., {Zhang} Y., {Abel} T., {Norman} M.~L., 1997, New Astronomy, 2,
  209

\bibitem[{{Baek} {et~al.}(2009){Baek}, {di Matteo}, {Semelin}, {Combes}, \&
  {Revaz}}]{2009A&A...495..389B}
{Baek} S., {di Matteo} P., {Semelin} B., {Combes} F., {Revaz} Y., 2009, \aap,
  495, 389

\bibitem[{{Bertoldi}(1989)}]{1989ApJ...346..735B}
{Bertoldi} F., 1989, \apj, 346, 735

\bibitem[{{Ciardi} {et~al.}(2003){Ciardi}, {Stoehr}, \&
  {White}}]{2003MNRAS.343.1101C}
{Ciardi} B., {Stoehr} F., {White} S.~D.~M., 2003, \mnras, 343, 1101

\bibitem[{{Colella} \& {Woodward}(1984)}]{1984JCoPh..54..174C}
{Colella} P., {Woodward} P.~R., 1984, Journal of Computational Physics, 54, 174

\bibitem[{{Dale} {et~al.}(2007{\natexlab{a}}){Dale}, {Bonnell}, \&
  {Whitworth}}]{2007MNRAS.375.1291D}
{Dale} J.~E., {Bonnell} I.~A., {Whitworth} A.~P., 2007{\natexlab{a}}, \mnras,
  375, 1291

\bibitem[{{Dale} {et~al.}(2007{\natexlab{b}}){Dale}, {Clark}, \&
  {Bonnell}}]{2007MNRAS.377..535D}
{Dale} J.~E., {Clark} P.~C., {Bonnell} I.~A., 2007{\natexlab{b}}, \mnras, 377,
  535

\bibitem[{{Dor{\'e}} {et~al.}(2007){Dor{\'e}}, {Holder}, {Alvarez}, {Iliev},
  {Mellema}, {Pen}, \& {Shapiro}}]{cmbpol}
{Dor{\'e}} O., {Holder} G., {Alvarez} M.~A., {Iliev} I.~T., {Mellema} G., {Pen}
  U.-L., {Shapiro} P.~R., 2007, \prd, 76, 043002

\bibitem[{{Draine} \& {Bertoldi}(1996)}]{1996ApJ...468..269D}
{Draine} B.~T., {Bertoldi} F., 1996, \apj, 468, 269

\bibitem[{{Franco} {et~al.}(1990){Franco}, {Tenorio-Tagle}, \&
  {Bodenheimer}}]{1990ApJ...349..126F}
{Franco} J., {Tenorio-Tagle} G., {Bodenheimer} P., 1990, \apj, 349, 126

\bibitem[{{Frank} \& {Mellema}(1994)}]{Frank1994}
{Frank} A., {Mellema} G., 1994, \aap, 289, 937

\bibitem[{{Fryxell} {et~al.}(2000){Fryxell}, {Olson}, {Ricker}, {Timmes},
  {Zingale}, {Lamb}, {MacNeice}, {Rosner}, {Truran}, \& {Tufo}}]{Fryxell2000}
{Fryxell} B., {Olson} K., {Ricker} P., {Timmes} F.~X., {Zingale} M., {Lamb}
  D.~Q., {MacNeice} P., {Rosner} R., {Truran} J.~W., {Tufo} H., 2000, \apjs,
  131, 273

\bibitem[{{Galli} \& {Palla}(1998{\natexlab{a}})}]{1998A&A...335..403G}
{Galli} D., {Palla} F., 1998{\natexlab{a}}, \aap, 335, 403

\bibitem[{{Galli} \& {Palla}(1998{\natexlab{b}})}]{gp98}
---, 1998{\natexlab{b}}, \aap, 335, 403

\bibitem[{{Garcia-Segura} \& {Franco}(1996)}]{gsf96}
{Garcia-Segura} G., {Franco} J., 1996, \apj, 469, 171

\bibitem[{{Giuliani}(1979)}]{g79}
{Giuliani} Jr. J.~L., 1979, \apj, 233, 280

\bibitem[{{Glover} \& {Abel}(2008)}]{ga08}
{Glover} S.~C.~O., {Abel} T., 2008, \mnras, 388, 1627

\bibitem[{{Gnedin}(2000)}]{2000ApJ...535..530G}
{Gnedin} N.~Y., 2000, \apj, 535, 530

\bibitem[{{Gnedin} {et~al.}(2008){Gnedin}, {Tassis}, \&
  {Kravtsov}}]{2008arXiv0810.4148G}
{Gnedin} N.~Y., {Tassis} K., {Kravtsov} A.~V., 2008, ArXiv e-prints

\bibitem[{{Gritschneder} {et~al.}(2009){Gritschneder}, {Naab}, {Burkert},
  {Walch}, {Heitsch}, \& {Wetzstein}}]{2009MNRAS.393...21G}
{Gritschneder} M., {Naab} T., {Burkert} A., {Walch} S., {Heitsch} F.,
  {Wetzstein} M., 2009, \mnras, 393, 21

\bibitem[{{Hasegawa} {et~al.}(2009){Hasegawa}, {Umemura}, \&
  {Susa}}]{2009MNRAS.tmp..445H}
{Hasegawa} K., {Umemura} M., {Susa} H., 2009, \mnras, 445

\bibitem[{{Heinemann} {et~al.}(2006){Heinemann}, {Dobler}, {Nordlund}, \&
  {Brandenburg}}]{2006A&A...448..731H}
{Heinemann} T., {Dobler} W., {Nordlund} {\AA}., {Brandenburg} A., 2006, \aap,
  448, 731

\bibitem[{Hindmarsh(1980)}]{Hindmarsh1980}
Hindmarsh A.~C., 1980, SIGNUM Newsl., 15, 10

\bibitem[{{Holder} {et~al.}(2007){Holder}, {Iliev}, \& {Mellema}}]{pol21}
{Holder} G.~P., {Iliev} I.~T., {Mellema} G., 2007, \apjl, 663, L1

\bibitem[{{Hosokawa} \& {Inutsuka}(2005)}]{2005ApJ...623..917H}
{Hosokawa} T., {Inutsuka} S., 2005, \apj, 623, 917

\bibitem[{{Hummer}(1994)}]{1994MNRAS.268..109H}
{Hummer} D.~G., 1994, \mnras, 268, 109

\bibitem[{{Hummer} \& {Storey}(1998)}]{1998MNRAS.297.1073H}
{Hummer} D.~G., {Storey} P.~J., 1998, \mnras, 297, 1073

\bibitem[{{Iliev} \& {et al.}(2006)}]{comparison1}
{Iliev} I.~T., {et al.}, 2006, \mnras, 371, 1057

\bibitem[{{Iliev} {et~al.}(2006{\natexlab{a}}){Iliev}, {Hirashita}, \&
  {Ferrara}}]{2006MNRAS.368.1885I}
{Iliev} I.~T., {Hirashita} H., {Ferrara} A., 2006{\natexlab{a}}, \mnras, 368,
  1885

\bibitem[{{Iliev} {et~al.}(2008{\natexlab{a}}){Iliev}, {Mellema}, {Pen},
  {Bond}, \& {Shapiro}}]{wmap3}
{Iliev} I.~T., {Mellema} G., {Pen} U.-L., {Bond} J.~R., {Shapiro} P.~R.,
  2008{\natexlab{a}}, \mnras, 77

\bibitem[{{Iliev} {et~al.}(2006{\natexlab{b}}){Iliev}, {Mellema}, {Pen},
  {Merz}, {Shapiro}, \& {Alvarez}}]{2006MNRAS.369.1625I}
{Iliev} I.~T., {Mellema} G., {Pen} U.-L., {Merz} H., {Shapiro} P.~R., {Alvarez}
  M.~A., 2006{\natexlab{b}}, \mnras, 369, 1625

\bibitem[{{Iliev} {et~al.}(2007{\natexlab{a}}){Iliev}, {Mellema}, {Shapiro}, \&
  {Pen}}]{selfregulated}
{Iliev} I.~T., {Mellema} G., {Shapiro} P.~R., {Pen} U.-L., 2007{\natexlab{a}},
  \mnras, 376, 534

\bibitem[{{Iliev} {et~al.}(2007{\natexlab{b}}){Iliev}, {Pen}, {Bond},
  {Mellema}, \& {Shapiro}}]{kSZ}
{Iliev} I.~T., {Pen} U.-L., {Bond} J.~R., {Mellema} G., {Shapiro} P.~R.,
  2007{\natexlab{b}}, \apj, 660, 933

\bibitem[{{Iliev} {et~al.}(2008{\natexlab{b}}){Iliev}, {Shapiro}, {Mellema},
  {Merz}, \& {Pen}}]{2008arXiv0806.2887I}
{Iliev} I.~T., {Shapiro} P.~R., {Mellema} G., {Merz} H., {Pen} U.-L.,
  2008{\natexlab{b}}, in refereed proceedings of TeraGrid08, ArXiv e-prints
  (0806.2887)

\bibitem[{{Iliev} {et~al.}(2005){Iliev}, {Shapiro}, \&
  {Raga}}]{2005MNRAS...361..405I}
{Iliev} I.~T., {Shapiro} P.~R., {Raga} A.~C., 2005, \mnras, 361, 405

\bibitem[{{Kahn} \& {Dyson}(1965)}]{1965ARA&A...3...47K}
{Kahn} F.~D., {Dyson} J.~E., 1965, \araa, 3, 47

\bibitem[{Khokhlov(1998)}]{khokhlov98}
Khokhlov A.~M., 1998, JCPh, 143, 519

\bibitem[{{Knoll} \& {Keyes}(2004)}]{KnollKeyes2004}
{Knoll} D.~A., {Keyes} D.~E., 2004, Journal of Computational Physics, 193, 357

\bibitem[{{Kohler} {et~al.}(2007){Kohler}, {Gnedin}, \&
  {Hamilton}}]{2007ApJ...657...15K}
{Kohler} K., {Gnedin} N.~Y., {Hamilton} A.~J.~S., 2007, \apj, 657, 15

\bibitem[{{Kravtsov}(1999)}]{kravtsov99}
{Kravtsov} A.~V., 1999, PhD thesis, New Mexico State University

\bibitem[{{Kravtsov} {et~al.}(2002){Kravtsov}, {Klypin}, \&
  {Hoffman}}]{kravtsov_etal02}
{Kravtsov} A.~V., {Klypin} A., {Hoffman} Y., 2002, \apj, 571, 563

\bibitem[{Kravtsov {et~al.}(1997)Kravtsov, Klypin, \&
  Khokhlov}]{kravtsov_etal97}
Kravtsov A.~V., Klypin A.~A., Khokhlov A.~M., 1997, \apjs, 111, 73

\bibitem[{{Krumholz} {et~al.}(2007){Krumholz}, {Stone}, \&
  {Gardiner}}]{2007ApJ...671..518K}
{Krumholz} M.~R., {Stone} J.~M., {Gardiner} T.~A., 2007, \apj, 671, 518

\bibitem[{{Lepp} \& {Shull}(1983)}]{ls83}
{Lepp} S., {Shull} J.~M., 1983, \apj, 270, 578

\bibitem[{{Lim} \& {Mellema}(2003)}]{2003A&A...405..189L}
{Lim} A.~J., {Mellema} G., 2003, \aap, 405, 189

\bibitem[{{Mac Low} {et~al.}(2007){Mac Low}, {Toraskar}, {Oishi}, \&
  {Abel}}]{2007ApJ...668..980M}
{Mac Low} M.-M., {Toraskar} J., {Oishi} J.~S., {Abel} T., 2007, \apj, 668, 980

\bibitem[{{Maselli} {et~al.}(2009){Maselli}, {Ciardi}, \&
  {Kanekar}}]{2009MNRAS.393..171M}
{Maselli} A., {Ciardi} B., {Kanekar} A., 2009, \mnras, 393, 171

\bibitem[{{Maselli} {et~al.}(2003){Maselli}, {Ferrara}, \&
  {Ciardi}}]{2003MNRAS.345..379M}
{Maselli} A., {Ferrara} A., {Ciardi} B., 2003, \mnras, 345, 379

\bibitem[{{Mellema} {et~al.}(2006{\natexlab{a}}){Mellema}, {Arthur}, {Henney},
  {Iliev}, \& {Shapiro}}]{2006ApJ...647..397M}
{Mellema} G., {Arthur} S.~J., {Henney} W.~J., {Iliev} I.~T., {Shapiro} P.~R.,
  2006{\natexlab{a}}, \apj, 647, 397

\bibitem[{{Mellema} {et~al.}(2006{\natexlab{b}}){Mellema}, {Iliev}, {Alvarez},
  \& {Shapiro}}]{methodpaper}
{Mellema} G., {Iliev} I.~T., {Alvarez} M.~A., {Shapiro} P.~R.,
  2006{\natexlab{b}}, New Astronomy, 11, 374

\bibitem[{{Mellema} {et~al.}(2006{\natexlab{c}}){Mellema}, {Iliev}, {Pen}, \&
  {Shapiro}}]{21cmreionpaper}
{Mellema} G., {Iliev} I.~T., {Pen} U.-L., {Shapiro} P.~R., 2006{\natexlab{c}},
  \mnras, 372, 679

\bibitem[{{Mellema} {et~al.}(1998){Mellema}, {Raga}, {Canto}, {Lundqvist},
  {Balick}, {Steffen}, \& {Noriega-Crespo}}]{1998A&A...331..335M}
{Mellema} G., {Raga} A.~C., {Canto} J., {Lundqvist} P., {Balick} B., {Steffen}
  W., {Noriega-Crespo} A., 1998, \aap, 331, 335

\bibitem[{{Monaghan}(1992)}]{1992ARA&A..30..543M}
{Monaghan} J.~J., 1992, \araa, 30, 543

\bibitem[{{Nakamoto} {et~al.}(2001){Nakamoto}, {Umemura}, \&
  {Susa}}]{2001MNRAS.321..593N}
{Nakamoto} T., {Umemura} M., {Susa} H., 2001, \mnras, 321, 593

\bibitem[{{Norman} {et~al.}(2007){Norman}, {Bryan}, {Harkness}, {Bordner},
  {Reynolds}, {O'Shea}, \& {Wagner}}]{2007arXiv0705.1556N}
{Norman} M.~L., {Bryan} G.~L., {Harkness} R., {Bordner} J., {Reynolds} D.,
  {O'Shea} B., {Wagner} R., 2007, ArXiv e-prints, (0705.1556)

\bibitem[{{Quirk}(1994)}]{quirk94}
{Quirk} J.~J., 1994, International Journal for Numerical Methods in Fluids, 18,
  555

\bibitem[{{Raga} {et~al.}(1999){Raga}, {Mellema}, {Arthur}, {Binette},
  {Ferruit}, \& {Steffen}}]{1999RMxAA..35..123R}
{Raga} A.~C., {Mellema} G., {Arthur} S.~J., {Binette} L., {Ferruit} P.,
  {Steffen} W., 1999, Revista Mexicana de Astronomia y Astrofisica, 35, 123

\bibitem[{{Raga} {et~al.}(1997){Raga}, {Mellema}, \&
  {Lundqvist}}]{1997ApJS..109..517R}
{Raga} A.~C., {Mellema} G., {Lundqvist} P., 1997, \apjs, 109, 517

\bibitem[{{Razoumov} {et~al.}(2002){Razoumov}, {Norman}, {Abel}, \&
  {Scott}}]{2002ApJ...572..695R}
{Razoumov} A.~O., {Norman} M.~L., {Abel} T., {Scott} D., 2002, \apj, 572, 695

\bibitem[{{Razoumov} {et~al.}(2006){Razoumov}, {Norman}, {Prochaska}, \&
  {Wolfe}}]{2006ApJ...645...55R}
{Razoumov} A.~O., {Norman} M.~L., {Prochaska} J.~X., {Wolfe} A.~M., 2006, \apj,
  645, 55

\bibitem[{{Razoumov} \& {Scott}(1999)}]{1999MNRAS.309..287R}
{Razoumov} A.~O., {Scott} D., 1999, \mnras, 309, 287

\bibitem[{{Reynolds} {et~al.}(2009){Reynolds}, {Hayes}, {Paschos}, \&
  {Norman}}]{2009arXiv0901.1110R}
{Reynolds} D.~R., {Hayes} J.~C., {Paschos} P., {Norman} M.~L., 2009, ArXiv
  e-prints (0901.1110)

\bibitem[{{Ricotti} {et~al.}(2002){Ricotti}, {Gnedin}, \&
  {Shull}}]{2002ApJ...575...49R}
{Ricotti} M., {Gnedin} N.~Y., {Shull} J.~M., 2002, \apj, 575, 49

\bibitem[{{Rijkhorst}(2005)}]{Rijkhorst2005}
{Rijkhorst} E.-J., 2005, PhD thesis, Leiden Observatory, Leiden University,
  P.O.~Box 9513, 2300 RA Leiden, The Netherlands

\bibitem[{{Rijkhorst} {et~al.}(2006){Rijkhorst}, {Plewa}, {Dubey}, \&
  {Mellema}}]{Rijkhorst2006}
{Rijkhorst} E.-J., {Plewa} T., {Dubey} A., {Mellema} G., 2006, \aap, 452, 907

\bibitem[{{Semelin} \& {Combes}(2002)}]{2002A&A...388..826S}
{Semelin} B., {Combes} F., 2002, \aap, 388, 826

\bibitem[{{Semelin} \& {Combes}(2005)}]{2005A&A...441...55S}
---, 2005, \aap, 441, 55

\bibitem[{{Shapiro} {et~al.}(2006){Shapiro}, {Iliev}, {Alvarez}, \&
  {Scannapieco}}]{2006ApJ...648..922S}
{Shapiro} P.~R., {Iliev} I.~T., {Alvarez} M.~A., {Scannapieco} E., 2006, \apj,
  648, 922

\bibitem[{{Shapiro} {et~al.}(2004){Shapiro}, {Iliev}, \&
  {Raga}}]{2004MNRAS.348..753S}
{Shapiro} P.~R., {Iliev} I.~T., {Raga} A.~C., 2004, \mnras, 348, 753

\bibitem[{{Sokasian} {et~al.}(2003){Sokasian}, {Abel}, {Hernquist}, \&
  {Springel}}]{2003MNRAS.344..607S}
{Sokasian} A., {Abel} T., {Hernquist} L., {Springel} V., 2003, \mnras, 344, 607

\bibitem[{{Spitzer}(1978)}]{1978ppim.book.....S}
{Spitzer} L., 1978, {Physical processes in the interstellar medium}. New York
  Wiley-Interscience, 1978

\bibitem[{{Springel} \& {Hernquist}(2002)}]{2002MNRAS.333..649S}
{Springel} V., {Hernquist} L., 2002, \mnras, 333, 649

\bibitem[{{Steinmetz} \& {Mueller}(1993)}]{1993A&A...268..391S}
{Steinmetz} M., {Mueller} E., 1993, \aap, 268, 391

\bibitem[{{Susa}(2006)}]{2006PASJ...58..445S}
{Susa} H., 2006, \pasj, 58, 445

\bibitem[{{Susa}(2007)}]{2007ApJ...659..908S}
---, 2007, \apj, 659, 908

\bibitem[{{Susa}(2008)}]{2008ApJ...684..226S}
---, 2008, \apj, 684, 226

\bibitem[{{Susa} \& {Kitayama}(2000)}]{2000MNRAS.317..175S}
{Susa} H., {Kitayama} T., 2000, \mnras, 317, 175

\bibitem[{{Susa} \& {Umemura}(2004)}]{2004ApJ...600....1S}
{Susa} H., {Umemura} M., 2004, \apj, 600, 1

\bibitem[{{Susa} \& {Umemura}(2006)}]{2006ApJ...645L..93S}
---, 2006, \apjl, 645, L93

\bibitem[{{Tenorio-Tagle} {et~al.}(1986){Tenorio-Tagle}, {Bodenheimer}, {Lin},
  \& {Noriega-Crespo}}]{1986MNRAS.221..635T}
{Tenorio-Tagle} G., {Bodenheimer} P., {Lin} D.~N.~C., {Noriega-Crespo} A.,
  1986, \mnras, 221, 635

\bibitem[{{Thacker} {et~al.}(2000){Thacker}, {Tittley}, {Pearce}, {Couchman},
  \& {Thomas}}]{2000MNRAS.319..619T}
{Thacker} R.~J., {Tittley} E.~R., {Pearce} F.~R., {Couchman} H.~M.~P., {Thomas}
  P.~A., 2000, \mnras, 319, 619

\bibitem[{{Trac} \& {Pen}(2004)}]{2004NewA....9..443T}
{Trac} H., {Pen} U.-L., 2004, New Astronomy, 9, 443

\bibitem[{{Umemura}(1993)}]{1993ApJ...406..361U}
{Umemura} M., 1993, \apj, 406, 361

\bibitem[{{Vishniac}(1983)}]{v83}
{Vishniac} E.~T., 1983, \apj, 274, 152

\bibitem[{{Whalen} {et~al.}(2004){Whalen}, {Abel}, \&
  {Norman}}]{2004ApJ...610...14W}
{Whalen} D., {Abel} T., {Norman} M.~L., 2004, \apj, 610, 14

\bibitem[{{Whalen} \& {Norman}(2006)}]{2006ApJS..162..281W}
{Whalen} D., {Norman} M.~L., 2006, \apjs, 162, 281

\bibitem[{{Whalen} \& {Norman}(2008{\natexlab{a}})}]{wn08b}
---, 2008{\natexlab{a}}, \apj, 673, 664

\bibitem[{{Whalen} \& {Norman}(2008{\natexlab{b}})}]{wn08a}
---, 2008{\natexlab{b}}, \apj, 672, 287

\bibitem[{{Whalen} {et~al.}(2008{\natexlab{a}}){Whalen}, {O'Shea}, {Smidt}, \&
  {Norman}}]{wet08b}
{Whalen} D., {O'Shea} B.~W., {Smidt} J., {Norman} M.~L., 2008{\natexlab{a}},
  \apj, 679, 925

\bibitem[{{Whalen} {et~al.}(2008{\natexlab{b}}){Whalen}, {Prochaska}, {Heger},
  \& {Tumlinson}}]{2008ApJ...682.1114W}
{Whalen} D., {Prochaska} J.~X., {Heger} A., {Tumlinson} J., 2008{\natexlab{b}},
  \apj, 682, 1114

\bibitem[{{Whalen} {et~al.}(2008{\natexlab{c}}){Whalen}, {van Veelen},
  {O'Shea}, \& {Norman}}]{wet08c}
{Whalen} D., {van Veelen} B., {O'Shea} B.~W., {Norman} M.~L.,
  2008{\natexlab{c}}, \apj, 682, 49

\bibitem[{{Williams}(1999)}]{rjr99}
{Williams} R.~J.~R., 1999, \mnras, 310, 789

\bibitem[{{Williams}(2002)}]{rjr02}
---, 2002, \mnras, 331, 693

\end{thebibliography}
\end{document}